\documentclass[rmp,aps,twocolumn,showpacs]{revtex4} 
\usepackage{times}
\usepackage{graphicx}

\begin{document}

\title{Ice structures, patterns, and processes: A view across the ice-fields}

\author{Thorsten Bartels-Rausch}
\affiliation{Lab. Radiochemistry and Environmental Chemistry, Paul Scherrer Institut, CH-5232 Villigen PSI, Switzerland}
\author{Vance Bergeron}
\affiliation{Ecole Normale Sup\'erieure de Lyon, F-69007 Lyon, France}
\author{Julyan H. E. Cartwright}
\affiliation{Instituto Andaluz de Ciencias de la Tierra, CSIC--Universidad de Granada, E-18071 Granada, Spain}
\author{Rafael Escribano}
\affiliation{Instituto de Estructura de la Materia, CSIC, E-28006 Madrid, Spain}
\author{John L. Finney}
\affiliation{Department of Physics \& Astronomy and London Centre for Nanotechnology, University College London, London WC1E 6BT, UK}
\author{Hinrich Grothe}
\affiliation{Institute of Materials Chemistry, Vienna University of Technology, A-1060 Vienna, Austria}
\author{Pedro J. Guti\'errez}
\affiliation{Instituto de Astrof{\'{\i}}sica de Andaluc{\'{\i}}a, CSIC, E-18080 Granada, Spain}
\author{Jari Haapala}
\affiliation{Finnish Meteorological Institute, FIN-00100 Helsinki, Finland}
\author{Werner F. Kuhs}
\affiliation{GZG Crystallography, University of G\"ottingen, D-37077 G\"ottingen, Germany}
\author{Jan B. C. Pettersson}
\affiliation{Department of Chemistry,  {}Atmospheric Science, University of Gothenburg, SE-41296 G\"oteborg, Sweden}
\author{Stephen D. Price}
\affiliation{Department of Chemistry, University College London,  London WC1H 0AJ, UK}
\author{C. Ignacio Sainz-D{\'{\i}}az}
\affiliation{Instituto Andaluz de Ciencias de la Tierra, CSIC--Universidad de Granada, E-18002 Granada, Spain}
\author{Debbie Stokes}
\affiliation{FEI Company, Achtseweg Noord 5, 5651 GG, Eindhoven, The Netherlands}
\author{Giovanni Strazzulla}
\affiliation{INAF --- Osservatorio Astrofisico di Catania, I-95123 Catania, Italy}
\author{Erik S. Thomson }
\affiliation{Department of Chemistry, Atmospheric Science, University of Gothenburg, SE-41296 G\"oteborg, Sweden}
\author{Hauke Trinks}
\affiliation{Technical University Hamburg Harburg, D- 21079 Hamburg, Germany}
\author{Nevin Uras-Aytemiz}
\affiliation{Department of Chemistry, Suleyman Demirel University, TR-32260 Isparta, Turkey}

\date{\today: Version 8.05}


\begin{abstract}
We look ahead from the frontiers of research on ice dynamics in its broadest sense; on the structures of ice, the patterns or morphologies it may assume, and the physical and chemical processes in which it is involved. We highlight open questions in the various fields of ice research in nature; ranging from terrestrial and oceanic ice on Earth, to ice in the atmosphere, to ice on other solar system bodies and in interstellar space.
\end{abstract}

\addtolength{\textheight}{10pt}
\maketitle
\addtolength{\textheight}{-10pt}

\tableofcontents 

\vspace{1.7cm}

\section{Introduction}\label{introduction}

\begin{quote}
The ice was here, the ice was there, \\
The ice was all around

Samuel Taylor Coleridge, \\ \emph{The Rime of the Ancient Mariner}
\end{quote}

Ice is indeed all around us. As the cryosphere, ice or snow covers a small but significant part of the Earth's surface, both land and sea, and it plays a similarly important role in our atmosphere. Moreover ice is present on many other celestial bodies in our solar system and --- surely --- beyond, and it coats grains of dust in interstellar space. Ice is not a static medium but a dynamical one; it shows strong variations of its characteristics with time and place, as we may readily experience at a human scale on any ski slope.  A better understanding of ice structures, patterns, and processes is thus a topic of current research in physics. We shall show in the following how progress in understanding these questions is elemental in understanding current questions in astrophysics, atmospheric, cryospheric, and environmental science. 

Ice research questions are not only tackled separately within distinct fields  --- in terrestrial, oceanic, atmospheric, planetary, and interstellar ice research --- but also by researchers with disparate backgrounds: by modelers, field and laboratory experimentalists and theoreticians from both physics and chemistry.  We work in these different fields and come from a variety of backgrounds. We came together, some of us initially for a Spanish national project supported by the Spanish CSIC, and then the majority of us for a workshop, Euroice 2008, sponsored by the European Science Foundation,  
which was organized to connect people working on structures and those working in applied ice fields; to find common ground in the physical and chemical processes at icy surfaces and the physics and chemistry of ice structures from the molecular scale to the macroscale, and to explore whether some of the questions we were asking and some of the answers we were seeking are the same.

During the workshop we found that, despite the diversity of ice research, a number of key themes are indeed common between the different fields. 
The common ground in any field of ice research is the urge to understand better its structure and dynamics. For example: What are the ordering mechanisms of ice as it changes from one  of its phases into another? What is the structure at its surface, and how does this differ from the bulk? What is the structure and microenvironment at the contact area of ice crystals? How does ice structure form initially? Are there meta-stable phases present in the environment?  
This work focuses on this common ground. 
It thus does not aim to be a comprehensive review; such a review of ice physics and chemistry would be a book; indeed there are excellent books available \cite{hobbs1974,petrenko1999}. However, many of the issues raised in this article are issues of the 21st century that are not addressed in the textbooks. This article provides a view of the way ahead from some frontiers of research on ice. We set out the main physical and chemical open questions on ice structures, patterns and processes from the fields of ice research in nature: from ice on Earth, in the oceans and the atmosphere, to planetary and interstellar ice.

We begin in Section~\ref{structures} by introducing open questions in the molecular structures of ices; we then examine 
open issues on dynamical patterns and processes in ice. Following this we look first, in Section~\ref{astrophysical}, at astrophysical ice. We then focus on ice on Earth, beginning with Section~\ref{atmospheric} on atmospheric ice, whose precipitation leads to the subsequent formation of terrestrial ice, Section~\ref{terrestrial}, and, Section~\ref{oceanic}, sea ice. We conclude in Section~\ref{perspectives}, in which we attempt to list the important open questions on ice from the perspectives of these different fields of ice research.

\section{Ice Structures}\label{structures}

\begin{quote}
``Now suppose,'' chortled Dr.~Breed, enjoying himself, ``that there were many
possible ways in which water could crystallize, could freeze. Suppose that the
sort of ice we skate upon and put into highballs---what we might call ice-one---is 
only one of several types of ice. Suppose water always froze as ice-one on
Earth because it had never had a seed to teach it how to form  ice-two,  
ice-three,  ice-four ...? And suppose,'' he rapped on his desk with his old hand
again, ``that there were one form, which we will call ice-nine---a crystal as
hard as this desk---with a melting point of, let us say, one-hundred degrees
Fahrenheit, or, better still, a melting point of one-hundred-and-thirty
degrees.''

Kurt Vonnegut, \emph{Cat's Cradle}
\end{quote}

\subsection{Order and disorder in crystalline ice structures}\label{crystalline}

\begin{figure*}[t]
\centering\includegraphics*[width=0.76\textwidth,clip=true]{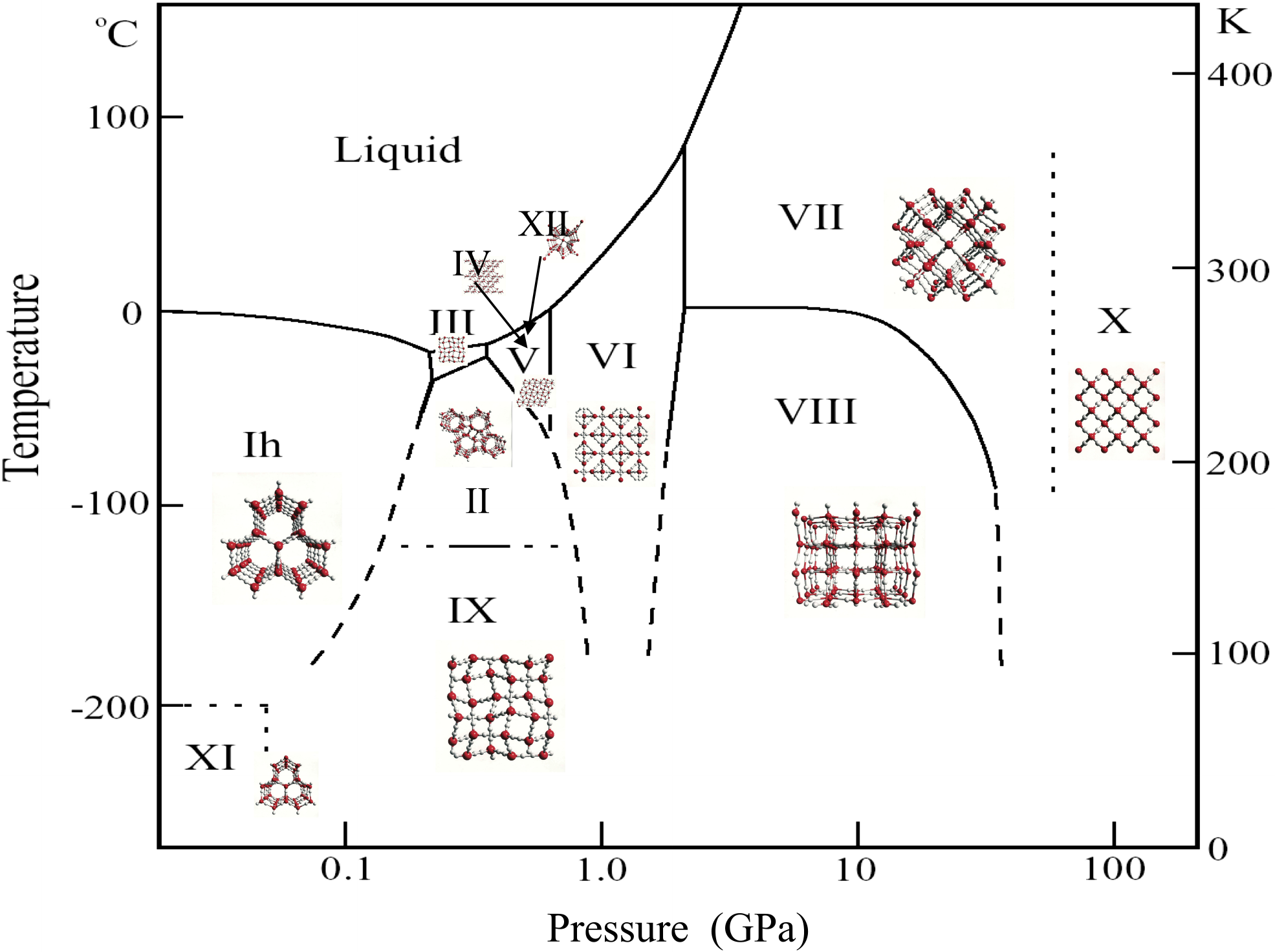} 
\caption{(Color online) The solid--liquid phase diagram of ice (the solid--liquid--gas triple point and liquid--gas coexistence line lie off the diagram to the left).
\label{john1}}
\end{figure*}

Unlike Vonnegut's fictional ice-nine, the real ice IX is not stable at ambient pressures and temperatures. But there are indeed many phases of ice. Although we normally experience only one of these --- the familiar ice Ih that forms in your freezer, and makes up snowflakes and icebergs --- changing pressure and temperature can cause changes of phase into other forms, as indicated in the phase diagram of Figure~\ref{john1}. Most of these phases are stable within a given range of temperature and pressure, but some are only metastable; for example, ice IV and XII, also indicated on Fig.~\ref{john1}, are found within the regions of stability of other phases. 

\subsubsection{Molecular structures of ices}

\begin{figure}[b]
\centering\includegraphics*[width=0.5\columnwidth,clip=true]{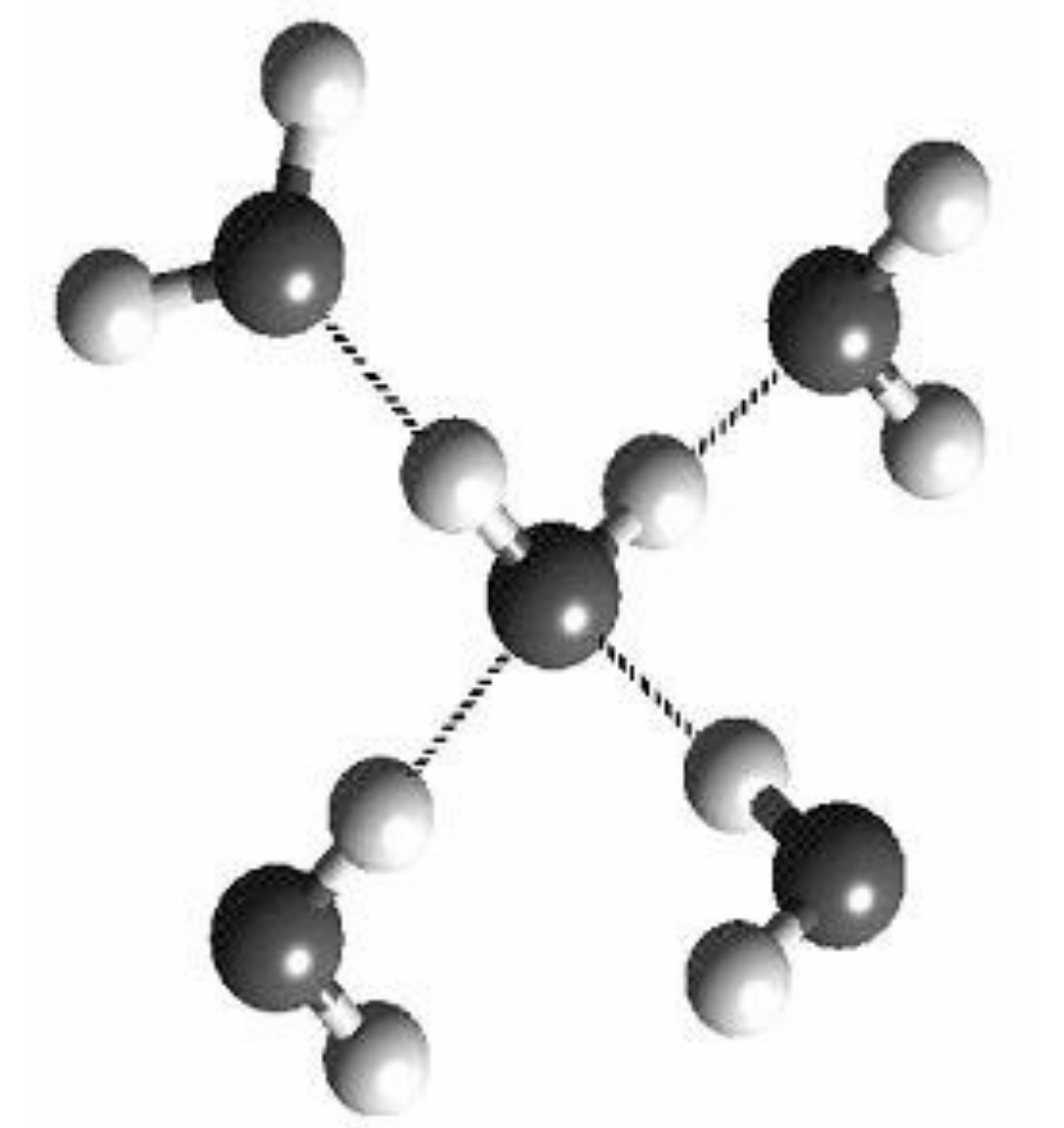} 
\caption{The local tetrahedral coordination of water molecules in ices. Each molecule accepts two hydrogen bonds from, and donates two to, its neighbors.
\label{john2}
}
\end{figure}

We now know the molecular structures of all of these phases, thanks largely to neutron diffraction crystallographic studies: these are indicated by the insets in Fig.~\ref{john1} --- further details and references can be found in \textcite{petrenko1999,finney2001,finney2004}. These structures can be simply rationalized in terms of fully connected tetrahedral networks of water molecules, with each molecule donating hydrogen bonds to two neighbors and accepting two hydrogen bonds from two others (Fig.~\ref{john2}). In the low-pressure ice Ih, the O--O--O angles are close to the ideal tetrahedral angle of  some 109$^o$. As pressure is increased, the molecules have to rearrange themselves to occupy less volume and this is done initially by both changes to the network structure (but still retaining four-coordination) and increased distortion of the O--O--O angles: for example, in ice II these angles vary between 80$^o$ and 129$^o$. As we increase pressure further, we come to a point at which the reduced volume available cannot be filled by merely increasing hydrogen bond distortion: the water molecules then form interpenetrating networks, as in ice VII which consists essentially of two interpenetrating diamond-type lattices. But each network remains four-coordinated.

\subsubsection{Hydrogen disorder}

\begin{figure}[b]
\centering\includegraphics*[width=\columnwidth,clip=true]{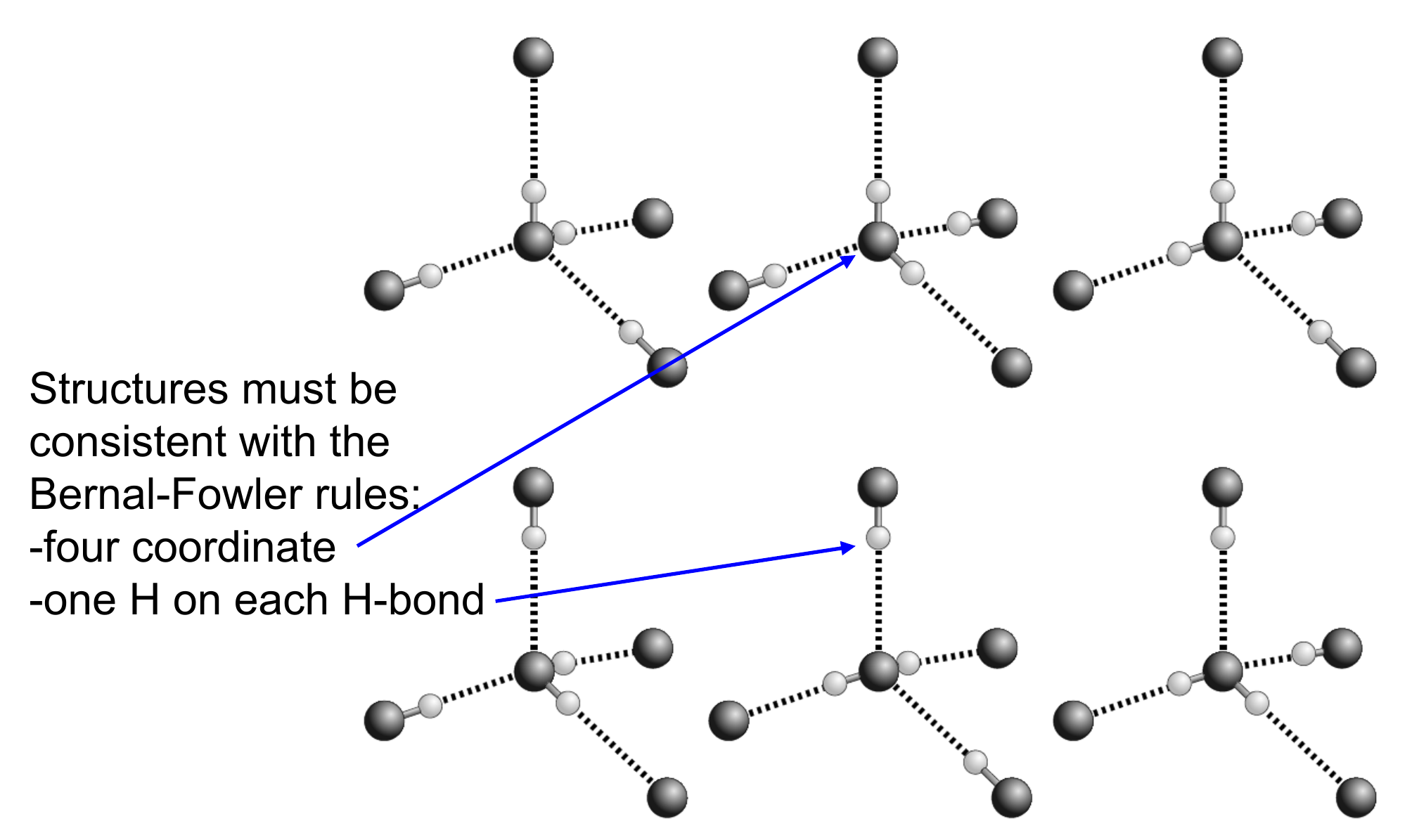} 
\caption{Rotational--hydrogen bond disorder in ices: the six possible orientations of a central water molecule and the hydrogen bonding consequences for its four neighbors.
\label{john3}}
\end{figure}

However, what we have said above refers only to the networks formed by the oxygen atoms of the water molecules. When we consider the hydrogens, things become a little more complicated. In fact, all the liquidus phases of ice --- those which share a phase boundary with the liquid (ices Ih, III, V, VI, and VII; see Fig.~\ref{john1}) --- are orientationally disordered. Thus, although the oxygen atoms are related to an underlying lattice, the hydrogens are not. To understand this, we can refer again to Fig.~\ref{john2} and note that the central water molecule is in one of six possible orientations that allow it to make four hydrogen bonds with its neighbors: these six possible orientations are shown schematically in Fig.~\ref{john3}. The only strong chemical constraints that a particular arrangement needs to fulfill are the Bernal--Fowler rules: each molecule must hydrogen bond to four neighbors, and there can be only one hydrogen participating in each hydrogen bond (again, see Fig.~\ref{john3}). Thus although the oxygen atoms in these ices are topologically ordered, the hydrogen atoms are not; they can be considered as decorating the underlying oxygen framework with one hydrogen along each hydrogen bond. This hydrogen disorder has another consequence that is sometimes neglected: the oxygen atoms too can show a positional disorder of several hundredths of an {\AA}  around their assigned high-symmetry crystallographic site as a response to the local hydrogen configuration \cite{kuhs1984,kuhs1986}. This leads in turn to local water-molecule geometries different from those obtained in a routine crystallographic structure analysis. Earlier contradictions between diffraction and spectroscopic results were removed by introducing this additional disorder. It has meanwhile been confirmed and quantified by quantum-chemical calculations considering the very many allowed hydrogen bond arrangements in a disordered ice phase using graph invariants \cite{kuo2001,kuo2003}; this method has been applied to ice VII--VIII \cite{kuo2004}, ice Ih \cite{kuo2005}, ice III--IX \cite{knight2006} and ice VI \cite{kuo2006}. 

As temperature is reduced, we would expect these disordered phases to order, becoming fully hydrogen ordered --- and hence of lower entropy --- at low temperature. This is indeed the case for ice III and ice VII, which order to ices IX and VIII respectively (Fig.~\ref{john1}). However, for all the other disordered phases, lowering the temperature does not lead to full hydrogen ordering: the molecular reorientations necessary for ordering to occur become more and more sluggish as temperature is reduced, freezing in the hydrogen --- or orientational --- disorder.

\begin{figure}[t]
\centering\includegraphics*[width=\columnwidth,clip=true]{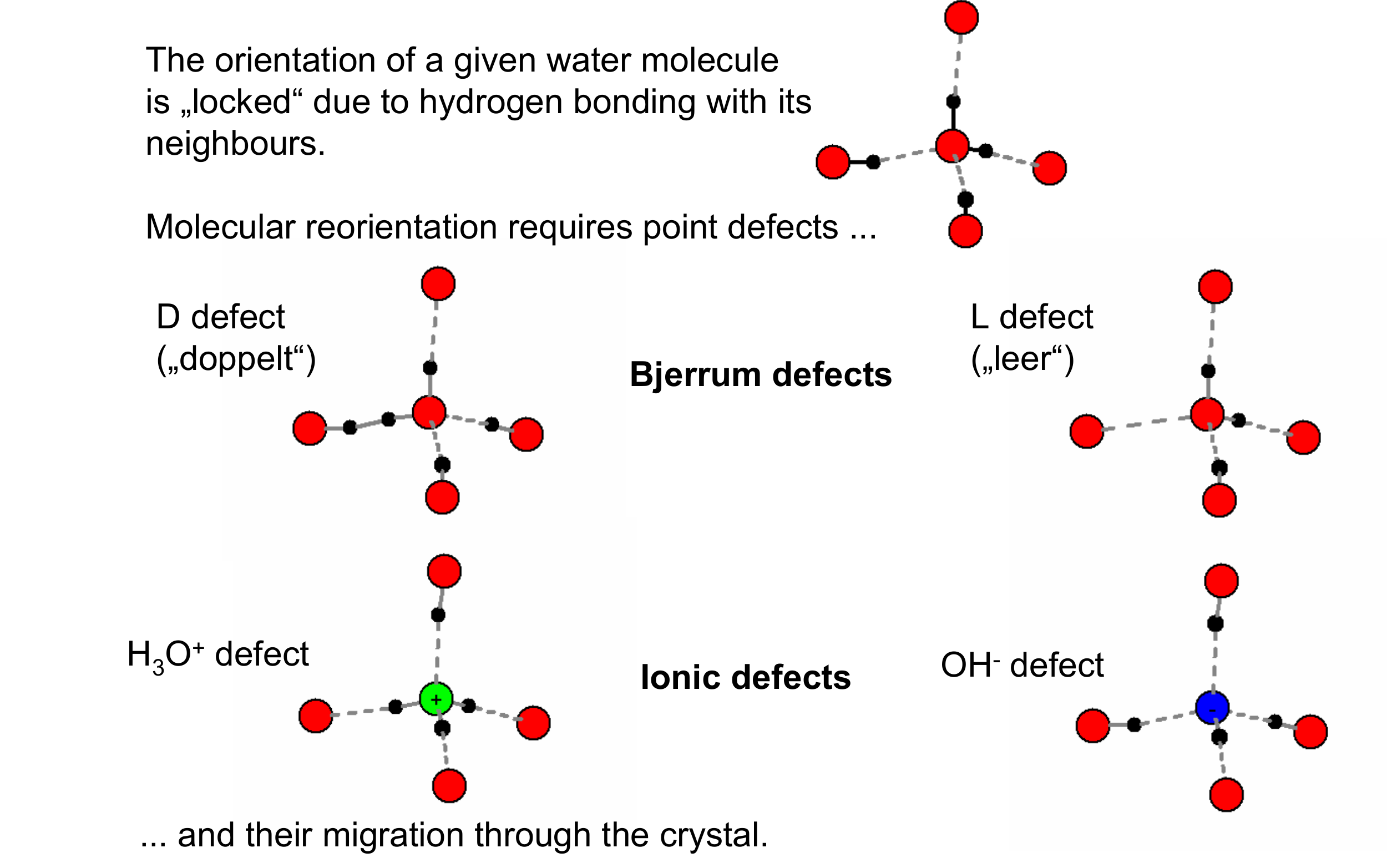} 
\caption{(Color online) Possible point defects in ice structures.
\label{john4}}
\end{figure}

If we want to obtain the ordered --- probably ground state --- structures of the other ices, we need to find some way of unlocking the motions that enable hydrogen atoms to move in the structure; essentially to ease the ability of the molecules to rotate and hence change the hydrogen locations. In principle this might be done through inducing appropriate defects, such as so-called L and D Bjerrum defects (where the Bernal--Fowler rules are violated by zero or double occupation of an O--O link respectively) and ionic defects (where a water molecule is ionized to either a hydronium (H$_3$O$^+$) or hydroxyl (OH$^-$) ion (see Fig.~\ref{john4}). In principle such defects should allow easier molecular rotations. In the case of ordering ices III and VII, presumably there is a sufficient intrinsic concentration of appropriate defects to allow ordering. In the cases of the other disordered ices that do not order on cooling, presumably the intrinsic defects are either insufficient in number or inoperative for some other reason.

\subsubsection{Ordering the familiar ice Ih}

Much effort was spent two decades ago to try to order the familiar ice Ih by forcing \emph{extrinsic} defects into the structure. This was done successfully by doping with KOH, which presumably created L and OH$^-$ defects. This doping was enough to allow careful, slow cooling partially to order the hydrogens so that the ideal structure of the ordered phase (ice XI) could be determined \cite{jackson1997}. The phase transition between the disordered (ice Ih) and ordered (ice XI) phases at 72~K (76~K for D$_2$O) was reported first by \textcite{kawada1972}, with more precise calorimetric measurements following \cite{tajima1984,matsuo1986} in which the doped ice was held a few degrees below the transition temperature for several days to allow the growth of the ordered material. Subsequently the ideal orthorhombic structure of the ordered phase (ice XI) was determined \cite{jackson1997}.

More recent time-resolved powder neutron diffraction studies have explored the nucleation and growth of ice XI. Working with KOD-doped deuterated ice Ih, \textcite{fukazawa2005} concluded that ice XI appears to nucleate at a temperature $57<T\leq62$~K, some 15 to 20~K lower than the transition temperature. Annealing the material at 68~K for three days encouraged growth of the ordered phase, but the fraction of ice XI achieved leveled off at about 6\%. Further work \cite{fukazawa2006} exploring annealing at higher temperatures to just below the transition temperature to ice Ih doubled the fraction of the ordered phase obtained to around 12\%. In addition, noting that the ice XI growth rate is accelerated if the ice is annealed at temperatures lower than 64~K for sufficient time in advance to produce ice XI, they concluded that the thermal history of the sample influences the growth mechanisms of ice XI. A very recent paper from the same group \cite{arakawa2010} takes this work further by taking annealed --- and hence ice XI containing --- samples through the transition at 76 to 100~K to transform the sample to the disordered Ice Ih. Reannealing this sample doubled the fraction of the ordered phase that was obtained. They suggest these interesting observations might be explained in terms of small hydrogen-ordered domains remaining in ice Ih, even when the ice temperature has been taken well above (here 24~K above) the ice Ih--XI transition temperature. This suggestion, that residual order may remain above the transition temperature, is an interesting one that may also be relevant to the ordering of other ice phases.

\subsubsection{Ordering high-pressure ices}

Early work on trying to order ice V by \textcite{handa1987} observed that an endothermic peak on heating ice V could be intensified by KOH doping. This transition was discussed, therefore, in terms of a hydrogen disorder-order transition. However, hydrogen ordering could not be confirmed by Raman spectroscopy \cite{minceva1988}. Attempts at ordering the remaining phases thus remained frozen --- like the molecular rotations themselves --- for nearly 20 years.

\begin{figure}[t]
\centering\includegraphics*[width=\columnwidth,clip=true]{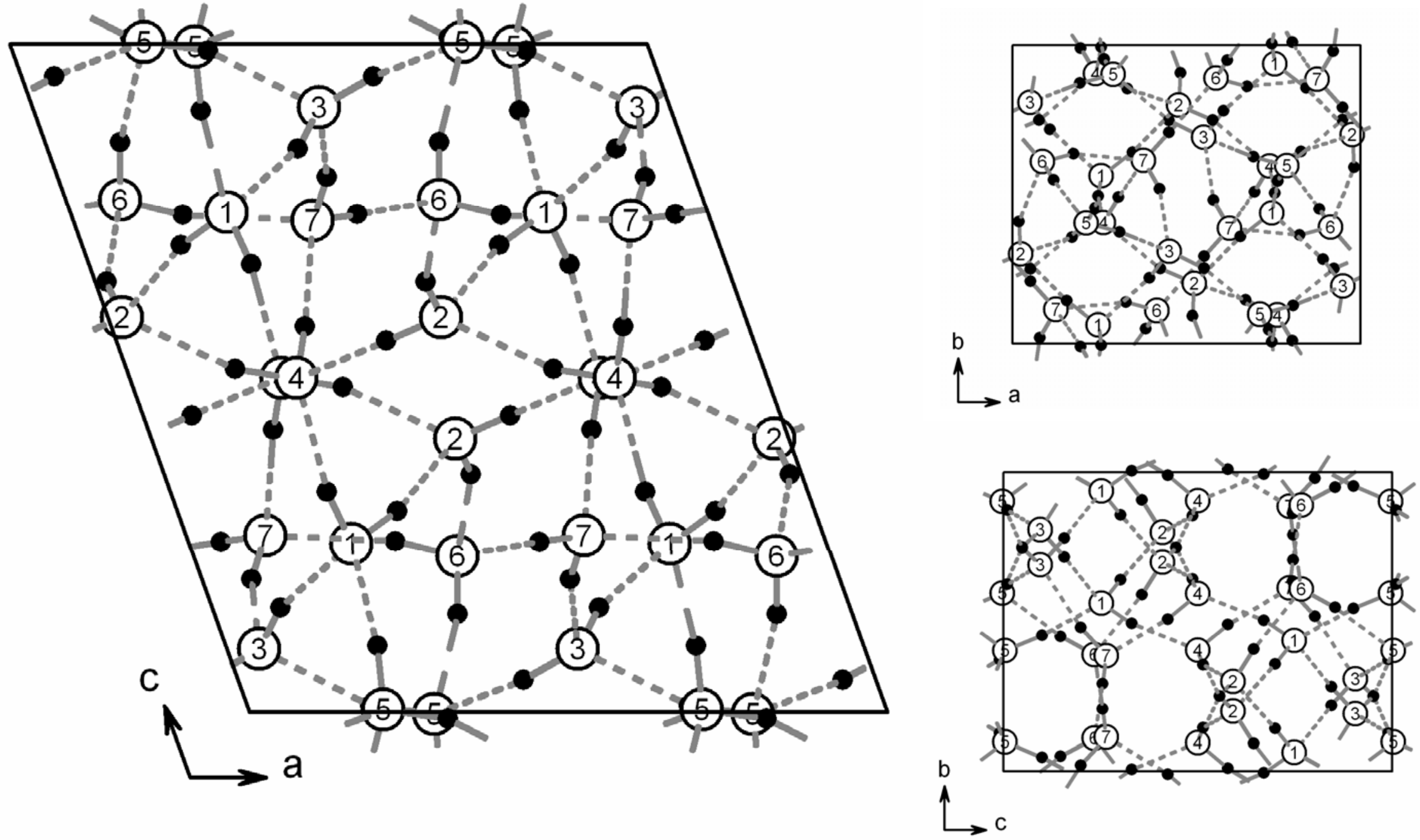} 
\caption{The three unit cell projections of the structure of ordered ice XIII; with 28 H$_2$O molecules per unit cell it is the most complicated structure of all crystalline phases of ice.
\label{john5}}
\end{figure}

The unfreezing of this problem occurred recently when the idea of trying acid rather than alkali doping was suggested by Salzmann and tested, initially, on ice V. This structure was thought to be a good candidate for ordering studies since Erwin Mayer in Innsbruck noted that up to 50\% ordering had been obtained in this structure by \textcite{lobban2000} without any doping. Accordingly, studies were initiated using both HF and HCl doping, coupled with careful controlled cooling procedures. The initial Raman data taken on these preparations suggested significant ordering had been achieved; an ordering that appeared to be much greater for the HCl-doped sample than the HF-doped one. So neutron powder diffraction measurements were made (on deuterated systems as neutrons see deuterium atoms much better than they do hydrogens), and significant ordering was confirmed for DCl doped samples. Not only was the ordered structure --- named ice XIII according to the convention that has developed in the ice field --- determined (Fig.~\ref{john5}) \cite{salzmann2006}; repeated thermal cycling showed that the ice V to ice XIII order--disorder transition is reversible. Interestingly, this transition occurs with no detectable volume change, though there are clear changes in the individual crystallographic unit cell parameters. 

Encouraged by this success, acid doping studies of disordered ices XII (a metastable phase) and ice VI were tried, resulting respectively in the new ordered phases XIV \cite{salzmann2006} and XV \cite{salzmann2009}. Both of these order--disorder phase transitions are also apparently reversible. So we now know the ordered structures of all the known disordered ices apart from (metastable) ice IV.

\subsubsection{Some outstanding questions}

So what are the outstanding questions to be answered about the ice phase diagram, in particular in the context of hydrogen ordering? We suggest the following to be of particular interest.

Firstly, can we order the remaining disordered phase, ice IV; and what is its structure? Being a metastable phase, this is a particularly difficult structure to make reliably. However, recently a reproducible procedure for its preparation has been found \cite{salzmann2002} which should facilitate future studies of its ordering.

Secondly, ice II is alone of all the known ice phases in occurring only as a hydrogen-ordered structure. Why is this? Can a disordered form be found, and if so what is its structure? If ice II cannot be disordered, why not? Can this be explained in simple energetic terms, with the ordered structure having the minimum free energy at this temperature, and if so, what is it about the actual structure that means that this is the case whereas ices Ih, V, and VI at similar temperatures are all disordered?

Thirdly, can we complete the phase diagram? This requires the resolution of the structure of ordered ice IV, and also probably the discovery of at least one other phase whose presence has been seen fleetingly in neutron crystallographic survey work; for example, the unidentified `yellow phase' found by \textcite{lobban1998} at about 5.5~kbar and 263~K --- a $(p,T)$ point close to where ices IV and V can also be found.

Fourthly, can we predict the topology of the phase diagram, either by computer simulation, or from first-principles quantum-chemistry calculations? For if we could, it would raise confidence in the potential functions used in simulations of aqueous systems. Recent work by \textcite{vega2008} has shown that some commonly used potentials are much better than others in reproducing the predominant features of the phase diagram up to medium pressures, though their work has not as yet considered the problems of the order--disorder transitions. Work by others has, however, attempted to predict the structures of ordered ices using density-functional theory approaches, a particularly impressive example being a correct ÔblindÕ prediction of the structure of ice XIV \cite{tribello2006}. Further successes in predicting the ordered ground-state structures would increase confidence in our understanding of the water--water interaction.

Fifthly, what are the ordering mechanisms for both acid (e.g., HCl for ices V, VI, and XII) and base (e.g., KOH for ice Ih) doping, and why is, for example, HCl much more effective than HF? And how might the ordering mechanism relate to the topology of the structure? Why, for example, does acid doping not work for ice Ih, whereas base doping does?

And finally, can we understand the existence of the metastable phases and their ranges of metastability? And can we perhaps use the metastable phases to try to improve our understanding of nucleation and growth mechanisms in ice?
Interestingly, ice XII was first observed within the range of stability of ice V, though later work found that it could also be obtained at much lower temperatures by crystallizing from high-density amorphous ice \cite{koza1999,koza2000,kohl2001}. Is there perhaps a region of the phase diagram where ice XII is the stable phase? The ice-V region is a particularly intriguing one in that two metastable phases --- ices IV and XII --- can also be found in this region. Moreover, all three phases have been observed at the same $(p,T)$ point \cite{lobban1998}. Furthermore, the free-energy difference between ice V and ice XII has been experimentally estimated to be about 67~J~mol$^{-1}$; a very small difference that is much less than the abilities of computer simulations and ab initio approaches to estimate \cite{lobban1998}. Lastly, in this region, as part of studies to map the stability--metastability ranges of the structures found in the ice V region, it was possible to observe a reversible transition between ice IV and ice XII (J.~L.~Finney and F.~Gotthardt, unpublished observation 2000) --- i.e., between two metastable phases.

Strange indeed.

\subsection{Cubic ice}\label{cubic}

Cubic ice, ice Ic, is a metastable hydrogen-disordered ambient-pressure phase of ice. Its ubiquitous occurrence, its importance for atmospheric science (see Section~\ref{atmospheric}) and for astrophysical issues (see Section~\ref{astrophysical}) merit a separate entry. Moreover, it turns out to be poorly crystallized and in this respect justifies a place between the crystalline and amorphous ices, possibly  --- as we shall see --- even in a literal sense. It can be formed from a multitude of starting phases and environments (see \textcite{hobbs1974,petrenko1999}), e.g., from amorphous ices, high-pressure ices, from water vapor or from the liquid in confined systems, or by hyper-quenching \cite{kohl2000}. More recently it has been established that it can also form from decomposing gas hydrates \cite{kuhs2004}. It may thus play a role in transitions within the complex phase diagram of water. Although ice Ih is the thermodynamically stable phase, the formation of ice Ic is understandable due to its lower nucleation barrier \cite{kobayashi1987} and can be seen as a manifestation of Ostwald's step rule.

Its idealized structure represents the crystalline water arrangement with the highest conceivable symmetry. The idealized structure is topologically related to the silicate structure of cristobalite (while common hexagonal ice is related to the structure of tridymite). Yet, this idealized structure has never been observed in experiments. After the first identification of ice Ic \cite{konig1943} it was realized by \textcite{shallcross1957} that the diffraction pattern of ice Ic had hexagonal components --- interpreted by them wrongly as an admixture of ice Ih. Many attempts were made to explain the origin of these hexagonal features and it was realized early on that there is a gradual transition from ill-defined ice Ic into well-crystallized ice Ih. Yet, the transition temperature (range) differed, partly owing to differences in the sensitivity of the method used for detection. It turns out that a suggestion by \textcite{kuhs1987} set the correct entry point for a final quantitative description of the real structure of ice Ic by ascribing the odd features in the diffraction pattern to stacking faults combined with particle size broadening --- the latter owing to the small crystallite size, typically of the order of a few hundred {\AA}. This idea was taken up by \textcite{morishige2005} and finally quantitatively treated by \textcite{hansen2008_1,hansen2008_2}. The model of the latter authors allows for a detailed description of differences in the stacking-fault pattern which appear to occur as a function of the formation procedure, of temperature and --- for isothermal systems --- also as a function of time. Thus, once formed, ice Ic can be considered as a material under ongoing reconstruction both in terms of stacking-fault pattern and of crystallite size. Its final destination is certainly ice Ih, yet to reach this state the material appears to need warming up to temperatures in the vicinity of 240~K \cite{kuhs2004}; at lower temperatures transformations may take longer than the usual laboratory time-scales and eventually may even get stuck with remaining faults. The energy difference between ice Ic and ice Ih is very small, some 50~J~mol$^{-1}$ \cite{handa1986}, as is the stacking fault energy, estimated to be on the order of 0.1--0.3~mJ~m$^{-2}$ \cite{oguro1988}; yet the annealing of stacking faults is linked to the migration of point defects (see, e.g., \textcite{petrenko1999}) with their much higher activation energies. Thus there is not only one ice Ic, but a range of cubic ices with varying degrees of (im)perfection and varying particles sizes, all annealing both as a function of time and temperature, yet seemingly doing so in different ways depending on the initial formation conditions. Modeling tools now allow for the study of processes involving ice Ic in any necessary detail \cite{hansen2008_1,hansen2008_2}.

However, not all the mysteries of ice Ic have been solved. It is at present unclear whether and to what extent the lattice spacing is affected by the stacking-fault sequences or by the local strain; the latter may well arise from incoherent interfaces between the ice Ic nanocrystallites \cite{johari1998}. The local molecular geometries arising from the crystallographic disorder of the oxygen atoms (see above) are considerably less well known than in ice Ih. On a nanoscopic scale it appears quite possible that between the stacking-faulty crystalline regions some interfacial, topologically disordered material, possibly with some enhanced mobility, persists in (sub-) nanometric patches, as has been suggested by recent molecular dynamics computer simulations \cite{moore2010,moore2011}. It will not be straightforward to confirm this experimentally, as small amounts of such disordered material are not easily detected in the presence of a dominant crystalline component. It should also be noted that these computer simulations reach `only' time-scales of several hundred ns and the material obtained would very likely undergo further rather rapid annealing and crystallization on time-scales beyond that limit.

As the ice microstructure, in particular its surface features, has implications for atmospheric and astrophysical considerations (see those Sections), it will be important to shed further light on this issue. Major contributions can be expected from in-situ X-ray and neutron scattering (diffraction and small angle scattering) combined with ex-situ (SEM) and in-situ (ESEM) electron-microscope work. We are only just beginning to see how this complex material behaves and are still far from any detailed understanding. Clearly, the multi-scalar microstructure and its complex transformation kinetics cannot be disentangled in a single experiment --- its complex nature rather needs a long-term commitment.

\subsection{Amorphous ices}\label{amorphous}

\subsubsection{Crystalline and amorphous structures}

In section~\ref{crystalline} we rationalized the structures of the crystalline ices phases in terms of the basic tetrahedral unit of Fig.~\ref{john2}, in which a central water molecule donates two hydrogen bonds to two of its four neighbors and accepts two hydrogen bonds from the other two. The ways in which these (varyingly distorted) units connect together give rise to the different ice structures that pepper the phase diagram shown in Fig.~\ref{john1}.

A necessary characteristic of all these structures --- as with all crystalline structures --- is that the molecules be arranged in a regular, repeating way. Formally, we can define a unit cell for each crystalline structure (see for example that of ice XIII  in Fig.~\ref{john5}) and then, by repeating this unit cell indefinitely in three dimensions generate the structure of the whole crystal. What happens when we remove this constraint of crystallinity? What kinds of structures can we obtain, and can such structures exist in reality? The answer to the second question is yes, and the resulting amorphous ices have been known for many years. Their structures have, however, only been definitively determined in the last decade, and as in the crystalline case, there remain major unanswered questions concerning some aspects of amorphous ices.

\begin{figure}[t]
\centering\includegraphics*[width=\columnwidth,clip=true]{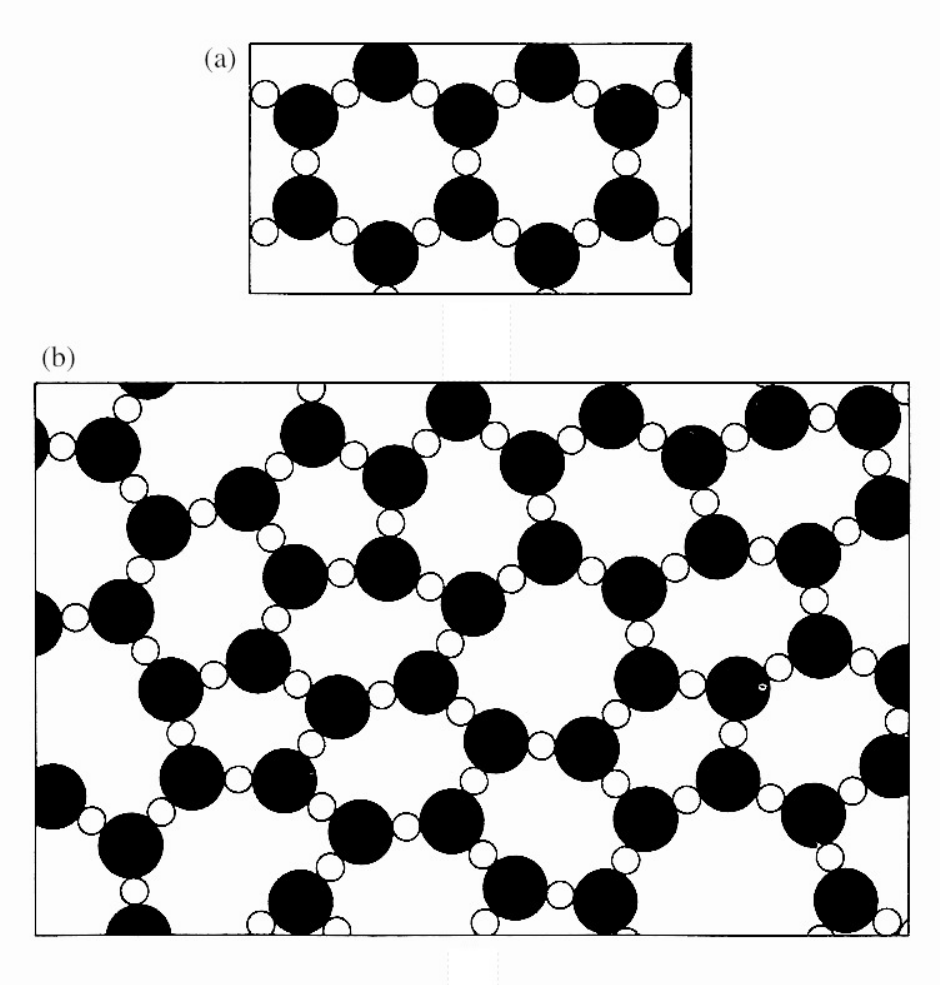} 
\caption{Two-dimensional analogs of (a) crystalline and (b) amorphous ``glassy'' solids made up of three-fold coordinated basic molecular units.
\label{john6}}
\end{figure}

Before looking at the structures of amorphous ices, it is worth making a short digression about amorphous solids in general. As we reduce the temperature of a liquid (which again has a non-crystalline arrangement of molecules), we usually come to a temperature (forgetting for the moment problems of supercooling) at which the liquid crystallizes and a crystalline phase is formed; just as ice Ih forms in our environment (again ignoring supercooling) at 0$^o$C. In many liquids, however, if we cool quickly enough, the molecules may not have sufficient time to rearrange themselves into the equilibrium crystalline structure and a non-crystalline glass may form; an obvious example being the cooling of silica-based liquids to form the glass we are all familiar with. Silica --- like water --- has a local tetrahedral four-coordinated molecular-level structure, and the atoms in the glass maintain the four-fold local coordination as found in the various crystal forms of silica, but without a regular repeating arrangement of the constituent atoms (see Fig.~\ref{john6} for a two-dimensional analog).

The same can happen with water: we can form a non-crystalline arrangement of water molecules, in which the basic four-fold local motif is expected to be retained, though the rates of cooling have to be very rapid indeed; and there are other ways apart from cooling the liquid in which a non-crystalline ice form can be made. Currently, it is generally recognized that there are five different kinds of amorphous ice, distinguished originally by their methods of preparation.

\subsubsection{Formation and structures of amorphous ices}

The first kind, amorphous solid water (ASW), was produced initially by  \textcite{burton1935} by depositing water vapor on a cold substrate. As formed, this material is microporous \cite{mayer1986}, but can be consolidated to a reproducible density by annealing or sintering at a temperature below about 110~K. Another non-crystalline form, hyperquenched glassy water (HGW), can be prepared by very rapid cooling of liquid droplets by either injecting a fine water spray into a cryoliquid or supersonically impacting micron-sized droplets onto a substrate at 77~K \cite{mayer1982,mayer1985}. Both the above forms have the same atomic density of 0.0937 atoms \AA$^{-3}$ (approximately 0.94~g~cm$^{-3}$; there is good reason to use atomic densities as the density depends on the isotopic composition of the sample, and the experimental neutron scattering work that is used to determine the structures uses different isotopic compositions). Pressurizing crystalline ice Ih to about 1~GPa at 77~K results in another apparently amorphous form \cite{mishima1984}. This was labelled high-density amorphous ice (HDA) on account of its significantly higher density of 0.117 atoms \AA$^{-3}$ ($\sim$1.15~g~cm$^{-3}$). Controlled heating of HDA recovered to ambient pressure produced a ``low density amorph'' (LDA) above about 120~K, which has the same density as ASW and HGW \cite{mishima1985}. A very high density form (VHDA; atomic density 0.125 atoms \AA$^{-3}$; $\sim$1.26~g~cm$^{-3}$) was discovered more recently by isobaric heating of HDA under pressure \cite{loerting2001}.

Although there has been an expectation for many years that the atomic-level structures of the three forms of amorphous ice produced differently but having the same density (ASW, HGW, and LDA) are similar, it is only in the last few years that neutron-diffraction techniques have established this structural identity \cite{bowron2006}. The water molecules in these three amorphous forms are almost perfectly four-fold coordinated, with each water molecule surrounded on average by four hydrogen-bonded neighboring water molecules at distances of between 2.5 to 3.3~\AA. The neutron scattering results additionally tell us that the mean O--O--O angle in the LDA network is $\sim$111$^o$, close to the ideal tetrahedral angle of some 109$^o$. We can thus conclude that the short range structures of LDA, ASW, and HGW all relate closely to a hydrogen-bonded tetrahedral network of water molecules.

The structures of the denser forms, HDA and VHDA, are somewhat more complex. A coordination-number analysis for these ices again shows that each oxygen atom is immediately coordinated by approximately four neighboring water oxygen atoms, situated in the distance range from 2.5 to 3.1~\AA. In contrast to the lower-density forms such as LDA, however, where the number of first neighbors remains at 4 up to a distance of 3.4~\AA, by this point the water-molecule oxygen coordination number in HDA rises to $\sim$5 and to a value of $\sim$6 in VHDA. Again the neutron data tell us that each water oxygen is hydrogen bonded to two hydrogen atoms, confirming that we have a four-coordinated hydrogen-bonded network. The higher density of these two forms thus appears to be obtained by molecules that are further away from a typical `central' molecule moving further in towards that central molecule, though not being hydrogen bonded to that central molecule. Although these `closer in' molecules were labelled ``interstitial'' molecules when they were first identified \cite{finney2002,finney2002_2}, it is important to emphasize that they themselves still appear to be hydrogen bonded to four neighbors in (distorted) tetrahedral geometries, and not are not ``interstitial'' in the common understanding of the term of being embedded separately in the main matrix structure (which would be a crystalline lattice in the crystalline case). 

So, just as with the crystalline ices, all these amorphous ice forms have structures that are essentially made up of the four-coordinated basic motif of Fig.~\ref{john2}, although in the amorphous case these units are connected together in a non-crystalline, non-repeating way. Moreover, as in some of the higher-pressure crystalline phases, the degree of distortion of this basic structural unit appears to be increased in the higher-density forms HDA and VHDA. In these ices, this distortion appears to be high enough to allow molecules that would in the case of, for example, LDA be non-hydrogen-bonded second neighbors to be `pushed in' to distances from the central molecule that are of the same order as those of the hydrogen-bonded first neighbors, yet which remain non-hydrogen bonded to that central molecule.

Thus, from a situation a decade ago in which we were unsure of the structures of the four amorphous forms ASW, HGW, LDA, and HDA, we have moved to one in which the essential structural identity of the first three, equal-density, forms has been established as being essentially a random network of four-coordinated water molecules. Moreover, we know that the higher density form, HDA, although still having a four-coordinated structure, has on average an additional molecule almost as close in as each of the hydrogen-bonded first neighbors. The more recently discovered very-high-density form, VHDA, appears to have a structure similar to that of HDA, though with additional occupation of the so-called interstitial location.

\subsubsection{Some outstanding questions}

However, although the amorphous ice situation has been simplified in some ways, in other respects it has become more complex, as a cursory look at the literature since 2002 will quickly demonstrate. A number of very interesting questions remain to be addressed.

The first of these concerns detailed structural characterization: what is the detailed relationship between the networks of the low-density and the high-density forms? How are the water molecules in the `interstitial' sites connected through the hydrogen-bonded network to the molecule at the tetrahedral `central' site? Is this connectivity the same as in, e.g., LDA with the interstitial being simply a previously more distant second neighbor being pushed in closer without any change in network connectivity, or has a change in the network topology occurred? If the first option is the case, then LDA, HDA, and VHDA can be seen as essentially the same topological structure, with just increasing distortion of the basic tetrahedral unit allowing us to move from the low-density structure to the higher-density ones. If the second is true, then how is the network connectivity changed? Are there instances in which hydrogen-bond linkages pass through ring structures formed by other water molecules, as is observed in the ÔinterpenetratingÕ networks found in the higher-pressure ices IV, VI, VII, and VIII? And if these higher-density amorphous forms do have a distinctly different network connectivity from the low-density forms, can this be considered a distinctly different water structure that could relate to the continuing debate \cite{mishima1998} about liquid water itself being able to exist in two distinctly different structural forms?

Whereas the above structural questions can be --- and, we hope, are being --- answered by a detailed examination of the network structures of the high- and low-density forms \cite{loerting2010}, there are further complexities that are more challenging to resolve. For example, structural annealing studies of the conversion of the high-density forms into the low-density forms by heating have been interpreted as suggesting the existence of a family of high-density structures  \cite{tulk2002,koza2003} and also the possibility of a range of low-density forms (\textcite{koza2003}; see also \textcite{koza2009}). However, this picture has been challenged by subsequent studies \cite{mishima2002,nelmes2006,winkel2008} that demonstrated that if care is taken to anneal the high-density amorphous ice forms to a structure that is thermodynamically metastable with respect to the applied pressure and temperature conditions of ice formation, the complexity of the high-density to low-density transition is simplified. Subsequently, the annealed form of HDA has been relabeled expanded HDA (eHDA) \cite{nelmes2006}. 

Furthermore, a recent computer-simulation study has raised the suggestion of a second distinct low-density form of amorphous ice \cite{guillot2003}. This challenges the prevailing experimental view of one low-density amorphous structure \cite{bowron2006} which is thought to be the metastable minimum structure. Within the framework of the simulation, a form labelled LDA$_\textrm{I}$ was produced by warming the model of HDA, while a second slightly more dense ice, LDA$_\textrm{II}$, was produced by warming the model of VHDA. Although interesting, these suggestions seemed to be at odds with existing experimental observations that the density of LDA produced by warming either eHDA or VHDA samples was found to be the same, at 0.0937 atoms \AA$^{-3}$ \cite{debenedetti2003,loerting2006}. Initial neutron-scattering experiments \cite{winkel2009} have however indicated that there exist at least two structural forms of low-density amorphous ice, where the primary difference in the structures occurs on an extended length scale in the hydrogen-bonded network of molecules. LDA$_\textrm{I}$ and LDA$_\textrm{II}$ are not seen to be distinct forms of ice in the way that LDA, HDA, and VHDA differ, but rather are two closely related, but kinetically trapped, forms of what could be considered the true metastable low-density amorphous ice characterized by an atomic density of 0.0937 atoms \AA$^{-3}$.

A further question that remains to be answered concerns the relationship between the structure(s) of amorphous ices and the structure of liquid water. The fact that LDA can be formed by rapid cooling of the liquid suggests a structural continuity with the liquid, and this is generally accepted to be the case. However, there has been much speculation concerning the possible existence of more than one liquid phase of composition H$_2$O, possible liquid--liquid immiscibility, and the existence of a second critical point (see, for example, \textcite{poole1992,debenedetti1996,mishima1998}). Noting the `polyamorphism' exhibited by the existence of more than one amorphous ice, the data currently available suggesting the existence of three structurally distinguishable amorphous ices (LDA, HDA, VHDA) \cite{loerting2011}, HDA (initially) \cite{mishima1998} and, since its discovery in 2001 \cite{loerting2001}, VHDA \cite{finney2002} have been proposed as possible glassy analogs of the putative `high-density liquid'. Such hypotheses do however remain hypotheses at the present time.

We concluded the section on crystalline ices with the phrase ``strange indeed''. The amorphous ices situation qualifies equally for this description --- and perhaps more so. Clearing up the current confusion in amorphous ices --- where we are of necessity dealing with non-equilibrium systems, complicated further by structural characterization problems made more difficult by the lack of an ordered crystalline structure --- is perhaps an even more challenging problem than that of finally sorting out the ice phase diagram.

\subsection{Polycrystallinity and interfaces}
\label{ssec:struct_sur}

A discussion of the structure and phase morphologies of ice is incomplete without addressing the important role that polycrystallinity, surfaces, and interfaces play.  Many of the important physical and chemical processes that are discussed in depth throughout this work are heavily influenced by ice grain structure and the character of ice surfaces and phase transitions. For example, in Section~\ref{sintering}, we discuss how the grain size of ice in glaciers changes with time, and in Section~\ref{snow_chemistry} the role of grain boundaries on the uptake of atmospheric trace gases is noted. 

Fundamentally, ice is very similar to a wide range of other polycrystalline materials, but is of particular interest because of its geophysical importance.  Furthermore, many interesting and important effects result from the fact that the ice around us exists at temperatures and pressures very close to where we expect phase coexistence.  Detailed and authoritative reviews have addressed many problems of grain structures, surfaces, and phase transitions in ice \cite{Dash2006}, metals \cite{Smith1964}, and other materials \cite{Luo2008}.  Here we simply summarize the most important aspects of these various behaviors.  

\subsubsection{Ice --- e pluribus unum}
\label{ssec:grains}

Naturally occurring ice, like most crystalline materials, is usually made up of many crystallites, or grains.  By the time it reaches the ground, even meteoric ice that begins as single crystals in the atmosphere is often polycrystalline. As a result, when we discuss the morphology of ice, in addition to the molecular structure of single crystals, the granular structure of the bulk ice can be very important.

\begin{figure} \centering
\includegraphics[width=\columnwidth,clip=true]{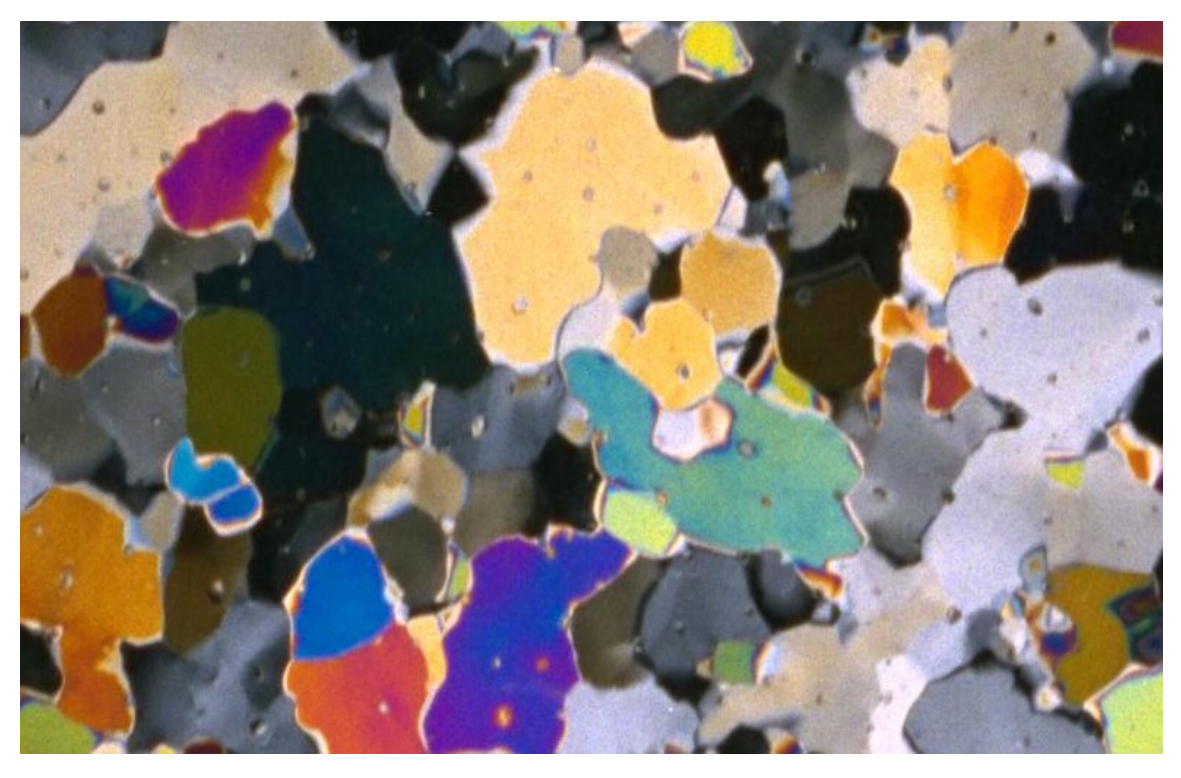}
\caption{(Color online) A thin 2~cm  by 3.7~cm polycrystalline ice section seen under cross polarized light. Individual crystals are seen in different shades/colors due to the birefringent nature of ice.}
\label{fig:thinsect}
\end{figure}

Ice grains can have a range of sizes that strongly depends upon the thermodynamic history of the particular sample. For example, within polar ice sheets crystal diameters range from less than 1~mm to more than 1~m \cite{Thorsteinsson1997}. 
When visualizing hexagonal ice crystals we may take advantage of their optical anisotropy; under cross-polarized light individual crystal domains are clearly distinguished; see Fig.~\ref{fig:thinsect}.  
With this technique generally even very small grains can be distinguished with a microscope.  In ice and other materials grain structure is observed either on exposed surfaces, or, as we see in Fig.~\ref{fig:thinsect}, by slicing thin sections of ice and examining the resulting two-dimensional structure.  At such length scales, in its two-dimensional form the underlying molecular structure of ice is not evident, and the individual crystals do not form standard polyhedra.  Instead, the discontinuities between crystals of different orientations stand out. 

When ice is warmed and the bulk melting temperature is approached, the atomic and molecular disorder present at low temperatures grows to produce macroscopic liquid layers at crystal interfaces. The planar structure between two single grains is referred to as a grain boundary.  Where three grains intersect \emph{veins} form, and where four or more crystals come together \emph{nodes} are formed.  These structures are themselves surfaces, internal to bulk polycrystals, and are discussed in more detail in the following Section~\ref{ssec:surf}.

In macroscopic terms, the hexagonal symmetry of molecular ice is described by prism facets surrounding an optical axis of symmetry, the $c$-axis, which is perpendicular to the basal plane.  The basal plane is characterized by a lower roughening temperature than the prism faces, and therefore is often observed as a smooth surface \cite{Wilen1995a,furukawa1997}.  In the atmosphere, crystal shape is generally determined by supersaturation (see Section~\ref{atmospheric}), but in bulk polycrystals 
many post-nucleation processes determine the evolution of crystal shape and size. In systems under stress $c$-axes are known to rotate in order to align themselves in the direction of deformation \cite{Paterson1994}. In polycrystalline systems with a distribution of grain sizes, grain boundary migration leads larger crystals to absorb smaller crystals, and over time the average grain size increases \cite{Alley1992}.  On the other hand, when dislocations are readily produced, they may accumulate to form new grain boundaries, a process called \emph{polygonalization}.  In systems where large strains exist, completely new grains may nucleate in a process termed \emph{recrystallization} \cite{Alley1992,Paterson1994}.  Bubbles, impurities, and other discontinuities within the crystal matrix will affect each of these processes in complicated and non-trivial ways.  The geometry and sponge-like nature of fallen snow add complexity due to the presence of air and water vapor.  Just as surface and strain gradients drive grain growth in solid ice, snow responds to vapor gradients.  Fundamentally the same morphological processes affecting ice crystal character are important to snow as it compacts and densifies, forming firn and eventually glacial ice (see Section~\ref{terrestrial}).  

Often in the laboratory idealized single crystals with measured or controlled size, shape, and crystal orientation are examined.  However, when working with polycrystalline ice samples, like those described in Sections \ref{astrophysical} and  \ref{atmospheric}, the effect of the polycrystalline morphology of ice must always be acknowledged.

\subsubsection{Surfaces and interfaces}
\label{ssec:surf}

In addition to the free surface, the grain boundary, vein, and node interfaces internal to bulk polycrystalline ice play important physical and chemical roles.  Surfaces are where phase change is initiated and are regions of enhanced impurity concentration.  Below the bulk melting temperature of ice, these interfaces can be disordered, even liquid, at equilibrium.  

With increasing temperature the outermost molecular layers of an ice surface in contact with a substrate or vapor can become increasingly mobile, and under the right circumstances form a \emph{premelted} liquid layer  \cite[for a complete review see][]{Dash2006}, which is of importance to the human activities of skiing, snowballing, skating, and so on.  The morphology of the surface as it nears the melting temperature depends upon the specifics of the system and is rooted in the strength of the intermolecular interactions within and between the various materials.  The same intermolecular interactions link the physics of melting with wetting behavior \cite{Schick1990,French2010}.  The existence of distinct droplets versus spreading liquid layers on surfaces is related to the relative cohesion versus adhesion of the material.  Droplets form when cohesion dominates, but when adhesion wins the liquid thins and spreads over the solid.  The balance between these effects represents the range of possible melting behaviors from partial to complete melting.  
Likewise the microscopic details of melting change with increasing temperature as surface structure breaks down and disorder and mobility increase.  In this transition liquid-like structures that retain some of the underlying crystal order, called \emph{quasi-liquid layers}, represent an intermediate state between the solid and bulk liquid.  In practice, the distinction between liquid and \emph{quasi-liquid} can be difficult to make and often experiment and theory require us to make assumptions concerning the physical properties (e.g., density, viscosity, index of refraction) of such layers.  In these cases, assuming bulk liquid properties represents one limiting case, which for many properties has been shown to be representative even for very thin layers \cite[e.g., ][]{Raviv2001}.

Thus a liquid or quasi-liquid surface layer will exist, in the bulk-solid region of the phase diagram, in thermodynamic equilibrium, on ice if by being there the total free energy of the system is reduced.  \textcite{Elbaum1993a} demonstrated that the surface melting of ice in equilibrium with its vapor can vary between partial and complete, but for systems with even small impurity levels complete melting is observed.  Ice surface and interfacial melting can be well-modeled theoretically using unretarded van der Waals forces, whose long-range potential falls off quadratically with distance \cite{Dash2006}.  This yields an expression for the thickness $d$ of the liquid layer as a function of temperature $T$ relative to the bulk melting temperature $T_\mathrm{m}$, 
\begin{equation}
\label{eq:d_sm}  d=\left( \frac{-2\Delta\gamma\sigma^2}{\rho_\mathrm{l}q_\mathrm{m}}\right)^{1/3} \left(\frac{T_\mathrm{m}-T}{T_\mathrm{m}}\right)^{-1/3},
\end{equation}  
where $\rho_\mathrm{l}$ is the density of the bulk liquid, $q_\mathrm{m}$ is the latent heat of transformation, and $\sigma$ is a constant on the order of a molecular diameter.  The difference between the wet and the dry interfacial free energy is $\Delta\gamma=\gamma_\mathrm{sl}+\gamma_\mathrm{lv}-\gamma_\mathrm{sv}$, where each $\gamma$ is an interfacial free energy per unit area with the subscripted letters defining the type of interface ($\mathrm{s}$, solid;  $\mathrm{l}$, liquid; $\mathrm{v}$, vapor).  In the case of interfacial melting the vapor phase is replaced by a substrate.  Modeled and observed thicknesses range from 1 to 20~nm and tend to diverge as the melting temperature is approached \cite{Elbaum1991, Elbaum1993a, Sadtchenko2002a}.

\begin{figure}
\centering
\includegraphics[width=\columnwidth,clip=true]{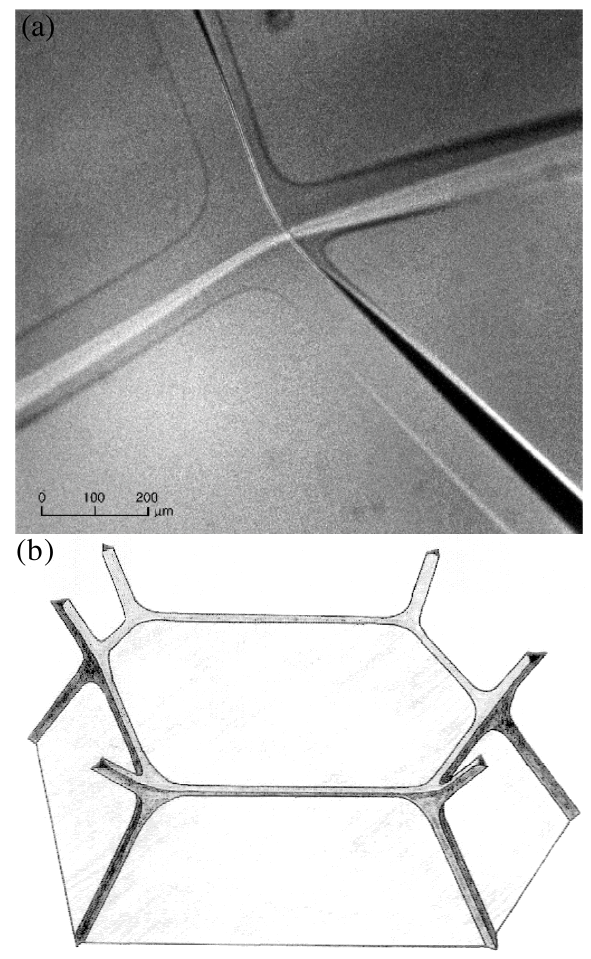}
\caption{The vein--node network in polycrystalline ice: (a) Micrograph published originally in \textcite{Mader1992}; (b) schematic from \textcite{Nye1992}, modified from \textcite{Smith1948}. Together they appear as Fig.~11 in \textcite{Dash2006}.}
\label{fig:veins}
\end{figure}

Like the free surface, within ice, veins and nodes of liquid water form in equilibrium below the bulk melting temperature; see Fig.~\ref{fig:veins}.  
However, in this case they exist largely because these interfaces have a high degree of curvature.  Thermodynamically, the energy cost of a curved solid interface is related to the surface energy per unit area $\gamma_\mathrm{sl}$  and interfacial curvature $\mathcal{K}=1/r$, vis-\`a-vis the Gibbs--Thomson effect \cite{Thomson1871}, which is also important in nucleation (Section~\ref{atmospheric}) and the bubble and crystal coarsening of Ostwald ripening \cite{Avron1992, Hillert1965}.  In the vein/node network the Gibbs--Thomson effect results in a depression of the equilibrium melting temperature that can be written as
\begin{equation}
T_\mathrm{m}-T= \frac{\gamma_\mathrm{sl}T_\mathrm{m}}{\rho_\mathrm{s} q_\mathrm{m}}\frac{2}{r},
\end{equation}
where for simplicity the solid is modeled as a sphere of radius $r$ (positive when the interface is convex into the melt), and $\rho_\mathrm{s}$ is the bulk density of the solid.  As $r$ increases the magnitude of $T_\mathrm{m}-T$ decreases, and approaches zero for flat interfaces.  Again the existence of liquid within the bulk solid polycrystalline ice is the thermodynamically favorable state.  In ice the liquid veins and nodes threading throughout the volume can be tens to hundreds of micrometers wide (Fig.~\ref{fig:veins}).

Grain boundaries are particularly important to understanding polycrystalline morphology.  In most materials they are challenging to access, yet they represent the dominant surface area of a polycrystal.  At these boundaries curvature disappears and the dispersion forces are attractive for systems of like materials straddling an interface.  However, the crystalline discontinuity suggests that some level of molecular or atomic disorder must exist between differently oriented grains.  For a range of materials and interaction potentials, grain-boundary disorder has been theoretically and numerically modeled \cite[e.g.,][]{Schick1987, Broughton1986,Mellenthin2008}.  \textcite{Thomson2010a} and \textcite{Benatov2004} suggest that wetted grain boundaries in impurity-doped ice may be measurable using optical light-scattering techniques \cite{Thomson2009b}.  Their experimental observations seem to support the idea that in impurity-rich systems grain-boundary disorder can be stabilized with thicknesses of tens to hundreds of monolayers.  In general, experiments on, and models of, grain boundaries show agreement that grain-boundary morphology is very sensitive to the specific interactions within a system.  In particular, in ice, surface charge and impurity loading strongly effect grain-boundary equilibrium.    

All ice surface effects are strongly influenced by the presence of impurities.  Ice is a particularly efficient segregator of impurities, and as a result impurities in water are strongly concentrated at ice surfaces, rather than being incorporated by the solid lattice, during freezing.  In addition to the classical colligative effect of depressing the equilibrium melting temperature, enhanced impurities can drive instabilities in freezing fronts \cite{Wettlaufer1994}, alter surface charge densities \cite{jackson1997}, and effect the intermolecular interactions of a system \cite{Dash2006}.  Impurities interacting with ice are particularly important in geophysical environments, where they are ubiquitous, and much of the discussion in this work centers on their physical and chemical importance in various settings. The fact that biological processes also take place at snow and ice surfaces has been emphasized in a recent paper \cite{ariya2011}.

Ice is an interesting material on many levels.  Its molecular structure and phase morphologies inspire a multitude of science.  In our natural environment ice characteristically exists as a polycrystalline material near its triple point, resulting in diverse behavior with macroscopic importance.  As stress, temperature, composition, and pressure of the environment vary so too do the structural and surface responses of ice.  From the molecular to the macroscopic, the structure of ice is an important variable underlying behavior in the cryosphere.

\section{Astrophysical Ice}\label{astrophysical}

\begin{quote}
The ice didn't have to be quarried. It existed in proper chunks in the rings of Saturn. That's all the rings were---pieces of nearly pure ice, circling Saturn. So spectroscopy stated and so it had turned out to be. He was standing on one such piece now, over two miles long, nearly one mile thick. It was almost half a billion tons of water, all in one piece, and he was standing on it.

Isaac Asimov, \emph{The Martian Way}
\end{quote}

In astrophysics ``ices'' are defined as those molecular species that are liquid or gaseous at room pressure and temperature (1~atm and 25$^o$C) but are solids under the physical conditions of the astrophysical environment in question. Water ice is the most abundant such ``ice'', and here, naturally, we shall concentrate on it, but in astrophysics one may often be interested in its abundance relative to other species and in mixtures of water ice with other ``ices''.

\begin{table}
\caption{Objects in the Solar System on which ice has been detected, or for which (with $^*$) its presence has been hypothesized. 
\label{astrophys_solarsys}}
\centering\begin{tabular}{ll}
\hline 
\textbf{Planet} & Observed species  \\
Satellite & in order of abundance \\
\hline
\textbf{Mercury} & H$_2$O in polar craters$^*$ \\
\multicolumn{2}{c}{\textcite{black2010,schilling2011}} \\
\textbf{Earth} &  \\
Moon & H$_2$O at poles and in craters$^*$ \\
\multicolumn{2}{c}{\textcite{colaprete2010}} \\
\textbf{Mars} & CO$_2$, H$_2$O surface, subsurface,  and \\
\multicolumn{2}{r}{stratospheric aerosols$^*$} \\
\multicolumn{2}{c}{\textcite{bibring2004,shean2007,cull2010}} \\
\multicolumn{2}{c}{\textcite{mccleese2010}} \\
\textbf{Asteroids} &  H$_2$O detected on 24 Themis, 65 Cybele \\
\multicolumn{2}{c}{\textcite{rivkin2010,campins2010};} \\
\multicolumn{2}{c}{\textcite{licandro2011}}
\\
\textbf{Jupiter} & NH$_4$SH, H$_2$O \\
Io & SO$_2$, H$_2$S, H$_2$O \\
Europa & H$_2$O, SO$_2$, CO$_2$, H$_2$O$_2$ \\
Ganymede & H$_2$O, O$_2$, O$_3$, CO$_2$ \\
Callisto & H$_2$O, SO$_2$, CO$_2$ \\
\multicolumn{2}{c}{\textcite{calvin1995,nash1995};} \\
\multicolumn{2}{c}{\textcite{mousis2006}} \\
\textbf{Saturn} & \\
Main rings & H$_2$O \\
Titan & H$_2$O on surface; CH$_4$ clathrates \\
\multicolumn{2}{r}{buried?; HCN, nitriles as icy stratospheric aerosols} \\
Mimas & H$_2$O \\
Enceladus & H$_2$O \\
Tetis & H$_2$O \\
Dione & H$_2$O, O$_3$ \\
Rhea & H$_2$O, O$_3$ \\
Hyperion & H$_2$O \\
Iapetus & H$_2$O \\
\multicolumn{2}{c}{\textcite{mastrapa2009,cruikshank1984,cruikshank2010}} \\
\multicolumn{2}{c}{\textcite{thomas1986,morrison1984,sohl2010};} \\
\textbf{Uranus} & \\
Miranda & H$_2$O \\
Ariel & H$_2$O \\
Umbriel & H$_2$O \\
Titania & H$_2$O \\
Oberon & H$_2$O \\
\multicolumn{2}{c}{\textcite{cruikshank1995,showalter2006}} \\
\textbf{Neptune} & \\
Triton & N$_2$, CH$_4$, CO, CO$_2$, H$_2$O \\
\multicolumn{2}{c}{\textcite{brown1995,coustenis2008}} \\
\textbf{Trans-Neptunian objects} & \\
Pluto & N$_2$, CH$_4$, CO, H$_2$O\\
Charon & H$_2$O \\
Haumea & H$_2$O \\
\multicolumn{2}{c}{\textcite{barkume2008, cruikshank1995,brown2010}} \\
\multicolumn{2}{c}{\textcite{gil-hutton2009,dotto2003,trujillo2007}} \\
\hline
\end{tabular}
\end{table}

Water ice is present on many objects in the Solar System (\textcite{cartwright2007}; \textcite{bell2009} and references therein). In addition, the occurrence of frozen volatile molecules in these objects is a matter of current discussion. We present in Table~\ref{astrophys_solarsys}  a summary of the ice species detected and hypothesized to exist on different bodies in the Solar System. At present the Messenger probe orbiting Mercury is providing indications that ice may be be present in polar craters there \cite{schilling2011}. Perhaps one of the most controversial issues is the possible presence of water on the Moon. Many articles have provided pros and cons for this case based on different interpretations of the available evidence through time. The latest results based on the analysis of debris following an impact on the lunar surface provide strong evidence for the presence of water \cite{colaprete2010}, but ice has not been detected yet.  The possibility that Earth may possess a faint ring, responsible for slight climate alterations, has even been raised in the past \cite{okeefe1980} and in very recent investigations \cite{hancock2010}, although the existence of the ring, and of its being composed of ice, seems rather dubious. The polar regions of Mars are rich in ices (mainly carbon dioxide and water), but frozen aerosols containing these species are also detected in the Martian stratosphere \cite{mccleese2010}. Very recently, two teams \cite{campins2010,rivkin2010} have provided evidence of water ice on the asteroid 24 Themis,
thus rekindling the discussion of the possible water delivery to Earth from these objects, abundant in the Main Asteroid Belt, between Mars and Jupiter. Even more recently, \textcite{licandro2011} have contributed to this argument with findings
of water and organics on the asteroid 65 Cybele.
The satellites of the outer planets --- Jupiter and beyond --- are dominated by water ice. Europa and Enceladus show fractured surfaces formed by blocks of water ice, which bear some similarity to icy surfaces on Earth. 
As Asimov anticipated, the rings of Saturn and other bodies of the outer Solar System, including the ring systems of Jupiter, Uranus, and Neptune, are composed of 95\% water ice \cite{heisselmann2010} (however the size of the fragments seems to be rather less than Asimov surmised \cite{cuzzi2009}). 
 On Pluto and the other so called trans-Neptunian objects --- a class of numerous small objects not yet well investigated --- some ice species have been detected. The composition of these far-off objects is not fully understood, but infrared reflectance data point to the existence of water and possibly methane ices. Water ice is found both in crystalline and amorphous phases, and models are employed to simulate the observations based on laboratory optical indices of these systems \cite{mastrapa2009}. 
Comets contain a large inventory of ices that are observed after their sublimation during comet passages near the Sun, but their presence is often masked by devolatilized surface materials (organics and/or silicates).  Experiments related to the sublimation of ice mixtures are then particularly relevant in this context; see Section~\ref{astrophys-lab}.
The existence of ice on the newly detected exoplanets is already being considered; observations are being carried
out to investigate this possibility, for instance through surface albedo and color \cite{cahoy2010}.

Ices in the Solar System suffer from energetic processing by solar photons, by solar wind and solar flares ions, and galactic cosmic rays. Furthermore satellites moving in the magnetosphere of their planet are exposed to the effects of magnetospheric ions (e.g., \textcite{johnson1990,strazzulla1991}). Just one example is the case of hydrogen peroxide (H$_2$O$_2$) that has been detected on Europa \cite{carlson1999}. Comparison with laboratory spectra indicates a surface concentration of about 0.13\%, by number of molecules, relative to water ice. Europa is immersed in a highly energetic plasma environment originating in Jupiter's magnetosphere \cite{cooper2001}. Based on the results of several experiments, radiolysis is supposed to be the dominant formation mechanism of hydrogen peroxide on Europa \cite{gomis2004,moore2000}. Laboratory experiments also suggest a patchy distribution of H$_2$O$_2$ on Europa. This result could be useful to support the hypothesis of a radiation-driven ecosystem on Europa based on the availability of organic molecules and oxidants such as hydrogen peroxide \cite{chyba2000}. More examples are available through the numerous missions currently deployed to several planets, especially Mars; the interested reader is advised to visit the webpages of NASA, ESA and other space agencies. 

Beyond the Solar System, the interstellar medium contains about 10\% of our galaxy's mass. It is mainly composed of atomic and molecular gas species (99\% by mass, of which 70\% is H, 28\% He, and 2\% are heavier elements), and dust (1\% by mass) and is pervaded by photon and particle radiation. The gas is characterized by two main parameters: its temperature ($T$) and its number density, largely dominated by the abundance of atomic and/or molecular hydrogen: $n_0=n(\mathrm{H})+2 n(\mathrm{H}_2)$~cm$^{-3}$. The average density of the dust is correlated with the gas density by $n_d=10^{-12}n_0$. The gas and dust together form either diffuse clouds that are characterized by low density ($n_0=10$--$10^3$~cm$^{-3}$), cold dust  (10--30~K) and hotter gas (100~K), in which hydrogen is essentially in atomic form, or dense clouds that are characterized by higher  densities ($n_0>10^4$~cm$^{-3}$), in which dust and gas are cold (10--30~K) and the hydrogen is mostly molecular. In the dense cold clouds gas phase species freeze out on cold grain surfaces ($\sim$10~K) and may chemically react to form new molecules (see section~\ref{price}); in this way the formation of polar (i.e., water-rich) and/or apolar (i.e., water-poor) ice mantles occurs.  The interaction with photons and cosmic particles drives the formation of other molecules, as has been simulated by experiments performed in several laboratories around the world. Occasional heating in the interstellar medium or the heating induced by star formation inside the cloud lead to segregation, crystallization and sublimation of ice mantles. Released species contribute to the inventory of the more than 120 chemical species identified in the gas phase. In contrast only some dozen molecules --- the most abundant being water ice --- have been identified so far in the solid phase; see the inventory in Table~\ref{astrophys_clouds}. This discrepancy in number is owing to the different sensitivities of radio and infrared techniques; in the gas phase molecules rotate and can be observed with an extremely high sensitivity at radio wavelengths, but in the solid phase the vibrational transitions are detected in the infrared with a much lower sensitivity.
	
\begin{table}
\caption{Molecular species detected on icy mantles in dense interstellar clouds and their relative abundance.
\label{astrophys_clouds}}
\centering\begin{tabular}{lcl}
\hline 
Species & Relative abundance & Reported by \\
\hline
H$_2$O & 100 & \\
CO & 0--40 & \textcite{chiar1994} \\
&& \textcite{boogert2004} \\
CH$_3$OH & 2--31 &  \textcite{allamandola1992}; \\
&& \textcite{dartois1999_2}; \\
&& \textcite{boogert2008} \\
NH$_4^+$ & 1--26 & \textcite{boogert2004}; \\
&& \textcite{boogert2008} \\
CO$_2$ & 10--25 & \textcite{gerakines1999}; \\
&& \textcite{zasowski2009} \\
CH$_4$ & 0.3--23 & \textcite{lacy1991}; \\
&& \textcite{boogert1997}; \\
&& \textcite{zasowski2009} \\
HCOOH &  1--16  &   \textcite{boogert2008} \\
NH$_3$ & 1--10 & \textcite{lacy1998}; \\
&& \textcite{williams2007} \\
OCN$^-$ & 1--8 & \textcite{tegler1995} \\
H$_2$CO & 0--7 & \textcite{schutte1996}; \\
&& \textcite{zasowski2009} \\
SO$_2$ &  0.3--1.7 & \textcite{boogert1997}; \\
&& \textcite{zasowski2009} \\
OCS & 0.04--0.4 & \textcite{palumbo1997}; \\
&& \textcite{williams2007}  \\
\hline
\end{tabular}
\end{table}

\subsection{Laboratory studies of astrophysical ices}
\label{astrophys-lab}

One way to gain insight into the icy fraction in interstellar medium particles and grains, as well as the composition and physical properties of cometary ice, is to perform laboratory experiments in which models are developed and investigated under circumstances that may resemble, or that would allow extrapolating the achieved results to, the harsher conditions in astrophysical media. Several laboratories in the world have been dedicated to such studies for some time (see for instance, \textcite{ehrenfreund2001,fraser2002,ehrenfreund2003,gerakines2005,munoz2006,hibbitts2007,hodyss2009,patel2009}, and references therein). Typical experiments consist of the preparation of ice samples, mainly by vapor deposition of individual substances or mixtures of several components, under controlled conditions of pressure and temperature. The solid fraction may be studied using infrared or Raman spectroscopy, and by transmission \cite{jenniskens1996b} and scanning (see below) electron microscopy. Similarly, the gas fraction in the chamber may be monitored by mass spectrometry. In some instances, the ices may be exposed to different forms of irradiation that may simulate solar or stellar emissions, to study the physico-chemical processes induced in the solid. The heating cycle of comets in their solar approximations is usually simulated by heating the samples, often in temperature programmed desorption experiments, which also provide information on the adsorption and surface interactions of gas and solid phases.

\subsubsection{Ice morphology on interstellar grain surfaces}

\begin{figure*}[tbp]
\begin{center}
\includegraphics*[width= \textwidth,clip=true]{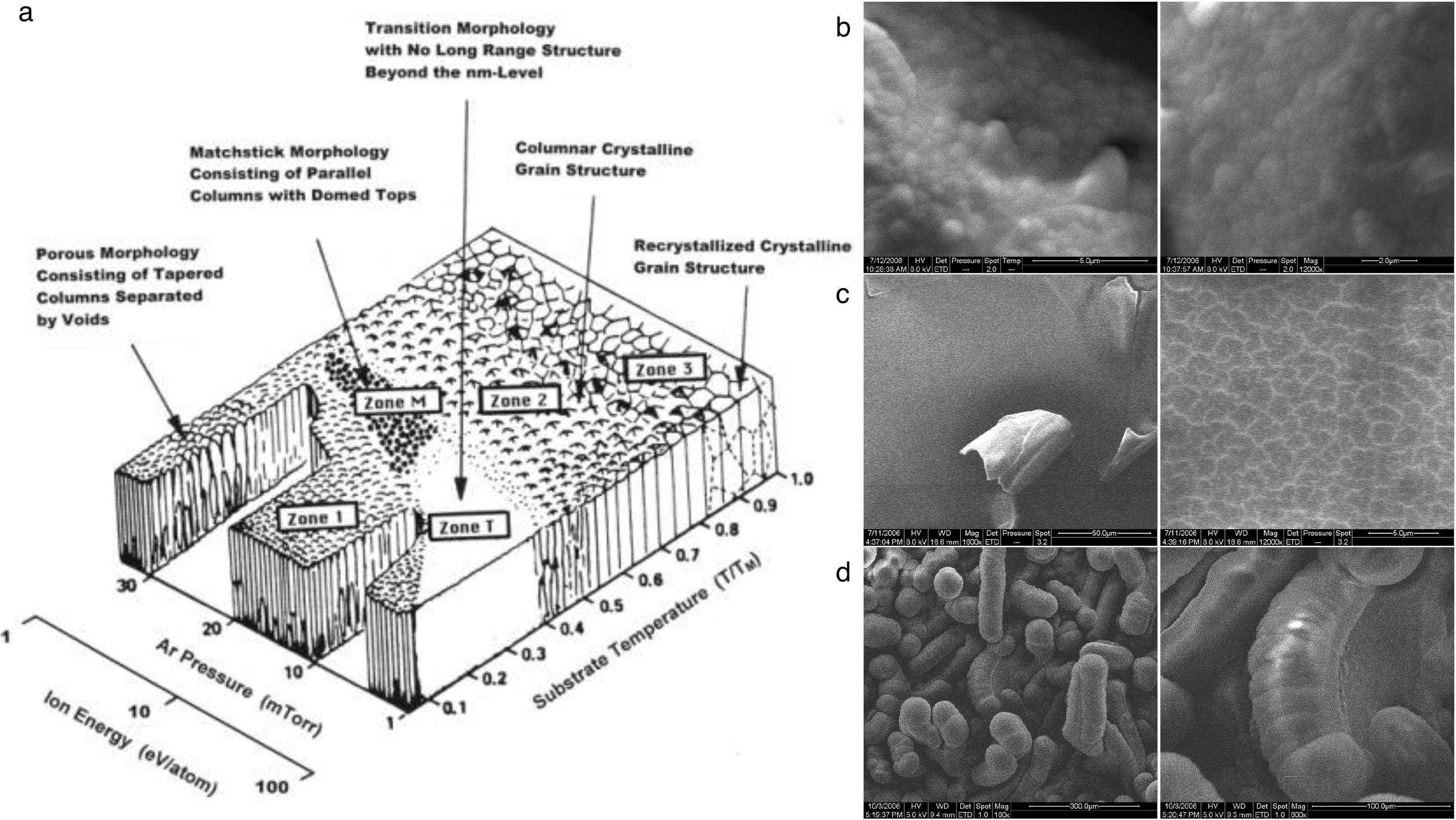}
\end{center}
\caption{\label{esem}
Film morphologies:
(a) A recent version of the structure zone model of solid film deposition \cite{lakhtakia2005};
ice films produced in situ in a liquid-helium cooled ESEM apparatus in water vapor deposition experiments \cite{cartwright2008}:
(b) zone 1, cauliflower morphology;
(c) zone T, transition morphology;
(d) zone M, matchstick morphology.
}
\end{figure*}

Dust grains in interstellar space are typically covered with an icy film formed at the very low temperatures of the interstellar medium \cite{tielens2005}; it has been estimated that most of the ice in the universe is to be found on these particles \cite{jenniskens1995}. Larger icy bodies are supposed to have formed through the accretion and reprocessing of this material \cite{ehrenfreund2003,vandishoeck2004}. Thus, icy films are found in the interstellar medium, in the Oort cloud, in planetesimals, asteroids, comets, and Kuiper-belt objects, on the surfaces of icy planets and moons, and in planetary ring systems and atmospheres, as mentioned above for the Solar System. Much astrochemistry is presumed to take place on these icy surfaces.  A fundamental question, on which hinge many astrophysical and astrochemical processes, is: in what forms --- with what morphologies --- may such ice films be deposited?

For many years films of numerous different materials --- ceramics, semiconductors, and metals --- have been deposited onto substrates from the vapor phase. The field is driven by a huge number of technological applications, but also has much scientific interest \cite{lakhtakia2005}.  One of the key differences between films and  bulk materials is in their morphologies, and from the 1960s on efforts have been made to construct a classification of the morphology  of a film depending on the conditions of its deposition. This has culminated in the structure zone model. A recent version, reproduced in Fig.~\ref{esem}(a), shows the distinct film morphologies obtained as a function of the renormalized substrate temperature $T/T_M$, where $T_M$ is the melting point of the material being deposited, and of the ion energy, which is inversely related to the argon gas pressure in the sputtering technique used in that research \cite{messier2000,lakhtakia2005}.  The structure zone model describes well the trends seen in experiments with ceramics, semiconductors, and metals. It has recently been found that the same morphologies are found in films of ice, despite the very different chemical bonding in ice compared to the materials from which the structure zone model was devised \cite{cartwright2010}.

An environmental scanning electron microscope (ESEM) equipped with a liquid helium cold stage may be used to grow ice films {\em in situ} at low pressures and temperatures down to 6~K appropriate to astrophysical conditions \cite{cartwright2008}. The region of lowest substrate temperature in the structure zone model is occupied by zone 1. Zone 1, or cauliflower morphology, consists of competing void-separated tapered columns whose diameters expand with the film depth according to a power-law \cite{lakhtakia2005}. The surface resembles a cauliflower, showing self-similarity over a range of scales; Fig.~\ref{esem}(b). At a higher water vapor pressure, there is obtained zone T, or transition morphology, in which there is no long-range structure above the nanoscale. Figure~\ref{esem}(c) shows the featureless surface characteristic of zone~T as well as  the boundaries of the individual densely-packed grains making up the surface, made visible as charge accumulated on the grain boundaries of the poorly conducting ice surface during imaging. In qualitative terms, the morphology of zone~1 is clearly driven by a competitive process of growth of clusters at all scales, leading to a fractal geometry, while greater gas-induced mobility of admolecules in zone~T than in zone~1 smoothes out the surface. As they occur at the lowest temperatures, both zone 1 and zone T morphologies can be presumed to be composed of a high-density amorphous ice (see Section~\ref{amorph_crystal}).  At substrate temperatures above those corresponding to cauliflower morphology (zone 1), and for relatively high gas-induced admolecule mobility there appears zone M, or matchstick morphology. In the example of this morphology we reproduce in Fig.~\ref{esem}(d) we see very large columns tens of micrometers in diameter. The columns display the domed tips characteristic of matchstick morphology, and also show interesting substructures both on the tip and along their length, presumably from their growth process; their shape is biomimetic, like an icy worm. With regard to the temperatures involved for zone M, it should typically be composed of low-density amorphous ice.

We see that structure zone morphologies do appear in ice films. Can this knowledge contribute to understanding the physics of the structure zone model and of solid ice films? The two axes of the structure zone model shown in Fig.~\ref{esem}(a) are the renormalized substrate temperature and the ion energy, which is inversely proportional to the gas pressure in sputtering experiments. On the other hand the deposition technique employed here is simply evaporation, involving thermal energies ($3/2\,kT< 1$~eV) for the admolecules. However, the energy required to sputter a water molecule is $\sim0.2$~eV \cite{bukowski2007}, much lower than the 10--30~eV sputtering threshold for ceramics, semiconductors or metals \cite{lakhtakia2005}. So we should renormalize the scale of the second axis by the sputtering threshold, in the same way as the substrate temperature is renormalized by the melting point for the first axis. It becomes clear then that the physical basis for the two axes is admolecule mobility induced in one case by the temperature of the film itself, and in the other by the impinging admolecules from the vapor. The latter mechanism is related in these experiments to the temperature of the vapor, and to the presence or absence of helium as an auxiliary gas in the chamber that will thermalize with the chamber walls. The different effects of the two sources for admolecule mobility leading to more or less compact structures surely arise because, unlike thermal mobility, admolecule movement induced by bombardment is highly directional. This directionality is quantifiable in terms of the sputtering yield as a function of the angle of incidence of the bombardment, which has a maximum at 45--60$^o$ \cite{lakhtakia2005}.  This is an avenue to explore in moving towards a physical understanding of the structure zone model as a consequence of the competition between the spatially disordered deposition of particles on the growing film surface and the ordering effect of activated particle mobility processes. The ESEM technique, considering the $p$--$T$ range it covers, is applicable also to atmospheric ices. In one way it is even more appropriate as the time-scales covered by ESEM ---minutes to hours--- are very similar to the actual time scale of atmospheric processes. This is not true for many astrophysical processes, which often take place over much longer times (deposition and aging effects, etc) and can thus be modeled in ESEM experiments only by rescaling time.

A recognition that these mesoscale structures exist, together with a knowledge of their morphologies, ought to aid understanding of fundamental astophysical and astrochemical processes involving surfaces coated with icy films and should lead to a reconsideration in terms of the structure zone model morphologies of what is at present in astrophysics often placed under the catch-all label of ice porosity \cite{cartwright2008}.

\subsubsection{Mixed ices}

\begin{figure}[tp]
\centering\includegraphics*[width=\columnwidth,clip=true]{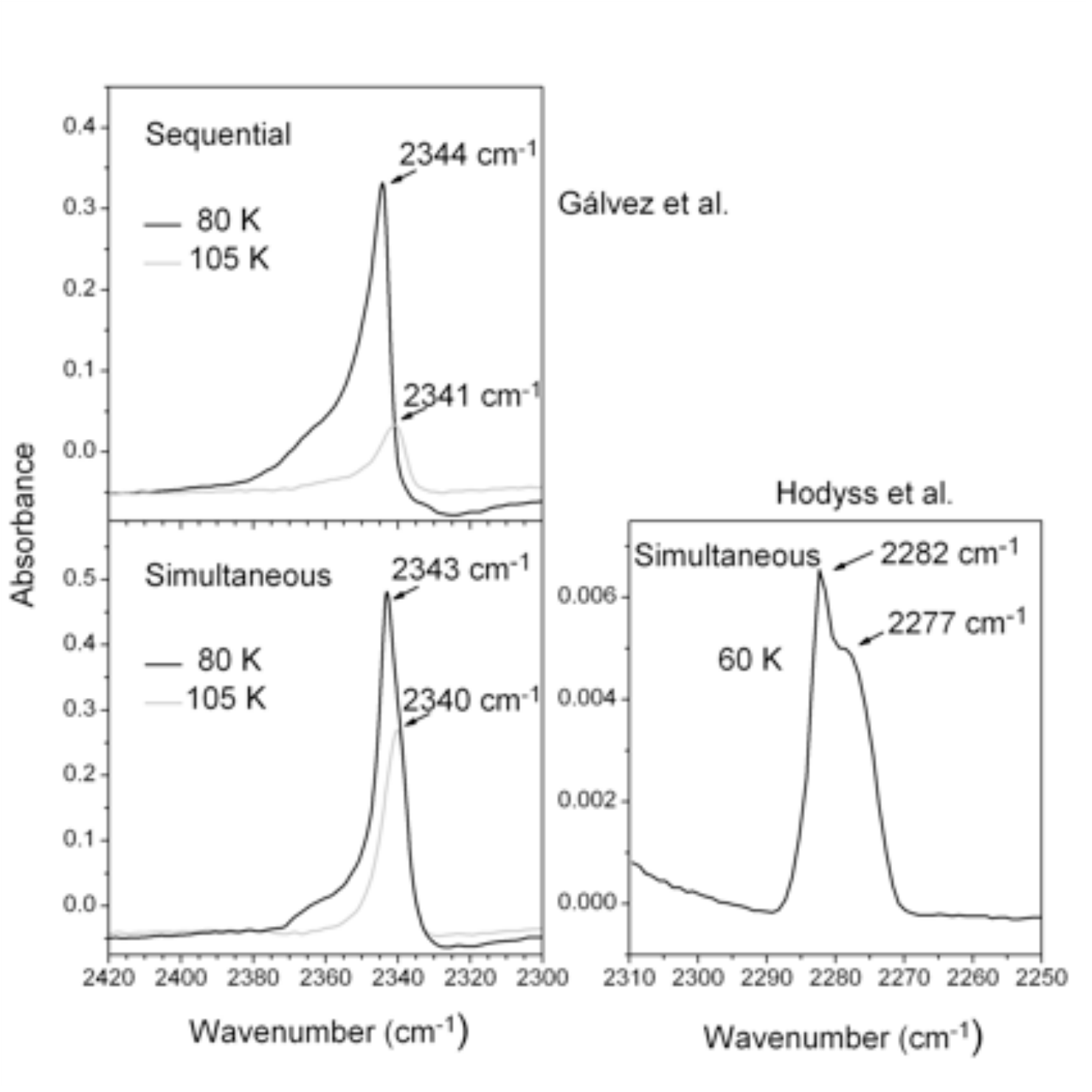} 
\caption{Transmission IR spectra of mixtures of CO$_2$ and water in the region of the $\nu_3$ band of CO$_2$. Left: spectra of two samples obtained by sequential deposition of water and CO$_2$ (above) and simultaneous deposition (below) of the vapours at 80~K. After warming the sample, most CO$_2$-norm has sublimated and CO$_2$-dist is better appreciated (from \textcite{galvez2007}). Right: spectrum focused on the same band for the $^{13}$CO$_2$ isotopomer. The two CO$_2$ structures are characterized by a band deconvolution analysis (from \textcite{hodyss2008}).
\label{rafa1}}
\end{figure}

Ices composed of mixtures of water, carbon dioxide and methanol are frequently studied as models for the ice envelope of cometary nuclei and other astrophysical objects \cite{bernstein2005,malyk2007,oberg2007,dartois1999}. In very recent publications dealing with these systems, evidence of two different CO$_2$ structures, characterized by their different infrared spectral signatures, has been clearly exposed \cite{mate2009, galvez2007, hodyss2008}. The spectrum of one of the structures is closer to that of pure CO$_2$, whereas in the other the main bands are shifted to lower frequencies and do not present the typical features of solid CO$_2$ like the crystal splitting in the $\nu_2$ band, or the sharp peaks in the combination band region. These two structures have been referred to as normal or pure and distorted, or CO$_2$-norm and CO$_2$-dist, respectively. Figure~\ref{rafa1}a presents two examples taken from \textcite{galvez2007} and \textcite{hodyss2008} that show the spectral region of the asymmetric C-O stretch $\nu_3$ band for the $^{12}$C and $^{13}$C isotopomers. 
Whereas CO$_2$-norm is assumed to be formed by small crystallites of CO$_2$ adsorbed at the surface (in sequentially deposited mixtures) or segregated in larger pores (in simultaneous deposition experiments), the structure of CO$_2$-dist is much less clear. The red-shifted vibrations of the latter may indicate some interaction with the surrounding host molecules that weakens the C-O bonding in CO$_2$ by $\sim$0.3 \%. The formation of clathrates retaining CO$_2$ in their interior is not excluded. Although these two CO$_2$ structures are well characterized from the spectroscopic point of view, there are some differences in their behavior when CO$_2$ is mixed with water or methanol. Some of these differences could be explained based on the various porosities of the host ices, but again a full understanding is not yet available. 
The internal structure of CO$_2$ ices in water and methanol mixtures thus remains an open question in this field.
This topic has been further extended in a very recent investigation \cite{cruikshank2010}. Theoretical calculations of small clusters have been performed \cite{chaban2007}, but the results are not conclusive, and in some cases seem to point to frequency shifts for the CO$_2$ bands in the opposite sense to the observations. On the other hand, calculations for larger models analogous to solid samples are difficult and have not been reported yet, but they would provide a powerful tool to understand this problem. The eventual spectroscopic observation in astrophysical measurements of the two kinds of CO$_2$--H$_2$O association revealed by these studies would provide a clue to the thermophysical evolution and growth process of the corresponding object. Other results discussed by \textcite{oberg2007,hodyss2008,cruikshank2010}, which include evidence for the adsorption of CO$_2$ on crystalline water ice under certain conditions \cite{galvez2007} underline the relevance of laboratory experiments on samples containing mixed constituents, as frequently found in astrophysical media.

A special mention should be made of the topic of astrophysical clathrate hydrates, as being a medium to store volatile species within crystalline water cages. Whereas most clathrates of terrestrial interest are formed under high pressure conditions, it has been proposed \cite{buch2009} that dipolar species such as acetone, formaldehyde, H$_2$S, HCN can become trapped in water cages at pressures in the 0.1 to 100~Pa range and temperatures as low as 100~K, and in some cases species like CH$_4$ or CO$_2$ can also form mixed clathrates with ethers in similar conditions of astronomical relevance. Whereas on Earth most clathrates exist under high-pressure conditions, this requirement is not essential on celestial bodies where the temperature is much lower. \textcite{dartois2008} have discussed the presence of clathrate species in comets and planets, in conjunction with laboratory experiments aimed to register infrared spectra of CH$_4$ clathrate hydrates. The Cassini mission has contributed with information that points to a vast reservoir of these hydrates beneath the surface of Titan \cite{niemann2010}.

\subsection{Cometary ice}

Cometary and trans-Neptunian objects (TNOs) research constitutes an important topic of Solar System ice studies.  Comets and TNOs contain invaluable information on the processes acting during the early stages of Solar System formation. As these objects have a high content of frozen material, the physics of ice is of fundamental importance in order to extract relevant information from them. In these complex systems, thermal evolution is strongly dependent on structural characteristics, and therefore it is not possible to extract dependable information without appropriate thermophysical models (both theoretical and experimental), including a complete description of both thermal and structural behavior. This is the main problem today in understanding the nature of comets: we do not yet know, for example, either the micro- or the macrostructure of cometary nuclei and of their constituents.

Cometary nuclei are made up of refractory material (generally described as dust) and a mixture of different frozen volatiles. Water ice is the major constituent, mixed with small quantities of more than 25 other volatiles. The second volatile is CO, with an abundance ranging from 0.01\% up to 10\% that of water, depending on the comet. Other minor volatiles, with an abundance smaller than 1\% of that of water, are CO$_2$, NH$_3$, and HCN.  One of the most discussed, but yet unsolved, problems regarding comets is the quasi-simultaneous sublimation of those volatiles, which have different saturation pressures. If it is accepted that sublimation of volatiles is triggered by insolation, it would be expected that the different volatiles should appear at different heliocentric distances. Nevertheless it is not possible to find a correspondence from observations between the corresponding temperature of sublimation of the different volatiles and heliocentric distance; see, e.g., \textcite{capria2000}.  This circumstance suggests that highly volatile constituents, or at least part of them, may be trapped within the water-ice structure, although the mechanisms involved in their trapping and storage are unknown or, at least, controversial. Two different mechanisms have been proposed to describe the trapping of highly volatile molecules: adsorption of volatiles on amorphous water ice and trapping of volatiles in crystalline water ice in the form of clathrate hydrates. There are arguments for and against both mechanisms, and the following illustrates not only the difficulties related to Solar System ice research, but also that progress can only be made with close cooperation among groups with different expertise.

\subsubsection{Amorphous versus crystalline ice}\label{amorph_crystal}

It is thought that comets formed at very low temperatures. This argument is founded, for example, on estimates of spin temperatures of, mainly, water and ammonia (see e.g., \textcite{crovisier2005} and references therein) which point towards a formation temperature of about 30~K.  At that temperature, water ice condenses in an amorphous phase, displaying an open structure and therefore being able to trap other volatile compounds within it. This idea is supported with observations of icy grains in the comae of some comets such as C/2002 C7 and Hale--Bopp. Near-infrared spectra of  detected icy grains do not show the characteristic 1.65~$\mu$m absorption feature of crystalline ice (see, e.g., \textcite{davies1997,kawakita2004}). Nevertheless, the non-detection of the absorption feature cannot be considered a definitive proof of amorphous ice since that spectral feature depends strongly on the temperature. Actually, there are laboratory experiments indicating that initial ice in the grains of the presolar nebula (considered seeds of cometary nuclei) is not necessarily amorphous. \textcite{moore1994} performed  a series of experiments to characterize spectroscopically water ices condensed on amorphous silicate smokes on an aluminum substrate at temperatures less than 20~K, to simulate the first stages of comet formation. These authors found that, for smoke particles with a diameter of about 5--10 nm, water ice condensed in a crystalline phase, suffering what they named a low-temperature crystallization (LTC). The authors suggested that this LTC occurs due to molecular interactions between the condensing gas and reactive silicate surface sites. Leaving the door open to amorphous ice formation, \textcite{moore1994} also mentioned that LTC was not observed for comparatively large grains (of the order of microns and larger), suggesting that size of the grains may be an important variable. Later, \textcite{schutte2002} extensively argued that the LTC reported by Moore et al could be caused by an experimental effect, mainly due to the poor thermal conductivity of the smoke layer. 
In any case, the most used argument against the existence, or at least, persistence of amorphous ice in comets is based on the results of thermophysical models (see \textcite{klinger1981,prialnik2003}). Considering that crystallization is an exothermic transformation, releasing an important quantity of energy, models predict that this energy would trigger a run-away process, producing the crystallization of cometary ice in a comparatively small time scale. Again, this argument has been questioned recently on the basis of laboratory experiments with impure water ice. \textcite{kouchi2001} performed experiments of differential thermal analysis on amorphous samples doped with various volatiles and found that crystallization becomes endothermic for CO ``impurities'' larger than 3\% by mass, which is a proportion plausible for comets. This result is in line with those of \textcite{sandford1990} and \textcite{manca2001}. \textcite{gonzalez2008} showed that for an endothermic crystallization, amorphous ice could survive for a long time, with a crystallization front comparatively close to the cometary surface.

The crystallization of amorphous ice is a much-studied phenomenon in ice physics; numerous laboratory studies have been devoted to this topic; to cite but a few, \textcite{dohnalek1999,dohnalek2000,hage1994,hage1995,lofgren1996,safarik2003,safarik2004,smith1996}. All these papers generally agree that the transition is gradual as a function of time and the higher the temperature the faster, leading eventually to cubic ice, ice Ic (see Section~\ref{cubic}). The beginning of the transition can be described in the well-known Johnson--Mehl--Avrami--Kolmogorov approach \cite{fanfoni1998} for nucleation and growth with an activation energy in the range of ~70 kJ/mol
 \begin{equation}
 f(t) = \frac{\pi}{3}Jv^3t^4,
 \end{equation}
where $f(t)$ is the fraction of transformed phase at a given time $t$, $J$ is the constant nucleation rate, and $v$ is the growth velocity of the crystal nuclei (NB, not cometary nuclei). Additionally, the reader is directed to \textcite{prialnik2003} for details on how the crystallization is dealt with in thermophysical models of comets. A paper by \textcite{jenniskens1996}, well known in the astrophysical communities and based on thin-film diffraction experiments, claims the co-existence of a more relaxed amorphous phase with crystallized ice Ic for temperatures up to 200~K. The existence of this viscous liquid-like or amorphous water is certainly not accepted in the ice-physics community as there is no other evidence for such a mixture of phases; the absorption spectroscopy work by \textcite{mitlin2002} is quite formal in that the measured spectra can perfectly well be described with optical constants of crystalline ice. Another finding of \textcite{jenniskens1996}, the release of adsorbed impurities upon crystallization, has been confirmed by later work \cite{kouchi2001}. 

A note on the terminology of amorphous (water) ice in the astrophysical communities is in order here. The multitude of amorphous ices discussed in Section~\ref{amorphous} has been reduced to a low-density form (LDA) and either one or two higher density forms, HDA (high density amorphous) and VHDA (very high density amorphous); it is unclear at present if HDA and VHDA should be considered as distinct forms or variants of the same form. They are characterized in great detail and LDA comprises older forms like amorphous solid water (ASW) and hyperquenched glassy water (HGW); see Section~\ref{amorphous}. Certainly LDA can be expected to exist under astrophysical conditions, in particular at temperatures higher than about 60~K. The situation at lower temperatures is more complex. \textcite{venkatesh1974} found evidence for an amorphous form with a density of $\sim$1.2~g~cm$^{-3}$ by X-ray diffraction work on samples prepared and measured at 10~K. A more detailed analysis and comparison with similar work at 77~K was subsequently presented by the same group \cite{narten1976}, correcting the density to $\sim$1.1~g~cm$^{-3}$ and introducing for the first time the term ``high-density amorphous ice'', yet still in parallel with the term ``low-temperature amorphous ice''. Referring to this work,  \textcite{jenniskens1994} took over the term ``high-density amorphous ice'' for water ice formed at low temperature, introducing it in the astrophysical communities. There it has stayed ever since, often referred to as ``low-temperature high-density amorphous ice'' or ``Ia(h)''. Whether this form is identical to HDA produced at higher pressures and described in Section~\ref{amorphous} is at present unclear and will be hard to elucidate on the basis of X-ray and electron diffraction data alone. It would be of interest to clarify this situation fully, in particular as a low-temperature phase transition may well have implications in the astrophysical context (see, e.g., \textcite{jenniskens1996}).

\subsubsection{Clathrates}

We return now to the mechanisms to describe the trapping of highly volatile molecules within cometary material. \textcite{delsemme1952} proposed the existence of clathrate hydrates in cometary nuclei to explain the quasi-simultaneous sublimation of different ices. This mechanism has been included in several thermophysical models (e.g., \textcite{houpis1985,flammer1998}) to describe cometary nuclei evolution. Nevertheless, \textcite{klinger1981} questioned the viability of such mechanisms with thermodynamical arguments, and, actually, of the plethora of volatiles detected in comets, formation of clathrate hydrates under solar nebula conditions has been proven only for CH$_3$OH \cite{blake1991} and H$_2$S \cite{richardson1985}. Clathrate hydrate formation has become an attractive explanation  in recent years, following the visit of comet Hale--Bopp and the analysis of the Stardust mission samples (e.g., \textcite{brownlee2006}) which may be consistent with large-scale circulation in the solar nebula. This large-scale circulation would suggest that clathrate hydrates could form at comparatively high pressures and temperatures. \textcite{gautier2005} proposed clathrate hydrate formation as a plausible mechanism for trapping volatiles that would explain the unexpectedly low value of the upper limit of the N$_2$--CO ratio inferred in several long-period comets, as determined by  \textcite{cochran2000,cochran2002}. \textcite{gautier2005} argued that, CH$_4$, CO, and N$_2$ being the most abundant species, they consume most of the available water in this order, given their relative affinities for clathrate formation. Since the amount of water is limited, some CO and N$_2$ molecules may not form clathrates, and they would not be incorporated into at least some cometary nuclei. But this circumstance cannot be used to exclude the possibility of amorphous ice. Laboratory experiments by \textcite{notesco1996}, work disregarded by the previous authors, show that the under-abundance of certain species can also be explained with amorphous water ice, which is more efficient at trapping CO than N$_2$.

Even if, or better, because, a complete picture of cometary ices structure is not yet available, numerical and laboratory experiments on the behavior of mixtures of dusty ices are still necessary, and they have to be designed bearing in mind observational implications. In this line it is worth mentioning the experiments of \textcite{notesco2000} and \textcite{bar-nun2007} focused on the effects of clathrate and other gas-trapping mechanisms on cometary ices, those of \textcite{gerakines2005}, \textcite{bernstein2006}, and \textcite{galvez2007}, etc.\ devoted to spectroscopic characterization of ice mixtures, and \textcite{trainer2009}, who discuss experiments showing that ethers (for which however there is not much evidence at all) may help to speed up the formation kinetics. From a numerical standpoint, a matter pending is to verify if the equations included in models are able to reproduce the complex patterns of sublimation observed in experiments with ice mixtures.

\subsection{Heterogeneous chemical processes on interstellar surfaces}\label{price}

Astrophysicists, particularly those interested in the synthesis of molecules in the interstellar medium, have for several decades recognized the importance of interstellar surface chemistry  (see the review of \textcite{williams2002}).  Indeed, in the last century, modeling clearly indicated that the formation of the most abundant interstellar molecule, molecular hydrogen, must occur efficiently on the surfaces of the abundant interstellar dust grains \cite{tielens1982}.

Molecules in the interstellar medium are present in giant interstellar clouds that are cold and tenuous \cite{williams2002,williams1999}.  As mentioned above, such clouds, which are composed of gas and dust, are broadly divided into two classes: diffuse and dense. The conditions in dense clouds allow the dust grains, which are carbonaceous or silicaceous in nature, to accumulate mantles of molecular ices \cite{williams2002,williams1999}.  Spectroscopic surveys show that water, methanol, carbon dioxide, carbon monoxide and ammonia molecules are often present in these mixed ice mantles \cite{gibb2000}.  Modeling of the chemistry in these interstellar clouds indicates that these ice mantles do not grow solely by the freeze-out of molecules from the gas phase. Heterogeneous processes must also be occurring, synthesizing molecules on grain surfaces which accumulate as ice layers \cite{oba2009}.

An important question linked to astrophysical ices is the origin of pre-biotic material \cite{meierhenrich2008}. \textcite{munoz2002} report the discovery of amino acids in ice made to mimic cometary ices from CO, CO$_2$, NH$_3$, and H$_2$O; another paper by the same group describes the finding of diamino acids in the Murchison meteorite \cite{meierhenrich2004}. Such diamino acids are the source material for the formation of RNA, which, it is generally believed,  made up the first living systems before the appearance of DNA. These works along with others like that of  \textcite{bernstein2006_2} all stress the importance of (inter)planetary and cometary ices as a likely place of formation of such pre-biotic material. We shall discuss this idea further in Section~\ref{life}.

\subsubsection{Laboratory experiments}

Heterogeneous astrochemistry, as detailed above, is currently of enormous interest in astrophysics.  However, there is an urgent demand for data from terrestrial experiments to allow the development of appropriate astrophysical models of molecule and ice formation on interstellar grain surfaces.  With regard to the formation of molecular hydrogen, recent experiments have shown that this process is possible at low temperatures on a variety of interstellar surfaces \cite{creighan2006,islam2007,williams2007,roser2002,vidali2004,perets2005,lemaire2010}.  Indeed, on carbonaceous surfaces, it appears that considerable internal energy is present in the nascent H$_2$ \cite{latimer2008} and perhaps such excitation may be observable spectroscopically.  

Recent results suggest that in the regions of protostars, where the gas is warmer, H atoms can chemisorb on carbonaceous grains. Pertinent to these regions, elegant scanning probe measurements indicate that
H$_2$ formation is rapid on graphitic materials \cite{hornekaer2005,hornekaer2006_1,hornekaer2006_2}. With regard to the synthesis of small molecules on interstellar grains, recent work has indicated, for example, that methanol can be generated from the reaction of CO with H atoms on surfaces at low temperatures \cite{watanabe2006}.

Much work, however, remains to be performed in the laboratory.  Modelers are in urgent need of surface diffusion rates at typical astrophysical temperatures for the accurate prediction of surface reaction rates on water ice and bare grain surfaces \cite{watanabe2010}.  Experiments are required to confirm that H$_2$O and other small molecules can be formed via heterogeneous reactions under pseudo-interstellar conditions, and some of these experiments have recently been performed \cite{hidaka2009,hiraoka2006,cuppen2010,ioppolo2010,dulieu2010,mokrane2009}.  Do such reactions generate molecules with sufficient excitation to desorb from the surface, or are the nascent molecules trapped on the surface and immediately incorporated into an ice layer?  How are the molecules in the ices processed by the cosmic rays and energetic photons present in the interstellar medium \cite{oberg2010,thrower2010,thrower2008}?  What are the morphologies of the water ice and the other ices present in the interstellar gas clouds \cite{collings2006,cuppen2007,cartwright2008}?  Questions of morphology are particularly pertinent; energized nascent molecules generated in surface reactions on highly porous surfaces are likely to thermalize with the surface before they can desorb \cite{hornekaer2003}.  If heterogeneous molecular synthesis generates a porous ice on a dust grain, how does the structure of the ice respond to the processing by photons and energetic particles that will occur in the interstellar medium?

\subsubsection{Desorption of ice-trapped molecules}

Already it seems clear that the desorption of mixed ices, as the dust clouds warm in the vicinity of a newly-born star, is more complex than astrophysicists first pictured.  Studies have shown that the desorption of astrophysically important species from water-rich analogs of interstellar ices does not occur at the single temperature previously assumed in many astrophysical models.  The formation of the icy mantles on the interstellar dust grains in the cold (10--30~K) environment of a dark interstellar cloud involves volatile molecules (e.g., CO, CH$_4$) and water being trapped on the grain surface.  Laboratory experiments show that during the process of stellar evolution, as the dust grains warm, some of the volatile molecules desorb \cite{collings2003,collings2004,burke2010,ayotte2001,malyk2007,galvez2007,barnun1985,hudson1991}.  
However, some volatile species may also diffuse into the pores of the ASW matrix that results from water accretion at the low temperatures of the dense cloud.  The structure and significant surface area of the ASW film gives both a range of suitable binding sites for the volatile molecules and a significant capacity for their retention.  As the water film warms, the collapse of the pores in the ASW results in the trapping of the volatile species in the water ice.  When the temperature of the dust grain increases the ASW may undergo a phase transition to form cubic ice. This transition to cubic ice occurs on time scales that are strongly dependent on the temperature. Upon warming over laboratory time scales, the transition to cubic ice is typically observed at temperatures of 140~K and above. At lower temperatures (10--20~K) the time for this recrystallization to form cubic ice is longer than the age of the universe \cite{schmitt1989}.
This crystallization of ASW upon warming allows the sudden release of the trapped volatile molecules into the gas phase: a so-called volcano desorption \cite{smith1997}.  Further warming eventually desorbs the cubic water ice and any remaining trapped volatiles.  These volcano and co-desorptions release volatile molecules into the gas phase at higher temperatures than had been assumed in simple models of ice mantle evolution.

An experimental survey shows that the desorption of a variety of different volatile molecules under these interstellar conditions can be broadly classified by the nature of their interaction with the water molecules that make up the ASW host \cite{collings2004}.  For example, ``water-like'' species, such as ammonia, bind very strongly to the water matrix and show a single co-desorption when the water ice evaporates,  whereas ``CO-like'' species (e.g., CO and CH$_4$), which interact less strongly with the water matrix, exhibit both volcano desorption and co-desorption, together with lower temperature ``monolayer desorption'' from the surface of the water ice.  These more accurate desorption data have been incorporated into astronomical models of the evolution of hot cores.  Hot cores are the ``clumps'' of warm gas which are remnants of the collapse of a gas cloud to form a nearby high-mass protostar.  Ignition of the protostar warms the surrounding gas and dust initiating desorption of the ice mantles from the dust grains to form the hot core.  Recent models of this desorption in the vicinity of a protostar show that correctly describing the desorption of the trapped molecules, and the water ice mantles, allows the gas-phase abundances of the desorbed species and be used as  a chemical clock, allowing a determination of the age of the nearby young star \cite{williams2007,viti2004}.

It is also interesting to note that as the dust temperature increases and the ice structure changes, the profile of the IR features of the icy species is also modified. This change in the IR profile is considered diagnostic of  the temperature of  the dust. For example, it has been suggested that the profile of the CO$_2$ ice bands observed in some young stellar objects can be reproduced by using different components having different temperatures \cite{ioppolo2009}. This in agreement with the existence of a temperature gradient in the circumstellar environment.

In summary, our understanding of the physics of molecular ices --- water and otherwise --- and their surface chemistry, under conditions relevant to interstellar medium, has developed rapidly in recent years.  However, a wealth of new data are required from the laboratory to improve astrophysical modeling of the physics and chemistry of this fundamental constituent of interstellar space.

\section{Atmospheric Ice}\label{atmospheric}

\begin{quote}
Every crystal was a masterpiece of design  \\
and no one design was ever repeated.

Wilson Bentley
\end{quote}

Like snowflakes, the clouds that generate them exhibit what seems to be endless variability.  They appear to form under a number of different atmospheric conditions and vary considerably in structures and patterns on relatively short spatial scales. Clouds help to regulate Earth's energy balance by scattering and absorbing radiation and by redistributing heat.  They are an essential part of the hydrologic cycle and a prerequisite for life on Earth, delivering the precipitation required for rivers to flow and lakes and glaciers to form. Although individual clouds may be short lived they are an important dynamic variable affecting conditions from the upper atmosphere to the biosphere.

\begin{figure}
\centering\includegraphics*[width=\columnwidth,clip=true]{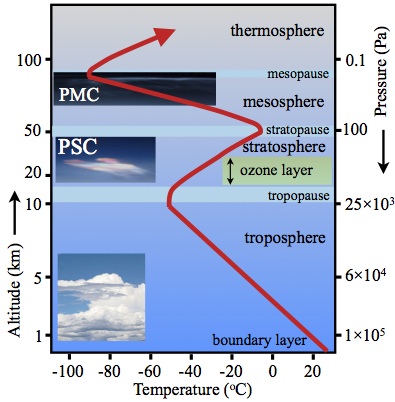}
\caption{(Color online) A schematic diagram of Earth's atmospheric stratigraphy, depicting the locations of icy cloud layers.}
\label{atmostrat}
\end{figure}

Clouds form when water vapor condenses into liquid water droplets, or directly into ice particles. The droplets or ice particles do not form from water alone; instead solid or liquid aerosol particles provide surfaces on which they may condense. These aerosol particles are omnipresent in the atmosphere and can act as cloud condensation nuclei (CCN) when an air mass is cooled to its saturation point.  In order for liquid droplets to nucleate homogeneously from pure water vapor an energy barrier must be overcome.  This energy barrier results from the competition between the free energy cost of adding a liquid surface $G_\mathrm{sur}$ versus the benefit of reducing the bulk free energy of the volume $G_\mathrm{bulk}$.  This competition is most simply treated using the edifice of classical nucleation theory (CNT):
\begin{equation}
G_\mathrm{tot} = G_\mathrm{sur}-G_\mathrm{bulk} =4\pi r^2 \gamma_\mathrm{lv}-\frac{4\pi}{3} r^3 g_\mathrm{v}. 
\end{equation}
Here we express the surface--volume competition in terms of Gibbs free energies of a spherical nucleus with radius $r$, surface free energy $\gamma_\mathrm{lv}$, and free energy per unit volume $g_\mathrm{v}$.  Thus the critical radius of a nucleus is found when $G_\mathrm{tot}$ reaches a maximum, or $\partial G_\mathrm{tot}/\partial r = 0$.  This yields a straightforward expression for the energy barrier which must be overcome for nucleation to occur
\begin{equation}
G_\mathrm{crit} = \frac{16\pi}{3} \frac{ \gamma_\mathrm{lv}^3}{g_\mathrm{v}^2}.
\label{G_crit}\end{equation}

Previously we used this surface--volume competition in the context of the Gibbs--Thomson effect (Section \ref{ssec:grains}), in order to explain the existence of the vein--node network in bulk ice, which is a thermodynamic consequence of the lower melting point of ice due to the curvature of the individual ice grains.  In the atmosphere the same effect, in this context sometimes known as the Kelvin effect, is responsible for a change in vapor pressure over curved surfaces, 
\begin{equation}
p=p_\mathrm{s} \exp\left({\kappa R_\mathrm{crit}}\right)\;\text{where}\;R_\mathrm{crit}=\frac{2 \gamma_\mathrm{lv}V_\mathrm{a}}{k_\mathrm{b}T}.
\end{equation}
Here we express the Gibbs--Thomson relation for a spherical droplet of curvature $\kappa=1/r$ in terms of the vapor pressure $p$ and the saturation vapor pressure $p_\mathrm{s}$.   The critical radius $R_\mathrm{crit}$ distinguishes between small droplets that evaporate and larger droplets that can continue to grow and depends on the atomic volume $V_\mathrm{a}$ and the temperature $T$.  As a system cools $R_\mathrm{crit}$ decreases, ultimately becoming small enough that droplets can homogeneously nucleate.  In CNT, when the energetic barrier to nucleation is overcome, only material diffusion limits the growth of water droplets or ice particles.  Thus the steady-state homogeneous volume nucleation rate can be written as
\begin{equation}
J_\mathrm{hom}(T) = J_\mathrm{o} \exp{\left( \frac{G_\mathrm{diff}+G_\mathrm{crit}}{k_b T}\right)},
\label{J_hom}\end{equation}
where $J_\mathrm{o}$ depends weakly on temperature and $G_\mathrm{diff}$ is the kinetic barrier to growth.  \citet{Vehkamaki2006} is an excellent reference for a complete discussion of CNT. 

At cold enough temperatures water droplets may nucleate and freeze essentially simultaneously.  In other cases, a second but analogous ice nucleation barrier must be overcome for liquid drops to become solid.  In the atmosphere this combination of effects means that in places very high supersaturations can exist and in others micrometer sized water droplets can be supercooled more than 35~K before they spontaneously freeze \citep{Pruppacher1997}.  Soluble impurities and other surface effects may further suppress freezing \citep[cf.,][]{Dash2006} while surfaces of solid particles may promote heterogeneous nucleation at higher temperatures \citep{cantrell2005}.

The majority of cloud formation occurs within the troposphere (Fig.~\ref{atmostrat}) between the Earth's surface and the tropopause.  Within the troposphere surface heating of the boundary layer (typically the first 1000 m) causes buoyant moist air to rise, cool, and mix with the dryer air aloft.  Internal heat sources and other mechanisms of transport, like latent heat release, advection, and aerosol adsorption are also important in influencing this process of rapid mixing.  Eventually clouds form as cooling air masses condense.  Throughout the troposphere these clouds, depending on their temperature and the local atmospheric composition, may contain ice.  Ice and mixed phase clouds are usually formed in cold regions and at high altitudes.  Above the troposphere the atmosphere is dry and clouds only exist when very low temperatures prevail. Polar stratospheric clouds (PSCs) form during wintertime when the temperature decreases to 195~K or lower in the polar regions. The highest clouds form in the polar mesopause region at an altitude of 80 to 90~km, where temperatures may drop to below 140~K during summertime, making it the coldest place in the atmosphere.

The development of a detailed understanding of icy clouds in the atmosphere relies on the combined use of field studies, modeling at a multitude of scales, and laboratory studies that provide a microscopic and molecular-level understanding. Atmospheric ice is studied by remote-sensing methods from the ground, and from airplanes and satellites, using passive spectroscopic and light-scattering methods and active methods like radar and lidar.  In the troposphere, and also, with greater difficulty, in the stratosphere, ice is studied \emph{in situ} using airborne platforms: aircraft and balloons. \emph{In situ} measurements in the mesopause region are achieved with rocket-borne instrumentation and are limited to brief sampling times as the rockets ascend and descend through cloud layers. These various methods typically lack sufficient access to fundamental physicochemical parameters of ice particles.  Furthermore, the representativeness of these types of studies is always an issue because of the transient character of the processes studied.  Off-line analysis of collected samples may clarify some aspects, but usually fails for metastable particles or when aging processes are important, because of sample degradation. In these cases laboratory studies may help.  Selected experiments can be performed under well-controlled conditions to achieve deeper understandings of underlying processes.  Theoretical and numerical models are then required to link laboratory and field studies and to transfer this knowledge into large-scale models using sensible parameterizations of key processes.

Here we focus on methods and open questions concerning the microscopic and molecular-level understanding of ice processes in the atmosphere. We begin with a discussion of experimental and theoretical methods used for detailed studies of atmospheric ices, and proceed to discuss the chemistry and microphysics of ice in the mesopause, stratosphere, and troposphere.

\subsection{Measurement and simulation methods }

\subsubsection{Experimental techniques}\label{atmos-lab}

\begin{figure}[t]
\centering\includegraphics*[width=\columnwidth,clip=true]{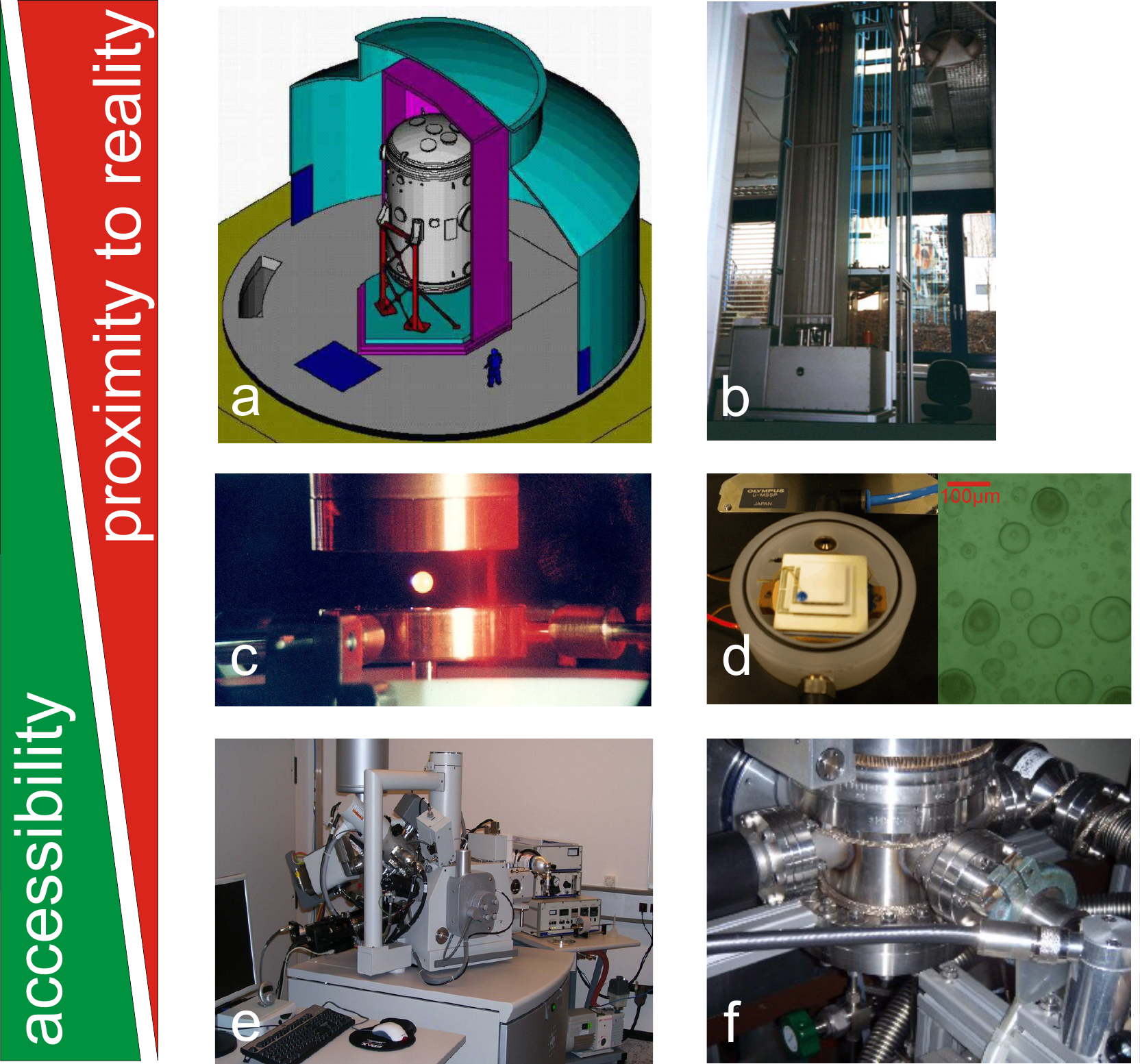} 
\caption{(Color online) Laboratory models sorted by their proximity to the atmospheric reality and the accessibility of physicochemical parameters:
(a) aerosol chamber AIDA built for ice nucleation experiments, 
(b) aerosol laminar flow tube, 
(c) electrodynamic balance with levitated ice particle, 
(d) cryo-chamber with Peltier element suited for light-microscopic observation of the freezing of water droplets in an oil emulsion, 
(e) Environmental scanning electron microscope (ESEM) with cryo-stage and cryo-transfer chamber, 
(f) UHV cryostat working at 6~K suited to isolate transient species in rare gas matrices. 
\label{hinrich1}}
\end{figure}

A multitude of experimental techniques have been developed and adapted to study atmospherically-relevant ice processes.  Broadly speaking, these techniques focus on specific chemical and physical phenomena using well constrained laboratory apparatuses.  An ensemble of such experiments may serve to simulate atmospherically-relevant conditions at a range of scales, from simulating clouds, to single particles, and idealized ice and substrate surfaces.  The benefit of any experiment can be reinforced if used in connection with field studies and modeling expertise (Fig.~\ref{hinrich1}).  Here we provide a brief overview of the range of experimental techniques applied to atmospheric ice.  In Sections \ref{sec:meso}--\ref{sec:tropo} we summarize the results of such experiments, and how they apply to ice within different regions of the atmosphere.

Aerosol chambers and flow tubes are considered to be the most realistic laboratory analogs to nature \citep{Callaghan1994, Khalizov2006, Mohler2006_2, Mohler2008, Stratmann2004, Hartmann2011}.  These relatively large devices contain real aerosol particle ensembles surrounded by gas phases.  Although flow tube designs vary, most seek to input laminar flows of sample aerosols into chambers with temperature and humidity controlled environments.  Often spectroscopic, size distribution, and chemical analyses are then made using the captured flow \citep[e.g.,][]{Khalizov2006}. The 84~m$^3$ cloud simulation chamber AIDA is one important example of a large-scale facility whose large volume makes it possible to carry out detailed cloud experiments extending over several hours \citep{Mohler2006_2}. A representative example is a recent study where the ice nucleation potential of mineral particles was observed to be substantially suppressed if they were coated with secondary organics \citep{Mohler2008}. In related work the ice nucleation ability of mineral dust particles coated with sulphuric acid was investigated at LACIS \citep{Reitz2011}, which is a thermostated laminar flow tube with a length of up to 10~m and residence times up to 60~s \citep{Stratmann2004, Hartmann2011}.

In the laboratory, macroscopic interfaces, micro- or nanoscopic particles, and clusters serve as analogs for atmospheric particles. Such investigations are motivated by the idea that laboratory studies of fundamental physicochemical processes can be used to illuminate complex atmospheric processes that are not easily accessible vis-\`{a}-vis field measurements (Fig.~\ref{hinrich1}). Individual laboratory analogs are each somewhat idealistic representations of reality, but can provide access to important parameters.  Furthermore, in the laboratory experimental conditions can be precisely chosen allowing sophisticated measurement techniques to be applied to outstanding problems. In principle any strategic research of complicated systems requires such simplifying approaches.

Laboratory-scale experimental techniques that focus on particles include levitation trapping \citep{Shaw1999, Swanson1999, Duft2004,Svensson2009}, powder diffraction \citep{Staykova2003}, and oil-matrix emulsions \citep{Murray2005a}. These tabletop-scale experiments allow researchers to focus on the properties of single ice particles or small ensembles of them.  These types of experiments are important for simulating realistic cloud-like environments, where particle interactions and behavior can be investigated. A recent example is high-precision measurements of homogeneous ice nucleation rates in individually charged microdroplets, levitated in an electrodynamic balance, which help to address the question of whether ice nucleation is initiated in the bulk liquid or near the surface of a droplet \citep{Duft2004}. As an alternative, studies may be carried out with a droplet resting on a substrate, which introduces the need to consider the effects of droplet--substrate interactions on the freezing process \citep{Gurganus2011}.

Still other techniques utilize substrate surfaces as particle surface analogs, typically to investigate fundamental surface-specific processes. Many modern surface techniques have been adapted from the fields of catalysis and materials science for atmospheric investigations. These include ultra-high vacuum (UHV) investigations of molecular complexes, clusters, nanoparticles and adsorbed films \cite{firanescu2006}. Examples are the use of secondary ion mass spectrometry to investigate interactions between condensed H$_2$O, NH$_3$, and HCOOH molecules \citep{Souda2003}, molecular beam studies of interactions between water molecules and water ice \citep{Gibson2011}, and the surface-science techniques applied to study molecular-level ice surface properties \citep{Li2007}. 

Several surface-science techniques that originally required UHV have also been adapted to higher experimental pressures. Examples include electron microscopy, mass spectroscopy, reflection IR spectroscopy, X-ray photoelectron spectroscopy (XPS), and molecular beam experiments. XPS and electron yield near edge X-ray absorption fine structure (NEXAFS) have recently been used to probe ice surfaces in the presence of HNO$_3$ at 230~K (\textcite{Krepelova2010a}; see also Section~\ref{snow_chem}), and molecular beam methods have been used to study ice nucleation in the deposition  mode at 215~K \citep{Kong2011}. These techniques continue to open new pathways for research in atmospheric science, particularly for models of particles in the upper troposphere. 
Light-scattering techniques are used to determine ice layer thickness \citep{Brown1996} and surface phase-transition behaviors \citep{Elbaum1993a,Thomson2009b}. In an exciting new development, laser confocal microscopy combined with differential interference contrast microscopy have been used to visualize the dynamic behavior of individual molecular layers at the air--ice interface \citep{Sazaki2010}.  At times multiple techniques, such as light microscopy and molecular beam measurements, can be applied coincidentally \citep{Andersson2007,Suter2007,Lejonthun2009,Kong2011}. For example, light reflection measurements can be made simultaneously with molecular beam scattering measurements to constrain adsorbate thickness and fractional coverage on substrate materials \citep{Thomson2011}. 

Comparing laboratory investigations of atmospherically relevant ice processes to their real world counterparts is particularly challenging.  In the atmosphere, temperature and partial pressures of important trace gases may have quite heterogeneous distributions and are susceptible to constant and sudden changes.  Often ice and aerosol particles are not in equilibrium with their environment, resulting in supersaturated gas phases, supercooled droplets, and inherently metastable particles.  Such common non-equilibrium conditions make it difficult to identify standard model substances and to compare results between different laboratories and investigators.  In many cases essential material parameters like structure, morphology, surface tension, density, viscosity, and dissociation equilibria vary owing to the non-equilibrium conditions.  This variability makes it particularly important for researchers to utilize the full array of laboratory techniques available to them.  Repeated measurements of simple systems allow for the most robust comparisons between experimental methods and observed phenomena.

\subsubsection{Theoretical methods}\label{atmos_theo}

Numerical simulation methods have dramatically improved the molecular-level understanding of many atmospheric phenomena. Like experimental design, theoretical calculations strive to simulate reality accurately.  Current state-of-the-art molecular-level models of atmospheric ice reaction dynamics utilize the full spectrum of molecular-dynamical, particle-interaction, and turbulence models.  Here we summarize the techniques most often used to simulate numerically the micro-physical processes associated with atmospherically relevant ice.  Such interaction models have largely focused on simulating conditions of the upper troposphere and stratosphere due to the global importance of chemical processes like ozone depletion \citep{lowe2008,molina1987}.  

Molecular-dynamics (MD) simulations that use field potentials are often referred to as ``classical" MD in chemical physics and are useful tools in ice research.  For example, using this class of models, preferential adsorption sites for molecular HCl on ice surfaces \citep{buch2002,devlin2002} or the sticking probability of HCl molecules on ice \citep{alhalabi1999,alhalabi2001} can be simulated.  Quantum-mechanical (QM) calculations are also very powerful and can be used for studying minima on potential energy surfaces (PES) but are more computationally expensive.  The advantages of both classical and quantum-mechanical calculations have been hybridized into techniques called quantum mechanical molecular mechanics (QMMM). These techniques have been applied by different groups to the problem of HCl solvation and ionization on and in ice surfaces  \citep{clary1997,estrin1997,svanberg2000}.  Developments of ab initio molecular dynamics (AIMD) computational techniques allow the dynamical aspect of structures and accompanying properties, such as the infrared spectrum, to be obtained from first-principles calculations \citep{car1985}.  \citet{mundy2006} provide a thorough review of first-principles theory and calculations and their application to atmospherically relevant interfaces.

Solvation and ionization of strong acids like HCl, HNO$_3$ and H$_2$SO$_4$ has become a particularly important field of atmospheric research because it strongly affects ozone depletion.   Furthermore these acids and others are known to be important chemical constituents of clouds in the stratosphere and the troposphere \citep{zondlo2000}.  Although the solvation and ionization of HCl have been studied theoretically in considerable detail, the same is not true for other strong acids such as HNO$_3$ and H$_2$SO$_4$. For these compounds some progress has been made using AIMD implementing electronic structure codes \citep{vandevondele2005}.  \citet{balci2011} studied HNO$_3$ solvation and ionization on and in water clusters and crystalline ice slabs.  Using a Kohn-Sham formulation of density functional theory (DFT) and the Gaussian plane wave method \citep{goedecker1996} samplings of the Born-Oppenheimer surface were calculated for each time step. This methodology is computationally advantageous, allowing for simulations with thousand of atoms by using first principles calculations \citep{mundy2006}.  However, as a consequence the simulations use relatively low level of electronic structure theory and classically treat the nuclear dynamics.

\begin{figure}[t]
\centering\includegraphics*[width=\columnwidth,clip=true]{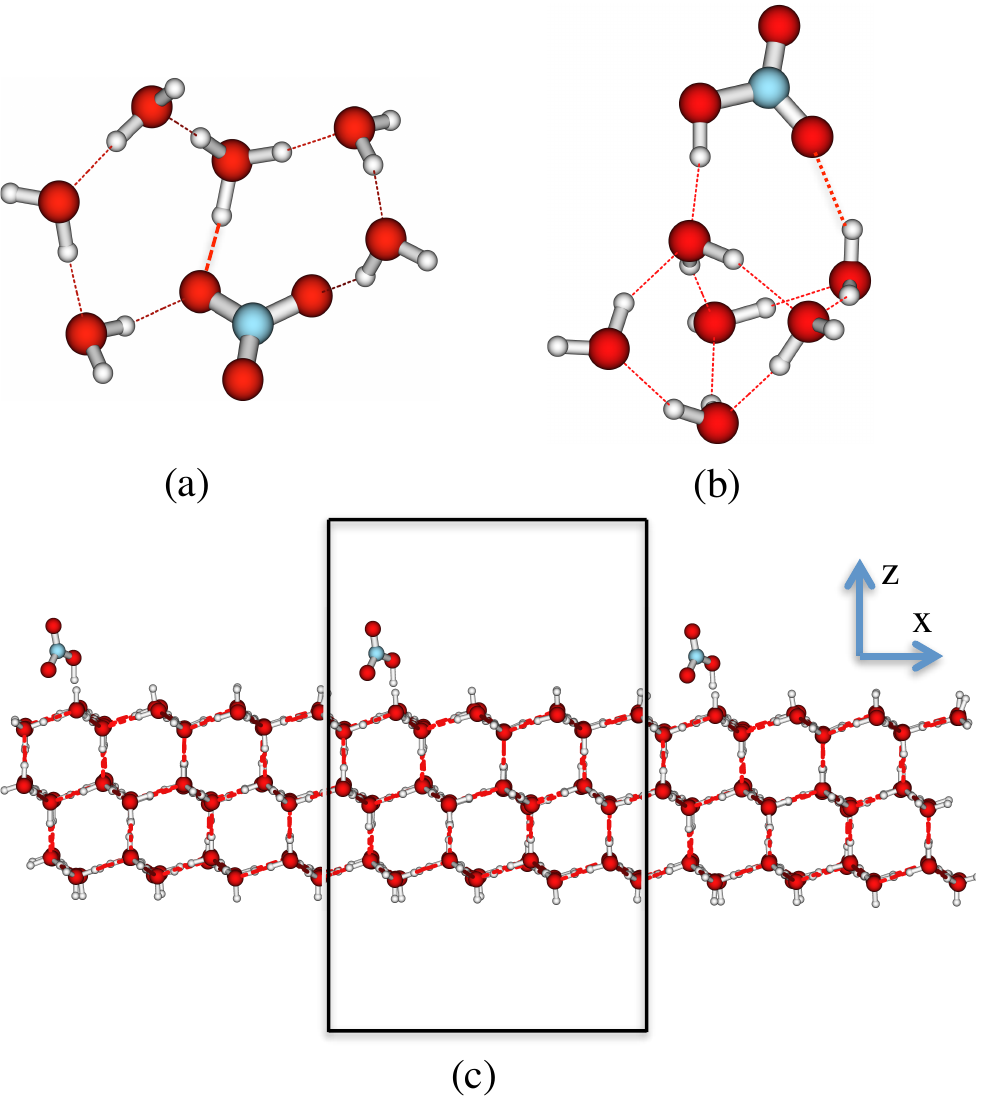} 
\caption{(Color online) Nitric acid solvation by water molecules: (a, b) HNO$_3$(H$_2$O)$_6$ clusters in two different configurations and (c) a crystalline cubic ice slab with one HNO$_3$ in the $x$--$z$ direction.
\label{nevin}}
\end{figure}

Uptake of HNO$_3$ on ice particles may play an important role for the NO$_x$ chemistry in the upper troposphere and lower stratosphere \citep{Leu2009}, and a detailed understanding of HNO$_3$--ice interactions is required to model HNO$_3$ uptake processes \citep{Karcher2009} and to constrain HNO$_3$ redistribution and removal mechanisms \citep{Scheuer2010}. Combined AIMD and DFT methodologies have been used to study single HNO$_3$ molecules interacting with up to eight water molecules  \citep{buch2007}. This is a model system for understanding microscopic solvation scenarios of HNO$_3$.  Water molecules in the clusters are linked either in chains forming a 2D network with respect to heavy atoms or a 3D network, as shown in Fig.~\ref{nevin}(a) and (b). Although the chains may not be representative of the condensed phase, the 3D network should be thought of as part of a condensed bulk. A periodic crystalline hexagonal ice slab containing 72 water molecules is illustrated in Fig.~\ref{nevin}(c).  Different adsorption positions for HNO$_3$ on the surface of the crystalline ice slab may be chosen for calculations run at temperatures below 130~K, and the infrared spectra of the clusters may be obtained from the Fourier transform of the dipole--dipole correlation function.  \citet{urasaytemiz2006} and \citet{buch2008} have extended this methodology to other condensed phase systems. An advantage of this approach is that anharmonicities of the system, like the proton sharing or transfer system, can be ignored because these effects are automatically included.

Numerical simulations of atmospherically relevant ice processes continue to improve as methods become more sophisticated and computation becomes faster.  Further advances will be made as the fundamental properties of ice and water (cf., Section~\ref{structures}) like density, heat capacity, and the boiling and freezing points are more accurately predicted using numerical models.  Numerical simulations are particularly convincing when coupled with validating experimental measurements, and it is only by fusing theory, simulation, and experiment that fundamental physical processes can be illuminated.

\subsection{Ice clouds in the mesopause region}
\label{sec:meso}

The highest clouds in the atmosphere form in the polar mesopause region, which can cool to below 140~K during the summer. The clouds appear as  narrow layers at altitudes between 80 and 86~km and consist of cloud particles with maximum sizes between 50 and 100~nm.  Because they become visible to ground observers after dusk when the high-altitude clouds remain sunlit they are known as noctilucent (``night-shining'') clouds.  When observed from satellite the clouds are known as polar mesospheric clouds (PMCs) \citep{Russell2009}. The historical record of their observations goes back to the 1880s, and their presence has been linked to climate change \citep{Fogle1966}, although there is ongoing debate on this point \citep{vonZahn2003, Thomas2003}.  Climate models predict that increasing greenhouse gas emissions should result in cooling of the mesosphere where methane oxidation leads to higher water concentrations, and both effects may contribute to an increased cloudiness \citep{Thomas1989, Thomas2001}.
 
The mechanism for ice-particle formation in the mesopause region is not yet fully understood. One hypothesis is that water condenses on existing aerosol particles. Meteoroids that enter the atmosphere vaporize as they reach the thicker atmosphere below an altitude of 120~km, and are visible as meteors from ground.  Vaporized material may recondense into nanometer-sized particles and serve as CCN within the mesopause region \citep{Bardeen2008}.  Although such dust particles have not yet been proven to exist, they are at present a likely hypothesis for the outcome of meteoroid vaporization, and modeling studies suggest that condensation on dust particles is likely \citep{Rapp2006}.  A recent study suggests that negatively-charged dust particles could play a significant role in the nucleation process \citep{Gumbel2009}.  A second hypothesis is that cloud particles are formed by ion-induced nucleation \citep{Sugiyama1994}. In the sunlit upper atmosphere, photo-ionization generates NO$^+$ and O$_2^+$ ions. If H$_2$O molecules condense on the ions and, by intra-cluster reactions, form a population of H$^+$(H$_2$O)$_n$, under favorable conditions these may grow into visible ice particles.  Still other recent work turns these arguments on their heads, returning to the idea that water may simply nucleate homogeneously under the conditions prevailing in the cold mesopause \citep{Murray2010a}.  Clearly, the interactions between water, dust, and ionic particles under the prevailing mesospheric conditions of low temperatures and pressures must continue to be studied. 

Further complicating our understanding of PMCs is that the ice particles in the mesopause exist in a weak plasma and are influenced by charged particles in the surrounding gas.   This may influence any formation process in a significant way. The evaluation of the ion-induced nucleation mechanism requires detailed knowledge of a large number of individual reaction steps that are required to model the formation of cloud particles, some of which are difficult to study experimentally. One example is the dissociative recombination (DR) process in which free electrons combine with water cluster ions \citep{Nagard2002, Ojekull2007, Ojekull2008, Thomas2010}. The neutralization process leads to energy release and partial or complete disintegration of the cluster; a process that competes with particle growth and may hinder the formation of large ice particles. It has only recently been possible to study the process in detail by applying methods usually used in accelerator physics.  Absolute rate coefficients for the DR process and the products of H$^+$(H$_2$O)$_n$ ($n = 1$--6) formed during DR  have been determined by the use of a heavy-ion storage ring \citep{Ojekull2007}.  Those results help to set stricter limits on models of ice-particle formation in the mesopause region.

Other open questions relate to the morphology and structure of the PMC particles. In particular, what is the molecular structure of PMC particles?  It is possible, but unlikely that they are amorphous, owing to their slow growth over tens of hours by water deposition and rare particle--particle collisions.  However, as discussed in Section~\ref{structures}, ice is only slowly transformed from amorphous to crystalline ice around and below 130~K.  Surface-specific effects may also play an important role as the particles grow from molecular scales to diameters of tens of nanometers. 

Under mesospheric conditions particle--particle collisions take place only rarely, raising a number of questions regarding particle growth and aggregation. Will particles formed by collisions merge into spherical particles? On what time scales do nanoparticles reorganize internally, and how is their shape influenced by further condensation of water? What is the effect of impurity components that may be incorporated into the growing ice particles?  These include a range of metals that are found within meteoroids.

Evidence suggests that gas uptake and heterogeneous processes on the surface of ice particles change the mesopause region's chemical composition. Measurements have shown that an element like sodium \citep{Thayer2006} is depleted when clouds are present. The same effect has also been observed for atomic oxygen \citep{Gumbel1998}, and model studies indicate that ice particles act as a source for odd hydrogen species that react with atomic oxygen and reduce the concentrations \citep{Murray2005b,kulikov2010}.

\subsection{Polar stratospheric clouds}
\label{sec:strato}

Lower in the atmosphere, polar stratospheric clouds (PSCs) are formed during wintertime when the air at an altitude of 15--25~km cools below 195~K. Polar stratospheric clouds are crucial to our understanding of polar ozone destruction. They allow for ozone depletion through denitrification of the stratosphere, and through the catalytic production of active chlorine species that subsequently attack ozone in early polar spring \citep{Hanson1991, Hanson1992, Leu2009}.

PSC particles consist of binary and ternary mixtures, mainly of H$_2$SO$_4$, HNO$_3$ and H$_2$O, but thorough analyses of their chemical compositions are difficult to perform \cite{lund2005}. Furthermore, the exact nucleation mechanisms of PSCs are still unknown \cite{tolbert2001}. In general it is uncertain which hydrates are in fact present in the aerosols. Several attempts have been made to perform in situ analyses of the chemical composition, phase composition and the surface structure of PSC particles, leading in some cases to contradictory results \cite{toon1995,voigt2000}. The metastability of particles causes particular problems for such measurements and the exact composition of supercooled liquids and metastable crystalline particles remains to be determined. In order to assign chemical spectra, precise optical indices of the compounds and phases in question are needed. However, the morphology, density, and dissociation equilibria are also important parameters for convincingly interpreting the gathered spectroscopic information.

In the stratosphere, acids (HNO$_3$, H$_2$SO$_4$, and HCl) form an array of hydrates and might undergo solid--solid phase transitions \cite{martin2000}. The related solids are polymorphs that exhibit a characteristic number of modifications. However, solid--solid phase changes have never been observed in situ in field experiments. The only hydrate determined so far is the nitric acid trihydrate (NAT) analyzed by space-borne infrared limb emission measurements of the Michelson Interferometer for Passive Atmospheric Sounding (MIPAS) on EnviSat \cite{hoepfner2006}. NAT exhibits a high- ($\beta$) and a low-temperature ($\alpha$) modification. The latter is metastable and is only known from laboratory experiments. The same is true also for both metastable modifications of nitric acid dihydrate (NAD) and for cubic ice (Ic). The nucleation rates \cite{koop2000}, phase transitions \cite{tizek2004}, transition kinetics \cite{grothe2006a}, and morphologies \cite{grothe2008} of these hydrates have primarily been studied in laboratories and the results are used in PSC microphysical models. Direct observation in the field is still hampered by the lack of suitable detection techniques. In principle, X-ray diffraction would be the method of choice but it is unsuitable in the field due to the sampling geometry. Mid-IR spectra have already proven to provide important results for understanding ice-cloud compositions. However, the interpretation of the spectra depends on high-quality reference spectra from the lab and has to take into account certain interlinked effects like sample texture (orientation), sample thickness, crystal morphologies, and phase composition. In principle, low-frequency Raman and/or terahertz spectroscopy for observing lattice mode vibrations would be the most appropriate methods. Recently, low-frequency Raman data have been measured in the lab \cite{escribano2007,grothe2006b}, but far-infrared spectroscopy data --- i.e., the THz range --- are still missing. THz observations are impossible from the ground due to water-vapor absorption lines, but might be obtainable directly in the upper atmosphere or from space. New missions are being deployed with the observation and analysis of PSCs, and also of PMCs, as part of their scientific objectives \cite{kadosaki2010}.

A specific PSC problem relates to the dissociation of acids in complexes with few water molecules, in environments that could be found in mixed icy particles. This is particularly important for sulfuric acid and nitric acid, which are believed to nucleate homogeneously from the gas phase \cite{givan2002,rozenberg2009} and potentially include additional trace gases in the nucleation process \cite{laaksonen1995}. Just how many water molecules are necessary to facilitate the dissociation of an acid? Several experiments and calculations have been carried out to determine whether a few molecules or larger clusters and condensed phases are required. These include molecular simulations of the nucleation process of sulfuric acid \cite{hale1996,kurdi1989,kusaka1996}. \textcite{arstila1998} presented a density functional study of sulfuric acid hydrate complexes containing 1 to 3 water molecules and were able to show that the proton transfer reaction is unlikely to occur for the mono- and dihydrate complexes. For the trihydrate complex the energy barrier is rather low. However, despite these theoretical efforts, proton transfer has not been observed and the mechanisms of solvation and ionization of sulfuric acid --- and for nitric acid too --- remain unclear owing to proton tunneling effects and the complexity of solution-phase reactions.

Recent AIMD simulations provide a detailed description of HNO$_3$ interactions with water ice \cite{balci2011}, but the influence of temperature and related effects on the dynamics of the system remain to be determined. Although the different hydrates of HNO$_3$ have been studied experimentally, computational studies are scarce. The hydrates themselves are interesting systems for exploring the dynamics of ions in crystalline environments. Furthermore, most theoretical studies of ice surface chemistry in PSCs have concerned the water-ice surface, and supplementary work should address the corresponding processes on NAT surfaces, which by comparison are poorly studied \cite{mantz2002}. Because the hydrates of HNO$_3$ are dominated by ions (NO$_3^-$ and H$_3$O$^+$) first-principle molecular dynamics simulations should be suitable to attack this problem. In addition, there is a general lack of information about ternary systems, which are likely to be present in atmospheric ices. One interesting question concerns the interactions of co-adsorbed HNO$_3$ and HCl on ice. Recent theoretical studies on HNO$_3$--HCl--H$_2$O clusters  \cite{gomez2009,gomez2010} could provide the seed to investigate systems of greater complexity leading into amorphous solids. 

Although our understanding of PSC particle microphysics and chemistry is improving, the empirical relationships currently used to describe ice particle microphysics and heterogeneous chemistry in models are based on incomplete data for one or more reaction steps. This drawback, combined with the uncertainties in atmospheric dynamics, causes the modeling of the propagation of PSC effects at global scales to remain of limited accuracy, in spite of the improvements implemented in the models \cite{kirner2010}.

\subsection{Ice-containing clouds in the troposphere}
\label{sec:tropo}

\subsubsection{Ice morphology}\label{snowflakes}

\begin{figure}[t]
\centering\includegraphics*[width=\columnwidth,clip=true]{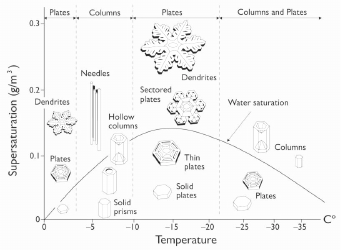}
\caption{The Nakaya diagram \cite{Nakaya1954} showing snowflake morphology as a function of supersaturation and temperature  \cite{libbrecht2005}.
\label{morph1}}
\end{figure}

The size and shape of tropospheric ice crystals --- cloud particles and snowflakes --- varies with atmospheric conditions, and in turn the ice morphology has a strong impact on its geophysical effects, from radiative properties to its tendency to agglomerate.  As demonstrated by Fig.~\ref{morph1}, crystals can have clean geometrical shapes such as hexagonal plates and columns, or more complex geometrical shapes like bullet rosettes and dendrites, or fractal-like configurations \citep{lynch2002}.  \citet{Nakaya1954} systematically mapped out the complex dependence of ice crystal habit as it varies with undercooling temperature and supersaturation (see \textcite{ball2004} for a good summary of the fascinating history of snowflake research). Plates and dendritic snowflake crystals are observed between 253 and 263~K.  Below 253~K habit is dominated by columnar crystals in addition to plates  where the particular morphology depends on supersaturation.  Above 263~K crystals form columns or needles before transitioning back to plates and dendrites close to the melting temperature.  Field and laboratory studies have continued through to the present in an attempt to assess more accurately the effect of varying thermodynamic parameters and the process of nucleation \citep[e.g.,][]{furukawa1997_2,bailey2004}.  Additionally, the effect of different types of seed aerosols on ice properties has been investigated \citep{bailey2002}.  Field studies suggest that columnar polycrystals dominate the cold upper regions of cirrus clouds, while planar ice crystals are more common in the lower warmer regions \citep{heymsfield1984, noel2002, noel2004, noel2006}.  There is also a considerable amount of work being done on the growth of ice crystals from the vapor phase \cite{libbrecht2005}.

Although much is known about how ice morphology depends upon thermodynamic parameters, linking this behavior to large-scale processes is often complex.  Clouds are dynamic and may contain irregularly shaped particles consisting of faceted polycrystalline particles or smooth sublimating particles, as has been observed in Arctic cirrus and stratiform clouds \citep{Korolev2000}.  Icy clouds can be highly inhomogeneous over a range of length scales and regions with high concentrations of small ice particles can be interspersed with regions of large particles \citep{lawson2001}. Much remains to be learned on the level of individual clouds, including the importance of particle--particle interactions.  Such interactions are already known to be significant for thunderstorm electrification \citep{vonnegut1994, dash2003,Dash2006,saunders2008, cartwright2010_2} but may also have other important chemical and physical effects.

\subsubsection{Ice nucleation in the troposphere}

Homogeneous freezing of the micrometer-sized droplets formed in the atmosphere usually takes place around 237~K. Although this process has been extensively studied, outstanding questions remain concerning the mechanism for homogeneous nucleation of ice.  Laboratory studies of homogeneous nucleation have primarily applied the concepts of CNT, which assumes that nucleation takes place within the volume of the droplet.  However, recent studies suggest that even for homogeneous nucleation the presence of surfaces cannot be ignored.   In contradiction to CNT, \citet{tabazadeh2002} suggested that homogeneous nucleation is initiated close to surfaces. There is an ongoing debate on the validity of these mechanisms and the currently available data may not be sufficiently detailed to distinguish between the two hypotheses \citep{tabazadeh2005, stetzer2006, mohler2006}.

According to CNT the freezing of a solution consists of two
consecutive steps: the formation of nuclei and the growth of
these nuclei into crystals (see Eq.~\ref{J_hom}). Only if the rates of both are sufficiently
high can the respective phase transition occur. If more
than one solid phase can nucleate under a set of given conditions,
then kinetics rather than thermodynamics controls the
phase transition. Thus, the phase with the lowest activation
barrier --- see Eq.~\ref{G_crit} --- will dominate the respective nucleation process
depending on the surface free energy and on supersaturation 
\citep[cf.,][]{tizek2002}. This explanation is an example of Ostwald's
step rule, which can also be visualized via locally ordered
structures in disordered systems: If this short-range order
also appears in one of the possible ordered crystalline systems,
then this crystallization process will preferentially occur
since the surface free energy stays low.
This is particularly true for systems containing hydrogen-bonded
constituents.

Several types of aerosol particles and dissolved components can act as ice nuclei above the homogeneous freezing temperature.  These can initiate heterogeneous freezing, which can be understood with a simple modification to the theoretical treatment of steady-state homogeneous nucleation,
\begin{equation}
J_\mathrm{het}(T) = n_\mathrm{s} J_\mathrm{o} \exp{\left( \frac{G_\mathrm{diff}+\Phi_\mathrm{het}G_\mathrm{crit}}{k_b T}\right)}.
\end{equation}
In this description of steady-state heterogeneous nucleation, $0<\Phi_\mathrm{het}<1$ quantifies the fractional reduction to the thermodynamic barrier to nucleation due to a change in material properties, external surfaces, etc., that may favor nucleation, and $n_\mathrm{s}$ is the number density of water molecules in contact with the catalyzing surface.  Heterogeneous freezing is traditionally categorized in four ways: deposition (ice deposits directly onto an ice nucleus), condensation (liquid water condenses and subsequently freezes), immersion (a solid particle within a cooling droplet initiates freezing) and contact (surface contact between a liquid and substrate results in freezing) \citep{cantrell2005}. Each freezing mode is active over a typical range of temperature and relative humidity (RH) \citep{Meyers1992, DeMott2010}. In particular, contact freezing is active at high heterogeneous freezing temperatures \citep{Pitter1973, vonBlohn2005}.  The freezing temperatures of evaporating droplets have been elevated up to four degrees as immersed solid particles come into contact with the surface \citep{Durant2005}.  The contact freezing process has been observed to be even more efficient when aerosol particles approach the liquid surface from the gas phase rather than from within the liquid \citep{cantrell2005}.  In another study of contact freezing on a silicon surface covered with different silanes, freezing was observed to occur preferentially at the contact lines where the solid, liquid, and air meet \citep{Suzuki2007}, while a recent study showed no preference for nucleation at the contact line \citep{Gurganus2011}. Further challenging simple understandings of phase behavior are new results identifying heterogeneous nucleation from glassy solution droplets \citep{zobrist2008, Murray2010b}.  

For the present, atmospheric ice researchers continue to build predictive fundamental physical models of nucleation processes.  A strong lattice match between ice and its substrate may be a good indicator of how effectively a given substrate material promotes ice nucleation, but it is not necessarily predictive \citep{saunders2010}.  Substrate defects that lead to surface charge effects \citep{Park2010}, the types of bonds formed by adsorbing water molecules, and the types of exposed substrate atoms all effect heterogeneous nucleation. The relative importance of these factors remains an important area of study.  

Biological materials serve as particularly interesting tropospheric ice nuclei, especially because they represent an important but elusive cryosphere--biosphere interconnection that may be strongly influenced by climate \cite{poschl2010}.  Airborne terrestrial biological particles like bacteria, lichen, and fungal spores \citep{szyrmer1997}, pollen \cite{pummer2011} and lofted marine species like diatoms \citep{Knopf2010, Alpert2011} can play a role in ice nucleation at altitudes of up to 70~km \citep{imshenetsky1978}.  However, the specific nucleation mechanisms involved are poorly understood and are complicated by the knowledge that biological materials can just as easily slow \citep{watanabe1995} or prevent \citep{Pertaya2007a, Pertaya2007b} ice-crystal growth.  The actual effect is unique for each type of biological material and can be sensitive to ice surface orientation, structure, and system thermodynamics. 

A recent airborne field study analyzed submicrometer ice-nucleating aerosol particles at an altitude of up to 8~km using aerosol mass spectroscopy \citep{pratt2009}. This study found that an important fraction of the ice-crystal residues in cirrus clouds consists of biological material.  This surprising result indicates that biological material may play an important role for the radiation budget of the atmosphere and has motivated more detailed investigations.  Aerosol-chamber experiments (AIDA) have been carried to investigate the ice nucleation efficiency of biological particles in clouds \citep{mohler2007}, and laboratory phase-composition studies of simplified systems have demonstrated that organic carbon proxies affect the nucleation kinetics \citep{murray2008}.  Continuing studies are required to relate the properties of relevant bacteria, pollen, marine diatoms, and other biological particles to their ice nucleation activity and to obtain a fundamental understanding of the processes in action.

\subsubsection{Cirrus clouds and the supersaturation puzzle}

Ice-containing clouds form in the cold regions of the troposphere, particularly the upper troposphere (UT) as warmer moist air is forced aloft.  As we have previously discussed, ice nucleation may occur either homogeneously from aqueous solution droplets \citep{koop2004}, or heterogeneously on aerosol particles \citep{cantrell2005}.  Once nucleated, ice particles continue to grow when the water vapor pressure is elevated and excess vapor condenses on existing particles.  The resultant decreasing vapor pressure allows the ice supersaturation to relax towards its equilibrium value.  In the UT and lower stratosphere (LS), homogeneous freezing is estimated to set in at a supersaturation of approximately 60\% \citep{koop2000}, while lower supersaturations may be sufficient for heterogeneous nucleation.  However, recent field studies suggest that we do not fully understand the conditions under which cloud particles form and grow in the UT \citep{Gao2004, Jensen2005, Lee2004}. Excessively large supersaturations have been observed using aircraft, balloon, and satellite measurements  \citep{ekstrom2008}.  Cloud-free regions with persistent supersaturation levels above 100\% have been observed \citep{Jensen2005} while supersaturations of 30\% have been measured within natural clouds and contrails \citep{Gao2004}.  Respectively, these values are above and below typical values we would expect for ice nucleation and cloud formation, suggesting our understanding is still incomplete. 
 
How can extreme supersaturations persist in cloud-free regions and why do ice particles not grow according to expectations?  Several explanations have been proposed to solve this climatically important ``supersaturation puzzle''  \citep{peter2006}.  The composition of aerosol particles might inhibit ice nucleation and ice particle growth. The 60\%  homogeneous ice nucleation threshold was established for salt solutions and sulfuric acid, but cloud-chamber data indicate that aerosols containing only organic and elemental carbon may almost completely inhibit ice nucleation \citep{Mohler2005}. Low ice-crystal numbers and high cloud humidity may also result from the presence of glassy aerosols rich in organic compounds \citep{Murray2010b}.   However, the mechanism of ice nucleation from such glassy droplets remains incompletely understood.  Other impurity species such as nitric acid  \citep{Gao2004} and sulfates \citep{Murray2007} can also effect the growth of ice crystals, and more laboratory studies need to be devoted to their freezing.

Previously, it had been assumed that all water molecules colliding with ice crystals would be incorporated into the solid lattice. However, recent laboratory studies suggest that we do not fully understand the basic interactions between water molecules and pure ice surfaces.  Laboratory data indicate that less than 10\% \citep{Pratte2006} or even less than 1\% \citep{Magee2006} of colliding water molecules are trapped by the solid surface.  The implication is that most molecules are trapped in a precursor state \citep{sadtchenko2004} on the ice surface and desorb before being incorporated into the lattice.  Recently, it has been shown that surface disorder begins to develop on the topmost layers of ice at temperatures as low as 180~K, and thus ice surfaces are potentially disordered under all conditions prevailing in the UT/LS \citep{suter2006}.  The effect of this surface disorder on the water accommodation is not currently known.

Experimentally observed equilibrium vapor pressures of supercooled water and ice show variations \citep{murphy2005} and \emph{in situ} atmospheric measurements are non-trivial to perform \citep{peter2006}.  However, observed supersaturations are too large to be explained solely by such uncertainties. The kinetically preferred formation of cubic ice (Ic) below 200~K may partially contribute to the explanation \citep{Demott2003, Murray2005a} because the equilibrium vapor pressure for cubic ice is about 10\% higher than that of the more stable hexagonal ice \citep{shilling2006}.  Other explanations may involve the large-scale dynamic development of clouds.  Conditions within clouds can vary faster than measurements can resolve, which may cause apparent supersaturation events due to the spatial or temporal averaging \citep{peter2006}.  On the other hand, observations of lower than expected ice-crystal concentrations are in agreement with measured high supersaturations \citep{kramer2009}.  This introduces a ``nucleation puzzle'', putting a different note on the open questions regarding supersaturation. Clearly, we lack understanding of the details of the freezing process of water at low temperature under atmospheric conditions. Not only it is not clear which type of ice is formed --- anything between ice Ih and a fully cubic ice Ic appears to be possible --- but the whole process leading to the observed low ice-crystal numbers is not fully understood.

The role of Ic, and its effect on supersaturation, is an open problem that deserves specific attention. It has been proposed that snow crystals generally begin with a nucleus of cubic symmetry before they transform into the well-known hexagonal forms \citep{kobayashi1987}. In particular, at large undercoolings the Ic form appears to have lower activation energy; i.e., the formation of Ic can be considered a manifestation of Ostwald's step rule.  Evidence for Ic in the atmosphere has been claimed from halo observations \citep{whalley1983, riikonen2000} but remains inconclusive \citep{weinheimer1987}.  However, laboratory observations suggest that, for atmospheric pressures, Ic is crystallized at temperatures below 200--210~K  \citep{kuhs2004, Murray2005a}.  There is no indication that this should be different in the atmosphere itself. Rather, the question is: what are the morphologies of Ic formed by homogeneous or heterogeneous nucleation for a wide range of atmospheric conditions \citep{hansen2008_1, hansen2008_2}?

The latter work clearly shows that the formation of ice Ic does not imply a cubic crystal morphology; rather, the expected morphology is trigonal based on the internal crystallographic symmetry of stacking-faulty ice Ic. The rare atmospheric optical phenomenon of Scheiner's halo \cite{petrenko1999} requires ice crystals of octahedral morphologies; thus a truly cubic crystal symmetry. Such crystals may indeed be much less frequent than ice crystals of trigonal shapes observed in cirrus clouds at low temperatures \citep[cf.][]{Heymsfield1986}. Such trigonal morphologies may well be indicative of the presence of ice Ic, yet this needs to be established more clearly by further investigations.  

The interaction of water with aerosol particles and ice particles in the UT/LS is a fairly open issue, which needs continued study due to its climatic importance. The currently available laboratory data and field observations are a good beginning but remain limited.  Further laboratory studies and \emph{in situ} atmospheric observations, combined with better molecular-level modeling of water interactions with surfaces \citep{kramer2009}, will be key factors for resolving these questions.

\subsubsection{Tropospheric chemistry}

In addition to their effects on the radiation budget, ice clouds also influence the chemical composition of the atmosphere. Cloud particles in the stratosphere are now well-known to play an important role in the chemistry of the stratosphere. In a similar way, cirrus clouds are likely to influence chemistry in the UT.  \citet{solomon1997} pointed out that the presence of high cirrus clouds near the tropical tropopause could decrease ozone in the stratosphere by activating chlorine constituents.  Nitric acid hydrates also form in the troposphere by nucleation on pre-existing ice particles \citep{voigt2003} or on meteor ablations \citep{voigt2005, curtius2005}.

At lower altitudes, ice chemistry is closely related with the snow chemistry to be discussed in Section~\ref{snow_chemistry}. Interesting research questions relate to the effects of the quasi-liquid layer that exists on ice below the melting point; see Section~\ref{ssec:grains} and references therein.  Trace gas adsorption and reactions on ice have been treated by \textcite{abbatt2003}, and open questions relating to gas uptake on ice and other surfaces have recently been covered by \textcite{Kolb2010}.  Several other important open questions relate to understanding organic chemistry on ice particles and the effects of organic compounds on ice processes \citep{Hallquist2009}. For example, the reversible uptake of small organic compounds on ice has been characterized in flow-tube experiments \citep{Winkler2002, vonHessberg2008} and the results have been implemented in models for gas uptake on cirrus cloud particles \citep{Marecal2010}.

The great number of open questions concerning the properties and action of clouds in the atmosphere introduces large uncertainties into models of the Earth climate system \citep{Lohmann2005}, and our limited scientific understanding is a major problem when we attempt to predict changes in climate over decades and longer. There are urgent needs for further studies that contribute to a detailed understanding of atmospheric ice on the microscopic and molecular levels.

\section{Terrestrial Ice}\label{terrestrial}

\begin{quote}
The abrupt sides of vast mountains were before me; the icy wall of the glacier overhung me; a few shattered pines were scattered around; and the solemn silence of this glorious presence-chamber of imperial nature was broken only by the brawling waves or the fall of some vast fragment, the thunder sound of the avalanche or the cracking, reverberated along the mountains, of the accumulated ice, which, through the silent working of immutable laws, was ever and anon rent and torn, as if it had been but a plaything in their hands. 

Mary Shelley, \emph{Frankenstein}
\end{quote}

The Earth's surface snow cover strongly affects our lives on a local and on a global scale. It is subject of great recent scientific interest with respect to its role in global warming; as a fresh-water supply; as an environmental archive; because of avalanches; for its ability to release pollution to the aqueous environment; and owing to its potential to modify the atmospheric composition. In the following we first discuss physical processes related to the above, followed by recent trends in laboratory experiments better to understand the physics and chemistry occurring in or on surface snow. Today, glaciers cover a little less than 10\% of Earth's land surface. Over geological timescales, this amount varies greatly. Were the ice sheets in Antarctica to melt entirely, global sea level would rise about 57~m \cite{DavidGVaughan:2007p26017}. At the last glacial maximum, some 18\,000 years ago, the quantity of water taken up in ice sheets and glaciers meant sea level was about 120~m lower than today. Most of that change was due to the formation of large ice sheets in northern North America and Europe; however, mountain glaciers, too, had their role.

Although we shall not discuss permafrost --- defined as soils with year-round temperatures below 273~K --- in detail in this review, it is an important cryospheric phenomenon and the focus of much research. The thawing of permafrost, which can be found over about a quarter of the global land area \cite{french1996} may lead to the release of carbon into the atmosphere, possibly increasing global warming \cite{Schiermeier2001};  thawing permafrost also changes the stability of bedrock, which is of particular importance in mountain areas; for reviews see \textcite{harris2009,margesin2008,haeberli2006}. Freeze--thaw cycles in permafrost and in soils that are affected by seasonal frost lead to frost heave and the striking phenomenon of patterned ground, often found in polar and high Alpine environments \cite{Dash2006,kessler2003}. 

The Antarctic and Greenland ice sheets, with 97\% of the glaciated area, or an estimated 99.8\% of the volume, contain the immense majority of the Earth's ice \cite{Ohmura:1996p26725}. Because of their tremendous mass, they react very slowly to environmental changes, with dynamic response times of up to 100\,000 years for the West Antarctic ice sheet, but at the same time have a great potential to affect sea level \cite{DavidGVaughan:2007p26017}. The much smaller glaciers and ice caps on mountain slopes --- also known as alpine, niche, or cirque glaciers --- like the Swiss one that Shelley described, appear to react very sensitively to environmental changes, which makes them, with their widespread distribution over the globe, good candidates for use as climate indicators. In many cases, air temperature, and in particular its variation with elevation, has a very high statistical correlation with the mass balance of glaciers \cite{GeorgKaser:2007p26005}; this correlation does not always hold, however, for high-latitude or high-altitude glaciers. Of particular interest with respect to short-term sea level rise are intermediate-scale ice fields and ice caps in Patagonia, Iceland, Alaska, and the Arctic that are small enough to respond to environmental changes on decadal time scales, but large enough to be important with regard to sea-level change \cite{ShawnJMarshall:2007p26023}.  
	
This direct connection between sea level and glacier volume has motivated intense research to understand and predict the fate of glaciers during global warming. In addition, ice and snow act as positive feedbacks in Earth's climate system: The extent of ice cover determines surface albedo and thus the radiative balance; the height of the ice sheets affects the geometry of upper atmosphere waves; and melt water may influence the global ocean circulation \cite{JohnTAndrews:2007p26015}. The most important removal processes of glacial ice are surface melting and ice transport, but it can also take place through sublimation \cite{GeorgKaser:2007p26005,ShawnJMarshall:2007p26023}.

\subsection{Glaciers}

\subsubsection{Glacier flow}

Glacier flow and consequent ice transport is the main ablation process in cold climates such as Antarctica where almost no surface melting occurs, and constitutes an important process for the Greenland ice sheet \cite{PhilippeHuybrechts:2007p26003,ShawnJMarshall:2007p26023}.  The flow of glaciers is a consequence of the weight and creep properties of ice \cite{MSchulson:2009p26776} and is successfully mathematically described for glaciers that are well coupled to their underlying beds  \cite{ShawnJMarshall:2007p26023}. For these, fluxes are dominated by internal shear stress leading to creep, or to plastic deformation. As ice tends to build up in the glacier, a surface slope is developed. This slope and the weight of the ice induce a shear stress throughout the mass. The two key factors that determine the stress driving the ice flow are ice thickness and the surface slope \cite{WendyLawson:2007p26018}. Each element of ice deforms according to the magnitude of the shear stress $\tau_\mathrm{xy}$ at a rate $\varepsilon_\mathrm{xy}$ determined by the Glen flow law
\begin{equation}
\label{eq:glensflowlaw} \varepsilon_{\mathrm{xy}}=A\, \tau_{\mathrm{xy}}.
\end{equation}
In this equation, $A$ is a parameter that is sensitive to temperature, water content, presence of impurities, and crystal properties; factors that also influence the flow velocity \cite{WendyLawson:2007p26018}. The effect of temperature on $A$ --- and thus the glacier flow --- is well understood, but the sensitivity of $A$ towards the other parameters, especially to the role of impurities, is much more uncertain. Summing or integrating the shear deformation of each element throughout the glacier thickness produces a velocity profile. In this simple case, velocity is approximately proportional to the fourth power of the depth. Therefore, if the thickness of a glacier is only slightly altered by changes in the net mass balance, there will be great changes in the rate of flow. In addition to the above factors, the nature of the bedrock has a great impact on a glacier's flow behavior, and this currently poses the greatest uncertainty in setting the appropriate factor $A$. Erodible bedrock favors rapid ice flow (\textcite{WendyLawson:2007p26018} and references therein) and on hard bedrock, the roughness affects sliding rates. Two mechanisms operate to permit sliding over a rough bed. Firstly, small protuberances on the bed cause stress concentrations in the ice, an increased amount of plastic flow, and ice streams around the protuberances. Secondly, ice on the upstream side of protuberances is subjected to higher pressure, which lowers the melting temperature and causes some of the ice to melt; on the downstream side the converse is true, and melt water freezes. This process, termed regelation, is controlled by the rate at which heat can be conducted through the bumps. The former process is more efficient with large knobs, and the latter process is more efficient with small bumps. These two processes together produce bed slip. Water-filled cavities may form in the lee of bedrock knobs, further complicating the phenomenon.  Although the process of glaciers sliding over bedrock is understood in a general way, more research is needed to account for the non-uniform viscosity of the glacier and the effect of the nature of the bedrock.

Climate scientists have been concerned by the observation of highly accelerated glacier flows  --- compared to the expectation based on the above considerations --- along the margins of the Greenland \cite{Rignot:2006p26032} and the Antarctic \cite{Pritchard:2009p26033}  ice sheets. This acceleration is thought to be a consequence of the retreat of ice shelves that act as barrier for the glaciers and of enhanced lubrication by melt water. The level of understanding of the loss  of ice sheets loss through this accelerated flow is presently too low to allow predictions on its impact on sea level during the 21st century \cite{DavidGVaughan:2007p26017}. 

To improve the assessment of the terrestrial ice fields in the context of global warming, mass balances of ice sheets and glaciers are current foci of research. Remote sensing has significantly contributed to our knowledge of the mass balance of polar ice sheets and has also confirmed much higher than expected mass wastage of glaciers in other regions such as Alaska or Patagonia \cite{JonathanBamber:2007p26021}. Satellite missions with a specific focus on the cryosphere are currently taking place. With these missions, some shortcomings of earlier work should be overcome; they should provide data with greater accuracy and with better coverage of the Antarctic continent.

\subsubsection{Glacial ablation}

The physics of snow and ice melt in glaciers and ice sheets are well understood, but owing to the spatial inhomogeneity are difficult to quantify in models \cite{ShawnJMarshall:2007p26023}. Also, not all melt water runs off a glacier; it can rather refreeze once it percolates into the deeper, and colder, snow pack. 
This refreezing can lead to a substantial growth of the ice sheet from below; a recent study found that up to half of the entire ice column thickness originated from this source in the Gumburtsev mountain range in Antarctica \cite{Bell2011}. 
The snow surface receives heat from short-wavelength solar radiation, long-wavelength radiation from clouds or water vapor, turbulent transfer from warm air, conduction upward from warmer lower layers, and the heat released by the condensation of dew or hoarfrost or by the freezing of liquid water. Heat is lost by outgoing long-wavelength radiation, turbulent transfer to colder air, the heat required for the evaporation, sublimation, or melting of ice, and conduction downward to lower layers.
Solar radiation is normally the greatest heat source, although much of the incoming radiation is reflected from a snow surface, and most of the heat loss goes to the melting of ice. It is incorrect to think of snow or ice melt as directly related to air temperature; it is the wind structure, the turbulent eddies near the surface, that determines most of the heat transfer from the atmosphere. In glaciers, as in seasonal snow pack, long periods of ablation can lead to cm-scale surface features termed suncups and penitentes. 

Suncups are characteristic quasiperiodic structures that form on snowfields after extended periods of ablation due to sun exposure.  The cups appear as shallow parabolic depressions separated by sharply peaked regions.  Qualitatively, these suncups form because the hollows trap sunlight more efficiently than the peaks.  A detailed radiative transfer model of ablating snow surfaces has suggested that the size, shape and time evolution of suncups can be theoretically constrained using the snow's physical parameters \cite{Tiedje2006}.  This work also suggests that unstable suncup surfaces may lead to the formation of penitentes, another ablation structure found in zones of high sublimation \cite{post1971}. 

\begin{figure}[t]
\begin{center}
\centering\includegraphics*[width=\columnwidth,clip=true]{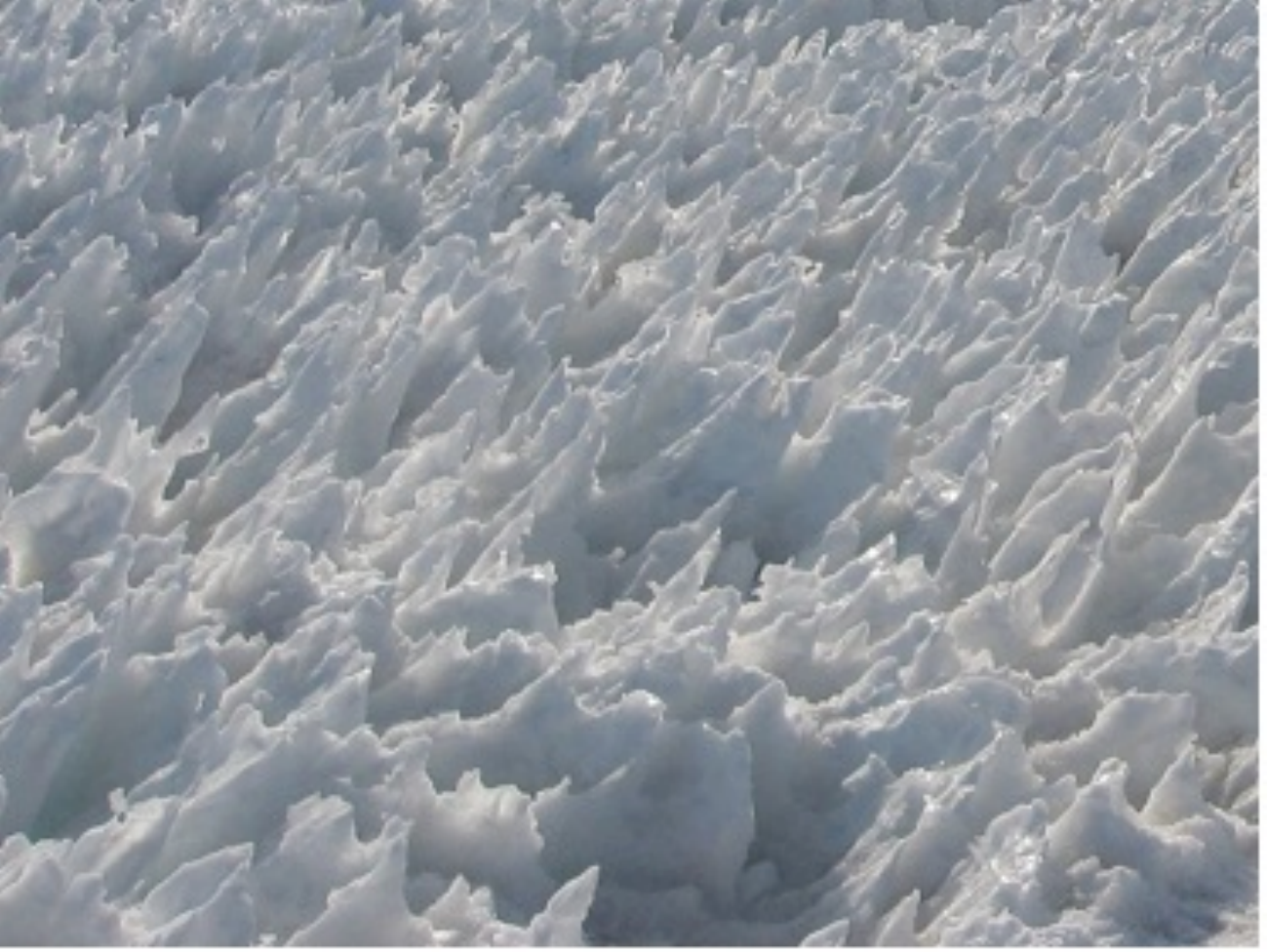}
\end{center}
\caption{(Color online) Penitentes tens of centimeters long tilt toward the predominant direction of solar radiation.
\label{vance3}
}
\end{figure}

A penitente is a column of snow, wider at the base and narrowing to a point at the tip, that takes its Spanish name from the resemblance of field of penitentes to a procession of penitents in white robes. Penitentes commonly form during the summer on glaciers or snowfields at high altitudes; Fig.~\ref{vance3} shows a typical penitente field.  The first observations leading to the present-day understanding came from \textcite{lliboutry1954} who noted that the key climatic condition for the differential ablation that leads to the formation of penitentes is that dew point is always below freezing. Thus, snow will sublimate, which requires a greater energy input than melting. Once the process of differential ablation starts, the surface geometry of the evolving penitente provides a positive feedback mechanism, and radiation is trapped by multiple reflections between the walls. The hollows become almost a black body for radiation, while decreased wind leads to air saturation, increasing dew-point temperature and the onset of melting. In this way peaks, where mass loss is only due to sublimation, will remain, as well as the steep walls, which intercept only a minimum of solar radiation. In the troughs ablation is enhanced, leading to the downward growth of penitentes. A mathematical model of the process has been developed \cite{betterton2001,bergeron2006}. The effect of penitentes on the energy balance of the snow surface, and therefore their effect on snow melt and water resources, has been described by \textcite{corripio2003,corripio2005}.

\subsubsection{Sintering processes}\label{sintering}

\begin{figure}[t]
\begin{center}
\centering\includegraphics*[width=\columnwidth,clip=true]{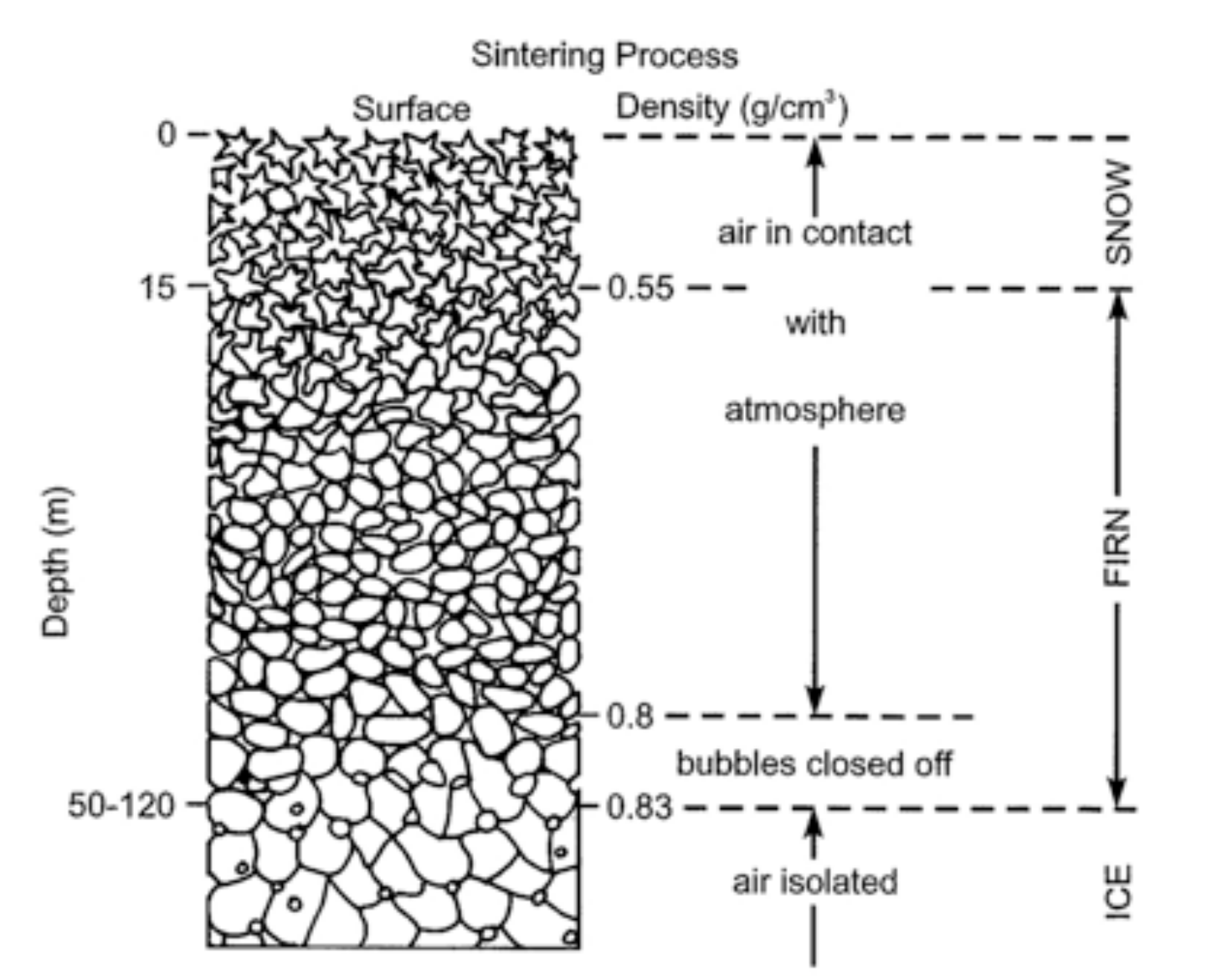}
\end{center}
\caption{Schematic cross-section of how snow transforms via firn into ice to create glaciers and ice sheets.
\label{vance1}
}
\end{figure}

Glaciers form where the accumulation of snow and ice exceeds ablation. The snow that eventually makes up ice sheets and glaciers undergoes a sequence of conversion stages, which are schematically represented in Fig~\ref{vance1}. Snowfall, in the form of crystals having tiny hexagonal plates, needles, stars, and other intricate shapes (Section~\ref{snowflakes}), first settles on the ground with a loose granular consistency.  As the crystals begin to collapse and break up  owing to the weight of the overlying snow, with partial melting and recrystallization the snow reaches a typical density of 0.2--0.3 g/cm$^3$. These processes proceed more rapidly at temperatures near the melting point and more slowly at colder temperatures, and result in a net densification of the snowpack. This snow densification proceeds more slowly after reaching a density of 0.5--0.6 g/cm$^3$, recrystallization under stress caused by the weight of the overlying snow becomes predominant, and grains change in size and shape in order to minimize the stress on them (Section~\ref{ssec:grains}). This change generally implies that large or favorably-oriented grains grow at the expense of others. Stresses due to glacier flow may cause further recrystallization. These processes thus lead to an increase in the density of the mass and in the size of the average grain. This stage is reached after one annual thaw--freeze cycle, and the resulting snow is referred to as firn or neve and appears as a white structureless mass with far less pore space than a fresh snowpack. Further sintering and fusion occurs over time and remaining air is squeezed out or trapped as tiny bubbles. The permeability change at a density of about 0.8 g/cm$^3$ marks the transition from firn to glacier ice. 
The transformation may take only three or four years and less than 10~m of burial in warm and wet environments, but on the high plateaux of Antarctica the same process takes several thousand years and burial to depths of up to 150~m. Most of the air transforms at a depth ranging from several hundred up to over 1000~m into air clathrate hydrates \cite{Shoji:1982kk} and the ice becomes bubble-free; however, rarely does the density exceed 0.9~g/cm$^3$. An open question is whether grain boundaries or even bulk ice also can take up molecules from the air. 
The initial densification can be observed by the naked eye as an gradual increase of the clear blue color of old ice along crevasses;  while snow is bright white, because of reflection and scattering by the air, ice is blue, because the red part of the optical spectrum is absorbed more efficiently --- just like in liquid water. 

The consequences of this metamorphism of the ice crystals are far ranging. The direct consequence is a vertical layering of the glacier and  --- as stress induces sintering --- local inhomogeneity of the snow and ice. Both affect the flow dynamics of the glacier, as we discussed above. Metamorphism in the snow surface layer also impacts the chemistry occurring within the snow pack, both directly due to release of trace gases from the shrinking snow surface area, and because snow properties modify the light penetration through snow \cite{Domine:2008p2711}; see below.
This metamorphism, which leads to an overall smoothing of the snow surface, is caused by the higher vapor pressure of small ice crystals making the larger ones thermodynamically more favorable (Ostwald ripening; the Gibbs--Thomson or Kelvin effect, Sections~\ref{ssec:surf} and \ref{atmospheric}). This process is rather fast; after a day structures of the size of several micrometers can disappear at 233~K \cite{kerbrat2008,legagneux2005}.

When ice cores drilled from ice sheets or glaciers are used as climate and environmental archives, the depth at which air is totally trapped by glacial ice is of great interest. Before this closure --- i.e., in firn --- the air is in contact with the atmosphere above, so that a time-lag between the age of the ice and the age of the air at any specific depths of the glacier evolves, which depends on the time required for complete trapping of the air bubbles in the glacial ice \cite{Hansen:2007p24301}. As both the composition of the enclosed air and of the ice are analyzed and used to reconstruct past environmental conditions, precise knowledge of this time-lag over the whole time-frame covered by the ice core would be of advantage to understand better leads and lags of, for example, temperature and greenhouse gases. The physics of the sintering is not well-enough known to model this process over time with high-enough accuracy \cite{Loulergue:2007p26030}.  Very recently \textcite{LemieuxDudon:2010p26029} have described a new approach to improve the ice--gas chronologies. In their work, they combine the information of several regional and global markers analyzed in the ice and gas of the  ice core with results from glaciological models to establish consistent gas and ice chronologies for several ice cores.  The results confirm an overestimation of the gas--ice time-lag purely based on densification firn models \cite{Loulergue:2007p26030} and allow the more precise comparison of phase relationships between ice cores from the northern and southern hemispheres \cite{LemieuxDudon:2010p26029}.

The layering of snow pack, which is a consequence of metamorphism or sintering, may also foster avalanche formation. One application of current research is to improve our capability to forecast the likelihood of avalanche formation in dry snow potentially triggered by recreationists \cite{Schweizer:2008p26034}. Such dry snow slab avalanches are released when two criteria are met \cite{Schweizer:2003p25880}: A weakness of the snowpack below one or more slabs is required and this failure needs to propagate on larger scales. The distinct differences in snow hardness and texture lead to initial failures between two snow slabs. Conditions for fracture propagation are less well known, but snowpack variability has a strong influence. New three-dimensional imaging techniques are currently being used to visualize snow microstructure with the goal to relate texture to mechanical properties~\cite{Pinzer:2009p25366}. Modeling will then be needed to close the gap between the micro-scale instabilities and macroscopic avalanche formation. Current modeling does include the sintering process that results in snow failure, but only in a parameterized fashion based on highly idealized structures.

\subsection{Snow chemistry}\label{snow_chemistry}

Earth's surface snow interacts vividly with the atmosphere. That the surface snow is thus not simply a passive cover was recognized starting in the 1980s when field measurements above snow-covered polar areas revealed concentrations of a number of atmospheric trace gases that were essentially unexpected based on known gas-phase chemistry (\textcite{Domine:2002p1662} and references therein). Rapid depletions of ozone and of gaseous mercury up to 1~km height during polar sunrise, elevated levels of nitrogen oxides, and of oxygenated organics within the snow-pack interstitial air, and emissions of halogens and hydrogen peroxides from the snow, are just a few examples of the observed snow--air interactions. The consequences of this reactivity, leading to a modified budget of atmospheric trace gases and of snow composition, are wide, ranging from changing the oxidative capacity and the climate feedbacks of the atmosphere, to complicating interpretation of ice-core data as a climate and environmental archive and fostering the transfer of pollutants to the aqueous phase during snow melt. To identify and to describe quantitatively the individual processes that contribute to the observed air--snow exchange is a current focus of research, a task that has turned out not always to be easy. Here, we pick some examples of well-controlled laboratory studies to explain recent developments and to identify some of the open issues. For a broader discussion including recent advances in field studies we recommend to the reader reviews on this topic by \textcite{Steffen:2008p24730,Domine:2008p2711,Simpson:2007p4813,Grannas:2007p2776,Huthwelker:2006p981, abbatt2003,Domine:2002p1662}.

\subsubsection{Uptake of trace gases}\label{snow_chem}

\begin{figure}[t]
\begin{center}
\centering\includegraphics*[width=0.7\columnwidth,clip=true]{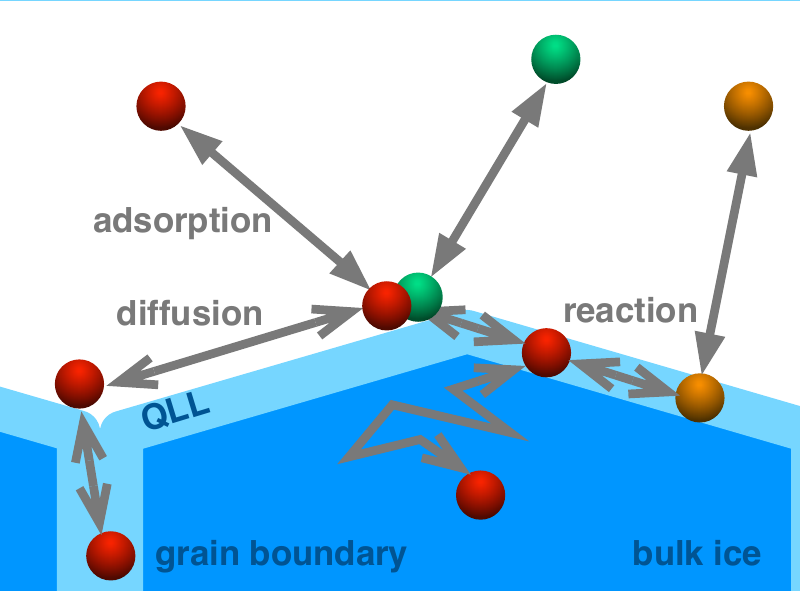}
\end{center}
\caption{(Color online) Illustration of individual uptake processes.
\label{thorsten1}
}
\end{figure}

Figure~\ref{thorsten1} illustrates processes that may contribute to changes in snow or in atmospheric composition: Physical processes such as the adsorption--desorption equilibrium, the competitive adsorption of several trace gases, and diffusion, as well as (photo-)reactions in the snow or dissociation of acidic trace gases. The distribution can further be influenced by dynamic snow processes such as snow metamorphism leading to the release of adsorbed and dissolved species or by the capture of trace gases into growing ice crystals.

For whatever process is governing the overall uptake (emission), the adsorption--desorption equilibrium is the initial (final) step and, as such, has garnered considerable attention during the last decades. It was found that, for a number of oxygenated organics, the interaction with ice surfaces can successfully be described based purely on the adsorption--desorption equilibrium \cite{abbatt2003}. The interest to study the interaction of volatile organics with ice was motivated by their detection throughout the troposphere, where they are an important source of the atmospheric oxidizers HO and HO$_2$.  In these studies the uptake was found to be fully reversible and to increase linearly with rising gas-phase concentration up to a saturation regime when a certain surface coverage is reached. Remarkably, for a wide range of oxygenated organics, the saturation was found to level off at similar surface concentrations of roughly $3 \times 10^{14}$ molecules cm$^{-2}$. Such an uptake behavior can be well described by the Langmuir adsorption isotherm with its simple linear relationship of the concentration of adsorbed molecules $[X_{\mathrm{s}}]$ and the gas-phase concentration$[X_{\mathrm{g}}]$ at low --- atmospherically relevant --- concentrations:
\begin{equation}
\label{eq:KlinC}  K_{\mathrm{linC}} = Ê\frac{[X_{\mathrm{s}}]}{[X_{\mathrm{g}}]}.
\end{equation}

The use of the linear relationship is recommended by \mbox{IUPAC} \cite{Ammann:2008p14103} and is supported by simulations, such as recent MD simulations \cite{Hantal:2008p25890}, confirming an uptake process where the organic adsorbs on the ice surface via hydrogen bonds. Knowledge of the temperature dependence of the partitioning allows one to extrapolate to environmental conditions at a specific temperature and snow-surface to air volume ratio. There are however observations that indicate that under environmental conditions, the uptake even of these well-behaved organics is more complicated. For example, \textcite{Burniston:2007p2458} found in a field study that the observed loss of semi-volatile organics from surface snow could not be well described taking into account the adsorption equilibrium and the evolution of total snow surface area with time, and concluded that an important factor is missing to describe the emissions, especially for wet snow. In line with this, recent laboratory studies on the uptake of organics at higher temperatures, above some 240~K, showed that the maximum uptake increases well above the $3 \times 10^{14}$ molecules cm$^{-2}$ with temperature \cite{Abbatt:2008p20344}. Clearly, there is a need for more experiments in this higher temperature region, most relevant to the Arctic snow pack. 

\begin{figure}[t]
\begin{center}
\centering\includegraphics*[width=\columnwidth,clip=true]{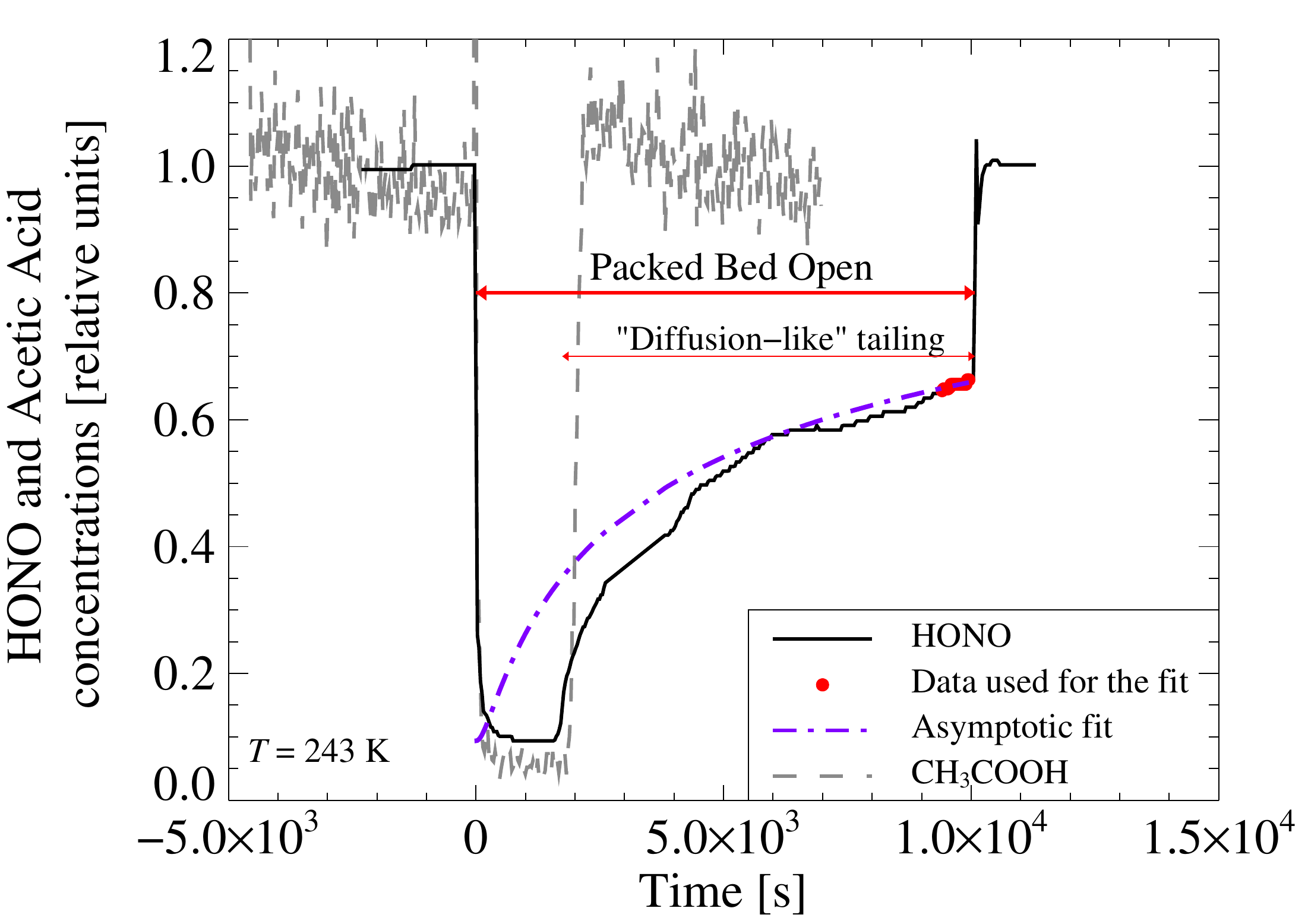}
\end{center}
\caption{(Color online) Comparison of breakthrough curves observed for nitrous acid (HONO; black solid line) and for acetic acid (CH$_3$COOH; gray dashed line). A packed ice bed flow tube at 243~K  was exposed to the trace gases at $t = 0$. For
the nitrous acid case, the ice was bypassed at $t = 1 \times 10^{4}$~s. The acetic
acid recovered back to its initial concentration before the ice was bypassed, while the nitrous acid trace shows a long-lasting tail. The blue dashed-dotted line represents the result of an asymptotic fit made to retrieve the diffusion parameters that describe the shape of the tail. Printed with permission from \textcite{Kerbrat:2010p25446}. Copyright 2010 American Chemical Society.
\label{thorsten2}
}
\end{figure}

While the laboratory experiments with organics at low temperature showed a fully reversible uptake that can be well described considering purely adsorptive processes, experiments with acidic gases --- typically HNO$_3$ and HCl --- reveal a more complex behavior. Research interest in HCl--ice interactions has been provoked by the discovery of the stratospheric ozone hole; uptake of HNO$_3$ plays a role both in the stratosphere and in the troposphere (see Section~\ref{atmos_theo}). The uptake experiments seem to show two regimes, an initial high uptake followed by a much slower uptake regime with diffusion-like kinetics. Figure~\ref{thorsten2} compares results from such uptake experiments for HONO and acetic acid, where the gas-phase concentration of the trace gas is shown. This study by \textcite{Kerbrat:2010p25446} was the first to observe this long-term behavior for the weak acid HONO (nitrous acid), which is an important precursor of the atmosphere's oxidizing agent OH. At $t = 0$ the trace gases are exposed to an ice surface leading to a sudden reduction of the gas-phase concentration. After some time the gas-phase concentration of acetic acid recovers just as fast, but for HONO the recovery is initially fast, but then dominated by a slow profile. By focusing on the initial part of the uptake process, information on the adsorption--desorption equilibrium can be derived.  Currently, there are several procedures used to disentangle the initial adsorptive process from the slower long-term process and a consistent picture of the amount of acidic gases adsorbed to ice surfaces begins now to emerge from some of the more recent laboratory studies \cite{Huthwelker:2006p981}.  As Table~\ref{surface_conc} summarizes, these results, as recommended by IUPAC \cite{Ammann:2008p14103}, show a more substantial uptake for acidic gases than observed for organics at identical partial pressures and temperatures. 

\begin{table}[b]
\caption{Partition coefficient ($K_\mathrm{linC}$) of some atmospheric trace gases at 220~K. The partition coefficient gives the concentration at the surface relative to the concentration in the gas phase at equilibrium. Data taken from \textcite{Ammann:2008p14103}.
\label{surface_conc}}
\centering
\begin{tabular}{ll}
& $K_\mathrm{linC}$ [cm$^{-1}$] \\
\hline  
Acetone & $3.5 \times 10^{0}$ 	 \\
HONO & $3.8 \times 10^{2}$  \\
HNO$_3$ & $8.4 \times 10^{4}$ \\
HCl & $9.6 \times 10^{3}$ \\
\hline
\end{tabular}
\end{table}

The physical nature of the subsequent slow uptake is an essentially open issue. Currently two different approaches to explain the observations are debated: either diffusion into reservoirs such as grain boundaries, or slow surface processes such as roughening or restructuring. On one hand, grain boundaries and junctions occur in natural ice (see Section~\ref{ssec:grains}; \textcite{Rempel:2001p26726}) and have been shown to host high concentrations of (acidic) impurities \cite{Mulvaney:1988p450}; thus it seems reasonable that trace gases might accumulate there after adsorption. Yet, based on this picture, the tail should not be observed on very thin ice films, as the diffusion should be fast enough to penetrate the entire film. Experimental observations of the influence of the thickness of the ice matrix on the observed tail are however not conclusive \cite{Huthwelker:2006p981}. One the other hand, there is strong theoretical and experimental evidence that impurities can influence the surface structure of ice (see Section~\ref{ssec:surf}; \textcite{Dash2006}) and, if the kinetics are slow and if this structural change results in a change of the uptake behavior, this might also explain the temporal behavior of the breakthrough curves. Yet, the depth of the restructured regions would need to be quite thick to explain fully the observed kinetics, as noted by \textcite{Huthwelker:2006p981}. 

There has been much effort spent recently to investigate how impurities impact the surface structure and in turn the uptake behavior. As discussed in detail in Section~\ref{ssec:surf}, the molecular structure even of clean ice surfaces is disordered at temperatures approaching the melting point. Because some properties of this disordered layer are similar to liquid water, this disorder is often called a quasi-liquid layer or, synonymously, a pre-melted layer. Distinct from this, high concentrations of impurities can lead to (partial) melting of the snow simply due to melting point depression, as observed in brine channels in sea ice. Currently, there are two questions of great importance and scientific interest in this context:  How does the presence of the quasi-liquid layer impact the interaction with trace gases? And how do impurities at concentrations too low to induce melting impact the surface structure and thus the uptake? Up to now the number of studies that have investigated the uptake of trace gases at temperatures approaching the melting point of ice, where pre-melting occurs, is limited \cite{Huthwelker:2006p981}. Recently, \textcite{Abbatt:2008p20344} observed a drastic increase in saturation surface coverage at such high temperatures and argued that this is due to a larger uptake to the quasi-liquid layer as compared to the solid ice surface. In the most direct way so far, \citet{McNeill:2006tk,mcneill2007} have combined results from ellipsometry and from classical uptake experiments to monitor the surface structure of ice. They showed that the long-lasting uptake of HCl on ice is enhanced when concentrations are high enough to induce surface disorder. \citet{Krepelova2010a} used electron-yield near-edge X-ray absorption fine-structure (NEXAFS) spectroscopy to probe ice surface properties in the presence of HNO$_{3}$ at 230~K. Their results show that nitrate on ice is hydrated by water molecules, just like it is in solution. The spectra further reveal the presence of solid ice at the surface region a few tens of water molecules thick. Likewise, neutron inelastic scattering on ice-covered nano-particles was recently used to get insight into the water mobility in the ice--HCl system \cite{demirdjian2002}. Interestingly, neutron experiments indicated that the translational diffusional mobility of the water molecules sets in at temperatures of some 250~K \cite{toubin2001, demirdjian2002}, putting in question the existence of a quasi-liquid-layer at temperatures below 200~K, suggested by \textcite{mcneill2007}. Clearly, the nature and importance of the long-lasting uptake regime and of changes at the ice surface during adsorption remain a top research priority.

The crucial point is that the temporal behavior, the temperature dependence, the sensitivity to ice morphology, and the reversibility of the uptake are very different in the two uptake regimes, which makes extrapolation of snow--air interactions from laboratory findings to environmental conditions uncertain without precise knowledge of the individual processes. For example, at long exposure times the overall uptake is increasingly dominated by diffusive processes and will essentially exceed the adsorptive uptake. Exposure times in laboratory experiments are typically in the seconds to minutes range, while in the environment they may vary from minutes, given by the lifetime of ice in clouds, to days or even weeks when pollutants are exposed to surface snow under stable meteorological conditions. Also, the long-lasting uptake becomes more dominant at higher temperatures, while adsorption increases at lower temperatures, as recently shown by \textcite{Kerbrat:2010p25446} for HONO. Yet, in a different study using a snow sample of significantly lower density --- as found in the environment --- no detectable contribution of long-term uptake of HONO to the snow was found, which was attributed to a lower grain-boundary density in the latter sample \cite{Pinzer:2010p25893}. This discrepancy underlines the need for a better characterization of ice and snow samples used in the laboratory and found in the field. Another difference of the two uptake regimes is that the initial uptake is mostly irreversible, while the long-lasting uptake has a reversible character, as shown for HNO$_3$ \cite{Ullerstam:2005p4622}. The irreversibility of the initial uptake calls for a new concept to describe the adsorptive uptake, as the current Langmuir isotherm and the linear relationship rely on equilibrium conditions that are in contradiction to the observed irreversible nature of the uptake process. 

In conclusion, experiments at rather low temperatures of $< 240$~K or so clearly show that the adsorption--desorption equilibrium plays an important role in the distribution of trace gases between the gas and the snow phase. Extrapolation to environmental conditions is still hampered by insufficient knowledge of the observed long-term uptake and scarce experiments at temperatures approaching the melting point. Whether this distribution will have an impact on a larger scale and thus be of atmospheric significance depends also on the transfer of trace gases between the interstitial, near surface, and boundary layer air and the height of the boundary layer.   

\subsubsection{Photochemistry of trace gases}

Surface snow hosts a rich photochemistry for two reasons: On one hand the high albedo of snow greatly enhances the path length of photons within a snow pack and increases the overall radiation intensity \cite{Grannas:2007p2776}; on the other, contaminants in the snow pack, such as H$_2$O$_2$, nitrate, and organics, to name the most dominant, may act as reactive chromophores. In filtered, molten samples from the Arctic and from Antarctica, H$_2$O$_2$ and nitrate were found to account for about half of sunlight absorption; the other half was attributed to unknown, presumably organic substances \cite{Anastasio:2007p6837}.  
Photolysis of H$_2$O$_2$ is a source of OH \cite{Chu:2005p2064} a strong oxidant, for example of organic matter possibly contributing to the observed release of volatile organic compounds from surface snow \cite{Grannas:2004p3348}. Nitrate photolysis is a source of nitrogen oxides that may be released to the atmosphere, where they impact the oxidative capacity \cite{Grannas:2007p2776}. In polluted environments this NO$_x$ release is of minor importance, but in remote areas it may be the primary source of gas-phase nitrogen oxides. At the South Pole the emissions of nitrogen oxides have even reached concentrations in the --- shallow --- boundary layer high enough for ozone production; an observation normally restricted to polluted environments \cite{Crawford:2001p}. 

The consequences of photochemistry occurring in snow are not restricted to modifications of the atmospheric trace gas budget, but may also alter concentrations within the snow pack.  Nitrate content in the surface snow at low-accumulation sites, such as in Antarctica, has been found to be highly influenced by post-depositional processes, one of which, although not the major process, is photochemistry \cite{Blunier:2005p25896}. This has consequences for the interpretation of concentration profiles gained from ice cores, where nitrate is an easy ion to measure, as an environmental archive.  Similar, ice cores provide records for reconstructing the mercury content preserved in snow from long before anthropogenic influences began \cite{Steffen:2008p24730}. They would thus be a valuable tool to study the natural atmospheric mercury cycle, provided that surface snow loss processes are well understood. Field studies have revealed that polar surface snow loses its mercury deposited from the atmosphere, the emission of which is enhanced with solar irradiation. In line with the known mercury chemistry in the (liquid) aqueous phase, \citet{BartelsRausch:2011p26430} have shown that organics can enhance photolytic mercury release also in the ice phase and at low temperatures. Some studies showed photolytic decomposition of anthropogenic organic compounds with products of increased toxic effect \cite{Grannas:2004p3348}. Furthermore, organic radical compounds, products of irradiation of organic chromophores, can participate in subsequent bimolecular reactions for example with mercury \cite{Steffen:2008p24730}. This chemistry directly impacts the amount of toxic substances in the snow pack, and thus the amount that can possibly be released and enter the biosphere during snow melting.
Field and laboratory studies have revealed that snow photochemistry affects both the composition of the boundary layer air and of the snow pack. Many relevant processes have been identified and a more quantitative picture starts to emerge, but we are far from understanding the fundamental chemistry, which hampers extrapolations to environmental conditions. In the following, we discuss current limitations and research topics.

Products from photochemistry in snow may differ from those found in aqueous environments.  For example, unusually high relative yields of HONO and NO are detected, both of which are secondary products of nitrate photolysis. The large temporal and spatial variations of the relative nitrogen oxides emissions found in field measurements underline the complexity of their production and release.  Identifying environmentally relevant production processes is a current focus of research and several ideas, such as the reaction of photolytically produced NO$_2$ with organics \cite{BartelsRausch:2010p25371} or hydrolysis of NO$_2$ \cite{Zhou:2001p3425} have been proposed. 
Studies on organic photochemistry in ice have also revealed some interesting chemistry: Klan et al have reported unusual photo-behavior of halobenzenes, the photolysis of which produced biphenyls that are more toxic than the educts. The rates of reactions may also be substantially different in ice as compared to in liquid solutions \cite{Grannas:2007p2776}.  Recent studies have shown reaction rates accelerated by up to a factor of 40 during the photochemistry of organics \cite{Grannas:2007p25490,Kahan2010}, but this acceleration is especially important for bimolecular reactions --- which often follow a photolysis reaction --- such as the oxidation of halogenides and nitrite, where rates of about $10^{5}$ times faster were determined in ice \cite{TAKENAKA:1992p2256,Grannas:2007p2776}. 

Yet, it is questionable whether results from laboratory studies that focus on identifying photolysis pathways can be applied to field studies, as the location and physical state of pollutants might be very different. Little is known about the chemical form and location of the reactive impurities in surface snow: Are they attached via hydrogen bonds on the surface or dissolved in a surface layer or in liquid-like micro-pockets, grain boundaries, or veins? Do they form supersaturated solid solutions or pockets of brine? Contaminants may enter the snow pack via nuclei for cloud condensation or ice formation, deposited aerosols, and adsorbed trace gases. A current limitation in this respect is that most laboratory studies use ice samples that have much higher contaminant concentrations than those found in natural snow samples. Segregation of solutes from the forming ice during the freezing process may depend on solute concentration, and possibly only occurs when a certain solubility limit of the ice is exceeded. So, even if wet metamorphism and refreezing occurs in the field, it is not certain whether segregation occurs in nature, as concentrations are so much lower. Experiments that more closely mimic the characteristics of natural snow are required. 

The microstructure of ice results in flow and sintering of glacial ice. Many open questions are currently tackled by addressing the microstructure of snow with its vast network of grain boundaries and veins. A better understanding of its properties will help to predict, for example, avalanche formation, the uptake and release of trace gases, and chemistry occurring in the snow pack.

\section {Sea Ice}\label{oceanic}

\begin{quote}
Pack-ice might be described as a gigantic and \\ interminable jigsaw-puzzle devised by nature. 

Ernest Shackleton, \emph{South!}
\end{quote}

\begin{figure*}[t]
\begin{center}
\centering\includegraphics*[width=\textwidth,clip=true]{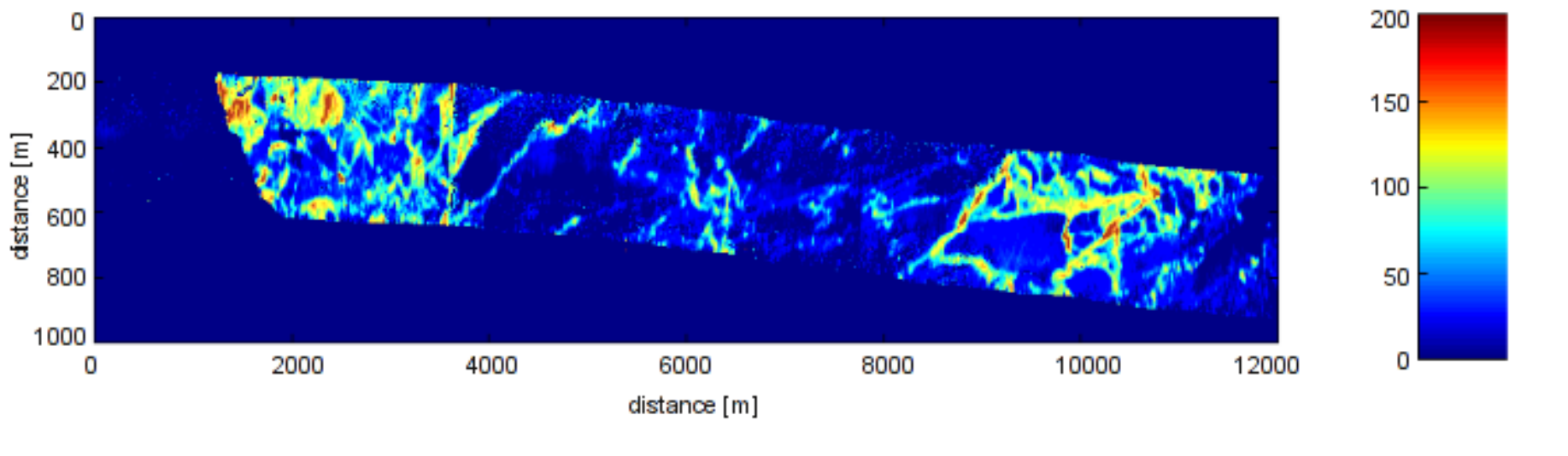}
\end{center}
\caption{(Color online) Lidar image of sea-ice topography in the Baltic Sea. The width of the swath is 400~m. Colors/shades indicate freeboard of ice (height of the surface relative to the ocean surface) in centimeters.
\label{jari1}}
\end{figure*}

A distinctive and crucial feature of water is that the solid --- in the form of ice Ih --- floats on the liquid, so sea ice generally forms at the ocean--atmosphere interface. Sea ice --- ice that originates as frozen ocean water ---  must be treated as a distinct genus of ice to that on land. It possesses a multiphase structure that, in addition to the water ice crystal lattice, includes also gas, liquid brine, solid salts and other impurities. Sea ice on the microscale is a so-called mushy layer \cite{Feltham:2006} and on that scale the presence of salt in the sea makes double-diffusive convection of heat and salt an important physical process.

In this section, we focus on morphological features of sea ice, those physical processes that create the landscape of sea ice, and how these processes are taken account in large-scale numerical models. On a geophysical scale, sea ice is considered to be a mixture of open water and several different classes of undeformed and deformed ices.  Each ice class has an inherent ice-thickness distribution, but ice classes also differ in their mechanical and thermodynamical properties. A typical floe of sea ice consists of undeformed ice, whose thickness varies from 0.5 to 5~m, and deformed ice or pressure ridges, whose thicknesses range from 2 to 30~m in the Arctic Ocean.

A typical feature of pack ice is the strong gradient of properties in a vertical space. Pack ice may consist of columnar and platelet ices, upon which may be superimposed through precipitation ice formed as a result of the sintering of snow and pure snow layers.
Density, brine content, crystal structure and prevailing orientation are different for these ice types. The top 1--10~cm of sea ice is usually composed of randomly-oriented ice crystals less than 5~mm in diameter, reflecting turbulent conditions during the initial freezing stage and freedom of growth in the horizontal directions. After this initial ice-cover formation, ice grows downward and ice crystals become larger. Deeper layers of ice are anisotropic because the crystals having the c-axis in a horizontal plane are favored to grow larger as heat is removed more efficiently in this orientation. This process ultimately leads to a uniform columnar ice type where the brine is trapped between the ice crystals \cite{Weeks:2010}. The bottom of the ice is porous and constantly interacting with the ocean water beneath.

Traditionally, large-scale sea-ice thickness characteristics have been mapped by profiling either the bottom or the surface of the ice with sonar or laser, which gives information about sea-ice draft or freeboard, respectively \cite{Wadhams:2000}. During the last few years, the utilization of side scanning sonars and lasers has extended these methods to fully three-dimensional mapping, enabling a much more detailed examination of the morphological characteristics of sea ice \cite{Hvidegaard:2006,Wadhams:2008}. Figure~\ref{jari1}, based on lidar measurements in the Baltic Sea in March 2005, depicts a highly deformed sea-ice field.  In this length scale of 1~m--1~km, dominant features are the pressure ridges and other deformed ice types, which can be identified from the image from their high freeboard and narrow elongated shape. At the time of the imaging, the mean ice thickness in the region was about a meter, but by purely thermodynamical means the ice would only have been 20--30~cm thick, which emphasizes the importance of a dynamical thickening of the ice due to deformation.  

Right from its beginnings --- in and even before Shackleton's time --- sea-ice research has been motivated by practical applications: shipping and off-shore activities in polar regions, exigencies of the Cold War, and climate-related questions. As the pack ice of the Arctic Ocean has experienced remarkable shrinking and thinning during the last decades, climate-related investigation has become the most important topic in sea-ice research. The most apparent changes are observed in the average and annual minimum sea-ice extents \cite{Comiso:2008} and in the mean sea-ice thickness \cite{rothrock:1999,Haas:2008}; but the drift speed and rate of deformation has also increased \cite{Rampal:2009}, and the age and residence time of the sea ice in the Arctic have decreased \cite{Maslanik:2007}. 

All climate models, which prognostically calculate the response of the pack ice to the increase of atmospheric CO$_2$ and other greenhouse gases, predict shrinking and thinning of the ice cover, and ultimately the total disappearance of the perennial ice cover in the Arctic Ocean within the next 30 years \cite{wang:2009}. Another result common to different climate models is that the increase of the surface air temperature will be two or three times larger in the Arctic than the average over the Earth \cite{Holland:2003}. Classically, polar amplification of climate change is understood to be caused by the positive feedback mechanism of snow and sea-ice changes, but changes in the atmospheric meridional transport of heat and moisture are also very significant driving factors \cite{Graversen:2008}.

Although climate models have projected similar changes to those that have occurred, modeled sea-ice changes are much smaller in magnitude than what has been observed during the last five years \cite{Stroeve:2007}. Some climate models project a similar accelerated disappearance of sea ice, but not until 10--20 years from now. It is well known that the natural variability of the climate is very large in the Arctic, which makes it difficult to assess to what extent the present changes are amplified by natural climate variability. In principle, climate models are able to reproduce internal climate variability when multiple simulations for the same climate periods are conducted. However, the ability of an ensemble of climate-model simulations to capture real natural variability is presently limited and thus it is as yet an open question whether the tipping point for accelerated melting has already been reached, or whether the present models are too conservative because of simplifications of some critical processes of ice physics.

\subsection {Processes determining the evolution of pack ice}

\begin{figure}[tb]
\begin{center}
\centering\includegraphics*[width=\columnwidth,clip=true]{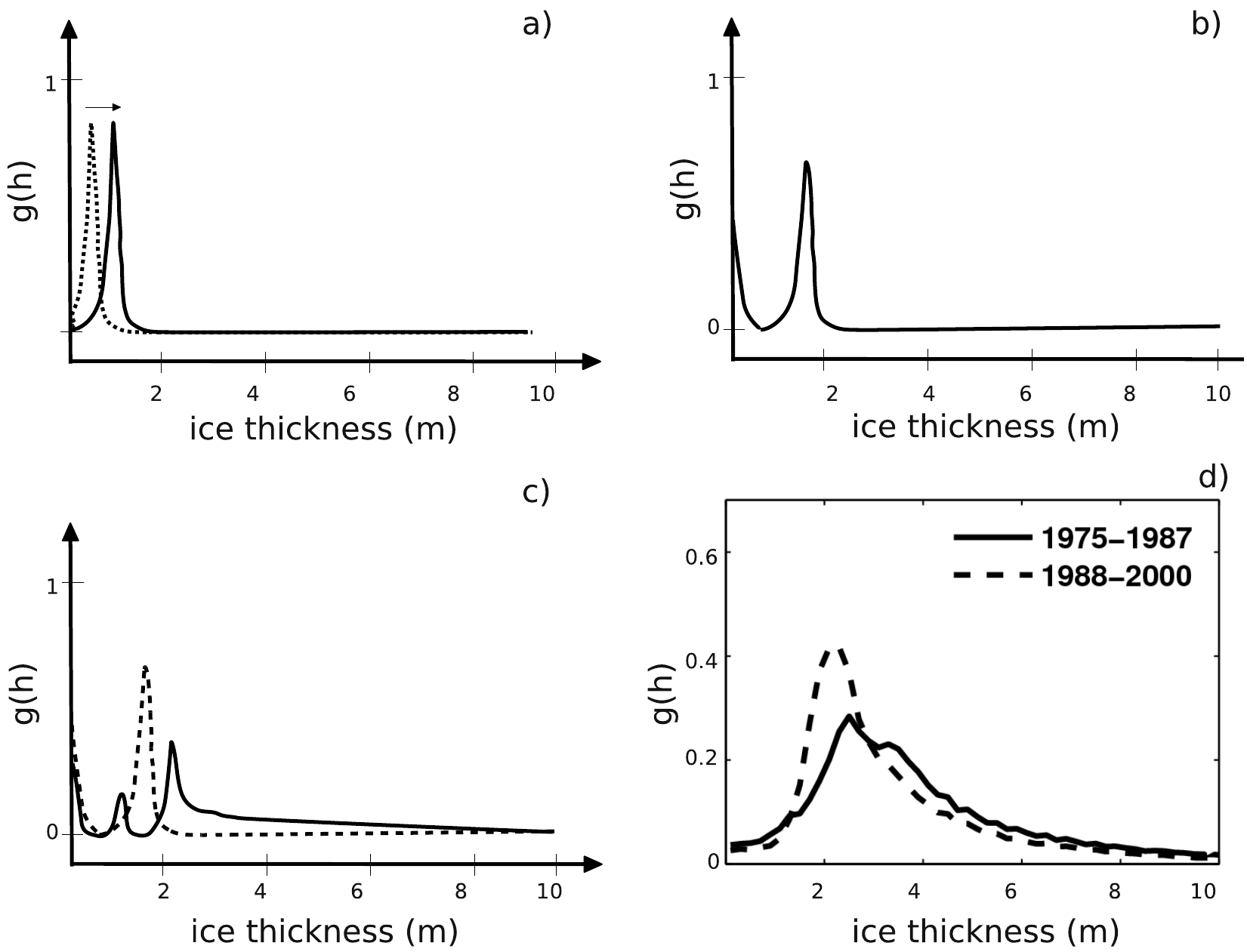}
\end{center}
\caption {\label{gh_evolution} Conceptual sketches illustrating the life cycle of pack ice by means of the ice-thickness distribution function $g(h)$. Panel (a) denotes the initial freezing of the ocean and the thermodynamical growth of sea ice; (b) depicts $g(h)$ after large-scale fracturing; and (c) illustrates the redistribution of pack ice. Panel (d) shows the observed distribution of sea-ice draft in the Canada basin in spring during two different climatological periods (solid line: 1975--1987, dashed line: 1988--2000)  \cite{Oikkonen:2011} .
\label{jari2}
}
\end{figure}

Pack-ice characteristics on a scale larger than the typical length scale of individual ridges and floes are described by the ice-thickness distribution function $g(h)$ \cite{Thorndike:1975} (Fig.~\ref{jari2}), defined as 
\begin{equation} 
\int_{h_1}^{h_2} g(h) dh = \frac{1}{R} A (h_1,h_2),
\end{equation} 
where $R$ is the region to be considered and $A (h_1, h_2)$ is the area within which the ice thickness varies between $h_1$ and $h_2$. The evolution equation for $g(h)$ is $ Dg(h)/Dt = \Theta + \Psi $, where the left-hand side of the equation describes the local change and advection, and $\Theta$ and $\Psi$ are the thermodynamic growth rate and the redistribution of ice thickness due to deformation, respectively.

The engine for sea-ice evolution is heat and momentum exchange between the ocean and atmosphere. Sea-ice dynamics is mainly driven by wind and ocean stresses. The thickness distribution and momentum balance are strongly coupled via the internal stress of pack ice. The material properties of sea ice depend very much on ice thickness; in particular the strength of thin ice is low and such an ice field moves easily, deforming and dynamically thickening under the action of relatively small external forces, but when the ice field becomes thicker, due to thermal or mechanical growth, the strength of ice increases, leading to large internal stresses and even to total stoppage of the motion. \textcite{Haapala:2005} showed that modeling of sea-ice dynamics is highly sensitive to the physical description of the sea-ice thickness distribution, especially of the thin fraction, and found that a multi-category sea-ice thickness model is more responsive to wind forcing than a commonly used two-level approach where ice concentration and mean ice thickness are the prognostic variables. 

For process studies, dissipation of the mechanical energy due to the floe flow,  floe--lead interaction, and deformation processes like ridging and rafting can be resolved explicitly by discrete models \cite{Shen1986,Hopkins1998,Hopkins2004,Herman2011}. For large-scale studies, continuum models of rheology need to be used \cite{Feltham2008}. Until recently, the most feasible approach has been using an isotropic plastic rheology \cite{Hibler:1979,Lepparanta:2005}, but satellite observations of large-scale fracturing of pack ice have inspired a reformulation of the plastic yield curve \cite{Wang2009}, the development of anisotropic rheological models \cite{Wilchinsky2006} and the formulation of a rheological model based on an elasto-brittle constitution law \cite{Weiss2008,Girard2011}.

Thermodynamic growth and decay of ice are a result of heat loss from the ocean to the atmosphere and absorption of the radiative fluxes inside the ice. The physics of thermodynamical ice growth is well captured by the general heat conduction equation. Having in mind that sea ice is a mixture of pure ice, brine and gas, its thermodynamics follows the theory of a mushy layer \cite{Feltham:2006}, which provides a theoretical framework for the interrelation of heat and salt.   

An apparent landscape of the ice pack exemplifies the life history of pack ice. Ridges and leads can be identified as generated during a particular event since most of the deformations are irreversible and the life cycle of pack ice can be separated into the following development stages: (1) initial formation of sea ice, (2) fracturing, (3) redistribution, and (4) aging.

\subsubsection{Initial formation of sea ice}

The life cycle of pack ice begins with the supercooling of the ocean and the formation of the first crystals of ice, termed frazil ice. Frazil ice can form very thin filaments of ice on the ocean surface or crystals can be mixed in the ocean surface layers by turbulent fluid motion as long as the buoyancy force is not large enough to lift them to the surface \cite{Weeks:2010}. As a result of an accumulation of frazil ice, the first 0.1--1~cm thick congealing layer, termed grease ice, is formed on the ocean surface. 

\begin{figure}[tp]
\begin{center}
\centering\includegraphics*[width=\columnwidth,clip=true]{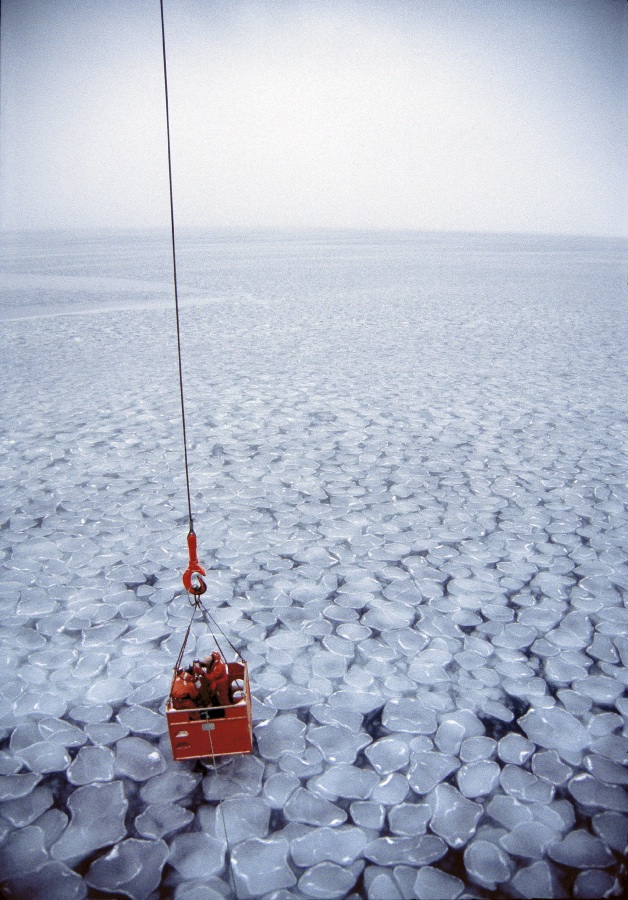}
\end{center}
\caption{(Color online) Pancake and frazil ice in the Weddell Sea. Pancake thickness here was 5--8~cm and their diameters around 1~m. The image shows the operation of taking samples of frazil ice between the pancake floes. 
\label{jari3}
}
\end{figure}

In shallow seas under very turbulent conditions, frazil ice can also accumulate on the sea floor, as anchor ice. Frazil-ice formation is not explicitly considered in large-scale models, but these processes can be parametrized in the models \cite{Walkington:2007}. The grease-ice layer effectively absorbs wave energy and decreases ocean turbulence, enhancing the accumulation of frazil ice at the surface. In calm conditions, grease ice becomes thicker and more solid due to downward crystal growth at the base of the ice. Otherwise, in addition to thermal growth, ice floes, typically 0.5--2~m in diameter, can be compressed together under wave action to form pancake ice \cite{Doble:2003}; Fig.~\ref{jari3}.

At an initial stage, the sea-ice-thickness distribution function, $g(h)$, exhibits only one peak. As the ice is thermodynamically growing, the peak shifts ahead in the ice-thickness space; Fig.~\ref{jari2}. Thickening leads to the direction of crystal growth becoming oriented mainly vertically and the size of the crystals becomes much larger. Also the temperature gradient between the top and bottom of the ice increases and the temperature profile changes from linear to nonlinear.  

\subsubsection{Fracturing}

\begin{figure}[tp]
\begin{center}
\centering\includegraphics*[width=\columnwidth,clip=true]{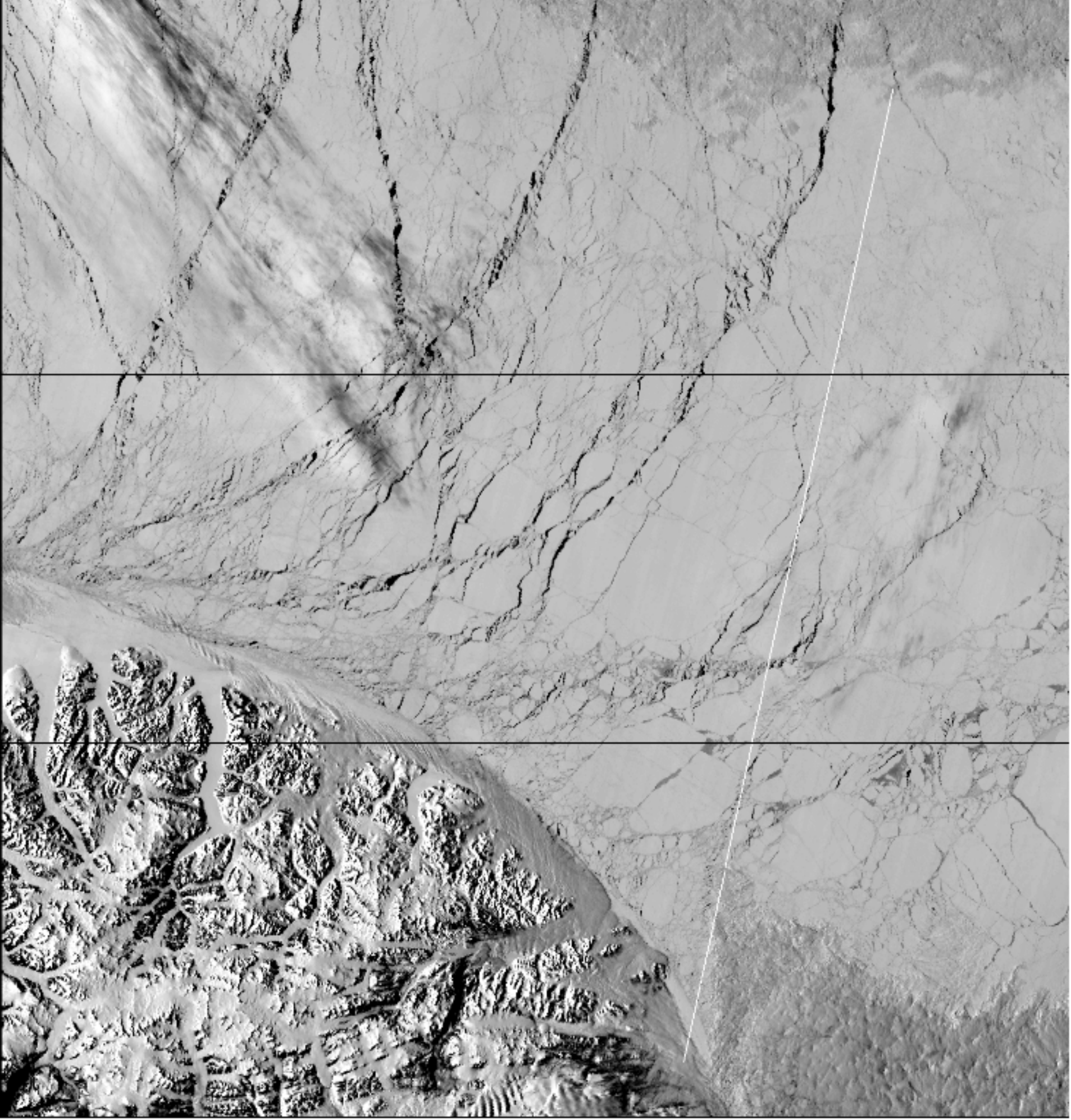}
\end{center}
\caption{$385\times 385$~km satellite (MODIS) image of pack ice in the Lincoln Sea.
\label{jari4}
}
\end{figure}

Fracturing can be regarded as a second stage in the life cycle of pack ice. Traditionally, pack ice has been considered to be a continuum of floes, but pack ice can also be understood to be a continuum of fractures \cite{Hibler:2001}. A single floe contains thousands of small-scale fractures: thermal cracks, which are caused by the contraction of the sea ice. Recently, \textcite{Marsan:2011} used a novel seismic monitoring system for recording the deformations of sea ice and found that during the spring, when large temperature gradients still exist, more than a thousand individual cracking events were observed in a day. Thermal cracking does not change the thickness or concentration of pack ice and hence $g(h)$ remains constant. Large fractures --- leads --- may extend hundreds of kilometers and are easily detectable from satellite images (Fig.~\ref{jari4}) and could also be detected by the seismic monitoring array \cite{Marsan:2011} These large leads are also called linear kinematic features (LKFs) since as well as the leads being discontinities in the ice thickness field, in these locations the sea-ice velocity field also has strong gradients \cite{kwok:2006}. The generation of leads changes the open-water fraction of $g(h)$ (Fig.~\ref{jari2}b). 

In contrast to initial ice formation in open-ocean conditions, freezing of ocean water in leads can occur in very calm and cold atmospheric conditions. In addition to the ice growth at the ice--ocean interface a particular form of ice, frost flowers, can be formed at the thin surface ice cover \cite{Perovich:1994}. The surface of thin ice ($\sim 1$--2~cm) in a refrozen lead often contains a liquid skin of brine ($\sim$1~mm). Initially pure fresh ice crystals are formed on the ice surface, but during the growth process such frost flowers become covered in brine and are thus extremely saline formations of ice \cite{Perovich:1994,Alvarez-Aviles:2008}. The presence of a skin of brine has been assumed to be essential for a development of frost flowers \cite{Weeks:2010}, but \textcite{Style:2009} showed that frost flowers could be formed by pure ice sublimation into an unsaturated atmosphere. In spite of their marginal fraction of the energy and mass balance of the pack ice, frost flowers are the dominant source of sea-salt aerosols and bromine monoxide, and influence tropospheric chemistry and ozone depletion \cite{Kaleschke:2004}.

\subsubsection{Redistribution}

The third stage is the redistribution of the ice pack. In this process some areas of pack ice are compressed together while open water is generated elsewhere in the ice pack. The total mass of the sea ice does not change, but redistribution changes the shape of $g(h)$; Figure~\ref{jari2}c. The most common process is that the thinnest ice experiences deformations and produces rafted and ridged ice. Rafted ice is particular form of deformed ice in which floes partially override each other; ridges are stripe-like formations having usually a triangular shape in the vertical dimension. 
It is known that rafted ice may be rafted many times, and rafted ice could form ridged ice. The deformed ice fraction depends very much on the location; in the Arctic, close to the North Pole the fraction is 20--40\%, but in the regions where ice drift is on average convergent the fraction of rafted and ridged ice could exceed 80\%.

The deformations begin when external forces cause ice to fail by flexure, buckling or crushing \cite{Mellor:1986}. Visual observation of the deformed ice field suggests that thin ice predominantly experiences rafting and thicker ice sheets are more often ridged, although thick ice ($h>1$~m) is also known to raft. The Parmeter model \cite{Parmeter:1975} suggests that rafting of an ice sheet is limited by the maximum ice thickness and that a cross-over thickness between rafting and ridging (e.g., 17~cm for the Arctic) relates to the mechanical parameters of the ice.

Laboratory experiments and numerical simulations \cite{Hopkins:1998,Hopkins:1999,Tuhkuri:2002} have given new insights into the deformations of sea ice. The major finding of ice-tank experiments, where an ice sheet is pushed against another, is that ice sheets of uniform thickness do not form ridges, but a nonuniform ice sheet generates both rafted and ridged ice. \textcite{Tuhkuri:2002} did not find any cross-over thickness that would determine whether an ice sheet is rafting or ridging, and noted that all ridging events began as rafting events. They distinguished four phases in the development of ridging events in their conceptual model of ridging. Firstly, a force exceeding a threshold value is needed for the initialization of deformation. Then an ice sheet begins to raft. The third phase is keel buildup, and the last phase is lateral growth of the ridge. 

\begin{figure}[tp]
\begin{center}
\centering\includegraphics*[width=\columnwidth,clip=true]{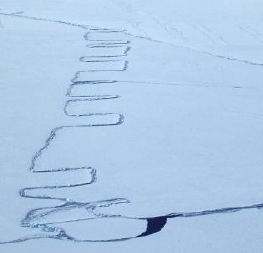}
\end{center}
\caption{(Color online) Finger rafting in the Lincoln Sea; the fingers are approximately 10--30~m  in length.
\label{finger_rafting}
}
\end{figure}

Field observations from the Baltic support the conclusion that all ridging events begins as rafting. \textcite{Babko:2002} have also highlighted the importance of rafting in the Arctic and have suggested that the cross-over thickness should be considerably larger than 17~cm. Recently, \textcite{Vella:2007,Vella:2008} have derived analytical solutions, based on thin plate theory, for determining the cross-over thickness for finger rafting, rafting and ridging. They found finger rafting --- Figure~\ref{finger_rafting} --- to be possible only with ice thicknesses less than 8~cm and rafting to occur with ice thickness less than 21~cm. 

\subsubsection{Aging}

The last stage in the life cycle of pack ice is aging. Although the life cycle of an individual ice floe can be considered as a chain of processes, in the ice pack as a whole all these processes are occurring simultaneously. The ice pack is opening and compressing, new ice is formed in the leads, and undeformed ice is thickening. The first-year ridges are weathered at the surface, consolidated internally \cite{Lepparanta:1995,Amundrad:2006}, but also melted at their bases due to larger oceanic heat fluxes at the depth of keels \cite{Wadhams:1992}.

In the perennial ice pack, the most prominent process during the summer period is the formation of melt ponds. Melt ponds effectively absorb solar radiation and accelerate the melting of ice since their albedo can be as low as 0.2 \cite{Curry:1995}; in comparison, the albedo of snow-cover pack ice is 0.77--0.87. \textcite{Pedersen:2009} have shown that the inclusion of parametrization of the melt ponds to climate models increases the sensitivity of the model and they suggest that this is the main reason why climate models are not capturing the present observed accelerated melting of the Arctic ice pack. 

Persistent drift of ice transports ice to other climatic regions. In the Antarctic, the flow is predominantly towards milder oceanic conditions, resulting in only 1--2 years lifetime for the ice. In the Arctic, the transpolar drift carries pack ice from the North American and Siberian shelf regions to the Greenland Sea via the North Pole, but part of the sea ice is recirculated in the Beaufort Sea gyre. In the 1980s the oldest individual floes were estimated to be 20--30 years old and most of the pack ice was over 10 years old \cite{Rigor:2004}.  Today, the lifetime of Arctic pack ice is considerably less; most of the ice in the Arctic is first- and second-year ice and older ice exists only in the Lincoln and Beaufort Seas \cite{Rigor:2004,Haas:2008}.

The state of the atmospheric circulation, which is the main driving force of sea-ice drift, is commonly described by empirical orthogonal functions (EOFs) of the surface air pressure. The two leading EOFs are the Arctic Oscillation (AO) \cite{Thompson:2000} and the Dipole Anomaly (DA) \cite{JiaWang:2009}. The variability of the AO/DA on a decadal scale affects to a great extent sea-ice circulation and mean sea-ice conditions \cite{Rigor:2002,Rothrock:2005,JiaWang:2009}, but also the form of the ice-thickness distribution function \cite{Oikkonen:2011}.

The simplifications to sea-ice physics made in different climate models have considerable implications. Many climate models still use rather simplified sea-ice physics, but great progress in the development of the sea-ice component of climate models has been made very recently. For example, none of the the models selected for the Arctic Climate Impact Assessment (ACIA) resolve sea-ice physics based on the full primitive equations, but now both the Community Climate System Model \cite{Holland:2006} and the Hadley Center climate model \cite{Johns:2006} use the explicitly calculated evolution of several ice categories based on the theory of the sea-ice thickness distribution \cite{Thorndike:1975,Bitz:2001}. \textcite{Holland:2006} concluded that the inclusion of the ice-thickness distribution model in climate simulations enhances the ice albedo feedback and the response of sea-ice thickness to external perturbations.

\subsection{A promoter of the emergence of the first life?}\label{life}

This review is on ice physics, not biology; it is not our intention to provide a review of physics--biogeochemistry interaction processes and we refer readers interested in that topic to \textcite{Thomas2009}.
However, the emergence of life from an abiotic environment is an interdisciplinary question of tremendous interest that involves important physical and chemical considerations in the self-organization necessary for life to arise; it is the physics and chemistry of biogenesis in ice that we concentrate on here.
 
Many different environments and processes facilitating the first steps to life have been proposed \cite{orgel1998,schopf2002,davies2003,schulze2008,bernstein2006_2}. As \textcite{orgel1998} has commented,
``the problem of the origin of life on the Earth has much in common with a well-constructed detective story. There is no shortage of clues pointing to the way in which the crime, the contamination of the pristine environment on the early Earth, was committed. On the contrary, there are far too many clues and far too many suspects. It would be hard to find two investigators who agree on even the broad outline of events.''
Among these many suspects, we may distinguish those that involve a hot origin of life on Earth from another group favoring a cold origin. One hypothesis that has been put forward within the context of a cold origin of life --- one that interests us in the present setting --- is that life may have originated in ice. 

\textcite{price2007} provides a very broad base of references defending the possibility of ice as promotor for life. While, as we mentioned in Section~\ref{astrophysical}, some researchers lean towards the extraterrestrial origins of prebiotic material  \cite{bernstein2006_2},  for which ``ices'' ---  astrophysical ices including but not limited to water ice --- are important, others consider the possible formation of pre-biotic material in sea ice \cite{trinks2005}. Here we discuss the arguments related to sea ice on the early Earth while remembering that these arguments pertain too to other similar astrophysical environments.

Owing to its being a multiphase system, sea ice reacts under small temperature variations with strong property changes. In this way sea ice renders many conditions and processes conducive to the generation of life, e.g., low temperature, the enrichment of compounds, the presence of small membranous compartments, chromatographic compound separation, temporal and spatial variations of pH, electrical potentials and ionic strength values together with catalytic effects on the surfaces of ice crystals and the presence of numerous tiny mineral particles always embedded in sea ice.  The above described properties of sea ice imply that it forms a complex system that may have some importance for some of the first steps towards the emergence of life.
 
Extant cells are autonomous: all properties and structures are an integral part of the cellular reproduction apparatus. But most researchers agree that the emergence of life was a process with many intermediate stages of increasing sophistication and complexity to reach this autonomous state. In the early steps to life, many features, such as the building of compartments and catalysis, presumably had to be provided by the environment, and in order to ensure the necessary continuity of the self-organization process that eventually led to life, had to be present in a quotidian manner. Investigations into the stability of RNA structures in cold environments \cite{moulton2000} and experiments performed at low temperatures under eutectic conditions \cite{stribling1991,liu1997,monnard2003,vlassov2004} indicate that  the sea-ice matrix might be an environment well suited to support processes that could finally lead to living systems. 

\begin{figure}[t]
\centering\includegraphics*[width=\columnwidth,clip=true]{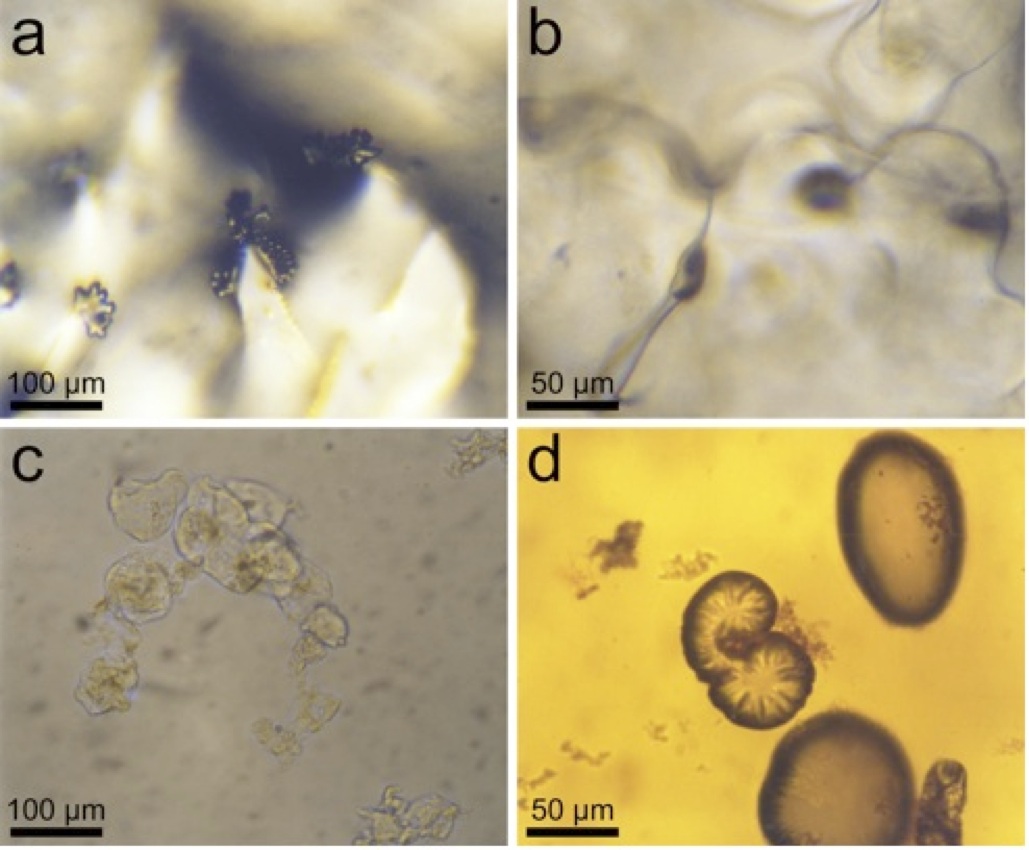} 
\caption{(Color online) Membranes and vesicles in ice: (a) Phosphatidylcholines distributed in sea water form membranes on ice surfaces and at the edges of mineral particles. (b) Formation of vesicles in sea ice by phosphatidylcholines during partial melting. (c) Peptide clusters of phenylalanine form vesicles in the ice. (d) Yeast RNA dissolved in sea water and frozen in sea ice forms vesicles under the influence of ultraviolet light. 
\label{trinks3}}
\end{figure}

Ice-reactor experiments have been carried out on artificially frozen sea water over time intervals of up to one year \cite{trinks2005}. Frozen sea water is a multiphase system which contains tiny ice clusters of frozen, salt-free water between which liquid, highly salty brine runs. Almost all ionic compounds have extremely low solubility in ice; as ice grows, ice polycrystals reject common ions into planar boundaries and, if their concentration is high enough, further concentrate them from two-grain boundaries into three-grain boundaries or veins (Section~\ref{ssec:struct_sur}). Small gas bubbles, and additionally, various mineral particles and salt crystals, are embedded in a huge number, approximately $10^{15}$~m$^{-3}$, of compartments with an overall surface area of about $10^5$~m$^2$ \cite{trinks2005}. The boundary layers between the solid clusters of ice and brine consists of flexible, less than 1~$\mu$m thick, membranes \cite{engemann2004}, which exist above temperatures of 200~K \cite{wei2001} and are stable for several minutes even after the ice has completely melted. Potential differences of up to 50~mV and local gradients of pH from 6 to 8 have been observed \cite{mazzega1976,bronshteyn1991}. The strong ultraviolet light that irradiated the young Earth \cite{zahnle1982} will have influenced the upper layer of sea ice to a depth of about 10~cm. Under typical local temperature gradients of up to 10~K/hour and 10~K/m, pressure differences provoke a steady movement of the brine, which leads to chromatographic effects due to the interaction of dissolved compounds with the ice surfaces. Small temperature variations have drastic effects on the constitution of the multiphase system of sea ice. They lead to sudden melting and freezing, de-gassing and dissolving of gas, crystallization and re-dissolving.

The concentration of prebiotic compounds in the oceans of the young Earth is estimated to have been less than a few mg/l \cite{stribling1986} and effective enrichment processes must have been indispensable steps towards self-organization. As we discussed above, when sea water freezes, dissolved compounds are enriched due to the growth of fresh-water ice, while salts and other ingredients concentrate in the liquid portion that fills the small cavioles. The concentration of this brine is enriched by a factor of 5 to 10, compared to unfrozen sea water, when temperatures of -15$^o$C to -20$^o$C are reached. At temperatures below -25$^o$C, and depending on their concentration related to the inorganic load, precipitation of organic ingredients (e.g., amino acids or RNA) starts. The dissolved ingredients have been observed to form aggregates in brine. 
Biomacromolecules like RNA spontaneously hydrolyze rather than polymerize in dilute aqueous solutions; however, because of their low water activity, veins in polycrystalline ice promote polymerization rather than hydrolysis. Due to the cyclical changes of temperature in natural sea ice these precipitates partially redissolve and precipitate once more, favoring the growth of more crystalline structures. In this way dissolved ingredients in the sea water, though low in concentration ($<1$~mg/l) under the freezing ice sheet, might form solid aggregates with typical local densities approaching approximately 1~kg per liter;  constituting an enrichment of compounds by a factor of $10^6$  \cite{trinks2005}.

Separating the melange of prebiotic compounds is essential for the synthesis of larger molecular structures, because some molecules inhibit the synthesis of others \cite{shapiro2000}. Sea ice is an environment with some separation power; it has the properties of a chromatograph in which the ice surface acts as the stationary phase and the liquid brine as the mobile phase \cite{dasgupta1997}. The experiments that have been conducted show that this is not enough to separate totally, for example, amino acids from carboxylic acids or amines; important for some prebiotic synthetic steps. Nevertheless, separation processes in sea ice, in addition to supposed similar other features in the environment of the young Earth, may have been instrumental in priming the generation of life. 

\begin{figure}[t]
\centering\includegraphics*[width=0.97\columnwidth,clip=true]{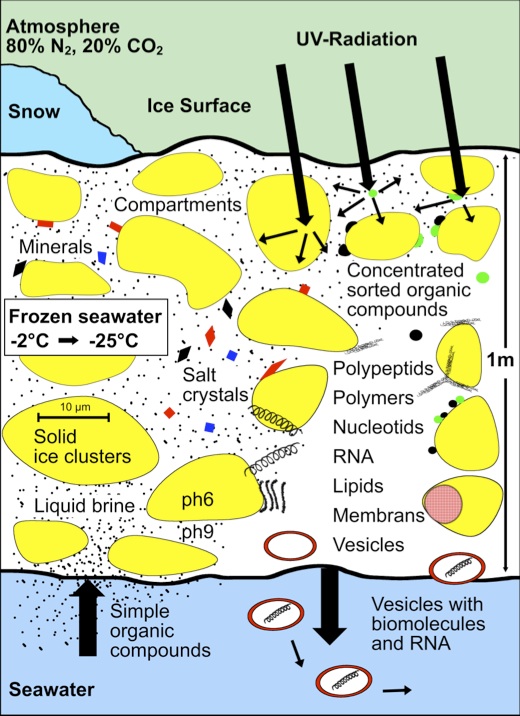} 
\caption{(Color online) Proposed life-promoting reactions in sea ice:
Simple prebiotic organic compounds were transported from sea water into the body of the ice (typical thickness 1--2~m); in the icy medium the reagents were sorted, separated and concentrated; chemical reactions led to polypeptides, RNA structures and membranes; the membranes formed vesicles embedding bio-molecules like RNA and others; and finally the vesicles left the ice and floated into the sea water beneath \cite{trinks2005}.
\label{trinks4}}
\end{figure}

The reaction of prebiotic molecules will have preferably taken place in an environment which supported synthetic processes. Condensation is a typical reaction in biological systems and should have played an important part in the prebiotic chemistry to build polymers from amino acids, RNA precursors or others. The growth of pure water crystals in sea ice during freezing can be regarded as a water consuming process possibly favoring such condensation reactions. In addition, a high concentration of NaCl has been found to initiate the condensation of amino acids into peptides, known as the salt-induced peptide formation (SIPF) reaction \cite{schwedinger1995,rode1999}. The surfaces of ice, as well as of mineral particles, always present in sea ice, are known for their catalytic properties \cite{ferris1998,graham2000}. Thus catalysis may also be considered to be involved in reactions in sea ice. 

Experiments in sea ice have been performed with solutions of alanine, phenylalanine, serine, tryptophan and isoleucine. After a few days of freezing under ultraviolet light, a material supposed to be peptides of high molecular weight was generated and formed a reticular structure that filled the cavioles and was partly attached to their walls. After complete melting of the ice the amino-acid condensate was found to remain as a reticular structure resistant to dissolution and even to hydrolysis in hydrochloric acid \cite{trinks2005}.

Biologically active molecules that are synthesized need to remain together in cellular structured compartments to be able to act in further evolutionary steps, e.g., to take part in self-organization during cyclic chemical processes, as proposed by \textcite{eigen1979}.  A great variety of cellular envelopes is contained in the sea-ice matrix. The channels at low temperatures ($< -20^o$C) reach typical dimensions of about 10--50~$\mu$m in diameter. The step from the wall-coated ice cells to the generation of autonomous vesicles composed of polymers from prebiotic molecules appears to take place surprisingly easily when ice melts \cite{trinks2005}.  During freezing, membrane-like coverings are constructed that follow the contours of the small ice channels (Fig.~\ref{trinks3}a, b). During melting these layers readily form clusters or vesicles (Fig.~\ref{trinks3}c) that float into the aqueous medium without being redissolved (Fig.~\ref{trinks3}d). 

Thus the freezing processes of sea water support chemical reactions, enrich and sort prebiotic compounds, and provide a medium for vesicle formation. Sea ice may be a matrix that is able to promote essential steps on the way to life; see Fig.~\ref{trinks4} for a sketch. Sea-ice vesicles, filled with the products of ice-matrix-supported prebiotic chemical processes, and released into the prebiotic aqueous environment underneath the ice sheet, may have floated in sea water to other locations on Earth until finally a cell fit for life had evolved.

As with several other themes we discuss in this paper, this issue could have appeared in another place, in the astrophysical ice section, Section~\ref{astrophysical}; as we mentioned there, similar ideas are being discussed in the astrophysical ice community with regard to the possible emergence of life on icy bodies in the solar system and beyond.  The presence of water is widely held to be crucial for the generation of life anywhere in the universe. While the role of ice in the emergence of life on Earth is not yet clear, there are many other places in the universe in which it may be of relevance. Planets and moons are, in many cases, thought to possess ice in their polar regions (e.g., Mars), whilst others seem to be covered by frozen oceans (e.g., Europa), and cometary ice may contain water from radioactive heating. Just as life is frequently preserved in the water beneath the icy crust of a frozen pond in winter, so perhaps there may be life in the ice-coated oceans of other worlds that can exist because of this quirk of ice physics that ice Ih floats on water. It may then be that ideas on ice as a cradle for life may be of interest in extraterrestrial contexts (see, e.g., \textcite{price2010}).

\section{Perspectives}\label{perspectives}

\begin{quote}
If you watch a glacier from a distance, and see the big rocks falling into the sea, and the way the ice moves, and so forth, it is not really essential to remember that it is made out of little hexagonal ice crystals. Yet if understood well enough the motion of the glacier is in fact a consequence of the character of the hexagonal ice crystals. But it takes quite a while to understand all the behavior of the glacier (in fact nobody knows enough about ice yet, no matter how much they've studied the crystal). However the hope is that if we do understand the ice crystal we shall ultimately understand the glacier.

Richard Feynman, \\ \emph{The Character of Physical Law}
\end{quote}

So which are the main open questions in research on ice today? 

If we begin by examining the issue of molecular structures, we find that there are even uncertainties remaining concerning some aspects of the ice phase diagram, especially in relation to understanding the mechanism of hydrogen ordering by acid or base doping and obtaining definitive information on the various ground-state structures. The possible existence of small ordered nuclei within the disordered phases is an interesting speculation. So-called cubic ice remains a particularly interesting problem, one which perhaps has some relation to the polytypes observed in systems such as silicon carbide. And understanding the details of both the molecular-level and intermediate-range structures of the amorphous ices is a particularly problematical issue that is far from being resolved.

In the field of astrophysical ice, the forthcoming years will be marked by the collection of new and exciting data from several space missions sent to explore Solar System bodies. Among them, it is worth mentioning the New Horizons mission (NASA) presently heading towards Pluto through the Kuiper belt and Rosetta (ESA), a spacecraft that will accompany comet Churyumov--Gerasimenko for more than a year and that will deploy a lander, loaded with nine instruments, on the surface of the comet; as we go to press, the Messenger mission to Mercury and the Dawn mission to 4 Vesta and Ceres will likely answer the questions of the presence of ice on these bodies. Data collected on such space missions will be added to the large amount of information already gathered by the very successful instruments used to explore space, like the space telescope Hubble and the Cassini--Huygens mission, to name but a couple. This wealth of knowledge should enable the astrophysical community to address some of the most intriguing questions related to astrophysical ice yet unanswered. For instance, what is the structure of cometary nuclei: do they contain crystalline as well as amorphous water ice, and are there other volatile species trapped in the ice?  Will the presence of clathrates be definitely confirmed, and if so, in which type of ice systems and astrophysical bodies? The unravelling of these mysteries will render the near future a very exciting time for research in astrophysical ices.

In the longer term, there is also the interesting question of the presence of the various ice phases in astrophysical environments. At present, almost everything remains speculation, as there is no significant difference between the infrared spectra of ice Ih and ice Ic, and high-pressure ices inside planets or moons are not accessible to remote-sensing IR spectroscopy. The conditions for the formation of any of the high-pressure phases of ice are unlikely to occur naturally anywhere on Earth; or perhaps that is too hasty a judgment? \textcite{bina2000} suggest that ice VII might be present in subduction zones of the Earth's crust. Elsewhere in the Solar System, many conditions could well exist under which these high-pressure phases could have formed. For example, it has been argued \cite{sohl2002} that the interior of Jupiter's moon Ganymede could consist of successive layers of ices Ih, III, V and VI. Similarly, the other largest icy Solar System moons, Callisto and Titan, might have ice VI in their depths, and layers of other ice phases in onion fashion towards the surface. The centers of somewhat smaller bodies, like the trans-Neptunian objects Pluto and Eris, might too consist of a mixture of rock and high-pressure ices. 
These high-pressure phases are all disordered phases. What about the ordered ones, whose presence could be astrophysically significant bearing in mind the general differences in physical (e.g., mechanical) properties of ordered and disordered phases? At present, it seems unlikely that the ordered phases will be found in the Solar System, given the ranges of temperature and pressure expected in icy Solar System bodies. So perhaps we shall have to wait for developments in exoplanets, but surely, somewhere in space, in some body orbiting some distant star or planet, the pressure and temperature conditions are right for ice XV?

An interesting related question to that of astrophysical ice phases, around which work is ongoing \cite{durham2001,greve2006}, is about the rheology or glaciology of ices in a planetary context; e.g., are there glaciers on Pluto? The deformation and flow of ice at low temperature is most likely qualitatively different to that on Earth, so that Glen's flow law cannot be extrapolated in a simple manner to low temperatures. As well as ice Ih, one has to consider flow in amorphous ices as well as in ice Ic and the high-pressure phases, and grain size may well affect flow to a considerable extent.  It will be very useful to have experimental results from the lab to compare with models, but at present there are no low-temperature data available, to say nothing of data at low temperature and high pressure, and they will be very difficult and time-consuming to obtain. Astrophysical ice rheology is important to tell us the minimum size at which an icy body becomes round under its own gravity; the figure is likely different than for a rocky body, and this question of roundness impinges on the current definition of a dwarf planet.

Further planetary geophysical issues include cryovolcanism and cryotectonics.  Cryovolcanoes, or ice volcanoes, are structures that are proposed to function analogously to terrestrial volcanoes, but their eruptions are of molten ice --- water --- rather than lava --- molten rock. These have been sought on various bodies in the solar system; at present, it is thought that features seen on Titan are good candidates, but unambiguous proof for cryovolcanism anywhere in the solar system is still lacking. Cryotectonics --- plate tectonics with ice --- is clearly a possibility for bodies that have an icy crust over a liquid water layer undergoing convection.  Might mixtures of ices permit there to be oceanic and continental plates as with terrestrial plate tectonics? That is as yet speculation, but cryotectonic activity has been observed on Enceladus, at ice fractures near its south pole. Europa too may have tectonic activity on what is believed to be an ice crust over an ocean of water; moreover, among many speculations, one is that some features on the Europan surface may be due to the ice crust itself convecting in the sort of slow flow we discussed above. Lastly, there is the puzzle of the ice ridge on Iapetus; this mountainous ridge, up to 20~km high, circles the equator on the Saturnian moon of diameter 1400~km, which consequently looks like a walnut. No theory for its formation seems completely satisfactory; it remains a mystery.

A great many open questions remain to be answered in the fascinating field of research on ices in the universe and their relationship with the evolution of matter and eventually life on Earth (and beyond?). Among them we may highlight:  Is the structure --- and thence the spectral properties --- of the interstellar water ice-rich mantles well reproduced in laboratory experiments? In the latter water ice is usually accreted from the gas phase onto cold substrates. In space oxygen and hydrogen atoms recombine by surface chemical reaction on the grain surfaces. Laboratory experiments focused on the study of the spectral characteristic of ices obtained by surface reactions are badly necessary. Has the water ice that dominates the crust of the bodies in the outer Solar System a terrestrial analog?; what we really observe is the outer skin (from micrometers to millimeters depending on the wavelength) of a given object. The structure and composition of those layers are continuously altered by ion irradiation and micrometeorite bombardment. Can we reproduce something similar on Earth? Comets are an aggregate of ices and dust. At which ice--dust ratio do the physical properties of ices change significantly? How to measure parameters such as thermal conductivity, compressibility, etc, for such a macroscopic (kilometer-scale) aggregate?   

The molecular structure sets the stage from which physical properties follow like the anisotropy of ice Ih with respect to deformation. What is the most stable structure? This is not so clear in ice; many competing solutions exist for minimizing free energy. This comes into play when nucleating the frequently occurring metastable structures. Ice nucleation is far from being understood; nor is the growth of ice crystals, in particular under atmospheric conditions. 
One of the main open questions in understanding Earth's radiation balance, and thus climate change, is related with ice nucleation. Atmospheric ice nuclei can trigger cloud formation and precipitation, having an impact on both the global radiation balance and the water cycle. The impact on the balance is still debated, as, on the one hand, ice clouds have a higher albedo than liquid water, so they should cool the climate \cite{mishcenko1996}, but, on the other, they are more likely to precipitate and so reduce total cloud albedo \cite{lohmann2002}. The light scattering and reflection of ice particles in the atmosphere is not yet well described, since these optical effects depend on the particle morphology (size and shape), which often varies with the atmospheric conditions and is difficult to model. 

In general, the entire microphysics of ice clouds is still under consideration. In particular heterogeneous ice nucleation is not well parametrized. For many nuclei, even the nucleation rates are not available. This is especially true for biological aerosols, which only recently have moved into the center of interest of the atmospheric ice community. Depending on the nucleation mechanism, morphology and phase composition might also change. The latter is also a problem for field measurements, since up to now there are no instruments available that can determine the crystallographic nature of an ice or hydrate phase beyond the well-known, but sometimes not unequivocal, spectroscopic specifications. Similar difficulties exist for obtaining nanometric images of atmospheric snow crystals; such microstructural information would be needed, for example, to confirm and quantify the suggested role of complex snow-crystal morphology on vapor pressure and chemical reactivity (e.g., \textcite{Gao2004}). We have good evidence of ice surfaces being important for various chemical reactions, but there remain many common atmospheric substances, from the biogenic to anthropogenic, whose chemical behavior remains to be fully quantified --- yet this leads us outside the field of our review.

Snow forms a wealth of different growth morphologies for which our understanding is mostly only qualitative. Morphologies are important regarding the sedimentation speed \cite{wang2002} of snow crystals and their interaction with atmospheric constituents during formation and fall, but also on the ground. The transfer functions from atmosphere to snow are often known only approximately while they are of primary importance for decoding the information contained in deep ice cores to reconstruct climate history. Related to this is the question of where exactly photo-reactions in environmental ice take place and how these can best be parameterized in models. Morphologies are also important for our understanding of processes in planetary science and astrophysics; again photochemistry on ice surfaces is one of the open fields in this context.

In terrestrial ice open questions range from the macroscopic behavior of large ice sheets down to the microscopic uptake of atmospheric trace gases: Generally, the flow of glaciers and its relation to microscopic ice properties is qualitatively well understood \cite{MSchulson:2009p26776}. However, important details on how the microscopic properties of single ice crystals link to the creep behavior of polycrystalline ice are still unclear. Self-consistent models developed in the field of materials science have now been applied to ice (e.g., \textcite{lebensohn2009}), yet dynamic recrystallization --- important for larger glaciers and ice sheets --- still awaits its full implementation. Theoretical advances in modeling are helped by new techniques like cryo-EBSD \cite{obbard2006} which are starting to provide access to microstructural details of recrystallization processes. Linking microstructural information to geophysical signals such as radar imaging and seismic data is another formidable challenge in ice-core research; if successful it would enable remote mapping of microstructural effects in large ice sheets \cite{drews2009}. Also open to debate are the reasons that explain the observed accelerated glacier flow along polar coasts and the role of sea ice. In this respect it has been proposed that thermal convection may play a role in the origin of ice streams \cite{Hughes:2009}. Also related to glaciers are open questions regarding the sintering process of glacial snow. A key question is to understand quantitatively the time offset of ice and neighboring closed-off air bubbles in deep ice cores. Deeper understanding of the dynamics during the transformation of glacial snow to ice --- where air bubbles are finally trapped --- would help to understand better the timing at which various possible climate forcings acted in the past. Small-scale dynamics are also of interest in understanding in detail the mechanism of icicle formation \cite{chen2011,neufeld2010}.

With regard to sea ice, the most important research questions are:  What mechanisms control variability and change of sea ice? What is the relationship between large- and small-scale processes? What governs the processes of interaction of snow--sea-ice--atmosphere? What is the nature of the interaction between the sea ice and radiation? How do biogeochemical processes depend on the physical properties of sea ice?  
Having in mind the well-documented changes now occurring in the ice pack of the Arctic Ocean, open questions remain on the differentiation of anthropogenic forced climate signals from natural variability, the separation of the relative importance of atmospheric and oceanic forced change, and on understanding what mechanisms control the variability and change of sea ice. Our lack of a comprehensive understanding is partly due to limited observational capabilities, and new satellite missions like CryoSat-2 and enhanced ocean monitoring activities will provide much more accurate and more numerous data to explore the physics of the Arctic climate system. Small-scale processes are known to be important also at the climate scale, but these processes are usually neglected or poorly parametrized in the large-scale models, and the  phenomena they generate have received hardly any attention.  To give just one example, consider sub-ice convection, during which ice may form around the draining cold brine plumes under floating ice, producing tapered hollow structures up to meters in length termed sea-ice stalactites  \cite{paige1970}.  
A fascinating open question of sea-ice physics is to establish a theoretical framework between the properties of ice, local-scale processes (e.g., fracturing, floe--floe friction, dissipation of kinetic energy during deformation), and large-scale ice motion. This kind of research could enable a statistical--dynamical parametrization of the sub-grid scale processes for climate models.

We should mention here the methane hydrates --- sometimes referred to as ``burning ice'' due to a number of similarities with ice Ih --- that exist in sediments of the ocean floor. In these crystalline compounds the water molecules form a three-dimensional structure consisting of cages containing mainly methane molecules, justifying the term ``clathrates". The amount of methane stored in hydrates is considered to be at least comparable to that in existing natural gas deposits, but it is unclear at present whether and how much of these gas hydrates are economically recoverable. These clathrates are stable only under the high pressures and at the fairly low temperatures of a few degrees Celsius prevailing in sediments on the sea floors of continental margins. The study and control of the stability of methane hydrates is thus likely of importance for their use as an energy resource. Due to their potential nuisance  --- upon decomposition methane gas may escape in large amounts into the atmosphere --- methane hydrates are also a concern in environmental studies and climate change predictions. Gas hydrates form a very large and independent field of research \cite{sloan2008}; in the marine context on Earth they belong to the cryosphere only marginally, via air clathrates in ice sheets and the likely occurrence of methane hydrates within permafrost; however in the planetological context the distinction between ice and clathrate hydrates is less strict. Conclusive evidence for the existence of gas hydrates on celestial bodies is still missing, since with infrared spectroscopy they are very hard to distinguish from a mixture of water ice and gases like CO$_2$. Moreover, the formation kinetics are very slow at low temperatures. A fair amount of discussion on the role of CO$_2$ clathrate hydrates in the planetological context can be found in \textcite{kargel2004}, but see also a more critical appreciation in \textcite{falenty2011}.

In our view across the ice-fields we have seen that there are many open questions in different aspects of our knowledge of the behavior of this small molecule, water, in its solid state as ice. What is more, many of these frontier questions on ice structure, patterns and processes are interrelated; multidisciplinary and interdisciplinary research is needed to resolve them. In this regard, an approach we commend, which may be applied fruitfully to ice (e.g., \textcite{faria2009}), is the study of the interaction of internal structure, form, and the environment.  \textcite{feynman}, whose words we set at the start of this section, was extremely prescient in highlighting this multiscalar approach from the atomic scale to geological scales. 

We conclude by reproducing the first verse of a poem \textcite{maxwell1871} composed for the meeting of the British Association for the Advancement of Science in 1871 \cite{booth1999}. In just a few lines he mentions three physical properties of this complex substance --- pressure melting or regelation, its tendency to form dendritic crystals, and veins and nodes within a polycrystal --- that still interest us.  Today, as in the 19th century, ice physics continues to fascinate.
\begin{quote}
I come from fields of fractured ice, \\
Whose wounds are cured by squeezing, \\
Melting they cool, but in a trice, \\
Get warm again by freezing. \\
Here, in the frosty air, the sprays \\
With fernlike hoar-frost bristle, \\
There, liquid stars their watery rays \\
Shoot through the solid crystal.

James Clerk Maxwell, \\ \emph{To The Chief Musician Upon Nabla: \\ A Tyndallic Ode}
\end{quote}

\begin{acknowledgments} 
We thank Christoph Salzmann for Figs~\ref{john4} and \ref{john5},  Larry Wilen for 
Fig.~\ref{fig:thinsect}, Heidy Mader for Fig.~\ref{fig:veins}a, Martin Doble for Fig.~\ref{jari4}, Mike Steele for Fig.~\ref{finger_rafting}, and Buford Price and David Thomas for useful discussions.  
This article is fruit of the European Science Foundation workshop Euroice 2008 held in Granada, Spain from 1--4 October 2008 attended by most of the present authors; we thank the ESF for its financial support. That workshop was itself born from an earlier Spanish national project, Hielocris, financed by the Consejo Superior de Investigaciones Cient\'{\i}ficas; we thank the CSIC for its financial support. We acknowledge funding in addition from: FWF, Austria (P23027); MINCINN, Spain (FIS2010-16455, PR2010-0012,  FIS2010-22322-528C02-02); SNSF, Switzerland (200021 121857). Neutron scattering studies have played a major role in molecular-level ice physics and we thank ILL (France) and ISIS (UK) for the neutron beam time that has made much of that work possible.
\end{acknowledgments}


\begin{thebibliography}{604}%
\makeatletter
\providecommand \@ifxundefined [1]{%
 \ifx #1\undefined \expandafter \@firstoftwo
 \else \expandafter \@secondoftwo
\fi
}%
\providecommand \@ifnum [1]{%
 \ifnum #1\expandafter \@firstoftwo
 \else \expandafter \@secondoftwo
\fi
}%
\providecommand \natexlab [1]{#1}%
\providecommand \enquote [1]{``#1''}%
\providecommand \bibnamefont  [1]{#1}%
\providecommand \bibfnamefont [1]{#1}%
\providecommand \citenamefont [1]{#1}%
\providecommand\href[0]{\@sanitize\@href}%
\providecommand\@href[1]{\endgroup\@@startlink{#1}\endgroup\@@href}%
\providecommand\@@href[1]{#1\@@endlink}%
\providecommand \@sanitize [0]{\begingroup\catcode`\&12\catcode`\#12\relax}%
\@ifxundefined \pdfoutput {\@firstoftwo}{%
 \@ifnum{\z@=\pdfoutput}{\@firstoftwo}{\@secondoftwo}%
}{%
 \providecommand\@@startlink[1]{\leavevmode\special{html:<a href="#1">}}%
 \providecommand\@@endlink[0]{\special{html:</a>}}%
}{%
 \providecommand\@@startlink[1]{%
  \leavevmode
  \pdfstartlink
   attr{/Border[0 0 1 ]/H/I/C[0 1 1]}%
   user{/Subtype/Link/A<</Type/Action/S/URI/URI(#1)>>}%
  \relax
 }%
 \providecommand\@@endlink[0]{\pdfendlink}%
}%
\providecommand \url  [0]{\begingroup\@sanitize \@url }%
\providecommand \@url [1]{\endgroup\@href {#1}{\urlprefix}}%
\providecommand \urlprefix [0]{URL }%
\providecommand \Eprint[0]{\href }%
\@ifxundefined \urlstyle {%
  \providecommand \doi [1]{doi:\discretionary{}{}{}#1}%
}{%
  \providecommand \doi [0]{doi:\discretionary{}{}{}\begingroup
  \urlstyle{rm}\Url }%
}%
\providecommand \doibase [0]{http://dx.doi.org/}%
\providecommand \Doi[1]{\href{\doibase#1}}%
\providecommand \bibAnnote [3]{%
  \BibitemShut{#1}%
  \begin{quotation}\noindent
    \textsc{Key:}\ #2\\\textsc{Annotation:}\ #3%
  \end{quotation}%
}%
\providecommand \bibAnnoteFile [2]{%
  \IfFileExists{#2}{\bibAnnote {#1} {#2} {\input{#2}}}{}%
}%
\providecommand \typeout [0]{\immediate \write \m@ne }%
\providecommand \selectlanguage [0]{\@gobble}%
\providecommand \bibinfo [0]{\@secondoftwo}%
\providecommand \bibfield [0]{\@secondoftwo}%
\providecommand \translation [1]{[#1]}%
\providecommand \BibitemOpen[0]{}%
\providecommand \bibitemStop [0]{}%
\providecommand \bibitemNoStop [0]{.\EOS\space}%
\providecommand \EOS [0]{\spacefactor3000\relax}%
\providecommand \BibitemShut [1]{\csname bibitem#1\endcsname}%
\bibitem[{\citenamefont{Abbatt}(2003)}]{abbatt2003}%
  \BibitemOpen
  \bibfield{author}{%
  \bibinfo {author} {\bibnamefont{Abbatt}, \bibfnamefont{J.~P.~D.}}}%
  , \bibinfo {year} {2003},\ \bibfield{title}{%
  \enquote{\bibinfo {title} {Interactions of atmospheric trace gases with ice
  surfaces: Adsorption and reaction},}\ }%
  \bibfield{journal}{%
  \bibinfo {journal} {Chem. Rev.}\ }%
  \textbf{\bibinfo {volume} {103}},\ \bibinfo {pages} {4783--4800}%
  \bibAnnoteFile{NoStop}{abbatt2003}%
\bibitem[{\citenamefont{Abbatt}\ \emph{et~al.}(2008)\citenamefont{Abbatt},
  \citenamefont{Bartels-Rausch}, \citenamefont{Ullerstam},\ and\
  \citenamefont{Ye}}]{Abbatt:2008p20344}%
  \BibitemOpen
  \bibfield{author}{%
  \bibinfo {author} {\bibnamefont{Abbatt}, \bibfnamefont{J.~P.~D.}}, \bibinfo
  {author} {\bibfnamefont{T.}~\bibnamefont{Bartels-Rausch}}, \bibinfo {author}
  {\bibfnamefont{M.}~\bibnamefont{Ullerstam}},\ and\ \bibinfo {author}
  {\bibfnamefont{T.~J.}\ \bibnamefont{Ye}}}%
  , \bibinfo {year} {2008},\ \bibfield{title}{%
  \enquote{\bibinfo {title} {Uptake of acetone, ethanol and benzene to snow and
  ice: effects of surface area and temperature},}\ }%
  \bibfield{journal}{%
  \bibinfo {journal} {Environ. Res. Lett.}\ }%
  \textbf{\bibinfo {volume} {3}},\ \bibinfo {pages} {045008}%
  \bibAnnoteFile{NoStop}{Abbatt:2008p20344}%
\bibitem[{\citenamefont{Al-Halabi}\
  \emph{et~al.}(1999)\citenamefont{Al-Halabi}, \citenamefont{Kleyn},\ and\
  \citenamefont{Kroes}}]{alhalabi1999}%
  \BibitemOpen
  \bibfield{author}{%
  \bibinfo {author} {\bibnamefont{Al-Halabi}, \bibfnamefont{A.}}, \bibinfo
  {author} {\bibfnamefont{A.~W.}\ \bibnamefont{Kleyn}},\ and\ \bibinfo {author}
  {\bibfnamefont{G.~J.}\ \bibnamefont{Kroes}}}%
  , \bibinfo {year} {1999},\ \bibfield{title}{%
  \enquote{\bibinfo {title} {New predictions on the sticking of {HCl} to ice at
  hyperthermal energies},}\ }%
  \bibfield{journal}{%
  \bibinfo {journal} {Chem. Phys. Lett.}\ }%
  \textbf{\bibinfo {volume} {307}},\ \bibinfo {pages} {505--510}%
  \bibAnnoteFile{NoStop}{alhalabi1999}%
\bibitem[{\citenamefont{Al-Halabi}\
  \emph{et~al.}(2001)\citenamefont{Al-Halabi}, \citenamefont{Kleyn},\ and\
  \citenamefont{Kroes}}]{alhalabi2001}%
  \BibitemOpen
  \bibfield{author}{%
  \bibinfo {author} {\bibnamefont{Al-Halabi}, \bibfnamefont{A.}}, \bibinfo
  {author} {\bibfnamefont{A.~W.}\ \bibnamefont{Kleyn}},\ and\ \bibinfo {author}
  {\bibfnamefont{G.~J.}\ \bibnamefont{Kroes}}}%
  , \bibinfo {year} {2001},\ \bibfield{title}{%
  \enquote{\bibinfo {title} {Sticking of {HCl} to ice at hyperthermal energies:
  Dependence on incidence energy, incidence angle, and surface temperature},}\
  }%
  \bibfield{journal}{%
  \bibinfo {journal} {J. Chem. Phys.}\ }%
  \textbf{\bibinfo {volume} {115}},\ \bibinfo {pages} {482--491}%
  \bibAnnoteFile{NoStop}{alhalabi2001}%
\bibitem[{\citenamefont{Allamandola}\
  \emph{et~al.}(1992)\citenamefont{Allamandola}, \citenamefont{Sandford},
  \citenamefont{Tielens},\ and\ \citenamefont{Herbst}}]{allamandola1992}%
  \BibitemOpen
  \bibfield{author}{%
  \bibinfo {author} {\bibnamefont{Allamandola}, \bibfnamefont{L.~J.}}, \bibinfo
  {author} {\bibfnamefont{S.~A.}\ \bibnamefont{Sandford}}, \bibinfo {author}
  {\bibfnamefont{A.~G. G.~M.}\ \bibnamefont{Tielens}},\ and\ \bibinfo {author}
  {\bibfnamefont{T.~M.}\ \bibnamefont{Herbst}}}%
  , \bibinfo {year} {1992},\ \bibfield{title}{%
  \enquote{\bibinfo {title} {Infrared spectroscopy of dense clouds in the {C-H}
  stretch region --- methanol and ``diamonds''},}\ }%
  \bibfield{journal}{%
  \bibinfo {journal} {Astrophys. J.}\ }%
  \textbf{\bibinfo {volume} {399}},\ \bibinfo {pages} {134--146}%
  \bibAnnoteFile{NoStop}{allamandola1992}%
\bibitem[{\citenamefont{Alley}(1992)}]{Alley1992}%
  \BibitemOpen
  \bibfield{author}{%
  \bibinfo {author} {\bibnamefont{Alley}, \bibfnamefont{R.~B.}}}%
  , \bibinfo {year} {1992},\ \bibfield{title}{%
  \enquote{\bibinfo {title} {Flow-law hypotheses for ice-sheet modeling},}\ }%
  \bibfield{journal}{%
  \bibinfo {journal} {J. Glaciol.}\ }%
  \textbf{\bibinfo {volume} {38}},\ \bibinfo {pages} {245--256}%
  \bibAnnoteFile{NoStop}{Alley1992}%
\bibitem[{\citenamefont{Alpert}\ \emph{et~al.}(2011)\citenamefont{Alpert},
  \citenamefont{Aller},\ and\ \citenamefont{Knopf}}]{Alpert2011}%
  \BibitemOpen
  \bibfield{author}{%
  \bibinfo {author} {\bibnamefont{Alpert}, \bibfnamefont{P.~A.}}, \bibinfo
  {author} {\bibfnamefont{J.~Y.}\ \bibnamefont{Aller}},\ and\ \bibinfo {author}
  {\bibfnamefont{D.~A.}\ \bibnamefont{Knopf}}}%
  , \bibinfo {year} {2011},\ \bibfield{title}{%
  \enquote{\bibinfo {title} {Initiation of the ice phase by marine biogenic
  surfaces in supersaturated gas and supercooled aqueous phases},}\ }%
  \bibfield{journal}{%
  \bibinfo {journal} {Phys. Chem. Chem. Phys.}\ }%
  \textbf{\bibinfo {volume} {13}},\ \bibinfo {pages} {19882--19894}%
  \bibAnnoteFile{NoStop}{Alpert2011}%
\bibitem[{\citenamefont{Alvarez-Aviles}\
  \emph{et~al.}(2008)\citenamefont{Alvarez-Aviles}, \citenamefont{Simpson},
  \citenamefont{Douglas}, \citenamefont{Sturm}, \citenamefont{Perovich},\ and\
  \citenamefont{Domine}}]{Alvarez-Aviles:2008}%
  \BibitemOpen
  \bibfield{author}{%
  \bibinfo {author} {\bibnamefont{Alvarez-Aviles}, \bibfnamefont{L.}}, \bibinfo
  {author} {\bibfnamefont{W.~R.}\ \bibnamefont{Simpson}}, \bibinfo {author}
  {\bibfnamefont{T.~A.}\ \bibnamefont{Douglas}}, \bibinfo {author}
  {\bibfnamefont{M.}~\bibnamefont{Sturm}}, \bibinfo {author}
  {\bibfnamefont{D.}~\bibnamefont{Perovich}},\ and\ \bibinfo {author}
  {\bibfnamefont{F.}~\bibnamefont{Domine}}}%
  , \bibinfo {year} {2008},\ \bibfield{title}{%
  \enquote{\bibinfo {title} {Frost flower chemical composition during growth
  and its implications for aerosol production and bromine activation},}\ }%
  \bibfield{journal}{%
  \bibinfo {journal} {J. Geophys. Res.}\ }%
  \textbf{\bibinfo {volume} {113}},\ \bibinfo {pages} {D21304}%
  \bibAnnoteFile{NoStop}{Alvarez-Aviles:2008}%
\bibitem[{\citenamefont{Amundrud}\ \emph{et~al.}(2006)\citenamefont{Amundrud},
  \citenamefont{Melling}, \citenamefont{Ingram},\ and\
  \citenamefont{Allen}}]{Amundrad:2006}%
  \BibitemOpen
  \bibfield{author}{%
  \bibinfo {author} {\bibnamefont{Amundrud}, \bibfnamefont{T.~L.}}, \bibinfo
  {author} {\bibfnamefont{H.}~\bibnamefont{Melling}}, \bibinfo {author}
  {\bibfnamefont{R.~G.}\ \bibnamefont{Ingram}},\ and\ \bibinfo {author}
  {\bibfnamefont{S.~E.}\ \bibnamefont{Allen}}}%
  , \bibinfo {year} {2006},\ \bibfield{title}{%
  \enquote{\bibinfo {title} {The effect of structural porosity on the ablation
  of sea ice ridges},}\ }%
  \bibfield{journal}{%
  \bibinfo {journal} {J.~Geophys.~Res.}\ }%
  \textbf{\bibinfo {volume} {111}},\ \bibinfo {pages} {C06004}%
  \bibAnnoteFile{NoStop}{Amundrad:2006}%
\bibitem[{\citenamefont{Anastasio}\ and\
  \citenamefont{Robles}(2007)}]{Anastasio:2007p6837}%
  \BibitemOpen
  \bibfield{author}{%
  \bibinfo {author} {\bibnamefont{Anastasio}, \bibfnamefont{C.}},\ and\
  \bibinfo {author} {\bibfnamefont{T.}~\bibnamefont{Robles}}}%
  , \bibinfo {year} {2007},\ \bibfield{title}{%
  \enquote{\bibinfo {title} {Light absorption by soluble chemical species in
  {Arctic} and {Antarctic} snow},}\ }%
  \bibfield{journal}{%
  \bibinfo {journal} {J. Geophys. Res.}\ }%
  \textbf{\bibinfo {volume} {112}},\ \bibinfo {pages} {D24304}%
  \bibAnnoteFile{NoStop}{Anastasio:2007p6837}%
\bibitem[{\citenamefont{Andersson}\
  \emph{et~al.}(2007)\citenamefont{Andersson}, \citenamefont{Suter},
  \citenamefont{Markovi{\'c}},\ and\
  \citenamefont{Pettersson}}]{Andersson2007}%
  \BibitemOpen
  \bibfield{author}{%
  \bibinfo {author} {\bibnamefont{Andersson}, \bibfnamefont{P.~U.}}, \bibinfo
  {author} {\bibfnamefont{M.~T.}\ \bibnamefont{Suter}}, \bibinfo {author}
  {\bibfnamefont{N.}~\bibnamefont{Markovi{\'c}}},\ and\ \bibinfo {author}
  {\bibfnamefont{J.~B.~C.}\ \bibnamefont{Pettersson}}}%
  , \bibinfo {year} {2007},\ \bibfield{title}{%
  \enquote{\bibinfo {title} {Water condensation on graphite studied by elastic
  helium scattering and molecular dynamics simulations},}\ }%
  \bibfield{journal}{%
  \bibinfo {journal} {J. Phys. Chem. A}\ }%
  \textbf{\bibinfo {volume} {111}},\ \bibinfo {pages} {15258--15266}%
  \bibAnnoteFile{NoStop}{Andersson2007}%
\bibitem[{\citenamefont{Andrews}(2007)}]{JohnTAndrews:2007p26015}%
  \BibitemOpen
  \bibfield{author}{%
  \bibinfo {author} {\bibnamefont{Andrews}, \bibfnamefont{J.~T.}}}%
  , \bibinfo {year} {2007},\ \bibfield{title}{%
  \enquote{\bibinfo {title} {Glaciers, oceans, atmosphere and climate},}\ }%
  \bibinfo {journal} {Glacier Science and Environmental Change (First
  Edition)},\ \bibinfo {pages} {95--113}%
  \bibAnnoteFile{NoStop}{JohnTAndrews:2007p26015}%
\bibitem[{\citenamefont{Arakawa}\ \emph{et~al.}(2010)\citenamefont{Arakawa},
  \citenamefont{Kagi},\ and\ \citenamefont{Fukazawa}}]{arakawa2010}%
  \BibitemOpen
\bibfield{journal}{%
    }%
  \bibfield{author}{%
  \bibinfo {author} {\bibnamefont{Arakawa}, \bibfnamefont{M.}}, \bibinfo
  {author} {\bibfnamefont{H.}~\bibnamefont{Kagi}},\ and\ \bibinfo {author}
  {\bibfnamefont{H.}~\bibnamefont{Fukazawa}}}%
  , \bibinfo {year} {2010},\ \bibfield{title}{%
  \enquote{\bibinfo {title} {Annealing effects on hydrogen ordering in
  {KOD}-doped ice observed using neutron diffraction},}\ }%
  \bibfield{journal}{%
  \bibinfo {journal} {J. Molec. Structure}\ }%
  \textbf{\bibinfo {volume} {972}},\ \bibinfo {pages} {111--114}%
  \bibAnnoteFile{NoStop}{arakawa2010}%
\bibitem[{\citenamefont{Ariya}\ \emph{et~al.}(2011)\citenamefont{Ariya},
  \citenamefont{Domine}, \citenamefont{Kos}, \citenamefont{Amyot},
  \citenamefont{C{\^o}t{\'e}}, \citenamefont{Vali}, \citenamefont{Lauzier},
  \citenamefont{Kuhs}, \citenamefont{Techmer}, \citenamefont{Heinrichs},\ and\
  \citenamefont{Mortazavi}}]{ariya2011}%
  \BibitemOpen
  \bibfield{author}{%
  \bibinfo {author} {\bibnamefont{Ariya}, \bibfnamefont{P.~A.}}, \bibinfo
  {author} {\bibfnamefont{F.}~\bibnamefont{Domine}}, \bibinfo {author}
  {\bibfnamefont{G.}~\bibnamefont{Kos}}, \bibinfo {author}
  {\bibfnamefont{M.}~\bibnamefont{Amyot}}, \bibinfo {author}
  {\bibfnamefont{V.}~\bibnamefont{C{\^o}t{\'e}}}, \bibinfo {author}
  {\bibfnamefont{H.}~\bibnamefont{Vali}}, \bibinfo {author}
  {\bibfnamefont{T.}~\bibnamefont{Lauzier}}, \bibinfo {author}
  {\bibfnamefont{W.~F.}\ \bibnamefont{Kuhs}}, \bibinfo {author}
  {\bibfnamefont{K.}~\bibnamefont{Techmer}}, \bibinfo {author}
  {\bibfnamefont{T.}~\bibnamefont{Heinrichs}},\ and\ \bibinfo {author}
  {\bibfnamefont{R.}~\bibnamefont{Mortazavi}}}%
  , \bibinfo {year} {2011},\ \bibfield{title}{%
  \enquote{\bibinfo {title} {Snow --- a photobiochemical exchange platform for
  volatile and semi-volatile organic compounds with the atmosphere},}\ }%
  \bibfield{journal}{%
  \bibinfo {journal} {Environ. Chem.}\ }%
  \textbf{\bibinfo {volume} {8}},\ \bibinfo {pages} {62--73}%
  \bibAnnoteFile{NoStop}{ariya2011}%
\bibitem[{\citenamefont{Arstila}\ \emph{et~al.}(1998)\citenamefont{Arstila},
  \citenamefont{Laasonen},\ and\ \citenamefont{Laaksonen}}]{arstila1998}%
  \BibitemOpen
  \bibfield{author}{%
  \bibinfo {author} {\bibnamefont{Arstila}, \bibfnamefont{H.}}, \bibinfo
  {author} {\bibfnamefont{K.}~\bibnamefont{Laasonen}},\ and\ \bibinfo {author}
  {\bibfnamefont{A.}~\bibnamefont{Laaksonen}}}%
  , \bibinfo {year} {1998},\ \bibfield{title}{%
  \enquote{\bibinfo {title} {Ab initio study of the gasphase sulphuric acid
  hydrates containing 1 to 3 water molecules},}\ }%
  \bibfield{journal}{%
  \bibinfo {journal} {J. Chem. Phys.}\ }%
  \textbf{\bibinfo {volume} {108}},\ \bibinfo {pages} {1031--1039}%
  \bibAnnoteFile{NoStop}{arstila1998}%
\bibitem[{\citenamefont{Avron}\ and\ \citenamefont{Levine}(1992)}]{Avron1992}%
  \BibitemOpen
  \bibfield{author}{%
  \bibinfo {author} {\bibnamefont{Avron}, \bibfnamefont{J.~E.}},\ and\ \bibinfo
  {author} {\bibfnamefont{D.}~\bibnamefont{Levine}}}%
  , \bibinfo {year} {1992},\ \bibfield{title}{%
  \enquote{\bibinfo {title} {Geometry and foams --- {2D} dynamics and {3D}
  statics},}\ }%
  \bibfield{journal}{%
  \bibinfo {journal} {Phys. Rev. Lett.}\ }%
  \textbf{\bibinfo {volume} {69}},\ \bibinfo {pages} {208--211}%
  \bibAnnoteFile{NoStop}{Avron1992}%
\bibitem[{\citenamefont{Ayotte}\ \emph{et~al.}(2001)\citenamefont{Ayotte},
  \citenamefont{Smith}, \citenamefont{Stevenson}, \citenamefont{Dohn{\'a}lek},
  \citenamefont{Kimmel},\ and\ \citenamefont{Kay}}]{ayotte2001}%
  \BibitemOpen
  \bibfield{author}{%
  \bibinfo {author} {\bibnamefont{Ayotte}, \bibfnamefont{P.}}, \bibinfo
  {author} {\bibfnamefont{R.~S.}\ \bibnamefont{Smith}}, \bibinfo {author}
  {\bibfnamefont{K.~P.}\ \bibnamefont{Stevenson}}, \bibinfo {author}
  {\bibfnamefont{Z.}~\bibnamefont{Dohn{\'a}lek}}, \bibinfo {author}
  {\bibfnamefont{G.~A.}\ \bibnamefont{Kimmel}},\ and\ \bibinfo {author}
  {\bibfnamefont{B.~D.}\ \bibnamefont{Kay}}}%
  , \bibinfo {year} {2001},\ \bibfield{title}{%
  \enquote{\bibinfo {title} {Effect of porosity on the adsorption, desorption,
  trapping, and release of volatile gases by amorphous solid water},}\ }%
  \bibfield{journal}{%
  \bibinfo {journal} {J. Geophys. Res. --- Planets}\ }%
  \textbf{\bibinfo {volume} {106}},\ \bibinfo {pages} {33387--33392}%
  \bibAnnoteFile{NoStop}{ayotte2001}%
\bibitem[{\citenamefont{Babko}\ \emph{et~al.}(2002)\citenamefont{Babko},
  \citenamefont{Rothrock},\ and\ \citenamefont{Maykut}}]{Babko:2002}%
  \BibitemOpen
  \bibfield{author}{%
  \bibinfo {author} {\bibnamefont{Babko}, \bibfnamefont{O.}}, \bibinfo {author}
  {\bibfnamefont{D.~A.}\ \bibnamefont{Rothrock}},\ and\ \bibinfo {author}
  {\bibfnamefont{G.~A.}\ \bibnamefont{Maykut}}}%
  , \bibinfo {year} {2002},\ \bibfield{title}{%
  \enquote{\bibinfo {title} {Role of rafting in the mechanical redistribution
  of sea ice},}\ }%
  \bibfield{journal}{%
  \bibinfo {journal} {J.~Geophys.~Res.}\ }%
  \textbf{\bibinfo {volume} {107}},\ \bibinfo {pages} {3113}%
  \bibAnnoteFile{NoStop}{Babko:2002}%
\bibitem[{\citenamefont{Bailey}\ and\
  \citenamefont{Hallett}(2002)}]{bailey2002}%
  \BibitemOpen
  \bibfield{author}{%
  \bibinfo {author} {\bibnamefont{Bailey}, \bibfnamefont{M.}},\ and\ \bibinfo
  {author} {\bibfnamefont{J.}~\bibnamefont{Hallett}}}%
  , \bibinfo {year} {2002},\ \bibfield{title}{%
  \enquote{\bibinfo {title} {Nucleation effects on the habit of vapour grown
  ice crystals from-18 to-42 degrees {C}},}\ }%
  \bibfield{journal}{%
  \bibinfo {journal} {Quart. J. Roy. Meteor. Soc.}\ }%
  \textbf{\bibinfo {volume} {128}},\ \bibinfo {pages} {1461--1483}%
  \bibAnnoteFile{NoStop}{bailey2002}%
\bibitem[{\citenamefont{Bailey}\ and\
  \citenamefont{Hallett}(2004)}]{bailey2004}%
  \BibitemOpen
  \bibfield{author}{%
  \bibinfo {author} {\bibnamefont{Bailey}, \bibfnamefont{M.}},\ and\ \bibinfo
  {author} {\bibfnamefont{J.}~\bibnamefont{Hallett}}}%
  , \bibinfo {year} {2004},\ \bibfield{title}{%
  \enquote{\bibinfo {title} {Growth rates and habits of ice crystals between
  -20 degrees and -70 degrees {C}},}\ }%
  \bibfield{journal}{%
  \bibinfo {journal} {J. Atmos. Sci.}\ }%
  \textbf{\bibinfo {volume} {61}},\ \bibinfo {pages} {514--544}%
  \bibAnnoteFile{NoStop}{bailey2004}%
\bibitem[{\citenamefont{Balci}\ and\
  \citenamefont{Uras-Aytemiz}(2011)}]{balci2011}%
  \BibitemOpen
  \bibfield{author}{%
  \bibinfo {author} {\bibnamefont{Balci}, \bibfnamefont{F.~M.}},\ and\ \bibinfo
  {author} {\bibfnamefont{N.}~\bibnamefont{Uras-Aytemiz}}}%
  , \bibinfo {year} {2011},\ \enquote{\bibinfo {title} {Auto-ionization of
  {HNO$_3$} on/in {H$_2$O} clusters: Structure, dynamics and spectroscopy},}\
  ,\ \bibinfo {pages} {submitted}%
  \bibAnnoteFile{NoStop}{balci2011}%
\bibitem[{\citenamefont{Ball}(2004)}]{ball2004}%
  \BibitemOpen
\bibfield{title}{%
    }%
  \bibfield{author}{%
  \bibinfo {author} {\bibnamefont{Ball}, \bibfnamefont{P.}}}%
  , \bibinfo {year} {2004},\ \bibfield{title}{%
  \enquote{\bibinfo {title} {The joy of six: the growth and form of snow
  crystals},}\ }%
  \bibfield{journal}{%
  \bibinfo {journal} {Interdisciplinary Sci. Rev.}\ }%
  \textbf{\bibinfo {volume} {29}},\ \bibinfo {pages} {353--365}%
  \bibAnnoteFile{NoStop}{ball2004}%
\bibitem[{\citenamefont{Bamber}(2007)}]{JonathanBamber:2007p26021}%
  \BibitemOpen
  \bibfield{author}{%
  \bibinfo {author} {\bibnamefont{Bamber}, \bibfnamefont{J.}}}%
  , \bibinfo {year} {2007},\ \bibfield{title}{%
  \enquote{\bibinfo {title} {Remote sensing in glaciology},}\ }%
  \bibinfo {journal} {Glacier Science and Environmental Change (First
  Edition)},\ \bibinfo {pages} {370--382}%
  \bibAnnoteFile{NoStop}{JonathanBamber:2007p26021}%
\bibitem[{\citenamefont{Bar-Nun}\ \emph{et~al.}(1985)\citenamefont{Bar-Nun},
  \citenamefont{Herman}, \citenamefont{Laufer},\ and\
  \citenamefont{Rappaport}}]{barnun1985}%
  \BibitemOpen
\bibfield{journal}{%
    }%
  \bibfield{author}{%
  \bibinfo {author} {\bibnamefont{Bar-Nun}, \bibfnamefont{A.}}, \bibinfo
  {author} {\bibfnamefont{G.}~\bibnamefont{Herman}}, \bibinfo {author}
  {\bibfnamefont{D.}~\bibnamefont{Laufer}},\ and\ \bibinfo {author}
  {\bibfnamefont{M.~L.}\ \bibnamefont{Rappaport}}}%
  , \bibinfo {year} {1985},\ \bibfield{title}{%
  \enquote{\bibinfo {title} {Trapping and release of gases by water ice and
  implications for icy bodies},}\ }%
  \bibfield{journal}{%
  \bibinfo {journal} {Icarus}\ }%
  \textbf{\bibinfo {volume} {63}},\ \bibinfo {pages} {317--332}%
  \bibAnnoteFile{NoStop}{barnun1985}%
\bibitem[{\citenamefont{Bar-Nun}\ \emph{et~al.}(2007)\citenamefont{Bar-Nun},
  \citenamefont{Notesco},\ and\ \citenamefont{Owen}}]{bar-nun2007}%
  \BibitemOpen
  \bibfield{author}{%
  \bibinfo {author} {\bibnamefont{Bar-Nun}, \bibfnamefont{A.}}, \bibinfo
  {author} {\bibfnamefont{G.}~\bibnamefont{Notesco}},\ and\ \bibinfo {author}
  {\bibfnamefont{T.}~\bibnamefont{Owen}}}%
  , \bibinfo {year} {2007},\ \bibfield{title}{%
  \enquote{\bibinfo {title} {Trapping of {N$_2$}, {CO} and {Ar} in amorphous
  ice --- application to comets},}\ }%
  \bibfield{journal}{%
  \bibinfo {journal} {Icarus}\ }%
  \textbf{\bibinfo {volume} {190}},\ \bibinfo {pages} {655--659}%
  \bibAnnoteFile{NoStop}{bar-nun2007}%
\bibitem[{\citenamefont{Bardeen}\ \emph{et~al.}(2008)\citenamefont{Bardeen},
  \citenamefont{Toon}, \citenamefont{Jensen}, \citenamefont{Marsh},\ and\
  \citenamefont{Harvey}}]{Bardeen2008}%
  \BibitemOpen
  \bibfield{author}{%
  \bibinfo {author} {\bibnamefont{Bardeen}, \bibfnamefont{C.~G.}}, \bibinfo
  {author} {\bibfnamefont{O.~B.}\ \bibnamefont{Toon}}, \bibinfo {author}
  {\bibfnamefont{E.~J.}\ \bibnamefont{Jensen}}, \bibinfo {author}
  {\bibfnamefont{D.~R.}\ \bibnamefont{Marsh}},\ and\ \bibinfo {author}
  {\bibfnamefont{V.~L.}\ \bibnamefont{Harvey}}}%
  , \bibinfo {year} {2008},\ \bibfield{title}{%
  \enquote{\bibinfo {title} {Numerical simulations of the three-dimensional
  distribution of meteoric dust in the mesosphere and upper stratosphere},}\ }%
  \bibfield{journal}{%
  \bibinfo {journal} {J. Geophy. Res.-Atmos.}\ }%
  \textbf{\bibinfo {volume} {113}},\ \bibinfo {pages} {D17202}%
  \bibAnnoteFile{NoStop}{Bardeen2008}%
\bibitem[{\citenamefont{{Barkume}}\
  \emph{et~al.}(2008)\citenamefont{{Barkume}}, \citenamefont{{Brown}},\ and\
  \citenamefont{{Schaller}}}]{barkume2008}%
  \BibitemOpen
  \bibfield{author}{%
  \bibinfo {author} {\bibnamefont{{Barkume}}, \bibfnamefont{K.~M.}}, \bibinfo
  {author} {\bibfnamefont{M.~E.}\ \bibnamefont{{Brown}}},\ and\ \bibinfo
  {author} {\bibfnamefont{E.~L.}\ \bibnamefont{{Schaller}}}}%
  , \bibinfo {year} {2008},\ \bibfield{title}{%
  \enquote{\bibinfo {title} {Near-infrared spectra of {Centaurs} and {Kuiper
  Belt Objects}},}\ }%
  \bibfield{journal}{%
  \bibinfo {journal} {Astronom. J.}\ }%
  \textbf{\bibinfo {volume} {135}},\ \bibinfo {pages} {55--67}%
  \bibAnnoteFile{NoStop}{barkume2008}%
\bibitem[{\citenamefont{Bartels-Rausch}\
  \emph{et~al.}(2010)\citenamefont{Bartels-Rausch}, \citenamefont{Brigante},
  \citenamefont{Elshorbany}, \citenamefont{Ammann}, \citenamefont{D'anna},
  \citenamefont{George}, \citenamefont{Stemmler}, \citenamefont{Ndour},\ and\
  \citenamefont{Kleffmann}}]{BartelsRausch:2010p25371}%
  \BibitemOpen
  \bibfield{author}{%
  \bibinfo {author} {\bibnamefont{Bartels-Rausch}, \bibfnamefont{T.}}, \bibinfo
  {author} {\bibfnamefont{M.}~\bibnamefont{Brigante}}, \bibinfo {author}
  {\bibfnamefont{Y.}~\bibnamefont{Elshorbany}}, \bibinfo {author}
  {\bibfnamefont{M.}~\bibnamefont{Ammann}}, \bibinfo {author}
  {\bibfnamefont{B.}~\bibnamefont{D'anna}}, \bibinfo {author}
  {\bibfnamefont{C.}~\bibnamefont{George}}, \bibinfo {author}
  {\bibfnamefont{K.}~\bibnamefont{Stemmler}}, \bibinfo {author}
  {\bibfnamefont{M.}~\bibnamefont{Ndour}},\ and\ \bibinfo {author}
  {\bibfnamefont{J.}~\bibnamefont{Kleffmann}}}%
  , \bibinfo {year} {2010},\ \bibfield{title}{%
  \enquote{\bibinfo {title} {Humic acid in ice: Photo-enhanced conversion of
  nitrogen dioxide into nitrous acid},}\ }%
  \bibfield{journal}{%
  \bibinfo {journal} {Atmos. Environ.}\ }%
  \textbf{\bibinfo {volume} {44}},\ \bibinfo {pages} {5443--5450}%
  \bibAnnoteFile{NoStop}{BartelsRausch:2010p25371}%
\bibitem[{\citenamefont{Bartels-Rausch}\
  \emph{et~al.}(2011)\citenamefont{Bartels-Rausch}, \citenamefont{Krysztofiak},
  \citenamefont{Bernhard}, \citenamefont{Schl{\"a}ppi},
  \citenamefont{Schwikowski},\ and\
  \citenamefont{Ammann}}]{BartelsRausch:2011p26430}%
  \BibitemOpen
  \bibfield{author}{%
  \bibinfo {author} {\bibnamefont{Bartels-Rausch}, \bibfnamefont{T.}}, \bibinfo
  {author} {\bibfnamefont{G.}~\bibnamefont{Krysztofiak}}, \bibinfo {author}
  {\bibfnamefont{A.}~\bibnamefont{Bernhard}}, \bibinfo {author}
  {\bibfnamefont{M.}~\bibnamefont{Schl{\"a}ppi}}, \bibinfo {author}
  {\bibfnamefont{M.}~\bibnamefont{Schwikowski}},\ and\ \bibinfo {author}
  {\bibfnamefont{M.}~\bibnamefont{Ammann}}}%
  , \bibinfo {year} {2011},\ \bibfield{title}{%
  \enquote{\bibinfo {title} {Photoinduced reduction of divalent mercury in ice
  by organic matter},}\ }%
  \bibfield{journal}{%
  \bibinfo {journal} {Chemosphere}\ }%
  \textbf{\bibinfo {volume} {82}},\ \bibinfo {pages} {199--203}%
  \bibAnnoteFile{NoStop}{BartelsRausch:2011p26430}%
\bibitem[{\citenamefont{Bell}\ \emph{et~al.}(2011)\citenamefont{Bell},
  \citenamefont{Ferraccioli}, \citenamefont{Creyts}, \citenamefont{Braaten},
  \citenamefont{Corr}, \citenamefont{Das}, \citenamefont{Damaske},
  \citenamefont{Frearson}, \citenamefont{Jordan}, \citenamefont{Rose},
  \citenamefont{Studinger},\ and\ \citenamefont{Wolovick}}]{Bell2011}%
  \BibitemOpen
  \bibfield{author}{%
  \bibinfo {author} {\bibnamefont{Bell}, \bibfnamefont{R.~E.}}, \bibinfo
  {author} {\bibfnamefont{F.}~\bibnamefont{Ferraccioli}}, \bibinfo {author}
  {\bibfnamefont{T.~T.}\ \bibnamefont{Creyts}}, \bibinfo {author}
  {\bibfnamefont{D.}~\bibnamefont{Braaten}}, \bibinfo {author}
  {\bibfnamefont{H.}~\bibnamefont{Corr}}, \bibinfo {author}
  {\bibfnamefont{I.}~\bibnamefont{Das}}, \bibinfo {author}
  {\bibfnamefont{D.}~\bibnamefont{Damaske}}, \bibinfo {author}
  {\bibfnamefont{N.}~\bibnamefont{Frearson}}, \bibinfo {author}
  {\bibfnamefont{T.}~\bibnamefont{Jordan}}, \bibinfo {author}
  {\bibfnamefont{K.}~\bibnamefont{Rose}}, \bibinfo {author}
  {\bibfnamefont{M.}~\bibnamefont{Studinger}},\ and\ \bibinfo {author}
  {\bibfnamefont{M.}~\bibnamefont{Wolovick}}}%
  , \bibinfo {year} {2011},\ \bibfield{title}{%
  \enquote{\bibinfo {title} {Widespread persistent thickening of the {East
  Antarctic} ice sheet by freezing from the base},}\ }%
  \bibfield{journal}{%
  \bibinfo {journal} {Science}\ }%
  \textbf{\bibinfo {volume} {331}},\ \bibinfo {pages} {1592--1595}%
  \bibAnnoteFile{NoStop}{Bell2011}%
\bibitem[{\citenamefont{{Bell III}}(2009)}]{bell2009}%
  \BibitemOpen
  \bibfield{author}{%
  \bibinfo {author} {\bibnamefont{{Bell III}}, \bibfnamefont{J.~F.}}}%
  , \bibinfo {year} {2009},\ \bibfield{title}{%
  \enquote{\bibinfo {title} {Water on planets},}\ }%
  \bibfield{journal}{%
  \bibinfo {journal} {Highlights of Astronomy}\ }%
  \textbf{\bibinfo {volume} {15}},\ \bibinfo {pages} {29--44}%
  \bibAnnoteFile{NoStop}{bell2009}%
\bibitem[{\citenamefont{Benatov}\ and\
  \citenamefont{Wettlaufer}(2004)}]{Benatov2004}%
  \BibitemOpen
  \bibfield{author}{%
  \bibinfo {author} {\bibnamefont{Benatov}, \bibfnamefont{L.}},\ and\ \bibinfo
  {author} {\bibfnamefont{J.~S.}\ \bibnamefont{Wettlaufer}}}%
  , \bibinfo {year} {2004},\ \bibfield{title}{%
  \enquote{\bibinfo {title} {Abrupt grain boundary melting in ice},}\ }%
  \bibfield{journal}{%
  \bibinfo {journal} {Phys. Rev. E}\ }%
  \textbf{\bibinfo {volume} {70}},\ \bibinfo {pages} {061606}%
  \bibAnnoteFile{NoStop}{Benatov2004}%
\bibitem[{\citenamefont{Bergeron}\ \emph{et~al.}(2006)\citenamefont{Bergeron},
  \citenamefont{Berger},\ and\ \citenamefont{Betterton}}]{bergeron2006}%
  \BibitemOpen
  \bibfield{author}{%
  \bibinfo {author} {\bibnamefont{Bergeron}, \bibfnamefont{V.}}, \bibinfo
  {author} {\bibfnamefont{C.}~\bibnamefont{Berger}},\ and\ \bibinfo {author}
  {\bibfnamefont{M.~D.}\ \bibnamefont{Betterton}}}%
  , \bibinfo {year} {2006},\ \bibfield{title}{%
  \enquote{\bibinfo {title} {Controlled irradiative formation of penitentes},}\
  }%
  \bibfield{journal}{%
  \bibinfo {journal} {Phys. Rev. Lett.}\ }%
  \textbf{\bibinfo {volume} {96}},\ \bibinfo {pages} {098502}%
  \bibAnnoteFile{NoStop}{bergeron2006}%
\bibitem[{\citenamefont{Bernstein}(2006)}]{bernstein2006_2}%
  \BibitemOpen
  \bibfield{author}{%
  \bibinfo {author} {\bibnamefont{Bernstein}, \bibfnamefont{M.}}}%
  , \bibinfo {year} {2006},\ \bibfield{title}{%
  \enquote{\bibinfo {title} {Prebiotic materials from on and off the early
  {Earth}},}\ }%
  \bibfield{journal}{%
  \bibinfo {journal} {Phil. Trans. R. Soc. B}\ }%
  \textbf{\bibinfo {volume} {361}},\ \bibinfo {pages} {1689--1702}%
  \bibAnnoteFile{NoStop}{bernstein2006_2}%
\bibitem[{\citenamefont{Bernstein}\
  \emph{et~al.}(2005)\citenamefont{Bernstein}, \citenamefont{Cruikshank},\ and\
  \citenamefont{Sandford}}]{bernstein2005}%
  \BibitemOpen
  \bibfield{author}{%
  \bibinfo {author} {\bibnamefont{Bernstein}, \bibfnamefont{M.~P.}}, \bibinfo
  {author} {\bibfnamefont{D.~P.}\ \bibnamefont{Cruikshank}},\ and\ \bibinfo
  {author} {\bibfnamefont{S.~A.}\ \bibnamefont{Sandford}}}%
  , \bibinfo {year} {2005},\ \bibfield{title}{%
  \enquote{\bibinfo {title} {Near infrared laboratory spectra of solid
  {H$_2$O/CO$_2$} and {CH$_3$OH/CO$_2$} ice mixtures},}\ }%
  \bibfield{journal}{%
  \bibinfo {journal} {Icarus}\ }%
  \textbf{\bibinfo {volume} {179}},\ \bibinfo {pages} {527--534}%
  \bibAnnoteFile{NoStop}{bernstein2005}%
\bibitem[{\citenamefont{Bernstein}\
  \emph{et~al.}(2006)\citenamefont{Bernstein}, \citenamefont{Cruikshank},\ and\
  \citenamefont{Sandford}}]{bernstein2006}%
  \BibitemOpen
  \bibfield{author}{%
  \bibinfo {author} {\bibnamefont{Bernstein}, \bibfnamefont{M.~P.}}, \bibinfo
  {author} {\bibfnamefont{D.~P.}\ \bibnamefont{Cruikshank}},\ and\ \bibinfo
  {author} {\bibfnamefont{S.A.}\ \bibnamefont{Sandford}}}%
  , \bibinfo {year} {2006},\ \bibfield{title}{%
  \enquote{\bibinfo {title} {Near-infrared spectra of laboratory {H$_2$O}
  {CH$_4$} ice mixtures},}\ }%
  \bibfield{journal}{%
  \bibinfo {journal} {Icarus}\ }%
  \textbf{\bibinfo {volume} {181}},\ \bibinfo {pages} {302--308}%
  \bibAnnoteFile{NoStop}{bernstein2006}%
\bibitem[{\citenamefont{Betterton}(2001)}]{betterton2001}%
  \BibitemOpen
  \bibfield{author}{%
  \bibinfo {author} {\bibnamefont{Betterton}, \bibfnamefont{M.~D.}}}%
  , \bibinfo {year} {2001},\ \bibfield{title}{%
  \enquote{\bibinfo {title} {Theory of structure formation in snowfields
  motivated by penitentes, suncups, and dirt cones},}\ }%
  \bibfield{journal}{%
  \bibinfo {journal} {Phys. Rev. E}\ }%
  \textbf{\bibinfo {volume} {63}},\ \bibinfo {pages} {056129}%
  \bibAnnoteFile{NoStop}{betterton2001}%
\bibitem[{\citenamefont{{Bibring}}\
  \emph{et~al.}(2004)\citenamefont{{Bibring}}, \citenamefont{{Langevin}},
  \citenamefont{{Poulet}}, \citenamefont{{Gendrin}}, \citenamefont{{Gondet}},
  \citenamefont{{Berth{\'e}}}, \citenamefont{{Soufflot}},
  \citenamefont{{Drossart}}, \citenamefont{{Combes}},
  \citenamefont{{Bellucci}}, \citenamefont{{Moroz}}, \citenamefont{{Mangold}},
  \citenamefont{{Schmitt}},\ and\ \citenamefont{{OMEGA team}}}]{bibring2004}%
  \BibitemOpen
  \bibfield{author}{%
  \bibinfo {author} {\bibnamefont{{Bibring}}, \bibfnamefont{{J.-P.}}}, \bibinfo
  {author} {\bibfnamefont{Y.}~\bibnamefont{{Langevin}}}, \bibinfo {author}
  {\bibfnamefont{F.}~\bibnamefont{{Poulet}}}, \bibinfo {author}
  {\bibfnamefont{A.}~\bibnamefont{{Gendrin}}}, \bibinfo {author}
  {\bibfnamefont{B.}~\bibnamefont{{Gondet}}}, \bibinfo {author}
  {\bibfnamefont{M.}~\bibnamefont{{Berth{\'e}}}}, \bibinfo {author}
  {\bibfnamefont{A.}~\bibnamefont{{Soufflot}}}, \bibinfo {author}
  {\bibfnamefont{P.}~\bibnamefont{{Drossart}}}, \bibinfo {author}
  {\bibfnamefont{M.}~\bibnamefont{{Combes}}}, \bibinfo {author}
  {\bibfnamefont{G.}~\bibnamefont{{Bellucci}}}, \bibinfo {author}
  {\bibfnamefont{V.}~\bibnamefont{{Moroz}}}, \bibinfo {author}
  {\bibfnamefont{N.}~\bibnamefont{{Mangold}}}, \bibinfo {author}
  {\bibfnamefont{B.}~\bibnamefont{{Schmitt}}},\ and\ \bibinfo {author}
  {\bibnamefont{{OMEGA team}}}}%
  , \bibinfo {year} {2004},\ \bibfield{title}{%
  \enquote{\bibinfo {title} {Perennial water ice identified in the south polar
  cap of {Mars}},}\ }%
  \bibfield{journal}{%
  \bibinfo {journal} {Nature}\ }%
  \textbf{\bibinfo {volume} {428}},\ \bibinfo {pages} {627--630}%
  \bibAnnoteFile{NoStop}{bibring2004}%
\bibitem[{\citenamefont{Bina}\ and\ \citenamefont{Navrotsky}(2000)}]{bina2000}%
  \BibitemOpen
  \bibfield{author}{%
  \bibinfo {author} {\bibnamefont{Bina}, \bibfnamefont{C.~R.}},\ and\ \bibinfo
  {author} {\bibfnamefont{A.}~\bibnamefont{Navrotsky}}}%
  , \bibinfo {year} {2000},\ \bibfield{title}{%
  \enquote{\bibinfo {title} {Possible presence of high-pressure ice in cold
  conducting slabs},}\ }%
  \bibfield{journal}{%
  \bibinfo {journal} {Nature}\ }%
  \textbf{\bibinfo {volume} {408}},\ \bibinfo {pages} {844--847}%
  \bibAnnoteFile{NoStop}{bina2000}%
\bibitem[{\citenamefont{Bitz}\ \emph{et~al.}(2001)\citenamefont{Bitz},
  \citenamefont{Holland}, \citenamefont{Eby},\ and\
  \citenamefont{Weaver}}]{Bitz:2001}%
  \BibitemOpen
  \bibfield{author}{%
  \bibinfo {author} {\bibnamefont{Bitz}, \bibfnamefont{C.~M.}}, \bibinfo
  {author} {\bibfnamefont{M.~M.}\ \bibnamefont{Holland}}, \bibinfo {author}
  {\bibfnamefont{M.}~\bibnamefont{Eby}},\ and\ \bibinfo {author}
  {\bibfnamefont{A.~J.}\ \bibnamefont{Weaver}}}%
  , \bibinfo {year} {2001},\ \bibfield{title}{%
  \enquote{\bibinfo {title} {Simulating the ice-thickness distribution in a
  coupled climate model},}\ }%
  \bibfield{journal}{%
  \bibinfo {journal} {J.~Geophys.~Res.}\ }%
  \textbf{\bibinfo {volume} {106}},\ \bibinfo {pages} {2441--2463}%
  \bibAnnoteFile{NoStop}{Bitz:2001}%
\bibitem[{\citenamefont{{Black}}\ \emph{et~al.}(2010)\citenamefont{{Black}},
  \citenamefont{{Campbell}},\ and\ \citenamefont{{Harmon}}}]{black2010}%
  \BibitemOpen
  \bibfield{author}{%
  \bibinfo {author} {\bibnamefont{{Black}}, \bibfnamefont{G.~J.}}, \bibinfo
  {author} {\bibfnamefont{D.~B.}\ \bibnamefont{{Campbell}}},\ and\ \bibinfo
  {author} {\bibfnamefont{J.~K.}\ \bibnamefont{{Harmon}}}}%
  , \bibinfo {year} {2010},\ \bibfield{title}{%
  \enquote{\bibinfo {title} {{Radar measurements of {Mercury's} north pole at
  70~cm wavelength}},}\ }%
  \bibfield{journal}{%
  \bibinfo {journal} {Icarus}\ }%
  \textbf{\bibinfo {volume} {209}},\ \bibinfo {pages} {224--229}%
  \bibAnnoteFile{NoStop}{black2010}%
\bibitem[{\citenamefont{Blake}\ \emph{et~al.}(1991)\citenamefont{Blake},
  \citenamefont{Allamandola}, \citenamefont{Sandford},\ and\
  \citenamefont{Freund}}]{blake1991}%
  \BibitemOpen
  \bibfield{author}{%
  \bibinfo {author} {\bibnamefont{Blake}, \bibfnamefont{D.~F.}}, \bibinfo
  {author} {\bibfnamefont{L.~J.}\ \bibnamefont{Allamandola}}, \bibinfo {author}
  {\bibfnamefont{S.}~\bibnamefont{Sandford}},\ and\ \bibinfo {author}
  {\bibfnamefont{F.}~\bibnamefont{Freund}}}%
  , \bibinfo {year} {1991},\ \bibfield{title}{%
  \enquote{\bibinfo {title} {Clathrate type {II} hydrate formation in vacuo
  under astrophysical conditions},}\ }%
  \bibfield{journal}{%
  \bibinfo {journal} {Meteoritics}\ }%
  \textbf{\bibinfo {volume} {26}},\ \bibinfo {pages} {319}%
  \bibAnnoteFile{NoStop}{blake1991}%
\bibitem[{\citenamefont{von Blohn}\ \emph{et~al.}(2005)\citenamefont{von
  Blohn}, \citenamefont{Mitra}, \citenamefont{Diehl},\ and\
  \citenamefont{Borrmann}}]{vonBlohn2005}%
  \BibitemOpen
  \bibfield{author}{%
  \bibinfo {author} {\bibnamefont{von Blohn}, \bibfnamefont{N.}}, \bibinfo
  {author} {\bibfnamefont{S.~K.}\ \bibnamefont{Mitra}}, \bibinfo {author}
  {\bibfnamefont{K.}~\bibnamefont{Diehl}},\ and\ \bibinfo {author}
  {\bibfnamefont{S.}~\bibnamefont{Borrmann}}}%
  , \bibinfo {year} {2005},\ \bibfield{title}{%
  \enquote{\bibinfo {title} {The ice nucleating ability of pollen: Part iii:
  New laboratory studies in immersion and contact freezing modes including more
  pollen types},}\ }%
  \bibfield{journal}{%
  \bibinfo {journal} {Atmos. Res.}\ }%
  \textbf{\bibinfo {volume} {78}},\ \bibinfo {pages} {182--189}%
  \bibAnnoteFile{NoStop}{vonBlohn2005}%
\bibitem[{\citenamefont{Blunier}\ \emph{et~al.}(2005)\citenamefont{Blunier},
  \citenamefont{Floch}, \citenamefont{Jacobi},\ and\
  \citenamefont{Quansah}}]{Blunier:2005p25896}%
  \BibitemOpen
  \bibfield{author}{%
  \bibinfo {author} {\bibnamefont{Blunier}, \bibfnamefont{T.}}, \bibinfo
  {author} {\bibfnamefont{G.}~\bibnamefont{Floch}}, \bibinfo {author}
  {\bibfnamefont{H.-W.}\ \bibnamefont{Jacobi}},\ and\ \bibinfo {author}
  {\bibfnamefont{E.}~\bibnamefont{Quansah}}}%
  , \bibinfo {year} {2005},\ \bibfield{title}{%
  \enquote{\bibinfo {title} {Isotopic view on nitrate loss in {Antarctic}
  surface snow},}\ }%
  \bibfield{journal}{%
  \bibinfo {journal} {Geophys. Res. Lett.}\ }%
  \textbf{\bibinfo {volume} {32}},\ \bibinfo {pages} {L13501}%
  \bibAnnoteFile{NoStop}{Blunier:2005p25896}%
\bibitem[{\citenamefont{Boogert}\ and\
  \citenamefont{Ehrenfreund}(2004)}]{boogert2004}%
  \BibitemOpen
  \bibfield{author}{%
  \bibinfo {author} {\bibnamefont{Boogert}, \bibfnamefont{A.~C.~A.}},\ and\
  \bibinfo {author} {\bibfnamefont{P.}~\bibnamefont{Ehrenfreund}}}%
  , \bibinfo {year} {2004},\ \enquote{\bibinfo {title} {Interstellar ices},}\
  in\ \emph{\bibinfo {booktitle} {Astrophysics of Dust}},\ \bibinfo {series}
  {ASP Conference Series}, Vol.\ \bibinfo {volume} {309},\ \bibinfo {editor}
  {edited by\ \bibinfo {editor} {\bibfnamefont{A.~N.}\ \bibnamefont{Witt}},
  \bibinfo {editor} {\bibfnamefont{G.C.}\ \bibnamefont{Clayton}},\ and\
  \bibinfo {editor} {\bibfnamefont{B.T.}\ \bibnamefont{Draine}}},\ p.\ \bibinfo
  {pages} {547}%
  \bibAnnoteFile{NoStop}{boogert2004}%
\bibitem[{\citenamefont{{Boogert}}\
  \emph{et~al.}(2008)\citenamefont{{Boogert}}, \citenamefont{{Pontoppidan}},
  \citenamefont{{Knez}}, \citenamefont{{Lahuis}},
  \citenamefont{{Kessler-Silacci}}, \citenamefont{{van Dishoeck}},
  \citenamefont{{Blake}}, \citenamefont{{Augereau}}, \citenamefont{{Bisschop}},
  \citenamefont{{Bottinelli}}, \citenamefont{{Brooke}}, \citenamefont{{Brown}},
  \citenamefont{{Crapsi}}, \citenamefont{{Evans}}, \citenamefont{{Fraser}},
  \citenamefont{{Geers}}, \citenamefont{{Huard}},
  \citenamefont{{J{\o}rgensen}}, \citenamefont{{{\"O}berg}},
  \citenamefont{{Allen}}, \citenamefont{{Harvey}}, \citenamefont{{Koerner}},
  \citenamefont{{Mundy}}, \citenamefont{{Padgett}}, \citenamefont{{Sargent}},\
  and\ \citenamefont{{Stapelfeldt}}}]{boogert2008}%
  \BibitemOpen
  \bibfield{author}{%
  \bibinfo {author} {\bibnamefont{{Boogert}}, \bibfnamefont{A.~C.~A.}},
  \bibinfo {author} {\bibfnamefont{K.~M.}\ \bibnamefont{{Pontoppidan}}},
  \bibinfo {author} {\bibfnamefont{C.}~\bibnamefont{{Knez}}}, \bibinfo {author}
  {\bibfnamefont{F.}~\bibnamefont{{Lahuis}}}, \bibinfo {author}
  {\bibfnamefont{J.}~\bibnamefont{{Kessler-Silacci}}}, \bibinfo {author}
  {\bibfnamefont{E.~F.}\ \bibnamefont{{van Dishoeck}}}, \bibinfo {author}
  {\bibfnamefont{G.~A.}\ \bibnamefont{{Blake}}}, \bibinfo {author}
  {\bibfnamefont{{J.-C.}}\ \bibnamefont{{Augereau}}}, \bibinfo {author}
  {\bibfnamefont{S.~E.}\ \bibnamefont{{Bisschop}}}, \bibinfo {author}
  {\bibfnamefont{S.}~\bibnamefont{{Bottinelli}}}, \bibinfo {author}
  {\bibfnamefont{T.~Y.}\ \bibnamefont{{Brooke}}}, \bibinfo {author}
  {\bibfnamefont{J.}~\bibnamefont{{Brown}}}, \bibinfo {author}
  {\bibfnamefont{A.}~\bibnamefont{{Crapsi}}}, \bibinfo {author}
  {\bibfnamefont{N.~J.}\ \bibnamefont{{Evans}}, \bibfnamefont{II}}, \bibinfo
  {author} {\bibfnamefont{H.~J.}\ \bibnamefont{{Fraser}}}, \bibinfo {author}
  {\bibfnamefont{V.}~\bibnamefont{{Geers}}}, \bibinfo {author}
  {\bibfnamefont{T.~L.}\ \bibnamefont{{Huard}}}, \bibinfo {author}
  {\bibfnamefont{J.~K.}\ \bibnamefont{{J{\o}rgensen}}}, \bibinfo {author}
  {\bibfnamefont{K.~I.}\ \bibnamefont{{{\"O}berg}}}, \bibinfo {author}
  {\bibfnamefont{L.~E.}\ \bibnamefont{{Allen}}}, \bibinfo {author}
  {\bibfnamefont{P.~M.}\ \bibnamefont{{Harvey}}}, \bibinfo {author}
  {\bibfnamefont{D.~W.}\ \bibnamefont{{Koerner}}}, \bibinfo {author}
  {\bibfnamefont{L.~G.}\ \bibnamefont{{Mundy}}}, \bibinfo {author}
  {\bibfnamefont{D.~L.}\ \bibnamefont{{Padgett}}}, \bibinfo {author}
  {\bibfnamefont{A.~I.}\ \bibnamefont{{Sargent}}},\ and\ \bibinfo {author}
  {\bibfnamefont{K.~R.}\ \bibnamefont{{Stapelfeldt}}}}%
  , \bibinfo {year} {2008},\ \bibfield{title}{%
  \enquote{\bibinfo {title} {The c2d spitzer spectroscopic survey of ices
  around low-mass young stellar objects. {I}. {H$_{2}$O} and the 5-8 $\mu$m
  bands},}\ }%
  \bibfield{journal}{%
  \bibinfo {journal} {Astrophys. J.}\ }%
  \textbf{\bibinfo {volume} {678}},\ \bibinfo {pages} {985--1004}%
  \bibAnnoteFile{NoStop}{boogert2008}%
\bibitem[{\citenamefont{Boogert}\ \emph{et~al.}(1997)\citenamefont{Boogert},
  \citenamefont{Schutte}, \citenamefont{Helmich}, \citenamefont{Tielens},\ and\
  \citenamefont{Wooden}}]{boogert1997}%
  \BibitemOpen
  \bibfield{author}{%
  \bibinfo {author} {\bibnamefont{Boogert}, \bibfnamefont{A.~C.~A.}}, \bibinfo
  {author} {\bibfnamefont{W.~A.}\ \bibnamefont{Schutte}}, \bibinfo {author}
  {\bibfnamefont{F.~P.}\ \bibnamefont{Helmich}}, \bibinfo {author}
  {\bibfnamefont{A.~G. G.~M.}\ \bibnamefont{Tielens}},\ and\ \bibinfo {author}
  {\bibfnamefont{D.~H.}\ \bibnamefont{Wooden}}}%
  , \bibinfo {year} {1997},\ \bibfield{title}{%
  \enquote{\bibinfo {title} {Infrared observations and laboratory simulations
  of interstellar {CH}$_4$ and {SO}$_2$},}\ }%
  \bibfield{journal}{%
  \bibinfo {journal} {Astron. Astrophys.}\ }%
  \textbf{\bibinfo {volume} {317}},\ \bibinfo {pages} {929--941}%
  \bibAnnoteFile{NoStop}{boogert1997}%
\bibitem[{\citenamefont{Booth}(1999)}]{booth1999}%
  \BibitemOpen
  \bibfield{author}{%
  \bibinfo {author} {\bibnamefont{Booth}, \bibfnamefont{G.~K.}}}%
  , \bibinfo {year} {1999},\ \emph{\bibinfo {title} {William Robertson Smith:
  The Scientific, Literary and Cultural Context from 1866 to 1881}},\ Ph.D.
  thesis\ (\bibinfo {school} {University of Aberdeen})%
  \bibAnnoteFile{NoStop}{booth1999}%
\bibitem[{\citenamefont{Bowron}\ \emph{et~al.}(2006)\citenamefont{Bowron},
  \citenamefont{Finney}, \citenamefont{Hallbrucker}, \citenamefont{Kohl},
  \citenamefont{Loerting}, \citenamefont{Mayer},\ and\
  \citenamefont{Soper}}]{bowron2006}%
  \BibitemOpen
  \bibfield{author}{%
  \bibinfo {author} {\bibnamefont{Bowron}, \bibfnamefont{D.~T.}}, \bibinfo
  {author} {\bibfnamefont{J.~L.}\ \bibnamefont{Finney}}, \bibinfo {author}
  {\bibfnamefont{A.}~\bibnamefont{Hallbrucker}}, \bibinfo {author}
  {\bibfnamefont{I.}~\bibnamefont{Kohl}}, \bibinfo {author}
  {\bibfnamefont{T.}~\bibnamefont{Loerting}}, \bibinfo {author}
  {\bibfnamefont{E.}~\bibnamefont{Mayer}},\ and\ \bibinfo {author}
  {\bibfnamefont{A.~K.}\ \bibnamefont{Soper}}}%
  , \bibinfo {year} {2006},\ \bibfield{title}{%
  \enquote{\bibinfo {title} {The local and intermediate range structures of the
  five amorphous ices at 80~{K} and ambient pressure: {A} {Faber--Ziman} and
  {Bhatia--Thornton} analysis},}\ }%
  \bibfield{journal}{%
  \bibinfo {journal} {J. Chem. Phys.}\ }%
  \textbf{\bibinfo {volume} {125}},\ \bibinfo {pages} {194502}%
  \bibAnnoteFile{NoStop}{bowron2006}%
\bibitem[{\citenamefont{Bronshteyn}\ and\
  \citenamefont{Chernov}(1991)}]{bronshteyn1991}%
  \BibitemOpen
  \bibfield{author}{%
  \bibinfo {author} {\bibnamefont{Bronshteyn}, \bibfnamefont{V.~K.}},\ and\
  \bibinfo {author} {\bibfnamefont{A.~A.}\ \bibnamefont{Chernov}}}%
  , \bibinfo {year} {1991},\ \bibfield{title}{%
  \enquote{\bibinfo {title} {Freezing potentials arising on solidification of
  dilute aqueous solutions of electrolytes},}\ }%
  \bibfield{journal}{%
  \bibinfo {journal} {J. Cryst. Growth}\ }%
  \textbf{\bibinfo {volume} {112}},\ \bibinfo {pages} {129--145}%
  \bibAnnoteFile{NoStop}{bronshteyn1991}%
\bibitem[{\citenamefont{Broughton}\ and\
  \citenamefont{Gilmer}(1986)}]{Broughton1986}%
  \BibitemOpen
  \bibfield{author}{%
  \bibinfo {author} {\bibnamefont{Broughton}, \bibfnamefont{J.~Q.}},\ and\
  \bibinfo {author} {\bibfnamefont{G.~H.}\ \bibnamefont{Gilmer}}}%
  , \bibinfo {year} {1986},\ \bibfield{title}{%
  \enquote{\bibinfo {title} {Thermodynamic criteria for grain-boundary melting
  --- a molecular-dynamics study},}\ }%
  \bibfield{journal}{%
  \bibinfo {journal} {Phys. Rev. Lett.}\ }%
  \textbf{\bibinfo {volume} {56}},\ \bibinfo {pages} {2692--2695}%
  \bibAnnoteFile{NoStop}{Broughton1986}%
\bibitem[{\citenamefont{Brown}\ \emph{et~al.}(1996)\citenamefont{Brown},
  \citenamefont{George}, \citenamefont{Huang}, \citenamefont{Wong},
  \citenamefont{Rider}, \citenamefont{Smith},\ and\
  \citenamefont{Kay}}]{Brown1996}%
  \BibitemOpen
  \bibfield{author}{%
  \bibinfo {author} {\bibnamefont{Brown}, \bibfnamefont{D.~E.}}, \bibinfo
  {author} {\bibfnamefont{S.~M.}\ \bibnamefont{George}}, \bibinfo {author}
  {\bibfnamefont{C.}~\bibnamefont{Huang}}, \bibinfo {author}
  {\bibfnamefont{E.~K.~L.}\ \bibnamefont{Wong}}, \bibinfo {author}
  {\bibfnamefont{K.~B.}\ \bibnamefont{Rider}}, \bibinfo {author}
  {\bibfnamefont{R.~S.}\ \bibnamefont{Smith}},\ and\ \bibinfo {author}
  {\bibfnamefont{B.~D.}\ \bibnamefont{Kay}}}%
  , \bibinfo {year} {1996},\ \bibfield{title}{%
  \enquote{\bibinfo {title} {H$_2$o condensation coefficient and refractive
  index for vapor-deposited ice from molecular beam and optical interference
  measurements},}\ }%
  \bibfield{journal}{%
  \bibinfo {journal} {J. Phys. Chem.}\ }%
  \textbf{\bibinfo {volume} {100}},\ \bibinfo {pages} {4988--4995}%
  \bibAnnoteFile{NoStop}{Brown1996}%
\bibitem[{\citenamefont{{Brown}}\ \emph{et~al.}(2010)\citenamefont{{Brown}},
  \citenamefont{{Ragozzine}}, \citenamefont{{Stansberry}},\ and\
  \citenamefont{{Fraser}}}]{brown2010}%
  \BibitemOpen
  \bibfield{author}{%
  \bibinfo {author} {\bibnamefont{{Brown}}, \bibfnamefont{M.~E.}}, \bibinfo
  {author} {\bibfnamefont{D.}~\bibnamefont{{Ragozzine}}}, \bibinfo {author}
  {\bibfnamefont{J.}~\bibnamefont{{Stansberry}}},\ and\ \bibinfo {author}
  {\bibfnamefont{W.~C.}\ \bibnamefont{{Fraser}}}}%
  , \bibinfo {year} {2010},\ \bibfield{title}{%
  \enquote{\bibinfo {title} {The size, density, and formation of the
  {Orcus-Vanth} system in the {Kuiper} belt},}\ }%
  \bibfield{journal}{%
  \bibinfo {journal} {Astrophys. J.}\ }%
  \textbf{\bibinfo {volume} {139}},\ \bibinfo {pages} {2700--2705}%
  \bibAnnoteFile{NoStop}{brown2010}%
\bibitem[{\citenamefont{Brown}\ \emph{et~al.}(1995)\citenamefont{Brown},
  \citenamefont{Cruikshank}, \citenamefont{Veverka},
  \citenamefont{Helfenstein},\ and\ \citenamefont{Eluszkiewicz}}]{brown1995}%
  \BibitemOpen
  \bibfield{author}{%
  \bibinfo {author} {\bibnamefont{Brown}, \bibfnamefont{R.~H.}}, \bibinfo
  {author} {\bibfnamefont{D.~P.}\ \bibnamefont{Cruikshank}}, \bibinfo {author}
  {\bibfnamefont{J.}~\bibnamefont{Veverka}}, \bibinfo {author}
  {\bibfnamefont{P.}~\bibnamefont{Helfenstein}},\ and\ \bibinfo {author}
  {\bibfnamefont{J.}~\bibnamefont{Eluszkiewicz}}}%
  , \bibinfo {year} {1995},\ \emph{\bibinfo {title} {Neptune and Triton}}\
  (\bibinfo {publisher} {University of Arizona Press})%
  \bibAnnoteFile{NoStop}{brown1995}%
\bibitem[{\citenamefont{{Brownlee}}\
  \emph{et~al.}(2006)\citenamefont{{Brownlee}}, \citenamefont{{Tsou}},
  \citenamefont{{Al{\'e}on}}, \citenamefont{{Alexander}},
  \citenamefont{{Araki}}, \citenamefont{{Bajt}}, \citenamefont{{Baratta}},
  \citenamefont{{Bastien}}, \citenamefont{{Bland}}, \citenamefont{{Bleuet}},
  \citenamefont{{Borg}}, \citenamefont{{Bradley}}, \citenamefont{{Brearley}},
  \citenamefont{{Brenker}}, \citenamefont{{Brennan}}, \citenamefont{{Bridges}},
  \citenamefont{{Browning}}, \citenamefont{{Brucato}},
  \citenamefont{{Bullock}}, \citenamefont{{Burchell}},
  \citenamefont{{Busemann}}, \citenamefont{{Butterworth}},
  \citenamefont{{Chaussidon}}, \citenamefont{{Cheuvront}},
  \citenamefont{{Chi}}, \citenamefont{{Cintala}}, \citenamefont{{Clark}},
  \citenamefont{{Clemett}}, \citenamefont{{Cody}}, \citenamefont{{Colangeli}},
  \citenamefont{{Cooper}}, \citenamefont{{Cordier}}, \citenamefont{{Daghlian}},
  \citenamefont{{Dai}}, \citenamefont{{D'Hendecourt}},
  \citenamefont{{Djouadi}}, \citenamefont{{Dominguez}},
  \citenamefont{{Duxbury}}, \citenamefont{{Dworkin}}, \citenamefont{{Ebel}},
  \citenamefont{{Economou}}, \citenamefont{{Fakra}}, \citenamefont{{Fairey}},
  \citenamefont{{Fallon}}, \citenamefont{{Ferrini}}, \citenamefont{{Ferroir}},
  \citenamefont{{Fleckenstein}}, \citenamefont{{Floss}},
  \citenamefont{{Flynn}}, \citenamefont{{Franchi}}, \citenamefont{{Fries}},
  \citenamefont{{Gainsforth}}, \citenamefont{{Gallien}},
  \citenamefont{{Genge}}, \citenamefont{{Gilles}}, \citenamefont{{Gillet}},
  \citenamefont{{Gilmour}}, \citenamefont{{Glavin}}, \citenamefont{{Gounelle}},
  \citenamefont{{Grady}}, \citenamefont{{Graham}}, \citenamefont{{Grant}},
  \citenamefont{{Green}}, \citenamefont{{Grossemy}}, \citenamefont{{Grossman}},
  \citenamefont{{Grossman}}, \citenamefont{{Guan}}, \citenamefont{{Hagiya}},
  \citenamefont{{Harvey}}, \citenamefont{{Heck}}, \citenamefont{{Herzog}},
  \citenamefont{{Hoppe}}, \citenamefont{{H{\"o}rz}}, \citenamefont{{Huth}},
  \citenamefont{{Hutcheon}}, \citenamefont{{Ignatyev}}, \citenamefont{{Ishii}},
  \citenamefont{{Ito}}, \citenamefont{{Jacob}}, \citenamefont{{Jacobsen}},
  \citenamefont{{Jacobsen}}, \citenamefont{{Jones}}, \citenamefont{{Joswiak}},
  \citenamefont{{Jurewicz}}, \citenamefont{{Kearsley}},
  \citenamefont{{Keller}}, \citenamefont{{Khodja}}, \citenamefont{{Kilcoyne}},
  \citenamefont{{Kissel}}, \citenamefont{{Krot}}, \citenamefont{{Langenhorst}},
  \citenamefont{{Lanzirotti}}, \citenamefont{{Le}}, \citenamefont{{Leshin}},
  \citenamefont{{Leitner}}, \citenamefont{{Lemelle}}, \citenamefont{{Leroux}},
  \citenamefont{{Liu}}, \citenamefont{{Luening}}, \citenamefont{{Lyon}},
  \citenamefont{{MacPherson}}, \citenamefont{{Marcus}},
  \citenamefont{{Marhas}}, \citenamefont{{Marty}}, \citenamefont{{Matrajt}},
  \citenamefont{{McKeegan}}, \citenamefont{{Meibom}},
  \citenamefont{{Mennella}}, \citenamefont{{Messenger}},
  \citenamefont{{Messenger}}, \citenamefont{{Mikouchi}},
  \citenamefont{{Mostefaoui}}, \citenamefont{{Nakamura}},
  \citenamefont{{Nakano}}, \citenamefont{{Newville}}, \citenamefont{{Nittler}},
  \citenamefont{{Ohnishi}}, \citenamefont{{Ohsumi}}, \citenamefont{{Okudaira}},
  \citenamefont{{Papanastassiou}}, \citenamefont{{Palma}},
  \citenamefont{{Palumbo}}, \citenamefont{{Pepin}}, \citenamefont{{Perkins}},
  \citenamefont{{Perronnet}}, \citenamefont{{Pianetta}}, \citenamefont{{Rao}},
  \citenamefont{{Rietmeijer}}, \citenamefont{{Robert}}, \citenamefont{{Rost}},
  \citenamefont{{Rotundi}}, \citenamefont{{Ryan}}, \citenamefont{{Sandford}},
  \citenamefont{{Schwandt}}, \citenamefont{{See}}, \citenamefont{{Schlutter}},
  \citenamefont{{Sheffield-Parker}}, \citenamefont{{Simionovici}},
  \citenamefont{{Simon}}, \citenamefont{{Sitnitsky}}, \citenamefont{{Snead}},
  \citenamefont{{Spencer}}, \citenamefont{{Stadermann}},
  \citenamefont{{Steele}}, \citenamefont{{Stephan}}, \citenamefont{{Stroud}},
  \citenamefont{{Susini}}, \citenamefont{{Sutton}}, \citenamefont{{Suzuki}},
  \citenamefont{{Taheri}}, \citenamefont{{Taylor}}, \citenamefont{{Teslich}},
  \citenamefont{{Tomeoka}}, \citenamefont{{Tomioka}}, \citenamefont{{Toppani}},
  \citenamefont{{Trigo-Rodr{\'{\i}}guez}}, \citenamefont{{Troadec}},
  \citenamefont{{Tsuchiyama}}, \citenamefont{{Tuzzolino}},
  \citenamefont{{Tyliszczak}}, \citenamefont{{Uesugi}},
  \citenamefont{{Velbel}}, \citenamefont{{Vellenga}}, \citenamefont{{Vicenzi}},
  \citenamefont{{Vincze}}, \citenamefont{{Warren}}, \citenamefont{{Weber}},
  \citenamefont{{Weisberg}}, \citenamefont{{Westphal}},
  \citenamefont{{Wirick}}, \citenamefont{{Wooden}}, \citenamefont{{Wopenka}},
  \citenamefont{{Wozniakiewicz}}, \citenamefont{{Wright}},
  \citenamefont{{Yabuta}}, \citenamefont{{Yano}}, \citenamefont{{Young}},
  \citenamefont{{Zare}}, \citenamefont{{Zega}}, \citenamefont{{Ziegler}},
  \citenamefont{{Zimmerman}}, \citenamefont{{Zinner}},\ and\
  \citenamefont{{Zolensky}}}]{brownlee2006}%
  \BibitemOpen
  \bibfield{author}{%
  \bibinfo {author} {\bibnamefont{{Brownlee}}, \bibfnamefont{D.}}, \bibinfo
  {author} {\bibfnamefont{P.}~\bibnamefont{{Tsou}}}, \bibinfo {author}
  {\bibfnamefont{J.}~\bibnamefont{{Al{\'e}on}}}, \bibinfo {author}
  {\bibfnamefont{C.~M.~O.~'.}\ \bibnamefont{{Alexander}}}, \bibinfo {author}
  {\bibfnamefont{T.}~\bibnamefont{{Araki}}}, \bibinfo {author}
  {\bibfnamefont{S.}~\bibnamefont{{Bajt}}}, \bibinfo {author}
  {\bibfnamefont{G.~A.}\ \bibnamefont{{Baratta}}}, \bibinfo {author}
  {\bibfnamefont{R.}~\bibnamefont{{Bastien}}}, \bibinfo {author}
  {\bibfnamefont{P.}~\bibnamefont{{Bland}}}, \bibinfo {author}
  {\bibfnamefont{P.}~\bibnamefont{{Bleuet}}}, \bibinfo {author}
  {\bibfnamefont{J.}~\bibnamefont{{Borg}}}, \bibinfo {author}
  {\bibfnamefont{J.~P.}\ \bibnamefont{{Bradley}}}, \bibinfo {author}
  {\bibfnamefont{A.}~\bibnamefont{{Brearley}}}, \bibinfo {author}
  {\bibfnamefont{F.}~\bibnamefont{{Brenker}}}, \bibinfo {author}
  {\bibfnamefont{S.}~\bibnamefont{{Brennan}}}, \bibinfo {author}
  {\bibfnamefont{J.~C.}\ \bibnamefont{{Bridges}}}, \bibinfo {author}
  {\bibfnamefont{N.~D.}\ \bibnamefont{{Browning}}}, \bibinfo {author}
  {\bibfnamefont{J.~R.}\ \bibnamefont{{Brucato}}}, \bibinfo {author}
  {\bibfnamefont{E.}~\bibnamefont{{Bullock}}}, \bibinfo {author}
  {\bibfnamefont{M.~J.}\ \bibnamefont{{Burchell}}}, \bibinfo {author}
  {\bibfnamefont{H.}~\bibnamefont{{Busemann}}}, \bibinfo {author}
  {\bibfnamefont{A.}~\bibnamefont{{Butterworth}}}, \bibinfo {author}
  {\bibfnamefont{M.}~\bibnamefont{{Chaussidon}}}, \bibinfo {author}
  {\bibfnamefont{A.}~\bibnamefont{{Cheuvront}}}, \bibinfo {author}
  {\bibfnamefont{M.}~\bibnamefont{{Chi}}}, \bibinfo {author}
  {\bibfnamefont{M.~J.}\ \bibnamefont{{Cintala}}}, \bibinfo {author}
  {\bibfnamefont{B.~C.}\ \bibnamefont{{Clark}}}, \bibinfo {author}
  {\bibfnamefont{S.~J.}\ \bibnamefont{{Clemett}}}, \bibinfo {author}
  {\bibfnamefont{G.}~\bibnamefont{{Cody}}}, \bibinfo {author}
  {\bibfnamefont{L.}~\bibnamefont{{Colangeli}}}, \bibinfo {author}
  {\bibfnamefont{G.}~\bibnamefont{{Cooper}}}, \bibinfo {author}
  {\bibfnamefont{P.}~\bibnamefont{{Cordier}}}, \bibinfo {author}
  {\bibfnamefont{C.}~\bibnamefont{{Daghlian}}}, \bibinfo {author}
  {\bibfnamefont{Z.}~\bibnamefont{{Dai}}}, \bibinfo {author}
  {\bibfnamefont{L.}~\bibnamefont{{D'Hendecourt}}}, \bibinfo {author}
  {\bibfnamefont{Z.}~\bibnamefont{{Djouadi}}}, \bibinfo {author}
  {\bibfnamefont{G.}~\bibnamefont{{Dominguez}}}, \bibinfo {author}
  {\bibfnamefont{T.}~\bibnamefont{{Duxbury}}}, \bibinfo {author}
  {\bibfnamefont{J.~P.}\ \bibnamefont{{Dworkin}}}, \bibinfo {author}
  {\bibfnamefont{D.~S.}\ \bibnamefont{{Ebel}}}, \bibinfo {author}
  {\bibfnamefont{T.~E.}\ \bibnamefont{{Economou}}}, \bibinfo {author}
  {\bibfnamefont{S.}~\bibnamefont{{Fakra}}}, \bibinfo {author}
  {\bibfnamefont{S.~A.~J.}\ \bibnamefont{{Fairey}}}, \bibinfo {author}
  {\bibfnamefont{S.}~\bibnamefont{{Fallon}}}, \bibinfo {author}
  {\bibfnamefont{G.}~\bibnamefont{{Ferrini}}}, \bibinfo {author}
  {\bibfnamefont{T.}~\bibnamefont{{Ferroir}}}, \bibinfo {author}
  {\bibfnamefont{H.}~\bibnamefont{{Fleckenstein}}}, \bibinfo {author}
  {\bibfnamefont{C.}~\bibnamefont{{Floss}}}, \bibinfo {author}
  {\bibfnamefont{G.}~\bibnamefont{{Flynn}}}, \bibinfo {author}
  {\bibfnamefont{I.~A.}\ \bibnamefont{{Franchi}}}, \bibinfo {author}
  {\bibfnamefont{M.}~\bibnamefont{{Fries}}}, \bibinfo {author}
  {\bibfnamefont{Z.}~\bibnamefont{{Gainsforth}}}, \bibinfo {author}
  {\bibfnamefont{{J.-P.}}\ \bibnamefont{{Gallien}}}, \bibinfo {author}
  {\bibfnamefont{M.}~\bibnamefont{{Genge}}}, \bibinfo {author}
  {\bibfnamefont{M.~K.}\ \bibnamefont{{Gilles}}}, \bibinfo {author}
  {\bibfnamefont{P.}~\bibnamefont{{Gillet}}}, \bibinfo {author}
  {\bibfnamefont{J.}~\bibnamefont{{Gilmour}}}, \bibinfo {author}
  {\bibfnamefont{D.~P.}\ \bibnamefont{{Glavin}}}, \bibinfo {author}
  {\bibfnamefont{M.}~\bibnamefont{{Gounelle}}}, \bibinfo {author}
  {\bibfnamefont{M.~M.}\ \bibnamefont{{Grady}}}, \bibinfo {author}
  {\bibfnamefont{G.~A.}\ \bibnamefont{{Graham}}}, \bibinfo {author}
  {\bibfnamefont{P.~G.}\ \bibnamefont{{Grant}}}, \bibinfo {author}
  {\bibfnamefont{S.~F.}\ \bibnamefont{{Green}}}, \bibinfo {author}
  {\bibfnamefont{F.}~\bibnamefont{{Grossemy}}}, \bibinfo {author}
  {\bibfnamefont{L.}~\bibnamefont{{Grossman}}}, \bibinfo {author}
  {\bibfnamefont{J.~N.}\ \bibnamefont{{Grossman}}}, \bibinfo {author}
  {\bibfnamefont{Y.}~\bibnamefont{{Guan}}}, \bibinfo {author}
  {\bibfnamefont{K.}~\bibnamefont{{Hagiya}}}, \bibinfo {author}
  {\bibfnamefont{R.}~\bibnamefont{{Harvey}}}, \bibinfo {author}
  {\bibfnamefont{P.}~\bibnamefont{{Heck}}}, \bibinfo {author}
  {\bibfnamefont{G.~F.}\ \bibnamefont{{Herzog}}}, \bibinfo {author}
  {\bibfnamefont{P.}~\bibnamefont{{Hoppe}}}, \bibinfo {author}
  {\bibfnamefont{F.}~\bibnamefont{{H{\"o}rz}}}, \bibinfo {author}
  {\bibfnamefont{J.}~\bibnamefont{{Huth}}}, \bibinfo {author}
  {\bibfnamefont{I.~D.}\ \bibnamefont{{Hutcheon}}}, \bibinfo {author}
  {\bibfnamefont{K.}~\bibnamefont{{Ignatyev}}}, \bibinfo {author}
  {\bibfnamefont{H.}~\bibnamefont{{Ishii}}}, \bibinfo {author}
  {\bibfnamefont{M.}~\bibnamefont{{Ito}}}, \bibinfo {author}
  {\bibfnamefont{D.}~\bibnamefont{{Jacob}}}, \bibinfo {author}
  {\bibfnamefont{C.}~\bibnamefont{{Jacobsen}}}, \bibinfo {author}
  {\bibfnamefont{S.}~\bibnamefont{{Jacobsen}}}, \bibinfo {author}
  {\bibfnamefont{S.}~\bibnamefont{{Jones}}}, \bibinfo {author}
  {\bibfnamefont{D.}~\bibnamefont{{Joswiak}}}, \bibinfo {author}
  {\bibfnamefont{A.}~\bibnamefont{{Jurewicz}}}, \bibinfo {author}
  {\bibfnamefont{A.~T.}\ \bibnamefont{{Kearsley}}}, \bibinfo {author}
  {\bibfnamefont{L.~P.}\ \bibnamefont{{Keller}}}, \bibinfo {author}
  {\bibfnamefont{H.}~\bibnamefont{{Khodja}}}, \bibinfo {author}
  {\bibfnamefont{A.~L.~D.}\ \bibnamefont{{Kilcoyne}}}, \bibinfo {author}
  {\bibfnamefont{J.}~\bibnamefont{{Kissel}}}, \bibinfo {author}
  {\bibfnamefont{A.}~\bibnamefont{{Krot}}}, \bibinfo {author}
  {\bibfnamefont{F.}~\bibnamefont{{Langenhorst}}}, \bibinfo {author}
  {\bibfnamefont{A.}~\bibnamefont{{Lanzirotti}}}, \bibinfo {author}
  {\bibfnamefont{L.}~\bibnamefont{{Le}}}, \bibinfo {author}
  {\bibfnamefont{L.~A.}\ \bibnamefont{{Leshin}}}, \bibinfo {author}
  {\bibfnamefont{J.}~\bibnamefont{{Leitner}}}, \bibinfo {author}
  {\bibfnamefont{L.}~\bibnamefont{{Lemelle}}}, \bibinfo {author}
  {\bibfnamefont{H.}~\bibnamefont{{Leroux}}}, \bibinfo {author}
  {\bibfnamefont{{M.-C.}}\ \bibnamefont{{Liu}}}, \bibinfo {author}
  {\bibfnamefont{K.}~\bibnamefont{{Luening}}}, \bibinfo {author}
  {\bibfnamefont{I.}~\bibnamefont{{Lyon}}}, \bibinfo {author}
  {\bibfnamefont{G.}~\bibnamefont{{MacPherson}}}, \bibinfo {author}
  {\bibfnamefont{M.~A.}\ \bibnamefont{{Marcus}}}, \bibinfo {author}
  {\bibfnamefont{K.}~\bibnamefont{{Marhas}}}, \bibinfo {author}
  {\bibfnamefont{B.}~\bibnamefont{{Marty}}}, \bibinfo {author}
  {\bibfnamefont{G.}~\bibnamefont{{Matrajt}}}, \bibinfo {author}
  {\bibfnamefont{K.}~\bibnamefont{{McKeegan}}}, \bibinfo {author}
  {\bibfnamefont{A.}~\bibnamefont{{Meibom}}}, \bibinfo {author}
  {\bibfnamefont{V.}~\bibnamefont{{Mennella}}}, \bibinfo {author}
  {\bibfnamefont{K.}~\bibnamefont{{Messenger}}}, \bibinfo {author}
  {\bibfnamefont{S.}~\bibnamefont{{Messenger}}}, \bibinfo {author}
  {\bibfnamefont{T.}~\bibnamefont{{Mikouchi}}}, \bibinfo {author}
  {\bibfnamefont{S.}~\bibnamefont{{Mostefaoui}}}, \bibinfo {author}
  {\bibfnamefont{T.}~\bibnamefont{{Nakamura}}}, \bibinfo {author}
  {\bibfnamefont{T.}~\bibnamefont{{Nakano}}}, \bibinfo {author}
  {\bibfnamefont{M.}~\bibnamefont{{Newville}}}, \bibinfo {author}
  {\bibfnamefont{L.~R.}\ \bibnamefont{{Nittler}}}, \bibinfo {author}
  {\bibfnamefont{I.}~\bibnamefont{{Ohnishi}}}, \bibinfo {author}
  {\bibfnamefont{K.}~\bibnamefont{{Ohsumi}}}, \bibinfo {author}
  {\bibfnamefont{K.}~\bibnamefont{{Okudaira}}}, \bibinfo {author}
  {\bibfnamefont{D.~A.}\ \bibnamefont{{Papanastassiou}}}, \bibinfo {author}
  {\bibfnamefont{R.}~\bibnamefont{{Palma}}}, \bibinfo {author}
  {\bibfnamefont{M.~E.}\ \bibnamefont{{Palumbo}}}, \bibinfo {author}
  {\bibfnamefont{R.~O.}\ \bibnamefont{{Pepin}}}, \bibinfo {author}
  {\bibfnamefont{D.}~\bibnamefont{{Perkins}}}, \bibinfo {author}
  {\bibfnamefont{M.}~\bibnamefont{{Perronnet}}}, \bibinfo {author}
  {\bibfnamefont{P.}~\bibnamefont{{Pianetta}}}, \bibinfo {author}
  {\bibfnamefont{W.}~\bibnamefont{{Rao}}}, \bibinfo {author}
  {\bibfnamefont{F.~J.~M.}\ \bibnamefont{{Rietmeijer}}}, \bibinfo {author}
  {\bibfnamefont{F.}~\bibnamefont{{Robert}}}, \bibinfo {author}
  {\bibfnamefont{D.}~\bibnamefont{{Rost}}}, \bibinfo {author}
  {\bibfnamefont{A.}~\bibnamefont{{Rotundi}}}, \bibinfo {author}
  {\bibfnamefont{R.}~\bibnamefont{{Ryan}}}, \bibinfo {author}
  {\bibfnamefont{S.~A.}\ \bibnamefont{{Sandford}}}, \bibinfo {author}
  {\bibfnamefont{C.~S.}\ \bibnamefont{{Schwandt}}}, \bibinfo {author}
  {\bibfnamefont{T.~H.}\ \bibnamefont{{See}}}, \bibinfo {author}
  {\bibfnamefont{D.}~\bibnamefont{{Schlutter}}}, \bibinfo {author}
  {\bibfnamefont{J.}~\bibnamefont{{Sheffield-Parker}}}, \bibinfo {author}
  {\bibfnamefont{A.}~\bibnamefont{{Simionovici}}}, \bibinfo {author}
  {\bibfnamefont{S.}~\bibnamefont{{Simon}}}, \bibinfo {author}
  {\bibfnamefont{I.}~\bibnamefont{{Sitnitsky}}}, \bibinfo {author}
  {\bibfnamefont{C.~J.}\ \bibnamefont{{Snead}}}, \bibinfo {author}
  {\bibfnamefont{M.~K.}\ \bibnamefont{{Spencer}}}, \bibinfo {author}
  {\bibfnamefont{F.~J.}\ \bibnamefont{{Stadermann}}}, \bibinfo {author}
  {\bibfnamefont{A.}~\bibnamefont{{Steele}}}, \bibinfo {author}
  {\bibfnamefont{T.}~\bibnamefont{{Stephan}}}, \bibinfo {author}
  {\bibfnamefont{R.}~\bibnamefont{{Stroud}}}, \bibinfo {author}
  {\bibfnamefont{J.}~\bibnamefont{{Susini}}}, \bibinfo {author}
  {\bibfnamefont{S.~R.}\ \bibnamefont{{Sutton}}}, \bibinfo {author}
  {\bibfnamefont{Y.}~\bibnamefont{{Suzuki}}}, \bibinfo {author}
  {\bibfnamefont{M.}~\bibnamefont{{Taheri}}}, \bibinfo {author}
  {\bibfnamefont{S.}~\bibnamefont{{Taylor}}}, \bibinfo {author}
  {\bibfnamefont{N.}~\bibnamefont{{Teslich}}}, \bibinfo {author}
  {\bibfnamefont{K.}~\bibnamefont{{Tomeoka}}}, \bibinfo {author}
  {\bibfnamefont{N.}~\bibnamefont{{Tomioka}}}, \bibinfo {author}
  {\bibfnamefont{A.}~\bibnamefont{{Toppani}}}, \bibinfo {author}
  {\bibfnamefont{J.~M.}\ \bibnamefont{{Trigo-Rodr{\'{\i}}guez}}}, \bibinfo
  {author} {\bibfnamefont{D.}~\bibnamefont{{Troadec}}}, \bibinfo {author}
  {\bibfnamefont{A.}~\bibnamefont{{Tsuchiyama}}}, \bibinfo {author}
  {\bibfnamefont{A.~J.}\ \bibnamefont{{Tuzzolino}}}, \bibinfo {author}
  {\bibfnamefont{T.}~\bibnamefont{{Tyliszczak}}}, \bibinfo {author}
  {\bibfnamefont{K.}~\bibnamefont{{Uesugi}}}, \bibinfo {author}
  {\bibfnamefont{M.}~\bibnamefont{{Velbel}}}, \bibinfo {author}
  {\bibfnamefont{J.}~\bibnamefont{{Vellenga}}}, \bibinfo {author}
  {\bibfnamefont{E.}~\bibnamefont{{Vicenzi}}}, \bibinfo {author}
  {\bibfnamefont{L.}~\bibnamefont{{Vincze}}}, \bibinfo {author}
  {\bibfnamefont{J.}~\bibnamefont{{Warren}}}, \bibinfo {author}
  {\bibfnamefont{I.}~\bibnamefont{{Weber}}}, \bibinfo {author}
  {\bibfnamefont{M.}~\bibnamefont{{Weisberg}}}, \bibinfo {author}
  {\bibfnamefont{A.~J.}\ \bibnamefont{{Westphal}}}, \bibinfo {author}
  {\bibfnamefont{S.}~\bibnamefont{{Wirick}}}, \bibinfo {author}
  {\bibfnamefont{D.}~\bibnamefont{{Wooden}}}, \bibinfo {author}
  {\bibfnamefont{B.}~\bibnamefont{{Wopenka}}}, \bibinfo {author}
  {\bibfnamefont{P.}~\bibnamefont{{Wozniakiewicz}}}, \bibinfo {author}
  {\bibfnamefont{I.}~\bibnamefont{{Wright}}}, \bibinfo {author}
  {\bibfnamefont{H.}~\bibnamefont{{Yabuta}}}, \bibinfo {author}
  {\bibfnamefont{H.}~\bibnamefont{{Yano}}}, \bibinfo {author}
  {\bibfnamefont{E.~D.}\ \bibnamefont{{Young}}}, \bibinfo {author}
  {\bibfnamefont{R.~N.}\ \bibnamefont{{Zare}}}, \bibinfo {author}
  {\bibfnamefont{T.}~\bibnamefont{{Zega}}}, \bibinfo {author}
  {\bibfnamefont{K.}~\bibnamefont{{Ziegler}}}, \bibinfo {author}
  {\bibfnamefont{L.}~\bibnamefont{{Zimmerman}}}, \bibinfo {author}
  {\bibfnamefont{E.}~\bibnamefont{{Zinner}}},\ and\ \bibinfo {author}
  {\bibfnamefont{M.}~\bibnamefont{{Zolensky}}}}%
  , \bibinfo {year} {2006},\ \bibfield{title}{%
  \enquote{\bibinfo {title} {Comet {81P/Wild} 2 under a microscope},}\ }%
  \bibfield{journal}{%
  \bibinfo {journal} {Science}\ }%
  \textbf{\bibinfo {volume} {314}},\ \bibinfo {pages} {1711--1716}%
  \bibAnnoteFile{NoStop}{brownlee2006}%
\bibitem[{\citenamefont{Buch}\ \emph{et~al.}(2009)\citenamefont{Buch},
  \citenamefont{Devlin}, \citenamefont{Abrrey~Monreal},
  \citenamefont{Jagoda-Cwiklik}, \citenamefont{Uras-Aytemiz},\ and\
  \citenamefont{Cwiklik}}]{buch2009}%
  \BibitemOpen
  \bibfield{author}{%
  \bibinfo {author} {\bibnamefont{Buch}, \bibfnamefont{V.}}, \bibinfo {author}
  {\bibfnamefont{J.~P.}\ \bibnamefont{Devlin}}, \bibinfo {author}
  {\bibfnamefont{I.}~\bibnamefont{Abrrey~Monreal}}, \bibinfo {author}
  {\bibfnamefont{B.}~\bibnamefont{Jagoda-Cwiklik}}, \bibinfo {author}
  {\bibfnamefont{N.}~\bibnamefont{Uras-Aytemiz}},\ and\ \bibinfo {author}
  {\bibfnamefont{L.}~\bibnamefont{Cwiklik}}}%
  , \bibinfo {year} {2009},\ \bibfield{title}{%
  \enquote{\bibinfo {title} {Clathrate hydrates with hydrogen-bonding
  guests},}\ }%
  \bibfield{journal}{%
  \bibinfo {journal} {Phys. Chem. Chem. Phys.}\ }%
  \textbf{\bibinfo {volume} {11}},\ \bibinfo {pages} {10245--10265}%
  \bibAnnoteFile{NoStop}{buch2009}%
\bibitem[{\citenamefont{Buch}\ \emph{et~al.}(2008)\citenamefont{Buch},
  \citenamefont{Dubrovskiy}, \citenamefont{Mohamed}, \citenamefont{Parrinello},
  \citenamefont{Sadlej}, \citenamefont{Hammerich},\ and\
  \citenamefont{Devlin}}]{buch2008}%
  \BibitemOpen
  \bibfield{author}{%
  \bibinfo {author} {\bibnamefont{Buch}, \bibfnamefont{V.}}, \bibinfo {author}
  {\bibfnamefont{A.}~\bibnamefont{Dubrovskiy}}, \bibinfo {author}
  {\bibfnamefont{F.}~\bibnamefont{Mohamed}}, \bibinfo {author}
  {\bibfnamefont{M.}~\bibnamefont{Parrinello}}, \bibinfo {author}
  {\bibfnamefont{J.}~\bibnamefont{Sadlej}}, \bibinfo {author}
  {\bibfnamefont{A.~D.}\ \bibnamefont{Hammerich}},\ and\ \bibinfo {author}
  {\bibfnamefont{J.~P.}\ \bibnamefont{Devlin}}}%
  , \bibinfo {year} {2008},\ \bibfield{title}{%
  \enquote{\bibinfo {title} {{HCl} hydrates as model systems for protonated
  water},}\ }%
  \bibfield{journal}{%
  \bibinfo {journal} {J. Phys. Chem. A}\ }%
  \textbf{\bibinfo {volume} {112}},\ \bibinfo {pages} {2144--2161}%
  \bibAnnoteFile{NoStop}{buch2008}%
\bibitem[{\citenamefont{Buch}\ \emph{et~al.}(2007)\citenamefont{Buch},
  \citenamefont{Mohamed}, \citenamefont{Parrinello},\ and\
  \citenamefont{Devlin}}]{buch2007}%
  \BibitemOpen
  \bibfield{author}{%
  \bibinfo {author} {\bibnamefont{Buch}, \bibfnamefont{V.}}, \bibinfo {author}
  {\bibfnamefont{F.}~\bibnamefont{Mohamed}}, \bibinfo {author}
  {\bibfnamefont{M.}~\bibnamefont{Parrinello}},\ and\ \bibinfo {author}
  {\bibfnamefont{J.~P.}\ \bibnamefont{Devlin}}}%
  , \bibinfo {year} {2007},\ \bibfield{title}{%
  \enquote{\bibinfo {title} {Elusive structure of {HCl} monohydrate},}\ }%
  \bibfield{journal}{%
  \bibinfo {journal} {J. Chem. Phys.}\ }%
  \textbf{\bibinfo {volume} {126}},\ \bibinfo {pages} {074503}%
  \bibAnnoteFile{NoStop}{buch2007}%
\bibitem[{\citenamefont{Buch}\ \emph{et~al.}(2002)\citenamefont{Buch},
  \citenamefont{Sadlej}, \citenamefont{Aytemiz-Uras},\ and\
  \citenamefont{Devlin}}]{buch2002}%
  \BibitemOpen
  \bibfield{author}{%
  \bibinfo {author} {\bibnamefont{Buch}, \bibfnamefont{V.}}, \bibinfo {author}
  {\bibfnamefont{J.}~\bibnamefont{Sadlej}}, \bibinfo {author}
  {\bibfnamefont{N.}~\bibnamefont{Aytemiz-Uras}},\ and\ \bibinfo {author}
  {\bibfnamefont{J.~P.}\ \bibnamefont{Devlin}}}%
  , \bibinfo {year} {2002},\ \bibfield{title}{%
  \enquote{\bibinfo {title} {Solvation and ionization stages of {HCl} on ice
  nanocrystals},}\ }%
  \bibfield{journal}{%
  \bibinfo {journal} {J. Phys. Chem. A}\ }%
  \textbf{\bibinfo {volume} {106}},\ \bibinfo {pages} {9374--9389}%
  \bibAnnoteFile{NoStop}{buch2002}%
\bibitem[{\citenamefont{Bukowski}\ \emph{et~al.}(2007)\citenamefont{Bukowski},
  \citenamefont{Szalewicz}, \citenamefont{Groenenboom},\ and\
  \citenamefont{{van der Avoird}}}]{bukowski2007}%
  \BibitemOpen
  \bibfield{author}{%
  \bibinfo {author} {\bibnamefont{Bukowski}, \bibfnamefont{R.}}, \bibinfo
  {author} {\bibfnamefont{K.}~\bibnamefont{Szalewicz}}, \bibinfo {author}
  {\bibfnamefont{G.~C.}\ \bibnamefont{Groenenboom}},\ and\ \bibinfo {author}
  {\bibfnamefont{A.}~\bibnamefont{{van der Avoird}}}}%
  , \bibinfo {year} {2007},\ \bibfield{title}{%
  \enquote{\bibinfo {title} {Predictions of the properties of water from first
  principles},}\ }%
  \bibfield{journal}{%
  \bibinfo {journal} {Science}\ }%
  \textbf{\bibinfo {volume} {315}},\ \bibinfo {pages} {1249--1252}%
  \bibAnnoteFile{NoStop}{bukowski2007}%
\bibitem[{\citenamefont{Burke}\ and\ \citenamefont{Brown}(2010)}]{burke2010}%
  \BibitemOpen
  \bibfield{author}{%
  \bibinfo {author} {\bibnamefont{Burke}, \bibfnamefont{D.~J.}},\ and\ \bibinfo
  {author} {\bibfnamefont{W.~A.}\ \bibnamefont{Brown}}}%
  , \bibinfo {year} {2010},\ \bibfield{title}{%
  \enquote{\bibinfo {title} {Ice in space: surface science investigations of
  the thermal desorption of model interstellar ices on dust grain analogue
  surfaces},}\ }%
  \bibfield{journal}{%
  \bibinfo {journal} {Phys. Chem. Chem. Phys.}\ }%
  \textbf{\bibinfo {volume} {12}},\ \bibinfo {pages} {5947--5969}%
  \bibAnnoteFile{NoStop}{burke2010}%
\bibitem[{\citenamefont{Burniston}\
  \emph{et~al.}(2007)\citenamefont{Burniston}, \citenamefont{Strachan},
  \citenamefont{Hoff},\ and\ \citenamefont{Wania}}]{Burniston:2007p2458}%
  \BibitemOpen
  \bibfield{author}{%
  \bibinfo {author} {\bibnamefont{Burniston}, \bibfnamefont{D.~A.}}, \bibinfo
  {author} {\bibfnamefont{W.~J.~M.}\ \bibnamefont{Strachan}}, \bibinfo {author}
  {\bibfnamefont{J.~T.}\ \bibnamefont{Hoff}},\ and\ \bibinfo {author}
  {\bibfnamefont{F.}~\bibnamefont{Wania}}}%
  , \bibinfo {year} {2007},\ \bibfield{title}{%
  \enquote{\bibinfo {title} {Changes in surface area and concentrations of
  semivolatile organic contaminants in aging snow},}\ }%
  \bibfield{journal}{%
  \bibinfo {journal} {Environ. Sci. Technol.}\ }%
  \textbf{\bibinfo {volume} {41}},\ \bibinfo {pages} {4932--4937}%
  \bibAnnoteFile{NoStop}{Burniston:2007p2458}%
\bibitem[{\citenamefont{Burton}\ and\
  \citenamefont{Oliver}(1935)}]{burton1935}%
  \BibitemOpen
  \bibfield{author}{%
  \bibinfo {author} {\bibnamefont{Burton}, \bibfnamefont{E.~F.}},\ and\
  \bibinfo {author} {\bibfnamefont{W.~F.}\ \bibnamefont{Oliver}}}%
  , \bibinfo {year} {1935},\ \bibfield{title}{%
  \enquote{\bibinfo {title} {The crystal structure of ice at low
  temperatures},}\ }%
  \bibfield{journal}{%
  \bibinfo {journal} {Proc. R. Soc. London A}\ }%
  \textbf{\bibinfo {volume} {153}},\ \bibinfo {pages} {166--172}%
  \bibAnnoteFile{NoStop}{burton1935}%
\bibitem[{\citenamefont{{Cahoy}}\ \emph{et~al.}(2010)\citenamefont{{Cahoy}},
  \citenamefont{{Marley}},\ and\ \citenamefont{{Fortney}}}]{cahoy2010}%
  \BibitemOpen
  \bibfield{author}{%
  \bibinfo {author} {\bibnamefont{{Cahoy}}, \bibfnamefont{K.~L.}}, \bibinfo
  {author} {\bibfnamefont{M.~S.}\ \bibnamefont{{Marley}}},\ and\ \bibinfo
  {author} {\bibfnamefont{J.~J.}\ \bibnamefont{{Fortney}}}}%
  , \bibinfo {year} {2010},\ \bibfield{title}{%
  \enquote{\bibinfo {title} {Exoplanet albedo spectra and colors as a function
  of planet phase, separation, and metallicity},}\ }%
  \bibfield{journal}{%
  \bibinfo {journal} {Astrophys. J.}\ }%
  \textbf{\bibinfo {volume} {724}},\ \bibinfo {pages} {189--214}%
  \bibAnnoteFile{NoStop}{cahoy2010}%
\bibitem[{\citenamefont{Callaghan}\
  \emph{et~al.}(1994)\citenamefont{Callaghan}, \citenamefont{Lim},
  \citenamefont{Murdock}, \citenamefont{Sloan},\ and\
  \citenamefont{Donaldson}}]{Callaghan1994}%
  \BibitemOpen
  \bibfield{author}{%
  \bibinfo {author} {\bibnamefont{Callaghan}, \bibfnamefont{R.}}, \bibinfo
  {author} {\bibfnamefont{I.~J.}\ \bibnamefont{Lim}}, \bibinfo {author}
  {\bibfnamefont{D.~E.}\ \bibnamefont{Murdock}}, \bibinfo {author}
  {\bibfnamefont{J.~J.}\ \bibnamefont{Sloan}},\ and\ \bibinfo {author}
  {\bibfnamefont{D.~J.}\ \bibnamefont{Donaldson}}}%
  , \bibinfo {year} {1994},\ \bibfield{title}{%
  \enquote{\bibinfo {title} {Laboratory simulation of polar stratospheric
  clouds},}\ }%
  \bibfield{journal}{%
  \bibinfo {journal} {Geophys. Res. Lett.}\ }%
  \textbf{\bibinfo {volume} {21}},\ \bibinfo {pages} {373--376}%
  \bibAnnoteFile{NoStop}{Callaghan1994}%
\bibitem[{\citenamefont{Calvin}\ \emph{et~al.}(1995)\citenamefont{Calvin},
  \citenamefont{Clark}, \citenamefont{Brown},\ and\
  \citenamefont{Spencer}}]{calvin1995}%
  \BibitemOpen
  \bibfield{author}{%
  \bibinfo {author} {\bibnamefont{Calvin}, \bibfnamefont{W.~M.}}, \bibinfo
  {author} {\bibfnamefont{R.~N.}\ \bibnamefont{Clark}}, \bibinfo {author}
  {\bibfnamefont{R.~H.}\ \bibnamefont{Brown}},\ and\ \bibinfo {author}
  {\bibfnamefont{J.~R.}\ \bibnamefont{Spencer}}}%
  , \bibinfo {year} {1995},\ \bibfield{title}{%
  \enquote{\bibinfo {title} {Spectra of the icy {Galilean} satellites from 0.2
  to 5$\mu$m: A compilation, new observations, and a recent summary},}\ }%
  \bibfield{journal}{%
  \bibinfo {journal} {J. Geophys. Res.}\ }%
  \textbf{\bibinfo {volume} {100}},\ \bibinfo {pages} {19041--19048}%
  \bibAnnoteFile{NoStop}{calvin1995}%
\bibitem[{\citenamefont{{Campins}}\
  \emph{et~al.}(2010)\citenamefont{{Campins}}, \citenamefont{{Hargrove}},
  \citenamefont{{Pinilla-Alonso}}, \citenamefont{{Howell}},
  \citenamefont{{Kelley}}, \citenamefont{{Licandro}},
  \citenamefont{{Moth{\'e}-Diniz}}, \citenamefont{{Fern{\'a}ndez}},\ and\
  \citenamefont{{Ziffer}}}]{campins2010}%
  \BibitemOpen
  \bibfield{author}{%
  \bibinfo {author} {\bibnamefont{{Campins}}, \bibfnamefont{H.}}, \bibinfo
  {author} {\bibfnamefont{K.}~\bibnamefont{{Hargrove}}}, \bibinfo {author}
  {\bibfnamefont{N.}~\bibnamefont{{Pinilla-Alonso}}}, \bibinfo {author}
  {\bibfnamefont{E.~S.}\ \bibnamefont{{Howell}}}, \bibinfo {author}
  {\bibfnamefont{M.~S.}\ \bibnamefont{{Kelley}}}, \bibinfo {author}
  {\bibfnamefont{J.}~\bibnamefont{{Licandro}}}, \bibinfo {author}
  {\bibfnamefont{T.}~\bibnamefont{{Moth{\'e}-Diniz}}}, \bibinfo {author}
  {\bibfnamefont{Y.}~\bibnamefont{{Fern{\'a}ndez}}},\ and\ \bibinfo {author}
  {\bibfnamefont{J.}~\bibnamefont{{Ziffer}}}}%
  , \bibinfo {year} {2010},\ \bibfield{title}{%
  \enquote{\bibinfo {title} {{Water ice and organics on the surface of the
  asteroid 24 Themis}},}\ }%
  \bibfield{journal}{%
  \bibinfo {journal} {Nature}\ }%
  \textbf{\bibinfo {volume} {464}},\ \bibinfo {pages} {1320--1321}%
  \bibAnnoteFile{NoStop}{campins2010}%
\bibitem[{\citenamefont{Cantrell}\ and\
  \citenamefont{Heymsfield}(2005)}]{cantrell2005}%
  \BibitemOpen
  \bibfield{author}{%
  \bibinfo {author} {\bibnamefont{Cantrell}, \bibfnamefont{W.}},\ and\ \bibinfo
  {author} {\bibfnamefont{A.}~\bibnamefont{Heymsfield}}}%
  , \bibinfo {year} {2005},\ \bibfield{title}{%
  \enquote{\bibinfo {title} {Production of ice in tropospheric clouds: A
  review},}\ }%
  \bibfield{journal}{%
  \bibinfo {journal} {Bull. Am. Meteor. Soc.}\ }%
  \textbf{\bibinfo {volume} {86}},\ \bibinfo {pages} {795--807}%
  \bibAnnoteFile{NoStop}{cantrell2005}%
\bibitem[{\citenamefont{Capria}(2000)}]{capria2000}%
  \BibitemOpen
  \bibfield{author}{%
  \bibinfo {author} {\bibnamefont{Capria}, \bibfnamefont{M.T.}}}%
  , \bibinfo {year} {2000},\ \bibfield{title}{%
  \enquote{\bibinfo {title} {Sublimation mechanisms of comet nuclei},}\ }%
  \bibfield{journal}{%
  \bibinfo {journal} {Earth, Moon and Planets}\ }%
  \textbf{\bibinfo {volume} {89}},\ \bibinfo {pages} {161--178}%
  \bibAnnoteFile{NoStop}{capria2000}%
\bibitem[{\citenamefont{Car}\ and\ \citenamefont{Parrinello}(1985)}]{car1985}%
  \BibitemOpen
  \bibfield{author}{%
  \bibinfo {author} {\bibnamefont{Car}, \bibfnamefont{R.}},\ and\ \bibinfo
  {author} {\bibfnamefont{M.}~\bibnamefont{Parrinello}}}%
  , \bibinfo {year} {1985},\ \bibfield{title}{%
  \enquote{\bibinfo {title} {Unified approach for molecular dynamics and
  density-functional theory},}\ }%
  \bibfield{journal}{%
  \bibinfo {journal} {Phys. Rev. Lett.}\ }%
  \textbf{\bibinfo {volume} {55}},\ \bibinfo {pages} {2471--2474}%
  \bibAnnoteFile{NoStop}{car1985}%
\bibitem[{\citenamefont{Carlson}\ \emph{et~al.}(1999)\citenamefont{Carlson},
  \citenamefont{Anderson}, \citenamefont{Johnson}, \citenamefont{Smythe},
  \citenamefont{Hendrix}, \citenamefont{Barth}, \citenamefont{Soderblom},
  \citenamefont{Hansen}, \citenamefont{McCord}, \citenamefont{Dalton},
  \citenamefont{Clark}, \citenamefont{Shirley}, \citenamefont{Ocampo},\ and\
  \citenamefont{Matson}}]{carlson1999}%
  \BibitemOpen
  \bibfield{author}{%
  \bibinfo {author} {\bibnamefont{Carlson}, \bibfnamefont{R.~W.}}, \bibinfo
  {author} {\bibfnamefont{M.~S.}\ \bibnamefont{Anderson}}, \bibinfo {author}
  {\bibfnamefont{R.~E.}\ \bibnamefont{Johnson}}, \bibinfo {author}
  {\bibfnamefont{W.~D.}\ \bibnamefont{Smythe}}, \bibinfo {author}
  {\bibfnamefont{A.~R}\ \bibnamefont{Hendrix}}, \bibinfo {author}
  {\bibfnamefont{C.~A.}\ \bibnamefont{Barth}}, \bibinfo {author}
  {\bibfnamefont{L.~A.}\ \bibnamefont{Soderblom}}, \bibinfo {author}
  {\bibfnamefont{G.~B.}\ \bibnamefont{Hansen}}, \bibinfo {author}
  {\bibfnamefont{T.~B.}\ \bibnamefont{McCord}}, \bibinfo {author}
  {\bibfnamefont{J.~B.}\ \bibnamefont{Dalton}}, \bibinfo {author}
  {\bibfnamefont{R.~N.}\ \bibnamefont{Clark}}, \bibinfo {author}
  {\bibfnamefont{J.~H.}\ \bibnamefont{Shirley}}, \bibinfo {author}
  {\bibfnamefont{A.~C.}\ \bibnamefont{Ocampo}},\ and\ \bibinfo {author}
  {\bibfnamefont{D.~L.}\ \bibnamefont{Matson}}}%
  , \bibinfo {year} {1999},\ \bibfield{title}{%
  \enquote{\bibinfo {title} {Hydrogen peroxide on the surface of {Europa}},}\
  }%
  \bibfield{journal}{%
  \bibinfo {journal} {Science}\ }%
  \textbf{\bibinfo {volume} {283}},\ \bibinfo {pages} {2062--2064}%
  \bibAnnoteFile{NoStop}{carlson1999}%
\bibitem[{\citenamefont{Cartwright}(2007)}]{cartwright2007}%
  \BibitemOpen
  \bibfield{author}{%
  \bibinfo {author} {\bibnamefont{Cartwright}, \bibfnamefont{J.~H.~E.}}}%
  , \bibinfo {year} {2007},\ \enquote{\bibinfo {title} {Ice in the solar
  system},}\ in\ \emph{\bibinfo {booktitle} {Solar and Stellar Physics through
  Eclipses}},\ \bibinfo {series} {ASP Conference Series}, Vol.\ \bibinfo
  {volume} {370},\ \bibinfo {editor} {edited by\ \bibinfo {editor}
  {\bibfnamefont{O.}~\bibnamefont{Demircan}}, \bibinfo {editor}
  {\bibfnamefont{S.~O.}\ \bibnamefont{Selam}},\ and\ \bibinfo {editor}
  {\bibfnamefont{B.}~\bibnamefont{Albayrak}}},\ pp.\ \bibinfo {pages}
  {265--69}%
  \bibAnnoteFile{NoStop}{cartwright2007}%
\bibitem[{\citenamefont{Cartwright}\
  \emph{et~al.}(2011)\citenamefont{Cartwright}, \citenamefont{Escribano},
  \citenamefont{Grothe}, \citenamefont{Piro}, \citenamefont{Sainz-D\'{\i}az},\
  and\ \citenamefont{Tuval}}]{cartwright2010_2}%
  \BibitemOpen
  \bibfield{author}{%
  \bibinfo {author} {\bibnamefont{Cartwright}, \bibfnamefont{J.~H.~E.}},
  \bibinfo {author} {\bibfnamefont{B.}~\bibnamefont{Escribano}}, \bibinfo
  {author} {\bibfnamefont{H.}~\bibnamefont{Grothe}}, \bibinfo {author}
  {\bibfnamefont{O.}~\bibnamefont{Piro}}, \bibinfo {author}
  {\bibfnamefont{C.~I.}\ \bibnamefont{Sainz-D\'{\i}az}},\ and\ \bibinfo
  {author} {\bibfnamefont{I.}~\bibnamefont{Tuval}}}%
  , \bibinfo {year} {2011},\ \bibfield{title}{%
  \enquote{\bibinfo {title} {Nonlinear dynamics of ice growth and charge
  production in thunderstorms},}\ }%
  \bibinfo {journal} {Proc. Roy. Soc. A},\ \bibinfo {pages} {submitted}%
  \bibAnnoteFile{NoStop}{cartwright2010_2}%
\bibitem[{\citenamefont{Cartwright}\
  \emph{et~al.}(2008)\citenamefont{Cartwright}, \citenamefont{Escribano},\ and\
  \citenamefont{Sainz-D\'{\i}az}}]{cartwright2008}%
  \BibitemOpen
\bibfield{journal}{%
    }%
  \bibfield{author}{%
  \bibinfo {author} {\bibnamefont{Cartwright}, \bibfnamefont{J.~H.~E.}},
  \bibinfo {author} {\bibfnamefont{B.}~\bibnamefont{Escribano}},\ and\ \bibinfo
  {author} {\bibfnamefont{C.~I.}\ \bibnamefont{Sainz-D\'{\i}az}}}%
  , \bibinfo {year} {2008},\ \bibfield{title}{%
  \enquote{\bibinfo {title} {The mesoscale morphologies of ice films: Porous
  and biomorphic forms of ice under astrophysical conditions},}\ }%
  \bibfield{journal}{%
  \bibinfo {journal} {Astrophys. J.}\ }%
  \textbf{\bibinfo {volume} {687}},\ \bibinfo {pages} {1406--1414}%
  \bibAnnoteFile{NoStop}{cartwright2008}%
\bibitem[{\citenamefont{Cartwright}\
  \emph{et~al.}(2010)\citenamefont{Cartwright}, \citenamefont{Escribano},\ and\
  \citenamefont{Sainz-D\'{\i}az}}]{cartwright2010}%
  \BibitemOpen
  \bibfield{author}{%
  \bibinfo {author} {\bibnamefont{Cartwright}, \bibfnamefont{J.~H.~E.}},
  \bibinfo {author} {\bibfnamefont{B.}~\bibnamefont{Escribano}},\ and\ \bibinfo
  {author} {\bibfnamefont{C.~I.}\ \bibnamefont{Sainz-D\'{\i}az}}}%
  , \bibinfo {year} {2010},\ \bibfield{title}{%
  \enquote{\bibinfo {title} {Ice films follow structure zone model
  morphologies},}\ }%
  \bibfield{journal}{%
  \bibinfo {journal} {Thin Solid Films}\ }%
  \textbf{\bibinfo {volume} {518}},\ \bibinfo {pages} {3422--3427}%
  \bibAnnoteFile{NoStop}{cartwright2010}%
\bibitem[{\citenamefont{Chaban}\ \emph{et~al.}(2007)\citenamefont{Chaban},
  \citenamefont{Bernstein},\ and\ \citenamefont{Cruikshank}}]{chaban2007}%
  \BibitemOpen
  \bibfield{author}{%
  \bibinfo {author} {\bibnamefont{Chaban}, \bibfnamefont{G.~M.}}, \bibinfo
  {author} {\bibfnamefont{M.}~\bibnamefont{Bernstein}},\ and\ \bibinfo {author}
  {\bibfnamefont{D.~P.}\ \bibnamefont{Cruikshank}}}%
  , \bibinfo {year} {2007},\ \bibfield{title}{%
  \enquote{\bibinfo {title} {Òcarbon dioxide on planetary bodies: Theoretical
  and experimental studies of molecular complexes},}\ }%
  \bibfield{journal}{%
  \bibinfo {journal} {Icarus}\ }%
  \textbf{\bibinfo {volume} {187}},\ \bibinfo {pages} {592--599}%
  \bibAnnoteFile{NoStop}{chaban2007}%
\bibitem[{\citenamefont{Chen}\ and\ \citenamefont{Morris}(2011)}]{chen2011}%
  \BibitemOpen
  \bibfield{author}{%
  \bibinfo {author} {\bibnamefont{Chen}, \bibfnamefont{A.~S.-H.}},\ and\
  \bibinfo {author} {\bibfnamefont{S.~W.}\ \bibnamefont{Morris}}}%
  , \bibinfo {year} {2011},\ \bibfield{title}{%
  \enquote{\bibinfo {title} {Experiments on the morphology of icicles},}\ }%
  \bibfield{journal}{%
  \bibinfo {journal} {Phys. Rev. E}\ }%
  \textbf{\bibinfo {volume} {83}},\ \bibinfo {pages} {026307}%
  \bibAnnoteFile{NoStop}{chen2011}%
\bibitem[{\citenamefont{Chiar}\ \emph{et~al.}(1994)\citenamefont{Chiar},
  \citenamefont{Adamson}, \citenamefont{Kerr},\ and\
  \citenamefont{Whittet}}]{chiar1994}%
  \BibitemOpen
  \bibfield{author}{%
  \bibinfo {author} {\bibnamefont{Chiar}, \bibfnamefont{J.~E.}}, \bibinfo
  {author} {\bibfnamefont{A.~J.}\ \bibnamefont{Adamson}}, \bibinfo {author}
  {\bibfnamefont{T.~H.}\ \bibnamefont{Kerr}},\ and\ \bibinfo {author}
  {\bibfnamefont{D.~C.~B.}\ \bibnamefont{Whittet}}}%
  , \bibinfo {year} {1994},\ \bibfield{title}{%
  \enquote{\bibinfo {title} {Solid carbon monoxide in the {Serpens} dark
  cloud},}\ }%
  \bibfield{journal}{%
  \bibinfo {journal} {Astrophys. J.}\ }%
  \textbf{\bibinfo {volume} {426}},\ \bibinfo {pages} {240--248}%
  \bibAnnoteFile{NoStop}{chiar1994}%
\bibitem[{\citenamefont{Chu}\ and\
  \citenamefont{Anastasio}(2005)}]{Chu:2005p2064}%
  \BibitemOpen
  \bibfield{author}{%
  \bibinfo {author} {\bibnamefont{Chu}, \bibfnamefont{L.}},\ and\ \bibinfo
  {author} {\bibfnamefont{C.}~\bibnamefont{Anastasio}}}%
  , \bibinfo {year} {2005},\ \bibfield{title}{%
  \enquote{\bibinfo {title} {Formation of hydroxyl radical from the photolysis
  of frozen hydrogen peroxide},}\ }%
  \bibfield{journal}{%
  \bibinfo {journal} {J. Phys. Chem. A}\ }%
  \textbf{\bibinfo {volume} {109}},\ \bibinfo {pages} {6264--6271}%
  \bibAnnoteFile{NoStop}{Chu:2005p2064}%
\bibitem[{\citenamefont{Chyba}(2000)}]{chyba2000}%
  \BibitemOpen
  \bibfield{author}{%
  \bibinfo {author} {\bibnamefont{Chyba}, \bibfnamefont{C.~F.}}}%
  , \bibinfo {year} {2000},\ \bibfield{title}{%
  \enquote{\bibinfo {title} {Energy for microbial life on {Europa}},}\ }%
  \bibfield{journal}{%
  \bibinfo {journal} {Nature}\ }%
  \textbf{\bibinfo {volume} {403}},\ \bibinfo {pages} {381--382}%
  \bibAnnoteFile{NoStop}{chyba2000}%
\bibitem[{\citenamefont{Clary}\ and\ \citenamefont{Wang}(1997)}]{clary1997}%
  \BibitemOpen
  \bibfield{author}{%
  \bibinfo {author} {\bibnamefont{Clary}, \bibfnamefont{D.~C.}},\ and\ \bibinfo
  {author} {\bibfnamefont{L.}~\bibnamefont{Wang}}}%
  , \bibinfo {year} {1997},\ \bibfield{title}{%
  \enquote{\bibinfo {title} {Influence of surface defects on the adsorption of
  {HCl} on ice},}\ }%
  \bibfield{journal}{%
  \bibinfo {journal} {J. Chem. Soc., Faraday Trans.}\ }%
  \textbf{\bibinfo {volume} {93}},\ \bibinfo {pages} {2763--2767}%
  \bibAnnoteFile{NoStop}{clary1997}%
\bibitem[{\citenamefont{Cochran}(2002)}]{cochran2002}%
  \BibitemOpen
  \bibfield{author}{%
  \bibinfo {author} {\bibnamefont{Cochran}, \bibfnamefont{A.~L.}}}%
  , \bibinfo {year} {2002},\ \bibfield{title}{%
  \enquote{\bibinfo {title} {A search for {N}$^{2+}$ in spectra of comet
  {C/2002 C1} ({Ikeya}--{Zhang})},}\ }%
  \bibfield{journal}{%
  \bibinfo {journal} {Astrophys. J.}\ }%
  \textbf{\bibinfo {volume} {576}},\ \bibinfo {pages} {165--168}%
  \bibAnnoteFile{NoStop}{cochran2002}%
\bibitem[{\citenamefont{Cochran}\ \emph{et~al.}(2000)\citenamefont{Cochran},
  \citenamefont{Cochran},\ and\ \citenamefont{Barker}}]{cochran2000}%
  \BibitemOpen
  \bibfield{author}{%
  \bibinfo {author} {\bibnamefont{Cochran}, \bibfnamefont{A.~L.}}, \bibinfo
  {author} {\bibfnamefont{W.~D.}\ \bibnamefont{Cochran}},\ and\ \bibinfo
  {author} {\bibfnamefont{E.~S.}\ \bibnamefont{Barker}}}%
  , \bibinfo {year} {2000},\ \bibfield{title}{%
  \enquote{\bibinfo {title} {{N}$^{2+}$ and {CO$^+$} in comets {122P/1995 S1}
  ({deVico}) and {C/1995 O1} ({Hale}--{Bopp})},}\ }%
  \bibinfo {journal} {Icarus},\ \bibinfo {pages} {583--593}%
  \bibAnnoteFile{NoStop}{cochran2000}%
\bibitem[{\citenamefont{{Colaprete}}\
  \emph{et~al.}(2010)\citenamefont{{Colaprete}}, \citenamefont{{Schultz}},
  \citenamefont{{Heldmann}}, \citenamefont{{Wooden}}, \citenamefont{{Shirley}},
  \citenamefont{{Ennico}}, \citenamefont{{Hermalyn}},
  \citenamefont{{Marshall}}, \citenamefont{{Ricco}}, \citenamefont{{Elphic}},
  \citenamefont{{Goldstein}}, \citenamefont{{Summy}}, \citenamefont{{Bart}},
  \citenamefont{{Asphaug}}, \citenamefont{{Korycansky}},
  \citenamefont{{Landis}},\ and\ \citenamefont{{Sollitt}}}]{colaprete2010}%
  \BibitemOpen
\bibfield{journal}{%
    }%
  \bibfield{author}{%
  \bibinfo {author} {\bibnamefont{{Colaprete}}, \bibfnamefont{A.}}, \bibinfo
  {author} {\bibfnamefont{P.}~\bibnamefont{{Schultz}}}, \bibinfo {author}
  {\bibfnamefont{J.}~\bibnamefont{{Heldmann}}}, \bibinfo {author}
  {\bibfnamefont{D.}~\bibnamefont{{Wooden}}}, \bibinfo {author}
  {\bibfnamefont{M.}~\bibnamefont{{Shirley}}}, \bibinfo {author}
  {\bibfnamefont{K.}~\bibnamefont{{Ennico}}}, \bibinfo {author}
  {\bibfnamefont{B.}~\bibnamefont{{Hermalyn}}}, \bibinfo {author}
  {\bibfnamefont{W.}~\bibnamefont{{Marshall}}}, \bibinfo {author}
  {\bibfnamefont{A.}~\bibnamefont{{Ricco}}}, \bibinfo {author}
  {\bibfnamefont{R.~C.}\ \bibnamefont{{Elphic}}}, \bibinfo {author}
  {\bibfnamefont{D.}~\bibnamefont{{Goldstein}}}, \bibinfo {author}
  {\bibfnamefont{D.}~\bibnamefont{{Summy}}}, \bibinfo {author}
  {\bibfnamefont{G.~D.}\ \bibnamefont{{Bart}}}, \bibinfo {author}
  {\bibfnamefont{E.}~\bibnamefont{{Asphaug}}}, \bibinfo {author}
  {\bibfnamefont{D.}~\bibnamefont{{Korycansky}}}, \bibinfo {author}
  {\bibfnamefont{D.}~\bibnamefont{{Landis}}},\ and\ \bibinfo {author}
  {\bibfnamefont{L.}~\bibnamefont{{Sollitt}}}}%
  , \bibinfo {year} {2010},\ \bibfield{title}{%
  \enquote{\bibinfo {title} {Detection of water in the {LCROSS} ejecta
  plume},}\ }%
  \bibfield{journal}{%
  \bibinfo {journal} {Science}\ }%
  \textbf{\bibinfo {volume} {330}},\ \bibinfo {pages} {463}%
  \bibAnnoteFile{NoStop}{colaprete2010}%
\bibitem[{\citenamefont{Collings}\ \emph{et~al.}(2004)\citenamefont{Collings},
  \citenamefont{Anderson}, \citenamefont{Chen}, \citenamefont{Dever},
  \citenamefont{Viti}, \citenamefont{Williams},\ and\
  \citenamefont{McCoustra}}]{collings2004}%
  \BibitemOpen
  \bibfield{author}{%
  \bibinfo {author} {\bibnamefont{Collings}, \bibfnamefont{M.~P.}}, \bibinfo
  {author} {\bibfnamefont{M.~A.}\ \bibnamefont{Anderson}}, \bibinfo {author}
  {\bibfnamefont{R.}~\bibnamefont{Chen}}, \bibinfo {author}
  {\bibfnamefont{J.~W.}\ \bibnamefont{Dever}}, \bibinfo {author}
  {\bibfnamefont{S.}~\bibnamefont{Viti}}, \bibinfo {author}
  {\bibfnamefont{D.~A.}\ \bibnamefont{Williams}},\ and\ \bibinfo {author}
  {\bibfnamefont{M.~R.~S.}\ \bibnamefont{McCoustra}}}%
  , \bibinfo {year} {2004},\ \bibfield{title}{%
  \enquote{\bibinfo {title} {A laboratory survey of the thermal desorption of
  astrophysically relevant molecules},}\ }%
  \bibfield{journal}{%
  \bibinfo {journal} {Mon. Not. Roy. Astron. Soc.}\ }%
  \textbf{\bibinfo {volume} {354}},\ \bibinfo {pages} {1133--1140}%
  \bibAnnoteFile{NoStop}{collings2004}%
\bibitem[{\citenamefont{Collings}\ \emph{et~al.}(2006)\citenamefont{Collings},
  \citenamefont{Chen},\ and\ \citenamefont{McCoustra}}]{collings2006}%
  \BibitemOpen
  \bibfield{author}{%
  \bibinfo {author} {\bibnamefont{Collings}, \bibfnamefont{M.~P.}}, \bibinfo
  {author} {\bibfnamefont{R.}~\bibnamefont{Chen}},\ and\ \bibinfo {author}
  {\bibfnamefont{M.~R.~S.}\ \bibnamefont{McCoustra}}}%
  , \bibinfo {year} {2006},\ in\ \emph{\bibinfo {booktitle} {Astrochemistry:
  From Laboratory Studies to Astronomical Observations}},\ Vol.\ \bibinfo
  {volume} {855},\ \bibinfo {editor} {edited by\ \bibinfo {editor}
  {\bibfnamefont{R.I.}\ \bibnamefont{Kaiser}}, \bibinfo {editor}
  {\bibfnamefont{P.}~\bibnamefont{Bernath}}, \bibinfo {editor}
  {\bibfnamefont{Y.}~\bibnamefont{Osamura}}, \bibinfo {editor}
  {\bibfnamefont{S.}~\bibnamefont{Petrie}},\ and\ \bibinfo {editor}
  {\bibfnamefont{A.~M.}\ \bibnamefont{Mebel}}}\ (\bibinfo {publisher} {AIP
  Conference Proceedings})\ pp.\ \bibinfo {pages} {62--68}%
  \bibAnnoteFile{NoStop}{collings2006}%
\bibitem[{\citenamefont{Collings}\ \emph{et~al.}(2003)\citenamefont{Collings},
  \citenamefont{Dever}, \citenamefont{Fraser},\ and\
  \citenamefont{McCoustra}}]{collings2003}%
  \BibitemOpen
  \bibfield{author}{%
  \bibinfo {author} {\bibnamefont{Collings}, \bibfnamefont{M.~P.}}, \bibinfo
  {author} {\bibfnamefont{J.~W.}\ \bibnamefont{Dever}}, \bibinfo {author}
  {\bibfnamefont{H.~J.}\ \bibnamefont{Fraser}},\ and\ \bibinfo {author}
  {\bibfnamefont{M.~R.~S.}\ \bibnamefont{McCoustra}}}%
  , \bibinfo {year} {2003},\ \bibfield{title}{%
  \enquote{\bibinfo {title} {Laboratory studies of the interaction of carbon
  monoxide with water ice},}\ }%
  \bibfield{journal}{%
  \bibinfo {journal} {Astrophys. Space Sci.}\ }%
  \textbf{\bibinfo {volume} {285}},\ \bibinfo {pages} {633--659}%
  \bibAnnoteFile{NoStop}{collings2003}%
\bibitem[{\citenamefont{Comiso}\ \emph{et~al.}(2008)\citenamefont{Comiso},
  \citenamefont{Parkinson}, \citenamefont{Gersten},\ and\
  \citenamefont{Stock}}]{Comiso:2008}%
  \BibitemOpen
  \bibfield{author}{%
  \bibinfo {author} {\bibnamefont{Comiso}, \bibfnamefont{J.~C.}}, \bibinfo
  {author} {\bibfnamefont{C.~L.}\ \bibnamefont{Parkinson}}, \bibinfo {author}
  {\bibfnamefont{R.}~\bibnamefont{Gersten}},\ and\ \bibinfo {author}
  {\bibfnamefont{L.}~\bibnamefont{Stock}}}%
  , \bibinfo {year} {2008},\ \bibfield{title}{%
  \enquote{\bibinfo {title} {Accelerated decline in the {Arctic} sea ice
  cover},}\ }%
  \bibfield{journal}{%
  \bibinfo {journal} {Geophys.~Res.~Lett.}\ }%
  \textbf{\bibinfo {volume} {35}},\ \bibinfo {pages} {L01703}%
  \bibAnnoteFile{NoStop}{Comiso:2008}%
\bibitem[{\citenamefont{Cooper}\ \emph{et~al.}(2001)\citenamefont{Cooper},
  \citenamefont{Johnson}, \citenamefont{Mauk}, \citenamefont{Garret},\ and\
  \citenamefont{Gehrels}}]{cooper2001}%
  \BibitemOpen
  \bibfield{author}{%
  \bibinfo {author} {\bibnamefont{Cooper}, \bibfnamefont{J.~F.}}, \bibinfo
  {author} {\bibfnamefont{R.~E.}\ \bibnamefont{Johnson}}, \bibinfo {author}
  {\bibfnamefont{B.~H.}\ \bibnamefont{Mauk}}, \bibinfo {author}
  {\bibfnamefont{H.~B.}\ \bibnamefont{Garret}},\ and\ \bibinfo {author}
  {\bibfnamefont{N.}~\bibnamefont{Gehrels}}}%
  , \bibinfo {year} {2001},\ \bibfield{title}{%
  \enquote{\bibinfo {title} {Energetic ion and electron irradiation of the icy
  galilean satellites},}\ }%
  \bibfield{journal}{%
  \bibinfo {journal} {Icarus}\ }%
  \textbf{\bibinfo {volume} {149}},\ \bibinfo {pages} {133--159}%
  \bibAnnoteFile{NoStop}{cooper2001}%
\bibitem[{\citenamefont{Corripio}(2003)}]{corripio2003}%
  \BibitemOpen
  \bibfield{author}{%
  \bibinfo {author} {\bibnamefont{Corripio}, \bibfnamefont{J.~G.}}}%
  , \bibinfo {year} {2003},\ \emph{\bibinfo {title} {Modelling the energy
  balance of high altitude glacierised basins in the Central {Andes}}},\ Ph.D.
  thesis\ (\bibinfo {school} {University of Edinburgh})%
  \bibAnnoteFile{NoStop}{corripio2003}%
\bibitem[{\citenamefont{Corripio}\ and\
  \citenamefont{Purves}(2005)}]{corripio2005}%
  \BibitemOpen
  \bibfield{author}{%
  \bibinfo {author} {\bibnamefont{Corripio}, \bibfnamefont{J.~G.}},\ and\
  \bibinfo {author} {\bibfnamefont{R.~S.}\ \bibnamefont{Purves}}}%
  , \bibinfo {year} {2005},\ \enquote{\bibinfo {title} {Surface energy balance
  of high altitude glaciers in the central {Andes}: the effect of snow
  penitentes},}\ in\ \emph{\bibinfo {booktitle} {Climate and Hydrology in
  Mountain Areas}},\ \bibinfo {editor} {edited by\ \bibinfo {editor}
  {\bibfnamefont{C.}~\bibnamefont{de~Jong}}, \bibinfo {editor}
  {\bibfnamefont{D.}~\bibnamefont{Collins}},\ and\ \bibinfo {editor}
  {\bibfnamefont{R.}~\bibnamefont{Ranzi}}}\ (\bibinfo {publisher} {Wiley})%
  \bibAnnoteFile{NoStop}{corripio2005}%
\bibitem[{\citenamefont{{Coustenis}}\
  \emph{et~al.}(2008)\citenamefont{{Coustenis}}, \citenamefont{{Atreya}},
  \citenamefont{{Casavecchia}}, \citenamefont{{Castillo}},
  \citenamefont{{Dutuit}}, \citenamefont{{Hussmann}},\ and\
  \citenamefont{{Sohl}}}]{coustenis2008}%
  \BibitemOpen
  \bibfield{author}{%
  \bibinfo {author} {\bibnamefont{{Coustenis}}, \bibfnamefont{A.}}, \bibinfo
  {author} {\bibfnamefont{S.}~\bibnamefont{{Atreya}}}, \bibinfo {author}
  {\bibfnamefont{P.}~\bibnamefont{{Casavecchia}}}, \bibinfo {author}
  {\bibfnamefont{J.}~\bibnamefont{{Castillo}}}, \bibinfo {author}
  {\bibfnamefont{O.}~\bibnamefont{{Dutuit}}}, \bibinfo {author}
  {\bibfnamefont{H.}~\bibnamefont{{Hussmann}}},\ and\ \bibinfo {author}
  {\bibfnamefont{F.}~\bibnamefont{{Sohl}}}}%
  , \bibinfo {year} {2008},\ \bibfield{title}{%
  \enquote{\bibinfo {title} {Surfaces and atmospheres of the outer planets,
  their satellites and ring systems: Part {IV}},}\ }%
  \bibfield{journal}{%
  \bibinfo {journal} {Planet. Space Sci.}\ }%
  \textbf{\bibinfo {volume} {56}},\ \bibinfo {pages} {1571--1572}%
  \bibAnnoteFile{NoStop}{coustenis2008}%
\bibitem[{\citenamefont{{Crawford}}\
  \emph{et~al.}(2001)\citenamefont{{Crawford}}, \citenamefont{{Davis}},
  \citenamefont{{Chen}}, \citenamefont{{Buhr}}, \citenamefont{{Oltmans}},
  \citenamefont{{Weller}}, \citenamefont{{Mauldin}}, \citenamefont{{Eisele}},
  \citenamefont{{Shetter}}, \citenamefont{{Lefer}}, \citenamefont{{Arimoto}},\
  and\ \citenamefont{{Hogan}}}]{Crawford:2001p}%
  \BibitemOpen
  \bibfield{author}{%
  \bibinfo {author} {\bibnamefont{{Crawford}}, \bibfnamefont{J.~H.}}, \bibinfo
  {author} {\bibfnamefont{D.~D.}\ \bibnamefont{{Davis}}}, \bibinfo {author}
  {\bibfnamefont{G.}~\bibnamefont{{Chen}}}, \bibinfo {author}
  {\bibfnamefont{M.}~\bibnamefont{{Buhr}}}, \bibinfo {author}
  {\bibfnamefont{S.}~\bibnamefont{{Oltmans}}}, \bibinfo {author}
  {\bibfnamefont{R.}~\bibnamefont{{Weller}}}, \bibinfo {author}
  {\bibfnamefont{L.}~\bibnamefont{{Mauldin}}}, \bibinfo {author}
  {\bibfnamefont{F.}~\bibnamefont{{Eisele}}}, \bibinfo {author}
  {\bibfnamefont{R.}~\bibnamefont{{Shetter}}}, \bibinfo {author}
  {\bibfnamefont{B.}~\bibnamefont{{Lefer}}}, \bibinfo {author}
  {\bibfnamefont{R.}~\bibnamefont{{Arimoto}}},\ and\ \bibinfo {author}
  {\bibfnamefont{A.}~\bibnamefont{{Hogan}}}}%
  , \bibinfo {year} {2001},\ \bibfield{title}{%
  \enquote{\bibinfo {title} {{Evidence for photochemical production of ozone at
  the South Pole surface}},}\ }%
  \bibfield{journal}{%
  \bibinfo {journal} {Geophys. Res. Lett.}\ }%
  \textbf{\bibinfo {volume} {28}},\ \bibinfo {pages} {3641--3644}%
  \bibAnnoteFile{NoStop}{Crawford:2001p}%
\bibitem[{\citenamefont{Creighan}\ \emph{et~al.}(2006)\citenamefont{Creighan},
  \citenamefont{Perry},\ and\ \citenamefont{Price}}]{creighan2006}%
  \BibitemOpen
  \bibfield{author}{%
  \bibinfo {author} {\bibnamefont{Creighan}, \bibfnamefont{S.~C.}}, \bibinfo
  {author} {\bibfnamefont{J.~S.~A.}\ \bibnamefont{Perry}},\ and\ \bibinfo
  {author} {\bibfnamefont{S.~D.}\ \bibnamefont{Price}}}%
  , \bibinfo {year} {2006},\ \bibfield{title}{%
  \enquote{\bibinfo {title} {The rovibrational distribution of {H}$_2$ and {HD}
  formed on a graphite surface at 15-50~{K}},}\ }%
  \bibfield{journal}{%
  \bibinfo {journal} {J. Chem. Phys.}\ }%
  \textbf{\bibinfo {volume} {124}},\ \bibinfo {pages} {114701}%
  \bibAnnoteFile{NoStop}{creighan2006}%
\bibitem[{\citenamefont{Crovisier}(2005)}]{crovisier2005}%
  \BibitemOpen
  \bibfield{author}{%
  \bibinfo {author} {\bibnamefont{Crovisier}, \bibfnamefont{J.}}}%
  , \bibinfo {year} {2005},\ \enquote{\bibinfo {title} {Cometary diversity and
  cometary families},}\ in\ \emph{\bibinfo {booktitle} {Proc. of the
  {XVIII}emes Rencontres de Blois}},\ \bibinfo {note}
  {arXiv:astro-ph/0703785v1}%
  \bibAnnoteFile{NoStop}{crovisier2005}%
\bibitem[{\citenamefont{Crowley}\ \emph{et~al.}(2010)\citenamefont{Crowley},
  \citenamefont{Ammann}, \citenamefont{Cox}, \citenamefont{Hynes},
  \citenamefont{Jenkin}, \citenamefont{Mellouki}, \citenamefont{Rossi},
  \citenamefont{Troe},\ and\ \citenamefont{Wallington}}]{Ammann:2008p14103}%
  \BibitemOpen
  \bibfield{author}{%
  \bibinfo {author} {\bibnamefont{Crowley}, \bibfnamefont{J.~N.}}, \bibinfo
  {author} {\bibfnamefont{M.}~\bibnamefont{Ammann}}, \bibinfo {author}
  {\bibfnamefont{R.~A.}\ \bibnamefont{Cox}}, \bibinfo {author}
  {\bibfnamefont{R.~G.}\ \bibnamefont{Hynes}}, \bibinfo {author}
  {\bibfnamefont{M.~E.}\ \bibnamefont{Jenkin}}, \bibinfo {author}
  {\bibfnamefont{A.}~\bibnamefont{Mellouki}}, \bibinfo {author}
  {\bibfnamefont{M.~J.}\ \bibnamefont{Rossi}}, \bibinfo {author}
  {\bibfnamefont{J.}~\bibnamefont{Troe}},\ and\ \bibinfo {author}
  {\bibfnamefont{T.~J.}\ \bibnamefont{Wallington}}}%
  , \bibinfo {year} {2010},\ \bibfield{title}{%
  \enquote{\bibinfo {title} {{Evaluated kinetic and photochemical data for
  atmospheric chemistry: Volume {V} --- heterogeneous reactions on solid
  substrates}},}\ }%
  \bibfield{journal}{%
  \bibinfo {journal} {Atmos. Chem. Phys.}\ }%
  \textbf{\bibinfo {volume} {10}},\ \bibinfo {pages} {9059--9223}%
  \bibAnnoteFile{NoStop}{Ammann:2008p14103}%
\bibitem[{\citenamefont{{Cruikshank}}(2010)}]{cruikshank2010}%
  \BibitemOpen
  \bibfield{author}{%
  \bibinfo {author} {\bibnamefont{{Cruikshank}}, \bibfnamefont{D.~P.}}}%
  , \bibinfo {year} {2010},\ \bibfield{title}{%
  \enquote{\bibinfo {title} {{Generating an Atmosphere}},}\ }%
  \bibfield{journal}{%
  \bibinfo {journal} {Science}\ }%
  \textbf{\bibinfo {volume} {330}},\ \bibinfo {pages} {1755}%
  \bibAnnoteFile{NoStop}{cruikshank2010}%
\bibitem[{\citenamefont{Cruikshank}\
  \emph{et~al.}(1995)\citenamefont{Cruikshank}, \citenamefont{Brown},
  \citenamefont{Calvin}, \citenamefont{Roush},\ and\
  \citenamefont{Bartholomew}}]{cruikshank1995}%
  \BibitemOpen
  \bibfield{author}{%
  \bibinfo {author} {\bibnamefont{Cruikshank}, \bibfnamefont{D.~P.}}, \bibinfo
  {author} {\bibfnamefont{R.~H.}\ \bibnamefont{Brown}}, \bibinfo {author}
  {\bibfnamefont{W.~M.}\ \bibnamefont{Calvin}}, \bibinfo {author}
  {\bibfnamefont{T.~L.}\ \bibnamefont{Roush}},\ and\ \bibinfo {author}
  {\bibfnamefont{M.~J.}\ \bibnamefont{Bartholomew}}}%
  , \bibinfo {year} {1995},\ \emph{\bibinfo {title} {Solar System Ices}},\
  Vol.\ \bibinfo {volume} {227}\ (\bibinfo {publisher} {Dordrecht: Kluwer
  Academic Publishers ASSL Series})%
  \bibAnnoteFile{NoStop}{cruikshank1995}%
\bibitem[{\citenamefont{Cruikshank}\
  \emph{et~al.}(1984)\citenamefont{Cruikshank}, \citenamefont{Veverka},\ and\
  \citenamefont{Lebofsky}}]{cruikshank1984}%
  \BibitemOpen
  \bibfield{author}{%
  \bibinfo {author} {\bibnamefont{Cruikshank}, \bibfnamefont{D.~P.}}, \bibinfo
  {author} {\bibfnamefont{J.}~\bibnamefont{Veverka}},\ and\ \bibinfo {author}
  {\bibfnamefont{L.~A.}\ \bibnamefont{Lebofsky}}}%
  , \bibinfo {year} {1984},\ \emph{\bibinfo {title} {Saturn}}\ (\bibinfo
  {publisher} {University of Arizona Press})%
  \bibAnnoteFile{NoStop}{cruikshank1984}%
\bibitem[{\citenamefont{{Cull}}\ \emph{et~al.}(2010)\citenamefont{{Cull}},
  \citenamefont{{Arvidson}}, \citenamefont{{Mellon}}, \citenamefont{{Skemer}},
  \citenamefont{{Shaw}},\ and\ \citenamefont{{Morris}}}]{cull2010}%
  \BibitemOpen
  \bibfield{author}{%
  \bibinfo {author} {\bibnamefont{{Cull}}, \bibfnamefont{S.}}, \bibinfo
  {author} {\bibfnamefont{R.~E.}\ \bibnamefont{{Arvidson}}}, \bibinfo {author}
  {\bibfnamefont{M.~T.}\ \bibnamefont{{Mellon}}}, \bibinfo {author}
  {\bibfnamefont{P.}~\bibnamefont{{Skemer}}}, \bibinfo {author}
  {\bibfnamefont{A.}~\bibnamefont{{Shaw}}},\ and\ \bibinfo {author}
  {\bibfnamefont{R.~V.}\ \bibnamefont{{Morris}}}}%
  , \bibinfo {year} {2010},\ \bibfield{title}{%
  \enquote{\bibinfo {title} {{Compositions of subsurface ices at the {Mars}
  {Phoenix} landing site}},}\ }%
  \bibfield{journal}{%
  \bibinfo {journal} {Geophys. Res. Lett.}\ }%
  \textbf{\bibinfo {volume} {37}},\ \bibinfo {pages} {24203}%
  \bibAnnoteFile{NoStop}{cull2010}%
\bibitem[{\citenamefont{Cuppen}\ and\
  \citenamefont{Herbst}(2007)}]{cuppen2007}%
  \BibitemOpen
  \bibfield{author}{%
  \bibinfo {author} {\bibnamefont{Cuppen}, \bibfnamefont{H.~M.}},\ and\
  \bibinfo {author} {\bibfnamefont{E.}~\bibnamefont{Herbst}}}%
  , \bibinfo {year} {2007},\ \bibfield{title}{%
  \enquote{\bibinfo {title} {Simulation of the formation and morphology of ice
  mantles on interstellar grains},}\ }%
  \bibfield{journal}{%
  \bibinfo {journal} {Astrophys. J.}\ }%
  \textbf{\bibinfo {volume} {668}},\ \bibinfo {pages} {294--309}%
  \bibAnnoteFile{NoStop}{cuppen2007}%
\bibitem[{\citenamefont{Cuppen}\ \emph{et~al.}(2010)\citenamefont{Cuppen},
  \citenamefont{Ioppolo}, \citenamefont{Romanzin},\ and\
  \citenamefont{Linnartz}}]{cuppen2010}%
  \BibitemOpen
  \bibfield{author}{%
  \bibinfo {author} {\bibnamefont{Cuppen}, \bibfnamefont{H.~M.}}, \bibinfo
  {author} {\bibfnamefont{S.}~\bibnamefont{Ioppolo}}, \bibinfo {author}
  {\bibfnamefont{C.}~\bibnamefont{Romanzin}},\ and\ \bibinfo {author}
  {\bibfnamefont{H.}~\bibnamefont{Linnartz}}}%
  , \bibinfo {year} {2010},\ \bibfield{title}{%
  \enquote{\bibinfo {title} {Water formation at low temperatures by surface
  {O}$_2$ hydrogenation {II}: the reaction network},}\ }%
  \bibfield{journal}{%
  \bibinfo {journal} {Phys. Chem. Chem. Phys.}\ }%
  \textbf{\bibinfo {volume} {12}},\ \bibinfo {pages} {12077--12088}%
  \bibAnnoteFile{NoStop}{cuppen2010}%
\bibitem[{\citenamefont{Curry}\ \emph{et~al.}(1995)\citenamefont{Curry},
  \citenamefont{Schramm},\ and\ \citenamefont{Ebert}}]{Curry:1995}%
  \BibitemOpen
  \bibfield{author}{%
  \bibinfo {author} {\bibnamefont{Curry}, \bibfnamefont{J.~A.}}, \bibinfo
  {author} {\bibfnamefont{J.~L.}\ \bibnamefont{Schramm}},\ and\ \bibinfo
  {author} {\bibfnamefont{E.~E.}\ \bibnamefont{Ebert}}}%
  , \bibinfo {year} {1995},\ \bibfield{title}{%
  \enquote{\bibinfo {title} {Sea ice-albedo climate feedback mechanism},}\ }%
  \bibfield{journal}{%
  \bibinfo {journal} {J.~Clim.}\ }%
  \textbf{\bibinfo {volume} {8}},\ \bibinfo {pages} {240--247}%
  \bibAnnoteFile{NoStop}{Curry:1995}%
\bibitem[{\citenamefont{Curtius}\ \emph{et~al.}(2005)\citenamefont{Curtius},
  \citenamefont{Weigel}, \citenamefont{Všssing}, \citenamefont{Wernli},
  \citenamefont{Werner}, \citenamefont{Volk}, \citenamefont{Konopka},
  \citenamefont{Krebsbach}, \citenamefont{Schiller}, \citenamefont{Roiger},
  \citenamefont{Schlager}, \citenamefont{Dreiling},\ and\
  \citenamefont{Borrmann}}]{curtius2005}%
  \BibitemOpen
  \bibfield{author}{%
  \bibinfo {author} {\bibnamefont{Curtius}, \bibfnamefont{J.}}, \bibinfo
  {author} {\bibfnamefont{R.}~\bibnamefont{Weigel}}, \bibinfo {author}
  {\bibfnamefont{H.-J.}\ \bibnamefont{Všssing}}, \bibinfo {author}
  {\bibfnamefont{H.}~\bibnamefont{Wernli}}, \bibinfo {author}
  {\bibfnamefont{A.}~\bibnamefont{Werner}}, \bibinfo {author}
  {\bibfnamefont{C.-M.}\ \bibnamefont{Volk}}, \bibinfo {author}
  {\bibfnamefont{P.}~\bibnamefont{Konopka}}, \bibinfo {author}
  {\bibfnamefont{M.}~\bibnamefont{Krebsbach}}, \bibinfo {author}
  {\bibfnamefont{C.}~\bibnamefont{Schiller}}, \bibinfo {author}
  {\bibfnamefont{A.}~\bibnamefont{Roiger}}, \bibinfo {author}
  {\bibfnamefont{H.}~\bibnamefont{Schlager}}, \bibinfo {author}
  {\bibfnamefont{V.}~\bibnamefont{Dreiling}},\ and\ \bibinfo {author}
  {\bibfnamefont{S.}~\bibnamefont{Borrmann}}}%
  , \bibinfo {year} {2005},\ \bibfield{title}{%
  \enquote{\bibinfo {title} {Observations of meteoric material and implications
  for aerosol nucleation in the winter {Arctic} lower stratosphere derived from
  in situ particle measurements},}\ }%
  \bibfield{journal}{%
  \bibinfo {journal} {Atmos. Chem. Phys.}\ }%
  \textbf{\bibinfo {volume} {5}},\ \bibinfo {pages} {3053--3069}%
  \bibAnnoteFile{NoStop}{curtius2005}%
\bibitem[{\citenamefont{Cuzzi}\ \emph{et~al.}(2009)\citenamefont{Cuzzi},
  \citenamefont{Clark}, \citenamefont{Filacchione}, \citenamefont{French},
  \citenamefont{Johnson}, \citenamefont{Marouf},\ and\
  \citenamefont{Spilker}}]{cuzzi2009}%
  \BibitemOpen
  \bibfield{author}{%
  \bibinfo {author} {\bibnamefont{Cuzzi}, \bibfnamefont{J.}}, \bibinfo {author}
  {\bibfnamefont{R.}~\bibnamefont{Clark}}, \bibinfo {author}
  {\bibfnamefont{G.}~\bibnamefont{Filacchione}}, \bibinfo {author}
  {\bibfnamefont{R.}~\bibnamefont{French}}, \bibinfo {author}
  {\bibfnamefont{R.}~\bibnamefont{Johnson}}, \bibinfo {author}
  {\bibfnamefont{E.}~\bibnamefont{Marouf}},\ and\ \bibinfo {author}
  {\bibfnamefont{L.}~\bibnamefont{Spilker}}}%
  , \bibinfo {year} {2009},\ \enquote{\bibinfo {title} {Ring particle
  composition and size distribution},}\ in\ \emph{\bibinfo {booktitle} {Saturn
  from {Cassini--Huygens}}},\ \bibinfo {editor} {edited by\ \bibinfo {editor}
  {\bibfnamefont{M.~K.}\ \bibnamefont{Dougherty}}, \bibinfo {editor}
  {\bibfnamefont{L.~W.}\ \bibnamefont{Esposito}},\ and\ \bibinfo {editor}
  {\bibfnamefont{S.~M.}\ \bibnamefont{Krimigis}}},\ Chap.~\bibinfo {chapter}
  {15}\ (\bibinfo {publisher} {Springer})%
  \bibAnnoteFile{NoStop}{cuzzi2009}%
\bibitem[{\citenamefont{{Dartois}}\ and\
  \citenamefont{{Deboffle}}(2008)}]{dartois2008}%
  \BibitemOpen
  \bibfield{author}{%
  \bibinfo {author} {\bibnamefont{{Dartois}}, \bibfnamefont{E.}},\ and\
  \bibinfo {author} {\bibfnamefont{D.}~\bibnamefont{{Deboffle}}}}%
  , \bibinfo {year} {2008},\ \bibfield{title}{%
  \enquote{\bibinfo {title} {{Methane clathrate hydrate {FTIR} spectrum.
  Implications for its cometary and planetary detection}},}\ }%
  \bibfield{journal}{%
  \bibinfo {journal} {Astron. Astrophys.}\ }%
  \textbf{\bibinfo {volume} {490}},\ \bibinfo {pages} {L19--L22}%
  \bibAnnoteFile{NoStop}{dartois2008}%
\bibitem[{\citenamefont{Dartois}\
  \emph{et~al.}(1999{\natexlab{a}})\citenamefont{Dartois},
  \citenamefont{Demyk}, \citenamefont{dÕHendecourt},\ and\
  \citenamefont{Ehrenfreund}}]{dartois1999}%
  \BibitemOpen
  \bibfield{author}{%
  \bibinfo {author} {\bibnamefont{Dartois}, \bibfnamefont{E.}}, \bibinfo
  {author} {\bibfnamefont{K.}~\bibnamefont{Demyk}}, \bibinfo {author}
  {\bibfnamefont{L.}~\bibnamefont{dÕHendecourt}},\ and\ \bibinfo {author}
  {\bibfnamefont{P.}~\bibnamefont{Ehrenfreund}}}%
  , \bibinfo {year} {1999}{\natexlab{a}},\ \bibfield{title}{%
  \enquote{\bibinfo {title} {Carbon dioxide--methanol intermolecular complexes
  in interstellar grain mantles},}\ }%
  \bibfield{journal}{%
  \bibinfo {journal} {Astron. Astrophys.}\ }%
  \textbf{\bibinfo {volume} {351}},\ \bibinfo {pages} {1066--1074}%
  \bibAnnoteFile{NoStop}{dartois1999}%
\bibitem[{\citenamefont{Dartois}\
  \emph{et~al.}(1999{\natexlab{b}})\citenamefont{Dartois},
  \citenamefont{Schutte}, \citenamefont{Geballe}, \citenamefont{Demyk},
  \citenamefont{Ehrenfreund},\ and\
  \citenamefont{D'Hendecourt}}]{dartois1999_2}%
  \BibitemOpen
  \bibfield{author}{%
  \bibinfo {author} {\bibnamefont{Dartois}, \bibfnamefont{E.}}, \bibinfo
  {author} {\bibfnamefont{W.}~\bibnamefont{Schutte}}, \bibinfo {author}
  {\bibfnamefont{T.~R.}\ \bibnamefont{Geballe}}, \bibinfo {author}
  {\bibfnamefont{K.}~\bibnamefont{Demyk}}, \bibinfo {author}
  {\bibfnamefont{P.}~\bibnamefont{Ehrenfreund}},\ and\ \bibinfo {author}
  {\bibfnamefont{L.}~\bibnamefont{D'Hendecourt}}}%
  , \bibinfo {year} {1999}{\natexlab{b}},\ \bibfield{title}{%
  \enquote{\bibinfo {title} {Methanol: The second most abundant ice species
  towards the high-mass protostars {RAFGL7009S} and {W 33A}},}\ }%
  \bibfield{journal}{%
  \bibinfo {journal} {Astron. Astrophys.}\ }%
  \textbf{\bibinfo {volume} {342}},\ \bibinfo {pages} {L32--L35}%
  \bibAnnoteFile{NoStop}{dartois1999_2}%
\bibitem[{\citenamefont{Dasgupta}\ and\
  \citenamefont{Mo}(1997)}]{dasgupta1997}%
  \BibitemOpen
  \bibfield{author}{%
  \bibinfo {author} {\bibnamefont{Dasgupta}, \bibfnamefont{P.~K.}},\ and\
  \bibinfo {author} {\bibfnamefont{Y.}~\bibnamefont{Mo}}}%
  , \bibinfo {year} {1997},\ \bibfield{title}{%
  \enquote{\bibinfo {title} {Chromatography on water-ice},}\ }%
  \bibfield{journal}{%
  \bibinfo {journal} {Anal. Chem.}\ }%
  \textbf{\bibinfo {volume} {69}},\ \bibinfo {pages} {4079--4081}%
  \bibAnnoteFile{NoStop}{dasgupta1997}%
\bibitem[{\citenamefont{Dash}\ \emph{et~al.}(2006)\citenamefont{Dash},
  \citenamefont{Rempel},\ and\ \citenamefont{Wettlaufer}}]{Dash2006}%
  \BibitemOpen
  \bibfield{author}{%
  \bibinfo {author} {\bibnamefont{Dash}, \bibfnamefont{J.~G.}}, \bibinfo
  {author} {\bibfnamefont{A.~W.}\ \bibnamefont{Rempel}},\ and\ \bibinfo
  {author} {\bibfnamefont{J.~S.}\ \bibnamefont{Wettlaufer}}}%
  , \bibinfo {year} {2006},\ \bibfield{title}{%
  \enquote{\bibinfo {title} {The physics of premelted ice and its geophysical
  consequences},}\ }%
  \bibfield{journal}{%
  \bibinfo {journal} {Rev. Mod. Phys.}\ }%
  \textbf{\bibinfo {volume} {78}},\ \bibinfo {pages} {695--741}%
  \bibAnnoteFile{NoStop}{Dash2006}%
\bibitem[{\citenamefont{Dash}\ and\
  \citenamefont{Wettlaufer}(2003)}]{dash2003}%
  \BibitemOpen
  \bibfield{author}{%
  \bibinfo {author} {\bibnamefont{Dash}, \bibfnamefont{J.~G.}},\ and\ \bibinfo
  {author} {\bibfnamefont{J.~S.}\ \bibnamefont{Wettlaufer}}}%
  , \bibinfo {year} {2003},\ \bibfield{title}{%
  \enquote{\bibinfo {title} {The surface physics of ice in thunderstorms},}\ }%
  \bibfield{journal}{%
  \bibinfo {journal} {Can. J. Phys.}\ }%
  \textbf{\bibinfo {volume} {81}},\ \bibinfo {pages} {201--207}%
  \bibAnnoteFile{NoStop}{dash2003}%
\bibitem[{\citenamefont{{Davies}}\ \emph{et~al.}(1997)\citenamefont{{Davies}},
  \citenamefont{{Roush}}, \citenamefont{{Cruikshank}},
  \citenamefont{{Bartholomew}}, \citenamefont{{Geballe}},
  \citenamefont{{Owen}},\ and\ \citenamefont{{de Bergh}}}]{davies1997}%
  \BibitemOpen
  \bibfield{author}{%
  \bibinfo {author} {\bibnamefont{{Davies}}, \bibfnamefont{J.~K.}}, \bibinfo
  {author} {\bibfnamefont{T.~L.}\ \bibnamefont{{Roush}}}, \bibinfo {author}
  {\bibfnamefont{D.~P.}\ \bibnamefont{{Cruikshank}}}, \bibinfo {author}
  {\bibfnamefont{M.~J.}\ \bibnamefont{{Bartholomew}}}, \bibinfo {author}
  {\bibfnamefont{T.~R.}\ \bibnamefont{{Geballe}}}, \bibinfo {author}
  {\bibfnamefont{T.}~\bibnamefont{{Owen}}},\ and\ \bibinfo {author}
  {\bibfnamefont{C.}~\bibnamefont{{de Bergh}}}}%
  , \bibinfo {year} {1997},\ \bibfield{title}{%
  \enquote{\bibinfo {title} {The detection of water ice in comet
  {Hale}--{Bopp}},}\ }%
  \bibfield{journal}{%
  \bibinfo {journal} {Icarus}\ }%
  \textbf{\bibinfo {volume} {127}},\ \bibinfo {pages} {238--245}%
  \bibAnnoteFile{NoStop}{davies1997}%
\bibitem[{\citenamefont{Davies}(2003)}]{davies2003}%
  \BibitemOpen
  \bibfield{author}{%
  \bibinfo {author} {\bibnamefont{Davies}, \bibfnamefont{P.}}}%
  , \bibinfo {year} {2003},\ \emph{\bibinfo {title} {The Origin of Life}}\
  (\bibinfo {publisher} {Penguin})%
  \bibAnnoteFile{NoStop}{davies2003}%
\bibitem[{\citenamefont{Debenedetti}(1996)}]{debenedetti1996}%
  \BibitemOpen
  \bibfield{author}{%
  \bibinfo {author} {\bibnamefont{Debenedetti}, \bibfnamefont{P.~G.}}}%
  , \bibinfo {year} {1996},\ \emph{\bibinfo {title} {Metastable Liquids}}\
  (\bibinfo {publisher} {Princeton University Press})%
  \bibAnnoteFile{NoStop}{debenedetti1996}%
\bibitem[{\citenamefont{Debenedetti}(2003)}]{debenedetti2003}%
  \BibitemOpen
  \bibfield{author}{%
  \bibinfo {author} {\bibnamefont{Debenedetti}, \bibfnamefont{P.~G.}}}%
  , \bibinfo {year} {2003},\ \bibfield{title}{%
  \enquote{\bibinfo {title} {Supercooled and glassy water},}\ }%
  \bibfield{journal}{%
  \bibinfo {journal} {J. Phys.: Condens. Matter}\ }%
  \textbf{\bibinfo {volume} {15}},\ \bibinfo {pages} {R1669--R1726}%
  \bibAnnoteFile{NoStop}{debenedetti2003}%
\bibitem[{\citenamefont{Delsemme}\ and\
  \citenamefont{Swings}(1952)}]{delsemme1952}%
  \BibitemOpen
  \bibfield{author}{%
  \bibinfo {author} {\bibnamefont{Delsemme}, \bibfnamefont{A.~H.}},\ and\
  \bibinfo {author} {\bibfnamefont{P.}~\bibnamefont{Swings}}}%
  , \bibinfo {year} {1952},\ \bibfield{title}{%
  \enquote{\bibinfo {title} {Hydrates de gaz dans les noyaux cometaires et les
  grains interstellairs},}\ }%
  \bibfield{journal}{%
  \bibinfo {journal} {Annales d'Astrophysique}\ }%
  \textbf{\bibinfo {volume} {15}},\ \bibinfo {pages} {1--6}%
  \bibAnnoteFile{NoStop}{delsemme1952}%
\bibitem[{\citenamefont{Demirdjian}\
  \emph{et~al.}(2002)\citenamefont{Demirdjian}, \citenamefont{Ferry},
  \citenamefont{Suzanne}, \citenamefont{Toubin}, \citenamefont{Picaud},
  \citenamefont{Hoang},\ and\ \citenamefont{Girardet}}]{demirdjian2002}%
  \BibitemOpen
  \bibfield{author}{%
  \bibinfo {author} {\bibnamefont{Demirdjian}, \bibfnamefont{B.}}, \bibinfo
  {author} {\bibfnamefont{D.}~\bibnamefont{Ferry}}, \bibinfo {author}
  {\bibfnamefont{J.}~\bibnamefont{Suzanne}}, \bibinfo {author}
  {\bibfnamefont{C.}~\bibnamefont{Toubin}}, \bibinfo {author}
  {\bibfnamefont{S.}~\bibnamefont{Picaud}}, \bibinfo {author}
  {\bibfnamefont{P.~N.~M.}\ \bibnamefont{Hoang}},\ and\ \bibinfo {author}
  {\bibfnamefont{C.}~\bibnamefont{Girardet}}}%
  , \bibinfo {year} {2002},\ \bibfield{title}{%
  \enquote{\bibinfo {title} {Structure and dynamics of ice {Ih} films upon
  {HCl} adsorption between 190 and 270 {K}. {I}. {N}eutron diffraction and
  quasielastic neutron scattering experiments},}\ }%
  \bibfield{journal}{%
  \bibinfo {journal} {J. Chem. Phys.}\ }%
  \textbf{\bibinfo {volume} {116}},\ \bibinfo {pages} {5143--5149}%
  \bibAnnoteFile{NoStop}{demirdjian2002}%
\bibitem[{\citenamefont{DeMott}\ \emph{et~al.}(2003)\citenamefont{DeMott},
  \citenamefont{Cziczo}, \citenamefont{Prenni}, \citenamefont{Murphy},
  \citenamefont{Kreidenweis}, \citenamefont{Thomson}, \citenamefont{Borys},\
  and\ \citenamefont{Rogers}}]{Demott2003}%
  \BibitemOpen
  \bibfield{author}{%
  \bibinfo {author} {\bibnamefont{DeMott}, \bibfnamefont{P.~J.}}, \bibinfo
  {author} {\bibfnamefont{D.~J.}\ \bibnamefont{Cziczo}}, \bibinfo {author}
  {\bibfnamefont{A.~J.}\ \bibnamefont{Prenni}}, \bibinfo {author}
  {\bibfnamefont{D.~M.}\ \bibnamefont{Murphy}}, \bibinfo {author}
  {\bibfnamefont{S.~M.}\ \bibnamefont{Kreidenweis}}, \bibinfo {author}
  {\bibfnamefont{D.~S.}\ \bibnamefont{Thomson}}, \bibinfo {author}
  {\bibfnamefont{R.}~\bibnamefont{Borys}},\ and\ \bibinfo {author}
  {\bibfnamefont{D.~C.}\ \bibnamefont{Rogers}}}%
  , \bibinfo {year} {2003},\ \bibfield{title}{%
  \enquote{\bibinfo {title} {Measurements of the concentration and composition
  of nuclei for cirrus formation},}\ }%
  \bibfield{journal}{%
  \bibinfo {journal} {Proc. Natl. Acad. Sci. U.S.A.}\ }%
  \textbf{\bibinfo {volume} {100}},\ \bibinfo {pages} {14655--14660}%
  \bibAnnoteFile{NoStop}{Demott2003}%
\bibitem[{\citenamefont{DeMott}\ \emph{et~al.}(2010)\citenamefont{DeMott},
  \citenamefont{Prenni}, \citenamefont{Liu}, \citenamefont{Kreidenweis},
  \citenamefont{Petters}, \citenamefont{Twohy}, \citenamefont{Richardson},
  \citenamefont{Eidhammer},\ and\ \citenamefont{Rogers}}]{DeMott2010}%
  \BibitemOpen
  \bibfield{author}{%
  \bibinfo {author} {\bibnamefont{DeMott}, \bibfnamefont{P.~J.}}, \bibinfo
  {author} {\bibfnamefont{A.~J.}\ \bibnamefont{Prenni}}, \bibinfo {author}
  {\bibfnamefont{X.}~\bibnamefont{Liu}}, \bibinfo {author}
  {\bibfnamefont{S.~M.}\ \bibnamefont{Kreidenweis}}, \bibinfo {author}
  {\bibfnamefont{M.~D.}\ \bibnamefont{Petters}}, \bibinfo {author}
  {\bibfnamefont{C.~H.}\ \bibnamefont{Twohy}}, \bibinfo {author}
  {\bibfnamefont{M.~S.}\ \bibnamefont{Richardson}}, \bibinfo {author}
  {\bibfnamefont{T.}~\bibnamefont{Eidhammer}},\ and\ \bibinfo {author}
  {\bibfnamefont{D.~C.}\ \bibnamefont{Rogers}}}%
  , \bibinfo {year} {2010},\ \bibfield{title}{%
  \enquote{\bibinfo {title} {Predicting global atmospheric ice nuclei
  distributions and their impacts on climate},}\ }%
  \bibfield{journal}{%
  \bibinfo {journal} {Proc. Natl Acad. Sci. USA}\ }%
  \textbf{\bibinfo {volume} {107}},\ \bibinfo {pages} {11217--11222}%
  \bibAnnoteFile{NoStop}{DeMott2010}%
\bibitem[{\citenamefont{Devlin}\ \emph{et~al.}(2002)\citenamefont{Devlin},
  \citenamefont{Uras}, \citenamefont{Sadlej},\ and\
  \citenamefont{Buch}}]{devlin2002}%
  \BibitemOpen
  \bibfield{author}{%
  \bibinfo {author} {\bibnamefont{Devlin}, \bibfnamefont{J.~P.}}, \bibinfo
  {author} {\bibfnamefont{N.}~\bibnamefont{Uras}}, \bibinfo {author}
  {\bibfnamefont{J.}~\bibnamefont{Sadlej}},\ and\ \bibinfo {author}
  {\bibfnamefont{V.}~\bibnamefont{Buch}}}%
  , \bibinfo {year} {2002},\ \bibfield{title}{%
  \enquote{\bibinfo {title} {Discrete stages in the solvation and ionization of
  hydrogen chloride adsorbed on ice particles},}\ }%
  \bibfield{journal}{%
  \bibinfo {journal} {Nature}\ }%
  \textbf{\bibinfo {volume} {417}},\ \bibinfo {pages} {269--271}%
  \bibAnnoteFile{NoStop}{devlin2002}%
\bibitem[{\citenamefont{van Dishoeck}(2004)}]{vandishoeck2004}%
  \BibitemOpen
  \bibfield{author}{%
  \bibinfo {author} {\bibnamefont{van Dishoeck}, \bibfnamefont{E.~F.}}}%
  , \bibinfo {year} {2004},\ \bibfield{title}{%
  \enquote{\bibinfo {title} {Iso spectroscopy of gas and dust: From molecular
  clouds to protoplanetary disks},}\ }%
  \bibfield{journal}{%
  \bibinfo {journal} {Annu. Rev. Astron. Astrophys.}\ }%
  \textbf{\bibinfo {volume} {42}},\ \bibinfo {pages} {119--167}%
  \bibAnnoteFile{NoStop}{vandishoeck2004}%
\bibitem[{\citenamefont{Doble}\ \emph{et~al.}(2003)\citenamefont{Doble},
  \citenamefont{Coon},\ and\ \citenamefont{Wadhams}}]{Doble:2003}%
  \BibitemOpen
  \bibfield{author}{%
  \bibinfo {author} {\bibnamefont{Doble}, \bibfnamefont{M.~J.}}, \bibinfo
  {author} {\bibfnamefont{M.~D.}\ \bibnamefont{Coon}},\ and\ \bibinfo {author}
  {\bibfnamefont{P.}~\bibnamefont{Wadhams}}}%
  , \bibinfo {year} {2003},\ \bibfield{title}{%
  \enquote{\bibinfo {title} {Pancake ice formation in the {Weddell Sea}},}\ }%
  \bibfield{journal}{%
  \bibinfo {journal} {J.~Geophys.~Res.}\ }%
  \textbf{\bibinfo {volume} {108}},\ \bibinfo {pages} {3209}%
  \bibAnnoteFile{NoStop}{Doble:2003}%
\bibitem[{\citenamefont{Dohn{\'a}lek}\
  \emph{et~al.}(1999)\citenamefont{Dohn{\'a}lek}, \citenamefont{Ciolli},
  \citenamefont{Kimmel}, \citenamefont{Stevenson}, \citenamefont{Smith},\ and\
  \citenamefont{Kay}}]{dohnalek1999}%
  \BibitemOpen
  \bibfield{author}{%
  \bibinfo {author} {\bibnamefont{Dohn{\'a}lek}, \bibfnamefont{Z.}}, \bibinfo
  {author} {\bibfnamefont{R.~L.}\ \bibnamefont{Ciolli}}, \bibinfo {author}
  {\bibfnamefont{G.~A.}\ \bibnamefont{Kimmel}}, \bibinfo {author}
  {\bibfnamefont{K.~P.}\ \bibnamefont{Stevenson}}, \bibinfo {author}
  {\bibfnamefont{R.~S.}\ \bibnamefont{Smith}},\ and\ \bibinfo {author}
  {\bibfnamefont{B.~D.}\ \bibnamefont{Kay}}}%
  , \bibinfo {year} {1999},\ \bibfield{title}{%
  \enquote{\bibinfo {title} {Substrate induced crystallization of amorphous
  solid water at low temperatures},}\ }%
  \bibfield{journal}{%
  \bibinfo {journal} {J. Chem. Phys.}\ }%
  \textbf{\bibinfo {volume} {110}},\ \bibinfo {pages} {5489--5492}%
  \bibAnnoteFile{NoStop}{dohnalek1999}%
\bibitem[{\citenamefont{Dohn{\'a}lek}\
  \emph{et~al.}(2000)\citenamefont{Dohn{\'a}lek}, \citenamefont{Kimmel},
  \citenamefont{Ciolli}, \citenamefont{Stevenson}, \citenamefont{Smith},\ and\
  \citenamefont{Kay}}]{dohnalek2000}%
  \BibitemOpen
  \bibfield{author}{%
  \bibinfo {author} {\bibnamefont{Dohn{\'a}lek}, \bibfnamefont{Z.}}, \bibinfo
  {author} {\bibfnamefont{G.~A.}\ \bibnamefont{Kimmel}}, \bibinfo {author}
  {\bibfnamefont{R.~L.}\ \bibnamefont{Ciolli}}, \bibinfo {author}
  {\bibfnamefont{K.~P.}\ \bibnamefont{Stevenson}}, \bibinfo {author}
  {\bibfnamefont{R.~S.}\ \bibnamefont{Smith}},\ and\ \bibinfo {author}
  {\bibfnamefont{B.~D.}\ \bibnamefont{Kay}}}%
  , \bibinfo {year} {2000},\ \bibfield{title}{%
  \enquote{\bibinfo {title} {The effect of the underlying substrate on the
  crystallization kinetics of dense amorphous solid water films},}\ }%
  \bibfield{journal}{%
  \bibinfo {journal} {J. Chem. Phys.}\ }%
  \textbf{\bibinfo {volume} {112}},\ \bibinfo {pages} {5932--5941}%
  \bibAnnoteFile{NoStop}{dohnalek2000}%
\bibitem[{\citenamefont{Domin{\'e}}\
  \emph{et~al.}(2008)\citenamefont{Domin{\'e}}, \citenamefont{Albert},
  \citenamefont{Huthwelker}, \citenamefont{Jacobi}, \citenamefont{Kokhanovsky},
  \citenamefont{Lehning}, \citenamefont{Picard},\ and\
  \citenamefont{Simpson}}]{Domine:2008p2711}%
  \BibitemOpen
  \bibfield{author}{%
  \bibinfo {author} {\bibnamefont{Domin{\'e}}, \bibfnamefont{F.}}, \bibinfo
  {author} {\bibfnamefont{M.}~\bibnamefont{Albert}}, \bibinfo {author}
  {\bibfnamefont{T.}~\bibnamefont{Huthwelker}}, \bibinfo {author}
  {\bibfnamefont{H.-W.}\ \bibnamefont{Jacobi}}, \bibinfo {author}
  {\bibfnamefont{A.~A.}\ \bibnamefont{Kokhanovsky}}, \bibinfo {author}
  {\bibfnamefont{M.}~\bibnamefont{Lehning}}, \bibinfo {author}
  {\bibfnamefont{G.}~\bibnamefont{Picard}},\ and\ \bibinfo {author}
  {\bibfnamefont{W.~R.}\ \bibnamefont{Simpson}}}%
  , \bibinfo {year} {2008},\ \bibfield{title}{%
  \enquote{\bibinfo {title} {Snow physics as relevant to snow
  photochemistry},}\ }%
  \bibfield{journal}{%
  \bibinfo {journal} {Atmos. Chem. Phys.}\ }%
  \textbf{\bibinfo {volume} {8}},\ \bibinfo {pages} {171--208}%
  \bibAnnoteFile{NoStop}{Domine:2008p2711}%
\bibitem[{\citenamefont{Domin{\'e}}\ and\
  \citenamefont{Shepson}(2002)}]{Domine:2002p1662}%
  \BibitemOpen
  \bibfield{author}{%
  \bibinfo {author} {\bibnamefont{Domin{\'e}}, \bibfnamefont{F.}},\ and\
  \bibinfo {author} {\bibfnamefont{P.}~\bibnamefont{Shepson}}}%
  , \bibinfo {year} {2002},\ \bibfield{title}{%
  \enquote{\bibinfo {title} {Air-snow interactions and atmospheric
  chemistry},}\ }%
  \bibfield{journal}{%
  \bibinfo {journal} {Science}\ }%
  \textbf{\bibinfo {volume} {297}},\ \bibinfo {pages} {1506--1510}%
  \bibAnnoteFile{NoStop}{Domine:2002p1662}%
\bibitem[{\citenamefont{{Dotto}}\ \emph{et~al.}(2003)\citenamefont{{Dotto}},
  \citenamefont{{Barucci}},\ and\ \citenamefont{{de Bergh}}}]{dotto2003}%
  \BibitemOpen
  \bibfield{author}{%
  \bibinfo {author} {\bibnamefont{{Dotto}}, \bibfnamefont{E.}}, \bibinfo
  {author} {\bibfnamefont{M.~A.}\ \bibnamefont{{Barucci}}},\ and\ \bibinfo
  {author} {\bibfnamefont{C.}~\bibnamefont{{de Bergh}}}}%
  , \bibinfo {year} {2003},\ \bibfield{title}{%
  \enquote{\bibinfo {title} {Surface composition of {TNO}s and {Centaurs}:
  visible and near-infrared spectroscopy},}\ }%
  \bibfield{journal}{%
  \bibinfo {journal} {Comptes Rendus Physique}\ }%
  \textbf{\bibinfo {volume} {4}},\ \bibinfo {pages} {775--782}%
  \bibAnnoteFile{NoStop}{dotto2003}%
\bibitem[{\citenamefont{Drews}\ \emph{et~al.}(2009)\citenamefont{Drews},
  \citenamefont{Eisen}, \citenamefont{Weikusat}, \citenamefont{Kipfstuhl},
  \citenamefont{Lambrecht}, \citenamefont{Steinhage}, \citenamefont{Wilhelms},\
  and\ \citenamefont{Miller}}]{drews2009}%
  \BibitemOpen
  \bibfield{author}{%
  \bibinfo {author} {\bibnamefont{Drews}, \bibfnamefont{R.}}, \bibinfo {author}
  {\bibfnamefont{O.}~\bibnamefont{Eisen}}, \bibinfo {author}
  {\bibfnamefont{I.}~\bibnamefont{Weikusat}}, \bibinfo {author}
  {\bibfnamefont{S.}~\bibnamefont{Kipfstuhl}}, \bibinfo {author}
  {\bibfnamefont{A.}~\bibnamefont{Lambrecht}}, \bibinfo {author}
  {\bibfnamefont{D.}~\bibnamefont{Steinhage}}, \bibinfo {author}
  {\bibfnamefont{F.}~\bibnamefont{Wilhelms}},\ and\ \bibinfo {author}
  {\bibfnamefont{H.}~\bibnamefont{Miller}}}%
  , \bibinfo {year} {2009},\ \bibfield{title}{%
  \enquote{\bibinfo {title} {Layer disturbances and the radio-echo free zone in
  ice sheets},}\ }%
  \bibfield{journal}{%
  \bibinfo {journal} {Cryosphere}\ }%
  \textbf{\bibinfo {volume} {3}},\ \bibinfo {pages} {195--203}%
  \bibAnnoteFile{NoStop}{drews2009}%
\bibitem[{\citenamefont{Duft}\ and\ \citenamefont{Leisner}(2004)}]{Duft2004}%
  \BibitemOpen
  \bibfield{author}{%
  \bibinfo {author} {\bibnamefont{Duft}, \bibfnamefont{D.}},\ and\ \bibinfo
  {author} {\bibfnamefont{T.}~\bibnamefont{Leisner}}}%
  , \bibinfo {year} {2004},\ \bibfield{title}{%
  \enquote{\bibinfo {title} {Laboratory evidence for volume-dominated
  nucleation of ice in supercooled water microdroplets},}\ }%
  \bibfield{journal}{%
  \bibinfo {journal} {Atmos. Chem. Phys.}\ }%
  \textbf{\bibinfo {volume} {4}},\ \bibinfo {pages} {1997--2000}%
  \bibAnnoteFile{NoStop}{Duft2004}%
\bibitem[{\citenamefont{Dulieu}\ \emph{et~al.}(2010)\citenamefont{Dulieu},
  \citenamefont{Amiaud}, \citenamefont{Congiu}, \citenamefont{Fillion},
  \citenamefont{Matar}, \citenamefont{Momeni}, \citenamefont{Pirronello},\ and\
  \citenamefont{Lemaire}}]{dulieu2010}%
  \BibitemOpen
  \bibfield{author}{%
  \bibinfo {author} {\bibnamefont{Dulieu}, \bibfnamefont{F.}}, \bibinfo
  {author} {\bibfnamefont{L.}~\bibnamefont{Amiaud}}, \bibinfo {author}
  {\bibfnamefont{E.}~\bibnamefont{Congiu}}, \bibinfo {author}
  {\bibfnamefont{J.~H.}\ \bibnamefont{Fillion}}, \bibinfo {author}
  {\bibfnamefont{E.}~\bibnamefont{Matar}}, \bibinfo {author}
  {\bibfnamefont{A.}~\bibnamefont{Momeni}}, \bibinfo {author}
  {\bibfnamefont{V.}~\bibnamefont{Pirronello}},\ and\ \bibinfo {author}
  {\bibfnamefont{J.~L.}\ \bibnamefont{Lemaire}}}%
  , \bibinfo {year} {2010},\ \bibfield{title}{%
  \enquote{\bibinfo {title} {Experimental evidence for water formation on
  interstellar dust grains by hydrogen and oxygen atoms},}\ }%
  \bibfield{journal}{%
  \bibinfo {journal} {Astron. Astrophys.}\ }%
  \textbf{\bibinfo {volume} {512}},\ \bibinfo {pages} {A30}%
  \bibAnnoteFile{NoStop}{dulieu2010}%
\bibitem[{\citenamefont{Durant}\ and\ \citenamefont{Shaw}(2005)}]{Durant2005}%
  \BibitemOpen
  \bibfield{author}{%
  \bibinfo {author} {\bibnamefont{Durant}, \bibfnamefont{A.~J.}},\ and\
  \bibinfo {author} {\bibfnamefont{R.~A.}\ \bibnamefont{Shaw}}}%
  , \bibinfo {year} {2005},\ \bibfield{title}{%
  \enquote{\bibinfo {title} {Evaporation freezing by contact nucleation
  inside-out},}\ }%
  \bibfield{journal}{%
  \bibinfo {journal} {Geophys. Res. Lett.}\ }%
  \textbf{\bibinfo {volume} {32}},\ \bibinfo {pages} {L20814}%
  \bibAnnoteFile{NoStop}{Durant2005}%
\bibitem[{\citenamefont{Durham}\ and\ \citenamefont{Stern}(2001)}]{durham2001}%
  \BibitemOpen
  \bibfield{author}{%
  \bibinfo {author} {\bibnamefont{Durham}, \bibfnamefont{W.~B.}},\ and\
  \bibinfo {author} {\bibfnamefont{L.~A.}\ \bibnamefont{Stern}}}%
  , \bibinfo {year} {2001},\ \bibfield{title}{%
  \enquote{\bibinfo {title} {Rheological properties of water ice ---
  applications to satellites of the outer planets},}\ }%
  \bibfield{journal}{%
  \bibinfo {journal} {Annu. Rev. Earth Planet. Sci.}\ }%
  \textbf{\bibinfo {volume} {29}},\ \bibinfo {pages} {295--330}%
  \bibAnnoteFile{NoStop}{durham2001}%
\bibitem[{\citenamefont{Ehrenfreund}\
  \emph{et~al.}(2001)\citenamefont{Ehrenfreund}, \citenamefont{d'Hendecourt},
  \citenamefont{Charnley},\ and\ \citenamefont{Ruiterkamp}}]{ehrenfreund2001}%
  \BibitemOpen
  \bibfield{author}{%
  \bibinfo {author} {\bibnamefont{Ehrenfreund}, \bibfnamefont{P.}}, \bibinfo
  {author} {\bibfnamefont{L.}~\bibnamefont{d'Hendecourt}}, \bibinfo {author}
  {\bibfnamefont{S.~B.}\ \bibnamefont{Charnley}},\ and\ \bibinfo {author}
  {\bibfnamefont{R.}~\bibnamefont{Ruiterkamp}}}%
  , \bibinfo {year} {2001},\ \bibfield{title}{%
  \enquote{\bibinfo {title} {Energetic and thermal processing of interstellar
  ices},}\ }%
  \bibfield{journal}{%
  \bibinfo {journal} {J. Geophys. Res.}\ }%
  \textbf{\bibinfo {volume} {106}},\ \bibinfo {pages} {33291--33301}%
  \bibAnnoteFile{NoStop}{ehrenfreund2001}%
\bibitem[{\citenamefont{Ehrenfreund}\
  \emph{et~al.}(2003)\citenamefont{Ehrenfreund}, \citenamefont{Fraser},
  \citenamefont{Blum}, \citenamefont{Cartwright},
  \citenamefont{Garc\'{\i}a-Ruiz}, \citenamefont{Hadamcik},
  \citenamefont{Levasseur-Regourd}, \citenamefont{Price},
  \citenamefont{Prodi},\ and\ \citenamefont{Sarkissian}}]{ehrenfreund2003}%
  \BibitemOpen
  \bibfield{author}{%
  \bibinfo {author} {\bibnamefont{Ehrenfreund}, \bibfnamefont{P.}}, \bibinfo
  {author} {\bibfnamefont{H.~J.}\ \bibnamefont{Fraser}}, \bibinfo {author}
  {\bibfnamefont{J.}~\bibnamefont{Blum}}, \bibinfo {author}
  {\bibfnamefont{J.~H.~E.}\ \bibnamefont{Cartwright}}, \bibinfo {author}
  {\bibfnamefont{J.~M.}\ \bibnamefont{Garc\'{\i}a-Ruiz}}, \bibinfo {author}
  {\bibfnamefont{E.}~\bibnamefont{Hadamcik}}, \bibinfo {author}
  {\bibfnamefont{A.~C.}\ \bibnamefont{Levasseur-Regourd}}, \bibinfo {author}
  {\bibfnamefont{S.}~\bibnamefont{Price}}, \bibinfo {author}
  {\bibfnamefont{F.}~\bibnamefont{Prodi}},\ and\ \bibinfo {author}
  {\bibfnamefont{A.}~\bibnamefont{Sarkissian}}}%
  , \bibinfo {year} {2003},\ \bibfield{title}{%
  \enquote{\bibinfo {title} {Physics and chemistry of icy particles in the
  universe: {A}nswers from microgravity},}\ }%
  \bibfield{journal}{%
  \bibinfo {journal} {Planet. Space Sci.}\ }%
  \textbf{\bibinfo {volume} {51}},\ \bibinfo {pages} {473--494}%
  \bibAnnoteFile{NoStop}{ehrenfreund2003}%
\bibitem[{\citenamefont{Eigen}\ and\
  \citenamefont{Schuster}(1979)}]{eigen1979}%
  \BibitemOpen
  \bibfield{author}{%
  \bibinfo {author} {\bibnamefont{Eigen}, \bibfnamefont{M.}},\ and\ \bibinfo
  {author} {\bibfnamefont{P.}~\bibnamefont{Schuster}}}%
  , \bibinfo {year} {1979},\ \emph{\bibinfo {title} {The Hypercycle --- A
  Principle of Natural Self-Organization}}\ (\bibinfo {publisher} {Springer,
  Berlin})%
  \bibAnnoteFile{NoStop}{eigen1979}%
\bibitem[{\citenamefont{Ekstr{\"o}m}\
  \emph{et~al.}(2008)\citenamefont{Ekstr{\"o}m}, \citenamefont{Eriksson},
  \citenamefont{Read}, \citenamefont{Milz},\ and\
  \citenamefont{Murtagh}}]{ekstrom2008}%
  \BibitemOpen
  \bibfield{author}{%
  \bibinfo {author} {\bibnamefont{Ekstr{\"o}m}, \bibfnamefont{M.}}, \bibinfo
  {author} {\bibfnamefont{P.}~\bibnamefont{Eriksson}}, \bibinfo {author}
  {\bibfnamefont{W.~G.}\ \bibnamefont{Read}}, \bibinfo {author}
  {\bibfnamefont{M.}~\bibnamefont{Milz}},\ and\ \bibinfo {author}
  {\bibfnamefont{D.~P.}\ \bibnamefont{Murtagh}}}%
  , \bibinfo {year} {2008},\ \bibfield{title}{%
  \enquote{\bibinfo {title} {Comparison of satellite limb-sounding humidity
  climatologies of the uppermost tropical troposphere},}\ }%
  \bibfield{journal}{%
  \bibinfo {journal} {Atmos. Chem. Phys.}\ }%
  \textbf{\bibinfo {volume} {8}},\ \bibinfo {pages} {309--320}%
  \bibAnnoteFile{NoStop}{ekstrom2008}%
\bibitem[{\citenamefont{Elbaum}\ \emph{et~al.}(1993)\citenamefont{Elbaum},
  \citenamefont{Lipson},\ and\ \citenamefont{Dash}}]{Elbaum1993a}%
  \BibitemOpen
  \bibfield{author}{%
  \bibinfo {author} {\bibnamefont{Elbaum}, \bibfnamefont{M.}}, \bibinfo
  {author} {\bibfnamefont{S.~G.}\ \bibnamefont{Lipson}},\ and\ \bibinfo
  {author} {\bibfnamefont{J.~G.}\ \bibnamefont{Dash}}}%
  , \bibinfo {year} {1993},\ \bibfield{title}{%
  \enquote{\bibinfo {title} {Optical study of surface melting on ice},}\ }%
  \bibfield{journal}{%
  \bibinfo {journal} {J. Crystal Growth}\ }%
  \textbf{\bibinfo {volume} {129}},\ \bibinfo {pages} {491--505}%
  \bibAnnoteFile{NoStop}{Elbaum1993a}%
\bibitem[{\citenamefont{Elbaum}\ and\
  \citenamefont{Schick}(1991)}]{Elbaum1991}%
  \BibitemOpen
  \bibfield{author}{%
  \bibinfo {author} {\bibnamefont{Elbaum}, \bibfnamefont{M.}},\ and\ \bibinfo
  {author} {\bibfnamefont{M.}~\bibnamefont{Schick}}}%
  , \bibinfo {year} {1991},\ \bibfield{title}{%
  \enquote{\bibinfo {title} {Application of the theory of dispersion forces to
  the surface melting of ice},}\ }%
  \bibfield{journal}{%
  \bibinfo {journal} {Phys. Rev. Lett.}\ }%
  \textbf{\bibinfo {volume} {66}},\ \bibinfo {pages} {1713--1716}%
  \bibAnnoteFile{NoStop}{Elbaum1991}%
\bibitem[{\citenamefont{Engemann}\ \emph{et~al.}(2004)\citenamefont{Engemann},
  \citenamefont{Reichert}, \citenamefont{Dosch}, \citenamefont{Bilgram},
  \citenamefont{Honkim{\"a}ki},\ and\ \citenamefont{Snigirev}}]{engemann2004}%
  \BibitemOpen
  \bibfield{author}{%
  \bibinfo {author} {\bibnamefont{Engemann}, \bibfnamefont{S.}}, \bibinfo
  {author} {\bibfnamefont{H.}~\bibnamefont{Reichert}}, \bibinfo {author}
  {\bibfnamefont{H.}~\bibnamefont{Dosch}}, \bibinfo {author}
  {\bibfnamefont{J.}~\bibnamefont{Bilgram}}, \bibinfo {author}
  {\bibfnamefont{V.}~\bibnamefont{Honkim{\"a}ki}},\ and\ \bibinfo {author}
  {\bibfnamefont{A.}~\bibnamefont{Snigirev}}}%
  , \bibinfo {year} {2004},\ \bibfield{title}{%
  \enquote{\bibinfo {title} {Interfacial melting of ice in contact with
  {SiO$_2$}},}\ }%
  \bibinfo {journal} {Phys. Rev. Lett.},\ \bibinfo {pages} {205701}%
  \bibAnnoteFile{NoStop}{engemann2004}%
\bibitem[{\citenamefont{Escribano}\
  \emph{et~al.}(2007)\citenamefont{Escribano},
  \citenamefont{Fern{\'a}ndez-Torre}, \citenamefont{Herrero},
  \citenamefont{Mart{\'{\i}}n-Llorente}, \citenamefont{Mat{\'e}},
  \citenamefont{Ortega},\ and\ \citenamefont{Grothe}}]{escribano2007}%
  \BibitemOpen
\bibfield{journal}{%
    }%
  \bibfield{author}{%
  \bibinfo {author} {\bibnamefont{Escribano}, \bibfnamefont{R.}}, \bibinfo
  {author} {\bibfnamefont{D.}~\bibnamefont{Fern{\'a}ndez-Torre}}, \bibinfo
  {author} {\bibfnamefont{V.~J.}\ \bibnamefont{Herrero}}, \bibinfo {author}
  {\bibfnamefont{B.}~\bibnamefont{Mart{\'{\i}}n-Llorente}}, \bibinfo {author}
  {\bibfnamefont{B.}~\bibnamefont{Mat{\'e}}}, \bibinfo {author}
  {\bibfnamefont{I.~K.}\ \bibnamefont{Ortega}},\ and\ \bibinfo {author}
  {\bibfnamefont{H.}~\bibnamefont{Grothe}}}%
  , \bibinfo {year} {2007},\ \bibfield{title}{%
  \enquote{\bibinfo {title} {The low-frequency {Raman} and {IR} spectra of
  nitric acid hydrates},}\ }%
  \bibfield{journal}{%
  \bibinfo {journal} {Vibr. Spectr.}\ }%
  \textbf{\bibinfo {volume} {43}},\ \bibinfo {pages} {254--259}%
  \bibAnnoteFile{NoStop}{escribano2007}%
\bibitem[{\citenamefont{Estrin}\ \emph{et~al.}(1997)\citenamefont{Estrin},
  \citenamefont{Kohanoff}, \citenamefont{Laria},\ and\
  \citenamefont{Weht}}]{estrin1997}%
  \BibitemOpen
  \bibfield{author}{%
  \bibinfo {author} {\bibnamefont{Estrin}, \bibfnamefont{D.~A.}}, \bibinfo
  {author} {\bibfnamefont{J.}~\bibnamefont{Kohanoff}}, \bibinfo {author}
  {\bibfnamefont{D.H.}\ \bibnamefont{Laria}},\ and\ \bibinfo {author}
  {\bibfnamefont{R.~O.}\ \bibnamefont{Weht}}}%
  , \bibinfo {year} {1997},\ \bibfield{title}{%
  \enquote{\bibinfo {title} {Hybrid quantum and classical mechanical {Monte
  Carlo} simulations of the interaction of hydrogen chloride with solid water
  clusters},}\ }%
  \bibfield{journal}{%
  \bibinfo {journal} {Chem. Phys. Lett.}\ }%
  \textbf{\bibinfo {volume} {280}},\ \bibinfo {pages} {280--286}%
  \bibAnnoteFile{NoStop}{estrin1997}%
\bibitem[{\citenamefont{Falenty}\ \emph{et~al.}(2011)\citenamefont{Falenty},
  \citenamefont{Genov}, \citenamefont{Hansen}, \citenamefont{Kuhs},\ and\
  \citenamefont{Salamatin}}]{falenty2011}%
  \BibitemOpen
  \bibfield{author}{%
  \bibinfo {author} {\bibnamefont{Falenty}, \bibfnamefont{A.}}, \bibinfo
  {author} {\bibfnamefont{G.}~\bibnamefont{Genov}}, \bibinfo {author}
  {\bibfnamefont{T.~C.}\ \bibnamefont{Hansen}}, \bibinfo {author}
  {\bibfnamefont{W.~F.}\ \bibnamefont{Kuhs}},\ and\ \bibinfo {author}
  {\bibfnamefont{A.~N.}\ \bibnamefont{Salamatin}}}%
  , \bibinfo {year} {2011},\ \bibfield{title}{%
  \enquote{\bibinfo {title} {Kinetics of {CO}$_2$ hydrate formation from water
  frost at low temperatures: Experimental results and theoretical model},}\ }%
  \bibfield{journal}{%
  \bibinfo {journal} {J. Phys. Chem. C}\ }%
  \textbf{\bibinfo {volume} {115}},\ \bibinfo {pages} {4022--4032}%
  \bibAnnoteFile{NoStop}{falenty2011}%
\bibitem[{\citenamefont{Fanfoni}\ and\
  \citenamefont{Tomellini}(1998)}]{fanfoni1998}%
  \BibitemOpen
  \bibfield{author}{%
  \bibinfo {author} {\bibnamefont{Fanfoni}, \bibfnamefont{M.}},\ and\ \bibinfo
  {author} {\bibfnamefont{M.}~\bibnamefont{Tomellini}}}%
  , \bibinfo {year} {1998},\ \bibfield{title}{%
  \enquote{\bibinfo {title} {The {Johnson--Mehl--Avrami--Kolmogorov} model: A
  brief review},}\ }%
  \bibfield{journal}{%
  \bibinfo {journal} {Il Nuovo Cimento}\ }%
  \textbf{\bibinfo {volume} {20D}},\ \bibinfo {pages} {1171--1181}%
  \bibAnnoteFile{NoStop}{fanfoni1998}%
\bibitem[{\citenamefont{Faria}\ \emph{et~al.}(2009)\citenamefont{Faria},
  \citenamefont{Kipfstuhl}, \citenamefont{Azuma}, \citenamefont{Freitag},
  \citenamefont{Weikusat}, \citenamefont{Murshed},\ and\
  \citenamefont{Kuhs}}]{faria2009}%
  \BibitemOpen
  \bibfield{author}{%
  \bibinfo {author} {\bibnamefont{Faria}, \bibfnamefont{S.~H.}}, \bibinfo
  {author} {\bibfnamefont{S.}~\bibnamefont{Kipfstuhl}}, \bibinfo {author}
  {\bibfnamefont{N.}~\bibnamefont{Azuma}}, \bibinfo {author}
  {\bibfnamefont{J.}~\bibnamefont{Freitag}}, \bibinfo {author}
  {\bibfnamefont{I.}~\bibnamefont{Weikusat}}, \bibinfo {author}
  {\bibfnamefont{M.~M.}\ \bibnamefont{Murshed}},\ and\ \bibinfo {author}
  {\bibfnamefont{W.~F.}\ \bibnamefont{Kuhs}}}%
  , \bibinfo {year} {2009},\ \bibfield{title}{%
  \enquote{\bibinfo {title} {The multiscale structure of {Antartica}. part {I}:
  Inland ice},}\ }%
  \bibfield{journal}{%
  \bibinfo {journal} {Low Temp. Sci.}\ }%
  \textbf{\bibinfo {volume} {68}},\ \bibinfo {pages} {39--59}%
  \bibAnnoteFile{NoStop}{faria2009}%
\bibitem[{\citenamefont{Feltham}(2008)}]{Feltham2008}%
  \BibitemOpen
  \bibfield{author}{%
  \bibinfo {author} {\bibnamefont{Feltham}, \bibfnamefont{D.~L.}}}%
  , \bibinfo {year} {2008},\ \bibfield{title}{%
  \enquote{\bibinfo {title} {Sea ice rheology},}\ }%
  \bibfield{journal}{%
  \bibinfo {journal} {Annu. Rev. Fluid Mech.}\ }%
  \textbf{\bibinfo {volume} {40}},\ \bibinfo {pages} {91--112}%
  \bibAnnoteFile{NoStop}{Feltham2008}%
\bibitem[{\citenamefont{Feltham}\ \emph{et~al.}(2006)\citenamefont{Feltham},
  \citenamefont{Untersteiner}, \citenamefont{Wettlaufer},\ and\
  \citenamefont{Worster}}]{Feltham:2006}%
  \BibitemOpen
  \bibfield{author}{%
  \bibinfo {author} {\bibnamefont{Feltham}, \bibfnamefont{D.~L.}}, \bibinfo
  {author} {\bibfnamefont{N.}~\bibnamefont{Untersteiner}}, \bibinfo {author}
  {\bibfnamefont{J.~S.}\ \bibnamefont{Wettlaufer}},\ and\ \bibinfo {author}
  {\bibfnamefont{M.~G.}\ \bibnamefont{Worster}}}%
  , \bibinfo {year} {2006},\ \bibfield{title}{%
  \enquote{\bibinfo {title} {Sea ice is a mushy layer},}\ }%
  \bibfield{journal}{%
  \bibinfo {journal} {Geophys.~Res.~Lett.}\ }%
  \textbf{\bibinfo {volume} {33}},\ \bibinfo {pages} {L14501}%
  \bibAnnoteFile{NoStop}{Feltham:2006}%
\bibitem[{\citenamefont{Ferris}(1998)}]{ferris1998}%
  \BibitemOpen
  \bibfield{author}{%
  \bibinfo {author} {\bibnamefont{Ferris}, \bibfnamefont{J.~P.}}}%
  , \bibinfo {year} {1998},\ \enquote{\bibinfo {title} {Catalyzed {RNA}
  synthesis for the {RNA} world},}\ in\ \emph{\bibinfo {booktitle} {The
  molecular origins of life}},\ \bibinfo {editor} {edited by\ \bibinfo {editor}
  {\bibfnamefont{A.}~\bibnamefont{Brack}}}\ (\bibinfo {publisher} {Cambridge
  University Press})\ pp.\ \bibinfo {pages} {255--268}%
  \bibAnnoteFile{NoStop}{ferris1998}%
\bibitem[{\citenamefont{Feynman}(1967)}]{feynman}%
  \BibitemOpen
  \bibfield{author}{%
  \bibinfo {author} {\bibnamefont{Feynman}, \bibfnamefont{R.~P.}}}%
  , \bibinfo {year} {1967},\ \emph{\bibinfo {title} {The Character of Physical
  Law}}\ (\bibinfo {publisher} {MIT Press})%
  \bibAnnoteFile{NoStop}{feynman}%
\bibitem[{\citenamefont{Finney}(2001)}]{finney2001}%
  \BibitemOpen
  \bibfield{author}{%
  \bibinfo {author} {\bibnamefont{Finney}, \bibfnamefont{J.~L.}}}%
  , \bibinfo {year} {2001},\ \enquote{\bibinfo {title} {Ice: {Structures}},}\
  in\ \emph{\bibinfo {booktitle} {Encyclopedia of Materials: Science and
  Technology}},\ Vol.~\bibinfo {volume} {5},\ \bibinfo {editor} {edited by\
  \bibinfo {editor} {\bibfnamefont{K.~H.~J.}\ \bibnamefont{Buschow}}, \bibinfo
  {editor} {\bibfnamefont{R.~W.}\ \bibnamefont{Cahn}}, \bibinfo {editor}
  {\bibfnamefont{M.~C.}\ \bibnamefont{Flemings}}, \bibinfo {editor}
  {\bibfnamefont{B.}~\bibnamefont{Ilschner}}, \bibinfo {editor}
  {\bibfnamefont{E.~J.}\ \bibnamefont{Kramer}},\ and\ \bibinfo {editor}
  {\bibfnamefont{S.}~\bibnamefont{Mahajan}}}\ (\bibinfo {publisher} {Elsevier
  Science, Oxford})\ pp.\ \bibinfo {pages} {4018--4027}%
  \bibAnnoteFile{NoStop}{finney2001}%
\bibitem[{\citenamefont{Finney}(2004)}]{finney2004}%
  \BibitemOpen
  \bibfield{author}{%
  \bibinfo {author} {\bibnamefont{Finney}, \bibfnamefont{J.~L.}}}%
  , \bibinfo {year} {2004},\ \bibfield{title}{%
  \enquote{\bibinfo {title} {Ice: the laboratory in your freezer},}\ }%
  \bibfield{journal}{%
  \bibinfo {journal} {Interdisciplinary Sci. Rev.}\ }%
  \textbf{\bibinfo {volume} {29}},\ \bibinfo {pages} {339--351}%
  \bibAnnoteFile{NoStop}{finney2004}%
\bibitem[{\citenamefont{Finney}\
  \emph{et~al.}(2002{\natexlab{a}})\citenamefont{Finney},
  \citenamefont{Bowron}, \citenamefont{Soper}, \citenamefont{Loerting},
  \citenamefont{Mayer},\ and\ \citenamefont{Hallbrucker}}]{finney2002}%
  \BibitemOpen
  \bibfield{author}{%
  \bibinfo {author} {\bibnamefont{Finney}, \bibfnamefont{J.~L.}}, \bibinfo
  {author} {\bibfnamefont{D.~T.}\ \bibnamefont{Bowron}}, \bibinfo {author}
  {\bibfnamefont{A.~K.}\ \bibnamefont{Soper}}, \bibinfo {author}
  {\bibfnamefont{T.}~\bibnamefont{Loerting}}, \bibinfo {author}
  {\bibfnamefont{E.}~\bibnamefont{Mayer}},\ and\ \bibinfo {author}
  {\bibfnamefont{A.}~\bibnamefont{Hallbrucker}}}%
  , \bibinfo {year} {2002}{\natexlab{a}},\ \bibfield{title}{%
  \enquote{\bibinfo {title} {Structure of a new dense amorphous ice},}\ }%
  \bibfield{journal}{%
  \bibinfo {journal} {Phys. Rev. Lett.}\ }%
  \textbf{\bibinfo {volume} {89}},\ \bibinfo {pages} {205503}%
  \bibAnnoteFile{NoStop}{finney2002}%
\bibitem[{\citenamefont{Finney}\
  \emph{et~al.}(2002{\natexlab{b}})\citenamefont{Finney},
  \citenamefont{Hallbrucker}, \citenamefont{Kohl}, \citenamefont{Soper},\ and\
  \citenamefont{Bowron}}]{finney2002_2}%
  \BibitemOpen
  \bibfield{author}{%
  \bibinfo {author} {\bibnamefont{Finney}, \bibfnamefont{J.~L.}}, \bibinfo
  {author} {\bibfnamefont{A.}~\bibnamefont{Hallbrucker}}, \bibinfo {author}
  {\bibfnamefont{I.}~\bibnamefont{Kohl}}, \bibinfo {author}
  {\bibfnamefont{A.~K.}\ \bibnamefont{Soper}},\ and\ \bibinfo {author}
  {\bibfnamefont{D.~T.}\ \bibnamefont{Bowron}}}%
  , \bibinfo {year} {2002}{\natexlab{b}},\ \bibfield{title}{%
  \enquote{\bibinfo {title} {Structures of high and low density amorphous ice
  by neutron diffraction},}\ }%
  \bibfield{journal}{%
  \bibinfo {journal} {Phys. Rev. Lett.}\ }%
  \textbf{\bibinfo {volume} {88}},\ \bibinfo {pages} {225503}%
  \bibAnnoteFile{NoStop}{finney2002_2}%
\bibitem[{\citenamefont{Firanescu}\
  \emph{et~al.}(2006)\citenamefont{Firanescu}, \citenamefont{Hermsdorf},
  \citenamefont{Ueberschaer},\ and\ \citenamefont{Signorell}}]{firanescu2006}%
  \BibitemOpen
  \bibfield{author}{%
  \bibinfo {author} {\bibnamefont{Firanescu}, \bibfnamefont{G.}}, \bibinfo
  {author} {\bibfnamefont{D.}~\bibnamefont{Hermsdorf}}, \bibinfo {author}
  {\bibfnamefont{R.}~\bibnamefont{Ueberschaer}},\ and\ \bibinfo {author}
  {\bibfnamefont{R.}~\bibnamefont{Signorell}}}%
  , \bibinfo {year} {2006},\ \bibfield{title}{%
  \enquote{\bibinfo {title} {Large molecular aggregates: from atmospheric
  aerosols to drug nanoparticles},}\ }%
  \bibfield{journal}{%
  \bibinfo {journal} {Phys. Chem. Chem. Phys.}\ }%
  \textbf{\bibinfo {volume} {8}},\ \bibinfo {pages} {4149--4165}%
  \bibAnnoteFile{NoStop}{firanescu2006}%
\bibitem[{\citenamefont{Flammer}\ \emph{et~al.}(1998)\citenamefont{Flammer},
  \citenamefont{Mendis},\ and\ \citenamefont{Houpis}}]{flammer1998}%
  \BibitemOpen
  \bibfield{author}{%
  \bibinfo {author} {\bibnamefont{Flammer}, \bibfnamefont{K.~R.}}, \bibinfo
  {author} {\bibfnamefont{D.~A.}\ \bibnamefont{Mendis}},\ and\ \bibinfo
  {author} {\bibfnamefont{H.~L.~F.}\ \bibnamefont{Houpis}}}%
  , \bibinfo {year} {1998},\ \bibfield{title}{%
  \enquote{\bibinfo {title} {On the outgassing profile of comet
  {H}ale--{B}opp},}\ }%
  \bibfield{journal}{%
  \bibinfo {journal} {Astrophys. J.}\ }%
  \textbf{\bibinfo {volume} {494}},\ \bibinfo {pages} {822--827}%
  \bibAnnoteFile{NoStop}{flammer1998}%
\bibitem[{\citenamefont{Fogle}\ and\
  \citenamefont{Haurwitz}(1966)}]{Fogle1966}%
  \BibitemOpen
  \bibfield{author}{%
  \bibinfo {author} {\bibnamefont{Fogle}, \bibfnamefont{B.}},\ and\ \bibinfo
  {author} {\bibfnamefont{B.}~\bibnamefont{Haurwitz}}}%
  , \bibinfo {year} {1966},\ \bibfield{title}{%
  \enquote{\bibinfo {title} {Noctilucent clouds},}\ }%
  \bibfield{journal}{%
  \bibinfo {journal} {Space Sci. Rev.}\ }%
  \textbf{\bibinfo {volume} {6}},\ \bibinfo {pages} {279--340}%
  \bibAnnoteFile{NoStop}{Fogle1966}%
\bibitem[{\citenamefont{Fraser}\ \emph{et~al.}(2002)\citenamefont{Fraser},
  \citenamefont{Collings},\ and\ \citenamefont{McCoustra}}]{fraser2002}%
  \BibitemOpen
  \bibfield{author}{%
  \bibinfo {author} {\bibnamefont{Fraser}, \bibfnamefont{H.~J.}}, \bibinfo
  {author} {\bibfnamefont{M.~P.}\ \bibnamefont{Collings}},\ and\ \bibinfo
  {author} {\bibfnamefont{M.~R.~S.}\ \bibnamefont{McCoustra}}}%
  , \bibinfo {year} {2002},\ \bibfield{title}{%
  \enquote{\bibinfo {title} {Laboratory surface astrophysics experiment},}\ }%
  \bibfield{journal}{%
  \bibinfo {journal} {Rev. Sci. Instrum.}\ }%
  \textbf{\bibinfo {volume} {73}},\ \bibinfo {pages} {2161--2170}%
  \bibAnnoteFile{NoStop}{fraser2002}%
\bibitem[{\citenamefont{French}\ \emph{et~al.}(1996)\citenamefont{French},
  \citenamefont{Kasischke}, \citenamefont{Bourgeau-Chavez},\ and\
  \citenamefont{Harrell}}]{french1996}%
  \BibitemOpen
  \bibfield{author}{%
  \bibinfo {author} {\bibnamefont{French}, \bibfnamefont{N.~H.~F.}}, \bibinfo
  {author} {\bibfnamefont{E.~S.}\ \bibnamefont{Kasischke}}, \bibinfo {author}
  {\bibfnamefont{L.~L.}\ \bibnamefont{Bourgeau-Chavez}},\ and\ \bibinfo
  {author} {\bibfnamefont{P.~A.}\ \bibnamefont{Harrell}}}%
  , \bibinfo {year} {1996},\ \bibfield{title}{%
  \enquote{\bibinfo {title} {Monitoring variations in soil moisture on fire
  disturbed sites in alaska using {ERS-1} {SAR} imagery},}\ }%
  \bibfield{journal}{%
  \bibinfo {journal} {Int. J. Remote Sensing}\ }%
  \textbf{\bibinfo {volume} {17}},\ \bibinfo {pages} {3037--3053}%
  \bibAnnoteFile{NoStop}{french1996}%
\bibitem[{\citenamefont{French}\ \emph{et~al.}(2010)\citenamefont{French},
  \citenamefont{Parsegian}, \citenamefont{Podgornik}, \citenamefont{Rajter},
  \citenamefont{Jagota}, \citenamefont{Luo}, \citenamefont{Asthagiri},
  \citenamefont{Chaudhury}, \citenamefont{Chiang}, \citenamefont{Granick},
  \citenamefont{Kalinin}, \citenamefont{Kardar}, \citenamefont{Kjellander},
  \citenamefont{Langreth}, \citenamefont{Lewis}, \citenamefont{Lustig},
  \citenamefont{Wesolowski}, \citenamefont{Wettlaufer}, \citenamefont{Ching},
  \citenamefont{Finnis}, \citenamefont{Houlihan}, \citenamefont{von
  Lilienfeld}, \citenamefont{van Oss},\ and\ \citenamefont{Zemb}}]{French2010}%
  \BibitemOpen
  \bibfield{author}{%
  \bibinfo {author} {\bibnamefont{French}, \bibfnamefont{R.~H.}}, \bibinfo
  {author} {\bibfnamefont{V.~A.}\ \bibnamefont{Parsegian}}, \bibinfo {author}
  {\bibfnamefont{R.}~\bibnamefont{Podgornik}}, \bibinfo {author}
  {\bibfnamefont{R.~F.}\ \bibnamefont{Rajter}}, \bibinfo {author}
  {\bibfnamefont{A.}~\bibnamefont{Jagota}}, \bibinfo {author}
  {\bibfnamefont{J.}~\bibnamefont{Luo}}, \bibinfo {author}
  {\bibfnamefont{D.}~\bibnamefont{Asthagiri}}, \bibinfo {author}
  {\bibfnamefont{M.~K.}\ \bibnamefont{Chaudhury}}, \bibinfo {author}
  {\bibfnamefont{Y.-M.}\ \bibnamefont{Chiang}}, \bibinfo {author}
  {\bibfnamefont{S.}~\bibnamefont{Granick}}, \bibinfo {author}
  {\bibfnamefont{S.}~\bibnamefont{Kalinin}}, \bibinfo {author}
  {\bibfnamefont{M.}~\bibnamefont{Kardar}}, \bibinfo {author}
  {\bibfnamefont{R.}~\bibnamefont{Kjellander}}, \bibinfo {author}
  {\bibfnamefont{D.~C.}\ \bibnamefont{Langreth}}, \bibinfo {author}
  {\bibfnamefont{J.}~\bibnamefont{Lewis}}, \bibinfo {author}
  {\bibfnamefont{S.}~\bibnamefont{Lustig}}, \bibinfo {author}
  {\bibfnamefont{D.}~\bibnamefont{Wesolowski}}, \bibinfo {author}
  {\bibfnamefont{J.~S.}\ \bibnamefont{Wettlaufer}}, \bibinfo {author}
  {\bibfnamefont{W.-Y.}\ \bibnamefont{Ching}}, \bibinfo {author}
  {\bibfnamefont{M.}~\bibnamefont{Finnis}}, \bibinfo {author}
  {\bibfnamefont{F.}~\bibnamefont{Houlihan}}, \bibinfo {author}
  {\bibfnamefont{O.~A.}\ \bibnamefont{von Lilienfeld}}, \bibinfo {author}
  {\bibfnamefont{C.~J.}\ \bibnamefont{van Oss}},\ and\ \bibinfo {author}
  {\bibfnamefont{T.}~\bibnamefont{Zemb}}}%
  , \bibinfo {year} {2010},\ \bibfield{title}{%
  \enquote{\bibinfo {title} {Long range interactions in nanoscale science},}\
  }%
  \bibfield{journal}{%
  \bibinfo {journal} {Rev. Mod. Phys.}\ }%
  \textbf{\bibinfo {volume} {82}},\ \bibinfo {pages} {1887--1944}%
  \bibAnnoteFile{NoStop}{French2010}%
\bibitem[{\citenamefont{Fukazawa}\ \emph{et~al.}(2006)\citenamefont{Fukazawa},
  \citenamefont{Hoshikawa}, \citenamefont{Yamauchi}, \citenamefont{Yamaguchi},
  \citenamefont{Igawa},\ and\ \citenamefont{Ishiib}}]{fukazawa2006}%
  \BibitemOpen
  \bibfield{author}{%
  \bibinfo {author} {\bibnamefont{Fukazawa}, \bibfnamefont{H.}}, \bibinfo
  {author} {\bibfnamefont{A.}~\bibnamefont{Hoshikawa}}, \bibinfo {author}
  {\bibfnamefont{H.}~\bibnamefont{Yamauchi}}, \bibinfo {author}
  {\bibfnamefont{Y.}~\bibnamefont{Yamaguchi}}, \bibinfo {author}
  {\bibfnamefont{N.}~\bibnamefont{Igawa}},\ and\ \bibinfo {author}
  {\bibfnamefont{Y.}~\bibnamefont{Ishiib}}}%
  , \bibinfo {year} {2006},\ \bibfield{title}{%
  \enquote{\bibinfo {title} {Deuteron ordering in ice containing impurities: A
  neutron diffraction study},}\ }%
  \bibfield{journal}{%
  \bibinfo {journal} {Physica B}\ }%
  \textbf{\bibinfo {volume} {385--386}},\ \bibinfo {pages} {113--115}%
  \bibAnnoteFile{NoStop}{fukazawa2006}%
\bibitem[{\citenamefont{Fukazawa}\ \emph{et~al.}(2005)\citenamefont{Fukazawa},
  \citenamefont{Hoshikawa}, \citenamefont{Yamauchi}, \citenamefont{Yamaguchi},\
  and\ \citenamefont{Ishiib}}]{fukazawa2005}%
  \BibitemOpen
  \bibfield{author}{%
  \bibinfo {author} {\bibnamefont{Fukazawa}, \bibfnamefont{H.}}, \bibinfo
  {author} {\bibfnamefont{A.}~\bibnamefont{Hoshikawa}}, \bibinfo {author}
  {\bibfnamefont{H.}~\bibnamefont{Yamauchi}}, \bibinfo {author}
  {\bibfnamefont{Y.}~\bibnamefont{Yamaguchi}},\ and\ \bibinfo {author}
  {\bibfnamefont{Y.}~\bibnamefont{Ishiib}}}%
  , \bibinfo {year} {2005},\ \bibfield{title}{%
  \enquote{\bibinfo {title} {Formation and growth of ice {XI}: A powder neutron
  diffraction study},}\ }%
  \bibfield{journal}{%
  \bibinfo {journal} {J.Cryst.Growth}\ }%
  \textbf{\bibinfo {volume} {282}},\ \bibinfo {pages} {251--259}%
  \bibAnnoteFile{NoStop}{fukazawa2005}%
\bibitem[{\citenamefont{Furukawa}\ and\
  \citenamefont{Nada}(1997{\natexlab{a}})}]{furukawa1997}%
  \BibitemOpen
  \bibfield{author}{%
  \bibinfo {author} {\bibnamefont{Furukawa}, \bibfnamefont{Y.}},\ and\ \bibinfo
  {author} {\bibfnamefont{H.}~\bibnamefont{Nada}}}%
  , \bibinfo {year} {1997}{\natexlab{a}},\ \bibfield{title}{%
  \enquote{\bibinfo {title} {Anisotropic surface melting of an ice crystal and
  its relationship to growth forms},}\ }%
  \bibfield{journal}{%
  \bibinfo {journal} {J. Phys. Chem. B}\ }%
  \textbf{\bibinfo {volume} {101}},\ \bibinfo {pages} {6167--6170}%
  \bibAnnoteFile{NoStop}{furukawa1997}%
\bibitem[{\citenamefont{Furukawa}\ and\
  \citenamefont{Nada}(1997{\natexlab{b}})}]{furukawa1997_2}%
  \BibitemOpen
  \bibfield{author}{%
  \bibinfo {author} {\bibnamefont{Furukawa}, \bibfnamefont{Y.}},\ and\ \bibinfo
  {author} {\bibfnamefont{H.}~\bibnamefont{Nada}}}%
  , \bibinfo {year} {1997}{\natexlab{b}},\ \enquote{\bibinfo {title}
  {Anisotropy in microscopic structures of ice-water and ice-vapor interfaces
  and its relation to growth kinetics},}\ in\ \emph{\bibinfo {booktitle}
  {Advances in the Understanding of Crystal Growth Mechanisms}},\ \bibinfo
  {editor} {edited by\ \bibinfo {editor}
  {\bibfnamefont{T.}~\bibnamefont{Nishinaga}}, \bibinfo {editor}
  {\bibfnamefont{K.}~\bibnamefont{Nishioka}}, \bibinfo {editor}
  {\bibfnamefont{J.}~\bibnamefont{Harada}}, \bibinfo {editor}
  {\bibfnamefont{A.}~\bibnamefont{Sasaki}},\ and\ \bibinfo {editor}
  {\bibfnamefont{H.}~\bibnamefont{Takei}}}\ (\bibinfo {publisher} {Elsevier})\
  pp.\ \bibinfo {pages} {559--573}%
  \bibAnnoteFile{NoStop}{furukawa1997_2}%
\bibitem[{\citenamefont{G{\'a}lvez}\
  \emph{et~al.}(2007)\citenamefont{G{\'a}lvez}, \citenamefont{Ortega},
  \citenamefont{Mat{\'e}}, \citenamefont{Moreno},
  \citenamefont{Mart{\'{\i}}n-Llorente}, \citenamefont{Herrero},
  \citenamefont{Escribano},\ and\ \citenamefont{Guti{\'e}rrez}}]{galvez2007}%
  \BibitemOpen
  \bibfield{author}{%
  \bibinfo {author} {\bibnamefont{G{\'a}lvez}, \bibfnamefont{O.}}, \bibinfo
  {author} {\bibfnamefont{I.~K.}\ \bibnamefont{Ortega}}, \bibinfo {author}
  {\bibfnamefont{B.}~\bibnamefont{Mat{\'e}}}, \bibinfo {author}
  {\bibfnamefont{M.~A.}\ \bibnamefont{Moreno}}, \bibinfo {author}
  {\bibfnamefont{B.}~\bibnamefont{Mart{\'{\i}}n-Llorente}}, \bibinfo {author}
  {\bibfnamefont{V.~J.}\ \bibnamefont{Herrero}}, \bibinfo {author}
  {\bibfnamefont{R.}~\bibnamefont{Escribano}},\ and\ \bibinfo {author}
  {\bibfnamefont{P.~J.}\ \bibnamefont{Guti{\'e}rrez}}}%
  , \bibinfo {year} {2007},\ \bibfield{title}{%
  \enquote{\bibinfo {title} {A study of the interaction of {CO$_2$} with water
  ice},}\ }%
  \bibfield{journal}{%
  \bibinfo {journal} {Astron. Astrophys.}\ }%
  \textbf{\bibinfo {volume} {472}},\ \bibinfo {pages} {691--698}%
  \bibAnnoteFile{NoStop}{galvez2007}%
\bibitem[{\citenamefont{Gao}\ \emph{et~al.}(2004)\citenamefont{Gao},
  \citenamefont{Popp}, \citenamefont{Fahey}, \citenamefont{Marcy},
  \citenamefont{Herman}, \citenamefont{Weinstock}, \citenamefont{Baumgardner},
  \citenamefont{Garrett}, \citenamefont{Rosenlof}, \citenamefont{Thompson},
  \citenamefont{Bui}, \citenamefont{Ridley}, \citenamefont{Wofsy},
  \citenamefont{Toon}, \citenamefont{Tolbert}, \citenamefont{K{\"a}rcher},
  \citenamefont{Peter}, \citenamefont{Hudson}, \citenamefont{Weinheimer},\ and\
  \citenamefont{Heymsfield}}]{Gao2004}%
  \BibitemOpen
  \bibfield{author}{%
  \bibinfo {author} {\bibnamefont{Gao}, \bibfnamefont{R.~S.}}, \bibinfo
  {author} {\bibfnamefont{P.~J.}\ \bibnamefont{Popp}}, \bibinfo {author}
  {\bibfnamefont{D.~W.}\ \bibnamefont{Fahey}}, \bibinfo {author}
  {\bibfnamefont{T.~P.}\ \bibnamefont{Marcy}}, \bibinfo {author}
  {\bibfnamefont{R.~L.}\ \bibnamefont{Herman}}, \bibinfo {author}
  {\bibfnamefont{E.~M.}\ \bibnamefont{Weinstock}}, \bibinfo {author}
  {\bibfnamefont{D.~G.}\ \bibnamefont{Baumgardner}}, \bibinfo {author}
  {\bibfnamefont{T.~J.}\ \bibnamefont{Garrett}}, \bibinfo {author}
  {\bibfnamefont{K.~H.}\ \bibnamefont{Rosenlof}}, \bibinfo {author}
  {\bibfnamefont{T.~L.}\ \bibnamefont{Thompson}}, \bibinfo {author}
  {\bibfnamefont{P.~T.}\ \bibnamefont{Bui}}, \bibinfo {author}
  {\bibfnamefont{B.~A.}\ \bibnamefont{Ridley}}, \bibinfo {author}
  {\bibfnamefont{S.~C.}\ \bibnamefont{Wofsy}}, \bibinfo {author}
  {\bibfnamefont{O.~B.}\ \bibnamefont{Toon}}, \bibinfo {author}
  {\bibfnamefont{M.~A.}\ \bibnamefont{Tolbert}}, \bibinfo {author}
  {\bibfnamefont{B.}~\bibnamefont{K{\"a}rcher}}, \bibinfo {author}
  {\bibfnamefont{T.}~\bibnamefont{Peter}}, \bibinfo {author}
  {\bibfnamefont{P.~K.}\ \bibnamefont{Hudson}}, \bibinfo {author}
  {\bibfnamefont{A.~J.}\ \bibnamefont{Weinheimer}},\ and\ \bibinfo {author}
  {\bibfnamefont{A.~J.}\ \bibnamefont{Heymsfield}}}%
  , \bibinfo {year} {2004},\ \bibfield{title}{%
  \enquote{\bibinfo {title} {Evidence that nitric acid increases relative
  humidity in low-temperature cirrus clouds},}\ }%
  \bibfield{journal}{%
  \bibinfo {journal} {Science}\ }%
  \textbf{\bibinfo {volume} {303}},\ \bibinfo {pages} {516--520}%
  \bibAnnoteFile{NoStop}{Gao2004}%
\bibitem[{\citenamefont{Gautier}\ and\
  \citenamefont{Hersant}(2005)}]{gautier2005}%
  \BibitemOpen
  \bibfield{author}{%
  \bibinfo {author} {\bibnamefont{Gautier}, \bibfnamefont{D.}},\ and\ \bibinfo
  {author} {\bibfnamefont{F.}~\bibnamefont{Hersant}}}%
  , \bibinfo {year} {2005},\ \bibfield{title}{%
  \enquote{\bibinfo {title} {Formation and composition of planetesimals},}\ }%
  \bibfield{journal}{%
  \bibinfo {journal} {Space Sci. Rev.}\ }%
  \textbf{\bibinfo {volume} {116}},\ \bibinfo {pages} {25--52}%
  \bibAnnoteFile{NoStop}{gautier2005}%
\bibitem[{\citenamefont{Gerakines}\
  \emph{et~al.}(2005)\citenamefont{Gerakines}, \citenamefont{Bray},
  \citenamefont{Davis},\ and\ \citenamefont{Richey}}]{gerakines2005}%
  \BibitemOpen
  \bibfield{author}{%
  \bibinfo {author} {\bibnamefont{Gerakines}, \bibfnamefont{P.~A.}}, \bibinfo
  {author} {\bibfnamefont{J.~J.}\ \bibnamefont{Bray}}, \bibinfo {author}
  {\bibfnamefont{A.}~\bibnamefont{Davis}},\ and\ \bibinfo {author}
  {\bibfnamefont{C.~R.}\ \bibnamefont{Richey}}}%
  , \bibinfo {year} {2005},\ \bibfield{title}{%
  \enquote{\bibinfo {title} {The strengths of near-infrared absorption features
  relevant to interstellar and planetary ices},}\ }%
  \bibfield{journal}{%
  \bibinfo {journal} {Astrophys. J.}\ }%
  \textbf{\bibinfo {volume} {620}},\ \bibinfo {pages} {1140--1150}%
  \bibAnnoteFile{NoStop}{gerakines2005}%
\bibitem[{\citenamefont{Gerakines}\
  \emph{et~al.}(1999)\citenamefont{Gerakines}, \citenamefont{Whittet},
  \citenamefont{Ehrenfreund}, \citenamefont{Boogert}, \citenamefont{Tielens},
  \citenamefont{Schutte}, \citenamefont{Chiar}, \citenamefont{vanDishoeck},
  \citenamefont{Prusti}, \citenamefont{Helmich},\ and\
  \citenamefont{De~Graauw}}]{gerakines1999}%
  \BibitemOpen
  \bibfield{author}{%
  \bibinfo {author} {\bibnamefont{Gerakines}, \bibfnamefont{P.~A.}}, \bibinfo
  {author} {\bibfnamefont{D.~C.~B.}\ \bibnamefont{Whittet}}, \bibinfo {author}
  {\bibfnamefont{P.}~\bibnamefont{Ehrenfreund}}, \bibinfo {author}
  {\bibfnamefont{A.~C.~A.}\ \bibnamefont{Boogert}}, \bibinfo {author}
  {\bibfnamefont{A.~G. G.~M.}\ \bibnamefont{Tielens}}, \bibinfo {author}
  {\bibfnamefont{W.~A.}\ \bibnamefont{Schutte}}, \bibinfo {author}
  {\bibfnamefont{J.~E.}\ \bibnamefont{Chiar}}, \bibinfo {author}
  {\bibfnamefont{E.~F.}\ \bibnamefont{vanDishoeck}}, \bibinfo {author}
  {\bibfnamefont{T.}~\bibnamefont{Prusti}}, \bibinfo {author}
  {\bibfnamefont{F.~P.}\ \bibnamefont{Helmich}},\ and\ \bibinfo {author}
  {\bibfnamefont{Th.}\ \bibnamefont{De~Graauw}}}%
  , \bibinfo {year} {1999},\ \bibfield{title}{%
  \enquote{\bibinfo {title} {Observations of solid carbon dioxide in molecular
  clouds with the infrared space observatory},}\ }%
  \bibfield{journal}{%
  \bibinfo {journal} {Astrophys. J.}\ }%
  \textbf{\bibinfo {volume} {522}},\ \bibinfo {pages} {357--377}%
  \bibAnnoteFile{NoStop}{gerakines1999}%
\bibitem[{\citenamefont{Gibb}\ \emph{et~al.}(2000)\citenamefont{Gibb},
  \citenamefont{Whittet}, \citenamefont{Schutte}, \citenamefont{Boogert},
  \citenamefont{Chiar}, \citenamefont{Ehrenfreund}, \citenamefont{Gerakines},
  \citenamefont{Keane}, \citenamefont{Tielens}, \citenamefont{van Dishoeck},\
  and\ \citenamefont{Kerkhof}}]{gibb2000}%
  \BibitemOpen
  \bibfield{author}{%
  \bibinfo {author} {\bibnamefont{Gibb}, \bibfnamefont{E.~L.}}, \bibinfo
  {author} {\bibfnamefont{D.~C.~B.}\ \bibnamefont{Whittet}}, \bibinfo {author}
  {\bibfnamefont{W.~A.}\ \bibnamefont{Schutte}}, \bibinfo {author}
  {\bibfnamefont{A.~C.~A.}\ \bibnamefont{Boogert}}, \bibinfo {author}
  {\bibfnamefont{J.~E.}\ \bibnamefont{Chiar}}, \bibinfo {author}
  {\bibfnamefont{P.}~\bibnamefont{Ehrenfreund}}, \bibinfo {author}
  {\bibfnamefont{P.~A.}\ \bibnamefont{Gerakines}}, \bibinfo {author}
  {\bibfnamefont{J.~V.}\ \bibnamefont{Keane}}, \bibinfo {author}
  {\bibfnamefont{A.}~\bibnamefont{Tielens}}, \bibinfo {author}
  {\bibfnamefont{E.~F.}\ \bibnamefont{van Dishoeck}},\ and\ \bibinfo {author}
  {\bibfnamefont{O.}~\bibnamefont{Kerkhof}}}%
  , \bibinfo {year} {2000},\ \bibfield{title}{%
  \enquote{\bibinfo {title} {An inventory of interstellar ices toward the
  embedded protostar {W33A}},}\ }%
  \bibfield{journal}{%
  \bibinfo {journal} {Astrophys. J.}\ }%
  \textbf{\bibinfo {volume} {536}},\ \bibinfo {pages} {347--356}%
  \bibAnnoteFile{NoStop}{gibb2000}%
\bibitem[{\citenamefont{Gibson}\ \emph{et~al.}(2011)\citenamefont{Gibson},
  \citenamefont{Killelea}, \citenamefont{Yuan}, \citenamefont{Becker},\ and\
  \citenamefont{Sibener}}]{Gibson2011}%
  \BibitemOpen
  \bibfield{author}{%
  \bibinfo {author} {\bibnamefont{Gibson}, \bibfnamefont{K.~D.}}, \bibinfo
  {author} {\bibfnamefont{D.~R.}\ \bibnamefont{Killelea}}, \bibinfo {author}
  {\bibfnamefont{H.~Q.}\ \bibnamefont{Yuan}}, \bibinfo {author}
  {\bibfnamefont{J.~S.}\ \bibnamefont{Becker}},\ and\ \bibinfo {author}
  {\bibfnamefont{S.~J.}\ \bibnamefont{Sibener}}}%
  , \bibinfo {year} {2011},\ \bibfield{title}{%
  \enquote{\bibinfo {title} {Determination of the sticking coefficient and
  scattering dynamics of water on ice using molecular beam techniques},}\ }%
  \bibfield{journal}{%
  \bibinfo {journal} {J. Chem. Phys.}\ }%
  \textbf{\bibinfo {volume} {134}},\ \bibinfo {pages} {034703}%
  \bibAnnoteFile{NoStop}{Gibson2011}%
\bibitem[{\citenamefont{{Gil-Hutton}}\
  \emph{et~al.}(2009)\citenamefont{{Gil-Hutton}}, \citenamefont{{Licandro}},
  \citenamefont{{Pinilla-Alonso}},\ and\
  \citenamefont{{Brunetto}}}]{gil-hutton2009}%
  \BibitemOpen
  \bibfield{author}{%
  \bibinfo {author} {\bibnamefont{{Gil-Hutton}}, \bibfnamefont{R.}}, \bibinfo
  {author} {\bibfnamefont{J.}~\bibnamefont{{Licandro}}}, \bibinfo {author}
  {\bibfnamefont{N.}~\bibnamefont{{Pinilla-Alonso}}},\ and\ \bibinfo {author}
  {\bibfnamefont{R.}~\bibnamefont{{Brunetto}}}}%
  , \bibinfo {year} {2009},\ \bibfield{title}{%
  \enquote{\bibinfo {title} {{The trans-Neptunian object size distribution at
  small sizes}},}\ }%
  \bibfield{journal}{%
  \bibinfo {journal} {Astron. Astrophys.}\ }%
  \textbf{\bibinfo {volume} {500}},\ \bibinfo {pages} {909--916}%
  \bibAnnoteFile{NoStop}{gil-hutton2009}%
\bibitem[{\citenamefont{Girard}\ \emph{et~al.}(2011)\citenamefont{Girard},
  \citenamefont{Bouillon}, \citenamefont{Weiss}, \citenamefont{Amitrano},
  \citenamefont{Fichefet},\ and\ \citenamefont{Legat}}]{Girard2011}%
  \BibitemOpen
  \bibfield{author}{%
  \bibinfo {author} {\bibnamefont{Girard}, \bibfnamefont{L.}}, \bibinfo
  {author} {\bibfnamefont{S.}~\bibnamefont{Bouillon}}, \bibinfo {author}
  {\bibfnamefont{J.}~\bibnamefont{Weiss}}, \bibinfo {author}
  {\bibfnamefont{D.}~\bibnamefont{Amitrano}}, \bibinfo {author}
  {\bibfnamefont{T.}~\bibnamefont{Fichefet}},\ and\ \bibinfo {author}
  {\bibfnamefont{V.}~\bibnamefont{Legat}}}%
  , \bibinfo {year} {2011},\ \bibfield{title}{%
  \enquote{\bibinfo {title} {A new modelling framework for sea-ice mechanics
  based on elasto-brittle rheology},}\ }%
  \bibfield{journal}{%
  \bibinfo {journal} {Ann. Glac.}\ }%
  \textbf{\bibinfo {volume} {52}},\ \bibinfo {pages} {123--132}%
  \bibAnnoteFile{NoStop}{Girard2011}%
\bibitem[{\citenamefont{Givan}\ \emph{et~al.}(2002)\citenamefont{Givan},
  \citenamefont{Grothe}, \citenamefont{Loewenschuss},\ and\
  \citenamefont{Nielsen}}]{givan2002}%
  \BibitemOpen
  \bibfield{author}{%
  \bibinfo {author} {\bibnamefont{Givan}, \bibfnamefont{A.}}, \bibinfo {author}
  {\bibfnamefont{H.}~\bibnamefont{Grothe}}, \bibinfo {author}
  {\bibfnamefont{A.}~\bibnamefont{Loewenschuss}},\ and\ \bibinfo {author}
  {\bibfnamefont{C.}~\bibnamefont{Nielsen}}}%
  , \bibinfo {year} {2002},\ \bibfield{title}{%
  \enquote{\bibinfo {title} {Infrared spectra and ab initio calculations of
  matrix isolated dimethyl sulfone and its water complex},}\ }%
  \bibfield{journal}{%
  \bibinfo {journal} {Phys. Chem. Chem. Phys.}\ }%
  \textbf{\bibinfo {volume} {4}},\ \bibinfo {pages} {255--263}%
  \bibAnnoteFile{NoStop}{givan2002}%
\bibitem[{\citenamefont{Goedecker}\
  \emph{et~al.}(1996)\citenamefont{Goedecker}, \citenamefont{Teter},\ and\
  \citenamefont{Hutter}}]{goedecker1996}%
  \BibitemOpen
  \bibfield{author}{%
  \bibinfo {author} {\bibnamefont{Goedecker}, \bibfnamefont{S.}}, \bibinfo
  {author} {\bibfnamefont{M.}~\bibnamefont{Teter}},\ and\ \bibinfo {author}
  {\bibfnamefont{J.}~\bibnamefont{Hutter}}}%
  , \bibinfo {year} {1996},\ \bibfield{title}{%
  \enquote{\bibinfo {title} {Separable dual-space gaussian pseudopotentials},}\
  }%
  \bibfield{journal}{%
  \bibinfo {journal} {Phys. Rev. B}\ }%
  \textbf{\bibinfo {volume} {54}},\ \bibinfo {pages} {1703--1710}%
  \bibAnnoteFile{NoStop}{goedecker1996}%
\bibitem[{\citenamefont{{G{\'o}mez}}\
  \emph{et~al.}(2009)\citenamefont{{G{\'o}mez}}, \citenamefont{{G{\'a}lvez}},\
  and\ \citenamefont{{Escribano}}}]{gomez2009}%
  \BibitemOpen
  \bibfield{author}{%
  \bibinfo {author} {\bibnamefont{{G{\'o}mez}}, \bibfnamefont{P.~C.}}, \bibinfo
  {author} {\bibfnamefont{O.}~\bibnamefont{{G{\'a}lvez}}},\ and\ \bibinfo
  {author} {\bibfnamefont{R.}~\bibnamefont{{Escribano}}}}%
  , \bibinfo {year} {2009},\ \bibfield{title}{%
  \enquote{\bibinfo {title} {{Theoretical study of atmospheric clusters:
  HNO$_3$-HCl-H$_2$O}},}\ }%
  \bibfield{journal}{%
  \bibinfo {journal} {Phys. Chem. Chem. Phys.}\ }%
  \textbf{\bibinfo {volume} {11}},\ \bibinfo {pages} {9710--9719}%
  \bibAnnoteFile{NoStop}{gomez2009}%
\bibitem[{\citenamefont{{G{\'o}mez}}\
  \emph{et~al.}(2010)\citenamefont{{G{\'o}mez}}, \citenamefont{{G{\'a}lvez}},
  \citenamefont{{Mosteo}}, \citenamefont{{Puzzarini}},\ and\
  \citenamefont{{Escribano}}}]{gomez2010}%
  \BibitemOpen
  \bibfield{author}{%
  \bibinfo {author} {\bibnamefont{{G{\'o}mez}}, \bibfnamefont{P.~C.}}, \bibinfo
  {author} {\bibfnamefont{O.}~\bibnamefont{{G{\'a}lvez}}}, \bibinfo {author}
  {\bibfnamefont{R.~G.}\ \bibnamefont{{Mosteo}}}, \bibinfo {author}
  {\bibfnamefont{C.}~\bibnamefont{{Puzzarini}}},\ and\ \bibinfo {author}
  {\bibfnamefont{R.}~\bibnamefont{{Escribano}}}}%
  , \bibinfo {year} {2010},\ \bibfield{title}{%
  \enquote{\bibinfo {title} {{Clusters of atmospheric relevance:
  H$_2$O/HCl/HNO$_3$. Prediction of IR \& MW spectra}},}\ }%
  \bibfield{journal}{%
  \bibinfo {journal} {Phys. Chem. Chem. Phys.}\ }%
  \textbf{\bibinfo {volume} {12}},\ \bibinfo {pages} {4617--4624}%
  \bibAnnoteFile{NoStop}{gomez2010}%
\bibitem[{\citenamefont{Gomis}\ \emph{et~al.}(2004)\citenamefont{Gomis},
  \citenamefont{Satorre}, \citenamefont{Strazzulla},\ and\
  \citenamefont{Leto}}]{gomis2004}%
  \BibitemOpen
  \bibfield{author}{%
  \bibinfo {author} {\bibnamefont{Gomis}, \bibfnamefont{O.}}, \bibinfo {author}
  {\bibfnamefont{M.~A.}\ \bibnamefont{Satorre}}, \bibinfo {author}
  {\bibfnamefont{G.}~\bibnamefont{Strazzulla}},\ and\ \bibinfo {author}
  {\bibfnamefont{G.}~\bibnamefont{Leto}}}%
  , \bibinfo {year} {2004},\ \bibfield{title}{%
  \enquote{\bibinfo {title} {Hydrogen peroxide formation by ion implantation in
  water ice and its relevance to the galilean satellites},}\ }%
  \bibfield{journal}{%
  \bibinfo {journal} {Planet. Space Sci.}\ }%
  \textbf{\bibinfo {volume} {52}},\ \bibinfo {pages} {371--378}%
  \bibAnnoteFile{NoStop}{gomis2004}%
\bibitem[{\citenamefont{Gonz{\'a}lez}\
  \emph{et~al.}(2008)\citenamefont{Gonz{\'a}lez}, \citenamefont{Guti{\'e}rrez},
  \citenamefont{Lara},\ and\ \citenamefont{Rodrigo}}]{gonzalez2008}%
  \BibitemOpen
  \bibfield{author}{%
  \bibinfo {author} {\bibnamefont{Gonz{\'a}lez}, \bibfnamefont{M.}}, \bibinfo
  {author} {\bibfnamefont{P.~J.}\ \bibnamefont{Guti{\'e}rrez}}, \bibinfo
  {author} {\bibfnamefont{L.~M.}\ \bibnamefont{Lara}},\ and\ \bibinfo {author}
  {\bibfnamefont{R.}~\bibnamefont{Rodrigo}}}%
  , \bibinfo {year} {2008},\ \bibfield{title}{%
  \enquote{\bibinfo {title} {Evolution of the crystallization front in cometary
  models. {E}ffect of the net energy released during crystallization},}\ }%
  \bibfield{journal}{%
  \bibinfo {journal} {Astron. Astrophys.}\ }%
  \textbf{\bibinfo {volume} {486}},\ \bibinfo {pages} {331--340}%
  \bibAnnoteFile{NoStop}{gonzalez2008}%
\bibitem[{\citenamefont{Graham}\ and\
  \citenamefont{Roberts}(2000)}]{graham2000}%
  \BibitemOpen
  \bibfield{author}{%
  \bibinfo {author} {\bibnamefont{Graham}, \bibfnamefont{J.~D.}},\ and\
  \bibinfo {author} {\bibfnamefont{J.~T.}\ \bibnamefont{Roberts}}}%
  , \bibinfo {year} {2000},\ \bibfield{title}{%
  \enquote{\bibinfo {title} {Chemical reactions of organic molecules adsorbed
  at ice: 2. {Chloride} substitution in 2-methyl-2-propanol},}\ }%
  \bibfield{journal}{%
  \bibinfo {journal} {Langmuir}\ }%
  \textbf{\bibinfo {volume} {16}},\ \bibinfo {pages} {3244--3248}%
  \bibAnnoteFile{NoStop}{graham2000}%
\bibitem[{\citenamefont{Grannas}\
  \emph{et~al.}(2007{\natexlab{a}})\citenamefont{Grannas},
  \citenamefont{Bausch},\ and\ \citenamefont{Mahanna}}]{Grannas:2007p25490}%
  \BibitemOpen
  \bibfield{author}{%
  \bibinfo {author} {\bibnamefont{Grannas}, \bibfnamefont{A.}}, \bibinfo
  {author} {\bibfnamefont{A.}~\bibnamefont{Bausch}},\ and\ \bibinfo {author}
  {\bibfnamefont{K.}~\bibnamefont{Mahanna}}}%
  , \bibinfo {year} {2007}{\natexlab{a}},\ \bibfield{title}{%
  \enquote{\bibinfo {title} {Enhanced aqueous photochemical reaction rates
  after freezing},}\ }%
  \bibfield{journal}{%
  \bibinfo {journal} {J. Phys. Chem. A}\ }%
  \textbf{\bibinfo {volume} {111}},\ \bibinfo {pages} {11043--11049}%
  \bibAnnoteFile{NoStop}{Grannas:2007p25490}%
\bibitem[{\citenamefont{Grannas}\
  \emph{et~al.}(2007{\natexlab{b}})\citenamefont{Grannas},
  \citenamefont{Jones}, \citenamefont{Dibb}, \citenamefont{Ammann},
  \citenamefont{Anastasio}, \citenamefont{Beine}, \citenamefont{Bergin},
  \citenamefont{Bottenheim}, \citenamefont{Boxe}, \citenamefont{Carver},
  \citenamefont{Chen}, \citenamefont{Crawford}, \citenamefont{Domin{\'e}},
  \citenamefont{Frey}, \citenamefont{Guzman}, \citenamefont{Heard},
  \citenamefont{Helmig}, \citenamefont{Hoffmann}, \citenamefont{Honrath},
  \citenamefont{Huey}, \citenamefont{Hutterli}, \citenamefont{Jacobi},
  \citenamefont{Kl{\'a}n}, \citenamefont{Lefer}, \citenamefont{McConnell},
  \citenamefont{Plane}, \citenamefont{Sander}, \citenamefont{Savarino},
  \citenamefont{Shepson}, \citenamefont{Simpson}, \citenamefont{Sodeau},
  \citenamefont{Glasow}, \citenamefont{Weller}, \citenamefont{Wolff},\ and\
  \citenamefont{Zhu}}]{Grannas:2007p2776}%
  \BibitemOpen
  \bibfield{author}{%
  \bibinfo {author} {\bibnamefont{Grannas}, \bibfnamefont{A.~M.}}, \bibinfo
  {author} {\bibfnamefont{A.~E.}\ \bibnamefont{Jones}}, \bibinfo {author}
  {\bibfnamefont{J.}~\bibnamefont{Dibb}}, \bibinfo {author}
  {\bibfnamefont{M.}~\bibnamefont{Ammann}}, \bibinfo {author}
  {\bibfnamefont{C.}~\bibnamefont{Anastasio}}, \bibinfo {author}
  {\bibfnamefont{H.}~\bibnamefont{Beine}}, \bibinfo {author}
  {\bibfnamefont{M.}~\bibnamefont{Bergin}}, \bibinfo {author}
  {\bibfnamefont{J.}~\bibnamefont{Bottenheim}}, \bibinfo {author}
  {\bibfnamefont{C.~S.}\ \bibnamefont{Boxe}}, \bibinfo {author}
  {\bibfnamefont{G.}~\bibnamefont{Carver}}, \bibinfo {author}
  {\bibfnamefont{G.}~\bibnamefont{Chen}}, \bibinfo {author}
  {\bibfnamefont{J.~H.}\ \bibnamefont{Crawford}}, \bibinfo {author}
  {\bibfnamefont{F.}~\bibnamefont{Domin{\'e}}}, \bibinfo {author}
  {\bibfnamefont{M.~M.}\ \bibnamefont{Frey}}, \bibinfo {author}
  {\bibfnamefont{M.~I.}\ \bibnamefont{Guzman}}, \bibinfo {author}
  {\bibfnamefont{D.~E.}\ \bibnamefont{Heard}}, \bibinfo {author}
  {\bibfnamefont{D.}~\bibnamefont{Helmig}}, \bibinfo {author}
  {\bibfnamefont{M.~R.}\ \bibnamefont{Hoffmann}}, \bibinfo {author}
  {\bibfnamefont{R.}~\bibnamefont{Honrath}}, \bibinfo {author}
  {\bibfnamefont{L.~G.}\ \bibnamefont{Huey}}, \bibinfo {author}
  {\bibfnamefont{M.}~\bibnamefont{Hutterli}}, \bibinfo {author}
  {\bibfnamefont{H.-W.}\ \bibnamefont{Jacobi}}, \bibinfo {author}
  {\bibfnamefont{P.}~\bibnamefont{Kl{\'a}n}}, \bibinfo {author}
  {\bibfnamefont{B.}~\bibnamefont{Lefer}}, \bibinfo {author}
  {\bibfnamefont{J.}~\bibnamefont{McConnell}}, \bibinfo {author}
  {\bibfnamefont{J.}~\bibnamefont{Plane}}, \bibinfo {author}
  {\bibfnamefont{R.}~\bibnamefont{Sander}}, \bibinfo {author}
  {\bibfnamefont{J.}~\bibnamefont{Savarino}}, \bibinfo {author}
  {\bibfnamefont{P.~B.}\ \bibnamefont{Shepson}}, \bibinfo {author}
  {\bibfnamefont{W.~R.}\ \bibnamefont{Simpson}}, \bibinfo {author}
  {\bibfnamefont{J.~R.}\ \bibnamefont{Sodeau}}, \bibinfo {author}
  {\bibfnamefont{R.~Von}\ \bibnamefont{Glasow}}, \bibinfo {author}
  {\bibfnamefont{R.}~\bibnamefont{Weller}}, \bibinfo {author}
  {\bibfnamefont{E.~W.}\ \bibnamefont{Wolff}},\ and\ \bibinfo {author}
  {\bibfnamefont{T.}~\bibnamefont{Zhu}}}%
  , \bibinfo {year} {2007}{\natexlab{b}},\ \bibfield{title}{%
  \enquote{\bibinfo {title} {An overview of snow photochemistry: {E}vidence,
  mechanisms and impacts},}\ }%
  \bibfield{journal}{%
  \bibinfo {journal} {Atmos. Chem. Phys.}\ }%
  \textbf{\bibinfo {volume} {7}},\ \bibinfo {pages} {4329--4373}%
  \bibAnnoteFile{NoStop}{Grannas:2007p2776}%
\bibitem[{\citenamefont{Grannas}\ \emph{et~al.}(2004)\citenamefont{Grannas},
  \citenamefont{Shepson},\ and\ \citenamefont{Filley}}]{Grannas:2004p3348}%
  \BibitemOpen
  \bibfield{author}{%
  \bibinfo {author} {\bibnamefont{Grannas}, \bibfnamefont{A.~M.}}, \bibinfo
  {author} {\bibfnamefont{P.~B.}\ \bibnamefont{Shepson}},\ and\ \bibinfo
  {author} {\bibfnamefont{T.~R.}\ \bibnamefont{Filley}}}%
  , \bibinfo {year} {2004},\ \bibfield{title}{%
  \enquote{\bibinfo {title} {Photochemistry and nature of organic matter in
  {Arctic} and {Antarctic} snow},}\ }%
  \bibfield{journal}{%
  \bibinfo {journal} {Global Biogeochem. Cycles}\ }%
  \textbf{\bibinfo {volume} {18}},\ \bibinfo {pages} {GB1006}%
  \bibAnnoteFile{NoStop}{Grannas:2004p3348}%
\bibitem[{\citenamefont{Graversen}\
  \emph{et~al.}(2008)\citenamefont{Graversen}, \citenamefont{Mauritsen},
  \citenamefont{Tjernstr\"om}, \citenamefont{K\"allen},\ and\
  \citenamefont{Svensson}}]{Graversen:2008}%
  \BibitemOpen
  \bibfield{author}{%
  \bibinfo {author} {\bibnamefont{Graversen}, \bibfnamefont{R.~G.}}, \bibinfo
  {author} {\bibfnamefont{T.}~\bibnamefont{Mauritsen}}, \bibinfo {author}
  {\bibfnamefont{M.}~\bibnamefont{Tjernstr\"om}}, \bibinfo {author}
  {\bibfnamefont{E.}~\bibnamefont{K\"allen}},\ and\ \bibinfo {author}
  {\bibfnamefont{G.}~\bibnamefont{Svensson}}}%
  , \bibinfo {year} {2008},\ \bibfield{title}{%
  \enquote{\bibinfo {title} {Vertical structure of recent {Arctic} warming},}\
  }%
  \bibfield{journal}{%
  \bibinfo {journal} {Nature}\ }%
  \textbf{\bibinfo {volume} {451}},\ \bibinfo {pages} {53--56}%
  \bibAnnoteFile{NoStop}{Graversen:2008}%
\bibitem[{\citenamefont{Greve}(2006)}]{greve2006}%
  \BibitemOpen
  \bibfield{author}{%
  \bibinfo {author} {\bibnamefont{Greve}, \bibfnamefont{R.}}}%
  , \bibinfo {year} {2006},\ \bibfield{title}{%
  \enquote{\bibinfo {title} {Fluid dynamics of planetary ices},}\ }%
  \bibfield{journal}{%
  \bibinfo {journal} {GAMM-Mitteilungen}\ }%
  \textbf{\bibinfo {volume} {29}},\ \bibinfo {pages} {29--51}%
  \bibAnnoteFile{NoStop}{greve2006}%
\bibitem[{\citenamefont{Grothe}(2008)}]{grothe2008}%
  \BibitemOpen
  \bibfield{author}{%
  \bibinfo {author} {\bibnamefont{Grothe}, \bibfnamefont{H.}}}%
  , \bibinfo {year} {2008},\ \bibfield{title}{%
  \enquote{\bibinfo {title} {Interactive comment on ``inhibition of ice
  crystallization in highly viscous aqueous organic acid droplets'' by b. j.
  murray},}\ }%
  \bibfield{journal}{%
  \bibinfo {journal} {Atmos. Chem. Phys. Discuss.}\ }%
  \textbf{\bibinfo {volume} {8}},\ \bibinfo {pages} {3992--3995}%
  \bibAnnoteFile{NoStop}{grothe2008}%
\bibitem[{\citenamefont{Grothe}\
  \emph{et~al.}(2006{\natexlab{a}})\citenamefont{Grothe}, \citenamefont{{Lund
  Myhre}},\ and\ \citenamefont{Nielsen}}]{grothe2006b}%
  \BibitemOpen
  \bibfield{author}{%
  \bibinfo {author} {\bibnamefont{Grothe}, \bibfnamefont{H.}}, \bibinfo
  {author} {\bibfnamefont{C.~E.}\ \bibnamefont{{Lund Myhre}}},\ and\ \bibinfo
  {author} {\bibfnamefont{C.~J.}\ \bibnamefont{Nielsen}}}%
  , \bibinfo {year} {2006}{\natexlab{a}},\ \bibfield{title}{%
  \enquote{\bibinfo {title} {Low-frequency {Raman} spectra of nitric acid
  hydrates},}\ }%
  \bibfield{journal}{%
  \bibinfo {journal} {J. Phys. Chem. A}\ }%
  \textbf{\bibinfo {volume} {110}},\ \bibinfo {pages} {171--176}%
  \bibAnnoteFile{NoStop}{grothe2006b}%
\bibitem[{\citenamefont{Grothe}\
  \emph{et~al.}(2006{\natexlab{b}})\citenamefont{Grothe}, \citenamefont{Tizek},
  \citenamefont{Waller},\ and\ \citenamefont{Stokes}}]{grothe2006a}%
  \BibitemOpen
  \bibfield{author}{%
  \bibinfo {author} {\bibnamefont{Grothe}, \bibfnamefont{H.}}, \bibinfo
  {author} {\bibfnamefont{H.}~\bibnamefont{Tizek}}, \bibinfo {author}
  {\bibfnamefont{D.}~\bibnamefont{Waller}},\ and\ \bibinfo {author}
  {\bibfnamefont{D.~J.}\ \bibnamefont{Stokes}}}%
  , \bibinfo {year} {2006}{\natexlab{b}},\ \bibfield{title}{%
  \enquote{\bibinfo {title} {The crystallization kinetics and morphology of
  nitric acid trihydrate},}\ }%
  \bibfield{journal}{%
  \bibinfo {journal} {Phys. Chem. Chem. Phys.}\ }%
  \textbf{\bibinfo {volume} {8}},\ \bibinfo {pages} {2232--2239}%
  \bibAnnoteFile{NoStop}{grothe2006a}%
\bibitem[{\citenamefont{Guillot}\ and\
  \citenamefont{Guissani}(2003)}]{guillot2003}%
  \BibitemOpen
  \bibfield{author}{%
  \bibinfo {author} {\bibnamefont{Guillot}, \bibfnamefont{B.}},\ and\ \bibinfo
  {author} {\bibfnamefont{Y.}~\bibnamefont{Guissani}}}%
  , \bibinfo {year} {2003},\ \bibfield{title}{%
  \enquote{\bibinfo {title} {Polyamorphism in low temperature water: A
  simulation study},}\ }%
  \bibfield{journal}{%
  \bibinfo {journal} {J. Chem. Phys.}\ }%
  \textbf{\bibinfo {volume} {119}},\ \bibinfo {pages} {11740--11752}%
  \bibAnnoteFile{NoStop}{guillot2003}%
\bibitem[{\citenamefont{Gumbel}\ and\
  \citenamefont{Megner}(2009)}]{Gumbel2009}%
  \BibitemOpen
  \bibfield{author}{%
  \bibinfo {author} {\bibnamefont{Gumbel}, \bibfnamefont{J.}},\ and\ \bibinfo
  {author} {\bibfnamefont{L.}~\bibnamefont{Megner}}}%
  , \bibinfo {year} {2009},\ \bibfield{title}{%
  \enquote{\bibinfo {title} {Charged meteoric smoke as ice nuclei in the
  mesosphere: Part 1-a review of basic concepts},}\ }%
  \bibfield{journal}{%
  \bibinfo {journal} {J. Atmos. Solar-Terrestrial Phys.}\ }%
  \textbf{\bibinfo {volume} {71}},\ \bibinfo {pages} {1225--1235}%
  \bibAnnoteFile{NoStop}{Gumbel2009}%
\bibitem[{\citenamefont{Gumbel}\ and\ \citenamefont{Witt}(1998)}]{Gumbel1998}%
  \BibitemOpen
  \bibfield{author}{%
  \bibinfo {author} {\bibnamefont{Gumbel}, \bibfnamefont{J.}},\ and\ \bibinfo
  {author} {\bibfnamefont{G.}~\bibnamefont{Witt}}}%
  , \bibinfo {year} {1998},\ \bibfield{title}{%
  \enquote{\bibinfo {title} {In situ measurements of the vertical structure of
  a noctilucent cloud},}\ }%
  \bibfield{journal}{%
  \bibinfo {journal} {Geophys. Res. Lett.}\ }%
  \textbf{\bibinfo {volume} {25}},\ \bibinfo {pages} {493--496}%
  \bibAnnoteFile{NoStop}{Gumbel1998}%
\bibitem[{\citenamefont{Gurganus}\ \emph{et~al.}(2011)\citenamefont{Gurganus},
  \citenamefont{Kostinski},\ and\ \citenamefont{Shaw}}]{Gurganus2011}%
  \BibitemOpen
  \bibfield{author}{%
  \bibinfo {author} {\bibnamefont{Gurganus}, \bibfnamefont{C.}}, \bibinfo
  {author} {\bibfnamefont{A.~B.}\ \bibnamefont{Kostinski}},\ and\ \bibinfo
  {author} {\bibfnamefont{R.~A.}\ \bibnamefont{Shaw}}}%
  , \bibinfo {year} {2011},\ \bibfield{title}{%
  \enquote{\bibinfo {title} {Fast imaging of freezing drops: No preference for
  nucleation at the contact line},}\ }%
  \bibfield{journal}{%
  \bibinfo {journal} {J. Phys. Chem. Lett.}\ }%
  \textbf{\bibinfo {volume} {2}},\ \bibinfo {pages} {1449--1454}%
  \bibAnnoteFile{NoStop}{Gurganus2011}%
\bibitem[{\citenamefont{Haapala}\ \emph{et~al.}(2005)\citenamefont{Haapala},
  \citenamefont{L\"onnroth},\ and\ \citenamefont{St\"ossel}}]{Haapala:2005}%
  \BibitemOpen
  \bibfield{author}{%
  \bibinfo {author} {\bibnamefont{Haapala}, \bibfnamefont{J.}}, \bibinfo
  {author} {\bibfnamefont{N.}~\bibnamefont{L\"onnroth}},\ and\ \bibinfo
  {author} {\bibfnamefont{A.}~\bibnamefont{St\"ossel}}}%
  , \bibinfo {year} {2005},\ \bibfield{title}{%
  \enquote{\bibinfo {title} {A numerical study of open water formation in sea
  ice},}\ }%
  \bibfield{journal}{%
  \bibinfo {journal} {J.~Geophys.~Res.}\ }%
  \textbf{\bibinfo {volume} {110}},\ \bibinfo {pages} {C09011}%
  \bibAnnoteFile{NoStop}{Haapala:2005}%
\bibitem[{\citenamefont{Haas}\ \emph{et~al.}(2008)\citenamefont{Haas},
  \citenamefont{Pfaffling}, \citenamefont{Hendricks},
  \citenamefont{Rabenstein}, \citenamefont{Etienne},\ and\
  \citenamefont{Rigor}}]{Haas:2008}%
  \BibitemOpen
  \bibfield{author}{%
  \bibinfo {author} {\bibnamefont{Haas}, \bibfnamefont{C.}}, \bibinfo {author}
  {\bibfnamefont{A.}~\bibnamefont{Pfaffling}}, \bibinfo {author}
  {\bibfnamefont{S.}~\bibnamefont{Hendricks}}, \bibinfo {author}
  {\bibfnamefont{L.}~\bibnamefont{Rabenstein}}, \bibinfo {author}
  {\bibfnamefont{J.~L.}\ \bibnamefont{Etienne}},\ and\ \bibinfo {author}
  {\bibfnamefont{I.}~\bibnamefont{Rigor}}}%
  , \bibinfo {year} {2008},\ \bibfield{title}{%
  \enquote{\bibinfo {title} {Reduced ice thickness in {Arctic} transpolar drift
  favors rapid ice retreat},}\ }%
  \bibfield{journal}{%
  \bibinfo {journal} {Geophys.~Res.~Lett.}\ }%
  \textbf{\bibinfo {volume} {35}},\ \bibinfo {pages} {L17501}%
  \bibAnnoteFile{NoStop}{Haas:2008}%
\bibitem[{\citenamefont{Haeberli}\ \emph{et~al.}(2006)\citenamefont{Haeberli},
  \citenamefont{Hallet}, \citenamefont{Arenson}, \citenamefont{Elconin},
  \citenamefont{Humlum}, \citenamefont{K{\"a}{\"a}b}, \citenamefont{Kaufmann},
  \citenamefont{Ladanyi}, \citenamefont{Matsuoka}, \citenamefont{Springman},\
  and\ \citenamefont{M{\"u}hll}}]{haeberli2006}%
  \BibitemOpen
  \bibfield{author}{%
  \bibinfo {author} {\bibnamefont{Haeberli}, \bibfnamefont{W.}}, \bibinfo
  {author} {\bibfnamefont{B.}~\bibnamefont{Hallet}}, \bibinfo {author}
  {\bibfnamefont{L.}~\bibnamefont{Arenson}}, \bibinfo {author}
  {\bibfnamefont{R.}~\bibnamefont{Elconin}}, \bibinfo {author}
  {\bibfnamefont{O.}~\bibnamefont{Humlum}}, \bibinfo {author}
  {\bibfnamefont{A.}~\bibnamefont{K{\"a}{\"a}b}}, \bibinfo {author}
  {\bibfnamefont{V.}~\bibnamefont{Kaufmann}}, \bibinfo {author}
  {\bibfnamefont{B.}~\bibnamefont{Ladanyi}}, \bibinfo {author}
  {\bibfnamefont{N.}~\bibnamefont{Matsuoka}}, \bibinfo {author}
  {\bibfnamefont{S.}~\bibnamefont{Springman}},\ and\ \bibinfo {author}
  {\bibfnamefont{D.~V.}\ \bibnamefont{M{\"u}hll}}}%
  , \bibinfo {year} {2006},\ \bibfield{title}{%
  \enquote{\bibinfo {title} {Permafrost creep and rock glacier dynamics},}\ }%
  \bibfield{journal}{%
  \bibinfo {journal} {Permafrost Periglac. Process.}\ }%
  \textbf{\bibinfo {volume} {17}},\ \bibinfo {pages} {189--214}%
  \bibAnnoteFile{NoStop}{haeberli2006}%
\bibitem[{\citenamefont{Hage}\ \emph{et~al.}(1995)\citenamefont{Hage},
  \citenamefont{Hallbrucker}, \citenamefont{Mayer},\ and\
  \citenamefont{Johari}}]{hage1995}%
  \BibitemOpen
  \bibfield{author}{%
  \bibinfo {author} {\bibnamefont{Hage}, \bibfnamefont{W.}}, \bibinfo {author}
  {\bibfnamefont{A.}~\bibnamefont{Hallbrucker}}, \bibinfo {author}
  {\bibfnamefont{E.}~\bibnamefont{Mayer}},\ and\ \bibinfo {author}
  {\bibfnamefont{G.~P.}\ \bibnamefont{Johari}}}%
  , \bibinfo {year} {1995},\ \bibfield{title}{%
  \enquote{\bibinfo {title} {Kinetics of crystallizing {D$_2$O} water near
  150~{K} by {Fourier} transform infrared spectroscopy and a comparison with
  the corresponding calorimetric studies on {H$_2$O} water},}\ }%
  \bibfield{journal}{%
  \bibinfo {journal} {J. Chem. Phys.}\ }%
  \textbf{\bibinfo {volume} {103}},\ \bibinfo {pages} {545--550}%
  \bibAnnoteFile{NoStop}{hage1995}%
\bibitem[{\citenamefont{Hage}\ \emph{et~al.}(1994)\citenamefont{Hage},
  \citenamefont{Hallbrucker}, \citenamefont{Mayer},\ and\
  \citenamefont{P.Johari}}]{hage1994}%
  \BibitemOpen
  \bibfield{author}{%
  \bibinfo {author} {\bibnamefont{Hage}, \bibfnamefont{W.}}, \bibinfo {author}
  {\bibfnamefont{A.}~\bibnamefont{Hallbrucker}}, \bibinfo {author}
  {\bibfnamefont{E.}~\bibnamefont{Mayer}},\ and\ \bibinfo {author}
  {\bibfnamefont{G.}~\bibnamefont{P.Johari}}}%
  , \bibinfo {year} {1994},\ \bibfield{title}{%
  \enquote{\bibinfo {title} {Crystallization kinetics of water below
  150~{K}},}\ }%
  \bibfield{journal}{%
  \bibinfo {journal} {J. Chem. Phys.}\ }%
  \textbf{\bibinfo {volume} {100}},\ \bibinfo {pages} {2743--2747}%
  \bibAnnoteFile{NoStop}{hage1994}%
\bibitem[{\citenamefont{Hale}\ and\ \citenamefont{Kathmann}(1996)}]{hale1996}%
  \BibitemOpen
  \bibfield{author}{%
  \bibinfo {author} {\bibnamefont{Hale}, \bibfnamefont{B.~N.}},\ and\ \bibinfo
  {author} {\bibfnamefont{S.~M.}\ \bibnamefont{Kathmann}}}%
  , \bibinfo {year} {1996},\ \enquote{\bibinfo {title} {Monte {C}arlo
  simulations of small {H$_2$SO$_4$--H$_2$O} clusters},}\ in\ \emph{\bibinfo
  {booktitle} {Nucleation and Atmospheric Aerosols. Proceedings of the 14th
  International Conference}}\ (\bibinfo {publisher} {Pergamon, Oxford})%
  \bibAnnoteFile{NoStop}{hale1996}%
\bibitem[{\citenamefont{Hallquist}\
  \emph{et~al.}(2009)\citenamefont{Hallquist}, \citenamefont{Wenger},
  \citenamefont{Baltensperger}, \citenamefont{Rudich}, \citenamefont{Simpson},
  \citenamefont{Claeys}, \citenamefont{Dommen}, \citenamefont{Donahue},
  \citenamefont{George}, \citenamefont{Goldstein}, \citenamefont{Hamilton},
  \citenamefont{Herrmann}, \citenamefont{Hoffmann}, \citenamefont{Iinuma},
  \citenamefont{Jang}, \citenamefont{Jenkin}, \citenamefont{Jimenez},
  \citenamefont{Kiendler-Scharr}, \citenamefont{Maenhaut},
  \citenamefont{McFiggans}, \citenamefont{Mentel}, \citenamefont{Monod},
  \citenamefont{Prevot}, \citenamefont{Seinfeld}, \citenamefont{Surratt},
  \citenamefont{Szmigielski},\ and\ \citenamefont{Wildt}}]{Hallquist2009}%
  \BibitemOpen
  \bibfield{author}{%
  \bibinfo {author} {\bibnamefont{Hallquist}, \bibfnamefont{M.}}, \bibinfo
  {author} {\bibfnamefont{J.~C.}\ \bibnamefont{Wenger}}, \bibinfo {author}
  {\bibfnamefont{U.}~\bibnamefont{Baltensperger}}, \bibinfo {author}
  {\bibfnamefont{Y.}~\bibnamefont{Rudich}}, \bibinfo {author}
  {\bibfnamefont{D.}~\bibnamefont{Simpson}}, \bibinfo {author}
  {\bibfnamefont{M.}~\bibnamefont{Claeys}}, \bibinfo {author}
  {\bibfnamefont{J.}~\bibnamefont{Dommen}}, \bibinfo {author}
  {\bibfnamefont{N.~M.}\ \bibnamefont{Donahue}}, \bibinfo {author}
  {\bibfnamefont{C.}~\bibnamefont{George}}, \bibinfo {author}
  {\bibfnamefont{A.~H.}\ \bibnamefont{Goldstein}}, \bibinfo {author}
  {\bibfnamefont{J.~F.}\ \bibnamefont{Hamilton}}, \bibinfo {author}
  {\bibfnamefont{H.}~\bibnamefont{Herrmann}}, \bibinfo {author}
  {\bibfnamefont{T.}~\bibnamefont{Hoffmann}}, \bibinfo {author}
  {\bibfnamefont{Y.}~\bibnamefont{Iinuma}}, \bibinfo {author}
  {\bibfnamefont{M.}~\bibnamefont{Jang}}, \bibinfo {author}
  {\bibfnamefont{M.~E.}\ \bibnamefont{Jenkin}}, \bibinfo {author}
  {\bibfnamefont{J.~L.}\ \bibnamefont{Jimenez}}, \bibinfo {author}
  {\bibfnamefont{A.}~\bibnamefont{Kiendler-Scharr}}, \bibinfo {author}
  {\bibfnamefont{W.}~\bibnamefont{Maenhaut}}, \bibinfo {author}
  {\bibfnamefont{G.}~\bibnamefont{McFiggans}}, \bibinfo {author}
  {\bibfnamefont{Th.~F.}\ \bibnamefont{Mentel}}, \bibinfo {author}
  {\bibfnamefont{A.}~\bibnamefont{Monod}}, \bibinfo {author}
  {\bibfnamefont{A.~S.~H.}\ \bibnamefont{Prevot}}, \bibinfo {author}
  {\bibfnamefont{J.~H.}\ \bibnamefont{Seinfeld}}, \bibinfo {author}
  {\bibfnamefont{J.~D.}\ \bibnamefont{Surratt}}, \bibinfo {author}
  {\bibfnamefont{R.}~\bibnamefont{Szmigielski}},\ and\ \bibinfo {author}
  {\bibfnamefont{J.}~\bibnamefont{Wildt}}}%
  , \bibinfo {year} {2009},\ \bibfield{title}{%
  \enquote{\bibinfo {title} {The formation, properties and impact of secondary
  organic aerosol: current and emerging issues},}\ }%
  \bibfield{journal}{%
  \bibinfo {journal} {Atmos. Chem. Phys.}\ }%
  \textbf{\bibinfo {volume} {9}},\ \bibinfo {pages} {5155--5236}%
  \bibAnnoteFile{NoStop}{Hallquist2009}%
\bibitem[{\citenamefont{{Hancock}}\ and\
  \citenamefont{{Povenmire}}(2010)}]{hancock2010}%
  \BibitemOpen
  \bibfield{author}{%
  \bibinfo {author} {\bibnamefont{{Hancock}}, \bibfnamefont{L.~O.}},\ and\
  \bibinfo {author} {\bibfnamefont{H.}~\bibnamefont{{Povenmire}}}}%
  , \bibinfo {year} {2010},\ \bibfield{title}{%
  \enquote{\bibinfo {title} {Earth: A ringed planet?}.}\ }%
  \bibinfo {journal} {AGU Fall Meeting Abstracts},\ \bibinfo {pages} {B1635+}%
  \bibAnnoteFile{NoStop}{hancock2010}%
\bibitem[{\citenamefont{Handa}\ \emph{et~al.}(1986)\citenamefont{Handa},
  \citenamefont{Klug},\ and\ \citenamefont{Whalley}}]{handa1986}%
  \BibitemOpen
\bibfield{journal}{%
    }%
  \bibfield{author}{%
  \bibinfo {author} {\bibnamefont{Handa}, \bibfnamefont{Y.~P.}}, \bibinfo
  {author} {\bibfnamefont{D.~D.}\ \bibnamefont{Klug}},\ and\ \bibinfo {author}
  {\bibfnamefont{E.}~\bibnamefont{Whalley}}}%
  , \bibinfo {year} {1986},\ \bibfield{title}{%
  \enquote{\bibinfo {title} {Difference in energy between cubic and hexagonal
  ice},}\ }%
  \bibfield{journal}{%
  \bibinfo {journal} {J. Chem. Phys.}\ }%
  \textbf{\bibinfo {volume} {84}},\ \bibinfo {pages} {7009--7010}%
  \bibAnnoteFile{NoStop}{handa1986}%
\bibitem[{\citenamefont{Handa}\ \emph{et~al.}(1987)\citenamefont{Handa},
  \citenamefont{Klug},\ and\ \citenamefont{Whalley}}]{handa1987}%
  \BibitemOpen
  \bibfield{author}{%
  \bibinfo {author} {\bibnamefont{Handa}, \bibfnamefont{Y.~P.}}, \bibinfo
  {author} {\bibfnamefont{D.~D.}\ \bibnamefont{Klug}},\ and\ \bibinfo {author}
  {\bibfnamefont{E.}~\bibnamefont{Whalley}}}%
  , \bibinfo {year} {1987},\ \bibfield{title}{%
  \enquote{\bibinfo {title} {Phase transitions of ice {V} and {VI}},}\ }%
  \bibfield{journal}{%
  \bibinfo {journal} {J. Phys. Colloq.}\ }%
  \textbf{\bibinfo {volume} {48}},\ \bibinfo {pages} {435--440}%
  \bibAnnoteFile{NoStop}{handa1987}%
\bibitem[{\citenamefont{Hansen}\ \emph{et~al.}(2007)\citenamefont{Hansen},
  \citenamefont{Sato}, \citenamefont{Kharecha}, \citenamefont{Russell},
  \citenamefont{Lea},\ and\ \citenamefont{Siddall}}]{Hansen:2007p24301}%
  \BibitemOpen
  \bibfield{author}{%
  \bibinfo {author} {\bibnamefont{Hansen}, \bibfnamefont{J.}}, \bibinfo
  {author} {\bibfnamefont{M.}~\bibnamefont{Sato}}, \bibinfo {author}
  {\bibfnamefont{P.}~\bibnamefont{Kharecha}}, \bibinfo {author}
  {\bibfnamefont{G.}~\bibnamefont{Russell}}, \bibinfo {author}
  {\bibfnamefont{D.}~\bibnamefont{Lea}},\ and\ \bibinfo {author}
  {\bibfnamefont{M.}~\bibnamefont{Siddall}}}%
  , \bibinfo {year} {2007},\ \bibfield{title}{%
  \enquote{\bibinfo {title} {Climate change and trace gases},}\ }%
  \bibfield{journal}{%
  \bibinfo {journal} {Phil. Trans. Roy. Soc. A}\ }%
  \textbf{\bibinfo {volume} {365}},\ \bibinfo {pages} {1925--1954}%
  \bibAnnoteFile{NoStop}{Hansen:2007p24301}%
\bibitem[{\citenamefont{Hansen}\
  \emph{et~al.}(2008{\natexlab{a}})\citenamefont{Hansen}, \citenamefont{Koza},\
  and\ \citenamefont{Kuhs}}]{hansen2008_1}%
  \BibitemOpen
  \bibfield{author}{%
  \bibinfo {author} {\bibnamefont{Hansen}, \bibfnamefont{T.~C.}}, \bibinfo
  {author} {\bibfnamefont{M.~M.}\ \bibnamefont{Koza}},\ and\ \bibinfo {author}
  {\bibfnamefont{W.~F.}\ \bibnamefont{Kuhs}}}%
  , \bibinfo {year} {2008}{\natexlab{a}},\ \bibfield{title}{%
  \enquote{\bibinfo {title} {Formation and annealing of cubic ice: {I}.
  {M}odelling of stacking faults},}\ }%
  \bibfield{journal}{%
  \bibinfo {journal} {J. Phys. Cond. Matt.}\ }%
  \textbf{\bibinfo {volume} {20}},\ \bibinfo {pages} {285104}%
  \bibAnnoteFile{NoStop}{hansen2008_1}%
\bibitem[{\citenamefont{Hansen}\
  \emph{et~al.}(2008{\natexlab{b}})\citenamefont{Hansen}, \citenamefont{Koza},
  \citenamefont{Lindner},\ and\ \citenamefont{Kuhs}}]{hansen2008_2}%
  \BibitemOpen
  \bibfield{author}{%
  \bibinfo {author} {\bibnamefont{Hansen}, \bibfnamefont{T.~C.}}, \bibinfo
  {author} {\bibfnamefont{M.~M.}\ \bibnamefont{Koza}}, \bibinfo {author}
  {\bibfnamefont{P.}~\bibnamefont{Lindner}},\ and\ \bibinfo {author}
  {\bibfnamefont{W.~F.}\ \bibnamefont{Kuhs}}}%
  , \bibinfo {year} {2008}{\natexlab{b}},\ \bibfield{title}{%
  \enquote{\bibinfo {title} {Formation and annealing of cubic ice: {II}.
  {K}inetic study},}\ }%
  \bibfield{journal}{%
  \bibinfo {journal} {J. Phys. Cond. Matt.}\ }%
  \textbf{\bibinfo {volume} {20}},\ \bibinfo {pages} {285105}%
  \bibAnnoteFile{NoStop}{hansen2008_2}%
\bibitem[{\citenamefont{Hanson}\ and\
  \citenamefont{Ravishankara}(1991)}]{Hanson1991}%
  \BibitemOpen
  \bibfield{author}{%
  \bibinfo {author} {\bibnamefont{Hanson}, \bibfnamefont{D.~R.}},\ and\
  \bibinfo {author} {\bibfnamefont{A.~R.}\ \bibnamefont{Ravishankara}}}%
  , \bibinfo {year} {1991},\ \bibfield{title}{%
  \enquote{\bibinfo {title} {The reaction probabilities of {CLONO$_2$} and
  {N$_2$O$_5$} on polar stratospheric cloud materials},}\ }%
  \bibfield{journal}{%
  \bibinfo {journal} {J. Geophys. Res.-Atmos.}\ }%
  \textbf{\bibinfo {volume} {96}},\ \bibinfo {pages} {5081--5090}%
  \bibAnnoteFile{NoStop}{Hanson1991}%
\bibitem[{\citenamefont{Hanson}\ and\
  \citenamefont{Ravishankara}(1992)}]{Hanson1992}%
  \BibitemOpen
  \bibfield{author}{%
  \bibinfo {author} {\bibnamefont{Hanson}, \bibfnamefont{D.~R.}},\ and\
  \bibinfo {author} {\bibfnamefont{A.~R.}\ \bibnamefont{Ravishankara}}}%
  , \bibinfo {year} {1992},\ \bibfield{title}{%
  \enquote{\bibinfo {title} {Investigation of the reactive and nonreactive
  processes involving {ClONO$_2$} and {HCl} on water and nitric-acid doped
  ice},}\ }%
  \bibfield{journal}{%
  \bibinfo {journal} {J. Phys. Chem.}\ }%
  \textbf{\bibinfo {volume} {96}},\ \bibinfo {pages} {2682--2691}%
  \bibAnnoteFile{NoStop}{Hanson1992}%
\bibitem[{\citenamefont{Hantal}\ \emph{et~al.}(2008)\citenamefont{Hantal},
  \citenamefont{Jedlovszky}, \citenamefont{Hoang},\ and\
  \citenamefont{Picaud}}]{Hantal:2008p25890}%
  \BibitemOpen
  \bibfield{author}{%
  \bibinfo {author} {\bibnamefont{Hantal}, \bibfnamefont{G.}}, \bibinfo
  {author} {\bibfnamefont{P.}~\bibnamefont{Jedlovszky}}, \bibinfo {author}
  {\bibfnamefont{P.~N.~M}\ \bibnamefont{Hoang}},\ and\ \bibinfo {author}
  {\bibfnamefont{S.}~\bibnamefont{Picaud}}}%
  , \bibinfo {year} {2008},\ \bibfield{title}{%
  \enquote{\bibinfo {title} {Investigation of the adsorption behaviour of
  acetone at the surface of ice. {A} grand canonical {Monte Carlo} simulation
  study},}\ }%
  \bibfield{journal}{%
  \bibinfo {journal} {Phys. Chem. Chem. Phys.}\ }%
  \textbf{\bibinfo {volume} {10}},\ \bibinfo {pages} {6369--6380}%
  \bibAnnoteFile{NoStop}{Hantal:2008p25890}%
\bibitem[{\citenamefont{Harris}\ \emph{et~al.}(2009)\citenamefont{Harris},
  \citenamefont{Arenson}, \citenamefont{Christiansen},
  \citenamefont{Etzelmuller}, \citenamefont{Frauenfelder},
  \citenamefont{Gruber}, \citenamefont{Haeberli}, \citenamefont{Hauck},
  \citenamefont{Holzle}, \citenamefont{Humlum}, \citenamefont{Isaksen},
  \citenamefont{Kaab}, \citenamefont{Kern-Lutschg}, \citenamefont{Lehning},
  \citenamefont{Matsuoka}, \citenamefont{Murton}, \citenamefont{Notzli},
  \citenamefont{Phillips}, \citenamefont{Ross}, \citenamefont{Seppala},
  \citenamefont{Springman},\ and\ \citenamefont{Muhll}}]{harris2009}%
  \BibitemOpen
  \bibfield{author}{%
  \bibinfo {author} {\bibnamefont{Harris}, \bibfnamefont{C.}}, \bibinfo
  {author} {\bibfnamefont{L.~U.}\ \bibnamefont{Arenson}}, \bibinfo {author}
  {\bibfnamefont{H.~H.}\ \bibnamefont{Christiansen}}, \bibinfo {author}
  {\bibfnamefont{B.}~\bibnamefont{Etzelmuller}}, \bibinfo {author}
  {\bibfnamefont{R.}~\bibnamefont{Frauenfelder}}, \bibinfo {author}
  {\bibfnamefont{S.}~\bibnamefont{Gruber}}, \bibinfo {author}
  {\bibfnamefont{W.}~\bibnamefont{Haeberli}}, \bibinfo {author}
  {\bibfnamefont{C.}~\bibnamefont{Hauck}}, \bibinfo {author}
  {\bibfnamefont{M.}~\bibnamefont{Holzle}}, \bibinfo {author}
  {\bibfnamefont{O.}~\bibnamefont{Humlum}}, \bibinfo {author}
  {\bibfnamefont{K.}~\bibnamefont{Isaksen}}, \bibinfo {author}
  {\bibfnamefont{A.}~\bibnamefont{Kaab}}, \bibinfo {author}
  {\bibfnamefont{M.~A.}\ \bibnamefont{Kern-Lutschg}}, \bibinfo {author}
  {\bibfnamefont{M.}~\bibnamefont{Lehning}}, \bibinfo {author}
  {\bibfnamefont{N.}~\bibnamefont{Matsuoka}}, \bibinfo {author}
  {\bibfnamefont{J.~B.}\ \bibnamefont{Murton}}, \bibinfo {author}
  {\bibfnamefont{J.}~\bibnamefont{Notzli}}, \bibinfo {author}
  {\bibfnamefont{M.}~\bibnamefont{Phillips}}, \bibinfo {author}
  {\bibfnamefont{N.}~\bibnamefont{Ross}}, \bibinfo {author}
  {\bibfnamefont{M.}~\bibnamefont{Seppala}}, \bibinfo {author}
  {\bibfnamefont{S.~M.}\ \bibnamefont{Springman}},\ and\ \bibinfo {author}
  {\bibfnamefont{D.~Vonder}\ \bibnamefont{Muhll}}}%
  , \bibinfo {year} {2009},\ \bibfield{title}{%
  \enquote{\bibinfo {title} {Permafrost and climate in {Europe}: Monitoring and
  modelling thermal, geomorphological and geotechnical responses},}\ }%
  \bibfield{journal}{%
  \bibinfo {journal} {Earth Sci. Revs}\ }%
  \textbf{\bibinfo {volume} {92}},\ \bibinfo {pages} {117--171}%
  \bibAnnoteFile{NoStop}{harris2009}%
\bibitem[{\citenamefont{Hartmann}\ \emph{et~al.}(2011)\citenamefont{Hartmann},
  \citenamefont{Niedermeier}, \citenamefont{Voigtlander},
  \citenamefont{Clauss}, \citenamefont{Shaw}, \citenamefont{Wex},
  \citenamefont{Kiselev},\ and\ \citenamefont{Stratmann}}]{Hartmann2011}%
  \BibitemOpen
  \bibfield{author}{%
  \bibinfo {author} {\bibnamefont{Hartmann}, \bibfnamefont{S.}}, \bibinfo
  {author} {\bibfnamefont{D.}~\bibnamefont{Niedermeier}}, \bibinfo {author}
  {\bibfnamefont{J.}~\bibnamefont{Voigtlander}}, \bibinfo {author}
  {\bibfnamefont{T.}~\bibnamefont{Clauss}}, \bibinfo {author}
  {\bibfnamefont{R.~A.}\ \bibnamefont{Shaw}}, \bibinfo {author}
  {\bibfnamefont{H.}~\bibnamefont{Wex}}, \bibinfo {author}
  {\bibfnamefont{A.}~\bibnamefont{Kiselev}},\ and\ \bibinfo {author}
  {\bibfnamefont{F.}~\bibnamefont{Stratmann}}}%
  , \bibinfo {year} {2011},\ \bibfield{title}{%
  \enquote{\bibinfo {title} {Homogeneous and heterogeneous ice nucleation at
  lacis: operating principle and theoretical studies},}\ }%
  \bibfield{journal}{%
  \bibinfo {journal} {Atmos. Chem. Phys.}\ }%
  \textbf{\bibinfo {volume} {11}},\ \bibinfo {pages} {1753--1767}%
  \bibAnnoteFile{NoStop}{Hartmann2011}%
\bibitem[{\citenamefont{{Hei{\ss}elmann}}\
  \emph{et~al.}(2010)\citenamefont{{Hei{\ss}elmann}}, \citenamefont{{Blum}},
  \citenamefont{{Fraser}},\ and\ \citenamefont{{Wolling}}}]{heisselmann2010}%
  \BibitemOpen
  \bibfield{author}{%
  \bibinfo {author} {\bibnamefont{{Hei{\ss}elmann}}, \bibfnamefont{D.}},
  \bibinfo {author} {\bibfnamefont{J.}~\bibnamefont{{Blum}}}, \bibinfo {author}
  {\bibfnamefont{H.~J.}\ \bibnamefont{{Fraser}}},\ and\ \bibinfo {author}
  {\bibfnamefont{K.}~\bibnamefont{{Wolling}}}}%
  , \bibinfo {year} {2010},\ \bibfield{title}{%
  \enquote{\bibinfo {title} {{Microgravity experiments on the collisional
  behavior of saturnian ring particles}},}\ }%
  \bibfield{journal}{%
  \bibinfo {journal} {Icarus}\ }%
  \textbf{\bibinfo {volume} {206}},\ \bibinfo {pages} {424--430}%
  \bibAnnoteFile{NoStop}{heisselmann2010}%
\bibitem[{\citenamefont{Herman}(2011)}]{Herman2011}%
  \BibitemOpen
  \bibfield{author}{%
  \bibinfo {author} {\bibnamefont{Herman}, \bibfnamefont{A.}}}%
  , \bibinfo {year} {2011},\ \bibfield{title}{%
  \enquote{\bibinfo {title} {Molecular-dynamics simulation of clustering
  processes in sea-ice floes},}\ }%
  \bibfield{journal}{%
  \bibinfo {journal} {Phys. Rev. E}\ }%
  \textbf{\bibinfo {volume} {84}},\ \bibinfo {pages} {056104}%
  \bibAnnoteFile{NoStop}{Herman2011}%
\bibitem[{\citenamefont{von Hessberg}\ \emph{et~al.}(2008)\citenamefont{von
  Hessberg}, \citenamefont{Pouvesle}, \citenamefont{Winkler},
  \citenamefont{Schuster},\ and\ \citenamefont{Crowley}}]{vonHessberg2008}%
  \BibitemOpen
  \bibfield{author}{%
  \bibinfo {author} {\bibnamefont{von Hessberg}, \bibfnamefont{P.}}, \bibinfo
  {author} {\bibfnamefont{N.}~\bibnamefont{Pouvesle}}, \bibinfo {author}
  {\bibfnamefont{A.~K.}\ \bibnamefont{Winkler}}, \bibinfo {author}
  {\bibfnamefont{G.}~\bibnamefont{Schuster}},\ and\ \bibinfo {author}
  {\bibfnamefont{J.~N.}\ \bibnamefont{Crowley}}}%
  , \bibinfo {year} {2008},\ \bibfield{title}{%
  \enquote{\bibinfo {title} {Interaction of formic and acetic acid with ice
  surfaces between 187 and 227~{K}. {Investigation} of single species- and
  competitive adsorption},}\ }%
  \bibfield{journal}{%
  \bibinfo {journal} {Phys. Chem. Chem. Phys.}\ }%
  \textbf{\bibinfo {volume} {10}},\ \bibinfo {pages} {2345--2355}%
  \bibAnnoteFile{NoStop}{vonHessberg2008}%
\bibitem[{\citenamefont{Heymsfield}(1986)}]{Heymsfield1986}%
  \BibitemOpen
  \bibfield{author}{%
  \bibinfo {author} {\bibnamefont{Heymsfield}, \bibfnamefont{A.~J.}}}%
  , \bibinfo {year} {1986},\ \bibfield{title}{%
  \enquote{\bibinfo {title} {Ice particles observed in a cirriform cloud at
  -83$^o${C} and implications for polar stratospheric clouds},}\ }%
  \bibfield{journal}{%
  \bibinfo {journal} {J. Atmos. Sci.}\ }%
  \textbf{\bibinfo {volume} {43}},\ \bibinfo {pages} {851--855}%
  \bibAnnoteFile{NoStop}{Heymsfield1986}%
\bibitem[{\citenamefont{Heymsfield}\ and\
  \citenamefont{Platt}(1984)}]{heymsfield1984}%
  \BibitemOpen
  \bibfield{author}{%
  \bibinfo {author} {\bibnamefont{Heymsfield}, \bibfnamefont{A.~J.}},\ and\
  \bibinfo {author} {\bibfnamefont{C.~M.~R.}\ \bibnamefont{Platt}}}%
  , \bibinfo {year} {1984},\ \bibfield{title}{%
  \enquote{\bibinfo {title} {A parameterization of the particle-size spectrum
  of ice clouds in terms of the ambient-temperature and the ice
  water-content},}\ }%
  \bibfield{journal}{%
  \bibinfo {journal} {J. Atmos. Sci.}\ }%
  \textbf{\bibinfo {volume} {41}},\ \bibinfo {pages} {846--855}%
  \bibAnnoteFile{NoStop}{heymsfield1984}%
\bibitem[{\citenamefont{{Hibbitts}}\ and\
  \citenamefont{{Szanyi}}(2007)}]{hibbitts2007}%
  \BibitemOpen
  \bibfield{author}{%
  \bibinfo {author} {\bibnamefont{{Hibbitts}}, \bibfnamefont{C.~A.}},\ and\
  \bibinfo {author} {\bibfnamefont{J.}~\bibnamefont{{Szanyi}}}}%
  , \bibinfo {year} {2007},\ \bibfield{title}{%
  \enquote{\bibinfo {title} {{Physisorption of CO$_{2}$ on non-ice materials
  relevant to icy satellites}},}\ }%
  \bibfield{journal}{%
  \bibinfo {journal} {Icarus}\ }%
  \textbf{\bibinfo {volume} {191}},\ \bibinfo {pages} {371--380}%
  \bibAnnoteFile{NoStop}{hibbitts2007}%
\bibitem[{\citenamefont{{Hibler III}}(1979)}]{Hibler:1979}%
  \BibitemOpen
  \bibfield{author}{%
  \bibinfo {author} {\bibnamefont{{Hibler III}}, \bibfnamefont{W.~D}}}%
  , \bibinfo {year} {1979},\ \bibfield{title}{%
  \enquote{\bibinfo {title} {A dynamic thermodynamic sea ice model},}\ }%
  \bibfield{journal}{%
  \bibinfo {journal} {J.~Phys.~Oceanogr.}\ }%
  \textbf{\bibinfo {volume} {9}},\ \bibinfo {pages} {815--846}%
  \bibAnnoteFile{NoStop}{Hibler:1979}%
\bibitem[{\citenamefont{{Hibler III}}(2001)}]{Hibler:2001}%
  \BibitemOpen
  \bibfield{author}{%
  \bibinfo {author} {\bibnamefont{{Hibler III}}, \bibfnamefont{W.~D.}}}%
  , \bibinfo {year} {2001},\ \bibfield{title}{%
  \enquote{\bibinfo {title} {Sea ice fracturing on the large scale},}\ }%
  \bibfield{journal}{%
  \bibinfo {journal} {Eng. Fracture Mech.}\ }%
  \textbf{\bibinfo {volume} {68}},\ \bibinfo {pages} {2013--2043}%
  \bibAnnoteFile{NoStop}{Hibler:2001}%
\bibitem[{\citenamefont{Hidaka}\ \emph{et~al.}(2009)\citenamefont{Hidaka},
  \citenamefont{Watanabe}, \citenamefont{Kouchi},\ and\
  \citenamefont{Watanabe}}]{hidaka2009}%
  \BibitemOpen
  \bibfield{author}{%
  \bibinfo {author} {\bibnamefont{Hidaka}, \bibfnamefont{H.}}, \bibinfo
  {author} {\bibfnamefont{M.}~\bibnamefont{Watanabe}}, \bibinfo {author}
  {\bibfnamefont{A.}~\bibnamefont{Kouchi}},\ and\ \bibinfo {author}
  {\bibfnamefont{N.}~\bibnamefont{Watanabe}}}%
  , \bibinfo {year} {2009},\ \bibfield{title}{%
  \enquote{\bibinfo {title} {Reaction routes in the
  {CO}-{H$_2$CO}-d$_n$-{CH$_3$OH}-d$_m$ system clarified from {H(D)} exposure
  of solid formaldehyde at low temperatures},}\ }%
  \bibfield{journal}{%
  \bibinfo {journal} {Astrophys. J.}\ }%
  \textbf{\bibinfo {volume} {702}},\ \bibinfo {pages} {291--300}%
  \bibAnnoteFile{NoStop}{hidaka2009}%
\bibitem[{\citenamefont{Hillert}(1965)}]{Hillert1965}%
  \BibitemOpen
  \bibfield{author}{%
  \bibinfo {author} {\bibnamefont{Hillert}, \bibfnamefont{M.}}}%
  , \bibinfo {year} {1965},\ \bibfield{title}{%
  \enquote{\bibinfo {title} {On theory of normal and abnormal grain growth},}\
  }%
  \bibfield{journal}{%
  \bibinfo {journal} {Acta Metallurgica}\ }%
  \textbf{\bibinfo {volume} {13}},\ \bibinfo {pages} {227--238}%
  \bibAnnoteFile{NoStop}{Hillert1965}%
\bibitem[{\citenamefont{Hiraoka}\ \emph{et~al.}(2006)\citenamefont{Hiraoka},
  \citenamefont{Mochizuki},\ and\ \citenamefont{Wada}}]{hiraoka2006}%
  \BibitemOpen
  \bibfield{author}{%
  \bibinfo {author} {\bibnamefont{Hiraoka}, \bibfnamefont{K.}}, \bibinfo
  {author} {\bibfnamefont{N.}~\bibnamefont{Mochizuki}},\ and\ \bibinfo {author}
  {\bibfnamefont{A.}~\bibnamefont{Wada}}}%
  , \bibinfo {year} {2006},\ in\ \emph{\bibinfo {booktitle} {Astrochemistry:
  From Laboratory Studies to Astronomical Observations}},\ Vol.\ \bibinfo
  {volume} {855},\ \bibinfo {editor} {edited by\ \bibinfo {editor}
  {\bibfnamefont{R.I.}\ \bibnamefont{Kaiser}}, \bibinfo {editor}
  {\bibfnamefont{P.}~\bibnamefont{Bernath}}, \bibinfo {editor}
  {\bibfnamefont{Y.}~\bibnamefont{Osamura}}, \bibinfo {editor}
  {\bibfnamefont{S.}~\bibnamefont{Petrie}},\ and\ \bibinfo {editor}
  {\bibfnamefont{A.~M.}\ \bibnamefont{Mebel}}}\ (\bibinfo {publisher} {AIP
  Conference Proceedings})\ pp.\ \bibinfo {pages} {86--99}%
  \bibAnnoteFile{NoStop}{hiraoka2006}%
\bibitem[{\citenamefont{Hobbs}(1974)}]{hobbs1974}%
  \BibitemOpen
  \bibfield{author}{%
  \bibinfo {author} {\bibnamefont{Hobbs}, \bibfnamefont{P.~V.}}}%
  , \bibinfo {year} {1974},\ \emph{\bibinfo {title} {Ice Physics}}\ (\bibinfo
  {publisher} {Oxford University Press})\ \bibinfo {note} {reprinted 2010}%
  \bibAnnoteFile{NoStop}{hobbs1974}%
\bibitem[{\citenamefont{{Hodyss}}\ \emph{et~al.}(2009)\citenamefont{{Hodyss}},
  \citenamefont{{Johnson}}, \citenamefont{{Stern}}, \citenamefont{{Goguen}},\
  and\ \citenamefont{{Kanik}}}]{hodyss2009}%
  \BibitemOpen
  \bibfield{author}{%
  \bibinfo {author} {\bibnamefont{{Hodyss}}, \bibfnamefont{R.}}, \bibinfo
  {author} {\bibfnamefont{P.~V.}\ \bibnamefont{{Johnson}}}, \bibinfo {author}
  {\bibfnamefont{J.~V.}\ \bibnamefont{{Stern}}}, \bibinfo {author}
  {\bibfnamefont{J.~D.}\ \bibnamefont{{Goguen}}},\ and\ \bibinfo {author}
  {\bibfnamefont{I.}~\bibnamefont{{Kanik}}}}%
  , \bibinfo {year} {2009},\ \bibfield{title}{%
  \enquote{\bibinfo {title} {{Photochemistry of methane water ices}},}\ }%
  \bibfield{journal}{%
  \bibinfo {journal} {Icarus}\ }%
  \textbf{\bibinfo {volume} {200}},\ \bibinfo {pages} {338--342}%
  \bibAnnoteFile{NoStop}{hodyss2009}%
\bibitem[{\citenamefont{Hodyss}\ \emph{et~al.}(2008)\citenamefont{Hodyss},
  \citenamefont{Jonson}, \citenamefont{Orzechowska}, \citenamefont{Goguen},\
  and\ \citenamefont{Kanik}}]{hodyss2008}%
  \BibitemOpen
  \bibfield{author}{%
  \bibinfo {author} {\bibnamefont{Hodyss}, \bibfnamefont{R.}}, \bibinfo
  {author} {\bibfnamefont{P.~V.}\ \bibnamefont{Jonson}}, \bibinfo {author}
  {\bibfnamefont{G.~E.}\ \bibnamefont{Orzechowska}}, \bibinfo {author}
  {\bibfnamefont{J.~D.}\ \bibnamefont{Goguen}},\ and\ \bibinfo {author}
  {\bibfnamefont{I.}~\bibnamefont{Kanik}}}%
  , \bibinfo {year} {2008},\ \bibfield{title}{%
  \enquote{\bibinfo {title} {Carbon dioxide segregation in 1:4 and 1:9
  {CO$_2$:H$_2$O} ices},}\ }%
  \bibfield{journal}{%
  \bibinfo {journal} {Icarus}\ }%
  \textbf{\bibinfo {volume} {194}},\ \bibinfo {pages} {836--842}%
  \bibAnnoteFile{NoStop}{hodyss2008}%
\bibitem[{\citenamefont{Holland}\ and\
  \citenamefont{Bitz}(2003)}]{Holland:2003}%
  \BibitemOpen
  \bibfield{author}{%
  \bibinfo {author} {\bibnamefont{Holland}, \bibfnamefont{M.~M.}},\ and\
  \bibinfo {author} {\bibfnamefont{C.~M.}\ \bibnamefont{Bitz}}}%
  , \bibinfo {year} {2003},\ \bibfield{title}{%
  \enquote{\bibinfo {title} {Polar amplification of climate change in the
  coupled model intercomparison project},}\ }%
  \bibfield{journal}{%
  \bibinfo {journal} {Climate~Dynamics}\ }%
  \textbf{\bibinfo {volume} {21}},\ \bibinfo {pages} {221--232}%
  \bibAnnoteFile{NoStop}{Holland:2003}%
\bibitem[{\citenamefont{Holland}\ \emph{et~al.}(2006)\citenamefont{Holland},
  \citenamefont{Bitz}, \citenamefont{Hunke}, \citenamefont{Lipscomb},\ and\
  \citenamefont{Schramm}}]{Holland:2006}%
  \BibitemOpen
  \bibfield{author}{%
  \bibinfo {author} {\bibnamefont{Holland}, \bibfnamefont{M.~M.}}, \bibinfo
  {author} {\bibfnamefont{C.~M.}\ \bibnamefont{Bitz}}, \bibinfo {author}
  {\bibfnamefont{E.~C.}\ \bibnamefont{Hunke}}, \bibinfo {author}
  {\bibfnamefont{W.~H.}\ \bibnamefont{Lipscomb}},\ and\ \bibinfo {author}
  {\bibfnamefont{J.~L.}\ \bibnamefont{Schramm}}}%
  , \bibinfo {year} {2006},\ \bibfield{title}{%
  \enquote{\bibinfo {title} {Influence of sea ice thickness distribution on
  polar climate in {CCSM3}},}\ }%
  \bibfield{journal}{%
  \bibinfo {journal} {J.~Clim.}\ }%
  \textbf{\bibinfo {volume} {19}},\ \bibinfo {pages} {2398--2414}%
  \bibAnnoteFile{NoStop}{Holland:2006}%
\bibitem[{\citenamefont{H{\"o}pfner}\
  \emph{et~al.}(2006)\citenamefont{H{\"o}pfner}, \citenamefont{Luo},
  \citenamefont{Massoli}, \citenamefont{Cairo}, \citenamefont{Spang},
  \citenamefont{Snels}, \citenamefont{Donfrancesco}, \citenamefont{Stiller},
  \citenamefont{von Clarmann}, \citenamefont{Fischer},\ and\
  \citenamefont{Biermann}}]{hoepfner2006}%
  \BibitemOpen
  \bibfield{author}{%
  \bibinfo {author} {\bibnamefont{H{\"o}pfner}, \bibfnamefont{M.}}, \bibinfo
  {author} {\bibfnamefont{B.~P.}\ \bibnamefont{Luo}}, \bibinfo {author}
  {\bibfnamefont{P.}~\bibnamefont{Massoli}}, \bibinfo {author}
  {\bibfnamefont{F.}~\bibnamefont{Cairo}}, \bibinfo {author}
  {\bibfnamefont{R.}~\bibnamefont{Spang}}, \bibinfo {author}
  {\bibfnamefont{M.}~\bibnamefont{Snels}}, \bibinfo {author}
  {\bibfnamefont{G.~Di}\ \bibnamefont{Donfrancesco}}, \bibinfo {author}
  {\bibfnamefont{G.}~\bibnamefont{Stiller}}, \bibinfo {author}
  {\bibfnamefont{T.}~\bibnamefont{von Clarmann}}, \bibinfo {author}
  {\bibfnamefont{H.}~\bibnamefont{Fischer}},\ and\ \bibinfo {author}
  {\bibfnamefont{U.}~\bibnamefont{Biermann}}}%
  , \bibinfo {year} {2006},\ \bibfield{title}{%
  \enquote{\bibinfo {title} {Spectroscopic evidence for {NAT}, {STS}, and ice
  in {MIPAS} infrared limb emission measurements of polar stratospheric
  clouds},}\ }%
  \bibfield{journal}{%
  \bibinfo {journal} {Atmos. Chem. Phys.}\ }%
  \textbf{\bibinfo {volume} {6}},\ \bibinfo {pages} {1201--1219}%
  \bibAnnoteFile{NoStop}{hoepfner2006}%
\bibitem[{\citenamefont{Hopkins}(1998{\natexlab{a}})}]{Hopkins1998}%
  \BibitemOpen
  \bibfield{author}{%
  \bibinfo {author} {\bibnamefont{Hopkins}, \bibfnamefont{M.~A.}}}%
  , \bibinfo {year} {1998}{\natexlab{a}},\ \bibfield{title}{%
  \enquote{\bibinfo {title} {Four stages of pressure ridging},}\ }%
  \bibfield{journal}{%
  \bibinfo {journal} {J. Geophys. Res.}\ }%
  \textbf{\bibinfo {volume} {103}},\ \bibinfo {pages} {21883--21891}%
  \bibAnnoteFile{NoStop}{Hopkins1998}%
\bibitem[{\citenamefont{Hopkins}(1998{\natexlab{b}})}]{Hopkins:1998}%
  \BibitemOpen
  \bibfield{author}{%
  \bibinfo {author} {\bibnamefont{Hopkins}, \bibfnamefont{M.~A.}}}%
  , \bibinfo {year} {1998}{\natexlab{b}},\ \bibfield{title}{%
  \enquote{\bibinfo {title} {Four stages of pressure ridging},}\ }%
  \bibfield{journal}{%
  \bibinfo {journal} {J.~Geophys.~Res.}\ }%
  \textbf{\bibinfo {volume} {103}},\ \bibinfo {pages} {21883--21891}%
  \bibAnnoteFile{NoStop}{Hopkins:1998}%
\bibitem[{\citenamefont{Hopkins}\ \emph{et~al.}(2004)\citenamefont{Hopkins},
  \citenamefont{Frankenstein},\ and\ \citenamefont{Thorndike}}]{Hopkins2004}%
  \BibitemOpen
  \bibfield{author}{%
  \bibinfo {author} {\bibnamefont{Hopkins}, \bibfnamefont{M.~A.}}, \bibinfo
  {author} {\bibfnamefont{S.}~\bibnamefont{Frankenstein}},\ and\ \bibinfo
  {author} {\bibfnamefont{A.~S.}\ \bibnamefont{Thorndike}}}%
  , \bibinfo {year} {2004},\ \bibfield{title}{%
  \enquote{\bibinfo {title} {Formation of an aggregate scale in {Arctic} sea
  ice},}\ }%
  \bibfield{journal}{%
  \bibinfo {journal} {J. Geophys. Res.}\ }%
  \textbf{\bibinfo {volume} {109}},\ \bibinfo {pages} {C01032}%
  \bibAnnoteFile{NoStop}{Hopkins2004}%
\bibitem[{\citenamefont{Hopkins}\ \emph{et~al.}(1999)\citenamefont{Hopkins},
  \citenamefont{Tuhkuri},\ and\ \citenamefont{Lensu}}]{Hopkins:1999}%
  \BibitemOpen
  \bibfield{author}{%
  \bibinfo {author} {\bibnamefont{Hopkins}, \bibfnamefont{M.~A.}}, \bibinfo
  {author} {\bibfnamefont{J.}~\bibnamefont{Tuhkuri}},\ and\ \bibinfo {author}
  {\bibfnamefont{M.}~\bibnamefont{Lensu}}}%
  , \bibinfo {year} {1999},\ \bibfield{title}{%
  \enquote{\bibinfo {title} {Rafting and ridging of thin ice sheets},}\ }%
  \bibfield{journal}{%
  \bibinfo {journal} {J.~Geophys.~Res.}\ }%
  \textbf{\bibinfo {volume} {104}},\ \bibinfo {pages} {13605--13613}%
  \bibAnnoteFile{NoStop}{Hopkins:1999}%
\bibitem[{\citenamefont{Hornekaer}\
  \emph{et~al.}(2003)\citenamefont{Hornekaer}, \citenamefont{Baurichter},
  \citenamefont{Petrunin}, \citenamefont{Field},\ and\
  \citenamefont{Luntz}}]{hornekaer2003}%
  \BibitemOpen
  \bibfield{author}{%
  \bibinfo {author} {\bibnamefont{Hornekaer}, \bibfnamefont{L.}}, \bibinfo
  {author} {\bibfnamefont{A.}~\bibnamefont{Baurichter}}, \bibinfo {author}
  {\bibfnamefont{V.~V.}\ \bibnamefont{Petrunin}}, \bibinfo {author}
  {\bibfnamefont{D.}~\bibnamefont{Field}},\ and\ \bibinfo {author}
  {\bibfnamefont{A.C.}\ \bibnamefont{Luntz}}}%
  , \bibinfo {year} {2003},\ \bibfield{title}{%
  \enquote{\bibinfo {title} {Importance of surface morphology in interstellar
  {H}$_2$ formation},}\ }%
  \bibfield{journal}{%
  \bibinfo {journal} {Science}\ }%
  \textbf{\bibinfo {volume} {302}},\ \bibinfo {pages} {1943--1946}%
  \bibAnnoteFile{NoStop}{hornekaer2003}%
\bibitem[{\citenamefont{Hornekaer}\
  \emph{et~al.}(2005)\citenamefont{Hornekaer}, \citenamefont{Baurichter},
  \citenamefont{Petrunin}, \citenamefont{Luntz}, \citenamefont{Kay},\ and\
  \citenamefont{Al-Halab}}]{hornekaer2005}%
  \BibitemOpen
  \bibfield{author}{%
  \bibinfo {author} {\bibnamefont{Hornekaer}, \bibfnamefont{L.}}, \bibinfo
  {author} {\bibfnamefont{A.}~\bibnamefont{Baurichter}}, \bibinfo {author}
  {\bibfnamefont{V.~V.}\ \bibnamefont{Petrunin}}, \bibinfo {author}
  {\bibfnamefont{A.~C.}\ \bibnamefont{Luntz}}, \bibinfo {author}
  {\bibfnamefont{B.~D.}\ \bibnamefont{Kay}},\ and\ \bibinfo {author}
  {\bibfnamefont{A.}~\bibnamefont{Al-Halab}}}%
  , \bibinfo {year} {2005},\ \bibfield{title}{%
  \enquote{\bibinfo {title} {'influence of surface morphology on {D}$_2$
  desorption kinetics from amorphous solid water},}\ }%
  \bibfield{journal}{%
  \bibinfo {journal} {J. Chem. Phys.}\ }%
  \textbf{\bibinfo {volume} {122}},\ \bibinfo {pages} {124701}%
  \bibAnnoteFile{NoStop}{hornekaer2005}%
\bibitem[{\citenamefont{Hornekaer}\
  \emph{et~al.}(2006{\natexlab{a}})\citenamefont{Hornekaer},
  \citenamefont{Rauls}, \citenamefont{Xu}, \citenamefont{Z.~Sljivancanin},
  \citenamefont{Stensgaard}, \citenamefont{Laegsgaard}, \citenamefont{Hammer},\
  and\ \citenamefont{Besenbacher}}]{hornekaer2006_2}%
  \BibitemOpen
  \bibfield{author}{%
  \bibinfo {author} {\bibnamefont{Hornekaer}, \bibfnamefont{L.}}, \bibinfo
  {author} {\bibfnamefont{E.}~\bibnamefont{Rauls}}, \bibinfo {author}
  {\bibfnamefont{W.}~\bibnamefont{Xu}}, \bibinfo {author}
  {\bibfnamefont{R.~Otero}\ \bibnamefont{Z.~Sljivancanin}}, \bibinfo {author}
  {\bibfnamefont{I.}~\bibnamefont{Stensgaard}}, \bibinfo {author}
  {\bibfnamefont{E.}~\bibnamefont{Laegsgaard}}, \bibinfo {author}
  {\bibfnamefont{B.}~\bibnamefont{Hammer}},\ and\ \bibinfo {author}
  {\bibfnamefont{F.}~\bibnamefont{Besenbacher}}}%
  , \bibinfo {year} {2006}{\natexlab{a}},\ \bibfield{title}{%
  \enquote{\bibinfo {title} {Clustering of chemisorbed {H(D)} atoms on the
  graphite (0001) surface due to preferential sticking},}\ }%
  \bibfield{journal}{%
  \bibinfo {journal} {Phys. Rev. Lett.}\ }%
  \textbf{\bibinfo {volume} {97}},\ \bibinfo {pages} {186102}%
  \bibAnnoteFile{NoStop}{hornekaer2006_2}%
\bibitem[{\citenamefont{Hornekaer}\
  \emph{et~al.}(2006{\natexlab{b}})\citenamefont{Hornekaer},
  \citenamefont{Sljivancanin}, \citenamefont{Xu}, \citenamefont{Otero},
  \citenamefont{Rauls}, \citenamefont{Stensgaard}, \citenamefont{Laegsgaard},
  \citenamefont{Hammer},\ and\ \citenamefont{Besenbacher}}]{hornekaer2006_1}%
  \BibitemOpen
  \bibfield{author}{%
  \bibinfo {author} {\bibnamefont{Hornekaer}, \bibfnamefont{L.}}, \bibinfo
  {author} {\bibfnamefont{Z.}~\bibnamefont{Sljivancanin}}, \bibinfo {author}
  {\bibfnamefont{W.}~\bibnamefont{Xu}}, \bibinfo {author}
  {\bibfnamefont{R.}~\bibnamefont{Otero}}, \bibinfo {author}
  {\bibfnamefont{E.}~\bibnamefont{Rauls}}, \bibinfo {author}
  {\bibfnamefont{I.}~\bibnamefont{Stensgaard}}, \bibinfo {author}
  {\bibfnamefont{E.}~\bibnamefont{Laegsgaard}}, \bibinfo {author}
  {\bibfnamefont{B.}~\bibnamefont{Hammer}},\ and\ \bibinfo {author}
  {\bibfnamefont{F.}~\bibnamefont{Besenbacher}}}%
  , \bibinfo {year} {2006}{\natexlab{b}},\ \bibfield{title}{%
  \enquote{\bibinfo {title} {Metastable structures and recombination pathways
  for atomic hydrogen on the graphite (0001) surface},}\ }%
  \bibfield{journal}{%
  \bibinfo {journal} {Phys. Rev. Lett.}\ }%
  \textbf{\bibinfo {volume} {96}},\ \bibinfo {pages} {156104}%
  \bibAnnoteFile{NoStop}{hornekaer2006_1}%
\bibitem[{\citenamefont{Houpis}\ \emph{et~al.}(1985)\citenamefont{Houpis},
  \citenamefont{Mendis},\ and\ \citenamefont{Ip}}]{houpis1985}%
  \BibitemOpen
  \bibfield{author}{%
  \bibinfo {author} {\bibnamefont{Houpis}, \bibfnamefont{H.~L.~F.}}, \bibinfo
  {author} {\bibfnamefont{D.~A.}\ \bibnamefont{Mendis}},\ and\ \bibinfo
  {author} {\bibfnamefont{W.-H.}\ \bibnamefont{Ip}}}%
  , \bibinfo {year} {1985},\ \bibfield{title}{%
  \enquote{\bibinfo {title} {The chemical differentiation of the cometary
  nucleus --- the process and its consequences},}\ }%
  \bibfield{journal}{%
  \bibinfo {journal} {Astrophys. J.}\ }%
  \textbf{\bibinfo {volume} {295}},\ \bibinfo {pages} {654--667}%
  \bibAnnoteFile{NoStop}{houpis1985}%
\bibitem[{\citenamefont{Hudson}\ and\ \citenamefont{Donn}(1991)}]{hudson1991}%
  \BibitemOpen
  \bibfield{author}{%
  \bibinfo {author} {\bibnamefont{Hudson}, \bibfnamefont{R.~L.}},\ and\
  \bibinfo {author} {\bibfnamefont{B.}~\bibnamefont{Donn}}}%
  , \bibinfo {year} {1991},\ \bibfield{title}{%
  \enquote{\bibinfo {title} {An experimental-study of the sublimation of water
  ice and the release of trapped gases},}\ }%
  \bibfield{journal}{%
  \bibinfo {journal} {Icarus}\ }%
  \textbf{\bibinfo {volume} {94}},\ \bibinfo {pages} {326--332}%
  \bibAnnoteFile{NoStop}{hudson1991}%
\bibitem[{\citenamefont{Hughes}(2009)}]{Hughes:2009}%
  \BibitemOpen
  \bibfield{author}{%
  \bibinfo {author} {\bibnamefont{Hughes}, \bibfnamefont{T.}}}%
  , \bibinfo {year} {2009},\ \bibfield{title}{%
  \enquote{\bibinfo {title} {Thermal convection and the origin of ice
  streams},}\ }%
  \bibfield{journal}{%
  \bibinfo {journal} {J. Glaciol.}\ }%
  \textbf{\bibinfo {volume} {55}},\ \bibinfo {pages} {524--536}%
  \bibAnnoteFile{NoStop}{Hughes:2009}%
\bibitem[{\citenamefont{Huthwelker}\
  \emph{et~al.}(2006)\citenamefont{Huthwelker}, \citenamefont{Ammann},\ and\
  \citenamefont{Peter}}]{Huthwelker:2006p981}%
  \BibitemOpen
  \bibfield{author}{%
  \bibinfo {author} {\bibnamefont{Huthwelker}, \bibfnamefont{T.}}, \bibinfo
  {author} {\bibfnamefont{M.}~\bibnamefont{Ammann}},\ and\ \bibinfo {author}
  {\bibfnamefont{T.}~\bibnamefont{Peter}}}%
  , \bibinfo {year} {2006},\ \bibfield{title}{%
  \enquote{\bibinfo {title} {The uptake of acidic gases on ice},}\ }%
  \bibfield{journal}{%
  \bibinfo {journal} {Chem. Rev.}\ }%
  \textbf{\bibinfo {volume} {106}},\ \bibinfo {pages} {1375--1444}%
  \bibAnnoteFile{NoStop}{Huthwelker:2006p981}%
\bibitem[{\citenamefont{Huybrechts}(2007)}]{PhilippeHuybrechts:2007p26003}%
  \BibitemOpen
  \bibfield{author}{%
  \bibinfo {author} {\bibnamefont{Huybrechts}, \bibfnamefont{P.}}}%
  , \bibinfo {year} {2007},\ \bibfield{title}{%
  \enquote{\bibinfo {title} {Numerical modelling of polar ice sheets through
  time},}\ }%
  \bibinfo {journal} {Glacier Science and Environmental Change (First
  Edition)},\ \bibinfo {pages} {406--412}%
  \bibAnnoteFile{NoStop}{PhilippeHuybrechts:2007p26003}%
\bibitem[{\citenamefont{Hvidegaard}\
  \emph{et~al.}(2006)\citenamefont{Hvidegaard}, \citenamefont{Forsberg},\ and\
  \citenamefont{Skourup}}]{Hvidegaard:2006}%
  \BibitemOpen
\bibfield{journal}{%
    }%
  \bibfield{author}{%
  \bibinfo {author} {\bibnamefont{Hvidegaard}, \bibfnamefont{S.~M.}}, \bibinfo
  {author} {\bibfnamefont{R.}~\bibnamefont{Forsberg}},\ and\ \bibinfo {author}
  {\bibfnamefont{H.}~\bibnamefont{Skourup}}}%
  , \bibinfo {year} {2006},\ \enquote{\bibinfo {title} {Sea ice thickness
  estimates from airborne laser scanning},}\ in\ \emph{\bibinfo {booktitle}
  {Arctic sea ice thickness (Climate Change and Natural Hazards series 10)}},\
  \bibinfo {editor} {edited by\ \bibinfo {editor}
  {\bibfnamefont{P.}~\bibnamefont{Wadhams}}\ and\ \bibinfo {editor}
  {\bibfnamefont{G.}~\bibnamefont{Amanatidis}}}\ (\bibinfo {publisher}
  {European Union})\ pp.\ \bibinfo {pages} {193--206}%
  \bibAnnoteFile{NoStop}{Hvidegaard:2006}%
\bibitem[{\citenamefont{Imshenetsky}\
  \emph{et~al.}(1978)\citenamefont{Imshenetsky}, \citenamefont{Lysenko},\ and\
  \citenamefont{Kazakov}}]{imshenetsky1978}%
  \BibitemOpen
  \bibfield{author}{%
  \bibinfo {author} {\bibnamefont{Imshenetsky}, \bibfnamefont{A.~A.}}, \bibinfo
  {author} {\bibfnamefont{S.~V.}\ \bibnamefont{Lysenko}},\ and\ \bibinfo
  {author} {\bibfnamefont{G.~A.}\ \bibnamefont{Kazakov}}}%
  , \bibinfo {year} {1978},\ \bibfield{title}{%
  \enquote{\bibinfo {title} {Upper boundary of the biosphere},}\ }%
  \bibfield{journal}{%
  \bibinfo {journal} {Appl. Environ. Microbiol.}\ }%
  \textbf{\bibinfo {volume} {35}},\ \bibinfo {pages} {1--5}%
  \bibAnnoteFile{NoStop}{imshenetsky1978}%
\bibitem[{\citenamefont{Ioppolo}\ \emph{et~al.}(2010)\citenamefont{Ioppolo},
  \citenamefont{Cuppen}, \citenamefont{Romanzin}, \citenamefont{van Dishoeck},\
  and\ \citenamefont{Linnartz}}]{ioppolo2010}%
  \BibitemOpen
  \bibfield{author}{%
  \bibinfo {author} {\bibnamefont{Ioppolo}, \bibfnamefont{S.}}, \bibinfo
  {author} {\bibfnamefont{H.~M.}\ \bibnamefont{Cuppen}}, \bibinfo {author}
  {\bibfnamefont{C.}~\bibnamefont{Romanzin}}, \bibinfo {author}
  {\bibfnamefont{E.~F.}\ \bibnamefont{van Dishoeck}},\ and\ \bibinfo {author}
  {\bibfnamefont{H.}~\bibnamefont{Linnartz}}}%
  , \bibinfo {year} {2010},\ \bibfield{title}{%
  \enquote{\bibinfo {title} {Water formation at low temperatures by surface
  {O}$_2$ hydrogenation {I}: characterization of ice penetration},}\ }%
  \bibfield{journal}{%
  \bibinfo {journal} {Phys. Chem. Chem. Phys.}\ }%
  \textbf{\bibinfo {volume} {12}},\ \bibinfo {pages} {12065--12076}%
  \bibAnnoteFile{NoStop}{ioppolo2010}%
\bibitem[{\citenamefont{Ioppolo}\ \emph{et~al.}(2009)\citenamefont{Ioppolo},
  \citenamefont{Palumbo}, \citenamefont{Baratta},\ and\
  \citenamefont{Mennella}}]{ioppolo2009}%
  \BibitemOpen
  \bibfield{author}{%
  \bibinfo {author} {\bibnamefont{Ioppolo}, \bibfnamefont{S.}}, \bibinfo
  {author} {\bibfnamefont{M.~E.}\ \bibnamefont{Palumbo}}, \bibinfo {author}
  {\bibfnamefont{G.~A.}\ \bibnamefont{Baratta}},\ and\ \bibinfo {author}
  {\bibfnamefont{V.}~\bibnamefont{Mennella}}}%
  , \bibinfo {year} {2009},\ \bibfield{title}{%
  \enquote{\bibinfo {title} {Formation of interstellar solid {CO$_2$} after
  energetic processing of icy grain mantles},}\ }%
  \bibfield{journal}{%
  \bibinfo {journal} {Astron. Astrophys.}\ }%
  \textbf{\bibinfo {volume} {493}},\ \bibinfo {pages} {1017--1028}%
  \bibAnnoteFile{NoStop}{ioppolo2009}%
\bibitem[{\citenamefont{Islam}\ \emph{et~al.}(2007)\citenamefont{Islam},
  \citenamefont{Latimer},\ and\ \citenamefont{Price}}]{islam2007}%
  \BibitemOpen
  \bibfield{author}{%
  \bibinfo {author} {\bibnamefont{Islam}, \bibfnamefont{F.}}, \bibinfo {author}
  {\bibfnamefont{E.~R.}\ \bibnamefont{Latimer}},\ and\ \bibinfo {author}
  {\bibfnamefont{S.~D.}\ \bibnamefont{Price}}}%
  , \bibinfo {year} {2007},\ \bibfield{title}{%
  \enquote{\bibinfo {title} {The formation of vibrationally excited {HD} from
  atomic recombination on cold graphite surfaces},}\ }%
  \bibfield{journal}{%
  \bibinfo {journal} {J. Chem. Phys.}\ }%
  \textbf{\bibinfo {volume} {127}},\ \bibinfo {pages} {064701}%
  \bibAnnoteFile{NoStop}{islam2007}%
\bibitem[{\citenamefont{Jackson}\ \emph{et~al.}(1997)\citenamefont{Jackson},
  \citenamefont{Nield}, \citenamefont{Whitworth}, \citenamefont{Oguro},\ and\
  \citenamefont{Wilson}}]{jackson1997}%
  \BibitemOpen
  \bibfield{author}{%
  \bibinfo {author} {\bibnamefont{Jackson}, \bibfnamefont{S.~M.}}, \bibinfo
  {author} {\bibfnamefont{V.~M.}\ \bibnamefont{Nield}}, \bibinfo {author}
  {\bibfnamefont{R.~W.}\ \bibnamefont{Whitworth}}, \bibinfo {author}
  {\bibfnamefont{M.}~\bibnamefont{Oguro}},\ and\ \bibinfo {author}
  {\bibfnamefont{C.~C.}\ \bibnamefont{Wilson}}}%
  , \bibinfo {year} {1997},\ \bibfield{title}{%
  \enquote{\bibinfo {title} {Single-crystal neutron diffraction studies of the
  structure of ice {XI}},}\ }%
  \bibfield{journal}{%
  \bibinfo {journal} {J. Phys. Chem. B}\ }%
  \textbf{\bibinfo {volume} {101}},\ \bibinfo {pages} {6142--6145}%
  \bibAnnoteFile{NoStop}{jackson1997}%
\bibitem[{\citenamefont{Jenniskens}\ and\
  \citenamefont{Blake}(1994)}]{jenniskens1994}%
  \BibitemOpen
  \bibfield{author}{%
  \bibinfo {author} {\bibnamefont{Jenniskens}, \bibfnamefont{P.}},\ and\
  \bibinfo {author} {\bibfnamefont{D.~F.}\ \bibnamefont{Blake}}}%
  , \bibinfo {year} {1994},\ \bibfield{title}{%
  \enquote{\bibinfo {title} {Structural transitions in amorphous water ice and
  astrophysical implications},}\ }%
  \bibfield{journal}{%
  \bibinfo {journal} {Science}\ }%
  \textbf{\bibinfo {volume} {265}},\ \bibinfo {pages} {753--756}%
  \bibAnnoteFile{NoStop}{jenniskens1994}%
\bibitem[{\citenamefont{Jenniskens}\ and\
  \citenamefont{Blake}(1996{\natexlab{a}})}]{jenniskens1996b}%
  \BibitemOpen
  \bibfield{author}{%
  \bibinfo {author} {\bibnamefont{Jenniskens}, \bibfnamefont{P.}},\ and\
  \bibinfo {author} {\bibfnamefont{D.~F.}\ \bibnamefont{Blake}}}%
  , \bibinfo {year} {1996}{\natexlab{a}},\ \bibfield{title}{%
  \enquote{\bibinfo {title} {Crystallization of amorphous water ice in the
  solar system},}\ }%
  \bibfield{journal}{%
  \bibinfo {journal} {Astrophys. J.}\ }%
  \textbf{\bibinfo {volume} {473}},\ \bibinfo {pages} {1104--1113}%
  \bibAnnoteFile{NoStop}{jenniskens1996b}%
\bibitem[{\citenamefont{Jenniskens}\ and\
  \citenamefont{Blake}(1996{\natexlab{b}})}]{jenniskens1996}%
  \BibitemOpen
  \bibfield{author}{%
  \bibinfo {author} {\bibnamefont{Jenniskens}, \bibfnamefont{P.}},\ and\
  \bibinfo {author} {\bibfnamefont{D.~F.}\ \bibnamefont{Blake}}}%
  , \bibinfo {year} {1996}{\natexlab{b}},\ \bibfield{title}{%
  \enquote{\bibinfo {title} {A mechanism for forming deep cracks in comets},}\
  }%
  \bibfield{journal}{%
  \bibinfo {journal} {Planet. Space Sci.}\ }%
  \textbf{\bibinfo {volume} {44}},\ \bibinfo {pages} {711--713}%
  \bibAnnoteFile{NoStop}{jenniskens1996}%
\bibitem[{\citenamefont{Jenniskens}\
  \emph{et~al.}(1995)\citenamefont{Jenniskens}, \citenamefont{Blake},
  \citenamefont{Wilson},\ and\ \citenamefont{Pohorille}}]{jenniskens1995}%
  \BibitemOpen
  \bibfield{author}{%
  \bibinfo {author} {\bibnamefont{Jenniskens}, \bibfnamefont{P.}}, \bibinfo
  {author} {\bibfnamefont{D.~F.}\ \bibnamefont{Blake}}, \bibinfo {author}
  {\bibfnamefont{M.~A.}\ \bibnamefont{Wilson}},\ and\ \bibinfo {author}
  {\bibfnamefont{A.}~\bibnamefont{Pohorille}}}%
  , \bibinfo {year} {1995},\ \bibfield{title}{%
  \enquote{\bibinfo {title} {High-density amorphous ice, the frost on
  interstellar grains},}\ }%
  \bibfield{journal}{%
  \bibinfo {journal} {Astrophys. J.}\ }%
  \textbf{\bibinfo {volume} {455}},\ \bibinfo {pages} {389--401}%
  \bibAnnoteFile{NoStop}{jenniskens1995}%
\bibitem[{\citenamefont{Jensen}\ \emph{et~al.}(2005)\citenamefont{Jensen},
  \citenamefont{Smith}, \citenamefont{Pfister}, \citenamefont{Pittman},
  \citenamefont{Weinstock}, \citenamefont{Sayres}, \citenamefont{Herman},
  \citenamefont{Troy}, \citenamefont{Rosenlof}, \citenamefont{Thompson},
  \citenamefont{Fridlind}, \citenamefont{Hudson}, \citenamefont{Cziczo},
  \citenamefont{Heymsfield}, \citenamefont{Schmitt},\ and\
  \citenamefont{Wilson}}]{Jensen2005}%
  \BibitemOpen
  \bibfield{author}{%
  \bibinfo {author} {\bibnamefont{Jensen}, \bibfnamefont{E.~J.}}, \bibinfo
  {author} {\bibfnamefont{J.~B.}\ \bibnamefont{Smith}}, \bibinfo {author}
  {\bibfnamefont{L.}~\bibnamefont{Pfister}}, \bibinfo {author}
  {\bibfnamefont{J.~V.}\ \bibnamefont{Pittman}}, \bibinfo {author}
  {\bibfnamefont{E.~M.}\ \bibnamefont{Weinstock}}, \bibinfo {author}
  {\bibfnamefont{D.~S.}\ \bibnamefont{Sayres}}, \bibinfo {author}
  {\bibfnamefont{R.~L.}\ \bibnamefont{Herman}}, \bibinfo {author}
  {\bibfnamefont{R.~F.}\ \bibnamefont{Troy}}, \bibinfo {author}
  {\bibfnamefont{K.}~\bibnamefont{Rosenlof}}, \bibinfo {author}
  {\bibfnamefont{T.~L.}\ \bibnamefont{Thompson}}, \bibinfo {author}
  {\bibfnamefont{A.~M.}\ \bibnamefont{Fridlind}}, \bibinfo {author}
  {\bibfnamefont{P.~K.}\ \bibnamefont{Hudson}}, \bibinfo {author}
  {\bibfnamefont{D.~J.}\ \bibnamefont{Cziczo}}, \bibinfo {author}
  {\bibfnamefont{A.~J.}\ \bibnamefont{Heymsfield}}, \bibinfo {author}
  {\bibfnamefont{C.}~\bibnamefont{Schmitt}},\ and\ \bibinfo {author}
  {\bibfnamefont{J.~C.}\ \bibnamefont{Wilson}}}%
  , \bibinfo {year} {2005},\ \bibfield{title}{%
  \enquote{\bibinfo {title} {Ice supersaturations exceeding 100{\%} at the cold
  tropical tropopause: implications for cirrus formation and dehydration},}\ }%
  \bibfield{journal}{%
  \bibinfo {journal} {Atmos. Chem. Phys.}\ }%
  \textbf{\bibinfo {volume} {5}},\ \bibinfo {pages} {851--862}%
  \bibAnnoteFile{NoStop}{Jensen2005}%
\bibitem[{\citenamefont{Johari}(1998)}]{johari1998}%
  \BibitemOpen
  \bibfield{author}{%
  \bibinfo {author} {\bibnamefont{Johari}, \bibfnamefont{G.~P.}}}%
  , \bibinfo {year} {1998},\ \bibfield{title}{%
  \enquote{\bibinfo {title} {On the coexistence of cubic and hexagonal ice
  between 160 and 240~{K}},}\ }%
  \bibfield{journal}{%
  \bibinfo {journal} {Phil. Mag. B}\ }%
  \textbf{\bibinfo {volume} {78}},\ \bibinfo {pages} {375--383}%
  \bibAnnoteFile{NoStop}{johari1998}%
\bibitem[{\citenamefont{{Johns}}\ \emph{et~al.}(2006)\citenamefont{{Johns}},
  \citenamefont{{Durman}}, \citenamefont{{Banks}}, \citenamefont{{Roberts}},
  \citenamefont{{McLaren}}, \citenamefont{{Ridley}}, \citenamefont{{Senior}},
  \citenamefont{{Williams}}, \citenamefont{{Jones}}, \citenamefont{{Rickard}},
  \citenamefont{{Cusack}}, \citenamefont{{Ingram}}, \citenamefont{{Crucifix}},
  \citenamefont{{Sexton}}, \citenamefont{{Joshi}}, \citenamefont{{Dong}},
  \citenamefont{{Spencer}}, \citenamefont{{Hill}}, \citenamefont{{Gregory}},
  \citenamefont{{Keen}}, \citenamefont{{Pardaens}}, \citenamefont{{Lowe}},
  \citenamefont{{Bodas-Salcedo}}, \citenamefont{{Stark}},\ and\
  \citenamefont{{Searl}}}]{Johns:2006}%
  \BibitemOpen
  \bibfield{author}{%
  \bibinfo {author} {\bibnamefont{{Johns}}, \bibfnamefont{T.~C.}}, \bibinfo
  {author} {\bibfnamefont{C.~F.}\ \bibnamefont{{Durman}}}, \bibinfo {author}
  {\bibfnamefont{H.~T.}\ \bibnamefont{{Banks}}}, \bibinfo {author}
  {\bibfnamefont{M.~J.}\ \bibnamefont{{Roberts}}}, \bibinfo {author}
  {\bibfnamefont{A.~J.}\ \bibnamefont{{McLaren}}}, \bibinfo {author}
  {\bibfnamefont{J.~K.}\ \bibnamefont{{Ridley}}}, \bibinfo {author}
  {\bibfnamefont{C.~A.}\ \bibnamefont{{Senior}}}, \bibinfo {author}
  {\bibfnamefont{K.~D.}\ \bibnamefont{{Williams}}}, \bibinfo {author}
  {\bibfnamefont{A.}~\bibnamefont{{Jones}}}, \bibinfo {author}
  {\bibfnamefont{G.~J.}\ \bibnamefont{{Rickard}}}, \bibinfo {author}
  {\bibfnamefont{S.}~\bibnamefont{{Cusack}}}, \bibinfo {author}
  {\bibfnamefont{W.~J.}\ \bibnamefont{{Ingram}}}, \bibinfo {author}
  {\bibfnamefont{M.}~\bibnamefont{{Crucifix}}}, \bibinfo {author}
  {\bibfnamefont{D.~M.~H.}\ \bibnamefont{{Sexton}}}, \bibinfo {author}
  {\bibfnamefont{M.~M.}\ \bibnamefont{{Joshi}}}, \bibinfo {author}
  {\bibfnamefont{B.-W.}\ \bibnamefont{{Dong}}}, \bibinfo {author}
  {\bibfnamefont{H.}~\bibnamefont{{Spencer}}}, \bibinfo {author}
  {\bibfnamefont{R.~S.~R.}\ \bibnamefont{{Hill}}}, \bibinfo {author}
  {\bibfnamefont{J.~M.}\ \bibnamefont{{Gregory}}}, \bibinfo {author}
  {\bibfnamefont{A.~B.}\ \bibnamefont{{Keen}}}, \bibinfo {author}
  {\bibfnamefont{A.~K.}\ \bibnamefont{{Pardaens}}}, \bibinfo {author}
  {\bibfnamefont{J.~A.}\ \bibnamefont{{Lowe}}}, \bibinfo {author}
  {\bibfnamefont{A.}~\bibnamefont{{Bodas-Salcedo}}}, \bibinfo {author}
  {\bibfnamefont{S.}~\bibnamefont{{Stark}}},\ and\ \bibinfo {author}
  {\bibfnamefont{Y.}~\bibnamefont{{Searl}}}}%
  , \bibinfo {year} {2006},\ \bibfield{title}{%
  \enquote{\bibinfo {title} {The new {Hadley Centre} climate model {(HadGEM1)}:
  {Evaluation} of coupled simulations},}\ }%
  \bibfield{journal}{%
  \bibinfo {journal} {J.~Clim.}\ }%
  \textbf{\bibinfo {volume} {19}},\ \bibinfo {pages} {1327--1353}%
  \bibAnnoteFile{NoStop}{Johns:2006}%
\bibitem[{\citenamefont{Johnson}(1990)}]{johnson1990}%
  \BibitemOpen
  \bibfield{author}{%
  \bibinfo {author} {\bibnamefont{Johnson}, \bibfnamefont{R.~E.}}}%
  , \bibinfo {year} {1990},\ \enquote{\bibinfo {title} {Energetic
  charged-particle interaction with atmospheres and surfaces},}\ in\
  \emph{\bibinfo {booktitle} {Physics and Chemistry in Space}},\ Vol.~\bibinfo
  {volume} {19}\ (\bibinfo {publisher} {Springer})%
  \bibAnnoteFile{NoStop}{johnson1990}%
\bibitem[{\citenamefont{{Kadosaki}}\
  \emph{et~al.}(2010)\citenamefont{{Kadosaki}}, \citenamefont{{Ichimaru}},
  \citenamefont{{Hirasawa}},\ and\ \citenamefont{{Yamanouchi}}}]{kadosaki2010}%
  \BibitemOpen
  \bibfield{author}{%
  \bibinfo {author} {\bibnamefont{{Kadosaki}}, \bibfnamefont{G.}}, \bibinfo
  {author} {\bibfnamefont{T.}~\bibnamefont{{Ichimaru}}}, \bibinfo {author}
  {\bibfnamefont{N.}~\bibnamefont{{Hirasawa}}},\ and\ \bibinfo {author}
  {\bibfnamefont{T.}~\bibnamefont{{Yamanouchi}}}}%
  , \bibinfo {year} {2010},\ \bibfield{title}{%
  \enquote{\bibinfo {title} {{The Information of PSC and PMC from GOSAT
  FTS}},}\ }%
  \bibinfo {journal} {AGU Fall Meeting Abstracts},\ \bibinfo {pages} {B210}%
  \bibAnnoteFile{NoStop}{kadosaki2010}%
\bibitem[{\citenamefont{Kahan}\ \emph{et~al.}(2010)\citenamefont{Kahan},
  \citenamefont{Zhao}, \citenamefont{Jumaa},\ and\
  \citenamefont{Donaldson}}]{Kahan2010}%
  \BibitemOpen
\bibfield{journal}{%
    }%
  \bibfield{author}{%
  \bibinfo {author} {\bibnamefont{Kahan}, \bibfnamefont{T.~F.}}, \bibinfo
  {author} {\bibfnamefont{R.}~\bibnamefont{Zhao}}, \bibinfo {author}
  {\bibfnamefont{K.~B.}\ \bibnamefont{Jumaa}},\ and\ \bibinfo {author}
  {\bibfnamefont{D.~J.}\ \bibnamefont{Donaldson}}}%
  , \bibinfo {year} {2010},\ \bibfield{title}{%
  \enquote{\bibinfo {title} {Anthracene photolysis in aqueous solution and ice:
  Photon flux dependence and comparison of kinetics in bulk ice and at the
  air--ice interface},}\ }%
  \bibfield{journal}{%
  \bibinfo {journal} {Environ. Sci. Technol.}\ }%
  \textbf{\bibinfo {volume} {44}},\ \bibinfo {pages} {1302--1306}%
  \bibAnnoteFile{NoStop}{Kahan2010}%
\bibitem[{\citenamefont{Kaleschke}\
  \emph{et~al.}(2004)\citenamefont{Kaleschke}, \citenamefont{Richter},
  \citenamefont{Burrows}, \citenamefont{Afe}, \citenamefont{Heygster},
  \citenamefont{Notholt}, \citenamefont{Rankin}, \citenamefont{Roscoe},
  \citenamefont{Hollwedel}, \citenamefont{Wagner},\ and\
  \citenamefont{Jacobi}}]{Kaleschke:2004}%
  \BibitemOpen
  \bibfield{author}{%
  \bibinfo {author} {\bibnamefont{Kaleschke}, \bibfnamefont{L.}}, \bibinfo
  {author} {\bibfnamefont{A.}~\bibnamefont{Richter}}, \bibinfo {author}
  {\bibfnamefont{J.}~\bibnamefont{Burrows}}, \bibinfo {author}
  {\bibfnamefont{O.}~\bibnamefont{Afe}}, \bibinfo {author}
  {\bibfnamefont{G.}~\bibnamefont{Heygster}}, \bibinfo {author}
  {\bibfnamefont{J.}~\bibnamefont{Notholt}}, \bibinfo {author}
  {\bibfnamefont{A.~M.}\ \bibnamefont{Rankin}}, \bibinfo {author}
  {\bibfnamefont{H.K.}\ \bibnamefont{Roscoe}}, \bibinfo {author}
  {\bibfnamefont{J.}~\bibnamefont{Hollwedel}}, \bibinfo {author}
  {\bibfnamefont{T.}~\bibnamefont{Wagner}},\ and\ \bibinfo {author}
  {\bibfnamefont{H.W.}\ \bibnamefont{Jacobi}}}%
  , \bibinfo {year} {2004},\ \bibfield{title}{%
  \enquote{\bibinfo {title} {Frost flowers on sea ice as a source of sea salt
  and their influence on tropospheric halogen chemistry},}\ }%
  \bibfield{journal}{%
  \bibinfo {journal} {Geophys.~Res.~Lett.}\ }%
  \textbf{\bibinfo {volume} {31}},\ \bibinfo {pages} {L16114}%
  \bibAnnoteFile{NoStop}{Kaleschke:2004}%
\bibitem[{\citenamefont{K{\"a}rcher}\
  \emph{et~al.}(2009)\citenamefont{K{\"a}rcher}, \citenamefont{Abbatt},
  \citenamefont{Cox}, \citenamefont{Popp},\ and\
  \citenamefont{Voigt}}]{Karcher2009}%
  \BibitemOpen
  \bibfield{author}{%
  \bibinfo {author} {\bibnamefont{K{\"a}rcher}, \bibfnamefont{B.}}, \bibinfo
  {author} {\bibfnamefont{J.~P.~D.}\ \bibnamefont{Abbatt}}, \bibinfo {author}
  {\bibfnamefont{R.~A.}\ \bibnamefont{Cox}}, \bibinfo {author}
  {\bibfnamefont{P.~J.}\ \bibnamefont{Popp}},\ and\ \bibinfo {author}
  {\bibfnamefont{C.}~\bibnamefont{Voigt}}}%
  , \bibinfo {year} {2009},\ \bibfield{title}{%
  \enquote{\bibinfo {title} {Trapping of trace gases by growing ice surfaces
  including surface-saturated adsorption},}\ }%
  \bibfield{journal}{%
  \bibinfo {journal} {J. Geophys. Res.-Atmos.}\ }%
  \textbf{\bibinfo {volume} {114}},\ \bibinfo {pages} {D13306}%
  \bibAnnoteFile{NoStop}{Karcher2009}%
\bibitem[{\citenamefont{Kargel}(2004)}]{kargel2004}%
  \BibitemOpen
  \bibfield{author}{%
  \bibinfo {author} {\bibnamefont{Kargel}, \bibfnamefont{J.~S.}}}%
  , \bibinfo {year} {2004},\ \emph{\bibinfo {title} {Mars --- A Warmer Wetter
  Planet}}\ (\bibinfo {publisher} {Springer})%
  \bibAnnoteFile{NoStop}{kargel2004}%
\bibitem[{\citenamefont{Kaser}(2007)}]{GeorgKaser:2007p26005}%
  \BibitemOpen
  \bibfield{author}{%
  \bibinfo {author} {\bibnamefont{Kaser}, \bibfnamefont{G.}}}%
  , \bibinfo {year} {2007},\ \bibfield{title}{%
  \enquote{\bibinfo {title} {Mountain glaciers},}\ }%
  \bibinfo {journal} {Glacier Science and Environmental Change (First
  Edition)},\ \bibinfo {pages} {268--271}%
  \bibAnnoteFile{NoStop}{GeorgKaser:2007p26005}%
\bibitem[{\citenamefont{Kawada}(1972)}]{kawada1972}%
  \BibitemOpen
\bibfield{journal}{%
    }%
  \bibfield{author}{%
  \bibinfo {author} {\bibnamefont{Kawada}, \bibfnamefont{S.}}}%
  , \bibinfo {year} {1972},\ \bibfield{title}{%
  \enquote{\bibinfo {title} {Dielectric dispersion and phase transition of
  {KOH} doped ice},}\ }%
  \bibfield{journal}{%
  \bibinfo {journal} {J. Phys. Soc. Japan}\ }%
  \textbf{\bibinfo {volume} {32}},\ \bibinfo {pages} {1442}%
  \bibAnnoteFile{NoStop}{kawada1972}%
\bibitem[{\citenamefont{Kawakita}\ \emph{et~al.}(2004)\citenamefont{Kawakita},
  \citenamefont{Watanabe}, \citenamefont{Ootsubo}, \citenamefont{Nakamura},
  \citenamefont{Fuse}, \citenamefont{Takato}, \citenamefont{Sasaki},\ and\
  \citenamefont{Sasaki}}]{kawakita2004}%
  \BibitemOpen
  \bibfield{author}{%
  \bibinfo {author} {\bibnamefont{Kawakita}, \bibfnamefont{H.}}, \bibinfo
  {author} {\bibfnamefont{J.}~\bibnamefont{Watanabe}}, \bibinfo {author}
  {\bibfnamefont{T.}~\bibnamefont{Ootsubo}}, \bibinfo {author}
  {\bibfnamefont{R.}~\bibnamefont{Nakamura}}, \bibinfo {author}
  {\bibfnamefont{T.}~\bibnamefont{Fuse}}, \bibinfo {author}
  {\bibfnamefont{N.}~\bibnamefont{Takato}}, \bibinfo {author}
  {\bibfnamefont{S.}~\bibnamefont{Sasaki}},\ and\ \bibinfo {author}
  {\bibfnamefont{T.}~\bibnamefont{Sasaki}}}%
  , \bibinfo {year} {2004},\ \bibfield{title}{%
  \enquote{\bibinfo {title} {Evidence of icy grains in comet {C/2002} {T7}
  (linear) at 3.52 {AU}},}\ }%
  \bibfield{journal}{%
  \bibinfo {journal} {Astrophys. J.}\ }%
  \textbf{\bibinfo {volume} {601}},\ \bibinfo {pages} {L191--L194}%
  \bibAnnoteFile{NoStop}{kawakita2004}%
\bibitem[{\citenamefont{Kerbrat}\ \emph{et~al.}(2010)\citenamefont{Kerbrat},
  \citenamefont{Huthwelker}, \citenamefont{G{\"a}ggeler},\ and\
  \citenamefont{Ammann}}]{Kerbrat:2010p25446}%
  \BibitemOpen
  \bibfield{author}{%
  \bibinfo {author} {\bibnamefont{Kerbrat}, \bibfnamefont{M.}}, \bibinfo
  {author} {\bibfnamefont{T.}~\bibnamefont{Huthwelker}}, \bibinfo {author}
  {\bibfnamefont{H.}~\bibnamefont{G{\"a}ggeler}},\ and\ \bibinfo {author}
  {\bibfnamefont{M.}~\bibnamefont{Ammann}}}%
  , \bibinfo {year} {2010},\ \bibfield{title}{%
  \enquote{\bibinfo {title} {Interaction of nitrous acid with polycrystalline
  ice: Adsorption on the surface and diffusion into the bulk},}\ }%
  \bibfield{journal}{%
  \bibinfo {journal} {J. Phys. Chem. C}\ }%
  \textbf{\bibinfo {volume} {114}},\ \bibinfo {pages} {2208--2219}%
  \bibAnnoteFile{NoStop}{Kerbrat:2010p25446}%
\bibitem[{\citenamefont{Kerbrat}\ \emph{et~al.}(2008)\citenamefont{Kerbrat},
  \citenamefont{Pinzer}, \citenamefont{Huthwelker},
  \citenamefont{G{\"a}ggeler}, \citenamefont{Ammann},\ and\
  \citenamefont{Schneebeli}}]{kerbrat2008}%
  \BibitemOpen
  \bibfield{author}{%
  \bibinfo {author} {\bibnamefont{Kerbrat}, \bibfnamefont{M.}}, \bibinfo
  {author} {\bibfnamefont{B.}~\bibnamefont{Pinzer}}, \bibinfo {author}
  {\bibfnamefont{T.}~\bibnamefont{Huthwelker}}, \bibinfo {author}
  {\bibfnamefont{H.~W.}\ \bibnamefont{G{\"a}ggeler}}, \bibinfo {author}
  {\bibfnamefont{M.}~\bibnamefont{Ammann}},\ and\ \bibinfo {author}
  {\bibfnamefont{M.}~\bibnamefont{Schneebeli}}}%
  , \bibinfo {year} {2008},\ \bibfield{title}{%
  \enquote{\bibinfo {title} {Measuring the specific surface area of snow with
  {X}-ray tomography and gas adsorption: comparison and implications for
  surface smoothness},}\ }%
  \bibfield{journal}{%
  \bibinfo {journal} {Atmos. Chem. Phys.}\ }%
  \textbf{\bibinfo {volume} {8}},\ \bibinfo {pages} {1261--1275}%
  \bibAnnoteFile{NoStop}{kerbrat2008}%
\bibitem[{\citenamefont{Kessler}\ and\
  \citenamefont{Werner}(2003)}]{kessler2003}%
  \BibitemOpen
  \bibfield{author}{%
  \bibinfo {author} {\bibnamefont{Kessler}, \bibfnamefont{M.~A.}},\ and\
  \bibinfo {author} {\bibfnamefont{B.~T.}\ \bibnamefont{Werner}}}%
  , \bibinfo {year} {2003},\ \bibfield{title}{%
  \enquote{\bibinfo {title} {Self-organization of sorted patterned ground},}\
  }%
  \bibfield{journal}{%
  \bibinfo {journal} {Science}\ }%
  \textbf{\bibinfo {volume} {299}},\ \bibinfo {pages} {380--383}%
  \bibAnnoteFile{NoStop}{kessler2003}%
\bibitem[{\citenamefont{Khalizov}\ \emph{et~al.}(2006)\citenamefont{Khalizov},
  \citenamefont{Earle}, \citenamefont{Johnson}, \citenamefont{Stubley},\ and\
  \citenamefont{Sloan}}]{Khalizov2006}%
  \BibitemOpen
  \bibfield{author}{%
  \bibinfo {author} {\bibnamefont{Khalizov}, \bibfnamefont{A.~F.}}, \bibinfo
  {author} {\bibfnamefont{M.~E.}\ \bibnamefont{Earle}}, \bibinfo {author}
  {\bibfnamefont{W.~J.~W.}\ \bibnamefont{Johnson}}, \bibinfo {author}
  {\bibfnamefont{G.~D.}\ \bibnamefont{Stubley}},\ and\ \bibinfo {author}
  {\bibfnamefont{J.~J.}\ \bibnamefont{Sloan}}}%
  , \bibinfo {year} {2006},\ \bibfield{title}{%
  \enquote{\bibinfo {title} {Development and characterization of a laminar
  aerosol flow tube},}\ }%
  \bibfield{journal}{%
  \bibinfo {journal} {Rev. Sci. Inst.}\ }%
  \textbf{\bibinfo {volume} {77}},\ \bibinfo {pages} {033102}%
  \bibAnnoteFile{NoStop}{Khalizov2006}%
\bibitem[{\citenamefont{{Kirner}}\ \emph{et~al.}(2010)\citenamefont{{Kirner}},
  \citenamefont{{Ruhnke}}, \citenamefont{{Buchholz-Dietsch}},
  \citenamefont{{J{\"o}ckel}}, \citenamefont{{Br{\"u}hl}},\ and\
  \citenamefont{{Steil}}}]{kirner2010}%
  \BibitemOpen
  \bibfield{author}{%
  \bibinfo {author} {\bibnamefont{{Kirner}}, \bibfnamefont{O.}}, \bibinfo
  {author} {\bibfnamefont{R.}~\bibnamefont{{Ruhnke}}}, \bibinfo {author}
  {\bibfnamefont{J.}~\bibnamefont{{Buchholz-Dietsch}}}, \bibinfo {author}
  {\bibfnamefont{P.}~\bibnamefont{{J{\"o}ckel}}}, \bibinfo {author}
  {\bibfnamefont{C.}~\bibnamefont{{Br{\"u}hl}}},\ and\ \bibinfo {author}
  {\bibfnamefont{B.}~\bibnamefont{{Steil}}}}%
  , \bibinfo {year} {2010},\ \bibfield{title}{%
  \enquote{\bibinfo {title} {{Simulation of polar stratospheric clouds in the
  chemistry-climate-model EMAC via the submodel PSC}},}\ }%
  \bibfield{journal}{%
  \bibinfo {journal} {Geoscientific Model Development Discussions}\ }%
  \textbf{\bibinfo {volume} {3}},\ \bibinfo {pages} {2071--2108}%
  \bibAnnoteFile{NoStop}{kirner2010}%
\bibitem[{\citenamefont{Klinger}(1981)}]{klinger1981}%
  \BibitemOpen
  \bibfield{author}{%
  \bibinfo {author} {\bibnamefont{Klinger}, \bibfnamefont{J.}}}%
  , \bibinfo {year} {1981},\ \bibfield{title}{%
  \enquote{\bibinfo {title} {Some consequences of a phase transition of water
  ice on the heat balance of comet nuclei},}\ }%
  \bibfield{journal}{%
  \bibinfo {journal} {Icarus}\ }%
  \textbf{\bibinfo {volume} {47}},\ \bibinfo {pages} {320--324}%
  \bibAnnoteFile{NoStop}{klinger1981}%
\bibitem[{\citenamefont{Knight}\ and\
  \citenamefont{Singer}(2006)}]{knight2006}%
  \BibitemOpen
  \bibfield{author}{%
  \bibinfo {author} {\bibnamefont{Knight}, \bibfnamefont{C.}},\ and\ \bibinfo
  {author} {\bibfnamefont{S.~J.}\ \bibnamefont{Singer}}}%
  , \bibinfo {year} {2006},\ \bibfield{title}{%
  \enquote{\bibinfo {title} {A re-examination of the ice {III/IX} hydrogen bond
  ordering phase transition},}\ }%
  \bibfield{journal}{%
  \bibinfo {journal} {J. Chem. Phys.}\ }%
  \textbf{\bibinfo {volume} {125}},\ \bibinfo {pages} {064506}%
  \bibAnnoteFile{NoStop}{knight2006}%
\bibitem[{\citenamefont{Knopf}\ \emph{et~al.}(2010)\citenamefont{Knopf},
  \citenamefont{Alpert}, \citenamefont{Want},\ and\
  \citenamefont{Y.}}]{Knopf2010}%
  \BibitemOpen
  \bibfield{author}{%
  \bibinfo {author} {\bibnamefont{Knopf}, \bibfnamefont{D.~A.}}, \bibinfo
  {author} {\bibfnamefont{P.~A.}\ \bibnamefont{Alpert}}, \bibinfo {author}
  {\bibfnamefont{B.}~\bibnamefont{Want}},\ and\ \bibinfo {author}
  {\bibfnamefont{Aller~J.}\ \bibnamefont{Y.}}}%
  , \bibinfo {year} {2010},\ \bibfield{title}{%
  \enquote{\bibinfo {title} {Stimulation of ice nucleation by marine
  diatoms},}\ }%
  \bibfield{journal}{%
  \bibinfo {journal} {Nature Geoscience}\ }%
  \textbf{\bibinfo {volume} {1037}},\ \bibinfo {pages} {1--3}%
  \bibAnnoteFile{NoStop}{Knopf2010}%
\bibitem[{\citenamefont{Kobayashi}\ and\
  \citenamefont{Kuroda}(1987)}]{kobayashi1987}%
  \BibitemOpen
  \bibfield{author}{%
  \bibinfo {author} {\bibnamefont{Kobayashi}, \bibfnamefont{T.}},\ and\
  \bibinfo {author} {\bibfnamefont{T.}~\bibnamefont{Kuroda}}}%
  , \bibinfo {year} {1987},\ \enquote{\bibinfo {title} {Snow crystals},}\ in\
  \emph{\bibinfo {booktitle} {Morphology of Crystals}},\ \bibinfo {editor}
  {edited by\ \bibinfo {editor} {\bibfnamefont{I.}~\bibnamefont{Sunagawa}}}\
  (\bibinfo {publisher} {Terra Scientific Publishing, Tokyo})\ pp.\ \bibinfo
  {pages} {649--743}%
  \bibAnnoteFile{NoStop}{kobayashi1987}%
\bibitem[{\citenamefont{Kohl}\ \emph{et~al.}(2000)\citenamefont{Kohl},
  \citenamefont{Mayer},\ and\ \citenamefont{Hallbrucker}}]{kohl2000}%
  \BibitemOpen
  \bibfield{author}{%
  \bibinfo {author} {\bibnamefont{Kohl}, \bibfnamefont{I.}}, \bibinfo {author}
  {\bibfnamefont{E.}~\bibnamefont{Mayer}},\ and\ \bibinfo {author}
  {\bibfnamefont{A.}~\bibnamefont{Hallbrucker}}}%
  , \bibinfo {year} {2000},\ \bibfield{title}{%
  \enquote{\bibinfo {title} {The glassy water -- cubic ice system: a
  comparative study by {X}-ray diffraction and differential scanning
  calorimetry},}\ }%
  \bibfield{journal}{%
  \bibinfo {journal} {Phys. Chem. Chem. Phys.}\ }%
  \textbf{\bibinfo {volume} {2}},\ \bibinfo {pages} {1579--1586}%
  \bibAnnoteFile{NoStop}{kohl2000}%
\bibitem[{\citenamefont{Kohl}\ \emph{et~al.}(2001)\citenamefont{Kohl},
  \citenamefont{Mayer},\ and\ \citenamefont{Hallbrucker}}]{kohl2001}%
  \BibitemOpen
  \bibfield{author}{%
  \bibinfo {author} {\bibnamefont{Kohl}, \bibfnamefont{I.}}, \bibinfo {author}
  {\bibfnamefont{E.}~\bibnamefont{Mayer}},\ and\ \bibinfo {author}
  {\bibfnamefont{A.}~\bibnamefont{Hallbrucker}}}%
  , \bibinfo {year} {2001},\ \bibfield{title}{%
  \enquote{\bibinfo {title} {Ice {XII} forms on compression of hexagonal ice at
  77~{K} via high-density amorphous water},}\ }%
  \bibfield{journal}{%
  \bibinfo {journal} {Phys. Chem. Chem. Phys.}\ }%
  \textbf{\bibinfo {volume} {3}},\ \bibinfo {pages} {602--605}%
  \bibAnnoteFile{NoStop}{kohl2001}%
\bibitem[{\citenamefont{Kolb}\ \emph{et~al.}(2010)\citenamefont{Kolb},
  \citenamefont{Cox}, \citenamefont{Abbatt}, \citenamefont{Ammann},
  \citenamefont{Davis}, \citenamefont{Donaldson}, \citenamefont{Garrett},
  \citenamefont{George}, \citenamefont{Griffiths}, \citenamefont{Hanson},
  \citenamefont{Kulmala}, \citenamefont{McFiggans}, \citenamefont{Poeschl},
  \citenamefont{Riipinen}, \citenamefont{Rossi}, \citenamefont{Rudich},
  \citenamefont{Wagner}, \citenamefont{Winkler}, \citenamefont{Worsnop},\ and\
  \citenamefont{O'~Dowd}}]{Kolb2010}%
  \BibitemOpen
  \bibfield{author}{%
  \bibinfo {author} {\bibnamefont{Kolb}, \bibfnamefont{C.~E.}}, \bibinfo
  {author} {\bibfnamefont{R.~A.}\ \bibnamefont{Cox}}, \bibinfo {author}
  {\bibfnamefont{J.~P.~D.}\ \bibnamefont{Abbatt}}, \bibinfo {author}
  {\bibfnamefont{M.}~\bibnamefont{Ammann}}, \bibinfo {author}
  {\bibfnamefont{E.~J.}\ \bibnamefont{Davis}}, \bibinfo {author}
  {\bibfnamefont{D.~J.}\ \bibnamefont{Donaldson}}, \bibinfo {author}
  {\bibfnamefont{B.~C.}\ \bibnamefont{Garrett}}, \bibinfo {author}
  {\bibfnamefont{C.}~\bibnamefont{George}}, \bibinfo {author}
  {\bibfnamefont{P.~T.}\ \bibnamefont{Griffiths}}, \bibinfo {author}
  {\bibfnamefont{D.~R.}\ \bibnamefont{Hanson}}, \bibinfo {author}
  {\bibfnamefont{M.}~\bibnamefont{Kulmala}}, \bibinfo {author}
  {\bibfnamefont{G.}~\bibnamefont{McFiggans}}, \bibinfo {author}
  {\bibfnamefont{U.}~\bibnamefont{Poeschl}}, \bibinfo {author}
  {\bibfnamefont{I.}~\bibnamefont{Riipinen}}, \bibinfo {author}
  {\bibfnamefont{M.~J.}\ \bibnamefont{Rossi}}, \bibinfo {author}
  {\bibfnamefont{Y.}~\bibnamefont{Rudich}}, \bibinfo {author}
  {\bibfnamefont{P.~E.}\ \bibnamefont{Wagner}}, \bibinfo {author}
  {\bibfnamefont{P.~M.}\ \bibnamefont{Winkler}}, \bibinfo {author}
  {\bibfnamefont{D.~R.}\ \bibnamefont{Worsnop}},\ and\ \bibinfo {author}
  {\bibfnamefont{C.~D.}\ \bibnamefont{O'~Dowd}}}%
  , \bibinfo {year} {2010},\ \bibfield{title}{%
  \enquote{\bibinfo {title} {An overview of current issues in the uptake of
  atmospheric trace gases by aerosols and clouds},}\ }%
  \bibfield{journal}{%
  \bibinfo {journal} {Atmos. Chem. and Phys.}\ }%
  \textbf{\bibinfo {volume} {10}},\ \bibinfo {pages} {10561--10605}%
  \bibAnnoteFile{NoStop}{Kolb2010}%
\bibitem[{\citenamefont{Kong}\ \emph{et~al.}(2011)\citenamefont{Kong},
  \citenamefont{Andersson}, \citenamefont{Markovi\'{c}},\ and\
  \citenamefont{Pettersson}}]{Kong2011}%
  \BibitemOpen
  \bibfield{author}{%
  \bibinfo {author} {\bibnamefont{Kong}, \bibfnamefont{X.}}, \bibinfo {author}
  {\bibfnamefont{P.~U.}\ \bibnamefont{Andersson}}, \bibinfo {author}
  {\bibfnamefont{N.}~\bibnamefont{Markovi\'{c}}},\ and\ \bibinfo {author}
  {\bibfnamefont{J.~B.~C.}\ \bibnamefont{Pettersson}}}%
  , \bibinfo {year} {2011},\ \bibfield{title}{%
  \enquote{\bibinfo {title} {Environmental molecular beam studies of ice
  surface processes},}\ }%
  \bibfield{journal}{%
  \bibinfo {journal} {Physics and Chemistry of Ice}\ }%
  \textbf{\bibinfo {volume} {in press}}%
  \bibAnnoteFile{NoStop}{Kong2011}%
\bibitem[{\citenamefont{K{\"o}nig}(1943)}]{konig1943}%
  \BibitemOpen
  \bibfield{author}{%
  \bibinfo {author} {\bibnamefont{K{\"o}nig}, \bibfnamefont{H.}}}%
  , \bibinfo {year} {1943},\ \bibfield{title}{%
  \enquote{\bibinfo {title} {Eine kubische {Eismodifikation}},}\ }%
  \bibfield{journal}{%
  \bibinfo {journal} {Z. Krist.}\ }%
  \textbf{\bibinfo {volume} {105}},\ \bibinfo {pages} {279--286}%
  \bibAnnoteFile{NoStop}{konig1943}%
\bibitem[{\citenamefont{Koop}(2004)}]{koop2004}%
  \BibitemOpen
  \bibfield{author}{%
  \bibinfo {author} {\bibnamefont{Koop}, \bibfnamefont{T.}}}%
  , \bibinfo {year} {2004},\ \bibfield{title}{%
  \enquote{\bibinfo {title} {Homogeneous ice nucleation in water and aqueous
  solutions},}\ }%
  \bibfield{journal}{%
  \bibinfo {journal} {Z. Phys. Chem.}\ }%
  \textbf{\bibinfo {volume} {218}},\ \bibinfo {pages} {1231--1238}%
  \bibAnnoteFile{NoStop}{koop2004}%
\bibitem[{\citenamefont{Koop}\ \emph{et~al.}(2000)\citenamefont{Koop},
  \citenamefont{Luo}, \citenamefont{Tsias},\ and\
  \citenamefont{Peter}}]{koop2000}%
  \BibitemOpen
  \bibfield{author}{%
  \bibinfo {author} {\bibnamefont{Koop}, \bibfnamefont{T.}}, \bibinfo {author}
  {\bibfnamefont{B.}~\bibnamefont{Luo}}, \bibinfo {author}
  {\bibfnamefont{A.}~\bibnamefont{Tsias}},\ and\ \bibinfo {author}
  {\bibfnamefont{T.}~\bibnamefont{Peter}}}%
  , \bibinfo {year} {2000},\ \bibfield{title}{%
  \enquote{\bibinfo {title} {Water activity as the determinant for homogeneous
  ice nucleation in aqueous solutions},}\ }%
  \bibfield{journal}{%
  \bibinfo {journal} {Nature}\ }%
  \textbf{\bibinfo {volume} {406}},\ \bibinfo {pages} {611--612}%
  \bibAnnoteFile{NoStop}{koop2000}%
\bibitem[{\citenamefont{Korolev}\ \emph{et~al.}(2000)\citenamefont{Korolev},
  \citenamefont{Isaac},\ and\ \citenamefont{Hallett}}]{Korolev2000}%
  \BibitemOpen
  \bibfield{author}{%
  \bibinfo {author} {\bibnamefont{Korolev}, \bibfnamefont{A.}}, \bibinfo
  {author} {\bibfnamefont{G.~A.}\ \bibnamefont{Isaac}},\ and\ \bibinfo {author}
  {\bibfnamefont{J.}~\bibnamefont{Hallett}}}%
  , \bibinfo {year} {2000},\ \bibfield{title}{%
  \enquote{\bibinfo {title} {Ice particle habits in stratiform clouds},}\ }%
  \bibfield{journal}{%
  \bibinfo {journal} {Quart. J. Roy. Meteor. Soc.}\ }%
  \textbf{\bibinfo {volume} {126}},\ \bibinfo {pages} {2873--2902}%
  \bibAnnoteFile{NoStop}{Korolev2000}%
\bibitem[{\citenamefont{Kouchi}\ and\
  \citenamefont{Sirono}(2001)}]{kouchi2001}%
  \BibitemOpen
  \bibfield{author}{%
  \bibinfo {author} {\bibnamefont{Kouchi}, \bibfnamefont{A.}},\ and\ \bibinfo
  {author} {\bibfnamefont{S.}~\bibnamefont{Sirono}}}%
  , \bibinfo {year} {2001},\ \bibfield{title}{%
  \enquote{\bibinfo {title} {Crystallization heat of impure amorphous {H$_2$O}
  ice},}\ }%
  \bibfield{journal}{%
  \bibinfo {journal} {Geophys. Ress. Lett.}\ }%
  \textbf{\bibinfo {volume} {28}},\ \bibinfo {pages} {827--830}%
  \bibAnnoteFile{NoStop}{kouchi2001}%
\bibitem[{\citenamefont{Koza}(2009)}]{koza2009}%
  \BibitemOpen
  \bibfield{author}{%
  \bibinfo {author} {\bibnamefont{Koza}, \bibfnamefont{M.~M.}}}%
  , \bibinfo {year} {2009},\ \bibfield{title}{%
  \enquote{\bibinfo {title} {Transient pronounced density variation in
  amorphous ice structures},}\ }%
  \bibfield{journal}{%
  \bibinfo {journal} {Z. Phys. Chem.}\ }%
  \textbf{\bibinfo {volume} {223}},\ \bibinfo {pages} {979--1000}%
  \bibAnnoteFile{NoStop}{koza2009}%
\bibitem[{\citenamefont{Koza}\ \emph{et~al.}(2003)\citenamefont{Koza},
  \citenamefont{Schober}, \citenamefont{Fischer}, \citenamefont{Hansen},\ and\
  \citenamefont{Fujara}}]{koza2003}%
  \BibitemOpen
  \bibfield{author}{%
  \bibinfo {author} {\bibnamefont{Koza}, \bibfnamefont{M.~M.}}, \bibinfo
  {author} {\bibfnamefont{H.}~\bibnamefont{Schober}}, \bibinfo {author}
  {\bibfnamefont{H.~E.}\ \bibnamefont{Fischer}}, \bibinfo {author}
  {\bibfnamefont{T.}~\bibnamefont{Hansen}},\ and\ \bibinfo {author}
  {\bibfnamefont{F.}~\bibnamefont{Fujara}}}%
  , \bibinfo {year} {2003},\ \bibfield{title}{%
  \enquote{\bibinfo {title} {Kinetics of the high- to low-density amorphous
  water transition},}\ }%
  \bibfield{journal}{%
  \bibinfo {journal} {J. Phys.: Condens. Matter}\ }%
  \textbf{\bibinfo {volume} {15}},\ \bibinfo {pages} {321--332}%
  \bibAnnoteFile{NoStop}{koza2003}%
\bibitem[{\citenamefont{Koza}\ \emph{et~al.}(2000)\citenamefont{Koza},
  \citenamefont{Schober}, \citenamefont{Hansen}, \citenamefont{T\"olle},\ and\
  \citenamefont{Fujara}}]{koza2000}%
  \BibitemOpen
  \bibfield{author}{%
  \bibinfo {author} {\bibnamefont{Koza}, \bibfnamefont{M.~M.}}, \bibinfo
  {author} {\bibfnamefont{H.}~\bibnamefont{Schober}}, \bibinfo {author}
  {\bibfnamefont{T.}~\bibnamefont{Hansen}}, \bibinfo {author}
  {\bibfnamefont{A.}~\bibnamefont{T\"olle}},\ and\ \bibinfo {author}
  {\bibfnamefont{F.}~\bibnamefont{Fujara}}}%
  , \bibinfo {year} {2000},\ \bibfield{title}{%
  \enquote{\bibinfo {title} {Ice {XII} in its second regime of
  metastability},}\ }%
  \bibfield{journal}{%
  \bibinfo {journal} {Phys. Rev. Lett.}\ }%
  \textbf{\bibinfo {volume} {84}},\ \bibinfo {pages} {4112--4115}%
  \bibAnnoteFile{NoStop}{koza2000}%
\bibitem[{\citenamefont{{Koza}}\ \emph{et~al.}(1999)\citenamefont{{Koza}},
  \citenamefont{{Schober}}, \citenamefont{{T{\"o}lle}},
  \citenamefont{{Fujara}},\ and\ \citenamefont{{Hansen}}}]{koza1999}%
  \BibitemOpen
  \bibfield{author}{%
  \bibinfo {author} {\bibnamefont{{Koza}}, \bibfnamefont{M.~M.}}, \bibinfo
  {author} {\bibfnamefont{H.}~\bibnamefont{{Schober}}}, \bibinfo {author}
  {\bibfnamefont{A.}~\bibnamefont{{T{\"o}lle}}}, \bibinfo {author}
  {\bibfnamefont{F.}~\bibnamefont{{Fujara}}},\ and\ \bibinfo {author}
  {\bibfnamefont{T.}~\bibnamefont{{Hansen}}}}%
  , \bibinfo {year} {1999},\ \bibfield{title}{%
  \enquote{\bibinfo {title} {{Formation of ice XII at different conditions}},}\
  }%
  \bibfield{journal}{%
  \bibinfo {journal} {Nature}\ }%
  \textbf{\bibinfo {volume} {397}},\ \bibinfo {pages} {660--661}%
  \bibAnnoteFile{NoStop}{koza1999}%
\bibitem[{\citenamefont{Kr{\"a}mer}\
  \emph{et~al.}(2009)\citenamefont{Kr{\"a}mer}, \citenamefont{Schiller},
  \citenamefont{Afchine}, \citenamefont{Bauer}, \citenamefont{Gensch},
  \citenamefont{Mangold}, \citenamefont{Schlicht}, \citenamefont{Spelten},
  \citenamefont{Ebert}, \citenamefont{M{\"o}hler}, \citenamefont{Saathoff},
  \citenamefont{Sitnikov}, \citenamefont{Borrman}, \citenamefont{deReus},\ and\
  \citenamefont{Spichtinger}}]{kramer2009}%
  \BibitemOpen
  \bibfield{author}{%
  \bibinfo {author} {\bibnamefont{Kr{\"a}mer}, \bibfnamefont{M.}}, \bibinfo
  {author} {\bibfnamefont{C.}~\bibnamefont{Schiller}}, \bibinfo {author}
  {\bibfnamefont{A.}~\bibnamefont{Afchine}}, \bibinfo {author}
  {\bibfnamefont{R.}~\bibnamefont{Bauer}}, \bibinfo {author}
  {\bibfnamefont{I.}~\bibnamefont{Gensch}}, \bibinfo {author}
  {\bibfnamefont{A.}~\bibnamefont{Mangold}}, \bibinfo {author}
  {\bibfnamefont{S.}~\bibnamefont{Schlicht}}, \bibinfo {author}
  {\bibfnamefont{N.}~\bibnamefont{Spelten}}, \bibinfo {author}
  {\bibfnamefont{V.}~\bibnamefont{Ebert}}, \bibinfo {author}
  {\bibfnamefont{O.}~\bibnamefont{M{\"o}hler}}, \bibinfo {author}
  {\bibfnamefont{H.}~\bibnamefont{Saathoff}}, \bibinfo {author}
  {\bibfnamefont{N.}~\bibnamefont{Sitnikov}}, \bibinfo {author}
  {\bibfnamefont{S.}~\bibnamefont{Borrman}}, \bibinfo {author}
  {\bibfnamefont{M.}~\bibnamefont{deReus}},\ and\ \bibinfo {author}
  {\bibfnamefont{P.}~\bibnamefont{Spichtinger}}}%
  , \bibinfo {year} {2009},\ \bibfield{title}{%
  \enquote{\bibinfo {title} {Ice supersaturations and cirrus cloud crystal
  numbers},}\ }%
  \bibfield{journal}{%
  \bibinfo {journal} {Atmos. Chem. Phys.}\ }%
  \textbf{\bibinfo {volume} {9}},\ \bibinfo {pages} {3505--3522}%
  \bibAnnoteFile{NoStop}{kramer2009}%
\bibitem[{\citenamefont{K{\v r}epelov{\'a}}\
  \emph{et~al.}(2010)\citenamefont{K{\v r}epelov{\'a}}, \citenamefont{Newberg},
  \citenamefont{Huthwelker}, \citenamefont{Bluhm},\ and\
  \citenamefont{Ammann}}]{Krepelova2010a}%
  \BibitemOpen
  \bibfield{author}{%
  \bibinfo {author} {\bibnamefont{K{\v r}epelov{\'a}}, \bibfnamefont{A.}},
  \bibinfo {author} {\bibfnamefont{J.}~\bibnamefont{Newberg}}, \bibinfo
  {author} {\bibfnamefont{T.}~\bibnamefont{Huthwelker}}, \bibinfo {author}
  {\bibfnamefont{H.}~\bibnamefont{Bluhm}},\ and\ \bibinfo {author}
  {\bibfnamefont{M.}~\bibnamefont{Ammann}}}%
  , \bibinfo {year} {2010},\ \bibfield{title}{%
  \enquote{\bibinfo {title} {The nature of nitrate at the ice surface studied
  by {XPS} and {NEXAFS}},}\ }%
  \bibfield{journal}{%
  \bibinfo {journal} {Phys. Chem. Chem. Phys.}\ }%
  \textbf{\bibinfo {volume} {12}},\ \bibinfo {pages} {8870--8880}%
  \bibAnnoteFile{NoStop}{Krepelova2010a}%
\bibitem[{\citenamefont{Kuhs}\ \emph{et~al.}(1987)\citenamefont{Kuhs},
  \citenamefont{Bliss},\ and\ \citenamefont{Finney}}]{kuhs1987}%
  \BibitemOpen
  \bibfield{author}{%
  \bibinfo {author} {\bibnamefont{Kuhs}, \bibfnamefont{W.~F.}}, \bibinfo
  {author} {\bibfnamefont{D.~V.}\ \bibnamefont{Bliss}},\ and\ \bibinfo {author}
  {\bibfnamefont{J.~L.}\ \bibnamefont{Finney}}}%
  , \bibinfo {year} {1987},\ \bibfield{title}{%
  \enquote{\bibinfo {title} {High-resolution neutron powder diffraction study
  of ice {Ic}},}\ }%
  \bibinfo {journal} {J. Physique. Colloque C1},\ \bibinfo {pages} {631--636}%
  \bibAnnoteFile{NoStop}{kuhs1987}%
\bibitem[{\citenamefont{Kuhs}\ \emph{et~al.}(1984)\citenamefont{Kuhs},
  \citenamefont{Finney}, \citenamefont{Vettier},\ and\
  \citenamefont{Bliss}}]{kuhs1984}%
  \BibitemOpen
\bibfield{journal}{%
    }%
  \bibfield{author}{%
  \bibinfo {author} {\bibnamefont{Kuhs}, \bibfnamefont{W.~F.}}, \bibinfo
  {author} {\bibfnamefont{J.~L.}\ \bibnamefont{Finney}}, \bibinfo {author}
  {\bibfnamefont{C.}~\bibnamefont{Vettier}},\ and\ \bibinfo {author}
  {\bibfnamefont{D.~V.}\ \bibnamefont{Bliss}}}%
  , \bibinfo {year} {1984},\ \bibfield{title}{%
  \enquote{\bibinfo {title} {Structure and hydrogen ordering in ices {VI},
  {VII} and {VIII} by neutron powder diffraction},}\ }%
  \bibfield{journal}{%
  \bibinfo {journal} {J. Chem. Phys.}\ }%
  \textbf{\bibinfo {volume} {81}},\ \bibinfo {pages} {3612--3623}%
  \bibAnnoteFile{NoStop}{kuhs1984}%
\bibitem[{\citenamefont{Kuhs}\ \emph{et~al.}(2004)\citenamefont{Kuhs},
  \citenamefont{Genov},\ and\ \citenamefont{Staykova}}]{kuhs2004}%
  \BibitemOpen
  \bibfield{author}{%
  \bibinfo {author} {\bibnamefont{Kuhs}, \bibfnamefont{W.~F.}}, \bibinfo
  {author} {\bibfnamefont{G.}~\bibnamefont{Genov}},\ and\ \bibinfo {author}
  {\bibfnamefont{D.~K.}\ \bibnamefont{Staykova}}}%
  , \bibinfo {year} {2004},\ \bibfield{title}{%
  \enquote{\bibinfo {title} {Ice perfection and the onset of anomalous
  preservation of gas hydrates},}\ }%
  \bibfield{journal}{%
  \bibinfo {journal} {Phys. Chem. Chem. Phys.}\ }%
  \textbf{\bibinfo {volume} {6}},\ \bibinfo {pages} {4917--4920}%
  \bibAnnoteFile{NoStop}{kuhs2004}%
\bibitem[{\citenamefont{Kuhs}\ and\ \citenamefont{Lehmann}(1986)}]{kuhs1986}%
  \BibitemOpen
  \bibfield{author}{%
  \bibinfo {author} {\bibnamefont{Kuhs}, \bibfnamefont{W.~F.}},\ and\ \bibinfo
  {author} {\bibfnamefont{M.~S.}\ \bibnamefont{Lehmann}}}%
  , \bibinfo {year} {1986},\ \bibfield{title}{%
  \enquote{\bibinfo {title} {The structure of ice {Ih}},}\ }%
  \bibfield{journal}{%
  \bibinfo {journal} {Water Sci. Rev.}\ }%
  \textbf{\bibinfo {volume} {2}},\ \bibinfo {pages} {1--65}%
  \bibAnnoteFile{NoStop}{kuhs1986}%
\bibitem[{\citenamefont{Kulikov}\ \emph{et~al.}(2010)\citenamefont{Kulikov},
  \citenamefont{Feigin}, \citenamefont{Ignatov}, \citenamefont{Sennikov},
  \citenamefont{Bluszcz},\ and\ \citenamefont{Schrems}}]{kulikov2010}%
  \BibitemOpen
  \bibfield{author}{%
  \bibinfo {author} {\bibnamefont{Kulikov}, \bibfnamefont{M.~Yu.}}, \bibinfo
  {author} {\bibfnamefont{A.~M.}\ \bibnamefont{Feigin}}, \bibinfo {author}
  {\bibfnamefont{S.~K.}\ \bibnamefont{Ignatov}}, \bibinfo {author}
  {\bibfnamefont{P.~G.}\ \bibnamefont{Sennikov}}, \bibinfo {author}
  {\bibfnamefont{Th.}\ \bibnamefont{Bluszcz}},\ and\ \bibinfo {author}
  {\bibfnamefont{O.}~\bibnamefont{Schrems}}}%
  , \bibinfo {year} {2010},\ \bibfield{title}{%
  \enquote{\bibinfo {title} {Technical note: {VUV} photodesorption rates from
  water ice in the 120--150~{K} temperature range --- significance for
  noctilucent clouds},}\ }%
  \bibfield{journal}{%
  \bibinfo {journal} {Atmos. Chem. Phys. Discuss.}\ }%
  \textbf{\bibinfo {volume} {10}},\ \bibinfo {pages} {22653--22668}%
  \bibAnnoteFile{NoStop}{kulikov2010}%
\bibitem[{\citenamefont{Kuo}\ \emph{et~al.}(2001)\citenamefont{Kuo},
  \citenamefont{Coe}, \citenamefont{Singer}, \citenamefont{Band},\ and\
  \citenamefont{{Ojam\"ae}}}]{kuo2001}%
  \BibitemOpen
  \bibfield{author}{%
  \bibinfo {author} {\bibnamefont{Kuo}, \bibfnamefont{J.~L}}, \bibinfo {author}
  {\bibfnamefont{J.~V.}\ \bibnamefont{Coe}}, \bibinfo {author}
  {\bibfnamefont{S.~J.}\ \bibnamefont{Singer}}, \bibinfo {author}
  {\bibfnamefont{Y.~B.}\ \bibnamefont{Band}},\ and\ \bibinfo {author}
  {\bibfnamefont{L.}~\bibnamefont{{Ojam\"ae}}}}%
  , \bibinfo {year} {2001},\ \bibfield{title}{%
  \enquote{\bibinfo {title} {On the use of graph invariants for efficiently
  generating hydrogen bond topologies and predicting physical properties of
  water clusters and ice},}\ }%
  \bibfield{journal}{%
  \bibinfo {journal} {J. Chem. Phys.}\ }%
  \textbf{\bibinfo {volume} {114}},\ \bibinfo {pages} {2527--2540}%
  \bibAnnoteFile{NoStop}{kuo2001}%
\bibitem[{\citenamefont{Kuo}\ and\ \citenamefont{Klein}(2004)}]{kuo2004}%
  \BibitemOpen
  \bibfield{author}{%
  \bibinfo {author} {\bibnamefont{Kuo}, \bibfnamefont{J.~L.}},\ and\ \bibinfo
  {author} {\bibfnamefont{M.~L.}\ \bibnamefont{Klein}}}%
  , \bibinfo {year} {2004},\ \bibfield{title}{%
  \enquote{\bibinfo {title} {Structure of ice {VII} and {VIII}: {A} quantum
  mechanical study},}\ }%
  \bibfield{journal}{%
  \bibinfo {journal} {J. Phys. Chem. B}\ }%
  \textbf{\bibinfo {volume} {109}},\ \bibinfo {pages} {19634--19639}%
  \bibAnnoteFile{NoStop}{kuo2004}%
\bibitem[{\citenamefont{Kuo}\ and\ \citenamefont{Kuhs}(2006)}]{kuo2006}%
  \BibitemOpen
  \bibfield{author}{%
  \bibinfo {author} {\bibnamefont{Kuo}, \bibfnamefont{J.~L.}},\ and\ \bibinfo
  {author} {\bibfnamefont{W.~F.}\ \bibnamefont{Kuhs}}}%
  , \bibinfo {year} {2006},\ \bibfield{title}{%
  \enquote{\bibinfo {title} {A first principles study on the structure of ice
  {VI}: {S}tatic distortion, molecular geometry and proton ordering},}\ }%
  \bibfield{journal}{%
  \bibinfo {journal} {J. Phys. Chem. B}\ }%
  \textbf{\bibinfo {volume} {110}},\ \bibinfo {pages} {3697--3703}%
  \bibAnnoteFile{NoStop}{kuo2006}%
\bibitem[{\citenamefont{Kuo}\ \emph{et~al.}(2005)\citenamefont{Kuo},
  \citenamefont{L.Klein},\ and\ \citenamefont{Kuhs}}]{kuo2005}%
  \BibitemOpen
  \bibfield{author}{%
  \bibinfo {author} {\bibnamefont{Kuo}, \bibfnamefont{J.~L.}}, \bibinfo
  {author} {\bibfnamefont{M.}~\bibnamefont{L.Klein}},\ and\ \bibinfo {author}
  {\bibfnamefont{W.~F.}\ \bibnamefont{Kuhs}}}%
  , \bibinfo {year} {2005},\ \bibfield{title}{%
  \enquote{\bibinfo {title} {The effect of proton disorder on the structure of
  ice {Ih}: {A} theoretical study},}\ }%
  \bibfield{journal}{%
  \bibinfo {journal} {J. Chem. Phys.}\ }%
  \textbf{\bibinfo {volume} {123}},\ \bibinfo {pages} {134505}%
  \bibAnnoteFile{NoStop}{kuo2005}%
\bibitem[{\citenamefont{Kuo}\ and\ \citenamefont{Singer}(2003)}]{kuo2003}%
  \BibitemOpen
  \bibfield{author}{%
  \bibinfo {author} {\bibnamefont{Kuo}, \bibfnamefont{J.~L.}},\ and\ \bibinfo
  {author} {\bibfnamefont{S.~J.}\ \bibnamefont{Singer}}}%
  , \bibinfo {year} {2003},\ \bibfield{title}{%
  \enquote{\bibinfo {title} {Graph invariants for periodic systems: {Towards}
  predicting physical properties from the hydrogen bond topology of ice},}\ }%
  \bibfield{journal}{%
  \bibinfo {journal} {Phys. Rev. E}\ }%
  \textbf{\bibinfo {volume} {67}},\ \bibinfo {pages} {016114}%
  \bibAnnoteFile{NoStop}{kuo2003}%
\bibitem[{\citenamefont{Kurdi}\ and\ \citenamefont{Ponche}(1989)}]{kurdi1989}%
  \BibitemOpen
  \bibfield{author}{%
  \bibinfo {author} {\bibnamefont{Kurdi}, \bibfnamefont{L.}},\ and\ \bibinfo
  {author} {\bibfnamefont{J.}~\bibnamefont{Ponche}}}%
  , \bibinfo {year} {1989},\ \bibfield{title}{%
  \enquote{\bibinfo {title} {Theoretical studies of sulphuric acid monohydrate:
  Neutral or ionic complex?}.}\ }%
  \bibfield{journal}{%
  \bibinfo {journal} {Chem. Phys. Lett.}\ }%
  \textbf{\bibinfo {volume} {158}},\ \bibinfo {pages} {111--115}%
  \bibAnnoteFile{NoStop}{kurdi1989}%
\bibitem[{\citenamefont{Kusaka}\ \emph{et~al.}(1996)\citenamefont{Kusaka},
  \citenamefont{Wang},\ and\ \citenamefont{Seinfeld}}]{kusaka1996}%
  \BibitemOpen
  \bibfield{author}{%
  \bibinfo {author} {\bibnamefont{Kusaka}, \bibfnamefont{I.}}, \bibinfo
  {author} {\bibfnamefont{Z.~G.}\ \bibnamefont{Wang}},\ and\ \bibinfo {author}
  {\bibfnamefont{J.}~\bibnamefont{Seinfeld}}}%
  , \bibinfo {year} {1996},\ \enquote{\bibinfo {title} {Monte {Carlo}
  simulation of homogeneous binary nucleation: Toward a theory of sulfuric
  acid-water system},}\ in\ \emph{\bibinfo {booktitle} {Nucleation and
  Atmospheric Aerosols. Proceedings of the 14th International Conference}}\
  (\bibinfo {publisher} {Pergamon, Oxford})%
  \bibAnnoteFile{NoStop}{kusaka1996}%
\bibitem[{\citenamefont{Kwok}(2006)}]{kwok:2006}%
  \BibitemOpen
  \bibfield{author}{%
  \bibinfo {author} {\bibnamefont{Kwok}, \bibfnamefont{R.}}}%
  , \bibinfo {year} {2006},\ \bibfield{title}{%
  \enquote{\bibinfo {title} {Contrasts in sea ice deformation and production in
  the {Arctic} seasonal and perennial ice zones},}\ }%
  \bibfield{journal}{%
  \bibinfo {journal} {J.~Geophys.~Res.}\ }%
  \textbf{\bibinfo {volume} {111}},\ \bibinfo {pages} {C11S22}%
  \bibAnnoteFile{NoStop}{kwok:2006}%
\bibitem[{\citenamefont{Laaksonen}\
  \emph{et~al.}(1995)\citenamefont{Laaksonen}, \citenamefont{Talanquer},\ and\
  \citenamefont{Oxtoby}}]{laaksonen1995}%
  \BibitemOpen
  \bibfield{author}{%
  \bibinfo {author} {\bibnamefont{Laaksonen}, \bibfnamefont{A.}}, \bibinfo
  {author} {\bibfnamefont{V.}~\bibnamefont{Talanquer}},\ and\ \bibinfo {author}
  {\bibfnamefont{D.~W.}\ \bibnamefont{Oxtoby}}}%
  , \bibinfo {year} {1995},\ \bibfield{title}{%
  \enquote{\bibinfo {title} {Nucleation: {M}easurements, theory, and
  atmospheric applications},}\ }%
  \bibfield{journal}{%
  \bibinfo {journal} {Annu. Rev. Phys. Chem.}\ }%
  \textbf{\bibinfo {volume} {46}},\ \bibinfo {pages} {489--524}%
  \bibAnnoteFile{NoStop}{laaksonen1995}%
\bibitem[{\citenamefont{Lacy}\ \emph{et~al.}(1991)\citenamefont{Lacy},
  \citenamefont{Carr}, \citenamefont{Evans}, \citenamefont{Baas},
  \citenamefont{Achtermann},\ and\ \citenamefont{Arens}}]{lacy1991}%
  \BibitemOpen
  \bibfield{author}{%
  \bibinfo {author} {\bibnamefont{Lacy}, \bibfnamefont{J.~H.}}, \bibinfo
  {author} {\bibfnamefont{J.~S.}\ \bibnamefont{Carr}}, \bibinfo {author}
  {\bibfnamefont{N.~J.~II}\ \bibnamefont{Evans}}, \bibinfo {author}
  {\bibfnamefont{F.}~\bibnamefont{Baas}}, \bibinfo {author}
  {\bibfnamefont{J.~M.}\ \bibnamefont{Achtermann}},\ and\ \bibinfo {author}
  {\bibfnamefont{J.~F.}\ \bibnamefont{Arens}}}%
  , \bibinfo {year} {1991},\ \bibfield{title}{%
  \enquote{\bibinfo {title} {Discovery of interstellar methane: Observations of
  gaseous and solid ch$_4$ absorption toward young stars in molecular
  clouds},}\ }%
  \bibfield{journal}{%
  \bibinfo {journal} {Astrophys. J.}\ }%
  \textbf{\bibinfo {volume} {376}},\ \bibinfo {pages} {556--560}%
  \bibAnnoteFile{NoStop}{lacy1991}%
\bibitem[{\citenamefont{Lacy}\ \emph{et~al.}(1998)\citenamefont{Lacy},
  \citenamefont{Faraji}, \citenamefont{Sandford},\ and\
  \citenamefont{Allamandola}}]{lacy1998}%
  \BibitemOpen
  \bibfield{author}{%
  \bibinfo {author} {\bibnamefont{Lacy}, \bibfnamefont{J.~H.}}, \bibinfo
  {author} {\bibfnamefont{H.}~\bibnamefont{Faraji}}, \bibinfo {author}
  {\bibfnamefont{S.~A.}\ \bibnamefont{Sandford}},\ and\ \bibinfo {author}
  {\bibfnamefont{L.~J.}\ \bibnamefont{Allamandola}}}%
  , \bibinfo {year} {1998},\ \bibfield{title}{%
  \enquote{\bibinfo {title} {Unraveling the 10 micron ``silicate'' feature of
  protostars: The detection of frozen interstellar ammonia},}\ }%
  \bibfield{journal}{%
  \bibinfo {journal} {Astrophys. J. Lett.}\ }%
  \textbf{\bibinfo {volume} {501}},\ \bibinfo {pages} {L105--L109}%
  \bibAnnoteFile{NoStop}{lacy1998}%
\bibitem[{\citenamefont{Lakhtakia}\ and\
  \citenamefont{Messier}(2005)}]{lakhtakia2005}%
  \BibitemOpen
  \bibfield{author}{%
  \bibinfo {author} {\bibnamefont{Lakhtakia}, \bibfnamefont{A.}},\ and\
  \bibinfo {author} {\bibfnamefont{R.}~\bibnamefont{Messier}}}%
  , \bibinfo {year} {2005},\ \emph{\bibinfo {title} {Sculptured Thin Films:
  Nanoengineered Morphology and Optics}}\ (\bibinfo {publisher} {SPIE Press})%
  \bibAnnoteFile{NoStop}{lakhtakia2005}%
\bibitem[{\citenamefont{Latimer}\ \emph{et~al.}(2008)\citenamefont{Latimer},
  \citenamefont{Islam},\ and\ \citenamefont{Price}}]{latimer2008}%
  \BibitemOpen
  \bibfield{author}{%
  \bibinfo {author} {\bibnamefont{Latimer}, \bibfnamefont{E.~R.}}, \bibinfo
  {author} {\bibfnamefont{F.}~\bibnamefont{Islam}},\ and\ \bibinfo {author}
  {\bibfnamefont{S.~D.}\ \bibnamefont{Price}}}%
  , \bibinfo {year} {2008},\ \bibfield{title}{%
  \enquote{\bibinfo {title} {Studies of {HD} formed in excited vibrational
  states from atomic recombination on cold graphite surfaces},}\ }%
  \bibfield{journal}{%
  \bibinfo {journal} {Chem. Phys. Lett.}\ }%
  \textbf{\bibinfo {volume} {455}},\ \bibinfo {pages} {174--177}%
  \bibAnnoteFile{NoStop}{latimer2008}%
\bibitem[{\citenamefont{Lawson}\ \emph{et~al.}(2001)\citenamefont{Lawson},
  \citenamefont{Baker}, \citenamefont{Schmitt},\ and\
  \citenamefont{Jensen}}]{lawson2001}%
  \BibitemOpen
  \bibfield{author}{%
  \bibinfo {author} {\bibnamefont{Lawson}, \bibfnamefont{R.~P.}}, \bibinfo
  {author} {\bibfnamefont{B.~A.}\ \bibnamefont{Baker}}, \bibinfo {author}
  {\bibfnamefont{C.~G.}\ \bibnamefont{Schmitt}},\ and\ \bibinfo {author}
  {\bibfnamefont{T.~L.}\ \bibnamefont{Jensen}}}%
  , \bibinfo {year} {2001},\ \bibfield{title}{%
  \enquote{\bibinfo {title} {An overview of microphysical properties of
  {Arctic} clouds observed in {May} and {July} 1998 during {FIRE} {ACE}},}\ }%
  \bibfield{journal}{%
  \bibinfo {journal} {J. Geophys. Res.-Atmos.}\ }%
  \textbf{\bibinfo {volume} {106}},\ \bibinfo {pages} {14989--15014}%
  \bibAnnoteFile{NoStop}{lawson2001}%
\bibitem[{\citenamefont{Lawson}(2007)}]{WendyLawson:2007p26018}%
  \BibitemOpen
  \bibfield{author}{%
  \bibinfo {author} {\bibnamefont{Lawson}, \bibfnamefont{W.}}}%
  , \bibinfo {year} {2007},\ \bibfield{title}{%
  \enquote{\bibinfo {title} {Environmental conditions, ice facies and glacier
  behaviour},}\ }%
  \bibinfo {journal} {Glacier Science and Environmental Change (First
  Edition)},\ \bibinfo {pages} {319--326}%
  \bibAnnoteFile{NoStop}{WendyLawson:2007p26018}%
\bibitem[{\citenamefont{Lebensohn}\
  \emph{et~al.}(2009)\citenamefont{Lebensohn}, \citenamefont{Montagnat},
  \citenamefont{Mansuy}, \citenamefont{Duval}, \citenamefont{Meysonnier},\ and\
  \citenamefont{Philip}}]{lebensohn2009}%
  \BibitemOpen
\bibfield{journal}{%
    }%
  \bibfield{author}{%
  \bibinfo {author} {\bibnamefont{Lebensohn}, \bibfnamefont{R.~A.}}, \bibinfo
  {author} {\bibfnamefont{M.}~\bibnamefont{Montagnat}}, \bibinfo {author}
  {\bibfnamefont{P.}~\bibnamefont{Mansuy}}, \bibinfo {author}
  {\bibfnamefont{P.}~\bibnamefont{Duval}}, \bibinfo {author}
  {\bibfnamefont{J.}~\bibnamefont{Meysonnier}},\ and\ \bibinfo {author}
  {\bibfnamefont{A.}~\bibnamefont{Philip}}}%
  , \bibinfo {year} {2009},\ \bibfield{title}{%
  \enquote{\bibinfo {title} {Modeling viscoplastic behavior and heterogeneous
  intracrystalline deformation of columnar ice polycrystals},}\ }%
  \bibfield{journal}{%
  \bibinfo {journal} {Acta Mater}\ }%
  \textbf{\bibinfo {volume} {57}},\ \bibinfo {pages} {1405--1415}%
  \bibAnnoteFile{NoStop}{lebensohn2009}%
\bibitem[{\citenamefont{Lee}\ \emph{et~al.}(2004)\citenamefont{Lee},
  \citenamefont{Wilson}, \citenamefont{Baumgardner}, \citenamefont{Herman},
  \citenamefont{Weinstock}, \citenamefont{LaFleur}, \citenamefont{Kok},
  \citenamefont{Anderson}, \citenamefont{Lawson}, \citenamefont{Baker},
  \citenamefont{Strawa}, \citenamefont{Pittman}, \citenamefont{Reeves},\ and\
  \citenamefont{Bui}}]{Lee2004}%
  \BibitemOpen
  \bibfield{author}{%
  \bibinfo {author} {\bibnamefont{Lee}, \bibfnamefont{S.-H.}}, \bibinfo
  {author} {\bibfnamefont{J.~C.}\ \bibnamefont{Wilson}}, \bibinfo {author}
  {\bibfnamefont{D.}~\bibnamefont{Baumgardner}}, \bibinfo {author}
  {\bibfnamefont{R.~L.}\ \bibnamefont{Herman}}, \bibinfo {author}
  {\bibfnamefont{E.~M.}\ \bibnamefont{Weinstock}}, \bibinfo {author}
  {\bibfnamefont{B.~G.}\ \bibnamefont{LaFleur}}, \bibinfo {author}
  {\bibfnamefont{G.}~\bibnamefont{Kok}}, \bibinfo {author}
  {\bibfnamefont{B.}~\bibnamefont{Anderson}}, \bibinfo {author}
  {\bibfnamefont{P.}~\bibnamefont{Lawson}}, \bibinfo {author}
  {\bibfnamefont{B.}~\bibnamefont{Baker}}, \bibinfo {author}
  {\bibfnamefont{A.}~\bibnamefont{Strawa}}, \bibinfo {author}
  {\bibfnamefont{J.~V.}\ \bibnamefont{Pittman}}, \bibinfo {author}
  {\bibfnamefont{J.~M.}\ \bibnamefont{Reeves}},\ and\ \bibinfo {author}
  {\bibfnamefont{T.~P.}\ \bibnamefont{Bui}}}%
  , \bibinfo {year} {2004},\ \bibfield{title}{%
  \enquote{\bibinfo {title} {New particle formation observed in the
  tropical/subtropical cirrus clouds},}\ }%
  \bibfield{journal}{%
  \bibinfo {journal} {J. Geophys. Res.}\ }%
  \textbf{\bibinfo {volume} {109}},\ \bibinfo {pages} {D20209}%
  \bibAnnoteFile{NoStop}{Lee2004}%
\bibitem[{\citenamefont{Legagneux}\ and\
  \citenamefont{Domine}(2005)}]{legagneux2005}%
  \BibitemOpen
  \bibfield{author}{%
  \bibinfo {author} {\bibnamefont{Legagneux}, \bibfnamefont{L.}},\ and\
  \bibinfo {author} {\bibfnamefont{F.}~\bibnamefont{Domine}}}%
  , \bibinfo {year} {2005},\ \bibfield{title}{%
  \enquote{\bibinfo {title} {A mean field model of the decrease of the specific
  surface area of dry snow during isothermal metamorphism},}\ }%
  \bibfield{journal}{%
  \bibinfo {journal} {J. Geophys. Res.}\ }%
  \textbf{\bibinfo {volume} {110}},\ \bibinfo {pages} {F04011}%
  \bibAnnoteFile{NoStop}{legagneux2005}%
\bibitem[{\citenamefont{Lejonthun}\
  \emph{et~al.}(2009)\citenamefont{Lejonthun}, \citenamefont{Svensson},
  \citenamefont{Andersson},\ and\ \citenamefont{Pettersson}}]{Lejonthun2009}%
  \BibitemOpen
  \bibfield{author}{%
  \bibinfo {author} {\bibnamefont{Lejonthun}, \bibfnamefont{L.~S. E.~R.}},
  \bibinfo {author} {\bibfnamefont{E.~A.}\ \bibnamefont{Svensson}}, \bibinfo
  {author} {\bibfnamefont{P.~U.}\ \bibnamefont{Andersson}},\ and\ \bibinfo
  {author} {\bibfnamefont{J.~B.~C.}\ \bibnamefont{Pettersson}}}%
  , \bibinfo {year} {2009},\ \bibfield{title}{%
  \enquote{\bibinfo {title} {Formation of adsorbed layers by deposition of
  dinitrogen pentoxide, nitric acid, and water on graphite},}\ }%
  \bibfield{journal}{%
  \bibinfo {journal} {J. Phys. Chem. C}\ }%
  \textbf{\bibinfo {volume} {113}},\ \bibinfo {pages} {7728--7734}%
  \bibAnnoteFile{NoStop}{Lejonthun2009}%
\bibitem[{\citenamefont{Lemaire}\ \emph{et~al.}(2010)\citenamefont{Lemaire},
  \citenamefont{Vidali}, \citenamefont{Baouche}, \citenamefont{Chehrouri},
  \citenamefont{Chaabouni},\ and\ \citenamefont{Mokrane}}]{lemaire2010}%
  \BibitemOpen
  \bibfield{author}{%
  \bibinfo {author} {\bibnamefont{Lemaire}, \bibfnamefont{J.~L.}}, \bibinfo
  {author} {\bibfnamefont{G.}~\bibnamefont{Vidali}}, \bibinfo {author}
  {\bibfnamefont{S.}~\bibnamefont{Baouche}}, \bibinfo {author}
  {\bibfnamefont{M.}~\bibnamefont{Chehrouri}}, \bibinfo {author}
  {\bibfnamefont{H.}~\bibnamefont{Chaabouni}},\ and\ \bibinfo {author}
  {\bibfnamefont{H.}~\bibnamefont{Mokrane}}}%
  , \bibinfo {year} {2010},\ \bibfield{title}{%
  \enquote{\bibinfo {title} {Competing mechanisms of molecular hydrogen
  formation in conditions relevant to the interstellar medium},}\ }%
  \bibfield{journal}{%
  \bibinfo {journal} {Astrophys. J. Lett.}\ }%
  \textbf{\bibinfo {volume} {725}},\ \bibinfo {pages} {L156--L160}%
  \bibAnnoteFile{NoStop}{lemaire2010}%
\bibitem[{\citenamefont{Lemieux-Dudon}\
  \emph{et~al.}(2010)\citenamefont{Lemieux-Dudon}, \citenamefont{Blayo},
  \citenamefont{Petit}, \citenamefont{Waelbroeck}, \citenamefont{Svensson},
  \citenamefont{Ritz}, \citenamefont{Barnola}, \citenamefont{Narcisi},\ and\
  \citenamefont{Parrenin}}]{LemieuxDudon:2010p26029}%
  \BibitemOpen
  \bibfield{author}{%
  \bibinfo {author} {\bibnamefont{Lemieux-Dudon}, \bibfnamefont{B.}}, \bibinfo
  {author} {\bibfnamefont{E.}~\bibnamefont{Blayo}}, \bibinfo {author}
  {\bibfnamefont{J.-R.}\ \bibnamefont{Petit}}, \bibinfo {author}
  {\bibfnamefont{C.}~\bibnamefont{Waelbroeck}}, \bibinfo {author}
  {\bibfnamefont{A.}~\bibnamefont{Svensson}}, \bibinfo {author}
  {\bibfnamefont{C.}~\bibnamefont{Ritz}}, \bibinfo {author}
  {\bibfnamefont{J.-M.}\ \bibnamefont{Barnola}}, \bibinfo {author}
  {\bibfnamefont{B.}~\bibnamefont{Narcisi}},\ and\ \bibinfo {author}
  {\bibfnamefont{F.}~\bibnamefont{Parrenin}}}%
  , \bibinfo {year} {2010},\ \bibfield{title}{%
  \enquote{\bibinfo {title} {Consistent dating for {Antarctic} and {Greenland}
  ice cores},}\ }%
  \bibfield{journal}{%
  \bibinfo {journal} {Quaternary Sci. Rev.}\ }%
  \textbf{\bibinfo {volume} {29}},\ \bibinfo {pages} {8--20}%
  \bibAnnoteFile{NoStop}{LemieuxDudon:2010p26029}%
\bibitem[{\citenamefont{Lepp\"aranta}(2005)}]{Lepparanta:2005}%
  \BibitemOpen
  \bibfield{author}{%
  \bibinfo {author} {\bibnamefont{Lepp\"aranta}, \bibfnamefont{M.}}}%
  , \bibinfo {year} {2005},\ \emph{\bibinfo {title} {The drift of sea ice}}\
  (\bibinfo {publisher} {Springer-Praxis, Heidelberg, Germany})%
  \bibAnnoteFile{NoStop}{Lepparanta:2005}%
\bibitem[{\citenamefont{Lepp\"aranta}\
  \emph{et~al.}(1995)\citenamefont{Lepp\"aranta}, \citenamefont{Lensu},
  \citenamefont{Kosloff},\ and\ \citenamefont{Veitch}}]{Lepparanta:1995}%
  \BibitemOpen
  \bibfield{author}{%
  \bibinfo {author} {\bibnamefont{Lepp\"aranta}, \bibfnamefont{M.}}, \bibinfo
  {author} {\bibfnamefont{M.}~\bibnamefont{Lensu}}, \bibinfo {author}
  {\bibfnamefont{P.}~\bibnamefont{Kosloff}},\ and\ \bibinfo {author}
  {\bibfnamefont{B.}~\bibnamefont{Veitch}}}%
  , \bibinfo {year} {1995},\ \bibfield{title}{%
  \enquote{\bibinfo {title} {The life story of a first-year sea ice ridge},}\
  }%
  \bibfield{journal}{%
  \bibinfo {journal} {Cold Regions Sci. Tech.}\ }%
  \textbf{\bibinfo {volume} {23}},\ \bibinfo {pages} {279--290}%
  \bibAnnoteFile{NoStop}{Lepparanta:1995}%
\bibitem[{\citenamefont{Leu}\ and\ \citenamefont{Keyser}(2009)}]{Leu2009}%
  \BibitemOpen
  \bibfield{author}{%
  \bibinfo {author} {\bibnamefont{Leu}, \bibfnamefont{M.~T.}},\ and\ \bibinfo
  {author} {\bibfnamefont{L.~F.}\ \bibnamefont{Keyser}}}%
  , \bibinfo {year} {2009},\ \bibfield{title}{%
  \enquote{\bibinfo {title} {Vapor-deposited water and nitric acid ices:
  Physical and chemical properties},}\ }%
  \bibfield{journal}{%
  \bibinfo {journal} {Int. Rev. Phys. Chem.}\ }%
  \textbf{\bibinfo {volume} {28}},\ \bibinfo {pages} {53--109}%
  \bibAnnoteFile{NoStop}{Leu2009}%
\bibitem[{\citenamefont{Li}\ and\ \citenamefont{Somorjai}(2007)}]{Li2007}%
  \BibitemOpen
  \bibfield{author}{%
  \bibinfo {author} {\bibnamefont{Li}, \bibfnamefont{Y.}},\ and\ \bibinfo
  {author} {\bibfnamefont{G.~A.}\ \bibnamefont{Somorjai}}}%
  , \bibinfo {year} {2007},\ \bibfield{title}{%
  \enquote{\bibinfo {title} {Surface premelting of ice},}\ }%
  \bibfield{journal}{%
  \bibinfo {journal} {J. Phys. Chem. C}\ }%
  \textbf{\bibinfo {volume} {111}},\ \bibinfo {pages} {9631--9637}%
  \bibAnnoteFile{NoStop}{Li2007}%
\bibitem[{\citenamefont{Libbrecht}(2005)}]{libbrecht2005}%
  \BibitemOpen
  \bibfield{author}{%
  \bibinfo {author} {\bibnamefont{Libbrecht}, \bibfnamefont{K.~G.}}}%
  , \bibinfo {year} {2005},\ \bibfield{title}{%
  \enquote{\bibinfo {title} {The physics of snow crystals},}\ }%
  \bibfield{journal}{%
  \bibinfo {journal} {Rep. Prog. Phys.}\ }%
  \textbf{\bibinfo {volume} {68}},\ \bibinfo {pages} {855--895}%
  \bibAnnoteFile{NoStop}{libbrecht2005}%
\bibitem[{\citenamefont{{Licandro}}\
  \emph{et~al.}(2011)\citenamefont{{Licandro}}, \citenamefont{{Campins}},
  \citenamefont{{Kelley}}, \citenamefont{{Hargrove}},
  \citenamefont{{Pinilla-Alonso}}, \citenamefont{{Cruikshank}},
  \citenamefont{{Rivkin}},\ and\ \citenamefont{{Emery}}}]{licandro2011}%
  \BibitemOpen
  \bibfield{author}{%
  \bibinfo {author} {\bibnamefont{{Licandro}}, \bibfnamefont{J.}}, \bibinfo
  {author} {\bibfnamefont{H.}~\bibnamefont{{Campins}}}, \bibinfo {author}
  {\bibfnamefont{M.}~\bibnamefont{{Kelley}}}, \bibinfo {author}
  {\bibfnamefont{K.}~\bibnamefont{{Hargrove}}}, \bibinfo {author}
  {\bibfnamefont{N.}~\bibnamefont{{Pinilla-Alonso}}}, \bibinfo {author}
  {\bibfnamefont{D.}~\bibnamefont{{Cruikshank}}}, \bibinfo {author}
  {\bibfnamefont{A.~S.}\ \bibnamefont{{Rivkin}}},\ and\ \bibinfo {author}
  {\bibfnamefont{J.}~\bibnamefont{{Emery}}}}%
  , \bibinfo {year} {2011},\ \bibfield{title}{%
  \enquote{\bibinfo {title} {{(65) Cybele: detection of small silicate grains,
  water-ice, and organics}},}\ }%
  \bibfield{journal}{%
  \bibinfo {journal} {Astron. Astrophys.}\ }%
  \textbf{\bibinfo {volume} {525}},\ \bibinfo {pages} {A34}%
  \bibAnnoteFile{NoStop}{licandro2011}%
\bibitem[{\citenamefont{Liu}\ and\ \citenamefont{Orgel}(1997)}]{liu1997}%
  \BibitemOpen
  \bibfield{author}{%
  \bibinfo {author} {\bibnamefont{Liu}, \bibfnamefont{R.}},\ and\ \bibinfo
  {author} {\bibfnamefont{L.~E.}\ \bibnamefont{Orgel}}}%
  , \bibinfo {year} {1997},\ \bibfield{title}{%
  \enquote{\bibinfo {title} {Efficient oligomerization of negatively-charged
  $\beta$-amino acids at 20$^o${C}},}\ }%
  \bibfield{journal}{%
  \bibinfo {journal} {J. Am. Chem. Soc.}\ }%
  \textbf{\bibinfo {volume} {119}},\ \bibinfo {pages} {4791--4792}%
  \bibAnnoteFile{NoStop}{liu1997}%
\bibitem[{\citenamefont{Lliboutry}(1954)}]{lliboutry1954}%
  \BibitemOpen
  \bibfield{author}{%
  \bibinfo {author} {\bibnamefont{Lliboutry}, \bibfnamefont{L.}}}%
  , \bibinfo {year} {1954},\ \bibfield{title}{%
  \enquote{\bibinfo {title} {The origin of penitentes},}\ }%
  \bibfield{journal}{%
  \bibinfo {journal} {J. Glaciol.}\ }%
  \textbf{\bibinfo {volume} {2}},\ \bibinfo {pages} {331--338}%
  \bibAnnoteFile{NoStop}{lliboutry1954}%
\bibitem[{\citenamefont{Lobban}(1998)}]{lobban1998}%
  \BibitemOpen
  \bibfield{author}{%
  \bibinfo {author} {\bibnamefont{Lobban}, \bibfnamefont{C.}}}%
  , \bibinfo {year} {1998},\ \emph{\bibinfo {title} {Neutron Diffraction
  Studies of Ices}},\ Ph.D. thesis\ (\bibinfo {school} {University of London})%
  \bibAnnoteFile{NoStop}{lobban1998}%
\bibitem[{\citenamefont{Lobban}\ \emph{et~al.}(2000)\citenamefont{Lobban},
  \citenamefont{Finney},\ and\ \citenamefont{Kuhs}}]{lobban2000}%
  \BibitemOpen
  \bibfield{author}{%
  \bibinfo {author} {\bibnamefont{Lobban}, \bibfnamefont{C.}}, \bibinfo
  {author} {\bibfnamefont{J.~L.}\ \bibnamefont{Finney}},\ and\ \bibinfo
  {author} {\bibfnamefont{W.~F.}\ \bibnamefont{Kuhs}}}%
  , \bibinfo {year} {2000},\ \bibfield{title}{%
  \enquote{\bibinfo {title} {The structure and ordering of ices {III} and
  {V}},}\ }%
  \bibfield{journal}{%
  \bibinfo {journal} {J. Chem. Phys.}\ }%
  \textbf{\bibinfo {volume} {112}},\ \bibinfo {pages} {7169--7180}%
  \bibAnnoteFile{NoStop}{lobban2000}%
\bibitem[{\citenamefont{Loerting}\ \emph{et~al.}(2010)\citenamefont{Loerting},
  \citenamefont{Bowron},\ and\ \citenamefont{Finney}}]{loerting2010}%
  \BibitemOpen
  \bibfield{author}{%
  \bibinfo {author} {\bibnamefont{Loerting}, \bibfnamefont{T.}}, \bibinfo
  {author} {\bibfnamefont{D.~T.}\ \bibnamefont{Bowron}},\ and\ \bibinfo
  {author} {\bibfnamefont{J.~L}\ \bibnamefont{Finney}}}%
  , \bibinfo {year} {2010},\ \bibinfo {journal} {unpublished work in progress}%
  \bibAnnoteFile{NoStop}{loerting2010}%
\bibitem[{\citenamefont{Loerting}\ and\
  \citenamefont{Giovambattista}(2006)}]{loerting2006}%
  \BibitemOpen
\bibfield{journal}{%
    }%
  \bibfield{author}{%
  \bibinfo {author} {\bibnamefont{Loerting}, \bibfnamefont{T.}},\ and\ \bibinfo
  {author} {\bibfnamefont{N.}~\bibnamefont{Giovambattista}}}%
  , \bibinfo {year} {2006},\ \bibfield{title}{%
  \enquote{\bibinfo {title} {Amorphous ices: experiments and numerical
  simulations},}\ }%
  \bibfield{journal}{%
  \bibinfo {journal} {J. Phys.: Condens. Matter}\ }%
  \textbf{\bibinfo {volume} {18}},\ \bibinfo {pages} {R919--R977}%
  \bibAnnoteFile{NoStop}{loerting2006}%
\bibitem[{\citenamefont{Loerting}\ \emph{et~al.}(2001)\citenamefont{Loerting},
  \citenamefont{Salzmann}, \citenamefont{Kohl}, \citenamefont{Mayer},\ and\
  \citenamefont{Hallbrucker}}]{loerting2001}%
  \BibitemOpen
  \bibfield{author}{%
  \bibinfo {author} {\bibnamefont{Loerting}, \bibfnamefont{T.}}, \bibinfo
  {author} {\bibfnamefont{C.}~\bibnamefont{Salzmann}}, \bibinfo {author}
  {\bibfnamefont{I.}~\bibnamefont{Kohl}}, \bibinfo {author}
  {\bibfnamefont{E.}~\bibnamefont{Mayer}},\ and\ \bibinfo {author}
  {\bibfnamefont{A.}~\bibnamefont{Hallbrucker}}}%
  , \bibinfo {year} {2001},\ \bibfield{title}{%
  \enquote{\bibinfo {title} {A second distinct structural ``state'' of
  high-density amorphous ice at 77~{K} and 1~bar},}\ }%
  \bibfield{journal}{%
  \bibinfo {journal} {Phys. Chem. Chem. Phys.}\ }%
  \textbf{\bibinfo {volume} {3}},\ \bibinfo {pages} {5355--5357}%
  \bibAnnoteFile{NoStop}{loerting2001}%
\bibitem[{\citenamefont{Loerting}\ \emph{et~al.}(2011)\citenamefont{Loerting},
  \citenamefont{Winkel}, \citenamefont{Seidl}, \citenamefont{Bauer},
  \citenamefont{Mitterdorfer}, \citenamefont{Handle}, \citenamefont{Mayer},
  \citenamefont{Finney},\ and\ \citenamefont{Bowron}}]{loerting2011}%
  \BibitemOpen
  \bibfield{author}{%
  \bibinfo {author} {\bibnamefont{Loerting}, \bibfnamefont{T.}}, \bibinfo
  {author} {\bibfnamefont{K.}~\bibnamefont{Winkel}}, \bibinfo {author}
  {\bibfnamefont{M.}~\bibnamefont{Seidl}}, \bibinfo {author}
  {\bibfnamefont{M.}~\bibnamefont{Bauer}}, \bibinfo {author}
  {\bibfnamefont{C.}~\bibnamefont{Mitterdorfer}}, \bibinfo {author}
  {\bibfnamefont{P.~H.}\ \bibnamefont{Handle}}, \bibinfo {author}
  {\bibfnamefont{E.}~\bibnamefont{Mayer}}, \bibinfo {author}
  {\bibfnamefont{J.~L.}\ \bibnamefont{Finney}},\ and\ \bibinfo {author}
  {\bibfnamefont{D.~T.}\ \bibnamefont{Bowron}}}%
  , \bibinfo {year} {2011},\ \bibfield{title}{%
  \enquote{\bibinfo {title} {How many amorphous ices are there?}.}\ }%
  \bibinfo {journal} {Phys. Chem. Chem. Phys.},\ \bibinfo {pages} {in press}%
  \bibAnnoteFile{NoStop}{loerting2011}%
\bibitem[{\citenamefont{L{\"o}fgren}\
  \emph{et~al.}(1996)\citenamefont{L{\"o}fgren}, \citenamefont{Ahlstr{\"o}m},
  \citenamefont{Chakarov}, \citenamefont{Lausmaa},\ and\
  \citenamefont{Kasemo}}]{lofgren1996}%
  \BibitemOpen
\bibfield{journal}{%
    }%
  \bibfield{author}{%
  \bibinfo {author} {\bibnamefont{L{\"o}fgren}, \bibfnamefont{P.}}, \bibinfo
  {author} {\bibfnamefont{P.}~\bibnamefont{Ahlstr{\"o}m}}, \bibinfo {author}
  {\bibfnamefont{D.~V.}\ \bibnamefont{Chakarov}}, \bibinfo {author}
  {\bibfnamefont{J.}~\bibnamefont{Lausmaa}},\ and\ \bibinfo {author}
  {\bibfnamefont{B.}~\bibnamefont{Kasemo}}}%
  , \bibinfo {year} {1996},\ \bibfield{title}{%
  \enquote{\bibinfo {title} {Substrate dependent sublimation kinetics of
  mesoscopic ice films},}\ }%
  \bibfield{journal}{%
  \bibinfo {journal} {Surf. Sci. Lett.}\ }%
  \textbf{\bibinfo {volume} {367}},\ \bibinfo {pages} {L19--L25}%
  \bibAnnoteFile{NoStop}{lofgren1996}%
\bibitem[{\citenamefont{Lohmann}(2002)}]{lohmann2002}%
  \BibitemOpen
  \bibfield{author}{%
  \bibinfo {author} {\bibnamefont{Lohmann}, \bibfnamefont{U.}}}%
  , \bibinfo {year} {2002},\ \bibfield{title}{%
  \enquote{\bibinfo {title} {A glaciation indirect aerosol effect caused by
  soot aerosols},}\ }%
  \bibfield{journal}{%
  \bibinfo {journal} {Geophys. Res. Lett.}\ }%
  \textbf{\bibinfo {volume} {29}},\ \bibinfo {pages} {1052}%
  \bibAnnoteFile{NoStop}{lohmann2002}%
\bibitem[{\citenamefont{Lohmann}\ and\
  \citenamefont{Feichter}(2005)}]{Lohmann2005}%
  \BibitemOpen
  \bibfield{author}{%
  \bibinfo {author} {\bibnamefont{Lohmann}, \bibfnamefont{U.}},\ and\ \bibinfo
  {author} {\bibfnamefont{J.}~\bibnamefont{Feichter}}}%
  , \bibinfo {year} {2005},\ \bibfield{title}{%
  \enquote{\bibinfo {title} {Global indirect aerosol effects: a review},}\ }%
  \bibfield{journal}{%
  \bibinfo {journal} {Atmos. Chem. Phys.}\ }%
  \textbf{\bibinfo {volume} {5}},\ \bibinfo {pages} {715--737}%
  \bibAnnoteFile{NoStop}{Lohmann2005}%
\bibitem[{\citenamefont{Loulergue}\
  \emph{et~al.}(2007)\citenamefont{Loulergue}, \citenamefont{Parrenin},
  \citenamefont{Blunier}, \citenamefont{Barnola}, \citenamefont{Spahni},
  \citenamefont{Schilt}, \citenamefont{Raisbeck},\ and\
  \citenamefont{Chappellaz}}]{Loulergue:2007p26030}%
  \BibitemOpen
  \bibfield{author}{%
  \bibinfo {author} {\bibnamefont{Loulergue}, \bibfnamefont{L.}}, \bibinfo
  {author} {\bibfnamefont{F.}~\bibnamefont{Parrenin}}, \bibinfo {author}
  {\bibfnamefont{T.}~\bibnamefont{Blunier}}, \bibinfo {author}
  {\bibfnamefont{J.-M.}\ \bibnamefont{Barnola}}, \bibinfo {author}
  {\bibfnamefont{R.}~\bibnamefont{Spahni}}, \bibinfo {author}
  {\bibfnamefont{A.}~\bibnamefont{Schilt}}, \bibinfo {author}
  {\bibfnamefont{G.}~\bibnamefont{Raisbeck}},\ and\ \bibinfo {author}
  {\bibfnamefont{J.}~\bibnamefont{Chappellaz}}}%
  , \bibinfo {year} {2007},\ \bibfield{title}{%
  \enquote{\bibinfo {title} {New constraints on the gas age-ice age difference
  along the epica ice cores, 0--50 kyr},}\ }%
  \bibfield{journal}{%
  \bibinfo {journal} {Clim. Past}\ }%
  \textbf{\bibinfo {volume} {3}},\ \bibinfo {pages} {527--540}%
  \bibAnnoteFile{NoStop}{Loulergue:2007p26030}%
\bibitem[{\citenamefont{Lowe}\ and\ \citenamefont{MacKenzie}(2008)}]{lowe2008}%
  \BibitemOpen
  \bibfield{author}{%
  \bibinfo {author} {\bibnamefont{Lowe}, \bibfnamefont{D.}},\ and\ \bibinfo
  {author} {\bibfnamefont{A.~R.}\ \bibnamefont{MacKenzie}}}%
  , \bibinfo {year} {2008},\ \bibfield{title}{%
  \enquote{\bibinfo {title} {Polar stratospheric cloud microphysics and
  chemistry},}\ }%
  \bibfield{journal}{%
  \bibinfo {journal} {J. Atmos. Solar-Terrestrial Phys.}\ }%
  \textbf{\bibinfo {volume} {70}},\ \bibinfo {pages} {13--40}%
  \bibAnnoteFile{NoStop}{lowe2008}%
\bibitem[{\citenamefont{Lund~Myhre}\
  \emph{et~al.}(2005)\citenamefont{Lund~Myhre}, \citenamefont{Grothe},
  \citenamefont{Gola},\ and\ \citenamefont{Nielsen}}]{lund2005}%
  \BibitemOpen
  \bibfield{author}{%
  \bibinfo {author} {\bibnamefont{Lund~Myhre}, \bibfnamefont{C.}}, \bibinfo
  {author} {\bibfnamefont{H.}~\bibnamefont{Grothe}}, \bibinfo {author}
  {\bibfnamefont{A.}~\bibnamefont{Gola}},\ and\ \bibinfo {author}
  {\bibfnamefont{C.}~\bibnamefont{Nielsen}}}%
  , \bibinfo {year} {2005},\ \bibfield{title}{%
  \enquote{\bibinfo {title} {Optical constants of {HNO}$_3$/{H}$_2${O} and
  {H}$_2${SO}$_4$/{HNO}$_3$/{H}$_2${O} at low temperatures in the infrared
  region},}\ }%
  \bibfield{journal}{%
  \bibinfo {journal} {J. Phys. Chem. A}\ }%
  \textbf{\bibinfo {volume} {109}},\ \bibinfo {pages} {7166--7171}%
  \bibAnnoteFile{NoStop}{lund2005}%
\bibitem[{\citenamefont{Luo}\ and\ \citenamefont{Chiang}(2008)}]{Luo2008}%
  \BibitemOpen
  \bibfield{author}{%
  \bibinfo {author} {\bibnamefont{Luo}, \bibfnamefont{J.}},\ and\ \bibinfo
  {author} {\bibfnamefont{Y.-M.}\ \bibnamefont{Chiang}}}%
  , \bibinfo {year} {2008},\ \bibfield{title}{%
  \enquote{\bibinfo {title} {Wetting and prewetting on ceramic surfaces},}\ }%
  \bibfield{journal}{%
  \bibinfo {journal} {Annu. Rev. Matls Res.}\ }%
  \textbf{\bibinfo {volume} {38}},\ \bibinfo {pages} {227--249}%
  \bibAnnoteFile{NoStop}{Luo2008}%
\bibitem[{\citenamefont{Lynch}\ \emph{et~al.}(2002)\citenamefont{Lynch},
  \citenamefont{Sassen}, \citenamefont{Starr},\ and\
  \citenamefont{Stephens}}]{lynch2002}%
  \BibitemOpen
  \bibfield{author}{%
  \bibinfo {author} {\bibnamefont{Lynch}, \bibfnamefont{D.~K.}}, \bibinfo
  {author} {\bibfnamefont{K.}~\bibnamefont{Sassen}}, \bibinfo {author}
  {\bibfnamefont{D.~{O'C.}}\ \bibnamefont{Starr}},\ and\ \bibinfo {author}
  {\bibfnamefont{G.}~\bibnamefont{Stephens}}}%
  , \bibinfo {year} {2002},\ \emph{\bibinfo {title} {Cirrus}}\ (\bibinfo
  {publisher} {Oxford University Press})%
  \bibAnnoteFile{NoStop}{lynch2002}%
\bibitem[{\citenamefont{Mader}(1992)}]{Mader1992}%
  \BibitemOpen
  \bibfield{author}{%
  \bibinfo {author} {\bibnamefont{Mader}, \bibfnamefont{H.~M.}}}%
  , \bibinfo {year} {1992},\ \bibfield{title}{%
  \enquote{\bibinfo {title} {Observations of the water-vein system in
  polycrystalline ice},}\ }%
  \bibfield{journal}{%
  \bibinfo {journal} {J. Glaciol.}\ }%
  \textbf{\bibinfo {volume} {38}},\ \bibinfo {pages} {333--347}%
  \bibAnnoteFile{NoStop}{Mader1992}%
\bibitem[{\citenamefont{Magee}\ \emph{et~al.}(2006)\citenamefont{Magee},
  \citenamefont{Moyle},\ and\ \citenamefont{Lamb}}]{Magee2006}%
  \BibitemOpen
  \bibfield{author}{%
  \bibinfo {author} {\bibnamefont{Magee}, \bibfnamefont{N.}}, \bibinfo {author}
  {\bibfnamefont{A.~M.}\ \bibnamefont{Moyle}},\ and\ \bibinfo {author}
  {\bibfnamefont{D.}~\bibnamefont{Lamb}}}%
  , \bibinfo {year} {2006},\ \bibfield{title}{%
  \enquote{\bibinfo {title} {Experimental determination of the deposition
  coefficient of small cirrus-like ice crystals near -50$^o${C}},}\ }%
  \bibfield{journal}{%
  \bibinfo {journal} {Geophys. Res. Lett.}\ }%
  \textbf{\bibinfo {volume} {33}},\ \bibinfo {pages} {L17813}%
  \bibAnnoteFile{NoStop}{Magee2006}%
\bibitem[{\citenamefont{Malyk}\ \emph{et~al.}(2007)\citenamefont{Malyk},
  \citenamefont{Kumi}, \citenamefont{Reisler},\ and\
  \citenamefont{Wittig}}]{malyk2007}%
  \BibitemOpen
  \bibfield{author}{%
  \bibinfo {author} {\bibnamefont{Malyk}, \bibfnamefont{S.}}, \bibinfo {author}
  {\bibfnamefont{G.}~\bibnamefont{Kumi}}, \bibinfo {author}
  {\bibfnamefont{H.}~\bibnamefont{Reisler}},\ and\ \bibinfo {author}
  {\bibfnamefont{C.}~\bibnamefont{Wittig}}}%
  , \bibinfo {year} {2007},\ \bibfield{title}{%
  \enquote{\bibinfo {title} {Trapping and release of {CO$_2$} guest molecules
  by amorphous ice},}\ }%
  \bibfield{journal}{%
  \bibinfo {journal} {J. Phys. Chem. A}\ }%
  \textbf{\bibinfo {volume} {111}},\ \bibinfo {pages} {13365--13370}%
  \bibAnnoteFile{NoStop}{malyk2007}%
\bibitem[{\citenamefont{Manca}\ \emph{et~al.}(2001)\citenamefont{Manca},
  \citenamefont{Martin}, \citenamefont{Allouche},\ and\
  \citenamefont{Roubin}}]{manca2001}%
  \BibitemOpen
  \bibfield{author}{%
  \bibinfo {author} {\bibnamefont{Manca}, \bibfnamefont{C.}}, \bibinfo {author}
  {\bibfnamefont{C.}~\bibnamefont{Martin}}, \bibinfo {author}
  {\bibfnamefont{A.}~\bibnamefont{Allouche}},\ and\ \bibinfo {author}
  {\bibfnamefont{P.}~\bibnamefont{Roubin}}}%
  , \bibinfo {year} {2001},\ \bibfield{title}{%
  \enquote{\bibinfo {title} {Experimental and theoretical reinvestigation of
  {CO} adsorption on amorphous ice},}\ }%
  \bibfield{journal}{%
  \bibinfo {journal} {J. Phys. Chem. B}\ }%
  \textbf{\bibinfo {volume} {105}},\ \bibinfo {pages} {12861--12869}%
  \bibAnnoteFile{NoStop}{manca2001}%
\bibitem[{\citenamefont{Mantz}\ \emph{et~al.}(2002)\citenamefont{Mantz},
  \citenamefont{Geiger}, \citenamefont{Molina}, \citenamefont{Molina},\ and\
  \citenamefont{Trout}}]{mantz2002}%
  \BibitemOpen
  \bibfield{author}{%
  \bibinfo {author} {\bibnamefont{Mantz}, \bibfnamefont{Y.~A.}}, \bibinfo
  {author} {\bibfnamefont{F.~M.}\ \bibnamefont{Geiger}}, \bibinfo {author}
  {\bibfnamefont{L.~T.}\ \bibnamefont{Molina}}, \bibinfo {author}
  {\bibfnamefont{M.~J.}\ \bibnamefont{Molina}},\ and\ \bibinfo {author}
  {\bibfnamefont{B.~L.}\ \bibnamefont{Trout}}}%
  , \bibinfo {year} {2002},\ \bibfield{title}{%
  \enquote{\bibinfo {title} {A theoretical study of the interaction of {HCl}
  with crystalline {NAT}},}\ }%
  \bibfield{journal}{%
  \bibinfo {journal} {J. Phys. Chem. A}\ }%
  \textbf{\bibinfo {volume} {106}},\ \bibinfo {pages} {6972}%
  \bibAnnoteFile{NoStop}{mantz2002}%
\bibitem[{\citenamefont{Marecal}\ \emph{et~al.}(2010)\citenamefont{Marecal},
  \citenamefont{Pirre}, \citenamefont{Riviere}, \citenamefont{Pouvesle},
  \citenamefont{Crowley}, \citenamefont{Freitas},\ and\
  \citenamefont{Longo}}]{Marecal2010}%
  \BibitemOpen
  \bibfield{author}{%
  \bibinfo {author} {\bibnamefont{Marecal}, \bibfnamefont{V.}}, \bibinfo
  {author} {\bibfnamefont{M.}~\bibnamefont{Pirre}}, \bibinfo {author}
  {\bibfnamefont{E.~D.}\ \bibnamefont{Riviere}}, \bibinfo {author}
  {\bibfnamefont{N.}~\bibnamefont{Pouvesle}}, \bibinfo {author}
  {\bibfnamefont{J.~N.}\ \bibnamefont{Crowley}}, \bibinfo {author}
  {\bibfnamefont{S.~R.}\ \bibnamefont{Freitas}},\ and\ \bibinfo {author}
  {\bibfnamefont{K.~M.}\ \bibnamefont{Longo}}}%
  , \bibinfo {year} {2010},\ \bibfield{title}{%
  \enquote{\bibinfo {title} {Modelling the reversible uptake of chemical
  species in the gas phase by ice particles formed in a convective cloud},}\ }%
  \bibfield{journal}{%
  \bibinfo {journal} {Atmos. Chem. Phys.}\ }%
  \textbf{\bibinfo {volume} {10}},\ \bibinfo {pages} {4977--5000}%
  \bibAnnoteFile{NoStop}{Marecal2010}%
\bibitem[{\citenamefont{Margesin}(2008)}]{margesin2008}%
  \BibitemOpen
  \bibfield{author}{%
  \bibinfo {author} {\bibnamefont{Margesin}, \bibfnamefont{R.}}}%
  , \bibinfo {year} {2008},\ \emph{\bibinfo {title} {Permafrost soils}}\
  (\bibinfo {publisher} {Springer})%
  \bibAnnoteFile{NoStop}{margesin2008}%
\bibitem[{\citenamefont{Marsan}\ \emph{et~al.}(2011)\citenamefont{Marsan},
  \citenamefont{Weiss}, \citenamefont{M{\'e}taxian}, \citenamefont{Grangeon},
  \citenamefont{Roux},\ and\ \citenamefont{Haapala}}]{Marsan:2011}%
  \BibitemOpen
  \bibfield{author}{%
  \bibinfo {author} {\bibnamefont{Marsan}, \bibfnamefont{D.}}, \bibinfo
  {author} {\bibfnamefont{J.}~\bibnamefont{Weiss}}, \bibinfo {author}
  {\bibfnamefont{J.~P.}\ \bibnamefont{M{\'e}taxian}}, \bibinfo {author}
  {\bibfnamefont{J.}~\bibnamefont{Grangeon}}, \bibinfo {author}
  {\bibfnamefont{P.-F.}\ \bibnamefont{Roux}},\ and\ \bibinfo {author}
  {\bibfnamefont{J.}~\bibnamefont{Haapala}}}%
  , \bibinfo {year} {2011},\ \bibfield{title}{%
  \enquote{\bibinfo {title} {Low frequency bursts of horizontally-polarized
  waves in the arctic sea-ice cover},}\ }%
  \bibfield{journal}{%
  \bibinfo {journal} {J.~Glaciology}\ }%
  \textbf{\bibinfo {volume} {57}},\ \bibinfo {pages} {231--237}%
  \bibAnnoteFile{NoStop}{Marsan:2011}%
\bibitem[{\citenamefont{Marshall}(2007)}]{ShawnJMarshall:2007p26023}%
  \BibitemOpen
  \bibfield{author}{%
  \bibinfo {author} {\bibnamefont{Marshall}, \bibfnamefont{S.~J.}}}%
  , \bibinfo {year} {2007},\ \bibfield{title}{%
  \enquote{\bibinfo {title} {Modelling glacier response to climate change},}\
  }%
  \bibinfo {journal} {Glacier Science and Environmental Change (First
  Edition)},\ \bibinfo {pages} {163--173}%
  \bibAnnoteFile{NoStop}{ShawnJMarshall:2007p26023}%
\bibitem[{\citenamefont{Martin}(2000)}]{martin2000}%
  \BibitemOpen
\bibfield{journal}{%
    }%
  \bibfield{author}{%
  \bibinfo {author} {\bibnamefont{Martin}, \bibfnamefont{S.~T.}}}%
  , \bibinfo {year} {2000},\ \bibfield{title}{%
  \enquote{\bibinfo {title} {Phase transitions of aqueous atmospheric
  particles},}\ }%
  \bibfield{journal}{%
  \bibinfo {journal} {Chem. Rev.}\ }%
  \textbf{\bibinfo {volume} {100}},\ \bibinfo {pages} {3403--3454}%
  \bibAnnoteFile{NoStop}{martin2000}%
\bibitem[{\citenamefont{Maslanik}\ \emph{et~al.}(2007)\citenamefont{Maslanik},
  \citenamefont{Fowler}, \citenamefont{Stroeve}, \citenamefont{Drobot},
  \citenamefont{Zwally}, \citenamefont{Yi},\ and\
  \citenamefont{Emery}}]{Maslanik:2007}%
  \BibitemOpen
  \bibfield{author}{%
  \bibinfo {author} {\bibnamefont{Maslanik}, \bibfnamefont{J.~A.}}, \bibinfo
  {author} {\bibfnamefont{C.}~\bibnamefont{Fowler}}, \bibinfo {author}
  {\bibfnamefont{J.}~\bibnamefont{Stroeve}}, \bibinfo {author}
  {\bibfnamefont{S.}~\bibnamefont{Drobot}}, \bibinfo {author}
  {\bibfnamefont{J.}~\bibnamefont{Zwally}}, \bibinfo {author}
  {\bibfnamefont{D.}~\bibnamefont{Yi}},\ and\ \bibinfo {author}
  {\bibfnamefont{W.}~\bibnamefont{Emery}}}%
  , \bibinfo {year} {2007},\ \bibfield{title}{%
  \enquote{\bibinfo {title} {A younger, thinner ice cover: {I}ncreased
  potential for rapid, extensive sea-ice loss},}\ }%
  \bibfield{journal}{%
  \bibinfo {journal} {Geophys.~Res.~Lett.}\ }%
  \textbf{\bibinfo {volume} {34}},\ \bibinfo {pages} {L24501}%
  \bibAnnoteFile{NoStop}{Maslanik:2007}%
\bibitem[{\citenamefont{{Mastrapa}}\
  \emph{et~al.}(2009)\citenamefont{{Mastrapa}}, \citenamefont{{Sandford}},
  \citenamefont{{Roush}}, \citenamefont{{Cruikshank}},\ and\
  \citenamefont{{Dalle Ore}}}]{mastrapa2009}%
  \BibitemOpen
  \bibfield{author}{%
  \bibinfo {author} {\bibnamefont{{Mastrapa}}, \bibfnamefont{R.~M.}}, \bibinfo
  {author} {\bibfnamefont{S.~A.}\ \bibnamefont{{Sandford}}}, \bibinfo {author}
  {\bibfnamefont{T.~L.}\ \bibnamefont{{Roush}}}, \bibinfo {author}
  {\bibfnamefont{D.~P.}\ \bibnamefont{{Cruikshank}}},\ and\ \bibinfo {author}
  {\bibfnamefont{C.~M.}\ \bibnamefont{{Dalle Ore}}}}%
  , \bibinfo {year} {2009},\ \bibfield{title}{%
  \enquote{\bibinfo {title} {Optical constants of amorphous and crystalline
  {H$_{2}$O}-ice: 2.5-22 $\mu$m (4000--455 cm$^{-1}$) optical constants of
  {H$_{2}$O}-ice},}\ }%
  \bibfield{journal}{%
  \bibinfo {journal} {Astrophys. J.}\ }%
  \textbf{\bibinfo {volume} {701}},\ \bibinfo {pages} {1347--1356}%
  \bibAnnoteFile{NoStop}{mastrapa2009}%
\bibitem[{\citenamefont{Mat{\'e}}\ \emph{et~al.}(2009)\citenamefont{Mat{\'e}},
  \citenamefont{G{\'a}lvez}, \citenamefont{Herrero},\ and\
  \citenamefont{Escribano}}]{mate2009}%
  \BibitemOpen
  \bibfield{author}{%
  \bibinfo {author} {\bibnamefont{Mat{\'e}}, \bibfnamefont{B.}}, \bibinfo
  {author} {\bibfnamefont{O.}~\bibnamefont{G{\'a}lvez}}, \bibinfo {author}
  {\bibfnamefont{V.~J.}\ \bibnamefont{Herrero}},\ and\ \bibinfo {author}
  {\bibfnamefont{R.}~\bibnamefont{Escribano}}}%
  , \bibinfo {year} {2009},\ \bibfield{title}{%
  \enquote{\bibinfo {title} {Infrared spectra and thermodynamic properties of
  {CO$_2$}/methanol ices},}\ }%
  \bibfield{journal}{%
  \bibinfo {journal} {Astrophys. J.}\ }%
  \textbf{\bibinfo {volume} {690}},\ \bibinfo {pages} {486--495}%
  \bibAnnoteFile{NoStop}{mate2009}%
\bibitem[{\citenamefont{Matsuo}\ \emph{et~al.}(1986)\citenamefont{Matsuo},
  \citenamefont{Tajima},\ and\ \citenamefont{Suga}}]{matsuo1986}%
  \BibitemOpen
  \bibfield{author}{%
  \bibinfo {author} {\bibnamefont{Matsuo}, \bibfnamefont{T.}}, \bibinfo
  {author} {\bibfnamefont{Y.}~\bibnamefont{Tajima}},\ and\ \bibinfo {author}
  {\bibfnamefont{H.}~\bibnamefont{Suga}}}%
  , \bibinfo {year} {1986},\ \bibfield{title}{%
  \enquote{\bibinfo {title} {Calorimetric study of a phase transition in
  {D$_2$O} ice {Ih} doped with {KOD}: ice {XI}},}\ }%
  \bibfield{journal}{%
  \bibinfo {journal} {J. Phys. Chem. Solids}\ }%
  \textbf{\bibinfo {volume} {47}},\ \bibinfo {pages} {165--173}%
  \bibAnnoteFile{NoStop}{matsuo1986}%
\bibitem[{\citenamefont{Maxwell}(1871)}]{maxwell1871}%
  \BibitemOpen
  \bibfield{author}{%
  \bibinfo {author} {\bibnamefont{Maxwell}, \bibfnamefont{J.~C.}}}%
  , \bibinfo {year} {1871},\ \bibfield{title}{%
  \enquote{\bibinfo {title} {To the chief musician upon nabla: A tyndallic
  ode},}\ }%
  \bibfield{journal}{%
  \bibinfo {journal} {Nature}\ }%
  \textbf{\bibinfo {volume} {4}},\ \bibinfo {pages} {291}%
  \bibAnnoteFile{NoStop}{maxwell1871}%
\bibitem[{\citenamefont{Mayer}(1985)}]{mayer1985}%
  \BibitemOpen
  \bibfield{author}{%
  \bibinfo {author} {\bibnamefont{Mayer}, \bibfnamefont{E.}}}%
  , \bibinfo {year} {1985},\ \bibfield{title}{%
  \enquote{\bibinfo {title} {New method for vitrifying water and other liquids
  by rapid cooling of their aerosols},}\ }%
  \bibfield{journal}{%
  \bibinfo {journal} {J. Appl. Phys.}\ }%
  \textbf{\bibinfo {volume} {58}},\ \bibinfo {pages} {663--667}%
  \bibAnnoteFile{NoStop}{mayer1985}%
\bibitem[{\citenamefont{Mayer}\ and\
  \citenamefont{Br{\"u}ggeller}(1982)}]{mayer1982}%
  \BibitemOpen
  \bibfield{author}{%
  \bibinfo {author} {\bibnamefont{Mayer}, \bibfnamefont{E.}},\ and\ \bibinfo
  {author} {\bibfnamefont{P.}~\bibnamefont{Br{\"u}ggeller}}}%
  , \bibinfo {year} {1982},\ \bibfield{title}{%
  \enquote{\bibinfo {title} {Vitrification of pure liquid water by high
  pressure jet freezing},}\ }%
  \bibfield{journal}{%
  \bibinfo {journal} {Nature}\ }%
  \textbf{\bibinfo {volume} {298}},\ \bibinfo {pages} {715--718}%
  \bibAnnoteFile{NoStop}{mayer1982}%
\bibitem[{\citenamefont{Mayer}\ and\ \citenamefont{Pletzer}(1986)}]{mayer1986}%
  \BibitemOpen
  \bibfield{author}{%
  \bibinfo {author} {\bibnamefont{Mayer}, \bibfnamefont{E.}},\ and\ \bibinfo
  {author} {\bibfnamefont{R.}~\bibnamefont{Pletzer}}}%
  , \bibinfo {year} {1986},\ \bibfield{title}{%
  \enquote{\bibinfo {title} {Astrophysical implications of amorphous ice --- a
  microporous solid},}\ }%
  \bibfield{journal}{%
  \bibinfo {journal} {Nature}\ }%
  \textbf{\bibinfo {volume} {319}},\ \bibinfo {pages} {298--301}%
  \bibAnnoteFile{NoStop}{mayer1986}%
\bibitem[{\citenamefont{Mazzega}\ \emph{et~al.}(1976)\citenamefont{Mazzega},
  \citenamefont{del Pennino}, \citenamefont{Loria},\ and\
  \citenamefont{Mantovani}}]{mazzega1976}%
  \BibitemOpen
  \bibfield{author}{%
  \bibinfo {author} {\bibnamefont{Mazzega}, \bibfnamefont{E.}}, \bibinfo
  {author} {\bibfnamefont{U.}~\bibnamefont{del Pennino}}, \bibinfo {author}
  {\bibfnamefont{A.}~\bibnamefont{Loria}},\ and\ \bibinfo {author}
  {\bibfnamefont{S.}~\bibnamefont{Mantovani}}}%
  , \bibinfo {year} {1976},\ \bibfield{title}{%
  \enquote{\bibinfo {title} {Volta effect and liquidlike layer at the ice
  surface},}\ }%
  \bibfield{journal}{%
  \bibinfo {journal} {J. Chem. Phys.}\ }%
  \textbf{\bibinfo {volume} {64}},\ \bibinfo {pages} {1028--1031}%
  \bibAnnoteFile{NoStop}{mazzega1976}%
\bibitem[{\citenamefont{{McCleese}}\
  \emph{et~al.}(2010)\citenamefont{{McCleese}}, \citenamefont{{Heavens}},
  \citenamefont{{Schofield}}, \citenamefont{{Abdou}},
  \citenamefont{{Bandfield}}, \citenamefont{{Calcutt}}, \citenamefont{{Irwin}},
  \citenamefont{{Kass}}, \citenamefont{{Kleinb{\"o}hl}},
  \citenamefont{{Lewis}}, \citenamefont{{Paige}}, \citenamefont{{Read}},
  \citenamefont{{Richardson}}, \citenamefont{{Shirley}},
  \citenamefont{{Taylor}}, \citenamefont{{Teanby}},\ and\
  \citenamefont{{Zurek}}}]{mccleese2010}%
  \BibitemOpen
  \bibfield{author}{%
  \bibinfo {author} {\bibnamefont{{McCleese}}, \bibfnamefont{D.~J.}}, \bibinfo
  {author} {\bibfnamefont{N.~G.}\ \bibnamefont{{Heavens}}}, \bibinfo {author}
  {\bibfnamefont{J.~T.}\ \bibnamefont{{Schofield}}}, \bibinfo {author}
  {\bibfnamefont{W.~A.}\ \bibnamefont{{Abdou}}}, \bibinfo {author}
  {\bibfnamefont{J.~L.}\ \bibnamefont{{Bandfield}}}, \bibinfo {author}
  {\bibfnamefont{S.~B.}\ \bibnamefont{{Calcutt}}}, \bibinfo {author}
  {\bibfnamefont{P.~G.~J.}\ \bibnamefont{{Irwin}}}, \bibinfo {author}
  {\bibfnamefont{D.~M.}\ \bibnamefont{{Kass}}}, \bibinfo {author}
  {\bibfnamefont{A.}~\bibnamefont{{Kleinb{\"o}hl}}}, \bibinfo {author}
  {\bibfnamefont{S.~R.}\ \bibnamefont{{Lewis}}}, \bibinfo {author}
  {\bibfnamefont{D.~A.}\ \bibnamefont{{Paige}}}, \bibinfo {author}
  {\bibfnamefont{P.~L.}\ \bibnamefont{{Read}}}, \bibinfo {author}
  {\bibfnamefont{M.~I.}\ \bibnamefont{{Richardson}}}, \bibinfo {author}
  {\bibfnamefont{J.~H.}\ \bibnamefont{{Shirley}}}, \bibinfo {author}
  {\bibfnamefont{F.~W.}\ \bibnamefont{{Taylor}}}, \bibinfo {author}
  {\bibfnamefont{N.}~\bibnamefont{{Teanby}}},\ and\ \bibinfo {author}
  {\bibfnamefont{R.~W.}\ \bibnamefont{{Zurek}}}}%
  , \bibinfo {year} {2010},\ \bibfield{title}{%
  \enquote{\bibinfo {title} {{Structure and dynamics of the Martian lower and
  middle atmosphere as observed by the Mars Climate Sounder: Seasonal
  variations in zonal mean temperature, dust, and water ice aerosols}},}\ }%
  \bibfield{journal}{%
  \bibinfo {journal} {J. Geophys. Res. (Planets)}\ }%
  \textbf{\bibinfo {volume} {115}},\ \bibinfo {pages} {12016}%
  \bibAnnoteFile{NoStop}{mccleese2010}%
\bibitem[{\citenamefont{McNeill}\ \emph{et~al.}(2007)\citenamefont{McNeill},
  \citenamefont{Geiger}, \citenamefont{Loerting}, \citenamefont{Trout},
  \citenamefont{Molina},\ and\ \citenamefont{Molina}}]{mcneill2007}%
  \BibitemOpen
  \bibfield{author}{%
  \bibinfo {author} {\bibnamefont{McNeill}, \bibfnamefont{V.~F.}}, \bibinfo
  {author} {\bibfnamefont{F.~M.}\ \bibnamefont{Geiger}}, \bibinfo {author}
  {\bibfnamefont{T.}~\bibnamefont{Loerting}}, \bibinfo {author}
  {\bibfnamefont{B.~L.}\ \bibnamefont{Trout}}, \bibinfo {author}
  {\bibfnamefont{L.~T.}\ \bibnamefont{Molina}},\ and\ \bibinfo {author}
  {\bibfnamefont{M.}~\bibnamefont{Molina}}}%
  , \bibinfo {year} {2007},\ \bibfield{title}{%
  \enquote{\bibinfo {title} {Interaction of hydrogen chloride with ice
  surfaces: The effect of grain size, surface roughness, and surface
  disorder},}\ }%
  \bibfield{journal}{%
  \bibinfo {journal} {J. Phys. Chem. A}\ }%
  \textbf{\bibinfo {volume} {111}},\ \bibinfo {pages} {6274--6284}%
  \bibAnnoteFile{NoStop}{mcneill2007}%
\bibitem[{\citenamefont{McNeill}\ \emph{et~al.}(2006)\citenamefont{McNeill},
  \citenamefont{Loerting}, \citenamefont{Geiger}, \citenamefont{Trout},\ and\
  \citenamefont{Molina}}]{McNeill:2006tk}%
  \BibitemOpen
  \bibfield{author}{%
  \bibinfo {author} {\bibnamefont{McNeill}, \bibfnamefont{V.~F.}}, \bibinfo
  {author} {\bibfnamefont{T.}~\bibnamefont{Loerting}}, \bibinfo {author}
  {\bibfnamefont{F.~M.}\ \bibnamefont{Geiger}}, \bibinfo {author}
  {\bibfnamefont{B.~L.}\ \bibnamefont{Trout}},\ and\ \bibinfo {author}
  {\bibfnamefont{M.~J.}\ \bibnamefont{Molina}}}%
  , \bibinfo {year} {2006},\ \bibfield{title}{%
  \enquote{\bibinfo {title} {Hydrogen chloride-induced surface disordering on
  ice},}\ }%
  \bibfield{journal}{%
  \bibinfo {journal} {Proc. Natl Acad. Sci. USA}\ }%
  \textbf{\bibinfo {volume} {103}},\ \bibinfo {pages} {9422--9427}%
  \bibAnnoteFile{NoStop}{McNeill:2006tk}%
\bibitem[{\citenamefont{Meierhenrich}(2008)}]{meierhenrich2008}%
  \BibitemOpen
  \bibfield{author}{%
  \bibinfo {author} {\bibnamefont{Meierhenrich}, \bibfnamefont{U.~J.}}}%
  , \bibinfo {year} {2008},\ \emph{\bibinfo {title} {Amino Acids and the
  Asymmetry of Life}}\ (\bibinfo {publisher} {Springer})%
  \bibAnnoteFile{NoStop}{meierhenrich2008}%
\bibitem[{\citenamefont{Meierhenrich}\
  \emph{et~al.}(2004)\citenamefont{Meierhenrich}, \citenamefont{Caro},
  \citenamefont{Bredeh{\"o}ft},\ and\
  \citenamefont{Jessberger}}]{meierhenrich2004}%
  \BibitemOpen
  \bibfield{author}{%
  \bibinfo {author} {\bibnamefont{Meierhenrich}, \bibfnamefont{U.~J.}},
  \bibinfo {author} {\bibfnamefont{G.~M.~M.}\ \bibnamefont{Caro}}, \bibinfo
  {author} {\bibfnamefont{J.~H.}\ \bibnamefont{Bredeh{\"o}ft}},\ and\ \bibinfo
  {author} {\bibfnamefont{E.~K.}\ \bibnamefont{Jessberger}}}%
  , \bibinfo {year} {2004},\ \bibfield{title}{%
  \enquote{\bibinfo {title} {Identification of diamino acids in the {Murchison}
  meteorite},}\ }%
  \bibfield{journal}{%
  \bibinfo {journal} {Proc. Natl Acad. Sci. U.S.A.}\ }%
  \textbf{\bibinfo {volume} {101}},\ \bibinfo {pages} {9182--9186}%
  \bibAnnoteFile{NoStop}{meierhenrich2004}%
\bibitem[{\citenamefont{Mellenthin}\
  \emph{et~al.}(2008)\citenamefont{Mellenthin}, \citenamefont{Karma},\ and\
  \citenamefont{Plapp}}]{Mellenthin2008}%
  \BibitemOpen
  \bibfield{author}{%
  \bibinfo {author} {\bibnamefont{Mellenthin}, \bibfnamefont{J.}}, \bibinfo
  {author} {\bibfnamefont{A.}~\bibnamefont{Karma}},\ and\ \bibinfo {author}
  {\bibfnamefont{M.}~\bibnamefont{Plapp}}}%
  , \bibinfo {year} {2008},\ \bibfield{title}{%
  \enquote{\bibinfo {title} {Phase-field crystal study of grain-boundary
  premelting},}\ }%
  \bibfield{journal}{%
  \bibinfo {journal} {Phys. Rev. B}\ }%
  \textbf{\bibinfo {volume} {78}},\ \bibinfo {pages} {184110}%
  \bibAnnoteFile{NoStop}{Mellenthin2008}%
\bibitem[{\citenamefont{Mellor}(1986)}]{Mellor:1986}%
  \BibitemOpen
  \bibfield{author}{%
  \bibinfo {author} {\bibnamefont{Mellor}, \bibfnamefont{M.~M.}}}%
  , \bibinfo {year} {1986},\ \enquote{\bibinfo {title} {Mechanical behavior of
  sea ice},}\ in\ \emph{\bibinfo {booktitle} {Geophysics of Sea Ice. NATO
  Advanced Science Institutes Series B, Physics, vol 146}},\ \bibinfo {editor}
  {edited by\ \bibinfo {editor}
  {\bibfnamefont{N.}~\bibnamefont{Untersteiner}}}\ (\bibinfo {publisher}
  {Plenium Press, New York})\ pp.\ \bibinfo {pages} {165--281}%
  \bibAnnoteFile{NoStop}{Mellor:1986}%
\bibitem[{\citenamefont{Messier}\ \emph{et~al.}(2000)\citenamefont{Messier},
  \citenamefont{Venugopal},\ and\ \citenamefont{Sunal}}]{messier2000}%
  \BibitemOpen
  \bibfield{author}{%
  \bibinfo {author} {\bibnamefont{Messier}, \bibfnamefont{R.}}, \bibinfo
  {author} {\bibfnamefont{V.~C.}\ \bibnamefont{Venugopal}},\ and\ \bibinfo
  {author} {\bibfnamefont{P.~D.}\ \bibnamefont{Sunal}}}%
  , \bibinfo {year} {2000},\ \bibfield{title}{%
  \enquote{\bibinfo {title} {Origin and evolution of sculptured thin films},}\
  }%
  \bibfield{journal}{%
  \bibinfo {journal} {J. Vac. Sci. Technol. A}\ }%
  \textbf{\bibinfo {volume} {18}},\ \bibinfo {pages} {1538--1545}%
  \bibAnnoteFile{NoStop}{messier2000}%
\bibitem[{\citenamefont{Meyers}\ \emph{et~al.}(1992)\citenamefont{Meyers},
  \citenamefont{Demott},\ and\ \citenamefont{Cotton}}]{Meyers1992}%
  \BibitemOpen
  \bibfield{author}{%
  \bibinfo {author} {\bibnamefont{Meyers}, \bibfnamefont{M.~P.}}, \bibinfo
  {author} {\bibfnamefont{P.~J.}\ \bibnamefont{Demott}},\ and\ \bibinfo
  {author} {\bibfnamefont{W.~R.}\ \bibnamefont{Cotton}}}%
  , \bibinfo {year} {1992},\ \bibfield{title}{%
  \enquote{\bibinfo {title} {New primary ice-nucleation parameterizations in an
  explicit cloud model},}\ }%
  \bibfield{journal}{%
  \bibinfo {journal} {J. Appl. Meteorol.}\ }%
  \textbf{\bibinfo {volume} {31}},\ \bibinfo {pages} {708--721}%
  \bibAnnoteFile{NoStop}{Meyers1992}%
\bibitem[{\citenamefont{Min{\v{c}}eva-{\v{S}}ukarova}\
  \emph{et~al.}(1988)\citenamefont{Min{\v{c}}eva-{\v{S}}ukarova},
  \citenamefont{Slark},\ and\ \citenamefont{Sherman}}]{minceva1988}%
  \BibitemOpen
  \bibfield{author}{%
  \bibinfo {author} {\bibnamefont{Min{\v{c}}eva-{\v{S}}ukarova},
  \bibfnamefont{B.}}, \bibinfo {author}
  {\bibfnamefont{G.}~\bibnamefont{Slark}},\ and\ \bibinfo {author}
  {\bibfnamefont{W.~F.}\ \bibnamefont{Sherman}}}%
  , \bibinfo {year} {1988},\ \bibfield{title}{%
  \enquote{\bibinfo {title} {The {Raman} spectra of the {KOH}-doped ice
  polymorphs: {V} and {VI}},}\ }%
  \bibfield{journal}{%
  \bibinfo {journal} {J. Mol. Struct.}\ }%
  \textbf{\bibinfo {volume} {175}},\ \bibinfo {pages} {289--293}%
  \bibAnnoteFile{NoStop}{minceva1988}%
\bibitem[{\citenamefont{Mishcenko}\
  \emph{et~al.}(1996)\citenamefont{Mishcenko}, \citenamefont{Rossow},
  \citenamefont{Macke},\ and\ \citenamefont{Lacis}}]{mishcenko1996}%
  \BibitemOpen
  \bibfield{author}{%
  \bibinfo {author} {\bibnamefont{Mishcenko}, \bibfnamefont{M.~I.}}, \bibinfo
  {author} {\bibfnamefont{W.~B.}\ \bibnamefont{Rossow}}, \bibinfo {author}
  {\bibfnamefont{A.}~\bibnamefont{Macke}},\ and\ \bibinfo {author}
  {\bibfnamefont{A.~A.}\ \bibnamefont{Lacis}}}%
  , \bibinfo {year} {1996},\ \bibfield{title}{%
  \enquote{\bibinfo {title} {Sensitivity of cirrus cloud albedo, bidirectional
  reflectance and optical thickness retrieval accuracy to ice particle
  shape},}\ }%
  \bibfield{journal}{%
  \bibinfo {journal} {J. Geophys. Res.}\ }%
  \textbf{\bibinfo {volume} {101}},\ \bibinfo {pages} {16973--16985}%
  \bibAnnoteFile{NoStop}{mishcenko1996}%
\bibitem[{\citenamefont{Mishima}\ \emph{et~al.}(1984)\citenamefont{Mishima},
  \citenamefont{Calvert},\ and\ \citenamefont{Whalley}}]{mishima1984}%
  \BibitemOpen
  \bibfield{author}{%
  \bibinfo {author} {\bibnamefont{Mishima}, \bibfnamefont{O.}}, \bibinfo
  {author} {\bibfnamefont{L.~D.}\ \bibnamefont{Calvert}},\ and\ \bibinfo
  {author} {\bibfnamefont{E.}~\bibnamefont{Whalley}}}%
  , \bibinfo {year} {1984},\ \bibfield{title}{%
  \enquote{\bibinfo {title} {`{M}elting ice' {I} at 77~{K} and 10~kbar: a new
  method of making amorphous solids},}\ }%
  \bibfield{journal}{%
  \bibinfo {journal} {Nature}\ }%
  \textbf{\bibinfo {volume} {310}},\ \bibinfo {pages} {393--395}%
  \bibAnnoteFile{NoStop}{mishima1984}%
\bibitem[{\citenamefont{Mishima}\ \emph{et~al.}(1985)\citenamefont{Mishima},
  \citenamefont{Calvert},\ and\ \citenamefont{Whalley}}]{mishima1985}%
  \BibitemOpen
  \bibfield{author}{%
  \bibinfo {author} {\bibnamefont{Mishima}, \bibfnamefont{O.}}, \bibinfo
  {author} {\bibfnamefont{L.~D.}\ \bibnamefont{Calvert}},\ and\ \bibinfo
  {author} {\bibfnamefont{E.}~\bibnamefont{Whalley}}}%
  , \bibinfo {year} {1985},\ \bibfield{title}{%
  \enquote{\bibinfo {title} {An apparently first-order transition between two
  amorphous phases of ice induced by pressure},}\ }%
  \bibfield{journal}{%
  \bibinfo {journal} {Nature}\ }%
  \textbf{\bibinfo {volume} {314}},\ \bibinfo {pages} {76--78}%
  \bibAnnoteFile{NoStop}{mishima1985}%
\bibitem[{\citenamefont{Mishima}\ and\
  \citenamefont{Stanley}(1998)}]{mishima1998}%
  \BibitemOpen
  \bibfield{author}{%
  \bibinfo {author} {\bibnamefont{Mishima}, \bibfnamefont{O.}},\ and\ \bibinfo
  {author} {\bibfnamefont{H.~E.}\ \bibnamefont{Stanley}}}%
  , \bibinfo {year} {1998},\ \bibfield{title}{%
  \enquote{\bibinfo {title} {The relationship between liquid, supercooled and
  glassy water},}\ }%
  \bibfield{journal}{%
  \bibinfo {journal} {Nature}\ }%
  \textbf{\bibinfo {volume} {396}},\ \bibinfo {pages} {329--335}%
  \bibAnnoteFile{NoStop}{mishima1998}%
\bibitem[{\citenamefont{Mishima}\ and\
  \citenamefont{Suzuki}(2002)}]{mishima2002}%
  \BibitemOpen
  \bibfield{author}{%
  \bibinfo {author} {\bibnamefont{Mishima}, \bibfnamefont{O.}},\ and\ \bibinfo
  {author} {\bibfnamefont{Y.}~\bibnamefont{Suzuki}}}%
  , \bibinfo {year} {2002},\ \bibfield{title}{%
  \enquote{\bibinfo {title} {Propagation of the polyamorphic transition of ice
  and the liquid--liquid critical point},}\ }%
  \bibfield{journal}{%
  \bibinfo {journal} {Nature}\ }%
  \textbf{\bibinfo {volume} {419}},\ \bibinfo {pages} {599--603}%
  \bibAnnoteFile{NoStop}{mishima2002}%
\bibitem[{\citenamefont{Mitlin}\ and\ \citenamefont{Leung}(2002)}]{mitlin2002}%
  \BibitemOpen
  \bibfield{author}{%
  \bibinfo {author} {\bibnamefont{Mitlin}, \bibfnamefont{S.}},\ and\ \bibinfo
  {author} {\bibfnamefont{K.~T.}\ \bibnamefont{Leung}}}%
  , \bibinfo {year} {2002},\ \bibfield{title}{%
  \enquote{\bibinfo {title} {Film growth of ice by vapor deposition at
  128--185{K} studied by {Fourier} transform infrared reflection-absorption
  spectroscopy: Evolution of the {OH} stretch and the dangling bond with film
  thickness},}\ }%
  \bibfield{journal}{%
  \bibinfo {journal} {J. Phys. Chem. B}\ }%
  \textbf{\bibinfo {volume} {106}},\ \bibinfo {pages} {6234--6247}%
  \bibAnnoteFile{NoStop}{mitlin2002}%
\bibitem[{\citenamefont{M{\"o}hler}\
  \emph{et~al.}(2008)\citenamefont{M{\"o}hler}, \citenamefont{Benz},
  \citenamefont{Saathoff}, \citenamefont{Schnaiter}, \citenamefont{Wagner},
  \citenamefont{Schneider}, \citenamefont{Walter}, \citenamefont{Ebert},\ and\
  \citenamefont{Wagner}}]{Mohler2008}%
  \BibitemOpen
  \bibfield{author}{%
  \bibinfo {author} {\bibnamefont{M{\"o}hler}, \bibfnamefont{O.}}, \bibinfo
  {author} {\bibfnamefont{S.}~\bibnamefont{Benz}}, \bibinfo {author}
  {\bibfnamefont{H.}~\bibnamefont{Saathoff}}, \bibinfo {author}
  {\bibfnamefont{M.}~\bibnamefont{Schnaiter}}, \bibinfo {author}
  {\bibfnamefont{R.}~\bibnamefont{Wagner}}, \bibinfo {author}
  {\bibfnamefont{J.}~\bibnamefont{Schneider}}, \bibinfo {author}
  {\bibfnamefont{S.}~\bibnamefont{Walter}}, \bibinfo {author}
  {\bibfnamefont{V.}~\bibnamefont{Ebert}},\ and\ \bibinfo {author}
  {\bibfnamefont{S.}~\bibnamefont{Wagner}}}%
  , \bibinfo {year} {2008},\ \bibfield{title}{%
  \enquote{\bibinfo {title} {The effect of organic coating on the heterogeneous
  ice nucleation efficiency of mineral dust aerosols},}\ }%
  \bibfield{journal}{%
  \bibinfo {journal} {Environ. Res. Lett.}\ }%
  \textbf{\bibinfo {volume} {3}},\ \bibinfo {pages} {025007}%
  \bibAnnoteFile{NoStop}{Mohler2008}%
\bibitem[{\citenamefont{M{\"o}hler}\
  \emph{et~al.}(2006{\natexlab{a}})\citenamefont{M{\"o}hler},
  \citenamefont{Bunz},\ and\ \citenamefont{Stetzer}}]{mohler2006}%
  \BibitemOpen
  \bibfield{author}{%
  \bibinfo {author} {\bibnamefont{M{\"o}hler}, \bibfnamefont{O.}}, \bibinfo
  {author} {\bibfnamefont{H.}~\bibnamefont{Bunz}},\ and\ \bibinfo {author}
  {\bibfnamefont{O.}~\bibnamefont{Stetzer}}}%
  , \bibinfo {year} {2006}{\natexlab{a}},\ \bibfield{title}{%
  \enquote{\bibinfo {title} {Homogeneous nucleation rates of nitric acid
  dihydrate ({NAD}) at simulated stratospheric conditions --- {P}art {II}:
  {M}odelling},}\ }%
  \bibfield{journal}{%
  \bibinfo {journal} {Atmos. Chem. Phys. Discuss.}\ }%
  \textbf{\bibinfo {volume} {6}},\ \bibinfo {pages} {2119--2149}%
  \bibAnnoteFile{NoStop}{mohler2006}%
\bibitem[{\citenamefont{M\"{o}hler}\
  \emph{et~al.}(2007)\citenamefont{M\"{o}hler}, \citenamefont{DeMott},
  \citenamefont{Vali},\ and\ \citenamefont{Levin}}]{mohler2007}%
  \BibitemOpen
  \bibfield{author}{%
  \bibinfo {author} {\bibnamefont{M\"{o}hler}, \bibfnamefont{O.}}, \bibinfo
  {author} {\bibfnamefont{P.~J.}\ \bibnamefont{DeMott}}, \bibinfo {author}
  {\bibfnamefont{G.}~\bibnamefont{Vali}},\ and\ \bibinfo {author}
  {\bibfnamefont{Z.}~\bibnamefont{Levin}}}%
  , \bibinfo {year} {2007},\ \bibfield{title}{%
  \enquote{\bibinfo {title} {Microbiology and atmospheric processes: the role
  of biological particles in cloud physics},}\ }%
  \bibfield{journal}{%
  \bibinfo {journal} {Biogeosci.}\ }%
  \textbf{\bibinfo {volume} {4}},\ \bibinfo {pages} {1059--1071}%
  \bibAnnoteFile{NoStop}{mohler2007}%
\bibitem[{\citenamefont{M{\"o}hler}\
  \emph{et~al.}(2006{\natexlab{b}})\citenamefont{M{\"o}hler},
  \citenamefont{Field}, \citenamefont{Connolly}, \citenamefont{Benz},
  \citenamefont{Saathoff}, \citenamefont{Schnaiter}, \citenamefont{Wagner},
  \citenamefont{Cotton}, \citenamefont{Kramer}, \citenamefont{Mangold},\ and\
  \citenamefont{Heymsfield}}]{Mohler2006_2}%
  \BibitemOpen
  \bibfield{author}{%
  \bibinfo {author} {\bibnamefont{M{\"o}hler}, \bibfnamefont{O.}}, \bibinfo
  {author} {\bibfnamefont{P.~R.}\ \bibnamefont{Field}}, \bibinfo {author}
  {\bibfnamefont{P.}~\bibnamefont{Connolly}}, \bibinfo {author}
  {\bibfnamefont{S.}~\bibnamefont{Benz}}, \bibinfo {author}
  {\bibfnamefont{H.}~\bibnamefont{Saathoff}}, \bibinfo {author}
  {\bibfnamefont{M.}~\bibnamefont{Schnaiter}}, \bibinfo {author}
  {\bibfnamefont{R.}~\bibnamefont{Wagner}}, \bibinfo {author}
  {\bibfnamefont{R.}~\bibnamefont{Cotton}}, \bibinfo {author}
  {\bibfnamefont{M.}~\bibnamefont{Kramer}}, \bibinfo {author}
  {\bibfnamefont{A.}~\bibnamefont{Mangold}},\ and\ \bibinfo {author}
  {\bibfnamefont{A.~J.}\ \bibnamefont{Heymsfield}}}%
  , \bibinfo {year} {2006}{\natexlab{b}},\ \bibfield{title}{%
  \enquote{\bibinfo {title} {Efficiency of the deposition mode ice nucleation
  on mineral dust particles},}\ }%
  \bibfield{journal}{%
  \bibinfo {journal} {Atmos. Chem. Phys.}\ }%
  \textbf{\bibinfo {volume} {6}},\ \bibinfo {pages} {3007--3021}%
  \bibAnnoteFile{NoStop}{Mohler2006_2}%
\bibitem[{\citenamefont{M{\"o}hler}\
  \emph{et~al.}(2005)\citenamefont{M{\"o}hler}, \citenamefont{Linke},
  \citenamefont{Saathoff}, \citenamefont{Schnaiter}, \citenamefont{Wagner},
  \citenamefont{Mangold}, \citenamefont{Kr{\"a}mer},\ and\
  \citenamefont{Schurath}}]{Mohler2005}%
  \BibitemOpen
  \bibfield{author}{%
  \bibinfo {author} {\bibnamefont{M{\"o}hler}, \bibfnamefont{O.}}, \bibinfo
  {author} {\bibfnamefont{C.}~\bibnamefont{Linke}}, \bibinfo {author}
  {\bibfnamefont{H.}~\bibnamefont{Saathoff}}, \bibinfo {author}
  {\bibfnamefont{M.}~\bibnamefont{Schnaiter}}, \bibinfo {author}
  {\bibfnamefont{R.}~\bibnamefont{Wagner}}, \bibinfo {author}
  {\bibfnamefont{A.}~\bibnamefont{Mangold}}, \bibinfo {author}
  {\bibfnamefont{M.}~\bibnamefont{Kr{\"a}mer}},\ and\ \bibinfo {author}
  {\bibfnamefont{U.}~\bibnamefont{Schurath}}}%
  , \bibinfo {year} {2005},\ \bibfield{title}{%
  \enquote{\bibinfo {title} {Ice nucleation on flame soot aerosol of different
  organic carbon content},}\ }%
  \bibfield{journal}{%
  \bibinfo {journal} {Meteorol. Z.}\ }%
  \textbf{\bibinfo {volume} {14}},\ \bibinfo {pages} {477--484}%
  \bibAnnoteFile{NoStop}{Mohler2005}%
\bibitem[{\citenamefont{Mokrane}\ \emph{et~al.}(2009)\citenamefont{Mokrane},
  \citenamefont{Chaabouni}, \citenamefont{Accolla}, \citenamefont{Congiu},
  \citenamefont{Dulieu}, \citenamefont{Chehrouri},\ and\
  \citenamefont{Lemaire}}]{mokrane2009}%
  \BibitemOpen
  \bibfield{author}{%
  \bibinfo {author} {\bibnamefont{Mokrane}, \bibfnamefont{H.}}, \bibinfo
  {author} {\bibfnamefont{H.}~\bibnamefont{Chaabouni}}, \bibinfo {author}
  {\bibfnamefont{M.}~\bibnamefont{Accolla}}, \bibinfo {author}
  {\bibfnamefont{E.}~\bibnamefont{Congiu}}, \bibinfo {author}
  {\bibfnamefont{F.}~\bibnamefont{Dulieu}}, \bibinfo {author}
  {\bibfnamefont{M.}~\bibnamefont{Chehrouri}},\ and\ \bibinfo {author}
  {\bibfnamefont{J.~L.}\ \bibnamefont{Lemaire}}}%
  , \bibinfo {year} {2009},\ \bibfield{title}{%
  \enquote{\bibinfo {title} {Experimental evidence for water formation via
  ozone hydrogenation on dust grains at 10~{K}},}\ }%
  \bibfield{journal}{%
  \bibinfo {journal} {Astrophys. J. Lett.}\ }%
  \textbf{\bibinfo {volume} {705}},\ \bibinfo {pages} {L195--L198}%
  \bibAnnoteFile{NoStop}{mokrane2009}%
\bibitem[{\citenamefont{Molina}\ \emph{et~al.}(1987)\citenamefont{Molina},
  \citenamefont{Tso}, \citenamefont{Molina},\ and\
  \citenamefont{Wang}}]{molina1987}%
  \BibitemOpen
  \bibfield{author}{%
  \bibinfo {author} {\bibnamefont{Molina}, \bibfnamefont{M.~J.}}, \bibinfo
  {author} {\bibfnamefont{T.~L.}\ \bibnamefont{Tso}}, \bibinfo {author}
  {\bibfnamefont{L.~T.}\ \bibnamefont{Molina}},\ and\ \bibinfo {author}
  {\bibfnamefont{F.~C.~Y.}\ \bibnamefont{Wang}}}%
  , \bibinfo {year} {1987},\ \bibfield{title}{%
  \enquote{\bibinfo {title} {Antarctic stratospheric chemistry of chlorine
  nitrate, hydrogen chloride, and ice: Release of active chlorine},}\ }%
  \bibfield{journal}{%
  \bibinfo {journal} {Science}\ }%
  \textbf{\bibinfo {volume} {238}},\ \bibinfo {pages} {1253--1257}%
  \bibAnnoteFile{NoStop}{molina1987}%
\bibitem[{\citenamefont{Monnard}\ \emph{et~al.}(2003)\citenamefont{Monnard},
  \citenamefont{Kanavarioti},\ and\ \citenamefont{Deamer}}]{monnard2003}%
  \BibitemOpen
  \bibfield{author}{%
  \bibinfo {author} {\bibnamefont{Monnard}, \bibfnamefont{P.}}, \bibinfo
  {author} {\bibfnamefont{A.}~\bibnamefont{Kanavarioti}},\ and\ \bibinfo
  {author} {\bibfnamefont{D.~W.}\ \bibnamefont{Deamer}}}%
  , \bibinfo {year} {2003},\ \bibfield{title}{%
  \enquote{\bibinfo {title} {Eutectic phase polymerization of activated
  ribonucleotide mixtures yields quasi-equimolar incorporation of purine and
  pyrimidine nucleobases},}\ }%
  \bibfield{journal}{%
  \bibinfo {journal} {J. Am. Chem. Soc.}\ }%
  \textbf{\bibinfo {volume} {125}},\ \bibinfo {pages} {13734--13740}%
  \bibAnnoteFile{NoStop}{monnard2003}%
\bibitem[{\citenamefont{Moore}\ \emph{et~al.}(2010)\citenamefont{Moore},
  \citenamefont{de~la Llave}, \citenamefont{Welke}, \citenamefont{Scherlis},\
  and\ \citenamefont{Molinero}}]{moore2010}%
  \BibitemOpen
  \bibfield{author}{%
  \bibinfo {author} {\bibnamefont{Moore}, \bibfnamefont{E.~B.}}, \bibinfo
  {author} {\bibfnamefont{E.}~\bibnamefont{de~la Llave}}, \bibinfo {author}
  {\bibfnamefont{K.}~\bibnamefont{Welke}}, \bibinfo {author}
  {\bibfnamefont{D.~A.}\ \bibnamefont{Scherlis}},\ and\ \bibinfo {author}
  {\bibfnamefont{V.}~\bibnamefont{Molinero}}}%
  , \bibinfo {year} {2010},\ \bibfield{title}{%
  \enquote{\bibinfo {title} {Freezing, melting and structure of ice in a
  hydrophilic nanopore},}\ }%
  \bibfield{journal}{%
  \bibinfo {journal} {Phys. Chem. Chem. Phys.}\ }%
  \textbf{\bibinfo {volume} {12}},\ \bibinfo {pages} {4124--4134}%
  \bibAnnoteFile{NoStop}{moore2010}%
\bibitem[{\citenamefont{Moore}\ and\
  \citenamefont{Molinero}(2011)}]{moore2011}%
  \BibitemOpen
  \bibfield{author}{%
  \bibinfo {author} {\bibnamefont{Moore}, \bibfnamefont{E.~B.}},\ and\ \bibinfo
  {author} {\bibfnamefont{V.}~\bibnamefont{Molinero}}}%
  , \bibinfo {year} {2011},\ \bibfield{title}{%
  \enquote{\bibinfo {title} {Is it cubic? {I}ce crystallization from deeply
  supercooled water},}\ }%
  \bibinfo {journal} {Phys. Chem. Chem. Phys.}%
  \bibAnnoteFile{Stop}{moore2011}%
\bibitem[{\citenamefont{Moore}\ \emph{et~al.}(1994)\citenamefont{Moore},
  \citenamefont{Ferrante}, \citenamefont{Hudson}, \citenamefont{{Nuth III}},\
  and\ \citenamefont{Donn}}]{moore1994}%
  \BibitemOpen
\bibfield{journal}{%
    }%
  \bibfield{author}{%
  \bibinfo {author} {\bibnamefont{Moore}, \bibfnamefont{M.~H.}}, \bibinfo
  {author} {\bibfnamefont{R.~F.}\ \bibnamefont{Ferrante}}, \bibinfo {author}
  {\bibfnamefont{R.~L.}\ \bibnamefont{Hudson}}, \bibinfo {author}
  {\bibfnamefont{J.~A.}\ \bibnamefont{{Nuth III}}},\ and\ \bibinfo {author}
  {\bibfnamefont{B.}~\bibnamefont{Donn}}}%
  , \bibinfo {year} {1994},\ \bibfield{title}{%
  \enquote{\bibinfo {title} {Infrared spectra of crystalline phase ices
  condensed on silicate smokes at {$T< 20$~K}},}\ }%
  \bibfield{journal}{%
  \bibinfo {journal} {Astrophys. J.}\ }%
  \textbf{\bibinfo {volume} {428}},\ \bibinfo {pages} {L81--L84}%
  \bibAnnoteFile{NoStop}{moore1994}%
\bibitem[{\citenamefont{Moore}\ and\ \citenamefont{Hudson}(2000)}]{moore2000}%
  \BibitemOpen
  \bibfield{author}{%
  \bibinfo {author} {\bibnamefont{Moore}, \bibfnamefont{M.~H.}},\ and\ \bibinfo
  {author} {\bibfnamefont{R.~L.}\ \bibnamefont{Hudson}}}%
  , \bibinfo {year} {2000},\ \bibfield{title}{%
  \enquote{\bibinfo {title} {{IR} detection of {H$_2$O$_2$} at 80~{K} in
  ion-irradiated laboratory ices relevant to {Europa}},}\ }%
  \bibfield{journal}{%
  \bibinfo {journal} {Icarus}\ }%
  \textbf{\bibinfo {volume} {145}},\ \bibinfo {pages} {282--288}%
  \bibAnnoteFile{NoStop}{moore2000}%
\bibitem[{\citenamefont{Morishige}\ and\
  \citenamefont{Uematsu}(2005)}]{morishige2005}%
  \BibitemOpen
  \bibfield{author}{%
  \bibinfo {author} {\bibnamefont{Morishige}, \bibfnamefont{K.}},\ and\
  \bibinfo {author} {\bibfnamefont{H.}~\bibnamefont{Uematsu}}}%
  , \bibinfo {year} {2005},\ \bibfield{title}{%
  \enquote{\bibinfo {title} {The proper structure of cubic ice confined in
  mesopores},}\ }%
  \bibfield{journal}{%
  \bibinfo {journal} {J. Chem. Phys.}\ }%
  \textbf{\bibinfo {volume} {122}},\ \bibinfo {pages} {044711}%
  \bibAnnoteFile{NoStop}{morishige2005}%
\bibitem[{\citenamefont{Morrison}\ \emph{et~al.}(1984)\citenamefont{Morrison},
  \citenamefont{Johnson}, \citenamefont{Shoemaker}, \citenamefont{Soderblom},
  \citenamefont{Thomas},\ and\ \citenamefont{Smith}}]{morrison1984}%
  \BibitemOpen
  \bibfield{author}{%
  \bibinfo {author} {\bibnamefont{Morrison}, \bibfnamefont{D.}}, \bibinfo
  {author} {\bibfnamefont{T.~V.}\ \bibnamefont{Johnson}}, \bibinfo {author}
  {\bibfnamefont{T.~M.}\ \bibnamefont{Shoemaker}}, \bibinfo {author}
  {\bibfnamefont{L.~A.}\ \bibnamefont{Soderblom}}, \bibinfo {author}
  {\bibfnamefont{J.}~\bibnamefont{Thomas}, \bibfnamefont{P.~Veverka}},\ and\
  \bibinfo {author} {\bibfnamefont{B.~A.}\ \bibnamefont{Smith}}}%
  , \bibinfo {year} {1984},\ \emph{\bibinfo {title} {Saturn}}\ (\bibinfo
  {publisher} {University of Arizona Press})%
  \bibAnnoteFile{NoStop}{morrison1984}%
\bibitem[{\citenamefont{Moulton}\ \emph{et~al.}(2000)\citenamefont{Moulton},
  \citenamefont{Gardner}, \citenamefont{Pointon}, \citenamefont{Creamer},
  \citenamefont{Jameson},\ and\ \citenamefont{Penny}}]{moulton2000}%
  \BibitemOpen
  \bibfield{author}{%
  \bibinfo {author} {\bibnamefont{Moulton}, \bibfnamefont{V.}}, \bibinfo
  {author} {\bibfnamefont{P.~P.}\ \bibnamefont{Gardner}}, \bibinfo {author}
  {\bibfnamefont{R.~F.}\ \bibnamefont{Pointon}}, \bibinfo {author}
  {\bibfnamefont{L.~K.}\ \bibnamefont{Creamer}}, \bibinfo {author}
  {\bibfnamefont{G.~B.}\ \bibnamefont{Jameson}},\ and\ \bibinfo {author}
  {\bibfnamefont{D.}~\bibnamefont{Penny}}}%
  , \bibinfo {year} {2000},\ \bibfield{title}{%
  \enquote{\bibinfo {title} {{RNA} folding argues against a hot-start origin of
  life},}\ }%
  \bibfield{journal}{%
  \bibinfo {journal} {J. Mol. Evol.}\ }%
  \textbf{\bibinfo {volume} {51}},\ \bibinfo {pages} {416--421}%
  \bibAnnoteFile{NoStop}{moulton2000}%
\bibitem[{\citenamefont{{Mousis}}\ and\
  \citenamefont{{Alibert}}(2006)}]{mousis2006}%
  \BibitemOpen
  \bibfield{author}{%
  \bibinfo {author} {\bibnamefont{{Mousis}}, \bibfnamefont{O.}},\ and\ \bibinfo
  {author} {\bibfnamefont{Y.}~\bibnamefont{{Alibert}}}}%
  , \bibinfo {year} {2006},\ \bibfield{title}{%
  \enquote{\bibinfo {title} {{Modeling the Jovian subnebula. II. Composition of
  regular satellite ices}},}\ }%
  \bibfield{journal}{%
  \bibinfo {journal} {Astron. Astrophys.}\ }%
  \textbf{\bibinfo {volume} {448}},\ \bibinfo {pages} {771--778}%
  \bibAnnoteFile{NoStop}{mousis2006}%
\bibitem[{\citenamefont{{Mu{\~n}oz Caro}}\
  \emph{et~al.}(2006)\citenamefont{{Mu{\~n}oz Caro}}, \citenamefont{Matrajt},
  \citenamefont{Dartois}, \citenamefont{Nuevo}, \citenamefont{d'Hendecourt},
  \citenamefont{Deboffle}, \citenamefont{Montagnac}, \citenamefont{Chauvin},
  \citenamefont{Boukari},\ and\ \citenamefont{Du}}]{munoz2006}%
  \BibitemOpen
  \bibfield{author}{%
  \bibinfo {author} {\bibnamefont{{Mu{\~n}oz Caro}}, \bibfnamefont{G.~M.}},
  \bibinfo {author} {\bibfnamefont{G.}~\bibnamefont{Matrajt}}, \bibinfo
  {author} {\bibfnamefont{E.}~\bibnamefont{Dartois}}, \bibinfo {author}
  {\bibfnamefont{M.}~\bibnamefont{Nuevo}}, \bibinfo {author}
  {\bibfnamefont{L.}~\bibnamefont{d'Hendecourt}}, \bibinfo {author}
  {\bibfnamefont{D.}~\bibnamefont{Deboffle}}, \bibinfo {author}
  {\bibfnamefont{G.}~\bibnamefont{Montagnac}}, \bibinfo {author}
  {\bibfnamefont{N.}~\bibnamefont{Chauvin}}, \bibinfo {author}
  {\bibfnamefont{C.}~\bibnamefont{Boukari}},\ and\ \bibinfo {author}
  {\bibfnamefont{D.~Le}\ \bibnamefont{Du}}}%
  , \bibinfo {year} {2006},\ \bibfield{title}{%
  \enquote{\bibinfo {title} {Nature and evolution of the dominant carbonaceous
  matter in interplanetary dust particles: effects of irradiation and
  identification with a type of amorphous carbon},}\ }%
  \bibfield{journal}{%
  \bibinfo {journal} {Astron. Astrophys.}\ }%
  \textbf{\bibinfo {volume} {459}},\ \bibinfo {pages} {147--159}%
  \bibAnnoteFile{NoStop}{munoz2006}%
\bibitem[{\citenamefont{{Mu{\~n}oz Caro}}\
  \emph{et~al.}(2002)\citenamefont{{Mu{\~n}oz Caro}},
  \citenamefont{Meierhenrich}, \citenamefont{Schutte}, \citenamefont{Barbier},
  \citenamefont{Arcones~Segovia}, \citenamefont{Rosenbauer},
  \citenamefont{Thiemann}, \citenamefont{Brack},\ and\
  \citenamefont{Greenberg}}]{munoz2002}%
  \BibitemOpen
  \bibfield{author}{%
  \bibinfo {author} {\bibnamefont{{Mu{\~n}oz Caro}}, \bibfnamefont{G.~M.}},
  \bibinfo {author} {\bibfnamefont{U.~J.}\ \bibnamefont{Meierhenrich}},
  \bibinfo {author} {\bibfnamefont{W.~A.}\ \bibnamefont{Schutte}}, \bibinfo
  {author} {\bibfnamefont{B.}~\bibnamefont{Barbier}}, \bibinfo {author}
  {\bibfnamefont{A.}~\bibnamefont{Arcones~Segovia}}, \bibinfo {author}
  {\bibfnamefont{H.}~\bibnamefont{Rosenbauer}}, \bibinfo {author}
  {\bibfnamefont{W.~H.-P.}\ \bibnamefont{Thiemann}}, \bibinfo {author}
  {\bibfnamefont{A.}~\bibnamefont{Brack}},\ and\ \bibinfo {author}
  {\bibfnamefont{J.~M.}\ \bibnamefont{Greenberg}}}%
  , \bibinfo {year} {2002},\ \bibfield{title}{%
  \enquote{\bibinfo {title} {Amino acids from ultraviolet irradiation of
  interstellar ice analogues},}\ }%
  \bibfield{journal}{%
  \bibinfo {journal} {Nature}\ }%
  \textbf{\bibinfo {volume} {416}},\ \bibinfo {pages} {403--406}%
  \bibAnnoteFile{NoStop}{munoz2002}%
\bibitem[{\citenamefont{Mulvaney}\ \emph{et~al.}(1988)\citenamefont{Mulvaney},
  \citenamefont{Wolff},\ and\ \citenamefont{Oates}}]{Mulvaney:1988p450}%
  \BibitemOpen
  \bibfield{author}{%
  \bibinfo {author} {\bibnamefont{Mulvaney}, \bibfnamefont{R.}}, \bibinfo
  {author} {\bibfnamefont{E.}~\bibnamefont{Wolff}},\ and\ \bibinfo {author}
  {\bibfnamefont{K.}~\bibnamefont{Oates}}}%
  , \bibinfo {year} {1988},\ \bibfield{title}{%
  \enquote{\bibinfo {title} {Sulphuric acid at grain boundaries in {Antarctic}
  ice},}\ }%
  \bibfield{journal}{%
  \bibinfo {journal} {Nature}\ }%
  \textbf{\bibinfo {volume} {331}},\ \bibinfo {pages} {247--249}%
  \bibAnnoteFile{NoStop}{Mulvaney:1988p450}%
\bibitem[{\citenamefont{Mundy}\ and\ \citenamefont{Kuo}(2006)}]{mundy2006}%
  \BibitemOpen
  \bibfield{author}{%
  \bibinfo {author} {\bibnamefont{Mundy}, \bibfnamefont{C.~J.}},\ and\ \bibinfo
  {author} {\bibfnamefont{I-F.~W.}\ \bibnamefont{Kuo}}}%
  , \bibinfo {year} {2006},\ \bibfield{title}{%
  \enquote{\bibinfo {title} {First-principles approaches to the structure and
  reactivity of atmospherically relevant aqueous interfaces},}\ }%
  \bibfield{journal}{%
  \bibinfo {journal} {Chem. Rev.}\ }%
  \textbf{\bibinfo {volume} {106}},\ \bibinfo {pages} {1282--1304}%
  \bibAnnoteFile{NoStop}{mundy2006}%
\bibitem[{\citenamefont{Murphy}\ and\ \citenamefont{Koop}(2005)}]{murphy2005}%
  \BibitemOpen
  \bibfield{author}{%
  \bibinfo {author} {\bibnamefont{Murphy}, \bibfnamefont{D.~M.}},\ and\
  \bibinfo {author} {\bibfnamefont{T.}~\bibnamefont{Koop}}}%
  , \bibinfo {year} {2005},\ \bibfield{title}{%
  \enquote{\bibinfo {title} {Review of the vapour pressures of ice and
  supercooled water for atmospheric applications},}\ }%
  \bibfield{journal}{%
  \bibinfo {journal} {Q. J. R. Meteorol. Soc.}\ }%
  \textbf{\bibinfo {volume} {131}},\ \bibinfo {pages} {1539--1565}%
  \bibAnnoteFile{NoStop}{murphy2005}%
\bibitem[{\citenamefont{Murray}\ and\
  \citenamefont{Bertram}(2007)}]{Murray2007}%
  \BibitemOpen
  \bibfield{author}{%
  \bibinfo {author} {\bibnamefont{Murray}, \bibfnamefont{B.~J.}},\ and\
  \bibinfo {author} {\bibfnamefont{A.~K.}\ \bibnamefont{Bertram}}}%
  , \bibinfo {year} {2007},\ \bibfield{title}{%
  \enquote{\bibinfo {title} {Strong dependence of cubic ice formation on
  droplet ammonium to sulfate ratio},}\ }%
  \bibfield{journal}{%
  \bibinfo {journal} {Geophys. Res. Lett.}\ }%
  \textbf{\bibinfo {volume} {34}},\ \bibinfo {pages} {L16810}%
  \bibAnnoteFile{NoStop}{Murray2007}%
\bibitem[{\citenamefont{Murray}\ and\
  \citenamefont{Bertram}(2008)}]{murray2008}%
  \BibitemOpen
  \bibfield{author}{%
  \bibinfo {author} {\bibnamefont{Murray}, \bibfnamefont{B.~J.}},\ and\
  \bibinfo {author} {\bibfnamefont{A.~K.}\ \bibnamefont{Bertram}}}%
  , \bibinfo {year} {2008},\ \bibfield{title}{%
  \enquote{\bibinfo {title} {Inhibition of solute crystallisation in aqueous
  {H}$^+$-{NH}$_4^+$-{SO}$_4^{2-}$-{H}$_2${O} droplets},}\ }%
  \bibfield{journal}{%
  \bibinfo {journal} {Phys. Chem. Chem. Phys.}\ }%
  \textbf{\bibinfo {volume} {10}},\ \bibinfo {pages} {3287--301}%
  \bibAnnoteFile{NoStop}{murray2008}%
\bibitem[{\citenamefont{Murray}\ and\
  \citenamefont{Jensen}(2010)}]{Murray2010a}%
  \BibitemOpen
  \bibfield{author}{%
  \bibinfo {author} {\bibnamefont{Murray}, \bibfnamefont{B.~J.}},\ and\
  \bibinfo {author} {\bibfnamefont{E.~J.}\ \bibnamefont{Jensen}}}%
  , \bibinfo {year} {2010},\ \bibfield{title}{%
  \enquote{\bibinfo {title} {Homogeneous nucleation of amorphous solid water
  particles in the upper mesosphere},}\ }%
  \bibfield{journal}{%
  \bibinfo {journal} {J. Atmos. Solar-Terrestrial Phys.}\ }%
  \textbf{\bibinfo {volume} {72}},\ \bibinfo {pages} {51--61}%
  \bibAnnoteFile{NoStop}{Murray2010a}%
\bibitem[{\citenamefont{Murray}\ \emph{et~al.}(2005)\citenamefont{Murray},
  \citenamefont{Knopf},\ and\ \citenamefont{Bertram}}]{Murray2005a}%
  \BibitemOpen
  \bibfield{author}{%
  \bibinfo {author} {\bibnamefont{Murray}, \bibfnamefont{B.~J.}}, \bibinfo
  {author} {\bibfnamefont{D.~A.}\ \bibnamefont{Knopf}},\ and\ \bibinfo {author}
  {\bibfnamefont{A.~K.}\ \bibnamefont{Bertram}}}%
  , \bibinfo {year} {2005},\ \bibfield{title}{%
  \enquote{\bibinfo {title} {The formation of cubic ice under conditions
  relevant to earth's atmosphere},}\ }%
  \bibfield{journal}{%
  \bibinfo {journal} {Nature}\ }%
  \textbf{\bibinfo {volume} {434}},\ \bibinfo {pages} {202--205}%
  \bibAnnoteFile{NoStop}{Murray2005a}%
\bibitem[{\citenamefont{Murray}\ and\
  \citenamefont{Plane}(2005)}]{Murray2005b}%
  \BibitemOpen
  \bibfield{author}{%
  \bibinfo {author} {\bibnamefont{Murray}, \bibfnamefont{B.~J.}},\ and\
  \bibinfo {author} {\bibfnamefont{J.~M.~C.}\ \bibnamefont{Plane}}}%
  , \bibinfo {year} {2005},\ \bibfield{title}{%
  \enquote{\bibinfo {title} {Modelling the impact of noctilucent cloud
  formation on atomic oxygen and other minor constituents of the summer
  mesosphere},}\ }%
  \bibfield{journal}{%
  \bibinfo {journal} {Atmos. Chem. Phys.}\ }%
  \textbf{\bibinfo {volume} {5}},\ \bibinfo {pages} {1027--1038}%
  \bibAnnoteFile{NoStop}{Murray2005b}%
\bibitem[{\citenamefont{Murray}\ \emph{et~al.}(2010)\citenamefont{Murray},
  \citenamefont{Wilson}, \citenamefont{Dobbie}, \citenamefont{Cui},
  \citenamefont{Al-Jumar}, \citenamefont{M{\"o}hler}, \citenamefont{Schnaiter},
  \citenamefont{Wagner}, \citenamefont{Benz}, \citenamefont{Niemand},
  \citenamefont{Saathoff}, \citenamefont{Ebert}, \citenamefont{Wagner},\ and\
  \citenamefont{K{\"a}rcher}}]{Murray2010b}%
  \BibitemOpen
  \bibfield{author}{%
  \bibinfo {author} {\bibnamefont{Murray}, \bibfnamefont{B.~J.}}, \bibinfo
  {author} {\bibfnamefont{T.~M.}\ \bibnamefont{Wilson}}, \bibinfo {author}
  {\bibfnamefont{S.}~\bibnamefont{Dobbie}}, \bibinfo {author}
  {\bibfnamefont{Z.}~\bibnamefont{Cui}}, \bibinfo {author} {\bibfnamefont{S.~M.
  R.~K.}\ \bibnamefont{Al-Jumar}}, \bibinfo {author}
  {\bibfnamefont{O.}~\bibnamefont{M{\"o}hler}}, \bibinfo {author}
  {\bibfnamefont{M.}~\bibnamefont{Schnaiter}}, \bibinfo {author}
  {\bibfnamefont{R.}~\bibnamefont{Wagner}}, \bibinfo {author}
  {\bibfnamefont{S.}~\bibnamefont{Benz}}, \bibinfo {author}
  {\bibfnamefont{M.}~\bibnamefont{Niemand}}, \bibinfo {author}
  {\bibfnamefont{H.}~\bibnamefont{Saathoff}}, \bibinfo {author}
  {\bibfnamefont{V.}~\bibnamefont{Ebert}}, \bibinfo {author}
  {\bibfnamefont{S.}~\bibnamefont{Wagner}},\ and\ \bibinfo {author}
  {\bibfnamefont{B.}~\bibnamefont{K{\"a}rcher}}}%
  , \bibinfo {year} {2010},\ \bibfield{title}{%
  \enquote{\bibinfo {title} {Heterogeneous nucleation of ice particles on
  glassy aerosols under cirrus conditions},}\ }%
  \bibfield{journal}{%
  \bibinfo {journal} {Nature Geosci.}\ }%
  \textbf{\bibinfo {volume} {3}},\ \bibinfo {pages} {233--237}%
  \bibAnnoteFile{NoStop}{Murray2010b}%
\bibitem[{\citenamefont{N{\aa}g{\aa}rd}\
  \emph{et~al.}(2002)\citenamefont{N{\aa}g{\aa}rd}, \citenamefont{Pettersson},
  \citenamefont{Derkatch}, \citenamefont{Al~Khalili}, \citenamefont{Neau},
  \citenamefont{Ros\'{e}n}, \citenamefont{Larsson}, \citenamefont{Semaniak},
  \citenamefont{Danared}, \citenamefont{K\"{a}llberg},
  \citenamefont{\"{O}sterdahl},\ and\ \citenamefont{af~Ugglas}}]{Nagard2002}%
  \BibitemOpen
  \bibfield{author}{%
  \bibinfo {author} {\bibnamefont{N{\aa}g{\aa}rd}, \bibfnamefont{M.~B.}},
  \bibinfo {author} {\bibfnamefont{J.~B.~C.}\ \bibnamefont{Pettersson}},
  \bibinfo {author} {\bibfnamefont{A.~M.}\ \bibnamefont{Derkatch}}, \bibinfo
  {author} {\bibfnamefont{A.}~\bibnamefont{Al~Khalili}}, \bibinfo {author}
  {\bibfnamefont{A.}~\bibnamefont{Neau}}, \bibinfo {author}
  {\bibfnamefont{S.}~\bibnamefont{Ros\'{e}n}}, \bibinfo {author}
  {\bibfnamefont{M.}~\bibnamefont{Larsson}}, \bibinfo {author}
  {\bibfnamefont{J.}~\bibnamefont{Semaniak}}, \bibinfo {author}
  {\bibfnamefont{H.}~\bibnamefont{Danared}}, \bibinfo {author}
  {\bibfnamefont{A.}~\bibnamefont{K\"{a}llberg}}, \bibinfo {author}
  {\bibfnamefont{F.}~\bibnamefont{\"{O}sterdahl}},\ and\ \bibinfo {author}
  {\bibfnamefont{M.}~\bibnamefont{af~Ugglas}}}%
  , \bibinfo {year} {2002},\ \bibfield{title}{%
  \enquote{\bibinfo {title} {Dissociative recombination of {D$^+$(D$_2$O)$_2$}
  water cluster ions with free electrons},}\ }%
  \bibfield{journal}{%
  \bibinfo {journal} {J. Chem. Phys.}\ }%
  \textbf{\bibinfo {volume} {117}},\ \bibinfo {pages} {5264--5270}%
  \bibAnnoteFile{NoStop}{Nagard2002}%
\bibitem[{\citenamefont{Nakaya}(1954)}]{Nakaya1954}%
  \BibitemOpen
  \bibfield{author}{%
  \bibinfo {author} {\bibnamefont{Nakaya}, \bibfnamefont{U.}}}%
  , \bibinfo {year} {1954},\ \emph{\bibinfo {title} {Snow Crystals: Natural and
  Artificial}}\ (\bibinfo {publisher} {Harvard University Press},\ \bibinfo
  {address} {Cambridge})%
  \bibAnnoteFile{NoStop}{Nakaya1954}%
\bibitem[{\citenamefont{Narten}\ \emph{et~al.}(1976)\citenamefont{Narten},
  \citenamefont{Venkatesh},\ and\ \citenamefont{Rice}}]{narten1976}%
  \BibitemOpen
  \bibfield{author}{%
  \bibinfo {author} {\bibnamefont{Narten}, \bibfnamefont{A.~H.}}, \bibinfo
  {author} {\bibfnamefont{C.~G.}\ \bibnamefont{Venkatesh}},\ and\ \bibinfo
  {author} {\bibfnamefont{S.~A.}\ \bibnamefont{Rice}}}%
  , \bibinfo {year} {1976},\ \bibfield{title}{%
  \enquote{\bibinfo {title} {Diffraction pattern and structure of amorphous
  solid water at 10 and 77$^o${K}},}\ }%
  \bibfield{journal}{%
  \bibinfo {journal} {J. Chem. Phys.}\ }%
  \textbf{\bibinfo {volume} {64}},\ \bibinfo {pages} {1106--1121}%
  \bibAnnoteFile{NoStop}{narten1976}%
\bibitem[{\citenamefont{Nash}\ and\ \citenamefont{Betts}(1995)}]{nash1995}%
  \BibitemOpen
  \bibfield{author}{%
  \bibinfo {author} {\bibnamefont{Nash}, \bibfnamefont{D.~B.}},\ and\ \bibinfo
  {author} {\bibfnamefont{B.~H.}\ \bibnamefont{Betts}}}%
  , \bibinfo {year} {1995},\ \emph{\bibinfo {title} {Solar System Ices}},\
  Vol.\ \bibinfo {volume} {227}\ (\bibinfo {publisher} {Kluwer})%
  \bibAnnoteFile{NoStop}{nash1995}%
\bibitem[{\citenamefont{Nelmes}\ \emph{et~al.}(2006)\citenamefont{Nelmes},
  \citenamefont{Loveday}, \citenamefont{Str{\"a}ssle},
  \citenamefont{C.~L.~Bull}, \citenamefont{Hamel},\ and\
  \citenamefont{Klotz}}]{nelmes2006}%
  \BibitemOpen
  \bibfield{author}{%
  \bibinfo {author} {\bibnamefont{Nelmes}, \bibfnamefont{R.~J.}}, \bibinfo
  {author} {\bibfnamefont{J.~S.}\ \bibnamefont{Loveday}}, \bibinfo {author}
  {\bibfnamefont{T.}~\bibnamefont{Str{\"a}ssle}}, \bibinfo {author}
  {\bibfnamefont{M.~Guthrie}\ \bibnamefont{C.~L.~Bull}}, \bibinfo {author}
  {\bibfnamefont{G.}~\bibnamefont{Hamel}},\ and\ \bibinfo {author}
  {\bibfnamefont{S.}~\bibnamefont{Klotz}}}%
  , \bibinfo {year} {2006},\ \bibfield{title}{%
  \enquote{\bibinfo {title} {Annealed high-density amorphous ice under
  pressure},}\ }%
  \bibfield{journal}{%
  \bibinfo {journal} {Nature Phys.}\ }%
  \textbf{\bibinfo {volume} {2}},\ \bibinfo {pages} {414--418}%
  \bibAnnoteFile{NoStop}{nelmes2006}%
\bibitem[{\citenamefont{Neufeld}\ \emph{et~al.}(2010)\citenamefont{Neufeld},
  \citenamefont{Goldstein},\ and\ \citenamefont{Worster}}]{neufeld2010}%
  \BibitemOpen
  \bibfield{author}{%
  \bibinfo {author} {\bibnamefont{Neufeld}, \bibfnamefont{J.~A.}}, \bibinfo
  {author} {\bibfnamefont{R.~E.}\ \bibnamefont{Goldstein}},\ and\ \bibinfo
  {author} {\bibfnamefont{M.~G.}\ \bibnamefont{Worster}}}%
  , \bibinfo {year} {2010},\ \bibfield{title}{%
  \enquote{\bibinfo {title} {On the mechanisms of icicle evolution},}\ }%
  \bibfield{journal}{%
  \bibinfo {journal} {J. Fluid Mech.}\ }%
  \textbf{\bibinfo {volume} {647}},\ \bibinfo {pages} {287--308}%
  \bibAnnoteFile{NoStop}{neufeld2010}%
\bibitem[{\citenamefont{{Niemann}}\
  \emph{et~al.}(2010)\citenamefont{{Niemann}}, \citenamefont{{Atreya}},
  \citenamefont{{Demick}}, \citenamefont{{Gautier}}, \citenamefont{{Haberman}},
  \citenamefont{{Harpold}}, \citenamefont{{Kasprzak}}, \citenamefont{{Lunine}},
  \citenamefont{{Owen}},\ and\ \citenamefont{{Raulin}}}]{niemann2010}%
  \BibitemOpen
  \bibfield{author}{%
  \bibinfo {author} {\bibnamefont{{Niemann}}, \bibfnamefont{H.~B.}}, \bibinfo
  {author} {\bibfnamefont{S.~K.}\ \bibnamefont{{Atreya}}}, \bibinfo {author}
  {\bibfnamefont{J.~E.}\ \bibnamefont{{Demick}}}, \bibinfo {author}
  {\bibfnamefont{D.}~\bibnamefont{{Gautier}}}, \bibinfo {author}
  {\bibfnamefont{J.~A.}\ \bibnamefont{{Haberman}}}, \bibinfo {author}
  {\bibfnamefont{D.~N.}\ \bibnamefont{{Harpold}}}, \bibinfo {author}
  {\bibfnamefont{W.~T.}\ \bibnamefont{{Kasprzak}}}, \bibinfo {author}
  {\bibfnamefont{J.~I.}\ \bibnamefont{{Lunine}}}, \bibinfo {author}
  {\bibfnamefont{T.~C.}\ \bibnamefont{{Owen}}},\ and\ \bibinfo {author}
  {\bibfnamefont{F.}~\bibnamefont{{Raulin}}}}%
  , \bibinfo {year} {2010},\ \bibfield{title}{%
  \enquote{\bibinfo {title} {{Composition of Titan's lower atmosphere and
  simple surface volatiles as measured by the Cassini--Huygens probe gas
  chromatograph mass spectrometer experiment}},}\ }%
  \bibfield{journal}{%
  \bibinfo {journal} {J. Geophys. Res. (Planets)}\ }%
  \textbf{\bibinfo {volume} {115}},\ \bibinfo {pages} {12006}%
  \bibAnnoteFile{NoStop}{niemann2010}%
\bibitem[{\citenamefont{Noel}\ \emph{et~al.}(2006)\citenamefont{Noel},
  \citenamefont{Chepfer}, \citenamefont{Haeffelin},\ and\
  \citenamefont{Morille}}]{noel2006}%
  \BibitemOpen
  \bibfield{author}{%
  \bibinfo {author} {\bibnamefont{Noel}, \bibfnamefont{V.}}, \bibinfo {author}
  {\bibfnamefont{E.}~\bibnamefont{Chepfer}}, \bibinfo {author}
  {\bibfnamefont{M.}~\bibnamefont{Haeffelin}},\ and\ \bibinfo {author}
  {\bibfnamefont{Y.}~\bibnamefont{Morille}}}%
  , \bibinfo {year} {2006},\ \bibfield{title}{%
  \enquote{\bibinfo {title} {Classification of ice crystal shapes in
  midlatitude ice clouds from three years of lidar observations over the
  {SIRTA} observatory},}\ }%
  \bibfield{journal}{%
  \bibinfo {journal} {J. Atmos. Sci.}\ }%
  \textbf{\bibinfo {volume} {63}},\ \bibinfo {pages} {2978--2991}%
  \bibAnnoteFile{NoStop}{noel2006}%
\bibitem[{\citenamefont{Noel}\ \emph{et~al.}(2002)\citenamefont{Noel},
  \citenamefont{Chepfer}, \citenamefont{Ledanois}, \citenamefont{Delaval},\
  and\ \citenamefont{Flamant}}]{noel2002}%
  \BibitemOpen
  \bibfield{author}{%
  \bibinfo {author} {\bibnamefont{Noel}, \bibfnamefont{V.}}, \bibinfo {author}
  {\bibfnamefont{H.}~\bibnamefont{Chepfer}}, \bibinfo {author}
  {\bibfnamefont{G.}~\bibnamefont{Ledanois}}, \bibinfo {author}
  {\bibfnamefont{A.}~\bibnamefont{Delaval}},\ and\ \bibinfo {author}
  {\bibfnamefont{P.~H.}\ \bibnamefont{Flamant}}}%
  , \bibinfo {year} {2002},\ \bibfield{title}{%
  \enquote{\bibinfo {title} {Classification of particle effective shape ratios
  in cirrus clouds based on the lidar depolarization ratio},}\ }%
  \bibfield{journal}{%
  \bibinfo {journal} {Appl. Opt.}\ }%
  \textbf{\bibinfo {volume} {41}},\ \bibinfo {pages} {4245--4257}%
  \bibAnnoteFile{NoStop}{noel2002}%
\bibitem[{\citenamefont{Noel}\ \emph{et~al.}(2004)\citenamefont{Noel},
  \citenamefont{Winker}, \citenamefont{McGill},\ and\
  \citenamefont{Lawson}}]{noel2004}%
  \BibitemOpen
  \bibfield{author}{%
  \bibinfo {author} {\bibnamefont{Noel}, \bibfnamefont{V.}}, \bibinfo {author}
  {\bibfnamefont{D.~M.}\ \bibnamefont{Winker}}, \bibinfo {author}
  {\bibfnamefont{M.}~\bibnamefont{McGill}},\ and\ \bibinfo {author}
  {\bibfnamefont{P.}~\bibnamefont{Lawson}}}%
  , \bibinfo {year} {2004},\ \bibfield{title}{%
  \enquote{\bibinfo {title} {Classification of particle shapes from lidar
  depolarization ratio in convective ice clouds compared to in situ
  observations during {CRYSTAL}-{FACE}},}\ }%
  \bibfield{journal}{%
  \bibinfo {journal} {J. of Geophys. Res.-Atmos.}\ }%
  \textbf{\bibinfo {volume} {109}}%
  \bibAnnoteFile{NoStop}{noel2004}%
\bibitem[{\citenamefont{Notesco}\ and\
  \citenamefont{Bar-Nun}(1996)}]{notesco1996}%
  \BibitemOpen
  \bibfield{author}{%
  \bibinfo {author} {\bibnamefont{Notesco}, \bibfnamefont{G.}},\ and\ \bibinfo
  {author} {\bibfnamefont{A.}~\bibnamefont{Bar-Nun}}}%
  , \bibinfo {year} {1996},\ \bibfield{title}{%
  \enquote{\bibinfo {title} {Enrichment of {CO} over {N$_2$} by their trapping
  in amorphous ice and implications to comet {P/Halley}},}\ }%
  \bibfield{journal}{%
  \bibinfo {journal} {Icarus}\ }%
  \textbf{\bibinfo {volume} {126}},\ \bibinfo {pages} {336--341}%
  \bibAnnoteFile{NoStop}{notesco1996}%
\bibitem[{\citenamefont{Notesco}\ and\
  \citenamefont{Bar-Nun}(2000)}]{notesco2000}%
  \BibitemOpen
  \bibfield{author}{%
  \bibinfo {author} {\bibnamefont{Notesco}, \bibfnamefont{G.}},\ and\ \bibinfo
  {author} {\bibfnamefont{A.}~\bibnamefont{Bar-Nun}}}%
  , \bibinfo {year} {2000},\ \bibfield{title}{%
  \enquote{\bibinfo {title} {The effect of methanol clathrate-hydrate formation
  and other gas-trapping mechanisms on the structure and dynamics of cometary
  ices},}\ }%
  \bibfield{journal}{%
  \bibinfo {journal} {Icarus}\ }%
  \textbf{\bibinfo {volume} {148}},\ \bibinfo {pages} {456--463}%
  \bibAnnoteFile{NoStop}{notesco2000}%
\bibitem[{\citenamefont{Nye}(1992)}]{Nye1992}%
  \BibitemOpen
  \bibfield{author}{%
  \bibinfo {author} {\bibnamefont{Nye}, \bibfnamefont{J~F}}}%
  , \bibinfo {year} {1992},\ \enquote{\bibinfo {title} {Water veins and lenses
  in polycrystalline ice},}\ in\ \emph{\bibinfo {booktitle} {Physics and
  Chemistry of Ice}}\ (\bibinfo {publisher} {Hokkaido University Press})\ pp.\
  \bibinfo {pages} {200--205}%
  \bibAnnoteFile{NoStop}{Nye1992}%
\bibitem[{\citenamefont{Oba}\ \emph{et~al.}(2009)\citenamefont{Oba},
  \citenamefont{Miyauchi}, \citenamefont{Hidaka}, \citenamefont{Chigai},
  \citenamefont{Watanabe},\ and\ \citenamefont{Kouchi}}]{oba2009}%
  \BibitemOpen
  \bibfield{author}{%
  \bibinfo {author} {\bibnamefont{Oba}, \bibfnamefont{Y.}}, \bibinfo {author}
  {\bibfnamefont{N.}~\bibnamefont{Miyauchi}}, \bibinfo {author}
  {\bibfnamefont{H.}~\bibnamefont{Hidaka}}, \bibinfo {author}
  {\bibfnamefont{T.}~\bibnamefont{Chigai}}, \bibinfo {author}
  {\bibfnamefont{N.}~\bibnamefont{Watanabe}},\ and\ \bibinfo {author}
  {\bibfnamefont{A.}~\bibnamefont{Kouchi}}}%
  , \bibinfo {year} {2009},\ \bibfield{title}{%
  \enquote{\bibinfo {title} {Formation of compact amorphous {H$_2$O} ice by
  codeposition of hydrogen atoms with oxygen molecules on grain surfaces},}\ }%
  \bibfield{journal}{%
  \bibinfo {journal} {Astrophys. J.}\ }%
  \textbf{\bibinfo {volume} {701}},\ \bibinfo {pages} {464--470}%
  \bibAnnoteFile{NoStop}{oba2009}%
\bibitem[{\citenamefont{Obbard}\ \emph{et~al.}(2006)\citenamefont{Obbard},
  \citenamefont{Baker},\ and\ \citenamefont{Sieg}}]{obbard2006}%
  \BibitemOpen
  \bibfield{author}{%
  \bibinfo {author} {\bibnamefont{Obbard}, \bibfnamefont{R.~I.}}, \bibinfo
  {author} {\bibfnamefont{I.}~\bibnamefont{Baker}},\ and\ \bibinfo {author}
  {\bibfnamefont{K.}~\bibnamefont{Sieg}}}%
  , \bibinfo {year} {2006},\ \bibfield{title}{%
  \enquote{\bibinfo {title} {Using electron backscatter diffraction patterns to
  examine recrystallization in polar ice sheet},}\ }%
  \bibfield{journal}{%
  \bibinfo {journal} {J. Glaciol.}\ }%
  \textbf{\bibinfo {volume} {52}},\ \bibinfo {pages} {546--557}%
  \bibAnnoteFile{NoStop}{obbard2006}%
\bibitem[{\citenamefont{{\"O}berg}\
  \emph{et~al.}(2010)\citenamefont{{\"O}berg}, \citenamefont{van Dishoeck},
  \citenamefont{Linnartz},\ and\ \citenamefont{Andersson}}]{oberg2010}%
  \BibitemOpen
  \bibfield{author}{%
  \bibinfo {author} {\bibnamefont{{\"O}berg}, \bibfnamefont{K.~I.}}, \bibinfo
  {author} {\bibfnamefont{E.~F.}\ \bibnamefont{van Dishoeck}}, \bibinfo
  {author} {\bibfnamefont{H.}~\bibnamefont{Linnartz}},\ and\ \bibinfo {author}
  {\bibfnamefont{S.}~\bibnamefont{Andersson}}}%
  , \bibinfo {year} {2010},\ \bibfield{title}{%
  \enquote{\bibinfo {title} {The effect of {H}$_2${O} on ice photochemistry},}\
  }%
  \bibfield{journal}{%
  \bibinfo {journal} {Astrophys. J.}\ }%
  \textbf{\bibinfo {volume} {718}},\ \bibinfo {pages} {832--840}%
  \bibAnnoteFile{NoStop}{oberg2010}%
\bibitem[{\citenamefont{{\"O}berg}\
  \emph{et~al.}(2007)\citenamefont{{\"O}berg}, \citenamefont{Fraser},
  \citenamefont{Boogert}, \citenamefont{Bisschop}, \citenamefont{Fuchs},
  \citenamefont{van Dishoeck},\ and\ \citenamefont{Linnartz}}]{oberg2007}%
  \BibitemOpen
  \bibfield{author}{%
  \bibinfo {author} {\bibnamefont{{\"O}berg}, \bibfnamefont{K.~I.}}, \bibinfo
  {author} {\bibfnamefont{H.~J.}\ \bibnamefont{Fraser}}, \bibinfo {author}
  {\bibfnamefont{A.~C.~A.}\ \bibnamefont{Boogert}}, \bibinfo {author}
  {\bibfnamefont{S.~E.}\ \bibnamefont{Bisschop}}, \bibinfo {author}
  {\bibfnamefont{G.~W.}\ \bibnamefont{Fuchs}}, \bibinfo {author}
  {\bibfnamefont{E.~F.}\ \bibnamefont{van Dishoeck}},\ and\ \bibinfo {author}
  {\bibfnamefont{H.}~\bibnamefont{Linnartz}}}%
  , \bibinfo {year} {2007},\ \bibfield{title}{%
  \enquote{\bibinfo {title} {Effects of {CO$_2$} on {H$_2$O} band profiles and
  band strengths in mixed {H$_2$O}:{CO$_2$} ices},}\ }%
  \bibfield{journal}{%
  \bibinfo {journal} {Astron. Astrophys.}\ }%
  \textbf{\bibinfo {volume} {462}},\ \bibinfo {pages} {1187--1198}%
  \bibAnnoteFile{NoStop}{oberg2007}%
\bibitem[{\citenamefont{Oguro}\ and\ \citenamefont{Hondoh}(1988)}]{oguro1988}%
  \BibitemOpen
  \bibfield{author}{%
  \bibinfo {author} {\bibnamefont{Oguro}, \bibfnamefont{M.}},\ and\ \bibinfo
  {author} {\bibfnamefont{T.}~\bibnamefont{Hondoh}}}%
  , \bibinfo {year} {1988},\ \enquote{\bibinfo {title} {Stacking faults in ice
  crystals},}\ in\ \emph{\bibinfo {booktitle} {Lattice Defects in Ice
  Crystals}},\ \bibinfo {editor} {edited by\ \bibinfo {editor}
  {\bibfnamefont{A.}~\bibnamefont{Higashi}}}\ (\bibinfo {publisher} {Hokkaido
  University Press, Sapporo})\ pp.\ \bibinfo {pages} {49--67}%
  \bibAnnoteFile{NoStop}{oguro1988}%
\bibitem[{\citenamefont{Ohmura}\ \emph{et~al.}(1996)\citenamefont{Ohmura},
  \citenamefont{Wild},\ and\ \citenamefont{Bengtsson}}]{Ohmura:1996p26725}%
  \BibitemOpen
  \bibfield{author}{%
  \bibinfo {author} {\bibnamefont{Ohmura}, \bibfnamefont{A.}}, \bibinfo
  {author} {\bibfnamefont{M.}~\bibnamefont{Wild}},\ and\ \bibinfo {author}
  {\bibfnamefont{L.}~\bibnamefont{Bengtsson}}}%
  , \bibinfo {year} {1996},\ \bibfield{title}{%
  \enquote{\bibinfo {title} {A possible change in mass balance of {Greenland}
  and {Antarctic} ice sheets in the coming century},}\ }%
  \bibfield{journal}{%
  \bibinfo {journal} {J. Climate}\ }%
  \textbf{\bibinfo {volume} {9}},\ \bibinfo {pages} {2124--2135}%
  \bibAnnoteFile{NoStop}{Ohmura:1996p26725}%
\bibitem[{\citenamefont{Oikkonen}\ and\
  \citenamefont{Haapala}(2011)}]{Oikkonen:2011}%
  \BibitemOpen
  \bibfield{author}{%
  \bibinfo {author} {\bibnamefont{Oikkonen}, \bibfnamefont{A.}},\ and\ \bibinfo
  {author} {\bibfnamefont{J.}~\bibnamefont{Haapala}}}%
  , \bibinfo {year} {2011},\ \bibfield{title}{%
  \enquote{\bibinfo {title} {Variability and changes of {Arctic} sea ice draft
  distribution --- submarine sonar measurements revisited},}\ }%
  \bibfield{journal}{%
  \bibinfo {journal} {The Cryosphere}\ }%
  \textbf{\bibinfo {volume} {5}},\ \bibinfo {pages} {917--929}%
  \bibAnnoteFile{NoStop}{Oikkonen:2011}%
\bibitem[{\citenamefont{\"{O}jekull}\
  \emph{et~al.}(2007)\citenamefont{\"{O}jekull}, \citenamefont{Andersson},
  \citenamefont{N{\aa}g{\aa}rd}, \citenamefont{Pettersson},
  \citenamefont{Markovi\'{c}}, \citenamefont{Derkatch}, \citenamefont{Neau},
  \citenamefont{Al~Khalili}, \citenamefont{Ros\'{e}n}, \citenamefont{Larsson},
  \citenamefont{Semaniak}, \citenamefont{Danared}, \citenamefont{K\"{a}llberg},
  \citenamefont{\"{O}sterdahl},\ and\ \citenamefont{af~Ugglas}}]{Ojekull2007}%
  \BibitemOpen
  \bibfield{author}{%
  \bibinfo {author} {\bibnamefont{\"{O}jekull}, \bibfnamefont{J.}}, \bibinfo
  {author} {\bibfnamefont{P.~U.}\ \bibnamefont{Andersson}}, \bibinfo {author}
  {\bibfnamefont{M.~B.}\ \bibnamefont{N{\aa}g{\aa}rd}}, \bibinfo {author}
  {\bibfnamefont{J.~B.~C.}\ \bibnamefont{Pettersson}}, \bibinfo {author}
  {\bibfnamefont{N.}~\bibnamefont{Markovi\'{c}}}, \bibinfo {author}
  {\bibfnamefont{A.~M.}\ \bibnamefont{Derkatch}}, \bibinfo {author}
  {\bibfnamefont{A.}~\bibnamefont{Neau}}, \bibinfo {author}
  {\bibfnamefont{A.}~\bibnamefont{Al~Khalili}}, \bibinfo {author}
  {\bibfnamefont{S.}~\bibnamefont{Ros\'{e}n}}, \bibinfo {author}
  {\bibfnamefont{M.}~\bibnamefont{Larsson}}, \bibinfo {author}
  {\bibfnamefont{J.}~\bibnamefont{Semaniak}}, \bibinfo {author}
  {\bibfnamefont{H.}~\bibnamefont{Danared}}, \bibinfo {author}
  {\bibfnamefont{A.}~\bibnamefont{K\"{a}llberg}}, \bibinfo {author}
  {\bibfnamefont{F.}~\bibnamefont{\"{O}sterdahl}},\ and\ \bibinfo {author}
  {\bibfnamefont{M.}~\bibnamefont{af~Ugglas}}}%
  , \bibinfo {year} {2007},\ \bibfield{title}{%
  \enquote{\bibinfo {title} {Dissociative recombination of {H$^+$(H$_2$O)$_3$}
  and {D$^+$(D$_2$O)$_3$} water cluster ions with electrons: Cross sections and
  branching ratios},}\ }%
  \bibfield{journal}{%
  \bibinfo {journal} {J. Chem. Phys.}\ }%
  \textbf{\bibinfo {volume} {127}},\ \bibinfo {pages} {194301}%
  \bibAnnoteFile{NoStop}{Ojekull2007}%
\bibitem[{\citenamefont{{\"O}jekull}\
  \emph{et~al.}(2008)\citenamefont{{\"O}jekull}, \citenamefont{Andersson},
  \citenamefont{Pettersson}, \citenamefont{Markovi{\'c}},
  \citenamefont{Thomas}, \citenamefont{Khalili}, \citenamefont{Ehlerding},
  \citenamefont{Larsson}, \citenamefont{Danared}, \citenamefont{K{\"a}llberg},
  \citenamefont{{\"O}sterdahl},\ and\ \citenamefont{af. Ugglas}}]{Ojekull2008}%
  \BibitemOpen
  \bibfield{author}{%
  \bibinfo {author} {\bibnamefont{{\"O}jekull}, \bibfnamefont{J.}}, \bibinfo
  {author} {\bibfnamefont{P.~U.}\ \bibnamefont{Andersson}}, \bibinfo {author}
  {\bibfnamefont{J.~B.~C.}\ \bibnamefont{Pettersson}}, \bibinfo {author}
  {\bibfnamefont{N.}~\bibnamefont{Markovi{\'c}}}, \bibinfo {author}
  {\bibfnamefont{R.~D.}\ \bibnamefont{Thomas}}, \bibinfo {author}
  {\bibfnamefont{A.~Al}\ \bibnamefont{Khalili}}, \bibinfo {author}
  {\bibfnamefont{A.}~\bibnamefont{Ehlerding}}, \bibinfo {author}
  {\bibfnamefont{M.}~\bibnamefont{Larsson}}, \bibinfo {author}
  {\bibfnamefont{H.}~\bibnamefont{Danared}}, \bibinfo {author}
  {\bibfnamefont{A.}~\bibnamefont{K{\"a}llberg}}, \bibinfo {author}
  {\bibfnamefont{F.}~\bibnamefont{{\"O}sterdahl}},\ and\ \bibinfo {author}
  {\bibfnamefont{M.}~\bibnamefont{af. Ugglas}}}%
  , \bibinfo {year} {2008},\ \bibfield{title}{%
  \enquote{\bibinfo {title} {Dissociative recombination of water cluster ions
  with free electrons: cross sections and branching ratios},}\ }%
  \bibfield{journal}{%
  \bibinfo {journal} {J. Chem. Phys.}\ }%
  \textbf{\bibinfo {volume} {128}},\ \bibinfo {pages} {044311}%
  \bibAnnoteFile{NoStop}{Ojekull2008}%
\bibitem[{\citenamefont{{O'Keefe}}(1980)}]{okeefe1980}%
  \BibitemOpen
  \bibfield{author}{%
  \bibinfo {author} {\bibnamefont{{O'Keefe}}, \bibfnamefont{J.~A.}}}%
  , \bibinfo {year} {1980},\ \bibfield{title}{%
  \enquote{\bibinfo {title} {{The terminal Eocene event: formation of a ring
  system around the Earth?}}.}\ }%
  \bibfield{journal}{%
  \bibinfo {journal} {Nature}\ }%
  \textbf{\bibinfo {volume} {285}},\ \bibinfo {pages} {309--311}%
  \bibAnnoteFile{NoStop}{okeefe1980}%
\bibitem[{\citenamefont{Orgel}(1998)}]{orgel1998}%
  \BibitemOpen
  \bibfield{author}{%
  \bibinfo {author} {\bibnamefont{Orgel}, \bibfnamefont{L.~E.}}}%
  , \bibinfo {year} {1998},\ \bibfield{title}{%
  \enquote{\bibinfo {title} {The origin of life --- a review of facts and
  speculations},}\ }%
  \bibfield{journal}{%
  \bibinfo {journal} {Trends Biochem. Sci.}\ }%
  \textbf{\bibinfo {volume} {23}},\ \bibinfo {pages} {491--495}%
  \bibAnnoteFile{NoStop}{orgel1998}%
\bibitem[{\citenamefont{Paige}(1970)}]{paige1970}%
  \BibitemOpen
  \bibfield{author}{%
  \bibinfo {author} {\bibnamefont{Paige}, \bibfnamefont{R.~A.}}}%
  , \bibinfo {year} {1970},\ \bibfield{title}{%
  \enquote{\bibinfo {title} {Stalactite growth beneath sea ice},}\ }%
  \bibfield{journal}{%
  \bibinfo {journal} {Science}\ }%
  \textbf{\bibinfo {volume} {167}},\ \bibinfo {pages} {171--172}%
  \bibAnnoteFile{NoStop}{paige1970}%
\bibitem[{\citenamefont{Palumbo}\ \emph{et~al.}(1997)\citenamefont{Palumbo},
  \citenamefont{Geballe},\ and\ \citenamefont{Tielens}}]{palumbo1997}%
  \BibitemOpen
  \bibfield{author}{%
  \bibinfo {author} {\bibnamefont{Palumbo}, \bibfnamefont{M.~E.}}, \bibinfo
  {author} {\bibfnamefont{T.~R.}\ \bibnamefont{Geballe}},\ and\ \bibinfo
  {author} {\bibfnamefont{A.~G. G.~M.}\ \bibnamefont{Tielens}}}%
  , \bibinfo {year} {1997},\ \bibfield{title}{%
  \enquote{\bibinfo {title} {Solid carbonyl sulfide ({OCS}) in dense molecular
  clouds},}\ }%
  \bibfield{journal}{%
  \bibinfo {journal} {Astrophys. J.}\ }%
  \textbf{\bibinfo {volume} {479}},\ \bibinfo {pages} {839--844}%
  \bibAnnoteFile{NoStop}{palumbo1997}%
\bibitem[{\citenamefont{Park}\ \emph{et~al.}(2010)\citenamefont{Park},
  \citenamefont{Moon},\ and\ \citenamefont{Kang}}]{Park2010}%
  \BibitemOpen
  \bibfield{author}{%
  \bibinfo {author} {\bibnamefont{Park}, \bibfnamefont{S.-C.}}, \bibinfo
  {author} {\bibfnamefont{E.-S.}\ \bibnamefont{Moon}},\ and\ \bibinfo {author}
  {\bibfnamefont{H.}~\bibnamefont{Kang}}}%
  , \bibinfo {year} {2010},\ \bibfield{title}{%
  \enquote{\bibinfo {title} {Some fundamental properties and reactions of ice
  surfaces at low temperatures},}\ }%
  \bibfield{journal}{%
  \bibinfo {journal} {Phys. Chem. Chem. Phys.}\ }%
  \textbf{\bibinfo {volume} {12}},\ \bibinfo {pages} {12000--12011}%
  \bibAnnoteFile{NoStop}{Park2010}%
\bibitem[{\citenamefont{Parmeter}(1975)}]{Parmeter:1975}%
  \BibitemOpen
  \bibfield{author}{%
  \bibinfo {author} {\bibnamefont{Parmeter}, \bibfnamefont{R.~R.}}}%
  , \bibinfo {year} {1975},\ \bibfield{title}{%
  \enquote{\bibinfo {title} {A model of simple rafting in sea ice},}\ }%
  \bibfield{journal}{%
  \bibinfo {journal} {J.~Geophys.~Res.}\ }%
  \textbf{\bibinfo {volume} {80}},\ \bibinfo {pages} {1948--1952}%
  \bibAnnoteFile{NoStop}{Parmeter:1975}%
\bibitem[{\citenamefont{{Pat-El}}\ \emph{et~al.}(2009)\citenamefont{{Pat-El}},
  \citenamefont{{Laufer}}, \citenamefont{{Notesco}},\ and\
  \citenamefont{{Bar-Nun}}}]{patel2009}%
  \BibitemOpen
  \bibfield{author}{%
  \bibinfo {author} {\bibnamefont{{Pat-El}}, \bibfnamefont{I.}}, \bibinfo
  {author} {\bibfnamefont{D.}~\bibnamefont{{Laufer}}}, \bibinfo {author}
  {\bibfnamefont{G.}~\bibnamefont{{Notesco}}},\ and\ \bibinfo {author}
  {\bibfnamefont{A.}~\bibnamefont{{Bar-Nun}}}}%
  , \bibinfo {year} {2009},\ \bibfield{title}{%
  \enquote{\bibinfo {title} {{An experimental study of the formation of an ice
  crust and migration of water vapor in a comet's upper layers}},}\ }%
  \bibfield{journal}{%
  \bibinfo {journal} {Icarus}\ }%
  \textbf{\bibinfo {volume} {201}},\ \bibinfo {pages} {406--411}%
  \bibAnnoteFile{NoStop}{patel2009}%
\bibitem[{\citenamefont{Paterson}(1994)}]{Paterson1994}%
  \BibitemOpen
  \bibfield{author}{%
  \bibinfo {author} {\bibnamefont{Paterson}, \bibfnamefont{W.~S.~B.}}}%
  , \bibinfo {year} {1994},\ \emph{\bibinfo {title} {The Physics of
  Glaciers}},\ \bibinfo {edition} {3rd}\ ed.\ (\bibinfo {publisher} {Reed})%
  \bibAnnoteFile{NoStop}{Paterson1994}%
\bibitem[{\citenamefont{Pedersen}\ \emph{et~al.}(2009)\citenamefont{Pedersen},
  \citenamefont{Roecker}, \citenamefont{L\"uthje},\ and\
  \citenamefont{Winter}}]{Pedersen:2009}%
  \BibitemOpen
  \bibfield{author}{%
  \bibinfo {author} {\bibnamefont{Pedersen}, \bibfnamefont{C.~A.}}, \bibinfo
  {author} {\bibfnamefont{E.}~\bibnamefont{Roecker}}, \bibinfo {author}
  {\bibfnamefont{M.}~\bibnamefont{L\"uthje}},\ and\ \bibinfo {author}
  {\bibfnamefont{J.-G.}\ \bibnamefont{Winter}}}%
  , \bibinfo {year} {2009},\ \bibfield{title}{%
  \enquote{\bibinfo {title} {A new sea ice albedo scheme including melt ponds
  for {ECHAM5} general circulation model},}\ }%
  \bibfield{journal}{%
  \bibinfo {journal} {J.~Geophys.~Res.}\ }%
  \textbf{\bibinfo {volume} {114}},\ \bibinfo {pages} {D08101}%
  \bibAnnoteFile{NoStop}{Pedersen:2009}%
\bibitem[{\citenamefont{Perets}\ \emph{et~al.}(2005)\citenamefont{Perets},
  \citenamefont{Biham}, \citenamefont{Manico}, \citenamefont{Pirronello},
  \citenamefont{Roser}, \citenamefont{Swords},\ and\
  \citenamefont{Vidali}}]{perets2005}%
  \BibitemOpen
  \bibfield{author}{%
  \bibinfo {author} {\bibnamefont{Perets}, \bibfnamefont{H.~B.}}, \bibinfo
  {author} {\bibfnamefont{O.}~\bibnamefont{Biham}}, \bibinfo {author}
  {\bibfnamefont{G.}~\bibnamefont{Manico}}, \bibinfo {author}
  {\bibfnamefont{V.}~\bibnamefont{Pirronello}}, \bibinfo {author}
  {\bibfnamefont{J.}~\bibnamefont{Roser}}, \bibinfo {author}
  {\bibfnamefont{S.}~\bibnamefont{Swords}},\ and\ \bibinfo {author}
  {\bibfnamefont{G.}~\bibnamefont{Vidali}}}%
  , \bibinfo {year} {2005},\ \bibfield{title}{%
  \enquote{\bibinfo {title} {Molecular hydrogen formation on ice under
  interstellar conditions},}\ }%
  \bibfield{journal}{%
  \bibinfo {journal} {Astrophys. J.}\ }%
  \textbf{\bibinfo {volume} {627}},\ \bibinfo {pages} {850--860}%
  \bibAnnoteFile{NoStop}{perets2005}%
\bibitem[{\citenamefont{Perovich}\ and\
  \citenamefont{Richter-Menge}(1994)}]{Perovich:1994}%
  \BibitemOpen
  \bibfield{author}{%
  \bibinfo {author} {\bibnamefont{Perovich}, \bibfnamefont{D.K.}},\ and\
  \bibinfo {author} {\bibfnamefont{J.~A.}\ \bibnamefont{Richter-Menge}}}%
  , \bibinfo {year} {1994},\ \bibfield{title}{%
  \enquote{\bibinfo {title} {Surface characteristics of lead ice},}\ }%
  \bibfield{journal}{%
  \bibinfo {journal} {J.~Geophys.~Res.}\ }%
  \textbf{\bibinfo {volume} {99}},\ \bibinfo {pages} {16341--16350}%
  \bibAnnoteFile{NoStop}{Perovich:1994}%
\bibitem[{\citenamefont{Pertaya}\
  \emph{et~al.}(2007{\natexlab{a}})\citenamefont{Pertaya},
  \citenamefont{Celik}, \citenamefont{DiPrinzio}, \citenamefont{Wettlaufer},
  \citenamefont{Davies},\ and\ \citenamefont{Braslavsky}}]{Pertaya2007b}%
  \BibitemOpen
  \bibfield{author}{%
  \bibinfo {author} {\bibnamefont{Pertaya}, \bibfnamefont{N.}}, \bibinfo
  {author} {\bibfnamefont{Y.}~\bibnamefont{Celik}}, \bibinfo {author}
  {\bibfnamefont{C.~L.}\ \bibnamefont{DiPrinzio}}, \bibinfo {author}
  {\bibfnamefont{J.~S.}\ \bibnamefont{Wettlaufer}}, \bibinfo {author}
  {\bibfnamefont{P.~L.}\ \bibnamefont{Davies}},\ and\ \bibinfo {author}
  {\bibfnamefont{I.}~\bibnamefont{Braslavsky}}}%
  , \bibinfo {year} {2007}{\natexlab{a}},\ \bibfield{title}{%
  \enquote{\bibinfo {title} {Growth--melt asymmetry in ice crystals under the
  influence of spruce budworm antifreeze protein},}\ }%
  \bibfield{journal}{%
  \bibinfo {journal} {J. Phys.: Cond. Matt.}\ }%
  \textbf{\bibinfo {volume} {19}},\ \bibinfo {pages} {412101}%
  \bibAnnoteFile{NoStop}{Pertaya2007b}%
\bibitem[{\citenamefont{Pertaya}\
  \emph{et~al.}(2007{\natexlab{b}})\citenamefont{Pertaya},
  \citenamefont{Marshall}, \citenamefont{DiPrinzio}, \citenamefont{Wilen},
  \citenamefont{Thomson}, \citenamefont{Wettlaufer}, \citenamefont{Davies},\
  and\ \citenamefont{Braslavsky}}]{Pertaya2007a}%
  \BibitemOpen
  \bibfield{author}{%
  \bibinfo {author} {\bibnamefont{Pertaya}, \bibfnamefont{N.}}, \bibinfo
  {author} {\bibfnamefont{C.~B.}\ \bibnamefont{Marshall}}, \bibinfo {author}
  {\bibfnamefont{C.~L.}\ \bibnamefont{DiPrinzio}}, \bibinfo {author}
  {\bibfnamefont{L.~A.}\ \bibnamefont{Wilen}}, \bibinfo {author}
  {\bibfnamefont{E.~S.}\ \bibnamefont{Thomson}}, \bibinfo {author}
  {\bibfnamefont{J.~S.}\ \bibnamefont{Wettlaufer}}, \bibinfo {author}
  {\bibfnamefont{P.~L~.}\ \bibnamefont{Davies}},\ and\ \bibinfo {author}
  {\bibfnamefont{I.}~\bibnamefont{Braslavsky}}}%
  , \bibinfo {year} {2007}{\natexlab{b}},\ \bibfield{title}{%
  \enquote{\bibinfo {title} {Fluorescence microscopy evidence for
  quasi-permanent attachment of antifreeze proteins to ice surfaces},}\ }%
  \bibfield{journal}{%
  \bibinfo {journal} {Biophys. J.}\ }%
  \textbf{\bibinfo {volume} {92}},\ \bibinfo {pages} {3663--3673}%
  \bibAnnoteFile{NoStop}{Pertaya2007a}%
\bibitem[{\citenamefont{Peter}\ \emph{et~al.}(2006)\citenamefont{Peter},
  \citenamefont{Marcolli}, \citenamefont{Spaichinger}, \citenamefont{Corti},
  \citenamefont{Baker},\ and\ \citenamefont{Koop}}]{peter2006}%
  \BibitemOpen
  \bibfield{author}{%
  \bibinfo {author} {\bibnamefont{Peter}, \bibfnamefont{T.}}, \bibinfo {author}
  {\bibfnamefont{C.}~\bibnamefont{Marcolli}}, \bibinfo {author}
  {\bibfnamefont{P.}~\bibnamefont{Spaichinger}}, \bibinfo {author}
  {\bibfnamefont{T.}~\bibnamefont{Corti}}, \bibinfo {author}
  {\bibfnamefont{M.~C.}\ \bibnamefont{Baker}},\ and\ \bibinfo {author}
  {\bibfnamefont{T.}~\bibnamefont{Koop}}}%
  , \bibinfo {year} {2006},\ \bibfield{title}{%
  \enquote{\bibinfo {title} {When dry air is too humid},}\ }%
  \bibfield{journal}{%
  \bibinfo {journal} {Science}\ }%
  \textbf{\bibinfo {volume} {314}},\ \bibinfo {pages} {1399--1402}%
  \bibAnnoteFile{NoStop}{peter2006}%
\bibitem[{\citenamefont{Petrenko}\ and\
  \citenamefont{Whitworth}(1999)}]{petrenko1999}%
  \BibitemOpen
  \bibfield{author}{%
  \bibinfo {author} {\bibnamefont{Petrenko}, \bibfnamefont{V.~F.}},\ and\
  \bibinfo {author} {\bibfnamefont{R.~W.}\ \bibnamefont{Whitworth}}}%
  , \bibinfo {year} {1999},\ \emph{\bibinfo {title} {Physics of Ice}}\
  (\bibinfo {publisher} {Oxford University Press})%
  \bibAnnoteFile{NoStop}{petrenko1999}%
\bibitem[{\citenamefont{Pinzer}\ \emph{et~al.}(2010)\citenamefont{Pinzer},
  \citenamefont{Kerbrat}, \citenamefont{Huthwelker},
  \citenamefont{G{\"a}ggeler}, \citenamefont{Schneebeli},\ and\
  \citenamefont{Ammann}}]{Pinzer:2010p25893}%
  \BibitemOpen
  \bibfield{author}{%
  \bibinfo {author} {\bibnamefont{Pinzer}, \bibfnamefont{B.}}, \bibinfo
  {author} {\bibfnamefont{M.}~\bibnamefont{Kerbrat}}, \bibinfo {author}
  {\bibfnamefont{T.}~\bibnamefont{Huthwelker}}, \bibinfo {author}
  {\bibfnamefont{H.~W.}\ \bibnamefont{G{\"a}ggeler}}, \bibinfo {author}
  {\bibfnamefont{M.}~\bibnamefont{Schneebeli}},\ and\ \bibinfo {author}
  {\bibfnamefont{M.}~\bibnamefont{Ammann}}}%
  , \bibinfo {year} {2010},\ \bibfield{title}{%
  \enquote{\bibinfo {title} {Diffusion of {NO$_x$} and {HONO} in snow: A
  laboratory study},}\ }%
  \bibfield{journal}{%
  \bibinfo {journal} {J. Geophys. Res.}\ }%
  \textbf{\bibinfo {volume} {115}},\ \bibinfo {pages} {D03304}%
  \bibAnnoteFile{NoStop}{Pinzer:2010p25893}%
\bibitem[{\citenamefont{Pinzer}\ and\
  \citenamefont{Schneebeli}(2009)}]{Pinzer:2009p25366}%
  \BibitemOpen
  \bibfield{author}{%
  \bibinfo {author} {\bibnamefont{Pinzer}, \bibfnamefont{B.}},\ and\ \bibinfo
  {author} {\bibfnamefont{M.}~\bibnamefont{Schneebeli}}}%
  , \bibinfo {year} {2009},\ \bibfield{title}{%
  \enquote{\bibinfo {title} {Snow metamorphism under alternating temperature
  gradients: Morphology and recrystallization in surface snow},}\ }%
  \bibfield{journal}{%
  \bibinfo {journal} {Geophys. Res. Lett.}\ }%
  \textbf{\bibinfo {volume} {36}},\ \bibinfo {pages} {L23503}%
  \bibAnnoteFile{NoStop}{Pinzer:2009p25366}%
\bibitem[{\citenamefont{Pitter}\ and\
  \citenamefont{Pruppacher}(1973)}]{Pitter1973}%
  \BibitemOpen
  \bibfield{author}{%
  \bibinfo {author} {\bibnamefont{Pitter}, \bibfnamefont{R.~L.}},\ and\
  \bibinfo {author} {\bibfnamefont{H.~R.}\ \bibnamefont{Pruppacher}}}%
  , \bibinfo {year} {1973},\ \bibfield{title}{%
  \enquote{\bibinfo {title} {Wind-tunnel investigation of freezing of small
  water drops falling at terminal velocity in air},}\ }%
  \bibfield{journal}{%
  \bibinfo {journal} {Quart. J. Roy. Meteorol. Soc.}\ }%
  \textbf{\bibinfo {volume} {99}},\ \bibinfo {pages} {540--550}%
  \bibAnnoteFile{NoStop}{Pitter1973}%
\bibitem[{\citenamefont{Poole}\ \emph{et~al.}(1992)\citenamefont{Poole},
  \citenamefont{Sciortino}, \citenamefont{Essman},\ and\
  \citenamefont{Stanley}}]{poole1992}%
  \BibitemOpen
  \bibfield{author}{%
  \bibinfo {author} {\bibnamefont{Poole}, \bibfnamefont{P.~H.}}, \bibinfo
  {author} {\bibfnamefont{F.}~\bibnamefont{Sciortino}}, \bibinfo {author}
  {\bibfnamefont{U.}~\bibnamefont{Essman}},\ and\ \bibinfo {author}
  {\bibfnamefont{H.~E.}\ \bibnamefont{Stanley}}}%
  , \bibinfo {year} {1992},\ \bibfield{title}{%
  \enquote{\bibinfo {title} {Phase behaviour of metastable water},}\ }%
  \bibfield{journal}{%
  \bibinfo {journal} {Nature}\ }%
  \textbf{\bibinfo {volume} {360}},\ \bibinfo {pages} {324--328}%
  \bibAnnoteFile{NoStop}{poole1992}%
\bibitem[{\citenamefont{{P{\"o}schl}}\
  \emph{et~al.}(2010)\citenamefont{{P{\"o}schl}}, \citenamefont{{Martin}},
  \citenamefont{{Sinha}}, \citenamefont{{Chen}}, \citenamefont{{Gunthe}},
  \citenamefont{{Huffman}}, \citenamefont{{Borrmann}}, \citenamefont{{Farmer}},
  \citenamefont{{Garland}}, \citenamefont{{Helas}}, \citenamefont{{Jimenez}},
  \citenamefont{{King}}, \citenamefont{{Manzi}}, \citenamefont{{Mikhailov}},
  \citenamefont{{Pauliquevis}}, \citenamefont{{Petters}},
  \citenamefont{{Prenni}}, \citenamefont{{Roldin}}, \citenamefont{{Rose}},
  \citenamefont{{Schneider}}, \citenamefont{{Su}}, \citenamefont{{Zorn}},
  \citenamefont{{Artaxo}},\ and\ \citenamefont{{Andreae}}}]{poschl2010}%
  \BibitemOpen
  \bibfield{author}{%
  \bibinfo {author} {\bibnamefont{{P{\"o}schl}}, \bibfnamefont{U.}}, \bibinfo
  {author} {\bibfnamefont{S.~T.}\ \bibnamefont{{Martin}}}, \bibinfo {author}
  {\bibfnamefont{B.}~\bibnamefont{{Sinha}}}, \bibinfo {author}
  {\bibfnamefont{Q.}~\bibnamefont{{Chen}}}, \bibinfo {author}
  {\bibfnamefont{S.~S.}\ \bibnamefont{{Gunthe}}}, \bibinfo {author}
  {\bibfnamefont{J.~A.}\ \bibnamefont{{Huffman}}}, \bibinfo {author}
  {\bibfnamefont{S.}~\bibnamefont{{Borrmann}}}, \bibinfo {author}
  {\bibfnamefont{D.~K.}\ \bibnamefont{{Farmer}}}, \bibinfo {author}
  {\bibfnamefont{R.~M.}\ \bibnamefont{{Garland}}}, \bibinfo {author}
  {\bibfnamefont{G.}~\bibnamefont{{Helas}}}, \bibinfo {author}
  {\bibfnamefont{J.~L.}\ \bibnamefont{{Jimenez}}}, \bibinfo {author}
  {\bibfnamefont{S.~M.}\ \bibnamefont{{King}}}, \bibinfo {author}
  {\bibfnamefont{A.}~\bibnamefont{{Manzi}}}, \bibinfo {author}
  {\bibfnamefont{E.}~\bibnamefont{{Mikhailov}}}, \bibinfo {author}
  {\bibfnamefont{T.}~\bibnamefont{{Pauliquevis}}}, \bibinfo {author}
  {\bibfnamefont{M.~D.}\ \bibnamefont{{Petters}}}, \bibinfo {author}
  {\bibfnamefont{A.~J.}\ \bibnamefont{{Prenni}}}, \bibinfo {author}
  {\bibfnamefont{P.}~\bibnamefont{{Roldin}}}, \bibinfo {author}
  {\bibfnamefont{D.}~\bibnamefont{{Rose}}}, \bibinfo {author}
  {\bibfnamefont{J.}~\bibnamefont{{Schneider}}}, \bibinfo {author}
  {\bibfnamefont{H.}~\bibnamefont{{Su}}}, \bibinfo {author}
  {\bibfnamefont{S.~R.}\ \bibnamefont{{Zorn}}}, \bibinfo {author}
  {\bibfnamefont{P.}~\bibnamefont{{Artaxo}}},\ and\ \bibinfo {author}
  {\bibfnamefont{M.~O.}\ \bibnamefont{{Andreae}}}}%
  , \bibinfo {year} {2010},\ \bibfield{title}{%
  \enquote{\bibinfo {title} {{Rainforest Aerosols as Biogenic Nuclei of Clouds
  and Precipitation in the Amazon}},}\ }%
  \bibfield{journal}{%
  \bibinfo {journal} {Science}\ }%
  \textbf{\bibinfo {volume} {329}},\ \bibinfo {pages} {1513--1516}%
  \bibAnnoteFile{NoStop}{poschl2010}%
\bibitem[{\citenamefont{Post}\ and\
  \citenamefont{LaChapelle}(1971)}]{post1971}%
  \BibitemOpen
  \bibfield{author}{%
  \bibinfo {author} {\bibnamefont{Post}, \bibfnamefont{A.}},\ and\ \bibinfo
  {author} {\bibfnamefont{E.~R.}\ \bibnamefont{LaChapelle}}}%
  , \bibinfo {year} {1971},\ \emph{\bibinfo {title} {Glacier Ice}}\ (\bibinfo
  {publisher} {University of Washinton Press})%
  \bibAnnoteFile{NoStop}{post1971}%
\bibitem[{\citenamefont{Pratt}\ \emph{et~al.}(2009)\citenamefont{Pratt},
  \citenamefont{DeMott}, \citenamefont{French}, \citenamefont{Wang},
  \citenamefont{Westphal}, \citenamefont{Heymsfield}, \citenamefont{Twohy},
  \citenamefont{Prenni},\ and\ \citenamefont{Prather}}]{pratt2009}%
  \BibitemOpen
  \bibfield{author}{%
  \bibinfo {author} {\bibnamefont{Pratt}, \bibfnamefont{K.~A.}}, \bibinfo
  {author} {\bibfnamefont{P.~J.}\ \bibnamefont{DeMott}}, \bibinfo {author}
  {\bibfnamefont{J.~R.}\ \bibnamefont{French}}, \bibinfo {author}
  {\bibfnamefont{Z.}~\bibnamefont{Wang}}, \bibinfo {author}
  {\bibfnamefont{D.~L.}\ \bibnamefont{Westphal}}, \bibinfo {author}
  {\bibfnamefont{A.~J.}\ \bibnamefont{Heymsfield}}, \bibinfo {author}
  {\bibfnamefont{C.~H.}\ \bibnamefont{Twohy}}, \bibinfo {author}
  {\bibfnamefont{A.~J.}\ \bibnamefont{Prenni}},\ and\ \bibinfo {author}
  {\bibfnamefont{K.~A.}\ \bibnamefont{Prather}}}%
  , \bibinfo {year} {2009},\ \bibfield{title}{%
  \enquote{\bibinfo {title} {In situ detection of biological particles in cloud
  ice-crystals},}\ }%
  \bibfield{journal}{%
  \bibinfo {journal} {Nature Geosci.}\ }%
  \textbf{\bibinfo {volume} {2}},\ \bibinfo {pages} {397--400}%
  \bibAnnoteFile{NoStop}{pratt2009}%
\bibitem[{\citenamefont{Pratte}\ \emph{et~al.}(2006)\citenamefont{Pratte},
  \citenamefont{van~den Bergh},\ and\ \citenamefont{Rossi}}]{Pratte2006}%
  \BibitemOpen
  \bibfield{author}{%
  \bibinfo {author} {\bibnamefont{Pratte}, \bibfnamefont{P.}}, \bibinfo
  {author} {\bibfnamefont{H.}~\bibnamefont{van~den Bergh}},\ and\ \bibinfo
  {author} {\bibfnamefont{M.~J.}\ \bibnamefont{Rossi}}}%
  , \bibinfo {year} {2006},\ \bibfield{title}{%
  \enquote{\bibinfo {title} {The kinetics of {H$_2$O} vapor condensation and
  evaporation on different types of ice in the range 130--210~{K}},}\ }%
  \bibinfo {journal} {J. Phys. Chem. A},\ \bibinfo {pages} {3042--3058}%
  \bibAnnoteFile{NoStop}{Pratte2006}%
\bibitem[{\citenamefont{Prialnik}\ \emph{et~al.}(2003)\citenamefont{Prialnik},
  \citenamefont{Benkhoff},\ and\ \citenamefont{Podolak}}]{prialnik2003}%
  \BibitemOpen
\bibfield{journal}{%
    }%
  \bibfield{author}{%
  \bibinfo {author} {\bibnamefont{Prialnik}, \bibfnamefont{D.}}, \bibinfo
  {author} {\bibfnamefont{J.}~\bibnamefont{Benkhoff}},\ and\ \bibinfo {author}
  {\bibfnamefont{M.}~\bibnamefont{Podolak}}}%
  , \bibinfo {year} {2003},\ \enquote{\bibinfo {title} {Modeling the structure
  and activity of comet nuclei},}\ in\ \emph{\bibinfo {booktitle} {Comets
  II}},\ \bibinfo {editor} {edited by\ \bibinfo {editor} {\bibfnamefont{M.~C.}\
  \bibnamefont{Festou}}, \bibinfo {editor} {\bibfnamefont{H.~U.}\
  \bibnamefont{Keller}},\ and\ \bibinfo {editor} {\bibfnamefont{H.~A.}\
  \bibnamefont{Weaver}}}\ (\bibinfo {publisher} {University of Arizona Press})\
  pp.\ \bibinfo {pages} {359--387}%
  \bibAnnoteFile{NoStop}{prialnik2003}%
\bibitem[{\citenamefont{Price}(2007)}]{price2007}%
  \BibitemOpen
  \bibfield{author}{%
  \bibinfo {author} {\bibnamefont{Price}, \bibfnamefont{P.~B.}}}%
  , \bibinfo {year} {2007},\ \bibfield{title}{%
  \enquote{\bibinfo {title} {Microbial life in glacial ice and implications for
  a cold origin of life},}\ }%
  \bibfield{journal}{%
  \bibinfo {journal} {FEMS Microbiol. Ecol.}\ }%
  \textbf{\bibinfo {volume} {59}},\ \bibinfo {pages} {217--231}%
  \bibAnnoteFile{NoStop}{price2007}%
\bibitem[{\citenamefont{Price}(2010)}]{price2010}%
  \BibitemOpen
  \bibfield{author}{%
  \bibinfo {author} {\bibnamefont{Price}, \bibfnamefont{P.~B.}}}%
  , \bibinfo {year} {2010},\ \bibfield{title}{%
  \enquote{\bibinfo {title} {Microbial life in martian ice: A biotic origin of
  methane on {Mars}?}.}\ }%
  \bibfield{journal}{%
  \bibinfo {journal} {Planet. Space Sci.}\ }%
  \textbf{\bibinfo {volume} {58}},\ \bibinfo {pages} {1199--1206}%
  \bibAnnoteFile{NoStop}{price2010}%
\bibitem[{\citenamefont{Pritchard}\
  \emph{et~al.}(2009)\citenamefont{Pritchard}, \citenamefont{Arthern},
  \citenamefont{Vaughan},\ and\ \citenamefont{Edwards}}]{Pritchard:2009p26033}%
  \BibitemOpen
  \bibfield{author}{%
  \bibinfo {author} {\bibnamefont{Pritchard}, \bibfnamefont{H.}}, \bibinfo
  {author} {\bibfnamefont{R.}~\bibnamefont{Arthern}}, \bibinfo {author}
  {\bibfnamefont{D.}~\bibnamefont{Vaughan}},\ and\ \bibinfo {author}
  {\bibfnamefont{L.}~\bibnamefont{Edwards}}}%
  , \bibinfo {year} {2009},\ \bibfield{title}{%
  \enquote{\bibinfo {title} {Extensive dynamic thinning on the margins of the
  {Greenland} and {Antarctic} ice sheets},}\ }%
  \bibfield{journal}{%
  \bibinfo {journal} {Nature}\ }%
  \textbf{\bibinfo {volume} {461}},\ \bibinfo {pages} {971--975}%
  \bibAnnoteFile{NoStop}{Pritchard:2009p26033}%
\bibitem[{\citenamefont{Pruppacher}\ and\
  \citenamefont{Klett}(1997)}]{Pruppacher1997}%
  \BibitemOpen
  \bibfield{author}{%
  \bibinfo {author} {\bibnamefont{Pruppacher}, \bibfnamefont{H.~R.}},\ and\
  \bibinfo {author} {\bibfnamefont{J.~D.}\ \bibnamefont{Klett}}}%
  , \bibinfo {year} {1997},\ \emph{\bibinfo {title} {Microphysics of clouds and
  precipitation}}\ (\bibinfo {publisher} {Reidel},\ \bibinfo {address}
  {Dordrecht, Holland})%
  \bibAnnoteFile{NoStop}{Pruppacher1997}%
\bibitem[{\citenamefont{Pummer}\ \emph{et~al.}(2011)\citenamefont{Pummer},
  \citenamefont{Bauer}, \citenamefont{Bernardi}, \citenamefont{Bleicher},\ and\
  \citenamefont{Grothe}}]{pummer2011}%
  \BibitemOpen
  \bibfield{author}{%
  \bibinfo {author} {\bibnamefont{Pummer}, \bibfnamefont{B.~G.}}, \bibinfo
  {author} {\bibfnamefont{H.}~\bibnamefont{Bauer}}, \bibinfo {author}
  {\bibfnamefont{J.}~\bibnamefont{Bernardi}}, \bibinfo {author}
  {\bibfnamefont{S.}~\bibnamefont{Bleicher}},\ and\ \bibinfo {author}
  {\bibfnamefont{H.}~\bibnamefont{Grothe}}}%
  , \bibinfo {year} {2011},\ \bibfield{title}{%
  \enquote{\bibinfo {title} {Birch and conifer pollen are efficient atmospheric
  ice nuclei},}\ }%
  \bibfield{journal}{%
  \bibinfo {journal} {Atmos. Chem. Phys. Disc.}\ }%
  \textbf{\bibinfo {volume} {11}},\ \bibinfo {pages} {27219--27241}%
  \bibAnnoteFile{NoStop}{pummer2011}%
\bibitem[{\citenamefont{Rampal}\ \emph{et~al.}(2009)\citenamefont{Rampal},
  \citenamefont{Weiss},\ and\ \citenamefont{Marsan}}]{Rampal:2009}%
  \BibitemOpen
  \bibfield{author}{%
  \bibinfo {author} {\bibnamefont{Rampal}, \bibfnamefont{P.}}, \bibinfo
  {author} {\bibfnamefont{J.}~\bibnamefont{Weiss}},\ and\ \bibinfo {author}
  {\bibfnamefont{D.}~\bibnamefont{Marsan}}}%
  , \bibinfo {year} {2009},\ \bibfield{title}{%
  \enquote{\bibinfo {title} {Positive trend in the mean speed and deformation
  rate of {Arctic} sea ice},}\ }%
  \bibfield{journal}{%
  \bibinfo {journal} {J.~Geophys.~Res.}\ }%
  \textbf{\bibinfo {volume} {114}},\ \bibinfo {pages} {C05013}%
  \bibAnnoteFile{NoStop}{Rampal:2009}%
\bibitem[{\citenamefont{Rapp}\ and\ \citenamefont{Thomas}(2006)}]{Rapp2006}%
  \BibitemOpen
  \bibfield{author}{%
  \bibinfo {author} {\bibnamefont{Rapp}, \bibfnamefont{M.}},\ and\ \bibinfo
  {author} {\bibfnamefont{G.~E.}\ \bibnamefont{Thomas}}}%
  , \bibinfo {year} {2006},\ \bibfield{title}{%
  \enquote{\bibinfo {title} {Modeling the microphysics of mesospheric ice
  particles: Assessment of current capabilities and basic sensitivities},}\ }%
  \bibfield{journal}{%
  \bibinfo {journal} {J. Atmos. Solar-Terrestrial Phys.}\ }%
  \textbf{\bibinfo {volume} {68}},\ \bibinfo {pages} {715--744}%
  \bibAnnoteFile{NoStop}{Rapp2006}%
\bibitem[{\citenamefont{Raviv}\ \emph{et~al.}(2001)\citenamefont{Raviv},
  \citenamefont{Laurat},\ and\ \citenamefont{Klein}}]{Raviv2001}%
  \BibitemOpen
  \bibfield{author}{%
  \bibinfo {author} {\bibnamefont{Raviv}, \bibfnamefont{U.}}, \bibinfo {author}
  {\bibfnamefont{P.}~\bibnamefont{Laurat}},\ and\ \bibinfo {author}
  {\bibfnamefont{J.}~\bibnamefont{Klein}}}%
  , \bibinfo {year} {2001},\ \bibfield{title}{%
  \enquote{\bibinfo {title} {Fluidity of water confined to subnanometre
  films},}\ }%
  \bibfield{journal}{%
  \bibinfo {journal} {Nature}\ }%
  \textbf{\bibinfo {volume} {413}},\ \bibinfo {pages} {51--54}%
  \bibAnnoteFile{NoStop}{Raviv2001}%
\bibitem[{\citenamefont{Reitz}\ \emph{et~al.}(2011)\citenamefont{Reitz},
  \citenamefont{Spindler}, \citenamefont{Mentel}, \citenamefont{Poulain},
  \citenamefont{Wex}, \citenamefont{Mildenberger}, \citenamefont{Niedermeier},
  \citenamefont{Hartmann}, \citenamefont{Clauss}, \citenamefont{Stratmann},
  \citenamefont{Sullivan}, \citenamefont{DeMott}, \citenamefont{Petters},
  \citenamefont{Sierau},\ and\ \citenamefont{Schneider}}]{Reitz2011}%
  \BibitemOpen
  \bibfield{author}{%
  \bibinfo {author} {\bibnamefont{Reitz}, \bibfnamefont{P.}}, \bibinfo {author}
  {\bibfnamefont{C.}~\bibnamefont{Spindler}}, \bibinfo {author}
  {\bibfnamefont{T.~F.}\ \bibnamefont{Mentel}}, \bibinfo {author}
  {\bibfnamefont{L.}~\bibnamefont{Poulain}}, \bibinfo {author}
  {\bibfnamefont{H.}~\bibnamefont{Wex}}, \bibinfo {author}
  {\bibfnamefont{K.}~\bibnamefont{Mildenberger}}, \bibinfo {author}
  {\bibfnamefont{D.}~\bibnamefont{Niedermeier}}, \bibinfo {author}
  {\bibfnamefont{S.}~\bibnamefont{Hartmann}}, \bibinfo {author}
  {\bibfnamefont{T.}~\bibnamefont{Clauss}}, \bibinfo {author}
  {\bibfnamefont{F.}~\bibnamefont{Stratmann}}, \bibinfo {author}
  {\bibfnamefont{R.~C.}\ \bibnamefont{Sullivan}}, \bibinfo {author}
  {\bibfnamefont{P.~J.}\ \bibnamefont{DeMott}}, \bibinfo {author}
  {\bibfnamefont{M.~D.}\ \bibnamefont{Petters}}, \bibinfo {author}
  {\bibfnamefont{B.}~\bibnamefont{Sierau}},\ and\ \bibinfo {author}
  {\bibfnamefont{J.}~\bibnamefont{Schneider}}}%
  , \bibinfo {year} {2011},\ \bibfield{title}{%
  \enquote{\bibinfo {title} {Surface modification of mineral dust particles by
  sulphuric acid processing: implications for ice nucleation abilities},}\ }%
  \bibfield{journal}{%
  \bibinfo {journal} {Atmos. Chem. Phys.}\ }%
  \textbf{\bibinfo {volume} {11}},\ \bibinfo {pages} {7839--7858}%
  \bibAnnoteFile{NoStop}{Reitz2011}%
\bibitem[{\citenamefont{Rempel}\ \emph{et~al.}(2001)\citenamefont{Rempel},
  \citenamefont{Waddington}, \citenamefont{Wettlaufer},\ and\
  \citenamefont{Worster}}]{Rempel:2001p26726}%
  \BibitemOpen
  \bibfield{author}{%
  \bibinfo {author} {\bibnamefont{Rempel}, \bibfnamefont{A.}}, \bibinfo
  {author} {\bibfnamefont{E.}~\bibnamefont{Waddington}}, \bibinfo {author}
  {\bibfnamefont{J.}~\bibnamefont{Wettlaufer}},\ and\ \bibinfo {author}
  {\bibfnamefont{M.}~\bibnamefont{Worster}}}%
  , \bibinfo {year} {2001},\ \bibfield{title}{%
  \enquote{\bibinfo {title} {Possible displacement of the climate signal in
  ancient ice by premelting and anomalous diffusion},}\ }%
  \bibfield{journal}{%
  \bibinfo {journal} {Nature}\ }%
  \textbf{\bibinfo {volume} {411}},\ \bibinfo {pages} {568--571}%
  \bibAnnoteFile{NoStop}{Rempel:2001p26726}%
\bibitem[{\citenamefont{Richardson}\
  \emph{et~al.}(1985)\citenamefont{Richardson}, \citenamefont{Wooldridge},\
  and\ \citenamefont{Devlin}}]{richardson1985}%
  \BibitemOpen
  \bibfield{author}{%
  \bibinfo {author} {\bibnamefont{Richardson}, \bibfnamefont{H.~H.}}, \bibinfo
  {author} {\bibfnamefont{P.~J.}\ \bibnamefont{Wooldridge}},\ and\ \bibinfo
  {author} {\bibfnamefont{J.~P.}\ \bibnamefont{Devlin}}}%
  , \bibinfo {year} {1985},\ \bibfield{title}{%
  \enquote{\bibinfo {title} {{FT-IR} spectra of vacuum deposited clathrate
  hydrate of oxirane {H$_2$S}, {THF}, and ethane},}\ }%
  \bibfield{journal}{%
  \bibinfo {journal} {J. Chem. Phys.}\ }%
  \textbf{\bibinfo {volume} {83}},\ \bibinfo {pages} {4387--4394}%
  \bibAnnoteFile{NoStop}{richardson1985}%
\bibitem[{\citenamefont{Rignot}\ and\
  \citenamefont{Kanagaratnam}(2006)}]{Rignot:2006p26032}%
  \BibitemOpen
  \bibfield{author}{%
  \bibinfo {author} {\bibnamefont{Rignot}, \bibfnamefont{E.}},\ and\ \bibinfo
  {author} {\bibfnamefont{P.}~\bibnamefont{Kanagaratnam}}}%
  , \bibinfo {year} {2006},\ \bibfield{title}{%
  \enquote{\bibinfo {title} {Changes in the velocity structure of the
  {Greenland} ice sheet},}\ }%
  \bibfield{journal}{%
  \bibinfo {journal} {Science}\ }%
  \textbf{\bibinfo {volume} {311}},\ \bibinfo {pages} {986--990}%
  \bibAnnoteFile{NoStop}{Rignot:2006p26032}%
\bibitem[{\citenamefont{Rigor}\ and\
  \citenamefont{Wallace}(2004)}]{Rigor:2004}%
  \BibitemOpen
  \bibfield{author}{%
  \bibinfo {author} {\bibnamefont{Rigor}, \bibfnamefont{I.~G.}},\ and\ \bibinfo
  {author} {\bibfnamefont{J.~M.}\ \bibnamefont{Wallace}}}%
  , \bibinfo {year} {2004},\ \bibfield{title}{%
  \enquote{\bibinfo {title} {Variations in the age of the {Arctic} sea-ice and
  summer sea-ice extent},}\ }%
  \bibfield{journal}{%
  \bibinfo {journal} {Geophys.~Res.~Lett.}\ }%
  \textbf{\bibinfo {volume} {31}},\ \bibinfo {pages} {L09401}%
  \bibAnnoteFile{NoStop}{Rigor:2004}%
\bibitem[{\citenamefont{Rigor}\ \emph{et~al.}(2002)\citenamefont{Rigor},
  \citenamefont{Wallace},\ and\ \citenamefont{Colony}}]{Rigor:2002}%
  \BibitemOpen
  \bibfield{author}{%
  \bibinfo {author} {\bibnamefont{Rigor}, \bibfnamefont{I.~G.}}, \bibinfo
  {author} {\bibfnamefont{J.~M.}\ \bibnamefont{Wallace}},\ and\ \bibinfo
  {author} {\bibfnamefont{R.~L.}\ \bibnamefont{Colony}}}%
  , \bibinfo {year} {2002},\ \bibfield{title}{%
  \enquote{\bibinfo {title} {Response of sea ice to the {A}rctic
  {O}scillation},}\ }%
  \bibfield{journal}{%
  \bibinfo {journal} {J.~Clim.}\ }%
  \textbf{\bibinfo {volume} {15}},\ \bibinfo {pages} {2648--2663}%
  \bibAnnoteFile{NoStop}{Rigor:2002}%
\bibitem[{\citenamefont{Riikonen}\ \emph{et~al.}(2000)\citenamefont{Riikonen},
  \citenamefont{M.Sillanp{\"a\"a}}, \citenamefont{Virta},
  \citenamefont{Sullivan}, \citenamefont{Moilanen},\ and\
  \citenamefont{Luukkonen}}]{riikonen2000}%
  \BibitemOpen
  \bibfield{author}{%
  \bibinfo {author} {\bibnamefont{Riikonen}, \bibfnamefont{M.}}, \bibinfo
  {author} {\bibnamefont{M.Sillanp{\"a\"a}}}, \bibinfo {author}
  {\bibfnamefont{L.}~\bibnamefont{Virta}}, \bibinfo {author}
  {\bibfnamefont{D.}~\bibnamefont{Sullivan}}, \bibinfo {author}
  {\bibfnamefont{J.}~\bibnamefont{Moilanen}},\ and\ \bibinfo {author}
  {\bibfnamefont{I.}~\bibnamefont{Luukkonen}}}%
  , \bibinfo {year} {2000},\ \bibfield{title}{%
  \enquote{\bibinfo {title} {Halo observations provide evidence of airborne
  cubic ice in the earthÕs atmosphere},}\ }%
  \bibfield{journal}{%
  \bibinfo {journal} {Appl. Optics}\ }%
  \textbf{\bibinfo {volume} {39}},\ \bibinfo {pages} {6080--6085}%
  \bibAnnoteFile{NoStop}{riikonen2000}%
\bibitem[{\citenamefont{{Rivkin}}\ and\
  \citenamefont{{Emery}}(2010)}]{rivkin2010}%
  \BibitemOpen
  \bibfield{author}{%
  \bibinfo {author} {\bibnamefont{{Rivkin}}, \bibfnamefont{A.~S.}},\ and\
  \bibinfo {author} {\bibfnamefont{J.~P.}\ \bibnamefont{{Emery}}}}%
  , \bibinfo {year} {2010},\ \bibfield{title}{%
  \enquote{\bibinfo {title} {{Detection of ice and organics on an asteroidal
  surface}},}\ }%
  \bibfield{journal}{%
  \bibinfo {journal} {Nature}\ }%
  \textbf{\bibinfo {volume} {464}},\ \bibinfo {pages} {1322--1323}%
  \bibAnnoteFile{NoStop}{rivkin2010}%
\bibitem[{\citenamefont{Rode}(1999)}]{rode1999}%
  \BibitemOpen
  \bibfield{author}{%
  \bibinfo {author} {\bibnamefont{Rode}, \bibfnamefont{B.~M.}}}%
  , \bibinfo {year} {1999},\ \bibfield{title}{%
  \enquote{\bibinfo {title} {Peptides and the origin of life},}\ }%
  \bibfield{journal}{%
  \bibinfo {journal} {Peptides}\ }%
  \textbf{\bibinfo {volume} {20}},\ \bibinfo {pages} {773--786}%
  \bibAnnoteFile{NoStop}{rode1999}%
\bibitem[{\citenamefont{Roser}\ \emph{et~al.}(2002)\citenamefont{Roser},
  \citenamefont{Manic\`o}, \citenamefont{Pirronello},\ and\
  \citenamefont{Vidali}}]{roser2002}%
  \BibitemOpen
  \bibfield{author}{%
  \bibinfo {author} {\bibnamefont{Roser}, \bibfnamefont{J.~E.}}, \bibinfo
  {author} {\bibfnamefont{G.}~\bibnamefont{Manic\`o}}, \bibinfo {author}
  {\bibfnamefont{V.}~\bibnamefont{Pirronello}},\ and\ \bibinfo {author}
  {\bibfnamefont{G.}~\bibnamefont{Vidali}}}%
  , \bibinfo {year} {2002},\ \bibfield{title}{%
  \enquote{\bibinfo {title} {Formation of molecular hydrogen on amorphous water
  ice: Influence of morphology and ultraviolet exposure},}\ }%
  \bibfield{journal}{%
  \bibinfo {journal} {Astrophys. J.}\ }%
  \textbf{\bibinfo {volume} {581}},\ \bibinfo {pages} {276--284}%
  \bibAnnoteFile{NoStop}{roser2002}%
\bibitem[{\citenamefont{Rothrock}\ \emph{et~al.}(1999)\citenamefont{Rothrock},
  \citenamefont{Yu},\ and\ \citenamefont{Maykut}}]{rothrock:1999}%
  \BibitemOpen
  \bibfield{author}{%
  \bibinfo {author} {\bibnamefont{Rothrock}, \bibfnamefont{D.~A.}}, \bibinfo
  {author} {\bibfnamefont{Y.}~\bibnamefont{Yu}},\ and\ \bibinfo {author}
  {\bibfnamefont{G.~A.}\ \bibnamefont{Maykut}}}%
  , \bibinfo {year} {1999},\ \bibfield{title}{%
  \enquote{\bibinfo {title} {Thinning of the {Arctic} sea-ice cover},}\ }%
  \bibfield{journal}{%
  \bibinfo {journal} {Geophys.~Res.~Lett.}\ }%
  \textbf{\bibinfo {volume} {26}},\ \bibinfo {pages} {3469--3472}%
  \bibAnnoteFile{NoStop}{rothrock:1999}%
\bibitem[{\citenamefont{Rothrock}\ and\
  \citenamefont{Zhang}(2005)}]{Rothrock:2005}%
  \BibitemOpen
  \bibfield{author}{%
  \bibinfo {author} {\bibnamefont{Rothrock}, \bibfnamefont{D.~A.}},\ and\
  \bibinfo {author} {\bibfnamefont{J.}~\bibnamefont{Zhang}}}%
  , \bibinfo {year} {2005},\ \bibfield{title}{%
  \enquote{\bibinfo {title} {Arctic ocean sea ice volume: {What} explains its
  recent depletion?}.}\ }%
  \bibfield{journal}{%
  \bibinfo {journal} {J.~Geophys.~Res.}\ }%
  \textbf{\bibinfo {volume} {110}},\ \bibinfo {pages} {C01002}%
  \bibAnnoteFile{NoStop}{Rothrock:2005}%
\bibitem[{\citenamefont{Rozenberg}\ and\
  \citenamefont{Loewenschuss}(2009)}]{rozenberg2009}%
  \BibitemOpen
  \bibfield{author}{%
  \bibinfo {author} {\bibnamefont{Rozenberg}, \bibfnamefont{M.}},\ and\
  \bibinfo {author} {\bibfnamefont{A.}~\bibnamefont{Loewenschuss}}}%
  , \bibinfo {year} {2009},\ \bibfield{title}{%
  \enquote{\bibinfo {title} {Matrix isolation infrared spectrum of the sulfuric
  acid--monohydrate complex: New assignments and resolution of the ``missing
  {H}-bonded $\nu$(oh) band'' issue},}\ }%
  \bibfield{journal}{%
  \bibinfo {journal} {J. Phys. Chem. A}\ }%
  \textbf{\bibinfo {volume} {113}},\ \bibinfo {pages} {4963--4971}%
  \bibAnnoteFile{NoStop}{rozenberg2009}%
\bibitem[{\citenamefont{Russell}\ \emph{et~al.}(2009)\citenamefont{Russell},
  \citenamefont{Bailey}, \citenamefont{Gordley}, \citenamefont{Rusch},
  \citenamefont{Horanyi}, \citenamefont{Hervig}, \citenamefont{Thomas},
  \citenamefont{Randall}, \citenamefont{Siskind}, \citenamefont{Stevens},
  \citenamefont{Summers}, \citenamefont{Taylor}, \citenamefont{Englert},
  \citenamefont{Espy}, \citenamefont{McClintock},\ and\
  \citenamefont{Merkel}}]{Russell2009}%
  \BibitemOpen
  \bibfield{author}{%
  \bibinfo {author} {\bibnamefont{Russell}, \bibfnamefont{III, J.~M.}},
  \bibinfo {author} {\bibfnamefont{S.~M.}\ \bibnamefont{Bailey}}, \bibinfo
  {author} {\bibfnamefont{L.~L.}\ \bibnamefont{Gordley}}, \bibinfo {author}
  {\bibfnamefont{D.~W.}\ \bibnamefont{Rusch}}, \bibinfo {author}
  {\bibfnamefont{M.}~\bibnamefont{Horanyi}}, \bibinfo {author}
  {\bibfnamefont{M.~E.}\ \bibnamefont{Hervig}}, \bibinfo {author}
  {\bibfnamefont{G.~E.}\ \bibnamefont{Thomas}}, \bibinfo {author}
  {\bibfnamefont{C.~E.}\ \bibnamefont{Randall}}, \bibinfo {author}
  {\bibfnamefont{D.~E.}\ \bibnamefont{Siskind}}, \bibinfo {author}
  {\bibfnamefont{M.~H.}\ \bibnamefont{Stevens}}, \bibinfo {author}
  {\bibfnamefont{M.~E.}\ \bibnamefont{Summers}}, \bibinfo {author}
  {\bibfnamefont{M.~J.}\ \bibnamefont{Taylor}}, \bibinfo {author}
  {\bibfnamefont{C.~R.}\ \bibnamefont{Englert}}, \bibinfo {author}
  {\bibfnamefont{P.~J.}\ \bibnamefont{Espy}}, \bibinfo {author}
  {\bibfnamefont{W.~E.}\ \bibnamefont{McClintock}},\ and\ \bibinfo {author}
  {\bibfnamefont{A.~W.}\ \bibnamefont{Merkel}}}%
  , \bibinfo {year} {2009},\ \bibfield{title}{%
  \enquote{\bibinfo {title} {The aeronomy of ice in the mesosphere ({AIM})
  mission: Overview and early science results},}\ }%
  \bibfield{journal}{%
  \bibinfo {journal} {J. Atmos. Solar-Terrestrial Phys.}\ }%
  \textbf{\bibinfo {volume} {71}},\ \bibinfo {pages} {289--299}%
  \bibAnnoteFile{NoStop}{Russell2009}%
\bibitem[{\citenamefont{Sadtchenko}\
  \emph{et~al.}(2004)\citenamefont{Sadtchenko}, \citenamefont{Brindza},
  \citenamefont{Chonde}, \citenamefont{Palmore},\ and\
  \citenamefont{Eom}}]{sadtchenko2004}%
  \BibitemOpen
  \bibfield{author}{%
  \bibinfo {author} {\bibnamefont{Sadtchenko}, \bibfnamefont{V.}}, \bibinfo
  {author} {\bibfnamefont{M.}~\bibnamefont{Brindza}}, \bibinfo {author}
  {\bibfnamefont{M.}~\bibnamefont{Chonde}}, \bibinfo {author}
  {\bibfnamefont{B.}~\bibnamefont{Palmore}},\ and\ \bibinfo {author}
  {\bibfnamefont{R.}~\bibnamefont{Eom}}}%
  , \bibinfo {year} {2004},\ \bibfield{title}{%
  \enquote{\bibinfo {title} {The vaporization rate of ice at temperatures near
  its melting point},}\ }%
  \bibfield{journal}{%
  \bibinfo {journal} {J. Chem. Phys.}\ }%
  \textbf{\bibinfo {volume} {121}},\ \bibinfo {pages} {11980--11992}%
  \bibAnnoteFile{NoStop}{sadtchenko2004}%
\bibitem[{\citenamefont{Sadtchenko}\ and\
  \citenamefont{Ewing}(2002)}]{Sadtchenko2002a}%
  \BibitemOpen
  \bibfield{author}{%
  \bibinfo {author} {\bibnamefont{Sadtchenko}, \bibfnamefont{V.}},\ and\
  \bibinfo {author} {\bibfnamefont{G.~E.}\ \bibnamefont{Ewing}}}%
  , \bibinfo {year} {2002},\ \bibfield{title}{%
  \enquote{\bibinfo {title} {Interfacial melting of thin ice films: An infrared
  study},}\ }%
  \bibfield{journal}{%
  \bibinfo {journal} {J. Chem. Phys.}\ }%
  \textbf{\bibinfo {volume} {116}},\ \bibinfo {pages} {4686--4697}%
  \bibAnnoteFile{NoStop}{Sadtchenko2002a}%
\bibitem[{\citenamefont{Safarik}\ \emph{et~al.}(2003)\citenamefont{Safarik},
  \citenamefont{Meyer},\ and\ \citenamefont{Mullins}}]{safarik2003}%
  \BibitemOpen
  \bibfield{author}{%
  \bibinfo {author} {\bibnamefont{Safarik}, \bibfnamefont{D.~J.}}, \bibinfo
  {author} {\bibfnamefont{R.~J.}\ \bibnamefont{Meyer}},\ and\ \bibinfo {author}
  {\bibfnamefont{C.B.}\ \bibnamefont{Mullins}}}%
  , \bibinfo {year} {2003},\ \bibfield{title}{%
  \enquote{\bibinfo {title} {Thickness dependent crystallization kinetics of
  sub-micron amorphous solid water films},}\ }%
  \bibfield{journal}{%
  \bibinfo {journal} {J. Chem. Phys.}\ }%
  \textbf{\bibinfo {volume} {118}},\ \bibinfo {pages} {4660--4671}%
  \bibAnnoteFile{NoStop}{safarik2003}%
\bibitem[{\citenamefont{Safarik}\ and\
  \citenamefont{Mullins}(2004)}]{safarik2004}%
  \BibitemOpen
  \bibfield{author}{%
  \bibinfo {author} {\bibnamefont{Safarik}, \bibfnamefont{D.~J.}},\ and\
  \bibinfo {author} {\bibfnamefont{C.~B.}\ \bibnamefont{Mullins}}}%
  , \bibinfo {year} {2004},\ \bibfield{title}{%
  \enquote{\bibinfo {title} {The nucleation rate of crystalline ice in
  amorphous solid water},}\ }%
  \bibfield{journal}{%
  \bibinfo {journal} {J. Chem. Phys.}\ }%
  \textbf{\bibinfo {volume} {121}},\ \bibinfo {pages} {6003--6010}%
  \bibAnnoteFile{NoStop}{safarik2004}%
\bibitem[{\citenamefont{Salzmann}\ \emph{et~al.}(2006)\citenamefont{Salzmann},
  \citenamefont{Radaelli}, \citenamefont{Hallbrucker}, \citenamefont{Mayer},\
  and\ \citenamefont{Finney}}]{salzmann2006}%
  \BibitemOpen
  \bibfield{author}{%
  \bibinfo {author} {\bibnamefont{Salzmann}, \bibfnamefont{C.}}, \bibinfo
  {author} {\bibfnamefont{P.~G.}\ \bibnamefont{Radaelli}}, \bibinfo {author}
  {\bibfnamefont{A.}~\bibnamefont{Hallbrucker}}, \bibinfo {author}
  {\bibfnamefont{E.}~\bibnamefont{Mayer}},\ and\ \bibinfo {author}
  {\bibfnamefont{J.~L.}\ \bibnamefont{Finney}}}%
  , \bibinfo {year} {2006},\ \bibfield{title}{%
  \enquote{\bibinfo {title} {The preparation and structures of hydrogen ordered
  phases of ice},}\ }%
  \bibfield{journal}{%
  \bibinfo {journal} {Science}\ }%
  \textbf{\bibinfo {volume} {311}},\ \bibinfo {pages} {1758--1761}%
  \bibAnnoteFile{NoStop}{salzmann2006}%
\bibitem[{\citenamefont{Salzmann}\ \emph{et~al.}(2009)\citenamefont{Salzmann},
  \citenamefont{Radaelli}, \citenamefont{Mayer},\ and\
  \citenamefont{Finney}}]{salzmann2009}%
  \BibitemOpen
  \bibfield{author}{%
  \bibinfo {author} {\bibnamefont{Salzmann}, \bibfnamefont{C.}}, \bibinfo
  {author} {\bibfnamefont{P.~G.}\ \bibnamefont{Radaelli}}, \bibinfo {author}
  {\bibfnamefont{E.}~\bibnamefont{Mayer}},\ and\ \bibinfo {author}
  {\bibfnamefont{J.~L.}\ \bibnamefont{Finney}}}%
  , \bibinfo {year} {2009},\ \bibfield{title}{%
  \enquote{\bibinfo {title} {Ice {XV}: {A} new thermodynamically stable phase
  of ice},}\ }%
  \bibfield{journal}{%
  \bibinfo {journal} {Phys. Rev. Lett.}\ }%
  \textbf{\bibinfo {volume} {103}},\ \bibinfo {pages} {105701}%
  \bibAnnoteFile{NoStop}{salzmann2009}%
\bibitem[{\citenamefont{Salzmann}\ \emph{et~al.}(2002)\citenamefont{Salzmann},
  \citenamefont{Loerting}, \citenamefont{Kohl}, \citenamefont{Mayer},\ and\
  \citenamefont{Hallbrucker}}]{salzmann2002}%
  \BibitemOpen
  \bibfield{author}{%
  \bibinfo {author} {\bibnamefont{Salzmann}, \bibfnamefont{C.~G.}}, \bibinfo
  {author} {\bibfnamefont{T.}~\bibnamefont{Loerting}}, \bibinfo {author}
  {\bibfnamefont{I.}~\bibnamefont{Kohl}}, \bibinfo {author}
  {\bibfnamefont{E.}~\bibnamefont{Mayer}},\ and\ \bibinfo {author}
  {\bibfnamefont{A.}~\bibnamefont{Hallbrucker}}}%
  , \bibinfo {year} {2002},\ \bibfield{title}{%
  \enquote{\bibinfo {title} {Pure ice {IV} from high-density amorphous ice},}\
  }%
  \bibfield{journal}{%
  \bibinfo {journal} {J. Phys. Chem. B}\ }%
  \textbf{\bibinfo {volume} {106}},\ \bibinfo {pages} {5587--5590}%
  \bibAnnoteFile{NoStop}{salzmann2002}%
\bibitem[{\citenamefont{Sandford}\ and\
  \citenamefont{Allamandola}(1990)}]{sandford1990}%
  \BibitemOpen
  \bibfield{author}{%
  \bibinfo {author} {\bibnamefont{Sandford}, \bibfnamefont{S.~A.}},\ and\
  \bibinfo {author} {\bibfnamefont{L.~J.}\ \bibnamefont{Allamandola}}}%
  , \bibinfo {year} {1990},\ \bibfield{title}{%
  \enquote{\bibinfo {title} {The volume- and surface-binding energies of ice
  systems containing {CO}, {CO$_2$} and {H$_2$O}},}\ }%
  \bibfield{journal}{%
  \bibinfo {journal} {Icarus}\ }%
  \textbf{\bibinfo {volume} {87}},\ \bibinfo {pages} {188--192}%
  \bibAnnoteFile{NoStop}{sandford1990}%
\bibitem[{\citenamefont{Saunders}(2008)}]{saunders2008}%
  \BibitemOpen
  \bibfield{author}{%
  \bibinfo {author} {\bibnamefont{Saunders}, \bibfnamefont{C.}}}%
  , \bibinfo {year} {2008},\ \bibfield{title}{%
  \enquote{\bibinfo {title} {Charge separation mechanisms in clouds},}\ }%
  \bibfield{journal}{%
  \bibinfo {journal} {Space Sci. Rev.}\ }%
  \textbf{\bibinfo {volume} {137}},\ \bibinfo {pages} {335--353}%
  \bibAnnoteFile{NoStop}{saunders2008}%
\bibitem[{\citenamefont{Saunders}\ \emph{et~al.}(2010)\citenamefont{Saunders},
  \citenamefont{M{\"o}hler}, \citenamefont{Schnaiter}, \citenamefont{Benz},
  \citenamefont{Wagner}, \citenamefont{Saathoff}, \citenamefont{Connolly},
  \citenamefont{Burgess}, \citenamefont{Murray}, \citenamefont{Gallagher},
  \citenamefont{Wills},\ and\ \citenamefont{Plane}}]{saunders2010}%
  \BibitemOpen
  \bibfield{author}{%
  \bibinfo {author} {\bibnamefont{Saunders}, \bibfnamefont{R.~W.}}, \bibinfo
  {author} {\bibfnamefont{O.}~\bibnamefont{M{\"o}hler}}, \bibinfo {author}
  {\bibfnamefont{M.}~\bibnamefont{Schnaiter}}, \bibinfo {author}
  {\bibfnamefont{S.}~\bibnamefont{Benz}}, \bibinfo {author}
  {\bibfnamefont{R.}~\bibnamefont{Wagner}}, \bibinfo {author}
  {\bibfnamefont{H.}~\bibnamefont{Saathoff}}, \bibinfo {author}
  {\bibfnamefont{P.~J.}\ \bibnamefont{Connolly}}, \bibinfo {author}
  {\bibfnamefont{R.}~\bibnamefont{Burgess}}, \bibinfo {author}
  {\bibfnamefont{B.~J.}\ \bibnamefont{Murray}}, \bibinfo {author}
  {\bibfnamefont{M.}~\bibnamefont{Gallagher}}, \bibinfo {author}
  {\bibfnamefont{R.}~\bibnamefont{Wills}},\ and\ \bibinfo {author}
  {\bibfnamefont{J.~M.~C.}\ \bibnamefont{Plane}}}%
  , \bibinfo {year} {2010},\ \bibfield{title}{%
  \enquote{\bibinfo {title} {An aerosol chamber investigation of the
  heterogeneous ice nucleating potential of refractory nanoparticles},}\ }%
  \bibfield{journal}{%
  \bibinfo {journal} {Atmos. Chem. Phys.}\ }%
  \textbf{\bibinfo {volume} {10}},\ \bibinfo {pages} {1227--1247}%
  \bibAnnoteFile{NoStop}{saunders2010}%
\bibitem[{\citenamefont{Sazaki}\ \emph{et~al.}(2010)\citenamefont{Sazaki},
  \citenamefont{Zepeda}, \citenamefont{Nakatsubo}, \citenamefont{Yokoyama},\
  and\ \citenamefont{Furukawa}}]{Sazaki2010}%
  \BibitemOpen
  \bibfield{author}{%
  \bibinfo {author} {\bibnamefont{Sazaki}, \bibfnamefont{G.}}, \bibinfo
  {author} {\bibfnamefont{S.}~\bibnamefont{Zepeda}}, \bibinfo {author}
  {\bibfnamefont{S.}~\bibnamefont{Nakatsubo}}, \bibinfo {author}
  {\bibfnamefont{E.}~\bibnamefont{Yokoyama}},\ and\ \bibinfo {author}
  {\bibfnamefont{Y.}~\bibnamefont{Furukawa}}}%
  , \bibinfo {year} {2010},\ \bibfield{title}{%
  \enquote{\bibinfo {title} {Elementary steps at the surface of ice crystals
  visualized by advanced optical microscopy},}\ }%
  \bibfield{journal}{%
  \bibinfo {journal} {Proc. Natl Acad. Sci. USA}\ }%
  \textbf{\bibinfo {volume} {107}},\ \bibinfo {pages} {19702--19707}%
  \bibAnnoteFile{NoStop}{Sazaki2010}%
\bibitem[{\citenamefont{Scheuer}\ \emph{et~al.}(2010)\citenamefont{Scheuer},
  \citenamefont{Dibb}, \citenamefont{Twohy}, \citenamefont{Rogers},
  \citenamefont{Heymsfield},\ and\ \citenamefont{Bansemer}}]{Scheuer2010}%
  \BibitemOpen
  \bibfield{author}{%
  \bibinfo {author} {\bibnamefont{Scheuer}, \bibfnamefont{E.}}, \bibinfo
  {author} {\bibfnamefont{J.~E.}\ \bibnamefont{Dibb}}, \bibinfo {author}
  {\bibfnamefont{C.}~\bibnamefont{Twohy}}, \bibinfo {author}
  {\bibfnamefont{D.~C.}\ \bibnamefont{Rogers}}, \bibinfo {author}
  {\bibfnamefont{A.~J.}\ \bibnamefont{Heymsfield}},\ and\ \bibinfo {author}
  {\bibfnamefont{A.}~\bibnamefont{Bansemer}}}%
  , \bibinfo {year} {2010},\ \bibfield{title}{%
  \enquote{\bibinfo {title} {Evidence of nitric acid uptake in warm cirrus
  anvil clouds during the {NASA} {TC4} campaign},}\ }%
  \bibfield{journal}{%
  \bibinfo {journal} {J. Geophys. Res.-Atmos.}\ }%
  \textbf{\bibinfo {volume} {115}},\ \bibinfo {pages} {D00J03}%
  \bibAnnoteFile{NoStop}{Scheuer2010}%
\bibitem[{\citenamefont{Schick}(1990)}]{Schick1990}%
  \BibitemOpen
  \bibfield{author}{%
  \bibinfo {author} {\bibnamefont{Schick}, \bibfnamefont{M.}}}%
  , \bibinfo {year} {1990},\ \enquote{\bibinfo {title} {Introduction to wetting
  phenomena},}\ in\ \emph{\bibinfo {booktitle} {Liquids at Interfaces}},\ Vol.\
  \bibinfo {volume} {{XLVIII} Les Houches Lectures 1988},\ \bibinfo {editor}
  {edited by\ \bibinfo {editor} {\bibfnamefont{J.}~\bibnamefont{Charvolin}},
  \bibinfo {editor} {\bibfnamefont{J.~F.}\ \bibnamefont{Joanny}},\ and\
  \bibinfo {editor} {\bibfnamefont{J.}~\bibnamefont{Zinn-Justin}}},\
  Chap.~\bibinfo {chapter} {9}\ (\bibinfo {publisher} {Elsevier})%
  \bibAnnoteFile{NoStop}{Schick1990}%
\bibitem[{\citenamefont{Schick}\ and\ \citenamefont{Shih}(1987)}]{Schick1987}%
  \BibitemOpen
  \bibfield{author}{%
  \bibinfo {author} {\bibnamefont{Schick}, \bibfnamefont{M.}},\ and\ \bibinfo
  {author} {\bibfnamefont{W.-H.}\ \bibnamefont{Shih}}}%
  , \bibinfo {year} {1987},\ \bibfield{title}{%
  \enquote{\bibinfo {title} {Z(n) model of grain-boundary wetting},}\ }%
  \bibfield{journal}{%
  \bibinfo {journal} {Phys. Rev. B}\ }%
  \textbf{\bibinfo {volume} {35}},\ \bibinfo {pages} {5030--5035}%
  \bibAnnoteFile{NoStop}{Schick1987}%
\bibitem[{\citenamefont{Schiermeier}(2001)}]{Schiermeier2001}%
  \BibitemOpen
  \bibfield{author}{%
  \bibinfo {author} {\bibnamefont{Schiermeier}, \bibfnamefont{Q.}}}%
  , \bibinfo {year} {2001},\ \bibfield{title}{%
  \enquote{\bibinfo {title} {Fears grow over melting permafrost},}\ }%
  \bibfield{journal}{%
  \bibinfo {journal} {Nature}\ }%
  \textbf{\bibinfo {volume} {409}},\ \bibinfo {pages} {751}%
  \bibAnnoteFile{NoStop}{Schiermeier2001}%
\bibitem[{\citenamefont{Schilling}(2011)}]{schilling2011}%
  \BibitemOpen
  \bibfield{author}{%
  \bibinfo {author} {\bibnamefont{Schilling}, \bibfnamefont{G.}}}%
  , \bibinfo {year} {2011},\ \bibfield{title}{%
  \enquote{\bibinfo {title} {Dark ice on a hot planet},}\ }%
  \bibfield{journal}{%
  \bibinfo {journal} {Science}\ }%
  \textbf{\bibinfo {volume} {334}},\ \bibinfo {pages} {160--162}%
  \bibAnnoteFile{NoStop}{schilling2011}%
\bibitem[{\citenamefont{Schmitt}\ \emph{et~al.}(1989)\citenamefont{Schmitt},
  \citenamefont{Greenberg},\ and\ \citenamefont{Grim}}]{schmitt1989}%
  \BibitemOpen
  \bibfield{author}{%
  \bibinfo {author} {\bibnamefont{Schmitt}, \bibfnamefont{B.}}, \bibinfo
  {author} {\bibfnamefont{J.~M.}\ \bibnamefont{Greenberg}},\ and\ \bibinfo
  {author} {\bibfnamefont{R.~J.~A.}\ \bibnamefont{Grim}}}%
  , \bibinfo {year} {1989},\ \enquote{\bibinfo {title} {Spectroscopy and
  physico-chemistry of {CO:H$_2$O} and {CO$_2$:H$_2$O} ices},}\ in\
  \emph{\bibinfo {booktitle} {Infrared Spectroscopy in Astronomy}}\ (\bibinfo
  {publisher} {ESA})\ pp.\ \bibinfo {pages} {213--219}%
  \bibAnnoteFile{NoStop}{schmitt1989}%
\bibitem[{\citenamefont{Schopf}(2002)}]{schopf2002}%
  \BibitemOpen
  \bibfield{author}{%
  \bibinfo {author} {\bibnamefont{Schopf}, \bibfnamefont{J.~W.}}}%
  , \bibinfo {year} {2002},\ \emph{\bibinfo {title} {Life's Origin: The
  Beginnings of Biological Evolution}}\ (\bibinfo {publisher} {University of
  California Press})%
  \bibAnnoteFile{NoStop}{schopf2002}%
\bibitem[{\citenamefont{Schulson}\ and\
  \citenamefont{Duval}(2009)}]{MSchulson:2009p26776}%
  \BibitemOpen
  \bibfield{author}{%
  \bibinfo {author} {\bibnamefont{Schulson}, \bibfnamefont{E.~M.}},\ and\
  \bibinfo {author} {\bibfnamefont{P.}~\bibnamefont{Duval}}}%
  , \bibinfo {year} {2009},\ \emph{\bibinfo {title} {Creep and Fracture of
  Ice}}\ (\bibinfo {publisher} {Cambridge University Press})%
  \bibAnnoteFile{NoStop}{MSchulson:2009p26776}%
\bibitem[{\citenamefont{Schulze-Makuch}\ and\
  \citenamefont{Irwin}(2008)}]{schulze2008}%
  \BibitemOpen
  \bibfield{author}{%
  \bibinfo {author} {\bibnamefont{Schulze-Makuch}, \bibfnamefont{D.}},\ and\
  \bibinfo {author} {\bibfnamefont{L.~N.}\ \bibnamefont{Irwin}}}%
  , \bibinfo {year} {2008},\ \emph{\bibinfo {title} {Life in the universe:
  expectations and constraints}},\ \bibinfo {edition} {2nd}\ ed.\ (\bibinfo
  {publisher} {Springer})%
  \bibAnnoteFile{NoStop}{schulze2008}%
\bibitem[{\citenamefont{Schutte}(2002)}]{schutte2002}%
  \BibitemOpen
  \bibfield{author}{%
  \bibinfo {author} {\bibnamefont{Schutte}, \bibfnamefont{W.~A.}}}%
  , \bibinfo {year} {2002},\ \bibfield{title}{%
  \enquote{\bibinfo {title} {The low temperature crystallization effect
  reevaluated},}\ }%
  \bibfield{journal}{%
  \bibinfo {journal} {Astron Astrophys.}\ }%
  \textbf{\bibinfo {volume} {386}},\ \bibinfo {pages} {1103--1105}%
  \bibAnnoteFile{NoStop}{schutte2002}%
\bibitem[{\citenamefont{Schutte}\ \emph{et~al.}(1996)\citenamefont{Schutte},
  \citenamefont{Gerakines}, \citenamefont{Geballe}, \citenamefont{van
  Dishoeck},\ and\ \citenamefont{Greenberg}}]{schutte1996}%
  \BibitemOpen
  \bibfield{author}{%
  \bibinfo {author} {\bibnamefont{Schutte}, \bibfnamefont{W.~A.}}, \bibinfo
  {author} {\bibfnamefont{P.~A.}\ \bibnamefont{Gerakines}}, \bibinfo {author}
  {\bibfnamefont{T.~R.}\ \bibnamefont{Geballe}}, \bibinfo {author}
  {\bibfnamefont{E.~F.}\ \bibnamefont{van Dishoeck}},\ and\ \bibinfo {author}
  {\bibfnamefont{J.~M.}\ \bibnamefont{Greenberg}}}%
  , \bibinfo {year} {1996},\ \bibfield{title}{%
  \enquote{\bibinfo {title} {Discovery of solid formaldehyde toward the
  protostar {GL}~2136: observations and laboratory simulation},}\ }%
  \bibfield{journal}{%
  \bibinfo {journal} {Astron. Astrophys.}\ }%
  \textbf{\bibinfo {volume} {309}},\ \bibinfo {pages} {633--647}%
  \bibAnnoteFile{NoStop}{schutte1996}%
\bibitem[{\citenamefont{Schweizer}(2008)}]{Schweizer:2008p26034}%
  \BibitemOpen
  \bibfield{author}{%
  \bibinfo {author} {\bibnamefont{Schweizer}, \bibfnamefont{J.}}}%
  , \bibinfo {year} {2008},\ \bibfield{title}{%
  \enquote{\bibinfo {title} {Snow avalanche formation and dynamics},}\ }%
  \bibfield{journal}{%
  \bibinfo {journal} {Cold Regions Sci. Tech.}\ }%
  \textbf{\bibinfo {volume} {54}},\ \bibinfo {pages} {153--154}%
  \bibAnnoteFile{NoStop}{Schweizer:2008p26034}%
\bibitem[{\citenamefont{Schweizer}\
  \emph{et~al.}(2003)\citenamefont{Schweizer}, \citenamefont{Jamieson},\ and\
  \citenamefont{Schneebeli}}]{Schweizer:2003p25880}%
  \BibitemOpen
  \bibfield{author}{%
  \bibinfo {author} {\bibnamefont{Schweizer}, \bibfnamefont{J.}}, \bibinfo
  {author} {\bibfnamefont{J.~B.}\ \bibnamefont{Jamieson}},\ and\ \bibinfo
  {author} {\bibfnamefont{M.}~\bibnamefont{Schneebeli}}}%
  , \bibinfo {year} {2003},\ \bibfield{title}{%
  \enquote{\bibinfo {title} {Snow avalanche formation},}\ }%
  \bibfield{journal}{%
  \bibinfo {journal} {Rev. Geophys.}\ }%
  \textbf{\bibinfo {volume} {41}},\ \bibinfo {pages} {1016}%
  \bibAnnoteFile{NoStop}{Schweizer:2003p25880}%
\bibitem[{\citenamefont{Schwendinger}\
  \emph{et~al.}(1995)\citenamefont{Schwendinger}, \citenamefont{Tattler},
  \citenamefont{Saetia}, \citenamefont{Liedl}, \citenamefont{Kroemer},\ and\
  \citenamefont{Rode}}]{schwedinger1995}%
  \BibitemOpen
  \bibfield{author}{%
  \bibinfo {author} {\bibnamefont{Schwendinger}, \bibfnamefont{M.~G.}},
  \bibinfo {author} {\bibfnamefont{R.}~\bibnamefont{Tattler}}, \bibinfo
  {author} {\bibfnamefont{S.}~\bibnamefont{Saetia}}, \bibinfo {author}
  {\bibfnamefont{K.~R.}\ \bibnamefont{Liedl}}, \bibinfo {author}
  {\bibfnamefont{R.T.}\ \bibnamefont{Kroemer}},\ and\ \bibinfo {author}
  {\bibfnamefont{B.~M.}\ \bibnamefont{Rode}}}%
  , \bibinfo {year} {1995},\ \bibfield{title}{%
  \enquote{\bibinfo {title} {Salt induced peptide formation: on the selectivity
  of the copper induced peptide formation under possible prebiotic
  conditions},}\ }%
  \bibfield{journal}{%
  \bibinfo {journal} {Inorg. Chim. Acta}\ }%
  \textbf{\bibinfo {volume} {228}},\ \bibinfo {pages} {207--214}%
  \bibAnnoteFile{NoStop}{schwedinger1995}%
\bibitem[{\citenamefont{Shallcross}\ and\
  \citenamefont{Carpenter}(1957)}]{shallcross1957}%
  \BibitemOpen
  \bibfield{author}{%
  \bibinfo {author} {\bibnamefont{Shallcross}, \bibfnamefont{F.~V.}},\ and\
  \bibinfo {author} {\bibfnamefont{G.~B.}\ \bibnamefont{Carpenter}}}%
  , \bibinfo {year} {1957},\ \bibfield{title}{%
  \enquote{\bibinfo {title} {X-ray diffraction study of the cubic phase of
  ice},}\ }%
  \bibfield{journal}{%
  \bibinfo {journal} {J. Chem. Phys.}\ }%
  \textbf{\bibinfo {volume} {26}},\ \bibinfo {pages} {782--784}%
  \bibAnnoteFile{NoStop}{shallcross1957}%
\bibitem[{\citenamefont{Shapiro}(2000)}]{shapiro2000}%
  \BibitemOpen
  \bibfield{author}{%
  \bibinfo {author} {\bibnamefont{Shapiro}, \bibfnamefont{R.}}}%
  , \bibinfo {year} {2000},\ \bibfield{title}{%
  \enquote{\bibinfo {title} {A replicator was not involved in the origin of
  life},}\ }%
  \bibfield{journal}{%
  \bibinfo {journal} {IUBMB Life}\ }%
  \textbf{\bibinfo {volume} {49}},\ \bibinfo {pages} {173--176}%
  \bibAnnoteFile{NoStop}{shapiro2000}%
\bibitem[{\citenamefont{Shaw}\ and\ \citenamefont{Lamb}(1999)}]{Shaw1999}%
  \BibitemOpen
  \bibfield{author}{%
  \bibinfo {author} {\bibnamefont{Shaw}, \bibfnamefont{R.~A.}},\ and\ \bibinfo
  {author} {\bibfnamefont{D.}~\bibnamefont{Lamb}}}%
  , \bibinfo {year} {1999},\ \bibfield{title}{%
  \enquote{\bibinfo {title} {Homogeneous freezing of evaporating cloud
  droplets},}\ }%
  \bibfield{journal}{%
  \bibinfo {journal} {Geophys. Res. Lett.}\ }%
  \textbf{\bibinfo {volume} {26}},\ \bibinfo {pages} {1181--1184}%
  \bibAnnoteFile{NoStop}{Shaw1999}%
\bibitem[{\citenamefont{{Shean}}\ \emph{et~al.}(2007)\citenamefont{{Shean}},
  \citenamefont{{Head}}, \citenamefont{{Fastook}},\ and\
  \citenamefont{{Marchant}}}]{shean2007}%
  \BibitemOpen
  \bibfield{author}{%
  \bibinfo {author} {\bibnamefont{{Shean}}, \bibfnamefont{D.~E.}}, \bibinfo
  {author} {\bibfnamefont{J.~W.}\ \bibnamefont{{Head}}}, \bibinfo {author}
  {\bibfnamefont{J.~L.}\ \bibnamefont{{Fastook}}},\ and\ \bibinfo {author}
  {\bibfnamefont{D.~R.}\ \bibnamefont{{Marchant}}}}%
  , \bibinfo {year} {2007},\ \bibfield{title}{%
  \enquote{\bibinfo {title} {Recent glaciation at high elevations on {Arsia
  Mons, Mars}: Implications for the formation and evolution of large tropical
  mountain glaciers},}\ }%
  \bibfield{journal}{%
  \bibinfo {journal} {J. Geophys. Res. (Planets)}\ }%
  \textbf{\bibinfo {volume} {112}},\ \bibinfo {pages} {3004}%
  \bibAnnoteFile{NoStop}{shean2007}%
\bibitem[{\citenamefont{Shen}\ \emph{et~al.}(1986)\citenamefont{Shen},
  \citenamefont{{Hibler III}},\ and\ \citenamefont{Lepp{\"a}ranta}}]{Shen1986}%
  \BibitemOpen
  \bibfield{author}{%
  \bibinfo {author} {\bibnamefont{Shen}, \bibfnamefont{H.~H.}}, \bibinfo
  {author} {\bibfnamefont{W.~D.}\ \bibnamefont{{Hibler III}}},\ and\ \bibinfo
  {author} {\bibfnamefont{M.}~\bibnamefont{Lepp{\"a}ranta}}}%
  , \bibinfo {year} {1986},\ \bibfield{title}{%
  \enquote{\bibinfo {title} {On applying granular flow theory to a deforming
  broken ice field},}\ }%
  \bibfield{journal}{%
  \bibinfo {journal} {Acta Mechanica}\ }%
  \textbf{\bibinfo {volume} {63}},\ \bibinfo {pages} {143--160}%
  \bibAnnoteFile{NoStop}{Shen1986}%
\bibitem[{\citenamefont{Shilling}\ \emph{et~al.}(2006)\citenamefont{Shilling},
  \citenamefont{Tolbert}, \citenamefont{Toon}, \citenamefont{Jensen},
  \citenamefont{Murray},\ and\ \citenamefont{Bertram}}]{shilling2006}%
  \BibitemOpen
  \bibfield{author}{%
  \bibinfo {author} {\bibnamefont{Shilling}, \bibfnamefont{J.~E.}}, \bibinfo
  {author} {\bibfnamefont{M.~A.}\ \bibnamefont{Tolbert}}, \bibinfo {author}
  {\bibfnamefont{O.~B.}\ \bibnamefont{Toon}}, \bibinfo {author}
  {\bibfnamefont{E.~J.}\ \bibnamefont{Jensen}}, \bibinfo {author}
  {\bibfnamefont{B.~J.}\ \bibnamefont{Murray}},\ and\ \bibinfo {author}
  {\bibfnamefont{A.~K.}\ \bibnamefont{Bertram}}}%
  , \bibinfo {year} {2006},\ \bibfield{title}{%
  \enquote{\bibinfo {title} {Measurements of the vapor pressure of cubic ice
  and their implications for atmospheric ice clouds},}\ }%
  \bibfield{journal}{%
  \bibinfo {journal} {Geophys. Res. Lett.}\ }%
  \textbf{\bibinfo {volume} {33}},\ \bibinfo {pages} {026671}%
  \bibAnnoteFile{NoStop}{shilling2006}%
\bibitem[{\citenamefont{Shoji}\ and\
  \citenamefont{Langway}(1982)}]{Shoji:1982kk}%
  \BibitemOpen
  \bibfield{author}{%
  \bibinfo {author} {\bibnamefont{Shoji}, \bibfnamefont{H.}},\ and\ \bibinfo
  {author} {\bibfnamefont{C.~C.}\ \bibnamefont{Langway}}}%
  , \bibinfo {year} {1982},\ \bibfield{title}{%
  \enquote{\bibinfo {title} {Air hydrate inclusions in fresh ice core},}\ }%
  \bibfield{journal}{%
  \bibinfo {journal} {Nature}\ }%
  \textbf{\bibinfo {volume} {298}},\ \bibinfo {pages} {548--550}%
  \bibAnnoteFile{NoStop}{Shoji:1982kk}%
\bibitem[{\citenamefont{{Showalter}}\ and\
  \citenamefont{{Lissauer}}(2006)}]{showalter2006}%
  \BibitemOpen
  \bibfield{author}{%
  \bibinfo {author} {\bibnamefont{{Showalter}}, \bibfnamefont{M.~R.}},\ and\
  \bibinfo {author} {\bibfnamefont{J.~J.}\ \bibnamefont{{Lissauer}}}}%
  , \bibinfo {year} {2006},\ \bibfield{title}{%
  \enquote{\bibinfo {title} {The second ring--moon system of {Uranus}:
  Discovery and dynamics},}\ }%
  \bibfield{journal}{%
  \bibinfo {journal} {Science}\ }%
  \textbf{\bibinfo {volume} {311}},\ \bibinfo {pages} {973--977}%
  \bibAnnoteFile{NoStop}{showalter2006}%
\bibitem[{\citenamefont{Simpson}\ \emph{et~al.}(2007)\citenamefont{Simpson},
  \citenamefont{Glasow}, \citenamefont{Riedel}, \citenamefont{Anderson},
  \citenamefont{Ariya}, \citenamefont{Bottenheim}, \citenamefont{Burrows},
  \citenamefont{Carpenter}, \citenamefont{Frie{\ss}}, \citenamefont{Goodsite},
  \citenamefont{Heard}, \citenamefont{Hutterli}, \citenamefont{Jacobi},
  \citenamefont{Kaleschke}, \citenamefont{Neff}, \citenamefont{Plane},
  \citenamefont{Platt}, \citenamefont{Richter}, \citenamefont{Roscoe},
  \citenamefont{Sander}, \citenamefont{Shepson}, \citenamefont{Sodeau},
  \citenamefont{Steffen}, \citenamefont{Wagner},\ and\
  \citenamefont{Wolff}}]{Simpson:2007p4813}%
  \BibitemOpen
  \bibfield{author}{%
  \bibinfo {author} {\bibnamefont{Simpson}, \bibfnamefont{W.}}, \bibinfo
  {author} {\bibfnamefont{R.~Von}\ \bibnamefont{Glasow}}, \bibinfo {author}
  {\bibfnamefont{K.}~\bibnamefont{Riedel}}, \bibinfo {author}
  {\bibfnamefont{P.}~\bibnamefont{Anderson}}, \bibinfo {author}
  {\bibfnamefont{P.}~\bibnamefont{Ariya}}, \bibinfo {author}
  {\bibfnamefont{J.}~\bibnamefont{Bottenheim}}, \bibinfo {author}
  {\bibfnamefont{J.}~\bibnamefont{Burrows}}, \bibinfo {author}
  {\bibfnamefont{L.}~\bibnamefont{Carpenter}}, \bibinfo {author}
  {\bibfnamefont{U.}~\bibnamefont{Frie{\ss}}}, \bibinfo {author}
  {\bibfnamefont{M.}~\bibnamefont{Goodsite}}, \bibinfo {author}
  {\bibfnamefont{D.}~\bibnamefont{Heard}}, \bibinfo {author}
  {\bibfnamefont{M.}~\bibnamefont{Hutterli}}, \bibinfo {author}
  {\bibfnamefont{H.-W.}\ \bibnamefont{Jacobi}}, \bibinfo {author}
  {\bibfnamefont{L.}~\bibnamefont{Kaleschke}}, \bibinfo {author}
  {\bibfnamefont{B.}~\bibnamefont{Neff}}, \bibinfo {author}
  {\bibfnamefont{J.}~\bibnamefont{Plane}}, \bibinfo {author}
  {\bibfnamefont{U.}~\bibnamefont{Platt}}, \bibinfo {author}
  {\bibfnamefont{A.}~\bibnamefont{Richter}}, \bibinfo {author}
  {\bibfnamefont{H.}~\bibnamefont{Roscoe}}, \bibinfo {author}
  {\bibfnamefont{R.}~\bibnamefont{Sander}}, \bibinfo {author}
  {\bibfnamefont{P.}~\bibnamefont{Shepson}}, \bibinfo {author}
  {\bibfnamefont{J.~R.}\ \bibnamefont{Sodeau}}, \bibinfo {author}
  {\bibfnamefont{A.}~\bibnamefont{Steffen}}, \bibinfo {author}
  {\bibfnamefont{T.}~\bibnamefont{Wagner}},\ and\ \bibinfo {author}
  {\bibfnamefont{E.}~\bibnamefont{Wolff}}}%
  , \bibinfo {year} {2007},\ \bibfield{title}{%
  \enquote{\bibinfo {title} {Halogens and their role in polar boundary-layer
  ozone depletion},}\ }%
  \bibfield{journal}{%
  \bibinfo {journal} {Atmos. Chem. Phys.}\ }%
  \textbf{\bibinfo {volume} {7}},\ \bibinfo {pages} {4375--4418}%
  \bibAnnoteFile{NoStop}{Simpson:2007p4813}%
\bibitem[{\citenamefont{Sloan}\ and\ \citenamefont{Koh}(2008)}]{sloan2008}%
  \BibitemOpen
  \bibfield{author}{%
  \bibinfo {author} {\bibnamefont{Sloan}, \bibfnamefont{E.~D.}},\ and\ \bibinfo
  {author} {\bibfnamefont{C.~A.}\ \bibnamefont{Koh}}}%
  , \bibinfo {year} {2008},\ \emph{\bibinfo {title} {Clathrate Hydrates of
  Natural Gases}}\ (\bibinfo {publisher} {CRC Press})%
  \bibAnnoteFile{NoStop}{sloan2008}%
\bibitem[{\citenamefont{Smith}(1948)}]{Smith1948}%
  \BibitemOpen
  \bibfield{author}{%
  \bibinfo {author} {\bibnamefont{Smith}, \bibfnamefont{C.~S.}}}%
  , \bibinfo {year} {1948},\ \bibfield{title}{%
  \enquote{\bibinfo {title} {Grains, phases, and interfaces --- an
  interpretation of microstructure},}\ }%
  \bibfield{journal}{%
  \bibinfo {journal} {Trans. Amer. Inst. Mining Metallurg. Eng.}\ }%
  \textbf{\bibinfo {volume} {175}},\ \bibinfo {pages} {15--51}%
  \bibAnnoteFile{NoStop}{Smith1948}%
\bibitem[{\citenamefont{Smith}(1964)}]{Smith1964}%
  \BibitemOpen
  \bibfield{author}{%
  \bibinfo {author} {\bibnamefont{Smith}, \bibfnamefont{C.~S.}}}%
  , \bibinfo {year} {1964},\ \bibfield{title}{%
  \enquote{\bibinfo {title} {Structure, substructure and superstructure},}\ }%
  \bibfield{journal}{%
  \bibinfo {journal} {Rev. Mod. Phys.}\ }%
  \textbf{\bibinfo {volume} {36}},\ \bibinfo {pages} {524--532}%
  \bibAnnoteFile{NoStop}{Smith1964}%
\bibitem[{\citenamefont{Smith}\ \emph{et~al.}(1996)\citenamefont{Smith},
  \citenamefont{Huang}, \citenamefont{Wong},\ and\
  \citenamefont{Kay}}]{smith1996}%
  \BibitemOpen
  \bibfield{author}{%
  \bibinfo {author} {\bibnamefont{Smith}, \bibfnamefont{R.~S.}}, \bibinfo
  {author} {\bibfnamefont{C.}~\bibnamefont{Huang}}, \bibinfo {author}
  {\bibfnamefont{E.~K.~L.}\ \bibnamefont{Wong}},\ and\ \bibinfo {author}
  {\bibfnamefont{B.~D.}\ \bibnamefont{Kay}}}%
  , \bibinfo {year} {1996},\ \bibfield{title}{%
  \enquote{\bibinfo {title} {Desorption and crystallization kinetics in
  nanoscale thin films of amorphous water ice},}\ }%
  \bibfield{journal}{%
  \bibinfo {journal} {Surf. Sci. Lett.}\ }%
  \textbf{\bibinfo {volume} {367}},\ \bibinfo {pages} {L13--L18}%
  \bibAnnoteFile{NoStop}{smith1996}%
\bibitem[{\citenamefont{Smith}\ \emph{et~al.}(1997)\citenamefont{Smith},
  \citenamefont{Huang}, \citenamefont{Wong},\ and\
  \citenamefont{Kay}}]{smith1997}%
  \BibitemOpen
  \bibfield{author}{%
  \bibinfo {author} {\bibnamefont{Smith}, \bibfnamefont{R.~S.}}, \bibinfo
  {author} {\bibfnamefont{C.}~\bibnamefont{Huang}}, \bibinfo {author}
  {\bibfnamefont{E.~K.~L.}\ \bibnamefont{Wong}},\ and\ \bibinfo {author}
  {\bibfnamefont{B.~D.}\ \bibnamefont{Kay}}}%
  , \bibinfo {year} {1997},\ \bibfield{title}{%
  \enquote{\bibinfo {title} {The molecular volcano: Abrupt {CCl$_4$} desorption
  driven by the crystallization of amorphous solid water},}\ }%
  \bibfield{journal}{%
  \bibinfo {journal} {Phys. Rev. Lett.}\ }%
  \textbf{\bibinfo {volume} {79}},\ \bibinfo {pages} {909--912}%
  \bibAnnoteFile{NoStop}{smith1997}%
\bibitem[{\citenamefont{Sohl}(2010)}]{sohl2010}%
  \BibitemOpen
  \bibfield{author}{%
  \bibinfo {author} {\bibnamefont{Sohl}, \bibfnamefont{F.}}}%
  , \bibinfo {year} {2010},\ \bibfield{title}{%
  \enquote{\bibinfo {title} {Revealing {T}itanÕs interior},}\ }%
  \bibfield{journal}{%
  \bibinfo {journal} {Science}\ }%
  \textbf{\bibinfo {volume} {327}},\ \bibinfo {pages} {1338--1339}%
  \bibAnnoteFile{NoStop}{sohl2010}%
\bibitem[{\citenamefont{Sohl}\ \emph{et~al.}(2002)\citenamefont{Sohl},
  \citenamefont{Spohn}, \citenamefont{Breuer},\ and\
  \citenamefont{Nagel}}]{sohl2002}%
  \BibitemOpen
  \bibfield{author}{%
  \bibinfo {author} {\bibnamefont{Sohl}, \bibfnamefont{F.}}, \bibinfo {author}
  {\bibfnamefont{T.}~\bibnamefont{Spohn}}, \bibinfo {author}
  {\bibfnamefont{D.}~\bibnamefont{Breuer}},\ and\ \bibinfo {author}
  {\bibfnamefont{K.}~\bibnamefont{Nagel}}}%
  , \bibinfo {year} {2002},\ \bibfield{title}{%
  \enquote{\bibinfo {title} {Implications from {Galileo} observations on the
  interior structure and chemistry of the {Galilean} satellites},}\ }%
  \bibfield{journal}{%
  \bibinfo {journal} {Icarus}\ }%
  \textbf{\bibinfo {volume} {157}},\ \bibinfo {pages} {104--119}%
  \bibAnnoteFile{NoStop}{sohl2002}%
\bibitem[{\citenamefont{Solomon}\ \emph{et~al.}(1997)\citenamefont{Solomon},
  \citenamefont{Borrmann}, \citenamefont{Garcia}, \citenamefont{Portmann},
  \citenamefont{Thomason}, \citenamefont{Poole}, \citenamefont{Winker},\ and\
  \citenamefont{McCormick}}]{solomon1997}%
  \BibitemOpen
  \bibfield{author}{%
  \bibinfo {author} {\bibnamefont{Solomon}, \bibfnamefont{S.}}, \bibinfo
  {author} {\bibfnamefont{S.}~\bibnamefont{Borrmann}}, \bibinfo {author}
  {\bibfnamefont{R.~R.}\ \bibnamefont{Garcia}}, \bibinfo {author}
  {\bibfnamefont{R.}~\bibnamefont{Portmann}}, \bibinfo {author}
  {\bibfnamefont{L.}~\bibnamefont{Thomason}}, \bibinfo {author}
  {\bibfnamefont{L.~R.}\ \bibnamefont{Poole}}, \bibinfo {author}
  {\bibfnamefont{D.}~\bibnamefont{Winker}},\ and\ \bibinfo {author}
  {\bibfnamefont{M.~wP.}\ \bibnamefont{McCormick}}}%
  , \bibinfo {year} {1997},\ \bibfield{title}{%
  \enquote{\bibinfo {title} {Heterogeneous chlorine chemistry in the tropopause
  region},}\ }%
  \bibfield{journal}{%
  \bibinfo {journal} {J. Geophys. Res.-Atmos.}\ }%
  \textbf{\bibinfo {volume} {102}},\ \bibinfo {pages} {21411--21429}%
  \bibAnnoteFile{NoStop}{solomon1997}%
\bibitem[{\citenamefont{Souda}(2003)}]{Souda2003}%
  \BibitemOpen
  \bibfield{author}{%
  \bibinfo {author} {\bibnamefont{Souda}, \bibfnamefont{R.}}}%
  , \bibinfo {year} {2003},\ \bibfield{title}{%
  \enquote{\bibinfo {title} {Temperature-programmed time of flight secondary
  ion mass spectrometry study of hydration of ammonia and formic acid adsorbed
  on the water-ice surface},}\ }%
  \bibfield{journal}{%
  \bibinfo {journal} {J. Chem. Phys.}\ }%
  \textbf{\bibinfo {volume} {119}},\ \bibinfo {pages} {2774--2779}%
  \bibAnnoteFile{NoStop}{Souda2003}%
\bibitem[{\citenamefont{Staykova}\ \emph{et~al.}(2003)\citenamefont{Staykova},
  \citenamefont{Kuhs}, \citenamefont{Salamatin},\ and\
  \citenamefont{Hansen}}]{Staykova2003}%
  \BibitemOpen
  \bibfield{author}{%
  \bibinfo {author} {\bibnamefont{Staykova}, \bibfnamefont{D.~K.}}, \bibinfo
  {author} {\bibfnamefont{W.~F.}\ \bibnamefont{Kuhs}}, \bibinfo {author}
  {\bibfnamefont{A.~N.}\ \bibnamefont{Salamatin}},\ and\ \bibinfo {author}
  {\bibfnamefont{T.}~\bibnamefont{Hansen}}}%
  , \bibinfo {year} {2003},\ \bibfield{title}{%
  \enquote{\bibinfo {title} {Formation of porous gas hydrates from ice powders:
  Diffraction experiments and multistage model},}\ }%
  \bibfield{journal}{%
  \bibinfo {journal} {J. Phys. Chem. B}\ }%
  \textbf{\bibinfo {volume} {107}},\ \bibinfo {pages} {10299--10311}%
  \bibAnnoteFile{NoStop}{Staykova2003}%
\bibitem[{\citenamefont{Steffen}\ \emph{et~al.}(2008)\citenamefont{Steffen},
  \citenamefont{Douglas}, \citenamefont{Amyot}, \citenamefont{Ariya},
  \citenamefont{Aspmo}, \citenamefont{Berg}, \citenamefont{Bottenheim},
  \citenamefont{Brooks}, \citenamefont{Cobbett}, \citenamefont{Dastoor},
  \citenamefont{Dommergue}, \citenamefont{Ebinghaus}, \citenamefont{Ferrari},
  \citenamefont{Gardfeldt}, \citenamefont{Goodsite}, \citenamefont{Lean},
  \citenamefont{Poulain}, \citenamefont{Scherz}, \citenamefont{Skov},
  \citenamefont{Sommar},\ and\ \citenamefont{Temme}}]{Steffen:2008p24730}%
  \BibitemOpen
  \bibfield{author}{%
  \bibinfo {author} {\bibnamefont{Steffen}, \bibfnamefont{A.}}, \bibinfo
  {author} {\bibfnamefont{T.}~\bibnamefont{Douglas}}, \bibinfo {author}
  {\bibfnamefont{M.}~\bibnamefont{Amyot}}, \bibinfo {author}
  {\bibfnamefont{P.}~\bibnamefont{Ariya}}, \bibinfo {author}
  {\bibfnamefont{K.}~\bibnamefont{Aspmo}}, \bibinfo {author}
  {\bibfnamefont{T.}~\bibnamefont{Berg}}, \bibinfo {author}
  {\bibfnamefont{J.}~\bibnamefont{Bottenheim}}, \bibinfo {author}
  {\bibfnamefont{S.}~\bibnamefont{Brooks}}, \bibinfo {author}
  {\bibfnamefont{F.}~\bibnamefont{Cobbett}}, \bibinfo {author}
  {\bibfnamefont{A.}~\bibnamefont{Dastoor}}, \bibinfo {author}
  {\bibfnamefont{A.}~\bibnamefont{Dommergue}}, \bibinfo {author}
  {\bibfnamefont{R.}~\bibnamefont{Ebinghaus}}, \bibinfo {author}
  {\bibfnamefont{C.}~\bibnamefont{Ferrari}}, \bibinfo {author}
  {\bibfnamefont{K.}~\bibnamefont{Gardfeldt}}, \bibinfo {author}
  {\bibfnamefont{M.}~\bibnamefont{Goodsite}}, \bibinfo {author}
  {\bibfnamefont{D.}~\bibnamefont{Lean}}, \bibinfo {author}
  {\bibfnamefont{A.}~\bibnamefont{Poulain}}, \bibinfo {author}
  {\bibfnamefont{C.}~\bibnamefont{Scherz}}, \bibinfo {author}
  {\bibfnamefont{H.}~\bibnamefont{Skov}}, \bibinfo {author}
  {\bibfnamefont{J.}~\bibnamefont{Sommar}},\ and\ \bibinfo {author}
  {\bibfnamefont{C.}~\bibnamefont{Temme}}}%
  , \bibinfo {year} {2008},\ \bibfield{title}{%
  \enquote{\bibinfo {title} {A synthesis of atmospheric mercury depletion event
  chemistry in the atmosphere and snow},}\ }%
  \bibfield{journal}{%
  \bibinfo {journal} {Atmos. Chem. Phys.}\ }%
  \textbf{\bibinfo {volume} {8}},\ \bibinfo {pages} {1445--1482}%
  \bibAnnoteFile{NoStop}{Steffen:2008p24730}%
\bibitem[{\citenamefont{Stetzer}\ \emph{et~al.}(2006)\citenamefont{Stetzer},
  \citenamefont{M{\"o}hler}, \citenamefont{Wagner}, \citenamefont{Benz},
  \citenamefont{Saathoff}, \citenamefont{Bunz},\ and\
  \citenamefont{Indris}}]{stetzer2006}%
  \BibitemOpen
  \bibfield{author}{%
  \bibinfo {author} {\bibnamefont{Stetzer}, \bibfnamefont{O.}}, \bibinfo
  {author} {\bibfnamefont{O.}~\bibnamefont{M{\"o}hler}}, \bibinfo {author}
  {\bibfnamefont{R.}~\bibnamefont{Wagner}}, \bibinfo {author}
  {\bibfnamefont{S.}~\bibnamefont{Benz}}, \bibinfo {author}
  {\bibfnamefont{H.}~\bibnamefont{Saathoff}}, \bibinfo {author}
  {\bibfnamefont{H.}~\bibnamefont{Bunz}},\ and\ \bibinfo {author}
  {\bibfnamefont{O.}~\bibnamefont{Indris}}}%
  , \bibinfo {year} {2006},\ \bibfield{title}{%
  \enquote{\bibinfo {title} {Homogeneous nucleation rates of nitric acid
  dihydrate ({NAD}) at simulated stratospheric conditions --- {P}art {I}:
  {E}xperimental results},}\ }%
  \bibfield{journal}{%
  \bibinfo {journal} {Atmos. Chem. Phys. Discuss.}\ }%
  \textbf{\bibinfo {volume} {6}},\ \bibinfo {pages} {2091--2117}%
  \bibAnnoteFile{NoStop}{stetzer2006}%
\bibitem[{\citenamefont{Stratmann}\
  \emph{et~al.}(2004)\citenamefont{Stratmann}, \citenamefont{Kiselev},
  \citenamefont{Wurzler}, \citenamefont{Wendisch}, \citenamefont{Heintzenberg},
  \citenamefont{Charlson}, \citenamefont{Diehl}, \citenamefont{Wex},\ and\
  \citenamefont{Schmidt}}]{Stratmann2004}%
  \BibitemOpen
  \bibfield{author}{%
  \bibinfo {author} {\bibnamefont{Stratmann}, \bibfnamefont{F.}}, \bibinfo
  {author} {\bibfnamefont{A.}~\bibnamefont{Kiselev}}, \bibinfo {author}
  {\bibfnamefont{S.}~\bibnamefont{Wurzler}}, \bibinfo {author}
  {\bibfnamefont{M.}~\bibnamefont{Wendisch}}, \bibinfo {author}
  {\bibfnamefont{J.}~\bibnamefont{Heintzenberg}}, \bibinfo {author}
  {\bibfnamefont{R.~J.}\ \bibnamefont{Charlson}}, \bibinfo {author}
  {\bibfnamefont{K.}~\bibnamefont{Diehl}}, \bibinfo {author}
  {\bibfnamefont{H.}~\bibnamefont{Wex}},\ and\ \bibinfo {author}
  {\bibfnamefont{S.}~\bibnamefont{Schmidt}}}%
  , \bibinfo {year} {2004},\ \bibfield{title}{%
  \enquote{\bibinfo {title} {Laboratory studies and numerical simulations of
  cloud droplet formation under realistic supersaturation conditions},}\ }%
  \bibfield{journal}{%
  \bibinfo {journal} {J. Atmos. Ocean. Technol.}\ }%
  \textbf{\bibinfo {volume} {21}},\ \bibinfo {pages} {876--887}%
  \bibAnnoteFile{NoStop}{Stratmann2004}%
\bibitem[{\citenamefont{Strazzulla}\ and\
  \citenamefont{Johnson}(1991)}]{strazzulla1991}%
  \BibitemOpen
  \bibfield{author}{%
  \bibinfo {author} {\bibnamefont{Strazzulla}, \bibfnamefont{G.}},\ and\
  \bibinfo {author} {\bibfnamefont{R.~E.}\ \bibnamefont{Johnson}}}%
  , \bibinfo {year} {1991},\ \emph{\bibinfo {title} {Comets in the
  post-{H}alley era}},\ Vol.\ \bibinfo {volume} {167}\ (\bibinfo {publisher}
  {Dordrecht: Kluwer Academic Publishers ASSL Series})%
  \bibAnnoteFile{NoStop}{strazzulla1991}%
\bibitem[{\citenamefont{Stribling}\ and\
  \citenamefont{Miller}(1986)}]{stribling1986}%
  \BibitemOpen
  \bibfield{author}{%
  \bibinfo {author} {\bibnamefont{Stribling}, \bibfnamefont{R.}},\ and\
  \bibinfo {author} {\bibfnamefont{S.~L.}\ \bibnamefont{Miller}}}%
  , \bibinfo {year} {1986},\ \bibfield{title}{%
  \enquote{\bibinfo {title} {Energy yields for hydrogen cyanide and
  formaldehyde syntheses: The {HCN} and amino acid concentrations in the
  primitive ocean},}\ }%
  \bibfield{journal}{%
  \bibinfo {journal} {Origins Life}\ }%
  \textbf{\bibinfo {volume} {17}},\ \bibinfo {pages} {261--273}%
  \bibAnnoteFile{NoStop}{stribling1986}%
\bibitem[{\citenamefont{Stribling}\ and\
  \citenamefont{Miller}(1991)}]{stribling1991}%
  \BibitemOpen
  \bibfield{author}{%
  \bibinfo {author} {\bibnamefont{Stribling}, \bibfnamefont{R.}},\ and\
  \bibinfo {author} {\bibfnamefont{S.~L.}\ \bibnamefont{Miller}}}%
  , \bibinfo {year} {1991},\ \bibfield{title}{%
  \enquote{\bibinfo {title} {Template-directed synthesis of oligonucleotides
  under eutectic conditions},}\ }%
  \bibfield{journal}{%
  \bibinfo {journal} {J. Mol. Evol.}\ }%
  \textbf{\bibinfo {volume} {32}},\ \bibinfo {pages} {289--295}%
  \bibAnnoteFile{NoStop}{stribling1991}%
\bibitem[{\citenamefont{Stroeve}\ \emph{et~al.}(2007)\citenamefont{Stroeve},
  \citenamefont{Holland}, \citenamefont{Meier}, \citenamefont{Scambos},\ and\
  \citenamefont{Serreze}}]{Stroeve:2007}%
  \BibitemOpen
  \bibfield{author}{%
  \bibinfo {author} {\bibnamefont{Stroeve}, \bibfnamefont{J.}}, \bibinfo
  {author} {\bibfnamefont{M.~M.}\ \bibnamefont{Holland}}, \bibinfo {author}
  {\bibfnamefont{W.}~\bibnamefont{Meier}}, \bibinfo {author}
  {\bibfnamefont{T.}~\bibnamefont{Scambos}},\ and\ \bibinfo {author}
  {\bibfnamefont{M.}~\bibnamefont{Serreze}}}%
  , \bibinfo {year} {2007},\ \bibfield{title}{%
  \enquote{\bibinfo {title} {Arctic sea ice decline: {Faster} than forecast},}\
  }%
  \bibfield{journal}{%
  \bibinfo {journal} {Geophys.~Res.~Lett.}\ }%
  \textbf{\bibinfo {volume} {34}},\ \bibinfo {pages} {L09501}%
  \bibAnnoteFile{NoStop}{Stroeve:2007}%
\bibitem[{\citenamefont{Style}\ and\
  \citenamefont{Worster}(2009)}]{Style:2009}%
  \BibitemOpen
  \bibfield{author}{%
  \bibinfo {author} {\bibnamefont{Style}, \bibfnamefont{R.~W.}},\ and\ \bibinfo
  {author} {\bibfnamefont{M.~G.}\ \bibnamefont{Worster}}}%
  , \bibinfo {year} {2009},\ \bibfield{title}{%
  \enquote{\bibinfo {title} {Frost flower formation on sea ice and lake ice},}\
  }%
  \bibfield{journal}{%
  \bibinfo {journal} {Geophys.~Res.~Lett.}\ }%
  \textbf{\bibinfo {volume} {36}},\ \bibinfo {pages} {L11501}%
  \bibAnnoteFile{NoStop}{Style:2009}%
\bibitem[{\citenamefont{Sugiyama}(1994)}]{Sugiyama1994}%
  \BibitemOpen
  \bibfield{author}{%
  \bibinfo {author} {\bibnamefont{Sugiyama}, \bibfnamefont{T.}}}%
  , \bibinfo {year} {1994},\ \bibfield{title}{%
  \enquote{\bibinfo {title} {Ion-recombination nucleation and growth of ice
  particles in noctilucent clouds},}\ }%
  \bibfield{journal}{%
  \bibinfo {journal} {J. Geophys. Res.-Space Phys.}\ }%
  \textbf{\bibinfo {volume} {99}},\ \bibinfo {pages} {3915--3929}%
  \bibAnnoteFile{NoStop}{Sugiyama1994}%
\bibitem[{\citenamefont{Suter}\ \emph{et~al.}(2006)\citenamefont{Suter},
  \citenamefont{Andersson},\ and\ \citenamefont{Pettersson}}]{suter2006}%
  \BibitemOpen
  \bibfield{author}{%
  \bibinfo {author} {\bibnamefont{Suter}, \bibfnamefont{M.~T.}}, \bibinfo
  {author} {\bibfnamefont{P.~U.}\ \bibnamefont{Andersson}},\ and\ \bibinfo
  {author} {\bibfnamefont{J.~B.~C.}\ \bibnamefont{Pettersson}}}%
  , \bibinfo {year} {2006},\ \bibfield{title}{%
  \enquote{\bibinfo {title} {Surface properties of water ice at 150--191~{K}
  studied by elastic helium scattering},}\ }%
  \bibfield{journal}{%
  \bibinfo {journal} {J. Chem. Phys.}\ }%
  \textbf{\bibinfo {volume} {125}},\ \bibinfo {pages} {174704}%
  \bibAnnoteFile{NoStop}{suter2006}%
\bibitem[{\citenamefont{Suter}\ \emph{et~al.}(2007)\citenamefont{Suter},
  \citenamefont{Andersson},\ and\ \citenamefont{Pettersson}}]{Suter2007}%
  \BibitemOpen
  \bibfield{author}{%
  \bibinfo {author} {\bibnamefont{Suter}, \bibfnamefont{M.~T.}}, \bibinfo
  {author} {\bibfnamefont{P.~U.}\ \bibnamefont{Andersson}},\ and\ \bibinfo
  {author} {\bibfnamefont{J.~B.~C.}\ \bibnamefont{Pettersson}}}%
  , \bibinfo {year} {2007},\ \bibfield{title}{%
  \enquote{\bibinfo {title} {Formation of water-ammonia ice on graphite studied
  by elastic helium scattering},}\ }%
  \bibfield{journal}{%
  \bibinfo {journal} {Chem. Phys. Lett.}\ }%
  \textbf{\bibinfo {volume} {445}},\ \bibinfo {pages} {208--212}%
  \bibAnnoteFile{NoStop}{Suter2007}%
\bibitem[{\citenamefont{Suzuki}\ \emph{et~al.}(2007)\citenamefont{Suzuki},
  \citenamefont{Nakajima}, \citenamefont{Yoshida}, \citenamefont{Sakai},
  \citenamefont{Hashimoto}, \citenamefont{Kameshima},\ and\
  \citenamefont{Okada}}]{Suzuki2007}%
  \BibitemOpen
  \bibfield{author}{%
  \bibinfo {author} {\bibnamefont{Suzuki}, \bibfnamefont{S.}}, \bibinfo
  {author} {\bibfnamefont{A.}~\bibnamefont{Nakajima}}, \bibinfo {author}
  {\bibfnamefont{N.}~\bibnamefont{Yoshida}}, \bibinfo {author}
  {\bibfnamefont{M.}~\bibnamefont{Sakai}}, \bibinfo {author}
  {\bibfnamefont{A.}~\bibnamefont{Hashimoto}}, \bibinfo {author}
  {\bibfnamefont{Y.}~\bibnamefont{Kameshima}},\ and\ \bibinfo {author}
  {\bibfnamefont{K.}~\bibnamefont{Okada}}}%
  , \bibinfo {year} {2007},\ \bibfield{title}{%
  \enquote{\bibinfo {title} {Freezing of water droplets on silicon surfaces
  coated with various silanes},}\ }%
  \bibfield{journal}{%
  \bibinfo {journal} {Chem. Phys. Lett.}\ }%
  \textbf{\bibinfo {volume} {445}},\ \bibinfo {pages} {37--41}%
  \bibAnnoteFile{NoStop}{Suzuki2007}%
\bibitem[{\citenamefont{Svanberg}\ \emph{et~al.}(2000)\citenamefont{Svanberg},
  \citenamefont{Pettersson},\ and\ \citenamefont{Bolton}}]{svanberg2000}%
  \BibitemOpen
  \bibfield{author}{%
  \bibinfo {author} {\bibnamefont{Svanberg}, \bibfnamefont{M.}}, \bibinfo
  {author} {\bibfnamefont{J.~B.~C.}\ \bibnamefont{Pettersson}},\ and\ \bibinfo
  {author} {\bibfnamefont{K.}~\bibnamefont{Bolton}}}%
  , \bibinfo {year} {2000},\ \bibfield{title}{%
  \enquote{\bibinfo {title} {Coupled {QM/MM} molecular dynamics simulations of
  {HCl} interacting with ice surfaces and water clusters --- evidence of rapid
  ionization},}\ }%
  \bibfield{journal}{%
  \bibinfo {journal} {J. Phys. Chem. A}\ }%
  \textbf{\bibinfo {volume} {104}},\ \bibinfo {pages} {5787--5798}%
  \bibAnnoteFile{NoStop}{svanberg2000}%
\bibitem[{\citenamefont{Svensson}\ \emph{et~al.}(2009)\citenamefont{Svensson},
  \citenamefont{Delval}, \citenamefont{von Hessberg}, \citenamefont{Johnson},\
  and\ \citenamefont{Pettersson}}]{Svensson2009}%
  \BibitemOpen
  \bibfield{author}{%
  \bibinfo {author} {\bibnamefont{Svensson}, \bibfnamefont{E.~A.}}, \bibinfo
  {author} {\bibfnamefont{C.}~\bibnamefont{Delval}}, \bibinfo {author}
  {\bibfnamefont{P.}~\bibnamefont{von Hessberg}}, \bibinfo {author}
  {\bibfnamefont{M.~S.}\ \bibnamefont{Johnson}},\ and\ \bibinfo {author}
  {\bibfnamefont{J.~B.~C.}\ \bibnamefont{Pettersson}}}%
  , \bibinfo {year} {2009},\ \bibfield{title}{%
  \enquote{\bibinfo {title} {Freezing of water droplets colliding with
  kaolinite particles},}\ }%
  \bibfield{journal}{%
  \bibinfo {journal} {Atmos. Chem. Phys.}\ }%
  \textbf{\bibinfo {volume} {9}},\ \bibinfo {pages} {4295--4300}%
  \bibAnnoteFile{NoStop}{Svensson2009}%
\bibitem[{\citenamefont{Swanson}\ \emph{et~al.}(1999)\citenamefont{Swanson},
  \citenamefont{Bacon}, \citenamefont{Davis},\ and\
  \citenamefont{Baker}}]{Swanson1999}%
  \BibitemOpen
  \bibfield{author}{%
  \bibinfo {author} {\bibnamefont{Swanson}, \bibfnamefont{B.~D.}}, \bibinfo
  {author} {\bibfnamefont{M.~J.}\ \bibnamefont{Bacon}}, \bibinfo {author}
  {\bibfnamefont{E.~J.}\ \bibnamefont{Davis}},\ and\ \bibinfo {author}
  {\bibfnamefont{M.~B.}\ \bibnamefont{Baker}}}%
  , \bibinfo {year} {1999},\ \bibfield{title}{%
  \enquote{\bibinfo {title} {Electrodynamic trapping and manipulation of ice
  crystals},}\ }%
  \bibfield{journal}{%
  \bibinfo {journal} {Quart. J. Roy. Meteor. Soc.}\ }%
  \textbf{\bibinfo {volume} {125}},\ \bibinfo {pages} {1039--1058}%
  \bibAnnoteFile{NoStop}{Swanson1999}%
\bibitem[{\citenamefont{Szyrmer}\ and\
  \citenamefont{Zawadzki}(1997)}]{szyrmer1997}%
  \BibitemOpen
  \bibfield{author}{%
  \bibinfo {author} {\bibnamefont{Szyrmer}, \bibfnamefont{W.}},\ and\ \bibinfo
  {author} {\bibfnamefont{I.}~\bibnamefont{Zawadzki}}}%
  , \bibinfo {year} {1997},\ \bibfield{title}{%
  \enquote{\bibinfo {title} {Biogenic and anthropogenic sources of ice-forming
  nuclei: A review},}\ }%
  \bibfield{journal}{%
  \bibinfo {journal} {Bull. Amer. Meteor. Soc.}\ }%
  \textbf{\bibinfo {volume} {78}},\ \bibinfo {pages} {209--228}%
  \bibAnnoteFile{NoStop}{szyrmer1997}%
\bibitem[{\citenamefont{Tabazadeh}(2005)}]{tabazadeh2005}%
  \BibitemOpen
  \bibfield{author}{%
  \bibinfo {author} {\bibnamefont{Tabazadeh}, \bibfnamefont{A.}}}%
  , \bibinfo {year} {2005},\ \bibfield{title}{%
  \enquote{\bibinfo {title} {Organic aggregate formation in aerosols and its
  impact on the physicochemical properties of atmospheric particles},}\ }%
  \bibfield{journal}{%
  \bibinfo {journal} {Atmos. Env.}\ }%
  \textbf{\bibinfo {volume} {39}},\ \bibinfo {pages} {5472--5480}%
  \bibAnnoteFile{NoStop}{tabazadeh2005}%
\bibitem[{\citenamefont{Tabazadeh}\
  \emph{et~al.}(2002)\citenamefont{Tabazadeh}, \citenamefont{Djikaev},
  \citenamefont{Hamill},\ and\ \citenamefont{Reiss}}]{tabazadeh2002}%
  \BibitemOpen
  \bibfield{author}{%
  \bibinfo {author} {\bibnamefont{Tabazadeh}, \bibfnamefont{A.}}, \bibinfo
  {author} {\bibfnamefont{Y.~S.}\ \bibnamefont{Djikaev}}, \bibinfo {author}
  {\bibfnamefont{P.}~\bibnamefont{Hamill}},\ and\ \bibinfo {author}
  {\bibfnamefont{H.~J.}\ \bibnamefont{Reiss}}}%
  , \bibinfo {year} {2002},\ \bibfield{title}{%
  \enquote{\bibinfo {title} {Laboratory evidence for surface nucleation of
  solid polar stratospheric cloud particles},}\ }%
  \bibfield{journal}{%
  \bibinfo {journal} {Phys. Chem. A}\ }%
  \textbf{\bibinfo {volume} {106}},\ \bibinfo {pages} {10238--10246}%
  \bibAnnoteFile{NoStop}{tabazadeh2002}%
\bibitem[{\citenamefont{Tajima}\ \emph{et~al.}(1984)\citenamefont{Tajima},
  \citenamefont{Matsuo},\ and\ \citenamefont{Suga}}]{tajima1984}%
  \BibitemOpen
  \bibfield{author}{%
  \bibinfo {author} {\bibnamefont{Tajima}, \bibfnamefont{Y.}}, \bibinfo
  {author} {\bibfnamefont{T.}~\bibnamefont{Matsuo}},\ and\ \bibinfo {author}
  {\bibfnamefont{H.}~\bibnamefont{Suga}}}%
  , \bibinfo {year} {1984},\ \bibfield{title}{%
  \enquote{\bibinfo {title} {Calorimetric studies of phase transition in
  hexagonal ice doped with alkali hydroxides},}\ }%
  \bibfield{journal}{%
  \bibinfo {journal} {J. Phys. Chem. Solids}\ }%
  \textbf{\bibinfo {volume} {45}},\ \bibinfo {pages} {1135--1144}%
  \bibAnnoteFile{NoStop}{tajima1984}%
\bibitem[{\citenamefont{Takenaka}\ \emph{et~al.}(1992)\citenamefont{Takenaka},
  \citenamefont{Ueda},\ and\ \citenamefont{Maeda}}]{TAKENAKA:1992p2256}%
  \BibitemOpen
  \bibfield{author}{%
  \bibinfo {author} {\bibnamefont{Takenaka}, \bibfnamefont{N.}}, \bibinfo
  {author} {\bibfnamefont{A.}~\bibnamefont{Ueda}},\ and\ \bibinfo {author}
  {\bibfnamefont{Y.}~\bibnamefont{Maeda}}}%
  , \bibinfo {year} {1992},\ \bibfield{title}{%
  \enquote{\bibinfo {title} {Acceleration of the rate of nitrite oxidation by
  freezing in aqueous solution},}\ }%
  \bibfield{journal}{%
  \bibinfo {journal} {Nature}\ }%
  \textbf{\bibinfo {volume} {358}},\ \bibinfo {pages} {736--738}%
  \bibAnnoteFile{NoStop}{TAKENAKA:1992p2256}%
\bibitem[{\citenamefont{Tegler}\ \emph{et~al.}(1995)\citenamefont{Tegler},
  \citenamefont{Weintraub}, \citenamefont{Rettig}, \citenamefont{Pendleton},
  \citenamefont{Whittet},\ and\ \citenamefont{Kulesa}}]{tegler1995}%
  \BibitemOpen
  \bibfield{author}{%
  \bibinfo {author} {\bibnamefont{Tegler}, \bibfnamefont{S.~C.}}, \bibinfo
  {author} {\bibfnamefont{D.~A.}\ \bibnamefont{Weintraub}}, \bibinfo {author}
  {\bibfnamefont{T.~W.}\ \bibnamefont{Rettig}}, \bibinfo {author}
  {\bibfnamefont{Y.~J.}\ \bibnamefont{Pendleton}}, \bibinfo {author}
  {\bibfnamefont{D.~C.~B.}\ \bibnamefont{Whittet}},\ and\ \bibinfo {author}
  {\bibfnamefont{C.~A.}\ \bibnamefont{Kulesa}}}%
  , \bibinfo {year} {1995},\ \bibfield{title}{%
  \enquote{\bibinfo {title} {Evidence for chemical processing of precometary
  icy grains in circumstellar environments of pre-main-sequence stars},}\ }%
  \bibfield{journal}{%
  \bibinfo {journal} {Astrophys. J.}\ }%
  \textbf{\bibinfo {volume} {439}},\ \bibinfo {pages} {279--287}%
  \bibAnnoteFile{NoStop}{tegler1995}%
\bibitem[{\citenamefont{Thayer}\ and\ \citenamefont{Pan}(2006)}]{Thayer2006}%
  \BibitemOpen
  \bibfield{author}{%
  \bibinfo {author} {\bibnamefont{Thayer}, \bibfnamefont{J.~P.}},\ and\
  \bibinfo {author} {\bibfnamefont{W.~L.}\ \bibnamefont{Pan}}}%
  , \bibinfo {year} {2006},\ \bibfield{title}{%
  \enquote{\bibinfo {title} {Lidar observations of sodium density depletions in
  the presence of polar mesospheric clouds},}\ }%
  \bibfield{journal}{%
  \bibinfo {journal} {J. Atmos. Solar-Terrestrial Phys.}\ }%
  \textbf{\bibinfo {volume} {68}},\ \bibinfo {pages} {85--92}%
  \bibAnnoteFile{NoStop}{Thayer2006}%
\bibitem[{\citenamefont{Thomas}\ and\
  \citenamefont{Dieckmann}(2010)}]{Thomas2009}%
  \BibitemOpen
  \bibfield{author}{%
  \bibinfo {author} {\bibnamefont{Thomas}, \bibfnamefont{D.~N.}},\ and\
  \bibinfo {author} {\bibfnamefont{G.~S.}\ \bibnamefont{Dieckmann}}}%
  , \bibinfo {year} {2010},\ \emph{\bibinfo {title} {Sea Ice}},\ \bibinfo
  {edition} {2nd}\ ed.\ (\bibinfo {publisher} {Wiley-Blackwell})%
  \bibAnnoteFile{NoStop}{Thomas2009}%
\bibitem[{\citenamefont{Thomas}\ and\
  \citenamefont{Olivero}(2001)}]{Thomas2001}%
  \BibitemOpen
  \bibfield{author}{%
  \bibinfo {author} {\bibnamefont{Thomas}, \bibfnamefont{G.~E.}},\ and\
  \bibinfo {author} {\bibfnamefont{J.}~\bibnamefont{Olivero}}}%
  , \bibinfo {year} {2001},\ \bibfield{title}{%
  \enquote{\bibinfo {title} {Noctilucent clouds as possible indicators of
  global change in the mesosphere},}\ }%
  \bibfield{journal}{%
  \bibinfo {journal} {Greenhouse Gases, Aerosols and Dust}\ }%
  \textbf{\bibinfo {volume} {28}},\ \bibinfo {pages} {937--946}%
  \bibAnnoteFile{NoStop}{Thomas2001}%
\bibitem[{\citenamefont{Thomas}\ \emph{et~al.}(2003)\citenamefont{Thomas},
  \citenamefont{Olivero}, \citenamefont{DeLand},\ and\
  \citenamefont{Shettle}}]{Thomas2003}%
  \BibitemOpen
  \bibfield{author}{%
  \bibinfo {author} {\bibnamefont{Thomas}, \bibfnamefont{G.~E.}}, \bibinfo
  {author} {\bibfnamefont{J.~J.}\ \bibnamefont{Olivero}}, \bibinfo {author}
  {\bibfnamefont{M.}~\bibnamefont{DeLand}},\ and\ \bibinfo {author}
  {\bibfnamefont{E.~P.}\ \bibnamefont{Shettle}}}%
  , \bibinfo {year} {2003},\ \bibfield{title}{%
  \enquote{\bibinfo {title} {Comment on ``are noctilucent clouds truly a
  `miner's canary' for global change?"},}\ }%
  \bibfield{journal}{%
  \bibinfo {journal} {Eos Trans. AGU}\ }%
  \textbf{\bibinfo {volume} {84}},\ \bibinfo {pages} {352}%
  \bibAnnoteFile{NoStop}{Thomas2003}%
\bibitem[{\citenamefont{Thomas}\ \emph{et~al.}(1989)\citenamefont{Thomas},
  \citenamefont{Olivero}, \citenamefont{Jensen}, \citenamefont{Schroeder},\
  and\ \citenamefont{Toon}}]{Thomas1989}%
  \BibitemOpen
  \bibfield{author}{%
  \bibinfo {author} {\bibnamefont{Thomas}, \bibfnamefont{G.~E.}}, \bibinfo
  {author} {\bibfnamefont{J.~J.}\ \bibnamefont{Olivero}}, \bibinfo {author}
  {\bibfnamefont{E.~J.}\ \bibnamefont{Jensen}}, \bibinfo {author}
  {\bibfnamefont{W.}~\bibnamefont{Schroeder}},\ and\ \bibinfo {author}
  {\bibfnamefont{O.~B.}\ \bibnamefont{Toon}}}%
  , \bibinfo {year} {1989},\ \bibfield{title}{%
  \enquote{\bibinfo {title} {Relation between increasing methane and the
  presence of ice clouds at the mesopause},}\ }%
  \bibfield{journal}{%
  \bibinfo {journal} {Nature}\ }%
  \textbf{\bibinfo {volume} {338}},\ \bibinfo {pages} {490--492}%
  \bibAnnoteFile{NoStop}{Thomas1989}%
\bibitem[{\citenamefont{Thomas}\ \emph{et~al.}(1986)\citenamefont{Thomas},
  \citenamefont{Veverka},\ and\ \citenamefont{Dermott}}]{thomas1986}%
  \BibitemOpen
  \bibfield{author}{%
  \bibinfo {author} {\bibnamefont{Thomas}, \bibfnamefont{P.}}, \bibinfo
  {author} {\bibfnamefont{J.}~\bibnamefont{Veverka}},\ and\ \bibinfo {author}
  {\bibfnamefont{S.}~\bibnamefont{Dermott}}}%
  , \bibinfo {year} {1986},\ \emph{\bibinfo {title} {Satellites}}\ (\bibinfo
  {publisher} {Tucson: University of Arizona Press})%
  \bibAnnoteFile{NoStop}{thomas1986}%
\bibitem[{\citenamefont{Thomas}\ \emph{et~al.}(2010)\citenamefont{Thomas},
  \citenamefont{Zhaunerchyk}, \citenamefont{Hellberg},
  \citenamefont{Ehlerding}, \citenamefont{Geppert}, \citenamefont{Bahati},
  \citenamefont{Bannister}, \citenamefont{Fogle}, \citenamefont{Vane},
  \citenamefont{Petrignani}, \citenamefont{Andersson},
  \citenamefont{\"{O}jekull}, \citenamefont{Pettersson}, \citenamefont{van~der
  Zande},\ and\ \citenamefont{Larsson}}]{Thomas2010}%
  \BibitemOpen
  \bibfield{author}{%
  \bibinfo {author} {\bibnamefont{Thomas}, \bibfnamefont{R.~D.}}, \bibinfo
  {author} {\bibfnamefont{V.}~\bibnamefont{Zhaunerchyk}}, \bibinfo {author}
  {\bibfnamefont{F.}~\bibnamefont{Hellberg}}, \bibinfo {author}
  {\bibfnamefont{A.}~\bibnamefont{Ehlerding}}, \bibinfo {author}
  {\bibfnamefont{W.~D.}\ \bibnamefont{Geppert}}, \bibinfo {author}
  {\bibfnamefont{E.}~\bibnamefont{Bahati}}, \bibinfo {author}
  {\bibfnamefont{M.~E.}\ \bibnamefont{Bannister}}, \bibinfo {author}
  {\bibfnamefont{M.~R.}\ \bibnamefont{Fogle}}, \bibinfo {author}
  {\bibfnamefont{C.~R.}\ \bibnamefont{Vane}}, \bibinfo {author}
  {\bibfnamefont{A.}~\bibnamefont{Petrignani}}, \bibinfo {author}
  {\bibfnamefont{P.~U.}\ \bibnamefont{Andersson}}, \bibinfo {author}
  {\bibfnamefont{J.}~\bibnamefont{\"{O}jekull}}, \bibinfo {author}
  {\bibfnamefont{J.~B.~C.}\ \bibnamefont{Pettersson}}, \bibinfo {author}
  {\bibfnamefont{W.~J.}\ \bibnamefont{van~der Zande}},\ and\ \bibinfo {author}
  {\bibfnamefont{M.}~\bibnamefont{Larsson}}}%
  , \bibinfo {year} {2010},\ \bibfield{title}{%
  \enquote{\bibinfo {title} {Hot water from cold. {The} dissociative
  recombination of water cluster ions},}\ }%
  \bibfield{journal}{%
  \bibinfo {journal} {J. Phys. Chem. A}\ }%
  \textbf{\bibinfo {volume} {114}},\ \bibinfo {pages} {4843--4846}%
  \bibAnnoteFile{NoStop}{Thomas2010}%
\bibitem[{\citenamefont{Thompson}\ and\
  \citenamefont{Wallace}(2000)}]{Thompson:2000}%
  \BibitemOpen
  \bibfield{author}{%
  \bibinfo {author} {\bibnamefont{Thompson}, \bibfnamefont{D.~W.}},\ and\
  \bibinfo {author} {\bibfnamefont{J.~M.}\ \bibnamefont{Wallace}}}%
  , \bibinfo {year} {2000},\ \bibfield{title}{%
  \enquote{\bibinfo {title} {Annular modes in extratropical circulation},}\ }%
  \bibfield{journal}{%
  \bibinfo {journal} {J.~Clim.}\ }%
  \textbf{\bibinfo {volume} {13}},\ \bibinfo {pages} {1000--1016}%
  \bibAnnoteFile{NoStop}{Thompson:2000}%
\bibitem[{\citenamefont{Thomson}\ \emph{et~al.}(2010)\citenamefont{Thomson},
  \citenamefont{Benatov},\ and\ \citenamefont{Wettlaufer}}]{Thomson2010a}%
  \BibitemOpen
  \bibfield{author}{%
  \bibinfo {author} {\bibnamefont{Thomson}, \bibfnamefont{E.~S.}}, \bibinfo
  {author} {\bibfnamefont{L.}~\bibnamefont{Benatov}},\ and\ \bibinfo {author}
  {\bibfnamefont{J.~S.}\ \bibnamefont{Wettlaufer}}}%
  , \bibinfo {year} {2010},\ \bibfield{title}{%
  \enquote{\bibinfo {title} {Erratum: Abrupt grain boundary melting in ice
  [{P}hys. {R}ev. {E} 70, 061606 (2004)]},}\ }%
  \bibfield{journal}{%
  \bibinfo {journal} {Phys. Rev. E}\ }%
  \textbf{\bibinfo {volume} {82}},\ \bibinfo {pages} {039907}%
  \bibAnnoteFile{NoStop}{Thomson2010a}%
\bibitem[{\citenamefont{Thomson}\ \emph{et~al.}(2011)\citenamefont{Thomson},
  \citenamefont{Kong}, \citenamefont{Andersson}, \citenamefont{Markovic},\ and\
  \citenamefont{Pettersson}}]{Thomson2011}%
  \BibitemOpen
  \bibfield{author}{%
  \bibinfo {author} {\bibnamefont{Thomson}, \bibfnamefont{E.~S.}}, \bibinfo
  {author} {\bibfnamefont{X.~R.}\ \bibnamefont{Kong}}, \bibinfo {author}
  {\bibfnamefont{P.~U.}\ \bibnamefont{Andersson}}, \bibinfo {author}
  {\bibfnamefont{N.}~\bibnamefont{Markovic}},\ and\ \bibinfo {author}
  {\bibfnamefont{J.~B.~C.}\ \bibnamefont{Pettersson}}}%
  , \bibinfo {year} {2011},\ \bibfield{title}{%
  \enquote{\bibinfo {title} {Collision dynamics and solvation of water
  molecules in a liquid methanol film},}\ }%
  \bibfield{journal}{%
  \bibinfo {journal} {J. Phys. Chem. Lett.}\ }%
  \textbf{\bibinfo {volume} {2}},\ \bibinfo {pages} {2174--2178}%
  \bibAnnoteFile{NoStop}{Thomson2011}%
\bibitem[{\citenamefont{Thomson}\ \emph{et~al.}(2009)\citenamefont{Thomson},
  \citenamefont{Wettlaufer},\ and\ \citenamefont{Wilen}}]{Thomson2009b}%
  \BibitemOpen
  \bibfield{author}{%
  \bibinfo {author} {\bibnamefont{Thomson}, \bibfnamefont{E.~S.}}, \bibinfo
  {author} {\bibfnamefont{J.~S.}\ \bibnamefont{Wettlaufer}},\ and\ \bibinfo
  {author} {\bibfnamefont{L.~A.}\ \bibnamefont{Wilen}}}%
  , \bibinfo {year} {2009},\ \bibfield{title}{%
  \enquote{\bibinfo {title} {A direct optical method for the study of grain
  boundary melting},}\ }%
  \bibfield{journal}{%
  \bibinfo {journal} {Rev. Sci. Inst.}\ }%
  \textbf{\bibinfo {volume} {80}},\ \bibinfo {pages} {103903}%
  \bibAnnoteFile{NoStop}{Thomson2009b}%
\bibitem[{\citenamefont{Thomson}(1871)}]{Thomson1871}%
  \BibitemOpen
  \bibfield{author}{%
  \bibinfo {author} {\bibnamefont{Thomson}, \bibfnamefont{W.}}}%
  , \bibinfo {year} {1871},\ \bibfield{title}{%
  \enquote{\bibinfo {title} {On the equilibrium of vapour at a curved surface
  of liquid},}\ }%
  \bibfield{journal}{%
  \bibinfo {journal} {Philos. Mag.}\ }%
  \textbf{\bibinfo {volume} {42}}~(\bibinfo {number} {282}),\ \bibinfo {pages}
  {448--452}%
  \bibAnnoteFile{NoStop}{Thomson1871}%
\bibitem[{\citenamefont{Thorndike}\
  \emph{et~al.}(1975)\citenamefont{Thorndike}, \citenamefont{Rothrock},
  \citenamefont{Maykut},\ and\ \citenamefont{Colony}}]{Thorndike:1975}%
  \BibitemOpen
  \bibfield{author}{%
  \bibinfo {author} {\bibnamefont{Thorndike}, \bibfnamefont{A.~S.}}, \bibinfo
  {author} {\bibfnamefont{D.~A.}\ \bibnamefont{Rothrock}}, \bibinfo {author}
  {\bibfnamefont{G.~A.}\ \bibnamefont{Maykut}},\ and\ \bibinfo {author}
  {\bibfnamefont{R.}~\bibnamefont{Colony}}}%
  , \bibinfo {year} {1975},\ \bibfield{title}{%
  \enquote{\bibinfo {title} {The thickness distribution of sea ice},}\ }%
  \bibfield{journal}{%
  \bibinfo {journal} {J.~Geophys.~Res.}\ }%
  \textbf{\bibinfo {volume} {80}},\ \bibinfo {pages} {4501--4513}%
  \bibAnnoteFile{NoStop}{Thorndike:1975}%
\bibitem[{\citenamefont{Thorsteinsson}\
  \emph{et~al.}(1997)\citenamefont{Thorsteinsson}, \citenamefont{Kipfstuhl},\
  and\ \citenamefont{Miller}}]{Thorsteinsson1997}%
  \BibitemOpen
  \bibfield{author}{%
  \bibinfo {author} {\bibnamefont{Thorsteinsson}, \bibfnamefont{T.}}, \bibinfo
  {author} {\bibfnamefont{J.}~\bibnamefont{Kipfstuhl}},\ and\ \bibinfo {author}
  {\bibfnamefont{H.}~\bibnamefont{Miller}}}%
  , \bibinfo {year} {1997},\ \bibfield{title}{%
  \enquote{\bibinfo {title} {Textures and fabrics in the grip ice core},}\ }%
  \bibfield{journal}{%
  \bibinfo {journal} {J. Geophys. Res.}\ }%
  \textbf{\bibinfo {volume} {102}},\ \bibinfo {pages} {26583--26599}%
  \bibAnnoteFile{NoStop}{Thorsteinsson1997}%
\bibitem[{\citenamefont{Thrower}\ \emph{et~al.}(2010)\citenamefont{Thrower},
  \citenamefont{Abdulgalil}, \citenamefont{Collings}, \citenamefont{McCoustra},
  \citenamefont{Burke}, \citenamefont{Brown}, \citenamefont{A.~Dawes},
  \citenamefont{Kendall}, \citenamefont{Mason}, \citenamefont{Jamme},
  \citenamefont{Fraser},\ and\ \citenamefont{Rutten}}]{thrower2010}%
  \BibitemOpen
  \bibfield{author}{%
  \bibinfo {author} {\bibnamefont{Thrower}, \bibfnamefont{J.~D.}}, \bibinfo
  {author} {\bibfnamefont{A.~G.~M.}\ \bibnamefont{Abdulgalil}}, \bibinfo
  {author} {\bibfnamefont{M.~P.}\ \bibnamefont{Collings}}, \bibinfo {author}
  {\bibfnamefont{M.~R.~S.}\ \bibnamefont{McCoustra}}, \bibinfo {author}
  {\bibfnamefont{D.~J.}\ \bibnamefont{Burke}}, \bibinfo {author}
  {\bibfnamefont{W.~A.}\ \bibnamefont{Brown}}, \bibinfo {author}
  {\bibfnamefont{P.~J.~Holtom}\ \bibnamefont{A.~Dawes}}, \bibinfo {author}
  {\bibfnamefont{P.}~\bibnamefont{Kendall}}, \bibinfo {author}
  {\bibfnamefont{N.~J.}\ \bibnamefont{Mason}}, \bibinfo {author}
  {\bibfnamefont{F.}~\bibnamefont{Jamme}}, \bibinfo {author}
  {\bibfnamefont{H.~J.}\ \bibnamefont{Fraser}},\ and\ \bibinfo {author}
  {\bibfnamefont{F.~J.~M.}\ \bibnamefont{Rutten}}}%
  , \bibinfo {year} {2010},\ \bibfield{title}{%
  \enquote{\bibinfo {title} {Photon- and electron-stimulated desorption from
  laboratory models of interstellar ice grains},}\ }%
  \bibfield{journal}{%
  \bibinfo {journal} {J. Vacuum Sci. Technol. A}\ }%
  \textbf{\bibinfo {volume} {28}},\ \bibinfo {pages} {799--806}%
  \bibAnnoteFile{NoStop}{thrower2010}%
\bibitem[{\citenamefont{Thrower}\ \emph{et~al.}(2008)\citenamefont{Thrower},
  \citenamefont{Collings}, \citenamefont{McCoustra}, \citenamefont{Burke},
  \citenamefont{Brown}, \citenamefont{Dawes}, \citenamefont{Holtom},
  \citenamefont{Kendall}, \citenamefont{Mason}, \citenamefont{Jamme},
  \citenamefont{Fraser}, \citenamefont{Clark},\ and\
  \citenamefont{Parker}}]{thrower2008}%
  \BibitemOpen
  \bibfield{author}{%
  \bibinfo {author} {\bibnamefont{Thrower}, \bibfnamefont{J.~D.}}, \bibinfo
  {author} {\bibfnamefont{M.~P.}\ \bibnamefont{Collings}}, \bibinfo {author}
  {\bibfnamefont{M.~R.~S.}\ \bibnamefont{McCoustra}}, \bibinfo {author}
  {\bibfnamefont{D.~J.}\ \bibnamefont{Burke}}, \bibinfo {author}
  {\bibfnamefont{W.~A.}\ \bibnamefont{Brown}}, \bibinfo {author}
  {\bibfnamefont{A.}~\bibnamefont{Dawes}}, \bibinfo {author}
  {\bibfnamefont{P.~D.}\ \bibnamefont{Holtom}}, \bibinfo {author}
  {\bibfnamefont{P.}~\bibnamefont{Kendall}}, \bibinfo {author}
  {\bibfnamefont{N.~J.}\ \bibnamefont{Mason}}, \bibinfo {author}
  {\bibfnamefont{F.}~\bibnamefont{Jamme}}, \bibinfo {author}
  {\bibfnamefont{H.~J.}\ \bibnamefont{Fraser}}, \bibinfo {author}
  {\bibfnamefont{I.~P.}\ \bibnamefont{Clark}},\ and\ \bibinfo {author}
  {\bibfnamefont{A.~W.}\ \bibnamefont{Parker}}}%
  , \bibinfo {year} {2008},\ \bibfield{title}{%
  \enquote{\bibinfo {title} {Surface science investigations of photoprocesses
  in model interstellar ices},}\ }%
  \bibfield{journal}{%
  \bibinfo {journal} {J. Vacuum Sci. Technol. A}\ }%
  \textbf{\bibinfo {volume} {26}},\ \bibinfo {pages} {919--924}%
  \bibAnnoteFile{NoStop}{thrower2008}%
\bibitem[{\citenamefont{Tiedje}\ \emph{et~al.}(2006)\citenamefont{Tiedje},
  \citenamefont{Mitchell}, \citenamefont{Lau}, \citenamefont{Ballestad},\ and\
  \citenamefont{Nodwell}}]{Tiedje2006}%
  \BibitemOpen
  \bibfield{author}{%
  \bibinfo {author} {\bibnamefont{Tiedje}, \bibfnamefont{T.}}, \bibinfo
  {author} {\bibfnamefont{K.~A.}\ \bibnamefont{Mitchell}}, \bibinfo {author}
  {\bibfnamefont{B.}~\bibnamefont{Lau}}, \bibinfo {author}
  {\bibfnamefont{A.}~\bibnamefont{Ballestad}},\ and\ \bibinfo {author}
  {\bibfnamefont{E.}~\bibnamefont{Nodwell}}}%
  , \bibinfo {year} {2006},\ \bibfield{title}{%
  \enquote{\bibinfo {title} {Radiation transport model for ablation hollows on
  snowfields},}\ }%
  \bibfield{journal}{%
  \bibinfo {journal} {J. Geophys. Res.-Earth Surface}\ }%
  \textbf{\bibinfo {volume} {111}},\ \bibinfo {pages} {F02015}%
  \bibAnnoteFile{NoStop}{Tiedje2006}%
\bibitem[{\citenamefont{Tielens}(2005)}]{tielens2005}%
  \BibitemOpen
  \bibfield{author}{%
  \bibinfo {author} {\bibnamefont{Tielens}, \bibfnamefont{A.~G. G.~M.}}}%
  , \bibinfo {year} {2005},\ \emph{\bibinfo {title} {The Physics and Chemistry
  of the Interstellar Medium}}\ (\bibinfo {publisher} {Cambridge University
  Press})%
  \bibAnnoteFile{NoStop}{tielens2005}%
\bibitem[{\citenamefont{Tielens}\ and\
  \citenamefont{Hagen}(1982)}]{tielens1982}%
  \BibitemOpen
  \bibfield{author}{%
  \bibinfo {author} {\bibnamefont{Tielens}, \bibfnamefont{A.~G. G.~M.}},\ and\
  \bibinfo {author} {\bibfnamefont{W.}~\bibnamefont{Hagen}}}%
  , \bibinfo {year} {1982},\ \bibfield{title}{%
  \enquote{\bibinfo {title} {Model calculations of the molecular composition of
  interstellar grain mantles},}\ }%
  \bibfield{journal}{%
  \bibinfo {journal} {Astron. Astrophys.}\ }%
  \textbf{\bibinfo {volume} {114}},\ \bibinfo {pages} {245--260}%
  \bibAnnoteFile{NoStop}{tielens1982}%
\bibitem[{\citenamefont{Tizek}\ \emph{et~al.}(2002)\citenamefont{Tizek},
  \citenamefont{E.},\ and\ \citenamefont{H.}}]{tizek2002}%
  \BibitemOpen
  \bibfield{author}{%
  \bibinfo {author} {\bibnamefont{Tizek}, \bibfnamefont{H.}}, \bibinfo {author}
  {\bibfnamefont{Kn{\"o}zinger}\ \bibnamefont{E.}},\ and\ \bibinfo {author}
  {\bibfnamefont{Grothe}\ \bibnamefont{H.}}}%
  , \bibinfo {year} {2002},\ \bibfield{title}{%
  \enquote{\bibinfo {title} {X-ray diffraction studies on nitric acid
  dihydrate},}\ }%
  \bibfield{journal}{%
  \bibinfo {journal} {Phys. Chem. Chem. Phys.}\ }%
  \textbf{\bibinfo {volume} {4}},\ \bibinfo {pages} {5128--5134}%
  \bibAnnoteFile{NoStop}{tizek2002}%
\bibitem[{\citenamefont{Tizek}\ \emph{et~al.}(2004)\citenamefont{Tizek},
  \citenamefont{Grothe},\ and\ \citenamefont{Kn{\"o}zinger}}]{tizek2004}%
  \BibitemOpen
  \bibfield{author}{%
  \bibinfo {author} {\bibnamefont{Tizek}, \bibfnamefont{H.}}, \bibinfo {author}
  {\bibfnamefont{H.}~\bibnamefont{Grothe}},\ and\ \bibinfo {author}
  {\bibfnamefont{E.}~\bibnamefont{Kn{\"o}zinger}}}%
  , \bibinfo {year} {2004},\ \bibfield{title}{%
  \enquote{\bibinfo {title} {Gas-phase deposition of acetonitrile: an attempt
  to understand {O}stwaldÕs step rule on a molecular basis},}\ }%
  \bibfield{journal}{%
  \bibinfo {journal} {Chem. Phys. Lett.}\ }%
  \textbf{\bibinfo {volume} {383}},\ \bibinfo {pages} {129--133}%
  \bibAnnoteFile{NoStop}{tizek2004}%
\bibitem[{\citenamefont{Tolbert}\ and\
  \citenamefont{Toon}(2001)}]{tolbert2001}%
  \BibitemOpen
  \bibfield{author}{%
  \bibinfo {author} {\bibnamefont{Tolbert}, \bibfnamefont{M.~A.}},\ and\
  \bibinfo {author} {\bibfnamefont{O.~B.}\ \bibnamefont{Toon}}}%
  , \bibinfo {year} {2001},\ \bibfield{title}{%
  \enquote{\bibinfo {title} {Atmospheric science. solving the {PSC} mystery},}\
  }%
  \bibfield{journal}{%
  \bibinfo {journal} {Science}\ }%
  \textbf{\bibinfo {volume} {292}},\ \bibinfo {pages} {61--63}%
  \bibAnnoteFile{NoStop}{tolbert2001}%
\bibitem[{\citenamefont{{Toon}}\ and\
  \citenamefont{{Tolbert}}(1995)}]{toon1995}%
  \BibitemOpen
  \bibfield{author}{%
  \bibinfo {author} {\bibnamefont{{Toon}}, \bibfnamefont{O.~B.}},\ and\
  \bibinfo {author} {\bibfnamefont{M.~A.}\ \bibnamefont{{Tolbert}}}}%
  , \bibinfo {year} {1995},\ \bibfield{title}{%
  \enquote{\bibinfo {title} {{Spectroscopic evidence against nitric acid
  trihydrate in polar stratospheric clouds}},}\ }%
  \bibfield{journal}{%
  \bibinfo {journal} {Nature}\ }%
  \textbf{\bibinfo {volume} {375}},\ \bibinfo {pages} {218--221}%
  \bibAnnoteFile{NoStop}{toon1995}%
\bibitem[{\citenamefont{Toubin}\ \emph{et~al.}(2001)\citenamefont{Toubin},
  \citenamefont{Picaud}, \citenamefont{Hoang}, \citenamefont{Girardet},
  \citenamefont{Demirdjian}, \citenamefont{Ferry},\ and\
  \citenamefont{Suzanne}}]{toubin2001}%
  \BibitemOpen
  \bibfield{author}{%
  \bibinfo {author} {\bibnamefont{Toubin}, \bibfnamefont{C.}}, \bibinfo
  {author} {\bibfnamefont{S.}~\bibnamefont{Picaud}}, \bibinfo {author}
  {\bibfnamefont{P.~N.~M.}\ \bibnamefont{Hoang}}, \bibinfo {author}
  {\bibfnamefont{C.}~\bibnamefont{Girardet}}, \bibinfo {author}
  {\bibfnamefont{B.}~\bibnamefont{Demirdjian}}, \bibinfo {author}
  {\bibfnamefont{D.}~\bibnamefont{Ferry}},\ and\ \bibinfo {author}
  {\bibfnamefont{J.}~\bibnamefont{Suzanne}}}%
  , \bibinfo {year} {2001},\ \bibfield{title}{%
  \enquote{\bibinfo {title} {Dynamics of ice layers deposited on {MgO}(001):
  Quasielastic neutron scattering experiments and molecular dynamics
  simulations},}\ }%
  \bibfield{journal}{%
  \bibinfo {journal} {J. Chem. Phys.}\ }%
  \textbf{\bibinfo {volume} {114}},\ \bibinfo {pages} {6371--6381}%
  \bibAnnoteFile{NoStop}{toubin2001}%
\bibitem[{\citenamefont{Trainer}\ \emph{et~al.}(2009)\citenamefont{Trainer},
  \citenamefont{Tolbert}, \citenamefont{McKay},\ and\
  \citenamefont{Toon}}]{trainer2009}%
  \BibitemOpen
  \bibfield{author}{%
  \bibinfo {author} {\bibnamefont{Trainer}, \bibfnamefont{M.~G.}}, \bibinfo
  {author} {\bibfnamefont{M.~A.}\ \bibnamefont{Tolbert}}, \bibinfo {author}
  {\bibfnamefont{C.~P.}\ \bibnamefont{McKay}},\ and\ \bibinfo {author}
  {\bibfnamefont{O.~B.}\ \bibnamefont{Toon}}}%
  , \bibinfo {year} {2009},\ \bibfield{title}{%
  \enquote{\bibinfo {title} {Enhanced {CO$_2$} trapping in water ice via
  atmospheric deposition with relevance to {Mars}},}\ }%
  \bibfield{journal}{%
  \bibinfo {journal} {Icarus}\ }%
  \textbf{\bibinfo {volume} {206}},\ \bibinfo {pages} {707--715}%
  \bibAnnoteFile{NoStop}{trainer2009}%
\bibitem[{\citenamefont{Tribello}\ \emph{et~al.}(2006)\citenamefont{Tribello},
  \citenamefont{Slater},\ and\ \citenamefont{Salzmann}}]{tribello2006}%
  \BibitemOpen
  \bibfield{author}{%
  \bibinfo {author} {\bibnamefont{Tribello}, \bibfnamefont{G.}}, \bibinfo
  {author} {\bibfnamefont{B.}~\bibnamefont{Slater}},\ and\ \bibinfo {author}
  {\bibfnamefont{C.~G.}\ \bibnamefont{Salzmann}}}%
  , \bibinfo {year} {2006},\ \bibfield{title}{%
  \enquote{\bibinfo {title} {A blind structure prediction of ice {XIV}},}\ }%
  \bibfield{journal}{%
  \bibinfo {journal} {J. Am. Chem. Soc.}\ }%
  \textbf{\bibinfo {volume} {128}},\ \bibinfo {pages} {12594--12595}%
  \bibAnnoteFile{NoStop}{tribello2006}%
\bibitem[{\citenamefont{Trinks}\ \emph{et~al.}(2005)\citenamefont{Trinks},
  \citenamefont{Schr{\"o}der},\ and\ \citenamefont{Bierbricher}}]{trinks2005}%
  \BibitemOpen
  \bibfield{author}{%
  \bibinfo {author} {\bibnamefont{Trinks}, \bibfnamefont{H.}}, \bibinfo
  {author} {\bibfnamefont{W.}~\bibnamefont{Schr{\"o}der}},\ and\ \bibinfo
  {author} {\bibfnamefont{C.~K.}\ \bibnamefont{Bierbricher}}}%
  , \bibinfo {year} {2005},\ \bibfield{title}{%
  \enquote{\bibinfo {title} {Sea ice as a promoter of the emergence of first
  life},}\ }%
  \bibfield{journal}{%
  \bibinfo {journal} {Origins of Life and Evolution of Biospheres}\ }%
  \textbf{\bibinfo {volume} {35}},\ \bibinfo {pages} {429--445}%
  \bibAnnoteFile{NoStop}{trinks2005}%
\bibitem[{\citenamefont{{Trujillo}}\
  \emph{et~al.}(2007)\citenamefont{{Trujillo}}, \citenamefont{{Brown}},
  \citenamefont{{Barkume}}, \citenamefont{{Schaller}},\ and\
  \citenamefont{{Rabinowitz}}}]{trujillo2007}%
  \BibitemOpen
  \bibfield{author}{%
  \bibinfo {author} {\bibnamefont{{Trujillo}}, \bibfnamefont{C.~A.}}, \bibinfo
  {author} {\bibfnamefont{M.~E.}\ \bibnamefont{{Brown}}}, \bibinfo {author}
  {\bibfnamefont{K.~M.}\ \bibnamefont{{Barkume}}}, \bibinfo {author}
  {\bibfnamefont{E.~L.}\ \bibnamefont{{Schaller}}},\ and\ \bibinfo {author}
  {\bibfnamefont{D.~L.}\ \bibnamefont{{Rabinowitz}}}}%
  , \bibinfo {year} {2007},\ \bibfield{title}{%
  \enquote{\bibinfo {title} {The surface of 2003 {EL}$_{61}$ in the
  near-infrared},}\ }%
  \bibfield{journal}{%
  \bibinfo {journal} {Astrophys. J.}\ }%
  \textbf{\bibinfo {volume} {655}},\ \bibinfo {pages} {1172--1178}%
  \bibAnnoteFile{NoStop}{trujillo2007}%
\bibitem[{\citenamefont{Tuhkuri}\ and\
  \citenamefont{Lensu}(2002)}]{Tuhkuri:2002}%
  \BibitemOpen
  \bibfield{author}{%
  \bibinfo {author} {\bibnamefont{Tuhkuri}, \bibfnamefont{J.}},\ and\ \bibinfo
  {author} {\bibfnamefont{M.}~\bibnamefont{Lensu}}}%
  , \bibinfo {year} {2002},\ \bibfield{title}{%
  \enquote{\bibinfo {title} {Laboratory tests on ridging and rafting of ice
  sheets},}\ }%
  \bibfield{journal}{%
  \bibinfo {journal} {J.~Geophys.~Res.}\ }%
  \textbf{\bibinfo {volume} {107}},\ \bibinfo {pages} {3125--3142}%
  \bibAnnoteFile{NoStop}{Tuhkuri:2002}%
\bibitem[{\citenamefont{Tulk}\ \emph{et~al.}(2002)\citenamefont{Tulk},
  \citenamefont{Benmore}, \citenamefont{Urquidi}, \citenamefont{Klug},
  \citenamefont{Neuefeind}, \citenamefont{Tomberli},\ and\
  \citenamefont{Egelstaff}}]{tulk2002}%
  \BibitemOpen
  \bibfield{author}{%
  \bibinfo {author} {\bibnamefont{Tulk}, \bibfnamefont{C.~A.}}, \bibinfo
  {author} {\bibfnamefont{C.~J.}\ \bibnamefont{Benmore}}, \bibinfo {author}
  {\bibfnamefont{J.}~\bibnamefont{Urquidi}}, \bibinfo {author}
  {\bibfnamefont{D.~D.}\ \bibnamefont{Klug}}, \bibinfo {author}
  {\bibfnamefont{J.}~\bibnamefont{Neuefeind}}, \bibinfo {author}
  {\bibfnamefont{B.}~\bibnamefont{Tomberli}},\ and\ \bibinfo {author}
  {\bibfnamefont{P.~A.}\ \bibnamefont{Egelstaff}}}%
  , \bibinfo {year} {2002},\ \bibfield{title}{%
  \enquote{\bibinfo {title} {Structural studies of several distinct metastable
  forms of amorphous ice},}\ }%
  \bibfield{journal}{%
  \bibinfo {journal} {Science}\ }%
  \textbf{\bibinfo {volume} {297}},\ \bibinfo {pages} {1320--1323}%
  \bibAnnoteFile{NoStop}{tulk2002}%
\bibitem[{\citenamefont{Ullerstam}\
  \emph{et~al.}(2005)\citenamefont{Ullerstam}, \citenamefont{Thornberry},\ and\
  \citenamefont{Abbatt}}]{Ullerstam:2005p4622}%
  \BibitemOpen
  \bibfield{author}{%
  \bibinfo {author} {\bibnamefont{Ullerstam}, \bibfnamefont{M.}}, \bibinfo
  {author} {\bibfnamefont{T.}~\bibnamefont{Thornberry}},\ and\ \bibinfo
  {author} {\bibfnamefont{J.}~\bibnamefont{Abbatt}}}%
  , \bibinfo {year} {2005},\ \bibfield{title}{%
  \enquote{\bibinfo {title} {Uptake of gas-phase nitric acid to ice at low
  partial pressures: evidence for unsaturated surface coverage},}\ }%
  \bibfield{journal}{%
  \bibinfo {journal} {Faraday Discuss.}\ }%
  \textbf{\bibinfo {volume} {130}},\ \bibinfo {pages} {211--226}%
  \bibAnnoteFile{NoStop}{Ullerstam:2005p4622}%
\bibitem[{\citenamefont{Uras-Aytemiz}\
  \emph{et~al.}(2006)\citenamefont{Uras-Aytemiz}, \citenamefont{Sadlej},
  \citenamefont{Devlin},\ and\ \citenamefont{Buch}}]{urasaytemiz2006}%
  \BibitemOpen
  \bibfield{author}{%
  \bibinfo {author} {\bibnamefont{Uras-Aytemiz}, \bibfnamefont{N.}}, \bibinfo
  {author} {\bibfnamefont{J.}~\bibnamefont{Sadlej}}, \bibinfo {author}
  {\bibfnamefont{J.~P.}\ \bibnamefont{Devlin}},\ and\ \bibinfo {author}
  {\bibfnamefont{V.}~\bibnamefont{Buch}}}%
  , \bibinfo {year} {2006},\ \bibfield{title}{%
  \enquote{\bibinfo {title} {{HCl} solvation at the surface and within methanol
  clusters/nanoparticles {II}: Evidence for molecular wires},}\ }%
  \bibfield{journal}{%
  \bibinfo {journal} {J. Phys. Chem. B}\ }%
  \textbf{\bibinfo {volume} {110}},\ \bibinfo {pages} {21751--21763}%
  \bibAnnoteFile{NoStop}{urasaytemiz2006}%
\bibitem[{\citenamefont{VandeVondele}\
  \emph{et~al.}(2005)\citenamefont{VandeVondele}, \citenamefont{Krack},
  \citenamefont{Mohamed}, \citenamefont{Parrinello}, \citenamefont{Chassaing},\
  and\ \citenamefont{Hutter}}]{vandevondele2005}%
  \BibitemOpen
  \bibfield{author}{%
  \bibinfo {author} {\bibnamefont{VandeVondele}, \bibfnamefont{J.}}, \bibinfo
  {author} {\bibfnamefont{M.}~\bibnamefont{Krack}}, \bibinfo {author}
  {\bibfnamefont{F.}~\bibnamefont{Mohamed}}, \bibinfo {author}
  {\bibfnamefont{M.}~\bibnamefont{Parrinello}}, \bibinfo {author}
  {\bibfnamefont{T.}~\bibnamefont{Chassaing}},\ and\ \bibinfo {author}
  {\bibfnamefont{J.}~\bibnamefont{Hutter}}}%
  , \bibinfo {year} {2005},\ \bibfield{title}{%
  \enquote{\bibinfo {title} {Quickstep: Fast and accurate density functional
  calculations using a mixed {Gaussian} and plane waves approach},}\ }%
  \bibfield{journal}{%
  \bibinfo {journal} {Comput. Phys. Commun.}\ }%
  \textbf{\bibinfo {volume} {167}},\ \bibinfo {pages} {103--128}%
  \bibAnnoteFile{NoStop}{vandevondele2005}%
\bibitem[{\citenamefont{Vaughan}(2007)}]{DavidGVaughan:2007p26017}%
  \BibitemOpen
  \bibfield{author}{%
  \bibinfo {author} {\bibnamefont{Vaughan}, \bibfnamefont{D.~G.}}}%
  , \bibinfo {year} {2007},\ \bibfield{title}{%
  \enquote{\bibinfo {title} {The {Antarctic} ice sheet},}\ }%
  \bibinfo {journal} {Glacier Science and Environmental Change (First
  Edition)},\ \bibinfo {pages} {209--220}%
  \bibAnnoteFile{NoStop}{DavidGVaughan:2007p26017}%
\bibitem[{\citenamefont{Vega}\ \emph{et~al.}(2009)\citenamefont{Vega},
  \citenamefont{Abascal}, \citenamefont{Conde},\ and\
  \citenamefont{Aragones}}]{vega2008}%
  \BibitemOpen
\bibfield{journal}{%
    }%
  \bibfield{author}{%
  \bibinfo {author} {\bibnamefont{Vega}, \bibfnamefont{C.}}, \bibinfo {author}
  {\bibfnamefont{J.~L.~F.}\ \bibnamefont{Abascal}}, \bibinfo {author}
  {\bibfnamefont{M.~M.}\ \bibnamefont{Conde}},\ and\ \bibinfo {author}
  {\bibfnamefont{J.~L.}\ \bibnamefont{Aragones}}}%
  , \bibinfo {year} {2009},\ \bibfield{title}{%
  \enquote{\bibinfo {title} {What ice can teach us about water interactions: a
  critical comparison of the performance of different water models},}\ }%
  \bibfield{journal}{%
  \bibinfo {journal} {Farad. Disc. Roy. Soc. Chem.}\ }%
  \textbf{\bibinfo {volume} {141}},\ \bibinfo {pages} {251--276}%
  \bibAnnoteFile{NoStop}{vega2008}%
\bibitem[{\citenamefont{Vehkam\"{a}ki}(2006)}]{Vehkamaki2006}%
  \BibitemOpen
  \bibfield{author}{%
  \bibinfo {author} {\bibnamefont{Vehkam\"{a}ki}, \bibfnamefont{H.}}}%
  , \bibinfo {year} {2006},\ \emph{\bibinfo {title} {Classical Nucleation
  Theory in Multicomponent Systems}}\ (\bibinfo {publisher} {Springer},\
  \bibinfo {address} {Germany})%
  \bibAnnoteFile{NoStop}{Vehkamaki2006}%
\bibitem[{\citenamefont{Vella}\ and\
  \citenamefont{Wettlaufer}(2007)}]{Vella:2007}%
  \BibitemOpen
  \bibfield{author}{%
  \bibinfo {author} {\bibnamefont{Vella}, \bibfnamefont{D.}},\ and\ \bibinfo
  {author} {\bibfnamefont{J.~S.}\ \bibnamefont{Wettlaufer}}}%
  , \bibinfo {year} {2007},\ \bibfield{title}{%
  \enquote{\bibinfo {title} {Finger rafting: a generic instability of floating
  elastic sheets},}\ }%
  \bibfield{journal}{%
  \bibinfo {journal} {Phys.~Rev.~Lett.}\ }%
  \textbf{\bibinfo {volume} {98}},\ \bibinfo {pages} {088303}%
  \bibAnnoteFile{NoStop}{Vella:2007}%
\bibitem[{\citenamefont{Vella}\ and\
  \citenamefont{Wettlaufer}(2008)}]{Vella:2008}%
  \BibitemOpen
  \bibfield{author}{%
  \bibinfo {author} {\bibnamefont{Vella}, \bibfnamefont{D.}},\ and\ \bibinfo
  {author} {\bibfnamefont{J.~S.}\ \bibnamefont{Wettlaufer}}}%
  , \bibinfo {year} {2008},\ \bibfield{title}{%
  \enquote{\bibinfo {title} {Explaining the patterns formed by ice floe
  interaction},}\ }%
  \bibfield{journal}{%
  \bibinfo {journal} {J.~Geophys.~Res.}\ }%
  \textbf{\bibinfo {volume} {113}},\ \bibinfo {pages} {C11011}%
  \bibAnnoteFile{NoStop}{Vella:2008}%
\bibitem[{\citenamefont{Venkatesh}\
  \emph{et~al.}(1974)\citenamefont{Venkatesh}, \citenamefont{Rice},\ and\
  \citenamefont{Narten}}]{venkatesh1974}%
  \BibitemOpen
  \bibfield{author}{%
  \bibinfo {author} {\bibnamefont{Venkatesh}, \bibfnamefont{C.~G.}}, \bibinfo
  {author} {\bibfnamefont{S.~A.}\ \bibnamefont{Rice}},\ and\ \bibinfo {author}
  {\bibfnamefont{A.~H.}\ \bibnamefont{Narten}}}%
  , \bibinfo {year} {1974},\ \bibfield{title}{%
  \enquote{\bibinfo {title} {Amorphous solid water: An {X}-ray diffraction
  study},}\ }%
  \bibfield{journal}{%
  \bibinfo {journal} {Science}\ }%
  \textbf{\bibinfo {volume} {186}},\ \bibinfo {pages} {927--928}%
  \bibAnnoteFile{NoStop}{venkatesh1974}%
\bibitem[{\citenamefont{Vidali}\ \emph{et~al.}(2004)\citenamefont{Vidali},
  \citenamefont{Roser}, \citenamefont{Manic\`o},\ and\
  \citenamefont{Pirronello}}]{vidali2004}%
  \BibitemOpen
  \bibfield{author}{%
  \bibinfo {author} {\bibnamefont{Vidali}, \bibfnamefont{G.}}, \bibinfo
  {author} {\bibfnamefont{J.~E.}\ \bibnamefont{Roser}}, \bibinfo {author}
  {\bibfnamefont{G.}~\bibnamefont{Manic\`o}},\ and\ \bibinfo {author}
  {\bibfnamefont{V.}~\bibnamefont{Pirronello}}}%
  , \bibinfo {year} {2004},\ \bibfield{title}{%
  \enquote{\bibinfo {title} {Experimental study of the formation of molecular
  hydrogen and carbon dioxide on dust grain analogues},}\ }%
  \bibfield{booktitle}{%
  \emph{\bibinfo {booktitle} {Space Life Sciences: Steps toward Origin(s) of
  Life}},\ }%
  \bibfield{journal}{%
  \bibinfo {journal} {Adv. Space Res.}\ }%
  \textbf{\bibinfo {volume} {33}},\ \bibinfo {pages} {6--13}%
  \bibAnnoteFile{NoStop}{vidali2004}%
\bibitem[{\citenamefont{Viti}\ \emph{et~al.}(2004)\citenamefont{Viti},
  \citenamefont{Collings}, \citenamefont{Dever}, \citenamefont{McCoustra},\
  and\ \citenamefont{Williams}}]{viti2004}%
  \BibitemOpen
  \bibfield{author}{%
  \bibinfo {author} {\bibnamefont{Viti}, \bibfnamefont{S.}}, \bibinfo {author}
  {\bibfnamefont{M.~P.}\ \bibnamefont{Collings}}, \bibinfo {author}
  {\bibfnamefont{J.~W.}\ \bibnamefont{Dever}}, \bibinfo {author}
  {\bibfnamefont{M.~R.~S.}\ \bibnamefont{McCoustra}},\ and\ \bibinfo {author}
  {\bibfnamefont{D.~A.}\ \bibnamefont{Williams}}}%
  , \bibinfo {year} {2004},\ \bibfield{title}{%
  \enquote{\bibinfo {title} {Evaporation of ices near massive stars: models
  based on laboratory temperature programmed desorption data},}\ }%
  \bibfield{journal}{%
  \bibinfo {journal} {Mon. Not. Roy. Astron. Soc.}\ }%
  \textbf{\bibinfo {volume} {354}},\ \bibinfo {pages} {1141--1145}%
  \bibAnnoteFile{NoStop}{viti2004}%
\bibitem[{\citenamefont{Vlassov}\ \emph{et~al.}(2004)\citenamefont{Vlassov},
  \citenamefont{Johnston}, \citenamefont{Landweber},\ and\
  \citenamefont{Kazakov}}]{vlassov2004}%
  \BibitemOpen
  \bibfield{author}{%
  \bibinfo {author} {\bibnamefont{Vlassov}, \bibfnamefont{A.~V.}}, \bibinfo
  {author} {\bibfnamefont{B.~H.}\ \bibnamefont{Johnston}}, \bibinfo {author}
  {\bibfnamefont{L.~F.}\ \bibnamefont{Landweber}},\ and\ \bibinfo {author}
  {\bibfnamefont{S.~A.}\ \bibnamefont{Kazakov}}}%
  , \bibinfo {year} {2004},\ \bibfield{title}{%
  \enquote{\bibinfo {title} {Ligation activity of fragmented ribozymes in
  frozen solution: implications for the {RNA} world},}\ }%
  \bibfield{journal}{%
  \bibinfo {journal} {Nucleic Acids Res.}\ }%
  \textbf{\bibinfo {volume} {32}},\ \bibinfo {pages} {2966--2974}%
  \bibAnnoteFile{NoStop}{vlassov2004}%
\bibitem[{\citenamefont{Voigt}\ \emph{et~al.}(2003)\citenamefont{Voigt},
  \citenamefont{Larsen}, \citenamefont{Deshler}, \citenamefont{Kršger},
  \citenamefont{Schreiner}, \citenamefont{Mauersberger}, \citenamefont{Luo},
  \citenamefont{Adriani}, \citenamefont{Cairo}, \citenamefont{Di~Donfrancesco},
  \citenamefont{Ovarlez}, \citenamefont{Ovarlez}, \citenamefont{D{\"o}rnbrack},
  \citenamefont{Knudsen},\ and\ \citenamefont{Rosen}}]{voigt2003}%
  \BibitemOpen
  \bibfield{author}{%
  \bibinfo {author} {\bibnamefont{Voigt}, \bibfnamefont{C.}}, \bibinfo {author}
  {\bibfnamefont{N.}~\bibnamefont{Larsen}}, \bibinfo {author}
  {\bibfnamefont{T.}~\bibnamefont{Deshler}}, \bibinfo {author}
  {\bibfnamefont{C.}~\bibnamefont{Kršger}}, \bibinfo {author}
  {\bibfnamefont{J.}~\bibnamefont{Schreiner}}, \bibinfo {author}
  {\bibfnamefont{K.}~\bibnamefont{Mauersberger}}, \bibinfo {author}
  {\bibfnamefont{B.~P.}\ \bibnamefont{Luo}}, \bibinfo {author}
  {\bibfnamefont{A.}~\bibnamefont{Adriani}}, \bibinfo {author}
  {\bibfnamefont{F.}~\bibnamefont{Cairo}}, \bibinfo {author}
  {\bibfnamefont{G.}~\bibnamefont{Di~Donfrancesco}}, \bibinfo {author}
  {\bibfnamefont{J.}~\bibnamefont{Ovarlez}}, \bibinfo {author}
  {\bibfnamefont{H.}~\bibnamefont{Ovarlez}}, \bibinfo {author}
  {\bibfnamefont{A.}~\bibnamefont{D{\"o}rnbrack}}, \bibinfo {author}
  {\bibfnamefont{B.}~\bibnamefont{Knudsen}},\ and\ \bibinfo {author}
  {\bibfnamefont{J.}~\bibnamefont{Rosen}}}%
  , \bibinfo {year} {2003},\ \bibfield{title}{%
  \enquote{\bibinfo {title} {In situ mountain-wave polar stratospheric cloud
  measurements: {I}mplications for nitric acid trihydrate formation},}\ }%
  \bibfield{journal}{%
  \bibinfo {journal} {J. Geophys. Res.}\ }%
  \textbf{\bibinfo {volume} {108}},\ \bibinfo {pages} {8331}%
  \bibAnnoteFile{NoStop}{voigt2003}%
\bibitem[{\citenamefont{Voigt}\ \emph{et~al.}(2005)\citenamefont{Voigt},
  \citenamefont{Schlager}, \citenamefont{Luo}, \citenamefont{D{\"o}rnbrack},
  \citenamefont{Roiger}, \citenamefont{Stock}, \citenamefont{Curtius},
  \citenamefont{V{\"o}ssing}, \citenamefont{Borrmann}, \citenamefont{Davies},
  \citenamefont{Konopka}, \citenamefont{Schiller}, \citenamefont{Shur},\ and\
  \citenamefont{Peter}}]{voigt2005}%
  \BibitemOpen
  \bibfield{author}{%
  \bibinfo {author} {\bibnamefont{Voigt}, \bibfnamefont{C.}}, \bibinfo {author}
  {\bibfnamefont{H.}~\bibnamefont{Schlager}}, \bibinfo {author}
  {\bibfnamefont{B.~P.}\ \bibnamefont{Luo}}, \bibinfo {author}
  {\bibfnamefont{A.~D.}\ \bibnamefont{D{\"o}rnbrack}}, \bibinfo {author}
  {\bibfnamefont{A.}~\bibnamefont{Roiger}}, \bibinfo {author}
  {\bibfnamefont{P.}~\bibnamefont{Stock}}, \bibinfo {author}
  {\bibfnamefont{J.}~\bibnamefont{Curtius}}, \bibinfo {author}
  {\bibfnamefont{H.}~\bibnamefont{V{\"o}ssing}}, \bibinfo {author}
  {\bibfnamefont{S.}~\bibnamefont{Borrmann}}, \bibinfo {author}
  {\bibfnamefont{S.}~\bibnamefont{Davies}}, \bibinfo {author}
  {\bibfnamefont{P.}~\bibnamefont{Konopka}}, \bibinfo {author}
  {\bibfnamefont{C.}~\bibnamefont{Schiller}}, \bibinfo {author}
  {\bibfnamefont{G.}~\bibnamefont{Shur}},\ and\ \bibinfo {author}
  {\bibfnamefont{T.}~\bibnamefont{Peter}}}%
  , \bibinfo {year} {2005},\ \bibfield{title}{%
  \enquote{\bibinfo {title} {Nitric acid trihydrate ({NAT}) formation at low
  {NAT} supersaturation in polar stratospheric clouds ({PSC}s)},}\ }%
  \bibfield{journal}{%
  \bibinfo {journal} {Atmos. Chem. Phys.}\ }%
  \textbf{\bibinfo {volume} {5}},\ \bibinfo {pages} {1371--1380}%
  \bibAnnoteFile{NoStop}{voigt2005}%
\bibitem[{\citenamefont{{Voigt}}\ \emph{et~al.}(2000)\citenamefont{{Voigt}},
  \citenamefont{{Schreiner}}, \citenamefont{{Kohlmann}}, \citenamefont{{Zink}},
  \citenamefont{{Mauersberger}}, \citenamefont{{Larsen}},
  \citenamefont{{Deshler}}, \citenamefont{{Kr{\"o}ger}},
  \citenamefont{{Rosen}}, \citenamefont{{Adriani}}, \citenamefont{{Cairo}},
  \citenamefont{{Di Donfrancesco}}, \citenamefont{{Viterbini}},
  \citenamefont{{Ovarlez}}, \citenamefont{{Ovarlez}}, \citenamefont{{David}},\
  and\ \citenamefont{{D{\"o}rnbrack}}}]{voigt2000}%
  \BibitemOpen
  \bibfield{author}{%
  \bibinfo {author} {\bibnamefont{{Voigt}}, \bibfnamefont{C.}}, \bibinfo
  {author} {\bibfnamefont{J.}~\bibnamefont{{Schreiner}}}, \bibinfo {author}
  {\bibfnamefont{A.}~\bibnamefont{{Kohlmann}}}, \bibinfo {author}
  {\bibfnamefont{P.}~\bibnamefont{{Zink}}}, \bibinfo {author}
  {\bibfnamefont{K.}~\bibnamefont{{Mauersberger}}}, \bibinfo {author}
  {\bibfnamefont{N.}~\bibnamefont{{Larsen}}}, \bibinfo {author}
  {\bibfnamefont{T.}~\bibnamefont{{Deshler}}}, \bibinfo {author}
  {\bibfnamefont{C.}~\bibnamefont{{Kr{\"o}ger}}}, \bibinfo {author}
  {\bibfnamefont{J.}~\bibnamefont{{Rosen}}}, \bibinfo {author}
  {\bibfnamefont{A.}~\bibnamefont{{Adriani}}}, \bibinfo {author}
  {\bibfnamefont{F.}~\bibnamefont{{Cairo}}}, \bibinfo {author}
  {\bibfnamefont{G.}~\bibnamefont{{Di Donfrancesco}}}, \bibinfo {author}
  {\bibfnamefont{M.}~\bibnamefont{{Viterbini}}}, \bibinfo {author}
  {\bibfnamefont{J.}~\bibnamefont{{Ovarlez}}}, \bibinfo {author}
  {\bibfnamefont{H.}~\bibnamefont{{Ovarlez}}}, \bibinfo {author}
  {\bibfnamefont{C.}~\bibnamefont{{David}}},\ and\ \bibinfo {author}
  {\bibfnamefont{A.}~\bibnamefont{{D{\"o}rnbrack}}}}%
  , \bibinfo {year} {2000},\ \bibfield{title}{%
  \enquote{\bibinfo {title} {{Nitric Acid Trihydrate (NAT) in Polar
  Stratospheric Clouds}},}\ }%
  \bibfield{journal}{%
  \bibinfo {journal} {Science}\ }%
  \textbf{\bibinfo {volume} {290}},\ \bibinfo {pages} {1756--1758}%
  \bibAnnoteFile{NoStop}{voigt2000}%
\bibitem[{\citenamefont{Vonnegut}(1994)}]{vonnegut1994}%
  \BibitemOpen
  \bibfield{author}{%
  \bibinfo {author} {\bibnamefont{Vonnegut}, \bibfnamefont{B.}}}%
  , \bibinfo {year} {1994},\ \bibfield{title}{%
  \enquote{\bibinfo {title} {The atmospheric electricity paradigm},}\ }%
  \bibfield{journal}{%
  \bibinfo {journal} {Bull. Am. Meteor. Soc.}\ }%
  \textbf{\bibinfo {volume} {75}},\ \bibinfo {pages} {53--61}%
  \bibAnnoteFile{NoStop}{vonnegut1994}%
\bibitem[{\citenamefont{Wadhams}(1992)}]{Wadhams:1992}%
  \BibitemOpen
  \bibfield{author}{%
  \bibinfo {author} {\bibnamefont{Wadhams}, \bibfnamefont{P.}}}%
  , \bibinfo {year} {1992},\ \bibfield{title}{%
  \enquote{\bibinfo {title} {Sea ice thickness distribution in the {Greenland
  Sea} and {Eurasian Basin}, {May} 1987},}\ }%
  \bibfield{journal}{%
  \bibinfo {journal} {J.~Geophys.~Res.}\ }%
  \textbf{\bibinfo {volume} {97}},\ \bibinfo {pages} {5331--5348}%
  \bibAnnoteFile{NoStop}{Wadhams:1992}%
\bibitem[{\citenamefont{Wadhams}(2000)}]{Wadhams:2000}%
  \BibitemOpen
  \bibfield{author}{%
  \bibinfo {author} {\bibnamefont{Wadhams}, \bibfnamefont{P.}}}%
  , \bibinfo {year} {2000},\ \emph{\bibinfo {title} {Ice in the ocean}}\
  (\bibinfo {publisher} {CRC Press})%
  \bibAnnoteFile{NoStop}{Wadhams:2000}%
\bibitem[{\citenamefont{Wadhams}\ and\
  \citenamefont{Doble}(2008)}]{Wadhams:2008}%
  \BibitemOpen
  \bibfield{author}{%
  \bibinfo {author} {\bibnamefont{Wadhams}, \bibfnamefont{P.}},\ and\ \bibinfo
  {author} {\bibfnamefont{M.}~\bibnamefont{Doble}}}%
  , \bibinfo {year} {2008},\ \bibfield{title}{%
  \enquote{\bibinfo {title} {Digital terrain mapping of the underside of sea
  ice from a small {AUV}},}\ }%
  \bibfield{journal}{%
  \bibinfo {journal} {Geophys.~Res.~Lett.}\ }%
  \textbf{\bibinfo {volume} {35}},\ \bibinfo {pages} {L01501}%
  \bibAnnoteFile{NoStop}{Wadhams:2008}%
\bibitem[{\citenamefont{Walkington}\
  \emph{et~al.}(2007)\citenamefont{Walkington},
  \citenamefont{Morales~Maqueda},\ and\
  \citenamefont{Willmott}}]{Walkington:2007}%
  \BibitemOpen
  \bibfield{author}{%
  \bibinfo {author} {\bibnamefont{Walkington}, \bibfnamefont{I.}}, \bibinfo
  {author} {\bibfnamefont{M.~A.}\ \bibnamefont{Morales~Maqueda}},\ and\
  \bibinfo {author} {\bibfnamefont{A.~J.}\ \bibnamefont{Willmott}}}%
  , \bibinfo {year} {2007},\ \bibfield{title}{%
  \enquote{\bibinfo {title} {A robust and computationally efficient model of a
  two-dimensional coastal polynya},}\ }%
  \bibfield{journal}{%
  \bibinfo {journal} {Ocean Modelling}\ }%
  \textbf{\bibinfo {volume} {17}},\ \bibinfo {pages} {140--152}%
  \bibAnnoteFile{NoStop}{Walkington:2007}%
\bibitem[{\citenamefont{Wang}\ \emph{et~al.}(2009)\citenamefont{Wang},
  \citenamefont{Zhang}, \citenamefont{Watanabe}, \citenamefont{Ikeda},
  \citenamefont{Mizobata}, \citenamefont{Walsh}, \citenamefont{Bai},\ and\
  \citenamefont{Wu}}]{JiaWang:2009}%
  \BibitemOpen
  \bibfield{author}{%
  \bibinfo {author} {\bibnamefont{Wang}, \bibfnamefont{J.}}, \bibinfo {author}
  {\bibfnamefont{J.}~\bibnamefont{Zhang}}, \bibinfo {author}
  {\bibfnamefont{E.}~\bibnamefont{Watanabe}}, \bibinfo {author}
  {\bibfnamefont{M.}~\bibnamefont{Ikeda}}, \bibinfo {author}
  {\bibfnamefont{K.}~\bibnamefont{Mizobata}}, \bibinfo {author}
  {\bibfnamefont{J.}~\bibnamefont{Walsh}}, \bibinfo {author}
  {\bibfnamefont{X.}~\bibnamefont{Bai}},\ and\ \bibinfo {author}
  {\bibfnamefont{B.}~\bibnamefont{Wu}}}%
  , \bibinfo {year} {2009},\ \bibfield{title}{%
  \enquote{\bibinfo {title} {Is the dipole anomaly a major driver to record
  lows in {Arctic} summer sea ice extent?}.}\ }%
  \bibfield{journal}{%
  \bibinfo {journal} {Geophys.~Res.~Lett.}\ }%
  \textbf{\bibinfo {volume} {36}},\ \bibinfo {pages} {L05706}%
  \bibAnnoteFile{NoStop}{JiaWang:2009}%
\bibitem[{\citenamefont{Wang}\ and\ \citenamefont{Wang}(2009)}]{Wang2009}%
  \BibitemOpen
  \bibfield{author}{%
  \bibinfo {author} {\bibnamefont{Wang}, \bibfnamefont{K.}},\ and\ \bibinfo
  {author} {\bibfnamefont{C.}~\bibnamefont{Wang}}}%
  , \bibinfo {year} {2009},\ \bibfield{title}{%
  \enquote{\bibinfo {title} {Modeling linear kinematic features in pack ice},}\
  }%
  \bibfield{journal}{%
  \bibinfo {journal} {J. Geophys. Res.}\ }%
  \textbf{\bibinfo {volume} {114}},\ \bibinfo {pages} {C12011}%
  \bibAnnoteFile{NoStop}{Wang2009}%
\bibitem[{\citenamefont{Wang}\ and\ \citenamefont{Overland}(2009)}]{wang:2009}%
  \BibitemOpen
  \bibfield{author}{%
  \bibinfo {author} {\bibnamefont{Wang}, \bibfnamefont{M.}},\ and\ \bibinfo
  {author} {\bibfnamefont{J.~E.}\ \bibnamefont{Overland}}}%
  , \bibinfo {year} {2009},\ \bibfield{title}{%
  \enquote{\bibinfo {title} {A sea ice free summer {Arctic} within 30
  years?}.}\ }%
  \bibfield{journal}{%
  \bibinfo {journal} {Geophys.~Res.~Lett.}\ }%
  \textbf{\bibinfo {volume} {36}},\ \bibinfo {pages} {L07502}%
  \bibAnnoteFile{NoStop}{wang:2009}%
\bibitem[{\citenamefont{Wang}(2002)}]{wang2002}%
  \BibitemOpen
  \bibfield{author}{%
  \bibinfo {author} {\bibnamefont{Wang}, \bibfnamefont{P.~K.}}}%
  , \bibinfo {year} {2002},\ \enquote{\bibinfo {title} {Ice microdynamics},}\
  in\ \emph{\bibinfo {booktitle} {Advances in Geophysics}},\ Vol.~\bibinfo
  {volume} {45},\ \bibinfo {editor} {edited by\ \bibinfo {editor}
  {\bibfnamefont{R.}~\bibnamefont{Dmowska}}\ and\ \bibinfo {editor}
  {\bibfnamefont{B.}~\bibnamefont{Saltzman}}}\ (\bibinfo {publisher} {Academic
  Press})%
  \bibAnnoteFile{NoStop}{wang2002}%
\bibitem[{\citenamefont{Watanabe}\ and\
  \citenamefont{Arai}(1995)}]{watanabe1995}%
  \BibitemOpen
  \bibfield{author}{%
  \bibinfo {author} {\bibnamefont{Watanabe}, \bibfnamefont{M.}},\ and\ \bibinfo
  {author} {\bibfnamefont{S.}~\bibnamefont{Arai}}}%
  , \bibinfo {year} {1995},\ \enquote{\bibinfo {title} {Applications of
  bacterial ice nucleation activity in food processing},}\ in\ \emph{\bibinfo
  {booktitle} {Biological Ice Nucleation and Its Applications}},\ \bibinfo
  {editor} {edited by\ \bibinfo {editor}
  {\bibfnamefont{Jr}~\bibnamefont{R.~E.~Lee}}, \bibinfo {editor}
  {\bibfnamefont{G.~J.}\ \bibnamefont{Warren}},\ and\ \bibinfo {editor}
  {\bibfnamefont{L.~V.}\ \bibnamefont{Gusta}}}\ (\bibinfo {publisher} {APS
  Press})\ pp.\ \bibinfo {pages} {299--313}%
  \bibAnnoteFile{NoStop}{watanabe1995}%
\bibitem[{\citenamefont{Watanabe}\ \emph{et~al.}(2010)\citenamefont{Watanabe},
  \citenamefont{Kimura}, \citenamefont{Kouchi}, \citenamefont{Chigai},
  \citenamefont{Hama},\ and\ \citenamefont{Pirronello}}]{watanabe2010}%
  \BibitemOpen
  \bibfield{author}{%
  \bibinfo {author} {\bibnamefont{Watanabe}, \bibfnamefont{N.}}, \bibinfo
  {author} {\bibfnamefont{Y.}~\bibnamefont{Kimura}}, \bibinfo {author}
  {\bibfnamefont{A.}~\bibnamefont{Kouchi}}, \bibinfo {author}
  {\bibfnamefont{T.}~\bibnamefont{Chigai}}, \bibinfo {author}
  {\bibfnamefont{T.}~\bibnamefont{Hama}},\ and\ \bibinfo {author}
  {\bibfnamefont{V.}~\bibnamefont{Pirronello}}}%
  , \bibinfo {year} {2010},\ \bibfield{title}{%
  \enquote{\bibinfo {title} {Direct measurements of hydrogen atom diffusion and
  the spin temperature of nascent {H}$_2$ molecule on amorphous solid water},}\
  }%
  \bibfield{journal}{%
  \bibinfo {journal} {Astrophys. J. Lett.}\ }%
  \textbf{\bibinfo {volume} {714}},\ \bibinfo {pages} {L233--L237}%
  \bibAnnoteFile{NoStop}{watanabe2010}%
\bibitem[{\citenamefont{Watanabe}\ \emph{et~al.}(2006)\citenamefont{Watanabe},
  \citenamefont{Nagaoka}, \citenamefont{Hidaka}, \citenamefont{Shiraki},
  \citenamefont{Chigai},\ and\ \citenamefont{Kouchi}}]{watanabe2006}%
  \BibitemOpen
  \bibfield{author}{%
  \bibinfo {author} {\bibnamefont{Watanabe}, \bibfnamefont{N.}}, \bibinfo
  {author} {\bibfnamefont{A.}~\bibnamefont{Nagaoka}}, \bibinfo {author}
  {\bibfnamefont{H.}~\bibnamefont{Hidaka}}, \bibinfo {author}
  {\bibfnamefont{T.}~\bibnamefont{Shiraki}}, \bibinfo {author}
  {\bibfnamefont{T.}~\bibnamefont{Chigai}},\ and\ \bibinfo {author}
  {\bibfnamefont{A.}~\bibnamefont{Kouchi}}}%
  , \bibinfo {year} {2006},\ \bibfield{title}{%
  \enquote{\bibinfo {title} {Dependence of the effective rate constants for the
  hydrogenation of {CO} on the temperature and composition of the surface},}\
  }%
  \bibfield{journal}{%
  \bibinfo {journal} {Planet Space Sci.}\ }%
  \textbf{\bibinfo {volume} {54}},\ \bibinfo {pages} {1107--1114}%
  \bibAnnoteFile{NoStop}{watanabe2006}%
\bibitem[{\citenamefont{Weeks}(2010)}]{Weeks:2010}%
  \BibitemOpen
  \bibfield{author}{%
  \bibinfo {author} {\bibnamefont{Weeks}, \bibfnamefont{W.~F.}}}%
  , \bibinfo {year} {2010},\ \emph{\bibinfo {title} {On Sea Ice}}\ (\bibinfo
  {publisher} {{U}niversity of {A}laska {P}ress})%
  \bibAnnoteFile{NoStop}{Weeks:2010}%
\bibitem[{\citenamefont{Wei}\ \emph{et~al.}(2001)\citenamefont{Wei},
  \citenamefont{Miranda},\ and\ \citenamefont{Shen}}]{wei2001}%
  \BibitemOpen
  \bibfield{author}{%
  \bibinfo {author} {\bibnamefont{Wei}, \bibfnamefont{X.}}, \bibinfo {author}
  {\bibfnamefont{P.~B.}\ \bibnamefont{Miranda}},\ and\ \bibinfo {author}
  {\bibfnamefont{Y.~R.}\ \bibnamefont{Shen}}}%
  , \bibinfo {year} {2001},\ \bibfield{title}{%
  \enquote{\bibinfo {title} {Surface vibrational spectroscopic study of surface
  melting of ice},}\ }%
  \bibinfo {journal} {Phys. Rev. Lett.},\ \bibinfo {pages} {1554--1557}%
  \bibAnnoteFile{NoStop}{wei2001}%
\bibitem[{\citenamefont{Weinheimer}\ and\
  \citenamefont{Knight}(1987)}]{weinheimer1987}%
  \BibitemOpen
\bibfield{journal}{%
    }%
  \bibfield{author}{%
  \bibinfo {author} {\bibnamefont{Weinheimer}, \bibfnamefont{A.~J.}},\ and\
  \bibinfo {author} {\bibfnamefont{C.~A.}\ \bibnamefont{Knight}}}%
  , \bibinfo {year} {1987},\ \bibfield{title}{%
  \enquote{\bibinfo {title} {ScheinerÕs halo: {C}ubic ice or polycrystalline
  hexagonal ice?}.}\ }%
  \bibfield{journal}{%
  \bibinfo {journal} {J. Atmos. Sci.}\ }%
  \textbf{\bibinfo {volume} {44}},\ \bibinfo {pages} {3304--3308}%
  \bibAnnoteFile{NoStop}{weinheimer1987}%
\bibitem[{\citenamefont{Weiss}(2008)}]{Weiss2008}%
  \BibitemOpen
  \bibfield{author}{%
  \bibinfo {author} {\bibnamefont{Weiss}, \bibfnamefont{J.}}}%
  , \bibinfo {year} {2008},\ \bibfield{title}{%
  \enquote{\bibinfo {title} {Intermittency of principal stress directions
  within {Arctic} sea ice},}\ }%
  \bibfield{journal}{%
  \bibinfo {journal} {Phys. Rev. E}\ }%
  \textbf{\bibinfo {volume} {77}},\ \bibinfo {pages} {056106}%
  \bibAnnoteFile{NoStop}{Weiss2008}%
\bibitem[{\citenamefont{Wettlaufer}(1994)}]{Wettlaufer1994}%
  \BibitemOpen
  \bibfield{author}{%
  \bibinfo {author} {\bibnamefont{Wettlaufer}, \bibfnamefont{J.~S.}}}%
  , \bibinfo {year} {1994},\ \enquote{\bibinfo {title} {Introduction to
  crystallization phenomena in natural and artificial sea ice},}\ in\
  \emph{\bibinfo {booktitle} {Physics of ice-covered seas}},\ \bibinfo {series}
  {Lecture notes from Savonlinna summer school}, Vol.~\bibinfo {volume} {1},\
  pp.\ \bibinfo {pages} {105--194}%
  \bibAnnoteFile{NoStop}{Wettlaufer1994}%
\bibitem[{\citenamefont{Whalley}(1983)}]{whalley1983}%
  \BibitemOpen
  \bibfield{author}{%
  \bibinfo {author} {\bibnamefont{Whalley}, \bibfnamefont{E.}}}%
  , \bibinfo {year} {1983},\ \bibfield{title}{%
  \enquote{\bibinfo {title} {Cubic ice in nature},}\ }%
  \bibfield{journal}{%
  \bibinfo {journal} {J. Phys. Chem.}\ }%
  \textbf{\bibinfo {volume} {87}},\ \bibinfo {pages} {4174--4179}%
  \bibAnnoteFile{NoStop}{whalley1983}%
\bibitem[{\citenamefont{Wilchinsky}\ and\
  \citenamefont{Feltham}(2006)}]{Wilchinsky2006}%
  \BibitemOpen
  \bibfield{author}{%
  \bibinfo {author} {\bibnamefont{Wilchinsky}, \bibfnamefont{A.~V.}},\ and\
  \bibinfo {author} {\bibfnamefont{D~.L.}\ \bibnamefont{Feltham}}}%
  , \bibinfo {year} {2006},\ \bibfield{title}{%
  \enquote{\bibinfo {title} {Anisotropic model for granular sea ice
  dynamics},}\ }%
  \bibfield{journal}{%
  \bibinfo {journal} {J. Mech. Phys. Solids}\ }%
  \textbf{\bibinfo {volume} {54}},\ \bibinfo {pages} {1147--1185}%
  \bibAnnoteFile{NoStop}{Wilchinsky2006}%
\bibitem[{\citenamefont{Wilen}\ and\ \citenamefont{Dash}(1995)}]{Wilen1995a}%
  \BibitemOpen
  \bibfield{author}{%
  \bibinfo {author} {\bibnamefont{Wilen}, \bibfnamefont{L.~A.}},\ and\ \bibinfo
  {author} {\bibfnamefont{J.~G.}\ \bibnamefont{Dash}}}%
  , \bibinfo {year} {1995},\ \bibfield{title}{%
  \enquote{\bibinfo {title} {Giant facets at grain boundary grooves},}\ }%
  \bibfield{journal}{%
  \bibinfo {journal} {Science}\ }%
  \textbf{\bibinfo {volume} {270}},\ \bibinfo {pages} {1184--1186}%
  \bibAnnoteFile{NoStop}{Wilen1995a}%
\bibitem[{\citenamefont{{Williams}}\
  \emph{et~al.}(2007)\citenamefont{{Williams}}, \citenamefont{{Brown}},
  \citenamefont{{Price}}, \citenamefont{{Rawlings}},\ and\
  \citenamefont{{Viti}}}]{williams2007}%
  \BibitemOpen
  \bibfield{author}{%
  \bibinfo {author} {\bibnamefont{{Williams}}, \bibfnamefont{D.~A.}}, \bibinfo
  {author} {\bibfnamefont{W.~A.}\ \bibnamefont{{Brown}}}, \bibinfo {author}
  {\bibfnamefont{S.~D.}\ \bibnamefont{{Price}}}, \bibinfo {author}
  {\bibfnamefont{J.~M.~C.}\ \bibnamefont{{Rawlings}}},\ and\ \bibinfo {author}
  {\bibfnamefont{S.}~\bibnamefont{{Viti}}}}%
  , \bibinfo {year} {2007},\ \bibfield{title}{%
  \enquote{\bibinfo {title} {Molecules, ices and astronomy},}\ }%
  \bibfield{journal}{%
  \bibinfo {journal} {Astron. Geophys.}\ }%
  \textbf{\bibinfo {volume} {48}},\ \bibinfo {pages} {25--34}%
  \bibAnnoteFile{NoStop}{williams2007}%
\bibitem[{\citenamefont{Williams}\ and\
  \citenamefont{Hartquist}(1999)}]{williams1999}%
  \BibitemOpen
  \bibfield{author}{%
  \bibinfo {author} {\bibnamefont{Williams}, \bibfnamefont{D.~A.}},\ and\
  \bibinfo {author} {\bibfnamefont{T.~W.}\ \bibnamefont{Hartquist}}}%
  , \bibinfo {year} {1999},\ \bibfield{title}{%
  \enquote{\bibinfo {title} {The chemistry of star-forming regions},}\ }%
  \bibinfo {journal} {Accounts Chem. Res.},\ \bibinfo {pages} {334--341}%
  \bibAnnoteFile{NoStop}{williams1999}%
\bibitem[{\citenamefont{Williams}\ and\
  \citenamefont{Herbst}(2002)}]{williams2002}%
  \BibitemOpen
\bibfield{journal}{%
    }%
  \bibfield{author}{%
  \bibinfo {author} {\bibnamefont{Williams}, \bibfnamefont{D.~A.}},\ and\
  \bibinfo {author} {\bibfnamefont{E.}~\bibnamefont{Herbst}}}%
  , \bibinfo {year} {2002},\ \bibfield{title}{%
  \enquote{\bibinfo {title} {It's a dusty universe: surface science in
  space},}\ }%
  \bibfield{journal}{%
  \bibinfo {journal} {Surface Sci.}\ }%
  \textbf{\bibinfo {volume} {500}},\ \bibinfo {pages} {823--837}%
  \bibAnnoteFile{NoStop}{williams2002}%
\bibitem[{\citenamefont{Winkel}\ \emph{et~al.}(2009)\citenamefont{Winkel},
  \citenamefont{Bowron}, \citenamefont{Loerting}, \citenamefont{Mayer},\ and\
  \citenamefont{Finney}}]{winkel2009}%
  \BibitemOpen
  \bibfield{author}{%
  \bibinfo {author} {\bibnamefont{Winkel}, \bibfnamefont{K.}}, \bibinfo
  {author} {\bibfnamefont{D.~T.}\ \bibnamefont{Bowron}}, \bibinfo {author}
  {\bibfnamefont{T.}~\bibnamefont{Loerting}}, \bibinfo {author}
  {\bibfnamefont{E.}~\bibnamefont{Mayer}},\ and\ \bibinfo {author}
  {\bibfnamefont{J.~L.}\ \bibnamefont{Finney}}}%
  , \bibinfo {year} {2009},\ \bibfield{title}{%
  \enquote{\bibinfo {title} {Relaxation effects in low density amorphous ice:
  Two distinct structural states observed by neutron diffraction},}\ }%
  \bibfield{journal}{%
  \bibinfo {journal} {J. Chem. Phys.}\ }%
  \textbf{\bibinfo {volume} {130}},\ \bibinfo {pages} {204502}%
  \bibAnnoteFile{NoStop}{winkel2009}%
\bibitem[{\citenamefont{Winkel}\ \emph{et~al.}(2008)\citenamefont{Winkel},
  \citenamefont{Elsaesser}, \citenamefont{Mayer},\ and\
  \citenamefont{Loerting}}]{winkel2008}%
  \BibitemOpen
  \bibfield{author}{%
  \bibinfo {author} {\bibnamefont{Winkel}, \bibfnamefont{K.}}, \bibinfo
  {author} {\bibfnamefont{M.~S.}\ \bibnamefont{Elsaesser}}, \bibinfo {author}
  {\bibfnamefont{E.}~\bibnamefont{Mayer}},\ and\ \bibinfo {author}
  {\bibfnamefont{T.}~\bibnamefont{Loerting}}}%
  , \bibinfo {year} {2008},\ \bibfield{title}{%
  \enquote{\bibinfo {title} {Water polyamorphism: Reversibility and
  (dis)continuity},}\ }%
  \bibfield{journal}{%
  \bibinfo {journal} {J. Chem. Phys.}\ }%
  \textbf{\bibinfo {volume} {128}},\ \bibinfo {pages} {044510}%
  \bibAnnoteFile{NoStop}{winkel2008}%
\bibitem[{\citenamefont{Winkler}\ \emph{et~al.}(2002)\citenamefont{Winkler},
  \citenamefont{Holmes},\ and\ \citenamefont{Crowley}}]{Winkler2002}%
  \BibitemOpen
  \bibfield{author}{%
  \bibinfo {author} {\bibnamefont{Winkler}, \bibfnamefont{A.~K.}}, \bibinfo
  {author} {\bibfnamefont{N.~S.}\ \bibnamefont{Holmes}},\ and\ \bibinfo
  {author} {\bibfnamefont{J.~N.}\ \bibnamefont{Crowley}}}%
  , \bibinfo {year} {2002},\ \bibfield{title}{%
  \enquote{\bibinfo {title} {Interaction of methanol, acetone and formaldehyde
  with ice surfaces between 198 and 223~{K}},}\ }%
  \bibfield{journal}{%
  \bibinfo {journal} {Phys. Chem. Chem. Phys.}\ }%
  \textbf{\bibinfo {volume} {4}},\ \bibinfo {pages} {5270--5275}%
  \bibAnnoteFile{NoStop}{Winkler2002}%
\bibitem[{\citenamefont{von Zahn}(2003)}]{vonZahn2003}%
  \BibitemOpen
  \bibfield{author}{%
  \bibinfo {author} {\bibnamefont{von Zahn}, \bibfnamefont{U.}}}%
  , \bibinfo {year} {2003},\ \bibfield{title}{%
  \enquote{\bibinfo {title} {Are noctilucent clouds truly a ``miner's canary''
  for global change?}.}\ }%
  \bibfield{journal}{%
  \bibinfo {journal} {Eos Trans. AGU}\ }%
  \textbf{\bibinfo {volume} {84}},\ \bibinfo {pages} {261}%
  \bibAnnoteFile{NoStop}{vonZahn2003}%
\bibitem[{\citenamefont{Zahnle}\ and\
  \citenamefont{Walker}(1982)}]{zahnle1982}%
  \BibitemOpen
  \bibfield{author}{%
  \bibinfo {author} {\bibnamefont{Zahnle}, \bibfnamefont{K.~L.}},\ and\
  \bibinfo {author} {\bibfnamefont{J.~C.~G.}\ \bibnamefont{Walker}}}%
  , \bibinfo {year} {1982},\ \bibfield{title}{%
  \enquote{\bibinfo {title} {The evolution of solar ultraviolet luminosity},}\
  }%
  \bibfield{journal}{%
  \bibinfo {journal} {Rev. Geophys.}\ }%
  \textbf{\bibinfo {volume} {20}},\ \bibinfo {pages} {280--292}%
  \bibAnnoteFile{NoStop}{zahnle1982}%
\bibitem[{\citenamefont{{Zasowski}}\
  \emph{et~al.}(2009)\citenamefont{{Zasowski}}, \citenamefont{{Kemper}},
  \citenamefont{{Watson}}, \citenamefont{{Furlan}}, \citenamefont{{Bohac}},
  \citenamefont{{Hull}},\ and\ \citenamefont{{Green}}}]{zasowski2009}%
  \BibitemOpen
  \bibfield{author}{%
  \bibinfo {author} {\bibnamefont{{Zasowski}}, \bibfnamefont{G.}}, \bibinfo
  {author} {\bibfnamefont{F.}~\bibnamefont{{Kemper}}}, \bibinfo {author}
  {\bibfnamefont{D.~M.}\ \bibnamefont{{Watson}}}, \bibinfo {author}
  {\bibfnamefont{E.}~\bibnamefont{{Furlan}}}, \bibinfo {author}
  {\bibfnamefont{C.~J.}\ \bibnamefont{{Bohac}}}, \bibinfo {author}
  {\bibfnamefont{C.}~\bibnamefont{{Hull}}},\ and\ \bibinfo {author}
  {\bibfnamefont{J.~D.}\ \bibnamefont{{Green}}}}%
  , \bibinfo {year} {2009},\ \bibfield{title}{%
  \enquote{\bibinfo {title} {Spitzer infrared spectrograph observations of
  class {I/II} objects in {Taurus}: Composition and thermal history of the
  circumstellar ices},}\ }%
  \bibfield{journal}{%
  \bibinfo {journal} {Astrophys. J.}\ }%
  \textbf{\bibinfo {volume} {694}},\ \bibinfo {pages} {459--478}%
  \bibAnnoteFile{NoStop}{zasowski2009}%
\bibitem[{\citenamefont{Zhou}\ \emph{et~al.}(2001)\citenamefont{Zhou},
  \citenamefont{Beine}, \citenamefont{Honrath}, \citenamefont{Fuentes},
  \citenamefont{Simpson}, \citenamefont{Shepson},\ and\
  \citenamefont{Bottenheim}}]{Zhou:2001p3425}%
  \BibitemOpen
  \bibfield{author}{%
  \bibinfo {author} {\bibnamefont{Zhou}, \bibfnamefont{X.~L.}}, \bibinfo
  {author} {\bibfnamefont{H.}~\bibnamefont{Beine}}, \bibinfo {author}
  {\bibfnamefont{R.}~\bibnamefont{Honrath}}, \bibinfo {author}
  {\bibfnamefont{J.~D.}\ \bibnamefont{Fuentes}}, \bibinfo {author}
  {\bibfnamefont{W.}~\bibnamefont{Simpson}}, \bibinfo {author}
  {\bibfnamefont{P.~B.}\ \bibnamefont{Shepson}},\ and\ \bibinfo {author}
  {\bibfnamefont{J.~W.}\ \bibnamefont{Bottenheim}}}%
  , \bibinfo {year} {2001},\ \bibfield{title}{%
  \enquote{\bibinfo {title} {Snowpack photochemical production of {HONO}: a
  major source of {OH} in the {Arctic} boundary layer in springtime},}\ }%
  \bibfield{journal}{%
  \bibinfo {journal} {Geophys. Res. Lett.}\ }%
  \textbf{\bibinfo {volume} {28}},\ \bibinfo {pages} {4087--4090}%
  \bibAnnoteFile{NoStop}{Zhou:2001p3425}%
\bibitem[{\citenamefont{Zobrist}\ \emph{et~al.}(2008)\citenamefont{Zobrist},
  \citenamefont{C.Marcolli}, \citenamefont{Pedernera},\ and\
  \citenamefont{Koop}}]{zobrist2008}%
  \BibitemOpen
  \bibfield{author}{%
  \bibinfo {author} {\bibnamefont{Zobrist}, \bibfnamefont{B.}}, \bibinfo
  {author} {\bibnamefont{C.Marcolli}}, \bibinfo {author} {\bibfnamefont{D.~A.}\
  \bibnamefont{Pedernera}},\ and\ \bibinfo {author}
  {\bibfnamefont{T.}~\bibnamefont{Koop}}}%
  , \bibinfo {year} {2008},\ \bibfield{title}{%
  \enquote{\bibinfo {title} {Do atmospheric aerosols form glasses?}.}\ }%
  \bibfield{journal}{%
  \bibinfo {journal} {Atmos. Chem. Phys.}\ }%
  \textbf{\bibinfo {volume} {8}},\ \bibinfo {pages} {5221--5244}%
  \bibAnnoteFile{NoStop}{zobrist2008}%
\bibitem[{\citenamefont{Zondlo}\ \emph{et~al.}(2000)\citenamefont{Zondlo},
  \citenamefont{Hudson}, \citenamefont{Prenni},\ and\
  \citenamefont{Tolbert}}]{zondlo2000}%
  \BibitemOpen
  \bibfield{author}{%
  \bibinfo {author} {\bibnamefont{Zondlo}, \bibfnamefont{M.~A.}}, \bibinfo
  {author} {\bibfnamefont{P.~K.}\ \bibnamefont{Hudson}}, \bibinfo {author}
  {\bibfnamefont{A.~J.}\ \bibnamefont{Prenni}},\ and\ \bibinfo {author}
  {\bibfnamefont{M.~A.}\ \bibnamefont{Tolbert}}}%
  , \bibinfo {year} {2000},\ \bibfield{title}{%
  \enquote{\bibinfo {title} {Chemistry and microphysics of polar stratospheric
  clouds and cirrus clouds},}\ }%
  \bibfield{journal}{%
  \bibinfo {journal} {Annu. Rev. Phys. Chem.}\ }%
  \textbf{\bibinfo {volume} {51}},\ \bibinfo {pages} {473--499}%
  \bibAnnoteFile{NoStop}{zondlo2000}%
\end{thebibliography}%

%

\end{document}